\documentclass[12pt,a4paper,dvips,fleqn]{report}
\usepackage{a4p}
\usepackage{PhysRep}
\usepackage{graphicx}
\usepackage{cite}
\usepackage{epsfig}
\usepackage{rotating}
\usepackage{hyperref}
\usepackage{mcite}

\newcommand{\Wdisco}      {Arnison:1983rp,*Banner:1983jy}
\newcommand{\Zdisco}      {Arnison:1983mk,*Bagnaia:1983zx}
\newcommand{\GWS}         {Glashow:1961tr,*Weinberg:1967tq,*Salam:1968rm,*Veltman:1968ki,*'tHooft:1971rn,*'tHooft:1972fi,*'tHooft:1972ue}
\newcommand{\LEPacc}      {LEP,*Myers:1990sk,*Brandt:2000xk,*Assmann:2002th}
\newcommand{\ALEPHdet}    {Decamp:1990jr,*Buskulic:1995wz}
\newcommand{\DELPHIdet}   {Aarnio:1991vx,*Abreu:1996uz}
\newcommand{\Ldet}        {L3:1990kx,*Acciarri:1994yk,*Chemarin:1994bp,*Adam:1996fj}
\newcommand{\OPALdet}     {Ahmet:1991eg,*Allport:1993kp,*Allport:1994ec,*Anderson:1994ve}
\newcommand{\SLDdet}      {SLD-CDC,*SLD-LAC,*SLD-CRID,*SLD-LUM}
\newcommand{\TOPAZref}    {Montagna:1993py,*Montagna:1993ai,*Montagna:1996ja,*Montagna:1998kp}
\newcommand{\ZFITTERref}  {Bardin:1989di,*Bardin:1990tq,*Bardin:1991fu,*Bardin:1991de,*Bardin:1992jc,*Bardin:1999yd,*Kobel:2000aw,*Arbuzov:2005ma}
\newcommand{\ALEPHls}     {Decamp:1990ky,*Decamp:1992aj,*Buskulic:1993gu,*Buskulic:1994ea,*Barate:1999ce}
\newcommand{\DELPHIls}    {Abreu:1991wj,*Abreu:1994wg,*Abreu:1994ds,*Abreu:2000mh}
\newcommand{\Lls}         {Adeva:1991jh,*Adriani:1993gk,*Acciarri:1994gx,*Acciarri:2000ai}
\newcommand{\OPALls}      {Alexander:1991qw,*Acton:1993yc,*Akers:1994is,*Abbiendi:2000hu}

\newcommand{\LUMImeas}    {Bederede:1995pc,*Brock:1996ty,*Abbiendi:1999zx}
\newcommand{\Smatrix}     {Adriani:1993gkS,*Acciarri:2000aiS,*Abbiendi:2000huS}
\newcommand{\MIZA}        {Martinez:1991ta,*Martinez:1995kp}
\newcommand{\Lpairs}      {Montagna:1998vb,*Montagna:1999eu}
\newcommand{\bibSMAT}     {bib-smat1,*bib-smat2,*bib-smat3}
\newcommand{\LEPIIsmatl}  {bib-LEP2smatl1,*bib-LEP2smatl2,*bib-LEP2smatl3}
\newcommand{\SLDalr}      {Abe:1993sh,*Abe:1994wx,*Abe:1997nj,*Abe:2000dq}

\newcommand{\SLDstrain}   {Maruyama:1992dv,*Maruyama:1991mk,*Nakanishi:1991}
\newcommand{\SLDqfc}      {Berridge:1998wy,*Onoprienko:2000qfc}
\newcommand{\SLDpgc}      {ref:sld-pgc}
\newcommand{\SLDespec}    {Levi:1989tn,*Kent:1989zc}
\newcommand{\ALEPHTAU}    {ALEPHTAU,*Buskulic:1996vx,*Buskulic:1993vk,*Decamp:1991vz}
\newcommand{\DELPHITAU}   {DELPHITAU,*Abreu:1995ku}
\newcommand{\LTAU}        {L3TAU,*Acciarri:1994qt,*Adriani:1992zn}
\newcommand{\OPALTAU}     {OPALTAU,*Alexander:1996ha,*Akers:1995db,*Alexander:1991am}
\newcommand{\KORALZ}      {Jadach:1990mz,*Jadach:1993hs}
\newcommand{\alasy}       {ref:alasy}
\newcommand{\dlasy}       {Abreu:1995vn,*Abdallah:2003gp}
\newcommand{\llasy}       {ref:llasy1,*ref:llasy2}

\newcommand{\SLDacbl}     {ref:SLD_AQL,*Abe:1999hb}
\newcommand{\SLDabj}      {Abe:1998wk,*Abe:2002fs}
\newcommand{\SLDabk}      {ref:SLD_ABK1}
\newcommand{\GBBmeas}     {gbb_meas_a,*gbb_meas_di,*gbb_meas_dii,*gbb_meas_o,*gbb_meas_s}
\newcommand{\bmulti}      {bmultid,*bmultil,*bmultio}
\newcommand{\cleoupver}   {cleoupver1,*cleoupver2,*cleoupver3}
\newcommand{\btotau}      {btotaua,*btotaud,*btotaula,*btotaulb}
\newcommand{\CDFtop}      {CDF-top:1994,*CDF-top:1995}
\newcommand{\bibGmu}      {bib-Gmu-1,*bib-Gmu-2,*bib-Gmu-3}
\newcommand{\CDFtopm}     {Mtop1-CDF-di-l-PRLa,*Mtop1-CDF-di-l-PRLb,*Mtop1-CDF-di-l-PRLb-E,*Mtop1-CDF-l+j-PRL,*Mtop1-CDF-l+j-PRD,*Mtop1-CDF-all-j-PRL}
\newcommand{\CDFWm}       {CDF-MW,*CDF-MW-PRL95,*CDF-MW-PRD95,*CDF-MW-PRL90,*CDF-MW-PRD90}
\newcommand{\Dztopm}      {D0-top:prl-ll,*D0-top:prd-ll,*D0-top:prl-lj,*D0-top:prd-lj,*Mtop1-D0-l+j-new1,*Mtop1-D0-l+j-new2,*Mtop1-D0-all-j-PRL}
\newcommand{\DzWm}        {D0-MW:central,*D0-MW:endcap,*D0-MW:edge,*D0-MW:large}
\newcommand{\STUpar}      {Peskin-Takeuchi:1990,*Peskin-Takeuchi:1992}
\newcommand{\SLDspin}     {ref:sld-arctests,*King:1994spin}
\newcommand{\epsilonpar}  {Altarelli:1991,*Altarelli:1992a,*Altarelli:1992b,*Altarelli:1993a,*Altarelli:1993b,*Altarelli:1995,*Altarelli:1997}
\newcommand{\dalphaQCD}   {bib-Swartz,*bib-Zeppe,*bib-Alemany,*bib-Davier,*bib-alphaKuhn,*bib-jeger99,*bib-Erler,*bib-ADMartin,*bib-Troconiz-Yndurain,*bib-Hagiwara:2003}
\newcommand{\LEPMW}       {ALEPH-MW,*DELPHI-MW,*L3-MW,*OPAL-MW}

\newcommand{\Zprime}      {Barate:1999qx,*Abreu:2000ap,*Adriani:1993ca,*Abbiendi:2003dh}
\newcommand{\fourferm}    {Buskulic:1996tx,*Acciarri:1998iw,*Abbiendi:2002vu}

\newcommand{\SLDvxdwic}   {SLD-VXD,*SLD-WIC}
\newcommand{\ECAL}        {bib-MZpaper,*bib-ECAL92,*bib-ECAL93}
\newcommand{\thonshell}   {Ross:1973fp,*Sirlin:1980nh}
\newcommand{\rhopar}      {Ross:1975fq,*Veltman:1977kh}
\newcommand{\cmplxpole}   {Grassi:2000dz,*Sirlin:1991rt,*Sirlin:1991fd}
\newcommand{\delewqcd}    {Czarnecki:1996ei,*Harlander:1997zb}
\newcommand{\BLUE}        {BLUE:1988,*BLUE:2003}

\begin{document}

\begin{center}
\Large {EUROPEAN ORGANIZATION FOR NUCLEAR RESEARCH}\\
\Large {STANFORD LINEAR ACCELERATOR CENTER}\\
\end{center}
\vspace*{0.4cm}
\begin{flushright}
       CERN-PH-EP/2005-041\\
       SLAC-R-774\\
       hep-ex/0509008\\
       {\bf 7 September 2005} \\
\end{flushright}

\vspace*{2cm}

\begin{center}

{\Huge  {\bf Precision Electroweak Measurements\\[4mm]
                on the Z Resonance\\
}}

\vspace*{4cm}

{\Large {\bf The ALEPH, DELPHI, L3, OPAL, SLD 
Collaborations,\footnote{ See Appendix~\ref{app:author-list}
for the lists of authors.}\\[1mm]
the LEP Electroweak Working Group,\footnote{ Web 
access at {\tt http://www.cern.ch/LEPEWWG}}\\[1mm]
the SLD Electroweak and Heavy Flavour Groups\\ 
}}

\vspace{\fill}

{\Large Accepted for publication in \emph{Physics Reports}}

\vskip 1cm
{\small Updated: 20 February 2006}
\end{center}

\clearpage

\begin{center}
{\bf Abstract}
\end{center}
  
We report on the final electroweak measurements performed with data
taken at the Z resonance by the experiments operating at the
electron-positron colliders SLC and LEP.  The data consist of 17
million Z decays accumulated by the ALEPH, DELPHI, L3 and OPAL
experiments at LEP, and 600 thousand Z decays by the SLD experiment
using a polarised beam at SLC.  The measurements include
cross-sections, forward-backward asymmetries and polarised
asymmetries.  The mass and width of the Z boson, $\MZ$ and $\GZ$, and
its couplings to fermions, for example the $\rho$ parameter and the
effective electroweak mixing angle for leptons, are precisely
measured:
\begin{eqnarray*}
\MZ       & = &           91.1875  \pm 0.0021~\GeV \\
\GZ       & = &            2.4952  \pm 0.0023~\GeV \\
\rho_\ell & = &            1.0050  \pm 0.0010      \\
\swsqeffl & =&             0.23153 \pm 0.00016     \,.
\end{eqnarray*}
The number of light neutrino species is determined to be
$2.9840\pm0.0082$, in agreement with the three observed generations of
fundamental fermions.

The results are compared to the predictions of the Standard Model.  At
the Z-pole, electroweak radiative corrections beyond the running of
the QED and QCD coupling constants are observed with a significance of
five standard deviations, and in agreement with the Standard Model.
Of the many Z-pole measurements, the forward-backward asymmetry in
b-quark production shows the largest difference with respect to its
Standard Model expectation, at the level of 2.8 standard deviations.

Through radiative corrections evaluated in the framework of the
Standard Model, the Z-pole data are also used to predict the mass of
the top quark, $\Mt=173^{+13}_{-10}~\GeV$, and the mass of the W
boson, $\MW=80.363\pm0.032~\GeV$.  These indirect constraints are
compared to the direct measurements, providing a stringent test of the
Standard Model.  Using in addition the direct measurements of $\Mt$
and $\MW$, the mass of the as yet unobserved Standard Model Higgs
boson is predicted with a relative uncertainty of about 50\% and found
to be less than $285~\GeV$ at 95\% confidence level.

\vskip 2cm



\clearpage

$ $
\vfill

\begin{center}
{\bf Acknowledgements\\}
\end{center}

We would like to thank the CERN accelerator divisions for the
efficient operation of the LEP accelerator, the precise information on
the absolute energy scale and their close cooperation with the four
experiments.  The SLD collaboration would like to thank the SLAC
accelerator department for the efficient operation of the SLC
accelerator.  We would also like to thank members of the CDF, D\O,
NuTeV and E-158 Collaborations for making results available to us and
for useful discussions concerning their combination.  Finally, the
results and their interpretation within the Standard Model would not
have been possible without the close collaboration of many theorists.

\vfill
$ $

\clearpage

\clearpage

\tableofcontents
\clearpage
\listoffigures
\clearpage
\listoftables
\clearpage

\chapter{Introduction}
\label{sec:intro}

With the observation of neutral current interactions in
neutrino-nucleon scattering in 1973~\cite{Hasert:1973ff} and the
discovery of the W and Z bosons in $\pp$ collisions ten years
later~\cite{\Wdisco,\Zdisco}, these key features of the Standard
Model~\cite{\GWS} ($\SM$) of electroweak interactions were well
established experimentally. The LEP and SLC accelerators were then
designed during the 1980s to produce copious numbers of Z bosons via
$\ee$ annihilation, allowing detailed studies of the properties of the
Z boson to be performed in a very clean environment.

The data accumulated by LEP and SLC in the 1990s are used to determine
the Z boson parameters with high precision: its mass, its partial and
total widths, and its couplings to fermion pairs.  These results are
compared to the predictions of the $\SM$ and found to be in
agreement.  From these measurements, the number of generations of
fermions with a light neutrino is determined.  Moreover, for the
first time, the experimental precision is sufficient to probe the
predictions of the $\SM$ at the loop level, demonstrating not only
that it is a good model at low energies but that as a quantum
field theory it gives an adequate description of experimental
observations up to much higher scales.  The significant constraints
which the data impose on the size of higher order electroweak
radiative corrections allow the effects of particles not produced at
LEP and SLC, most notably the top quark and the Higgs boson, to be
investigated.

\section{LEP and SLC Data\label{sec:intro_datataking}}

\begin{figure}[htbp]
\begin{center}
\includegraphics[width=\textwidth]{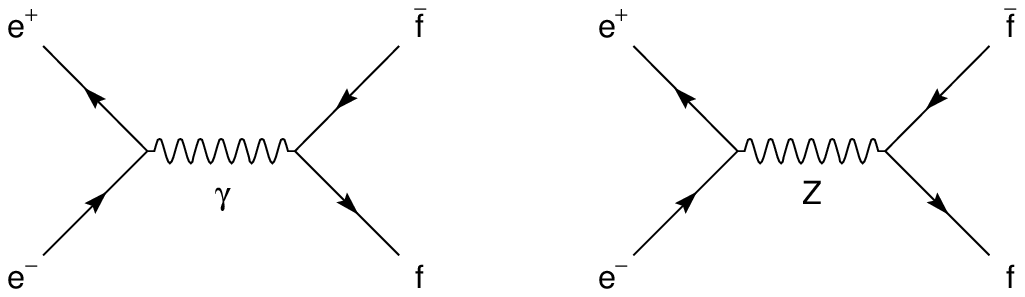}
\caption[The lowest-order $s$-channel Feynman diagrams for $\ee
\rightarrow \ff$.]{The lowest-order $s$-channel Feynman diagrams for
$\ee \rightarrow \ff$.  For $\ee$ final states, the photon and the Z
boson can also be exchanged via the $t$-channel.  The contribution
of Higgs boson exchange diagrams is negligible.}
\label{fig:intro_eeff}
\end{center}
\end{figure}

\begin{figure}[htbp]
\begin{center}
\includegraphics[width=0.9\textwidth]{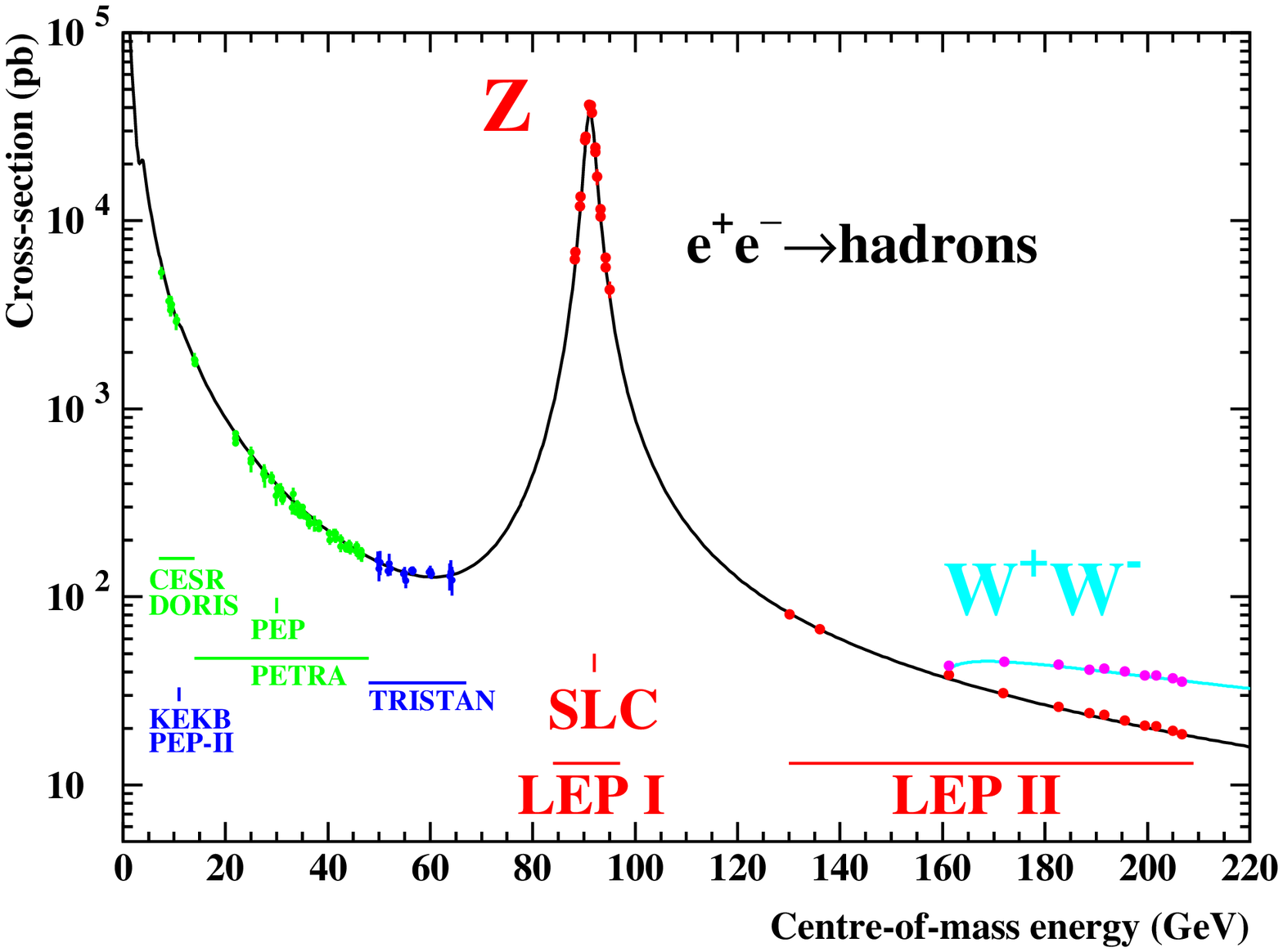}
\caption[The hadronic cross-section as a function of energy.]  {The
hadronic cross-section as a function of centre-of-mass energy.  The
solid line is the prediction of the $\SM$, and the points are the
experimental measurements.  Also indicated are the energy ranges of
various $\ee$ accelerators.  The cross-sections have been corrected
for the effects of photon radiation.}
\label{fig:intro_xhad}
\end{center}
\end{figure}

The process under study is $\ee \rightarrow \ff$, which proceeds in
lowest order via photon and Z boson exchange, as shown in
Figure~\ref{fig:intro_eeff}.  Here the fermion f is a quark, charged
lepton or neutrino.  All known fermions except the top quark are light
enough to be pair produced in Z decays.  The LEP~\cite{\LEPacc} and
SLC~\cite{SLC} $\ee$ accelerators were designed to operate at
centre-of-mass energies of approximately 91 \GeV, close to the mass of
the Z boson.\footnote{ In this report $\hbar=c=1$.}
Figure~\ref{fig:intro_xhad} illustrates two prominent features of the
hadronic cross-section as a function of the centre-of-mass energy.
The first is the $1/s$ fall-off, due to virtual photon exchange,
corresponding to the left-hand diagram in Figure~\ref{fig:intro_eeff},
which leads to the peak at low energies.  The second is the peak at 91
\GeV, due to Z exchange, which corresponds to the right-hand diagram
of Figure~\ref{fig:intro_eeff}, and allows LEP and SLC to function as
``Z factories''.

The LEP accelerator operated from 1989 to 2000, and until 1995, the
running was dedicated to the Z boson region.  From 1996 to 2000, the
centre-of-mass energy was increased to 161~\GeV\ and ultimately to
209~\GeV\, allowing the production of pairs of W bosons, $\ee
\rightarrow \mathrm{W}^+ \mathrm{W}^-$, as indicated in
Figure~\ref{fig:intro_xhad}.  Although some results from this later
running will be used in this report, the bulk of the data stems from
the Z period.  When needed, the Z period will be denoted ``\LEPI'',
and the period beginning in 1996 ``\LEPII''.  During the seven years
of running at \LEPI, the four experiments ALEPH~\cite{\ALEPHdet},
DELPHI~\cite{\DELPHIdet}, L3~\cite{\Ldet} and OPAL~\cite{\OPALdet}
collected approximately 17 million Z decays in total, distributed over
seven centre-of-mass energy points within plus or minus $3~\GeV$ of
the Z pole.

The SLC accelerator started running in 1989 and the Mark-II
collaboration published the first observations of Z production in
$\ee$ collisions~\cite{Abrams:1989aw}.  However, it was not until 1992
that longitudinal polarisation of the SLC electron beam was
established.  By then the SLD detector~\cite{\SLDdet,\SLDvxdwic}
had replaced Mark-II.  From 1992 until 1998, when the accelerator was
shut down, SLD accumulated approximately 600 thousand Z decays.
Although the data set is much smaller than that of LEP, the presence
of longitudinal polarisation allows complementary and competitive
measurements of the Z couplings.  Other properties of the accelerator
have been used to improve further the statistical power of the data.
For example, the extremely small luminous volume of the interaction
point improves the resolution in the measurement of the lifetimes of
heavy flavour hadrons, which are used to select b- and c-quark events.

\subsection{LEP\label{sec:intro_LEP-data}}
 
\begin{figure}[p]
\begin{center}
\includegraphics[width=0.9\textwidth]{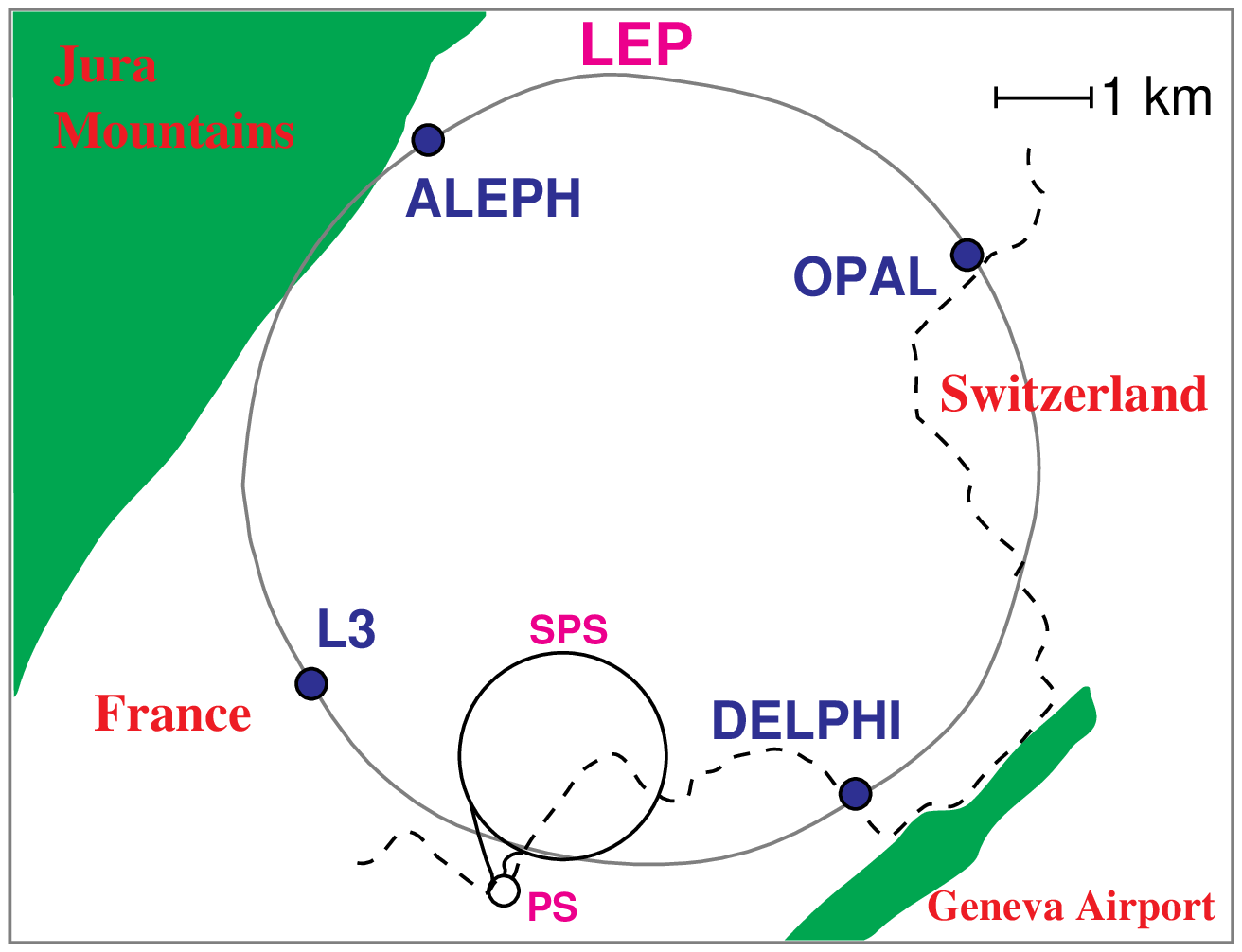}
\caption[The LEP storage ring] {The LEP storage ring, showing the
locations of the four experiments, and the PS and SPS accelerators
used to pre-accelerate the electron and positron bunches.}
\label{fig:LEPRING}
\end{center}
\end{figure}

LEP~\cite{\LEPacc} was an electron-positron collider ring with a
circumference of approximately 27 km, making it the largest particle
accelerator in the world.  The collider layout included eight straight
sections, with collisions between electron and positron bunches
allowed to take place in four of them.  The four interaction regions
were each instrumented with a multipurpose detector: L3, ALEPH, OPAL
and DELPHI, as indicated in Figure~\ref{fig:LEPRING}.

In the summer of 1989 the first $\Zzero$ bosons were produced at LEP and
observed by the four experiments.  Over the following years the
operation of the machine and its performance were steadily improved.
At the end of LEP data taking around the $\Zzero$ resonance in autumn
1995 the peak luminosity had reached $2\times 10^{31}
\mbox{cm}^{-2}\mbox{s}^{-1}$, above its design value of $1.6\times
10^{31} \mbox{cm}^{-2}\mbox{s}^{-1}$. At this luminosity, approximately
1000 Z bosons were recorded every hour by each of the four
experiments, making LEP a true Z factory.  Table~\ref{tab:lepop}
summarises the data taking periods, the approximate centre-of-mass
energies and the delivered integrated luminosities.

\begin{table}[p]
\begin{center}
\begin{tabular}{|c||c|c|}
\hline
 Year  & Centre-of-mass  & Integrated   \\
       & energy range    & luminosity  \\
       &  ~[$\GeV$]       & [pb$^{-1}$] \\
\hline
\hline
 1989  & 88.2 -- 94.2 &  \pz 1.7        \\
 1990  & 88.2 -- 94.2 &  \pz 8.6        \\
 1991  & 88.5 -- 93.7 & 18.9        \\
 1992  & 91.3            & 28.6        \\
 1993  & 89.4, 91.2, 93.0 & 40.0        \\
 1994  & 91.2            & 64.5        \\
 1995  & 89.4, 91.3, 93.0 & 39.8        \\ 
\hline
\end{tabular}
\caption[LEP beam energies and integrated luminosities] { Approximate
centre-of-mass energies and integrated luminosities delivered per LEP
experiment.  In 1990 and 1991, a total of about 7~\pb{} was taken at
off-peak energies, and 20~\pb{} per year in 1993 and in 1995. The
total luminosity used by the experiments in the analyses was smaller
by 10--15\% due to data taking inefficiencies and data quality cuts. }
\label{tab:lepop}
\end{center}
\end{table}

The data collected in 1989 constitute only a very small subset of the
total statistics and are of lower quality, and therefore these have
not been used in the final analyses.  In the years 1990 and 1991
``energy scans'' were performed at seven different centre-of-mass
energies around the peak of the $\Zzero$ resonance, placed about one
\GeV{} apart. In 1992 and 1994 there were high-statistics runs only at
the peak energy. In 1993 and 1995 data taking took place at three
centre-of-mass energies, about 1.8 \GeV{} below and above the peak and
at the peak.  The accumulated event statistics amount to about 17
million $\Zzero$ decays recorded by the four experiments.  A detailed
break-down is given in Table~\ref{tab:LSstat}.

\begin{table}[hbtp]
\begin{center}
\begin{tabular}{|r||rrrr|r||rrrr|r|}
\hline
   \multicolumn{11}{|c|}{Number of Events} \\
\hline
 & \multicolumn{ 5}{|c||}{$\Zzero\rightarrow\qq$} 
 & \multicolumn{ 5}{|c|}{$\Zzero\rightarrow\leptlept$} \\
\hline
  Year & A &   D  &  L  & O  & LEP &  A &  D &  L &  O & LEP\\
\hline\hline
1990/91& 433& 357 & 416 & 454& 1660& 53 & 36 & 39 & 58 & 186\\
1992   & 633& 697 & 678 & 733& 2741& 77 & 70 & 59 & 88 & 294\\
1993   & 630& 682 & 646 & 649& 2607& 78 & 75 & 64 & 79 & 296\\
1994   &1640&1310 &1359 &1601& 5910&202 &137 &127 &191 & 657\\
1995   & 735& 659 & 526 & 659& 2579& 90 & 66 & 54 & 81 & 291\\
\hline
 Total &4071&3705 &3625 &4096&15497&500 &384 &343 &497 &1724\\ 
\hline
\end{tabular}
\caption[Recorded event statistics for LEP]{\label{tab:LSstat} The
  $\qq$ and $\leptlept$ event statistics, in units of $10^3$, used for
  $\Zzero$ analyses by the experiments {ALEPH} (A), {DELPHI} (D), {L3}
  (L) and {OPAL} (O).}
\end{center}
\end{table}

Originally four bunches of electrons and four bunches of positrons
circulated in the ring, leading to a collision rate of 45~kHz.  The
luminosity was increased in later years by using eight equally spaced
bunches, or alternatively four trains of bunches with a spacing of
order a hundred meters between bunches in a train.  Electrons and
positrons were accelerated to about $20~\GeV$ in the
PS and SPS accelerators, then injected and accumulated in bunches in
the LEP ring.  When the desired bunch currents were achieved, the
beams were accelerated and only then brought into collision at the
interaction regions at the nominal centre-of-mass energy for that
``fill''. A fill would continue for up to about 10 hours before the
remaining beams were dumped and the machine refilled.  The main
bending field was provided by $3280$ concrete-loaded dipole magnets,
with hundreds of quadrupoles and sextupoles for focusing and
correcting the beams in the arcs and in the straight sections.  For
LEP-I running, the typical energy loss per turn of $125~\MeV$ was
compensated by a radio-frequency accelerating system comprised of
copper cavities installed in just two of the straight sections, to
either side of L3 and OPAL.

Much effort was dedicated to the determination of the energy of the
colliding beams. A precision of about $2~\MeV$ in the centre-of-mass
energy was achieved, corresponding to a relative uncertainty of about
$2\cdot10^{-5}$ on the absolute energy scale.  This level of accuracy
was vital for the precision of the measurements of the mass and width
of the $\Zzero$, as described in Chapter~\ref{chap:lsafb}.  In
particular the off-peak energies in the 1993 and 1995 scans were
carefully calibrated employing the technique of resonant
depolarisation of the transversely polarised
beams~\cite{\ECAL,bib-ECAL95}.  In order
to minimise the effects of any long-term instabilities during the
energy scans, the centre-of-mass energy was changed for every new fill
of the machine. As a result, the data samples taken above and below
the resonance are well balanced within each year, and the data at each
energy are spread evenly in time.  The data recorded within a year
around one centre-of-mass energy were combined to give one measurement
at this ``energy point''.

The build-up of transverse polarisation due to the emission of
synchrotron radiation~\cite{bib-transpol} was achieved with specially
smoothed beam trajectories.  Measurements with resonant depolarisation
were therefore only made outside normal data taking, and typically at
the ends of fills. Numerous potential causes of shifts in the
centre-of-mass energy were investigated, and some unexpected sources
identified.  These include the effects of earth tides generated by
the moon and sun, and local geological deformations following heavy
rainfall or changes in the level of Lake Geneva.  While the beam orbit
length was constrained by the RF accelerating system, the focusing
quadrupoles were fixed to the earth and moved with respect to the
beam, changing the effective total bending magnetic field and the beam
energy by $10~\MeV$ over several hours.  Leakage currents from
electric trains operating in the vicinity provoked a gradual change in
the bending field of the main dipoles, directly affecting the beam
energy.  The collision energy at each interaction point also depended
for example on the exact configuration of the RF accelerating system.
All these effects are large compared to the less than $2~\MeV$
systematic uncertainty on the centre-of-mass energy eventually
achieved through careful monitoring of the running conditions and
modelling of the beam energy.

\subsection{SLC\label{sec:intro_SLC-data}}

\begin{figure}[h]
\begin{center}
\includegraphics[width=0.9\textwidth]{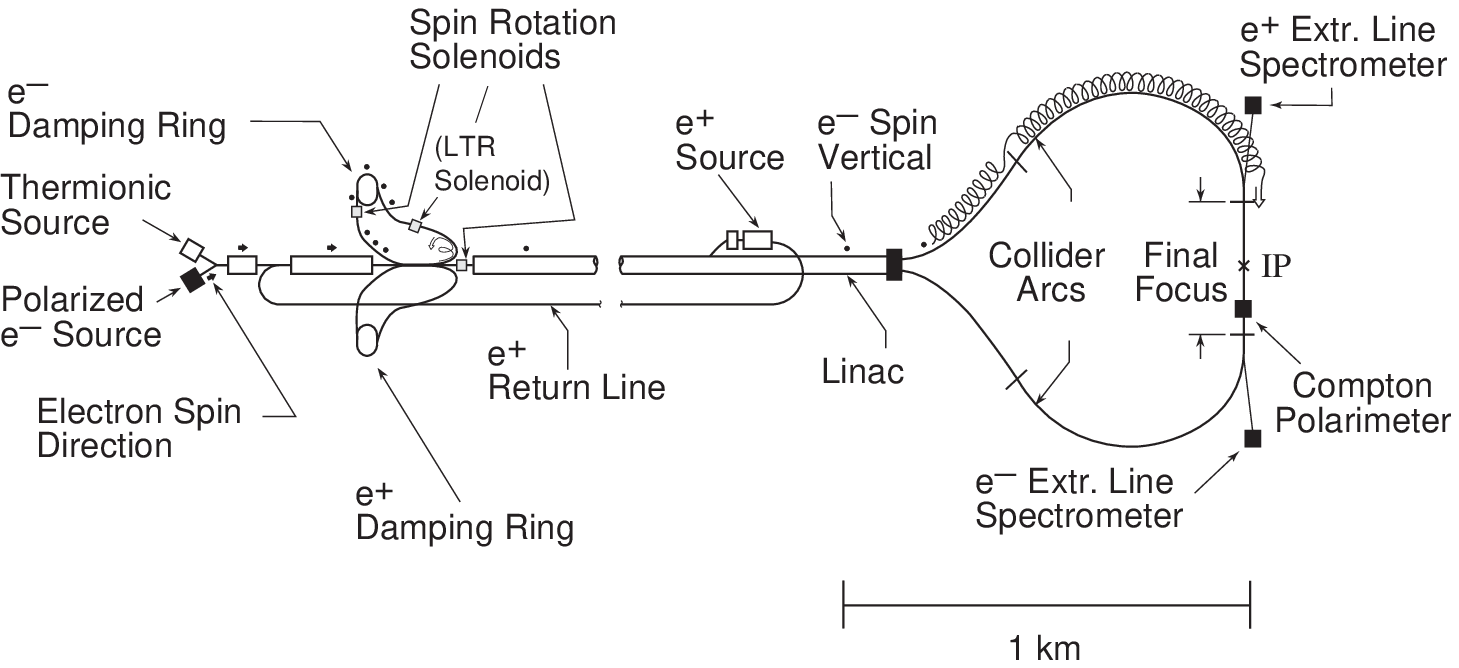}
\caption[The SLC linear collider] { The SLC linear collider complex,
showing the electron source, the damping rings, the positron source,
the 3~km long linac and arcs and the final focus. The helix and arrow
superimposed on the upper arc schematically indicate the electron spin
precession which occurs during transport.  }
\label{fig:SLC}
\end{center}
\end{figure}

The SLC~\cite{SLC} was the first $\ee$ linear collider.  As such, its
mode of operation was significantly different from that of LEP.  It
used the SLAC linear accelerator to accelerate alternate bunches of
electrons and positrons, a set of two damping rings to reduce the size
and energy spread of the electron and positron bunches, and two
separate arcs to guide the bunches to a single interaction region, as
shown in Figure~\ref{fig:SLC}.  The repetition rate was 120~Hz,
compared to either 45~kHz or 90~kHz, depending on the mode, for LEP.

The standard operating cycle began with the production of two closely
spaced
electron bunches, the first of which was longitudinally polarised.
These bunches were accelerated part way down the linac before being
stored in the electron damping rings at $1.2~\GeV$.  In the
linac-to-ring (LTR) transfer line, the longitudinal polarisation was
rotated first into a horizontal transverse orientation, and then,
using a spin rotator magnet, into a vertical orientation perpendicular
to the plane of the damping ring.  After damping,
the two bunches were extracted and accelerated in the linac.  At
$30~\GeV$, the second bunch was diverted to a target, where positrons
were created.  The positrons were captured, accelerated to $200~\MeV$
and sent back to the beginning of the linac, where they were then
stored in the positron damping ring.
The positron bunch was then extracted just before
the next two electron bunches, and accelerated.  The remaining positron
and electron bunches were accelerated to the final energy of $\approx
46.5~\GeV$ and then transported in the arcs to the final focus and
interaction point. Approximately $1~\GeV$ was lost in the arcs due to
synchrotron radiation, so the centre-of-mass energy of the $\ee$ collisions
was at the peak of the $\Zzero$ resonance.
The electron spins were manipulated
during transport in the arcs, so that the electrons arrived at the
interaction point with longitudinal polarisation.

\begin{figure}[tbp]
\begin{center}
\includegraphics[angle=-90,width=0.9\textwidth]{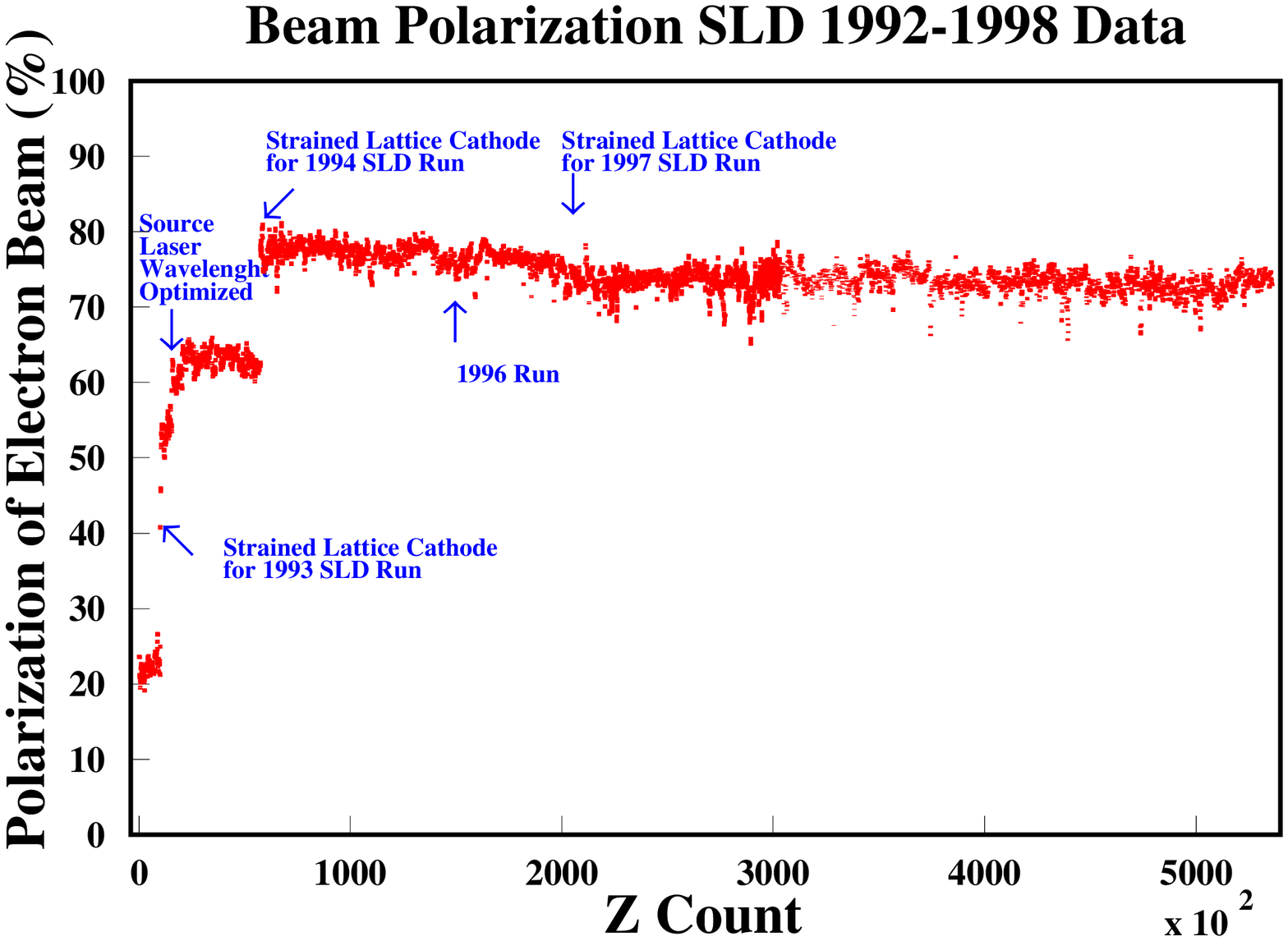}
\caption[Longitudinal polarisation at SLC]{The amount of
  longitudinal electron polarisation as a function of the number of
  recorded Z decays at SLD.}
\label{fig:slc_polar}
\end{center}
\end{figure}

The era of high-precision measurements at SLC started in 1992 with the
first longitudinally polarised beams.  The polarisation was achieved
by shining circularly polarised laser light on a gallium arsenide
photo-cathode at the electron source.  At that time, the electron
polarisation was only 22\%.  Shortly thereafter, ``strained lattice''
photocathodes were introduced, and the electron polarisation increased
significantly, as shown in Figure~\ref{fig:slc_polar}.
About 60\% of the data
were collected in the last two years of SLC running, from 1997 to
1998, with the second to last week of running producing more than
20000 Z bosons.  Much work was invested in the SLC machine to maintain
the electron polarisation at a very high value throughout the
production, damping, acceleration and transfer through the arcs.  In
addition, to avoid as much as possible any correlations in the SLC
machine or SLD detector, the electron helicity was randomly changed on
a pulse-to-pulse basis by changing the circular polarisation of the
laser.

The polarised beam physics programme at the SLC required additional
instrumentation beyond the main SLD detector, most notably, precision
polarimetry.  At the onset of the programme, it was hoped that the
Compton-scattering polarimeter installed near the beam interaction
point (IP) would reach a relative precision of 1\%.  In fact, an
ultimate precision of 0.5\% was achieved, which ensured that
polarimetry systematics were never the leading contributor to the
uncertainty of even the highest precision SLD measurements.  This
device employed a high-power circularly-polarised laser which was
brought into nearly head-on collision with the electron beam downstream from
the IP.  Compton scattered electrons were deflected by dipole magnets
and detected in a threshold Cherenkov counter, providing a beam
polarisation measurement with good statistical precision every few
minutes.  Over the course of SLC operation, significant time was
expended in a number of polarimetry cross-checks which served to
ensure confidence in the final polarimeter results. These took the
form of additional polarimeter detectors used at the IP and elsewhere
in the SLC (the more widely used but less precise M\o ller scattering
polarimeters), and specialized short-term accelerator experiments
designed to test polarised beam transport and to reveal, and mitigate,
unanticipated systematic effects.

Secondary in importance compared to the polarimeter, but essential to
the precision electroweak measurements, were two energy spectrometers
installed in the extraction lines for the electron and positron beams.
These instruments employed precisely calibrated analyzing bend
magnets, and were needed to accurately determine the centre-of-mass
collision energy.  The expected precision of this measurement was
about $20~\MeV$.  In 1998 SLD performed a scan of the $\Zzero$
resonance, which allowed recalibration of the SLC energy scale to the
precise value of $\MZ$ determined at LEP.  Further details of the SLC
operation, in particular concerning polarisation, are given in
Chapter~\ref{sec-ALR}.

\section{LEP/SLC Detectors\label{sec:intro_detectors}}

\begin{figure}[tbp]
\begin{center}
\includegraphics[width=0.9\textwidth]{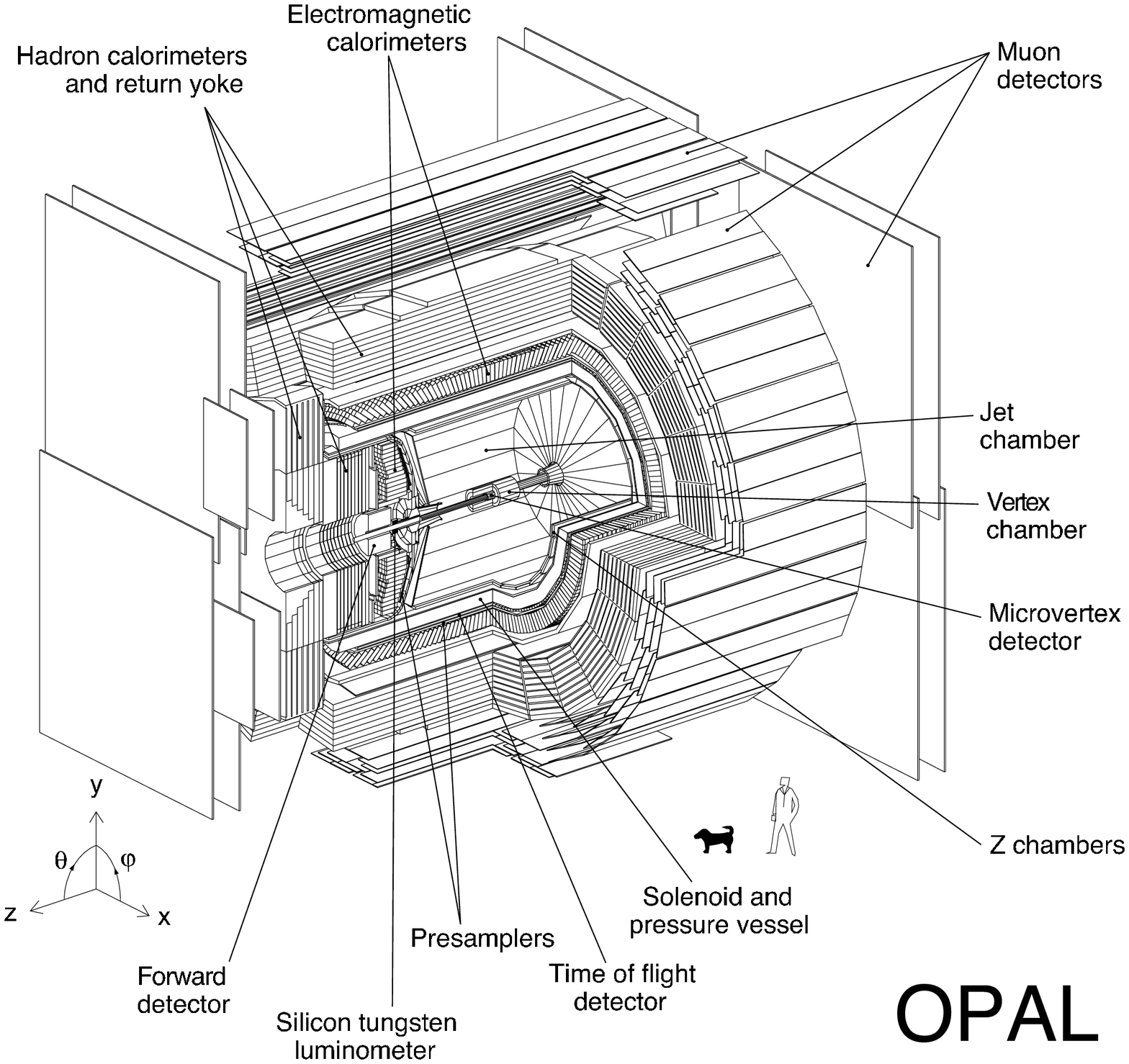}
\caption[A typical LEP/SLC detector.]{A cut-away view of the OPAL
  detector, as an example LEP/SLC detector. The $z$-axis points along
  the direction of the electron beam.}
\label{fig:intro_OPAL_det}
\end{center}
\end{figure}

The designs of the LEP and SLC detectors are quite similar, although
the details vary significantly among them.  As an example, the OPAL
detector is shown in Figure~\ref{fig:intro_OPAL_det}.  All five
detectors use the coordinate conventions indicated in this figure.
The polar angle $\theta$ is measured with respect to the electron
beam, which travels in the direction of the $z$-axis.  The azimuthal
angle $\phi$ is measured in the $x$-$y$ plane.  Starting radially from
the interaction point, there is first a vertex detector, followed by a
gas drift chamber to measure the parameters of charged particle
tracks.  Typically all tracks with transverse momenta greater than
$\sim 200\MeV$ resulting from each $\Zzero$ decay could be
reconstructed in three dimensions with high efficiency.  The momentum
resolution provided by the tracking chamber was also sufficient to
determine the sign of a single charged particle carrying the full beam
momentum.

Surrounding the tracking system is a calorimeter system, usually
divided into two sections.  The first section is designed to measure
the position and energy of electromagnetic showers from photons,
including those from $\pi^0$ decay, and electrons.  The
electromagnetic calorimeter is followed by a hadronic calorimeter to
measure the energy of hadronic particles.  Finally, an outer tracking
system designed to measure the parameters of penetrating particles
(muons) completes the system.

The central part of the detector (at least the tracking chamber) is
immersed in a solenoidal magnetic field to allow the measurement of
the momentum of charged particles.  In addition, particle
identification systems may be installed, including
$\mathrm{d}E/\mathrm{d}x$ ionisation loss measurements in the central
chamber, time-of-flight, and ring-imaging Cherenkov detectors.
These measurements can be used to determine the velocity of particles;
coupled with the momentum, they yield the particle masses.

Special detectors extending to polar angles of $\sim 25$mrad with
respect to the beam axis detect small-angle Bhabha scattering events.
The rate of these events was used for the luminosity determinations,
as the small-angle Bhabha process is due almost entirely to QED, and
the cross-section can be calculated precisely.  All the LEP
experiments replaced their first-generation luminosity detectors,
which had systematic uncertainties around the percent level, by
high-precision devices capable of pushing systematic errors on the
acceptance of small-angle Bhabha scattering events below one
per-mille.

Each LEP experiment also upgraded its original vertex detector with
multi-layer silicon devices, which significantly improved the ability
to measure impact parameters and to identify secondary vertices with a
resolution of approximately 300~$\mu$m.  As the typical B-hadron
produced in Z decays will move about 3 mm from the primary vertex
before decaying, the use of these detectors allowed the selection of a
heavy quark sample with high purity.  The typical beam spot size was
150~$\mu$m $\times$ 5~$\mu$m for LEP and 1.5~$\mu$m $\times$
0.7~$\mu$m for SLC, in the bending and non-bending planes,
respectively.

The smaller dimensions of the SLC beams and their low repetition rate
allowed SLD to place slow but very high-resolution CCD arrays at a
smaller radius than the micro-strip devices used at LEP.  Both features
resulted in SLD's superior vertex reconstruction.

As a consequence of the improvements to the detectors and also in the
understanding of the beam energy at \LEPI{}, and the production of high
beam polarisation at SLC, statistical and systematic errors are much
smaller for the later years of data taking, which hence dominate the
precision achieved on the $\Zzero$ parameters.

\begin{figure}[htbp]
\begin{center}
\mbox{\epsfig{file=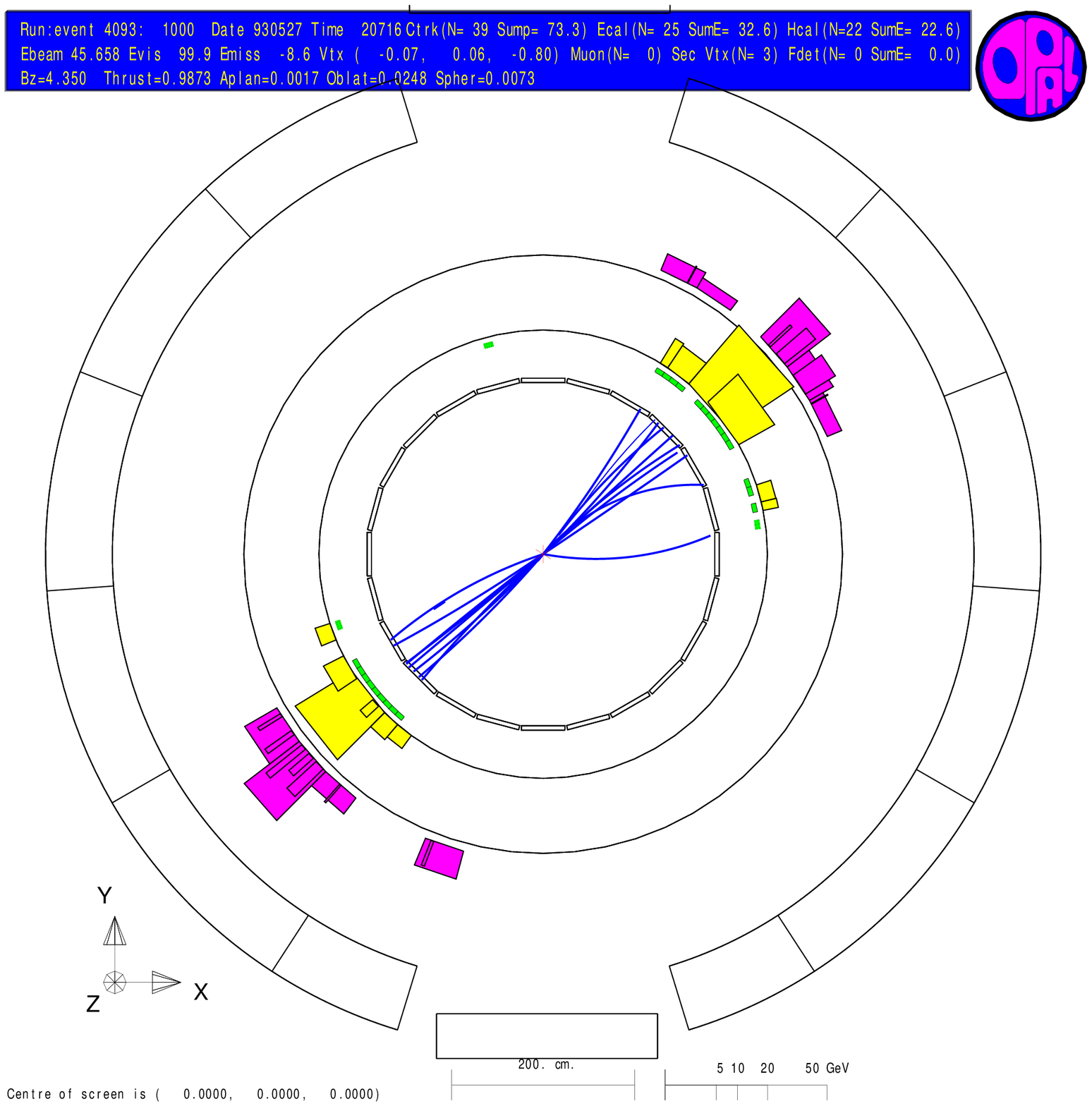,width=0.495\textwidth}}
\hfill
\mbox{\epsfig{file=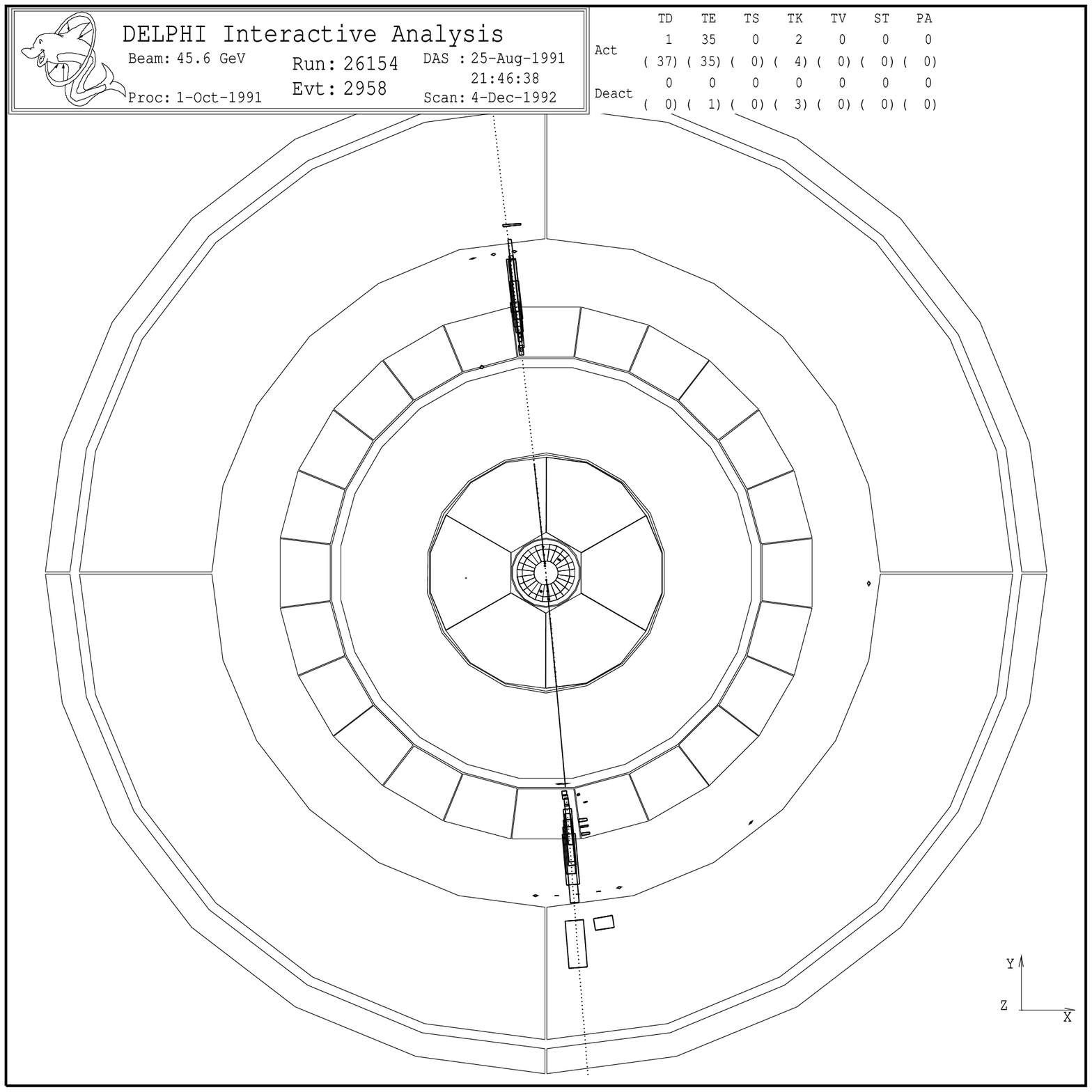,width=0.495\textwidth}}
\vskip 2cm
\mbox{\epsfig{file=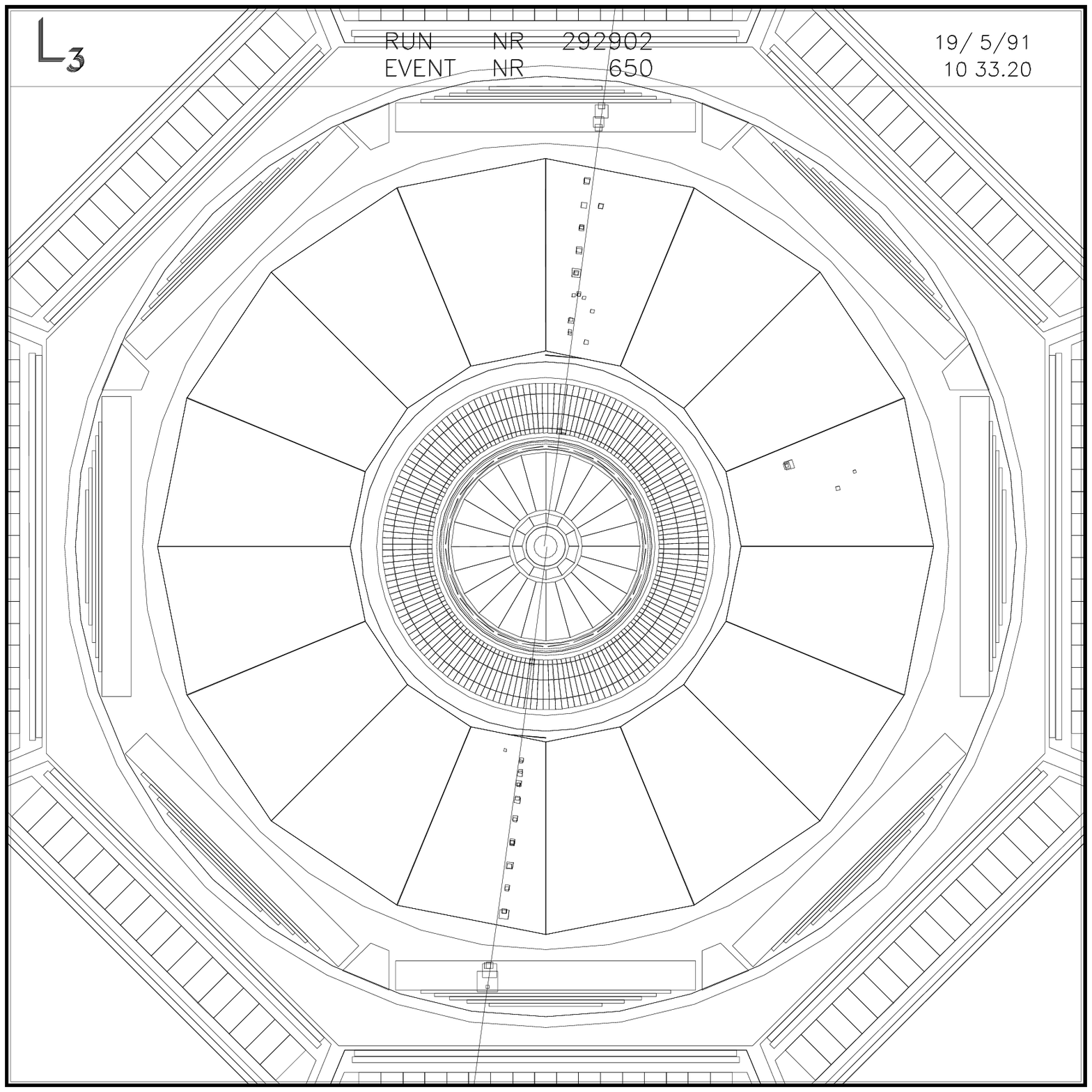,width=0.495\textwidth}}
\hfill
\includegraphics[angle=-90,origin=c,width=0.495\textwidth]{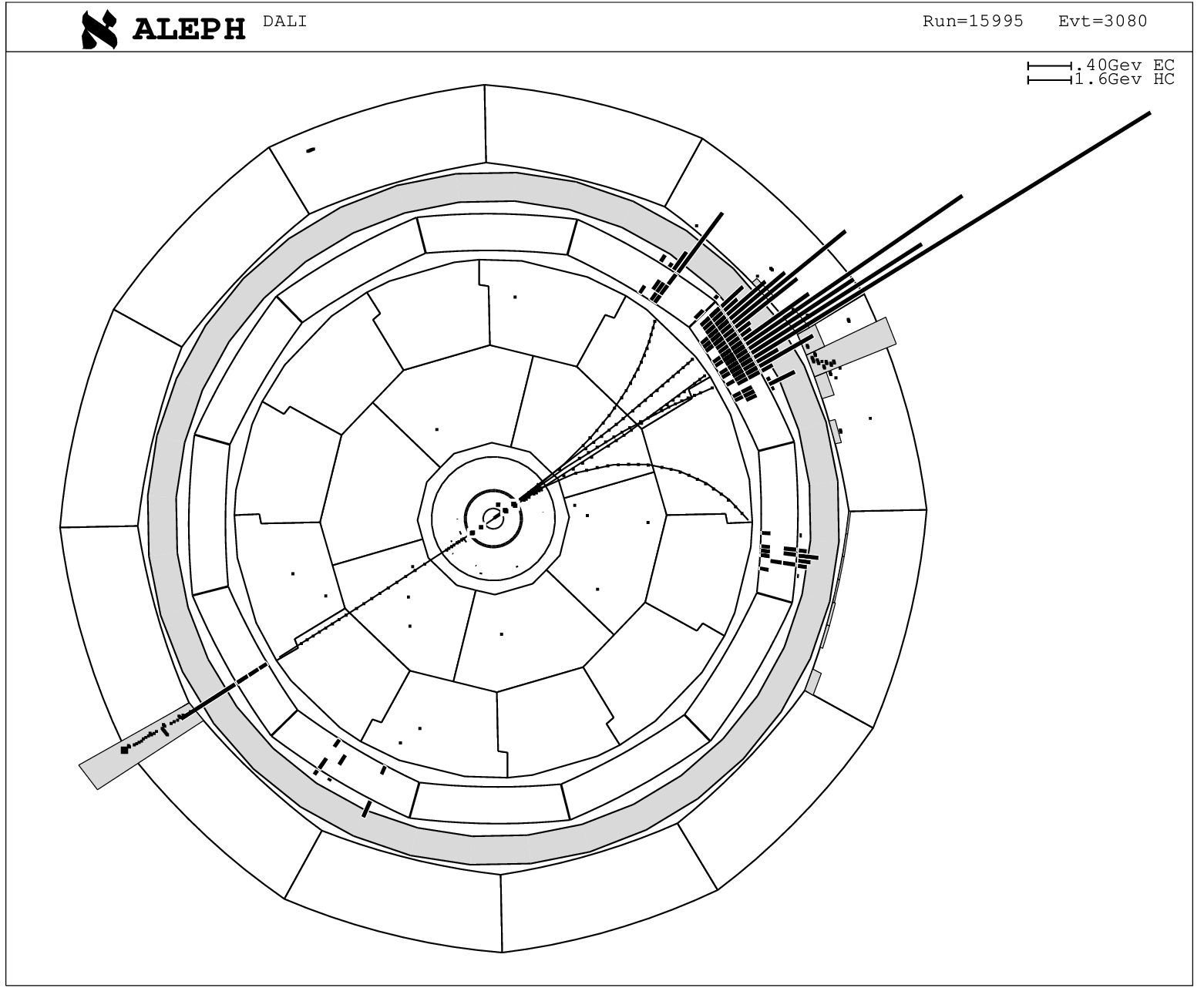}
\caption[Event display pictures of $\qq$, $\ee$, $\mumu$ and
$\tautau$] {\label{fig:intro_events} Pictures of $\qq$, $\ee$, $\mumu$
and $\tautau$ final states, visualised with the event displays of the
OPAL, DELPHI, L3 and ALEPH collaborations, respectively.  In all
views, the electron-positron beam axis is perpendicular to the plane
of the page.  The stability of the electron and the long lifetime of
the muon allow these fundamental $\Zzero$ decays to be directly
observed, while the low-multiplicity products of $\tau$ decays are
confined to well-isolated cones.  Hadronic $\Zzero$ decays result in
higher-multiplicity jets of particles produced in the QCD cascades
initiated by the initial $\qq$ pair.  }
\end{center}
\end{figure}

All five detectors had almost complete solid angle coverage; the only
holes being at polar angles below the coverage of the luminosity
detectors.  Thus, most events were fully contained in the active
elements of the detectors, allowing straight-forward identification.
A few typical $\Zzero$ decays, as seen in the detectors, are shown in
Figure~\ref{fig:intro_events}.  As can be seen, the events at LEP and
SLC were extremely clean, with practically no detector activity
unrelated to the products of the annihilation event, allowing
high-efficiency and high-purity selections to be made.  Shown in
Figure~\ref{fig:intro_sideview} is a side view of an SLD event
interpreted as the decay of a $\Zzero$ into $\bb$.  The displaced
vertex from the decay of a B hadron is clearly visible.

\section{Basic Measurements\label{sec:intro_measurements}}

As suggested by the event pictures, the decays of the $\Zzero$ to
charged leptons and to quarks are distinguished relatively easily, and
in addition some specific quark flavours can be identified.  Total
cross-sections for a given process are determined by counting selected
events, $N_\mathrm{sel}$, subtracting the expected background,
$N_\mathrm{bg}$, and normalising by the selection efficiency
(including acceptance), $\epsilon_\mathrm{sel}$, and the luminosity,
$\calL$:
\begin{equation}
  \label{eq:intro_xsec_def}
  \sigma ~ = ~ \frac{N_\mathrm{sel} -
  N_\mathrm{bg}}{\epsilon_\mathrm{sel}{\calL}} \,.
\end{equation}
The expected background and the selection efficiencies are determined
using Monte Carlo event generators (for
example~\cite{JETSET,HERWIG,ARIADNE,KORALZ,KK,BABAMC,BHLUMI4}).  The
generated events are typically passed through a program that
simulates the detector response, using packages such as
GEANT~\cite{GEANT}, and then processed by the same reconstruction
program as used for the data.

The cross-sections as a function of centre-of-mass energy around the
$\Zzero$ pole yield the $\Zzero$ mass, $\MZ$, and total width, $\GZ$,
together with a pole cross-section. The ratios of cross-sections for
different processes give the partial widths and information about the
relative strengths of the $\Zzero$ couplings to different final-state
fermions.

The $\Zzero$ couples with a mixture of vector and axial-vector
couplings.  This results in measurable asymmetries in the angular
distributions of the final-state fermions, the dependence of
$\Zzero$ production on the helicities of the colliding electrons and
positrons, and the polarisation of the produced particles.

\begin{figure}[thb]
\begin{center}
\includegraphics[width=0.9\textwidth]{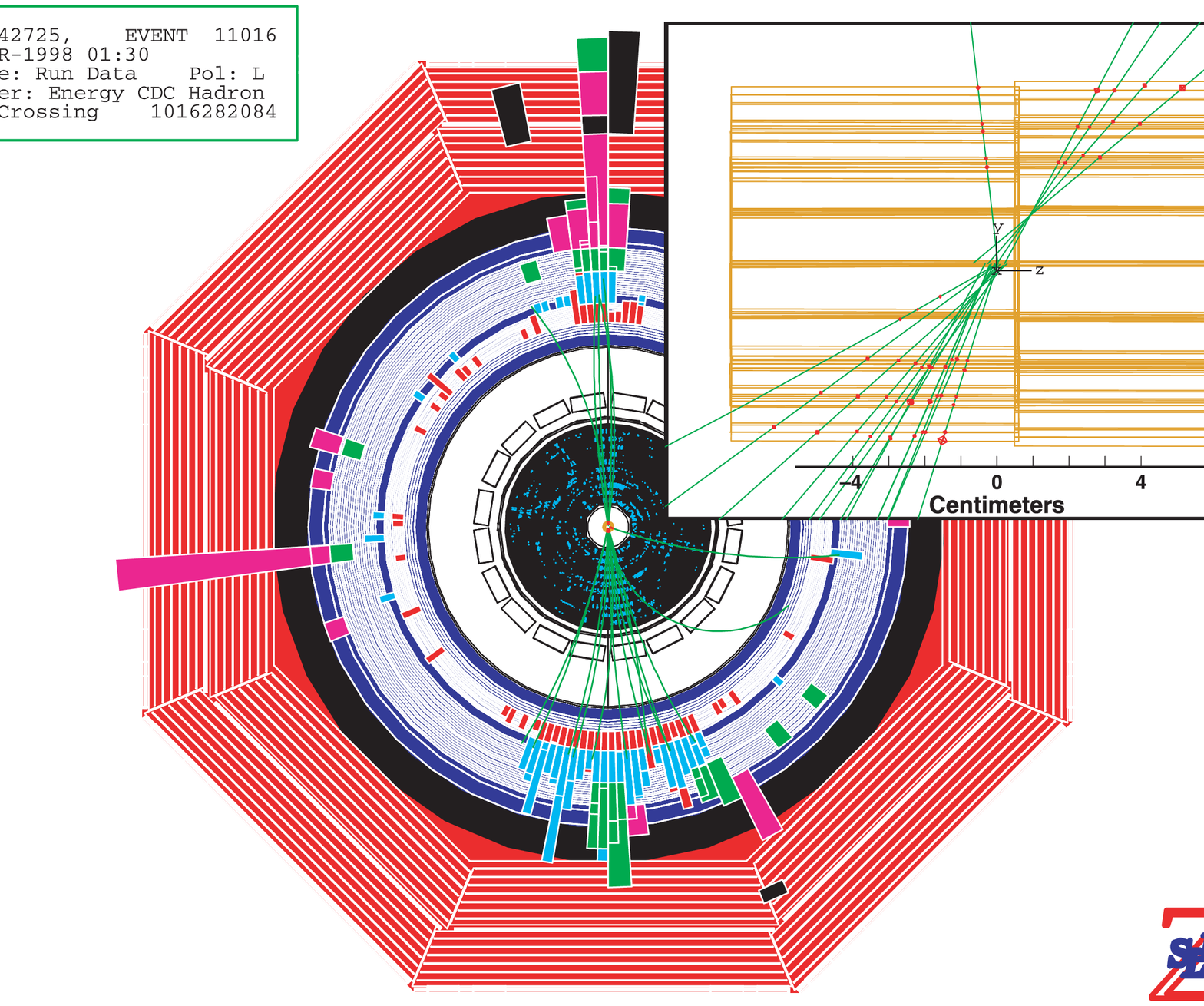}
\end{center}
\caption[Side view of $\Ztobb$] {\label{fig:intro_sideview} Front view
of an event classified as $\Ztobb$.  The displaced secondary vertex is
visible in the expanded side view ($r$-$z$ view) of the beam
interaction point.}
\end{figure}

One of the simplest such asymmetries to measure is the number of
forward events, $N_{\mathrm{F}}$, minus the number of backward events,
$N_{\mathrm{B}}$, divided by the total number of produced events:
\begin{eqnarray}
  \label{eq:intro_afb_count}
  \Afb &=& \frac{N_\mathrm{F} - N_\mathrm{B}}{N_\mathrm{F}+N_\mathrm{B}},
\end{eqnarray}
where ``forward" means that the produced fermion (as opposed to
anti-fermion) is in the hemisphere defined by the direction of the
electron beam (polar scattering angle $\theta < \pi/2$).
For example, the tagged jet with four tracks all emerging from a common
secondary vertex in Figure~\ref{fig:intro_sideview} is in the forward
part of the detector. If it is determined that this jet was generated by
the decay of a primary b-quark rather than
$\mathrm{\overline{b}}$-quark (see Section~\ref{sec:hqtag}),
it would be classified as a forward event.

The simple expression in terms of the numbers of forward
and backward events given in Equation~\ref{eq:intro_afb_count}
is only valid for full $4\pi$ acceptance.  The
forward-backward asymmetries are therefore usually derived from fits
to the differential distribution of events as a function of the polar
angle of the outgoing fermion with respect to the incoming electron
beam, see Section~\ref{sec:intro_zpar}.

This is the usual type of asymmetry measured at LEP.  Further
asymmetries, defined in Section~\ref{sec:intro_asymm}, can be measured
if information is available about the helicities of the incoming or
outgoing particles.  In particular, the polarised electron beam at the
SLC allowed the measurement of the left-right asymmetry:
\begin{equation}
  \label{eq:ALR_count}
    \ALR ~ = ~ \frac{N_{\mathrm{L}} - N_{\mathrm{R}}}%
              {N_{\mathrm{L}} + N_{\mathrm{R}}}%
         \frac{1}{\langle \pole \rangle}  \,,
\end{equation}
where, irrespective of the final state, $N_{\mathrm{L}}$ is the number
of Z bosons produced for left-handed electron bunches,
$N_{\mathrm{R}}$ is the corresponding number for right-handed bunches
and $\langle \pole \rangle$ is the magnitude of luminosity-weighted
electron polarisation.  This expression assumes that the luminosity
and the magnitude of the beam polarisation are helicity-symmetric (see
Chapter~\ref{sec-ALR}).  One attractive feature of the $\ALR$
measurement is the fact that it depends only on knowing the beam
polarisation, and not the acceptance of the detector.

When the $\Zzero$ decays to a pair of $\tau$ leptons, their polarisation
asymmetry is determined through the distribution of their decay products,
which are visible in the detectors.

The relationships between the cross-sections and asymmetries and the
$\Zzero$ couplings to fermions will be discussed further in
Section~\ref{sec:intro_zpar} after examining the underlying theory and
its implications for the process $\ee \rightarrow \ff$.

\section{Standard Model Relations\label{sec:intro_ew}}

In the $\SM$ at tree level, the relationship between the weak and
electromagnetic couplings is given by
\begin{equation}
  \label{eq:GF}
  \GF ~ = ~ \frac{\pi\alpha}{\sqrt{2}\MW^2\streesq},
\end{equation}
where $\GF$ is the Fermi constant determined in muon decay, $\alpha$
is the electromagnetic fine-structure constant, $\MW$ is the W boson
mass, and $\streesq$ is the electroweak mixing angle.  In addition, the
relationship between the neutral and charged weak couplings is fixed
by the ratio of the W and Z boson masses:
\begin{equation}
  \label{eq:rhotree}
  \rhoo ~ = ~ \frac{\MW^2}{\MZ^2\ctreesq}.
\end{equation}
The $\rhoo$ parameter~\cite{\rhopar} is determined
by the Higgs structure of the theory; in the Minimal Standard Model
containing only Higgs doublets, $\rhoo = 1$.

\begin{table}[t]
  \begin{center}
\renewcommand{\arraystretch}{1.3}
\begin{tabular}{|ccc||ccc|} \hline
\multicolumn{3}{|c||}{Family}& $T$ & $T_3$ & $Q$ \\
\hline
\hline
& & & & & \\[-4mm]
$\left(\begin{array}{c}
\nu_{\mathrm{e}} \\
\mathrm{e}
\end{array}\right)_L$ &
$\left(\begin{array}{c}
\nu_{\mu} \\
\mu
\end{array}\right)_L$ &
$\left(\begin{array}{c}
\nu_{\tau} \\
\tau
\end{array}\right)_L$ &
$1/2$&
$\begin{array}{c}
+1/2 \\
-1/2
\end{array}$ &
$\begin{array}{r}
0 \\
-1
\end{array}$ \\
$\nu_{\mathrm{e} R}$ & $\nu_{\mu R}$ & $\nu_{\tau R}$ & 0 & 0 & $\phantom{-}0$ \\
$\mathrm{e}_R$ & $\mu_R$ & $\tau_R$ & 0 & 0 & $-1$ \\
\hline
& & & & & \\[-4mm]
 $\left(\begin{array}{c}
\mathrm{u} \\
\mathrm{d}
\end{array}\right)_L$ &
 $\left(\begin{array}{c}
\mathrm{c} \\
\mathrm{s}
\end{array}\right)_L$ &
 $\left(\begin{array}{c}
\mathrm{t} \\
\mathrm{b}
\end{array}\right)_L$ &
$1/2$&
$\begin{array}{c}
+1/2 \\
-1/2
\end{array}$ &
$\begin{array}{r}
+2/3 \\
-1/3
\end{array}$ \\
$\mathrm{u}_R$ & $\mathrm{c}_R$ & $\mathrm{t}_R$ & 0 & 0 & $+2/3$ \\
$\mathrm{d}_R$ & $\mathrm{s}_R$ & $\mathrm{b}_R$ & 0 & 0 & $-1/3$ \\\hline
\end{tabular}
\caption[The weak-isospin structure of the fermions in the $\SM$.]{The
  weak-isospin structure of the fermions in the $\SM$.  ``L'' and
  ``R'' stand for left-handed and right-handed fermions, $T$ and $T_3$
  are the total weak-isospin and its third component, and $Q$ is the
  electric charge.  Note that the results presented in
  this report are insensitive to, and independent of, any small ($<
  \MeV$) neutrino masses.}
\label{tab:intro_SM}
\end{center}
\end{table}

\begin{figure}[htbp]
\begin{center}
\includegraphics[width=\textwidth]{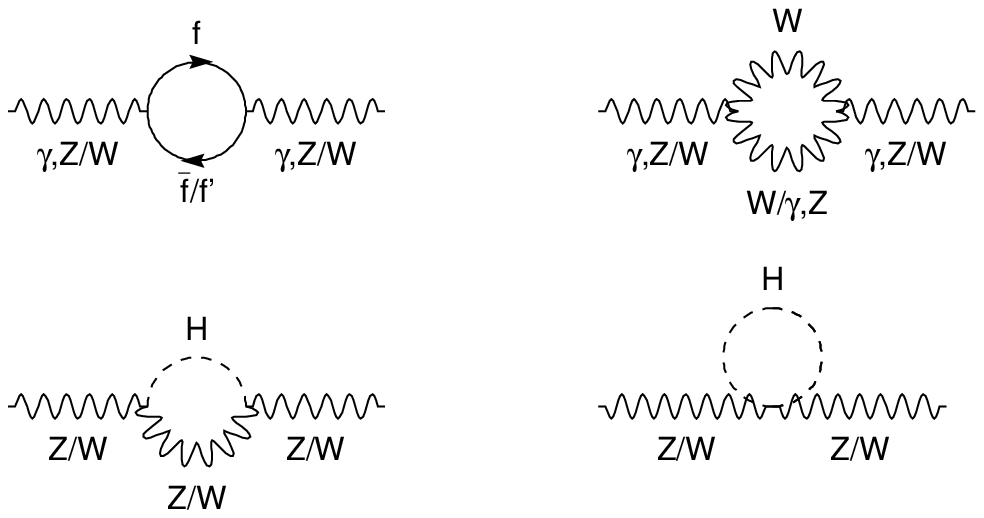}
\caption[Higher-order corrections to the gauge boson
propagators]{Higher-order corrections to the gauge boson propagators
  due to boson and fermion loops.}
\label{fig:intro_ewcor}
\end{center}
\end{figure}

\begin{figure}[tbp]
  \begin{center}
    \includegraphics[width=\textwidth]{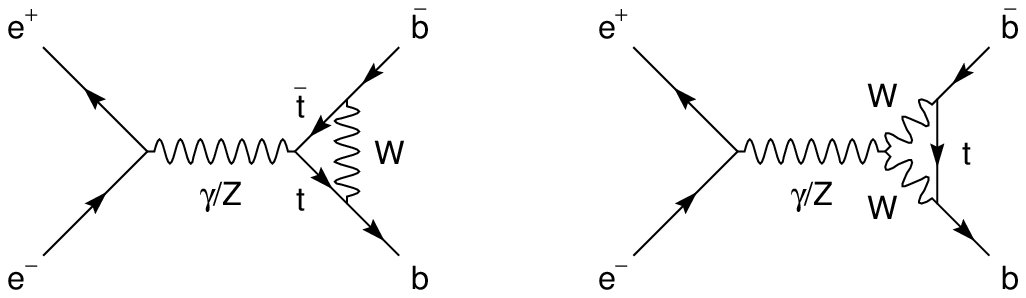}
    \caption{Vertex corrections to the process $\ee\to\bb$.}
    \label{fig:b_vertex}
  \end{center}
\end{figure}

The fermions are arranged in weak-isospin doublets for left-handed
particles and weak-isospin singlets for right-handed particles, as
shown in Table~\ref{tab:intro_SM}.  The interaction of the Z boson
with fermions depends on charge, $Q$, and the third component of
weak-isospin, $T_3$, and is given by the left- and right-handed
couplings:
\begin{eqnarray}
  \gltree &=&\sqrt{\rhoo}\,(\Tf - \Qf \streesq)\label{eq:gl}\\
  \grtree &=& - \sqrt{\rhoo}\, \Qf\streesq\,, \label{eq:gr}
\end{eqnarray}
or, equivalently in terms of vector and axial-vector couplings:
\begin{eqnarray}
 \gvtree & \equiv & \gltree + \grtree 
 ~=~ \sqrt{\rhoo}\,(\Tf - 2\Qf\streesq) \label{eq:gv}\\
 \gatree & \equiv & \gltree - \grtree 
 ~=~ \sqrt{\rhoo}\, \Tf \,.\label{eq:ga}
\end{eqnarray}
These tree-level quantities are modified by radiative corrections to
the propagators and vertices such as those shown in
Figures~\ref{fig:intro_ewcor} and~\ref{fig:b_vertex}.
When these corrections are renormalized in the ``on-shell''
scheme~\cite{\thonshell}, which we adopt here, the form of
Equation~\ref{eq:rhotree} is maintained, and taken to define the
on-shell electroweak mixing angle, $\thw$, to all orders, in terms of
the vector boson pole masses:
\begin{equation}
  \label{eq:rho}
  \rhoo ~ = ~ \frac{\MW^2}{\MZ^2\cwsq}.
\end{equation}
In the following, $\rhoo=1$ is assumed.

The bulk of the electroweak corrections~\cite{\rhopar} to the
couplings at the \Zzero-pole is absorbed into complex form factors,
$\calRf$ for the overall scale and $\calKf$ for the on-shell
electroweak mixing angle, resulting in complex effective couplings:
\begin{eqnarray}
  \cgvf &=& \sqrt{\calRf}\,(\Tf - 2\Qf\calKf\swsq) \\
  \cgaf &=& \sqrt{\calRf}\, \Tf\,.
\end{eqnarray}
In terms of the real parts of the complex form factors,
\begin{eqnarray}
 \rhof   & \equiv & \Re(\calRf)
 ~=~ 1 + \Delta\rhose   + \Delta\rhof    \label{eq:rhoeff} \\
 \kappaf & \equiv & \Re(\calKf)
 ~=~ 1 + \Delta\kappase + \Delta\kappaf  \label{eq:kappaeff}\,,
\end{eqnarray}
the effective electroweak mixing angle and the real effective
couplings are defined as:
\begin{eqnarray}
 \swsqefff & \equiv & \kappaf \swsq                       \label{eq:sin}  \\
 \gvf      & \equiv & \sqrt{\rhof}\,(\Tf-2\Qf\swsqefff) \label{eq:gveff}\\
 \gaf      & \equiv & \sqrt{\rhof}\, \Tf                \label{eq:gaeff}\,,
\end{eqnarray}
so that:
\begin{eqnarray}
\frac{\gvf}{\gaf} & = & \Re\left(\frac{\cgvf}{\cgaf}\right)
~=~ 1-4|\Qf|\swsqefff \,.
\end{eqnarray}

The quantities $\Delta\rhose$ and $\Delta\kappase$ are universal
corrections arising from the propagator self-energies, while
$\Delta\rhof$ and $\Delta\kappaf$ are flavour-specific vertex
corrections.  For simplicity we ignore the small imaginary components
of these corrections in most of the following discussion.
The leading order terms in $\Delta\rhose$ and $\Delta\kappase$ for
$\MH\gg\MW$ are~\cite{Burgers:LEP1YR89VOL1}:
\begin{eqnarray}
   \label{eq:deltarho}
    \Delta\rhose & = &
 \frac{3\GF\MW^2}{8\sqrt{2}\pi^2}\left[\frac{\Mt^2}{\MW^2} -
    \frac{\swsq}{\cwsq}\left(\ln\frac{\MH^2}{\MW^2} -\frac{5}{6}\right) + \cdots\right]  \\ 
   \label{eq:deltakappa}
 \Delta\kappase & = &
     \frac{3\GF\MW^2}{8\sqrt{2}\pi^2}\left[\frac{\Mt^2}{\MW^2}\frac{\cwsq}{\swsq} - 
     \frac{10}{9}\left(\ln\frac{\MH^2}{\MW^2} -\frac{5}{6}\right) + \cdots\right] 
\end{eqnarray}
For $\MH\ll\MW$, the Higgs terms are modified, for example:
\begin{eqnarray}
   \label{eq:deltarhomhsmall}
    \Delta\rhose & = &
\frac{3\GF\MW^2}{8\sqrt{2}\pi^2}\left[\frac{\Mt^2}{\MW^2} +
\frac{2}{3}\frac{\MZ^2}{\MW^2}\ln\frac{\MH^2}{\MZ^2} -  
\frac{7\pi}{3}\frac{\MH\MZ}{\MW^2} + \cdots\right] 
\end{eqnarray}
where only internal Higgs loops are considered.
Note the change of sign in the slope of the Higgs correction for
low $\MH$ seen in Equation~\ref{eq:deltarhomhsmall} compared to
Equation~\ref{eq:deltarho}, which is due to
contributions from the derivative of the $\Zzero$ self-energy with
respect to momentum transfer~\cite{Jegerlehner:1991dq}.
Existence of the process $\ee\rightarrow\Zzero^*\mathrm{H}$ (Higgs\-strahlung)
would tend to reduce the $\MH$ dependence in
Equation~\ref{eq:deltarhomhsmall}~\cite{Kawamoto:2004pi}.
The radiative corrections have a quadratic dependence on the top quark
mass and a weaker logarithmic dependence on the Higgs boson mass.
The flavour dependence is very small for all fermions, except for the
b-quark, where the effects of the diagrams shown in
Figure~\ref{fig:b_vertex} are significant, due to the large mass
splitting between the bottom and top quarks and the size of the
diagonal CKM matrix element $|V_{\mathrm{tb}}|\simeq 1$ , resulting in
a significant additional contribution for $\bb$
production~\cite{Jegerlehner:1991dq} (The effects of the off-diagonal
CKM matrix elements are here negligible.):
\begin{eqnarray}
  \label{eq:b_vertex}
  \Delta\kappa_{\mathrm{b}} &=& \frac{\GF\Mt^2}{4\sqrt{2}\pi^2} + \cdots\,,\\
  \Delta\rho_{\mathrm{b}}   &=& -2\Delta\kappa_{\mathrm{b}} + \cdots \,.
\end{eqnarray}

By interpreting the \Zzero-pole measurements in terms of these corrections,
the top quark mass can be determined indirectly, and compared to
the direct measurements.  The \Zzero-pole measurements, even when
taken alone, have sufficient power to separate the Higgs and top
corrections to some extent, and thus provide independent indications
of both $\Mt$, and, less sensitively, $\MH$.  The constraint on $\MH$
becomes more precise when additional results, in particular the direct
measurement of $\Mt$, are also considered (see Section~\ref{sec:mtdirect}).

The classic ``$\rho$ parameter"~\cite{\rhopar}, which describes the ratio of the
neutral to charged current couplings in neutrino interactions at low
momentum transfer, is also modified by radiative corrections:
\begin{equation}
  \label{eq:drho}
  \rho ~ = ~ 1 + \Delta\rho\,.
\end{equation}
Although $\rho$ displays a similar $\Mt$-dependence to that of $\rhof$,
its $\MH$-dependence specifically lacks the change in sign at low
$\MH$ which is evident in Equation~\ref{eq:deltarhomhsmall}.

The form of the fundamental $\SM$ relation derived from
Equations~\ref{eq:GF} and~\ref{eq:rho} is preserved in the presence of
radiative corrections for both low momentum transfer, and at the
\Zzero-pole~\cite{Burgers:LEP1YR89VOL1}:
\begin{eqnarray}
  \label{eq:GFmod}
 \cwsq\swsq & = & \frac{\pi\alpha(0)}{\sqrt{2}\MZ^2\GF}\frac{1}{1-\Dr} \\
  \label{eq:GFmodf}
  \cwsqefff\swsqefff & = & \frac{\pi\alpha(0)}{\sqrt{2}\MZ^2\GF}\frac{1}{1-\Drf} ,
\end{eqnarray}
where $\Dr$ and $\Drf$ are given by:
\begin{eqnarray}
  \label{eq:Dr}
  \Dr &=& \Delta\alpha  + \Drw \\
 \Drf &=& \Delta\alpha + \Drwf .
\end{eqnarray}
The $\Delta\alpha$ term arises from the running of the electromagnetic
coupling due to fermion loops in the photon propagator, and is usually
divided into three categories: from leptonic loops, top quark loops and
light quark (u/d/s/c/b) loops:
\begin{equation}
  \label{eq:Dalpha}
  \Delta\alpha(s) ~ = ~ 
                    \Delta\alpha_{\mathrm{e\mu\tau}}(s)
                   +\Delta\alpha_{\mathrm{top     }}(s)
                   +\Delta\alpha_{\mathrm{had     }}^{(5)}(s).
\end{equation}
The terms $\Delta\alpha_{\mathrm{e\mu\tau}}(s)$ and
$\Delta\alpha_{\mathrm{top }}(s)$ can be precisely calculated, whereas
the term $\Delta\alpha_{\mathrm{had }}^{(5)}(s)$ is best determined
by analysing low-energy $\ee$ data using a dispersion relation
(see Section~\ref{sec:msm:vacpol}).  These
effects are absorbed into $\alpha$ as:
\begin{equation}
  \label{eq:alpharun}
  \alpha(s) ~ = ~ \frac{\alpha(0)}{1-\Delta\alpha(s)}.
\end{equation}
At LEP/SLC energies, $\alpha$ is increased from the zero $q^2$ limit of
$1/137.036$ to $1/128.945$.

The weak part of the corrections contains $\Delta\rho$
(see Equation~\ref{eq:drho})
plus a remainder~\cite{Burgers:LEP1YR89VOL1}:
\begin{eqnarray}
  \label{eq:Drw}
  \Drw &=& -\frac{\cwsq}{\swsq}\Delta\rho + \cdots \\ %
  \Drwf &=& -\Delta\rho + \cdots . %
\end{eqnarray}

It should be noted that since $\GF$ and $\MZ$ are better determined
than $\MW$, Equations~\ref{eq:rho} and~\ref{eq:GFmod} are often used to
eliminate direct dependence on $\MW$~\cite{Burgers:LEP1YR89VOL1}:
\begin{eqnarray}
  \label{eq:MW_gone}
  \MW^2 &=& \frac{\MZ^2}{2}\left(1+\sqrt{1-4\frac{\pi\alpha}{\sqrt{2}\GF\MZ^2}\frac{1}{1-\Dr}}\,\,\right)\,.
\end{eqnarray}
This substitution introduces further significant $\Mt$ and $\MH$
dependencies through $\Delta r$.  For example, in
Equation~\ref{eq:sin} $\swsqeffl$ receives radiative corrections both
from $\Delta\kappase$ directly, and from $\Drw$ implicitly through
$\swsq$, as can be seen in Equation~\ref{eq:GFmodf}.
Here the implicit correction is of opposite sign, and in
fact dominates the direct correction, so that the $\Mt$ and $\MH$
dependences of $\swsqeffl$ are opposite in sign from the dependences
of $\Delta\kappase$ described in Equation~\ref{eq:deltakappa}.

The discussion of radiative corrections given here is leading order
only.  The actual calculations used in fits (\eg,
Chapters~\ref{chap:Z+coup} and~\ref{chap:MSM}) are performed to higher
order, using the programs TOPAZ0~\cite{\TOPAZref} and
ZFITTER~\cite{\ZFITTERref}.  The interested reader is encouraged to
consult the authoritative discussion in
Reference~\citen{BardinPassarinoBook}.

\section[The Process \protect$\ee\rightarrow\ff$]%
        {The Process \protect\boldmath$\ee\rightarrow\ff$}
\label{sec:intro_zpar}
       
The differential cross-sections for fermion pair production (see
Figure~\ref{fig:intro_eeff}) around the $\Zzero$ resonance can be cast
into a Born-type structure using the complex-valued effective coupling
constants given in the previous section.  Effects from photon vacuum
polarisation are taken into account by the running electromagnetic
coupling constant (Equation~\ref{eq:alpharun}), which also acquires a
small imaginary piece.  Neglecting initial and final state photon
radiation, final state gluon radiation and fermion masses, the
electroweak kernel cross-section for unpolarised beams can thus be
written as the sum of three contributions, from $s$-channel $\gamma$ and $\Zzero$
exchange and from their interference~\cite{BardinPassarinoBook},
\begin{equation} \begin{array}{ll}
\lefteqn{
\frac{2s}{\pi}\frac{1}{N_c^{\rm f}}\frac{d\sigma_{\rm ew}}{d\cost}(\eeff)~=}&\\
                                                              &    \\ [-2mm]
 & \underbrace{
\left| \alpha(s) \Qf \right|^{2} (1+\cos^{2}\theta) 
                }_{\textstyle \sigma^{\gamma}}                 \\
                                                              &    \\ [-2mm]
 & \underbrace{
 -8 \Re \left\{ \alpha^*(s) \Qf \chi (s)  
 \left[ \cgve\cgvf (1+\cos^{2}\theta)
 + 2 \cgae\cgaf \cost  \right] \hspace*{-0.2em}\right\} 
                      }_{\textstyle \gammaZ{\rm ~interference}}   \\
                                                               &  \\ [-2mm]
 &  +16|\chi(s)|^{2}
 \left[ (|\cgve|^2+|\cgae|^2)(|\cgvf|^2+|\cgaf|^2)(1+\cos^{2}\theta) \right.
 \\
                                                   
 & \underbrace{
 \left. \hspace*{13ex}+ 8\Re\left\{\cgve{\cgae}^*\right\}
                         \Re\left\{\cgvf{\cgaf}^*\right\}
                        \cos\theta \right] \hspace*{9ex}  
                        }_{\textstyle \sigma^{\rm Z}} \\

\end{array}\label{eq:intro_bigugly}\end{equation}
with:
\begin{equation}
\label{eq:intro_prop}
\chi(s) ~ = ~ \frac{\GF\MZ^{2}}{8\pi\sqrt{2}}
       \frac{s}{s-\MZ^{2} + is\GZ/\MZ} \, ,
\end{equation}
where $\theta$ is the scattering angle of the out-going fermion with
respect to the direction of the e$^-$.  The colour factor $N_c^{\rm
f}$ is one for leptons (f=$\nu_e$, $\nu_\mu$, $\nu_\tau$, e, $\mu$,
$\tau$) and three for quarks (f=d, u, s, c, b), and $\chi(s)$ is the
propagator term with a Breit-Wigner denominator with an $s$-dependent
width.

If the couplings are left free to depart from their $\SM$ values, the
above expression allows the resonance properties of the $\Zzero$ to be
parametrised in a very model-independent manner.  Essentially the only
assumptions imposed by Equation~\ref{eq:intro_bigugly} are that the
$\Zzero$ possesses vector and axial-vector couplings to fermions, has
spin 1, and interferes with the photon. Certain $\SM$ assumptions are
nevertheless employed when extracting and interpreting the couplings;
these are discussed in Sections~\ref{sec:intro_SM_remnants}
and~\ref{sec-jhad}.

The $1+\cos^2\theta$ terms in the above formula contribute to the
total cross-section, whereas the terms multiplying $\cost$ contribute
only to the forward-backward asymmetries for an experimental
acceptance symmetric in $\cost$. In the region of the $\Zzero$ peak,
the total cross-section is completely
dominated by $\Zzero$ exchange. The $\gammaZ$ interference determines
the energy dependence of the forward-backward asymmetries and
dominates them at off-peak energies, but its leading contribution,
from the real parts of the couplings, vanishes at $\sqrt{s}=\MZ$.

In Bhabha scattering, $\ee\rightarrow\ee$, the $t$-channel diagrams
also contribute to the cross-sections, with a very dominant photon
contribution at large $\cost$, \ie, in the forward direction. This
contribution, and its interference with the $s$-channel, add to the
pure $s$-channel cross-section for $\ee\rightarrow\ee$ (see
Section~\ref{sec-tcherr} for details).

The definition of the mass and width with an $s$-dependent width term
in the Breit-Wigner denominator is
suggested~\cite{Berends:LEP1YR89VOL1} by phase-space and the structure
of the electroweak radiative corrections within the $\SM$. It is
different from another commonly used definition, the real part of the
complex pole~\cite{\cmplxpole}, where the propagator term takes the
form $\chi(s)\propto s/(s- {\overline{m}_{\rm Z}}^2 +
i\overline{m}_{\rm Z} \overline{\Gamma}_{\rm Z})$.  However, under the
transformations $\MZbar = \MZ/\sqrt{1+\GZ^2/\MZ^2}$ and $\GZbar
=\GZ/\sqrt{1+\GZ^2/\MZ^2}$, and adjusting the scales of Z exchange and
$\gamma$/Z interference, the two formulations lead to exactly
equivalent resonance shapes, $\sigma(s)$.

\begin{figure}[htb]
\begin{center}
\includegraphics*[width=\textwidth]{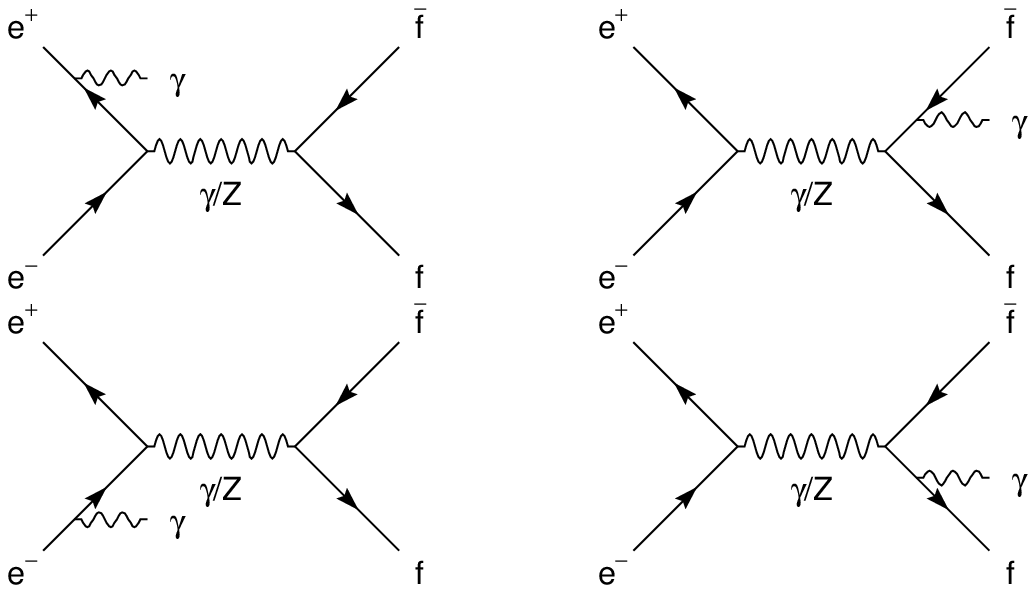}\\
\includegraphics*[width=\textwidth]{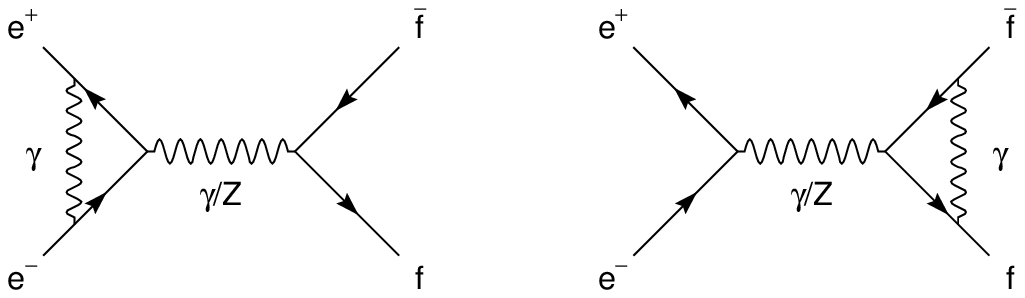}
\caption[QED corrections to fermion-pair production]
{\label{fig:intro_eeffrad} Some of the lowest order QED corrections to
  fermion-pair production. Together with photonic box diagrams, which give much
  smaller contributions, these form a gauge-invariant sub-set included
  in the radiator functions $H_{\rm QED}$.  Weak boxes are added explicitly
  to the kernel cross-section~\cite{BardinPassarinoBook}.
  }
\end{center}
\end{figure}

\begin{figure}[hbtp]
\begin{center}
  \includegraphics[width=0.6\textwidth]{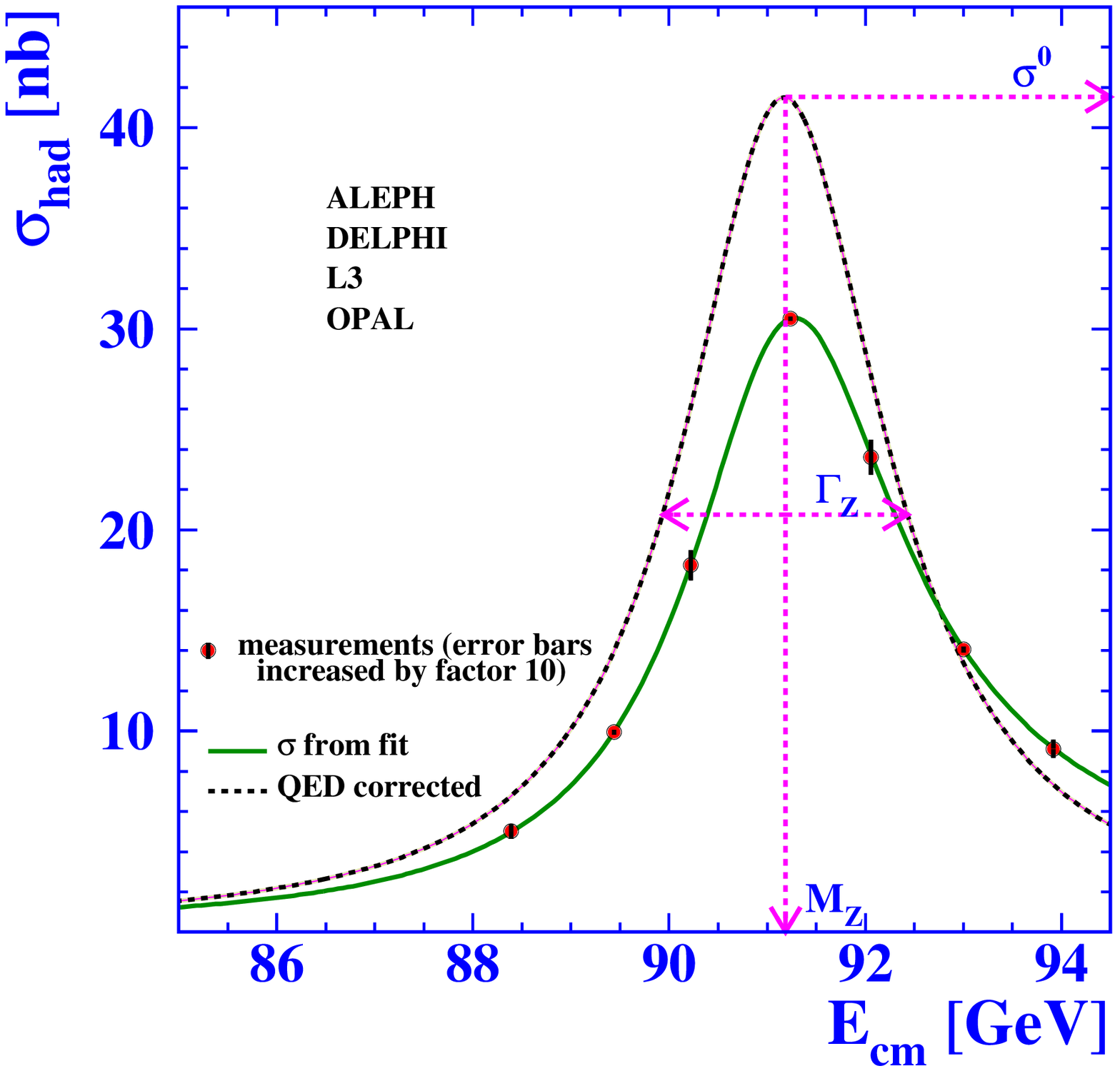}\\
  \includegraphics[width=0.6\textwidth]{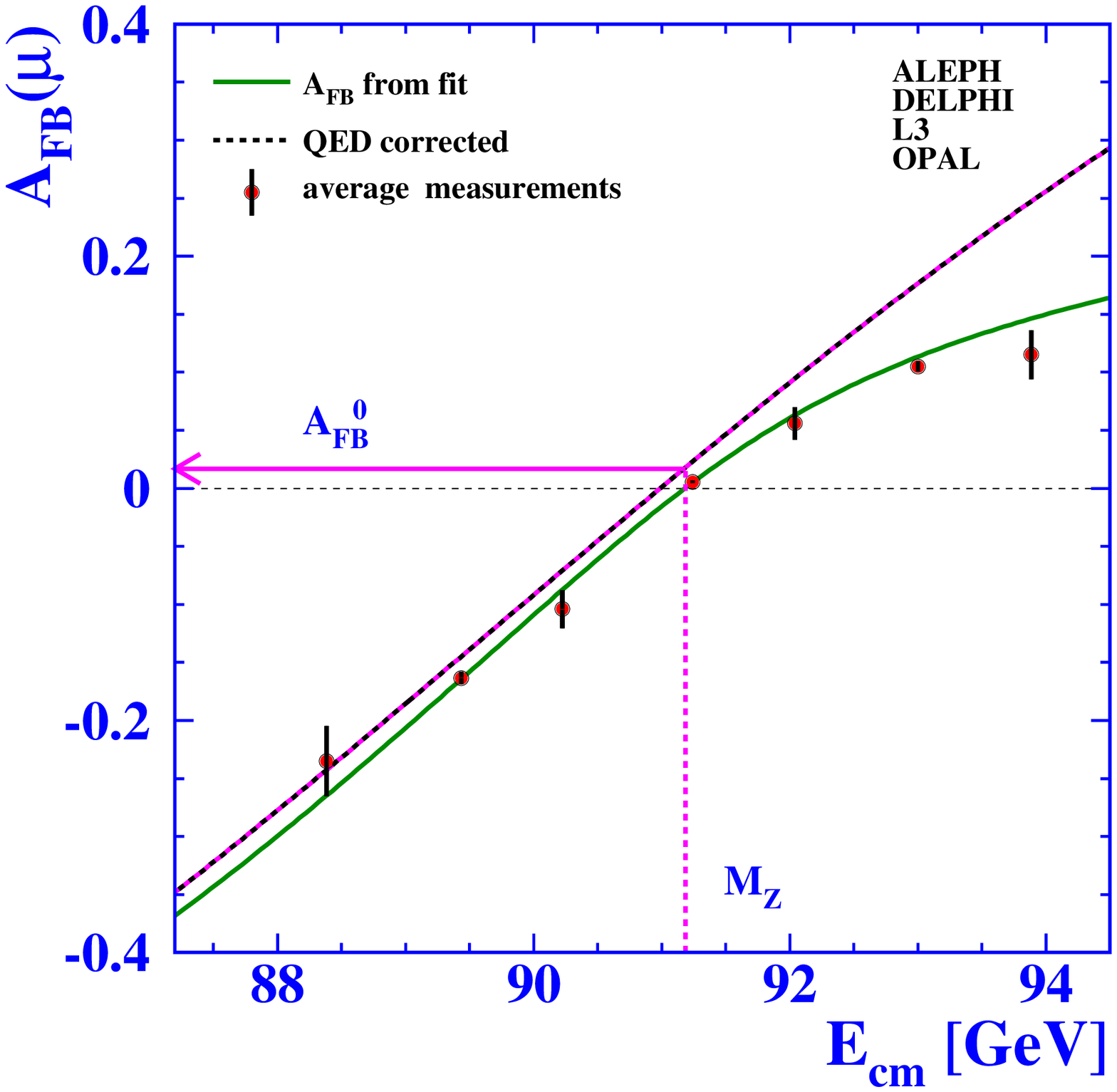} \\
\end{center}
\caption[The hadronic cross-section and muon
forward-backward asymmetry \emph{vs.} $\sqrt{s}$] {\label{fig:xshafb}
  Average over measurements of the hadronic cross-sections (top) and
  of the muon forward-backward asymmetry (bottom) by the four
  experiments, as a function of centre-of-mass energy. The full line
  represents the results of model-independent fits to the
  measurements, as outlined in Section~\ref{sec:intro_zpar}.
  Correcting for QED photonic effects yields the dashed curves, which
  define the $\Zzero$ parameters described in the text.}
\end{figure}

Photon radiation (Figure~\ref{fig:intro_eeffrad}) from the initial and
final states, and their interference, are conveniently treated by
convoluting the electroweak kernel cross-section, $\sigma_{\rm
  ew}(s)$, with a QED radiator, $H^{\rm tot}_{\rm QED}$,
\begin{equation} 
 \sigma(s) ~ = ~
   \int_{4m_{\rm f}^2/s}^{1} dz\,
   H^{\rm tot}_{\rm QED}(z,s) \sigma_{\rm ew}(zs).
\end{equation}
The difference between the forward and backward cross-sections
entering into the determination of the forward-backward asymmetries,
$\sigmaf-\sigmab$, is treated in the same way using a radiator
function $H^{\rm FB}_{\rm QED}$.  These QED corrections are calculated
to third order, and their effects on the cross-sections and
asymmetries are shown in Figure~\ref{fig:xshafb}.
At the peak the QED deconvoluted cross-section is 36\% larger than
the measured one, and the peak position is shifted downwards by about
$100~\MeV$.  At and below the peak $\Afbm$ and $\Afbt$ are 
offset by an amount about equal to their deconvoluted value of 0.017.
The estimated precision of these important corrections is
discussed in Section~\ref{sec-QEDerr}.
It is important to realize that these QED corrections are essentially
independent of the electroweak corrections discussed in
Section~\ref{sec:intro_ew}, and therefore allow the parameters of
Equation~\ref{eq:intro_bigugly} to be extracted from the data in
a model-independent manner.

\subsection{Cross-Sections and Partial Widths}
\label{sec:intro_xrpar}

The partial Z decay widths are defined inclusively, \emph{i.e.}, they
contain QED and QCD~\cite{bib-PCLI-QCD} final-state corrections and
contributions from the imaginary and non-factorisable
parts~\cite{\delewqcd} of the effective couplings,
\begin{equation}
 \Gff ~ = ~ N_c^{\rm f} \frac {\GF \MZ^3} {6\sqrt{2}\pi} 
      \left( |\cgaf|^2 R_{\mathrm{Af}} + |\cgvf|^2 R_{\mathrm{Vf}} \right) 
          +  \Delta_{\rm ew/QCD}. 
\label{eq:Gff}
\end{equation} 
The primary reason to define the partial widths including final state
corrections and the contribution of the complex non-factorisable terms
of the couplings is that the partial widths defined in this way add up
straightforwardly to yield the total width of the $\Zzero$ boson.  The
radiator factors $R_{\mathrm{Vf}}$ and $R_{\mathrm{Af}}$ take into
account final state QED and QCD corrections as well as non-zero
fermion masses; $\Delta_{\rm ew/QCD}$ accounts for small contributions
from non-factorisable electroweak/QCD corrections. The inclusion of
the complex parts of the couplings in the definition of the leptonic
width, $\Gll$, leads to changes of 0.15~per-mille corresponding to
only 15\% of the LEP-combined experimental error on $\Gll$.  The QCD
corrections only affect final states containing quarks.  To first
order in $\alfas$ for massless quarks, the QCD corrections are flavour
independent and the same for vector and axial-vector contributions:
\begin{equation}
  \label{eq:QCDcorr}
  R_{\mathrm{A,QCD}} ~ = ~ R_{\mathrm{V,QCD}} = 
                  R_{\mathrm{QCD}} = 1 + \frac{\alfmz}{\pi} + \cdots\,.
\end{equation}
The hadronic partial width therefore depends strongly on $\alfas$.
The final state QED correction is formally similar, but much smaller
due to the smaller size of the electromagnetic coupling:
\begin{eqnarray}
  \label{eq:QEDcorr}
  R_{\mathrm{A,QED}} & = & R_{\mathrm{V,QED}} = R_{\mathrm{QED}} =
        1 + \frac{3}{4}\Qf^2\frac{\alqed}{\pi} + \cdots\,.
\end{eqnarray}

The total cross-section arising from the $\cost$-symmetric $\Zzero$
production term can also be written in terms of the partial decay
widths of the initial and final states, $\Gee$ and $\Gff$,
\begin{equation}\label{eq:sigff}
\sigma_{\ff}^{\rm Z} ~ = ~ \sigma_{\ff}^{\rm peak}
     \frac{s\GZ^2}{(s-\MZ^2)^2+s^2\GZ^2/\MZ^2},
\end{equation}
where
\begin{equation}\label{eq:sigpeak}
\sigma_{\ff}^{\rm peak} ~ = ~ \frac{1}{R_{\mathrm{QED}}}\sigma_{\ff}^0
\end{equation} 
and
\begin{equation}\label{sig0}
\sigma_{\ff}^0 ~ = ~ \frac{12\pi}{\MZ^2}\ \frac{\Gee\Gff}{\GZ^2}.
\end{equation}
The term $1/R_{\mathrm{QED}}$ removes the final state QED correction
included in the definition of $\Gee$.

The overall hadronic cross-section is parametrised in terms of the
hadronic width given by the sum over all quark final states,
\begin{equation}\label{eq:ghad}
\Ghad~=~\sum_{{\rm q \neq t}}^{} \Gamma_{\qq}.
\end{equation}
The invisible width from $\Zzero$ decays to neutrinos,
$\Ginv=\Nnu\Gnn$, where $\Nnu$ is the number of light neutrino
species, is determined from the measurements of the decay widths
to all visible final states and the total width,
\begin{equation}
  \label{eq:def_GZ}
\GZ~=~\Gee+\Gmumu+\Gtautau+\Ghad+\Ginv.
\end{equation}

Because the measured cross-sections depend on products of the partial
widths and also on the total width, the widths constitute a highly
correlated parameter set. In order to reduce correlations among the
fit parameters, an experimentally-motivated set of six parameters is
used to describe the total hadronic and leptonic cross-sections around
the $\Zzero$ peak. These are
\begin{itemize}
\item the mass of the $\Zzero$, $\MZ$;
\item the $\Zzero$ total width, $\GZ$;
\item the ``hadronic pole cross-section'',
  \begin{equation}\label{eq:def_shad}
 \shad~\equiv~{12\pi\over\MZ^2}{\Gee\Ghad\over\GZ^2}; 
  \end{equation}
\item the three ratios
  \begin{equation}\label{eq:def_Rl}
 \Ree~\equiv~\Ghad/\Gee,~\Rmu\equiv\Ghad/\Gmumu 
  ~~{\rm and}~~ \Rtau\equiv\Ghad/\Gtautau.
\end{equation}
If lepton universality is assumed, the last three ratios reduce to a
single parameter:
\begin{equation}
  \label{eq:def_Rl_univ}
  \Rl~\equiv~\Ghad/\Gll,
\end{equation}
where $\Gll$ is the partial width of the Z into one massless charged
lepton flavour.  (Due to the mass of the tau lepton, even with the
assumption of lepton universality, $\Gtautau$ differs from $\Gll$ by
about $\delta_\tau=-0.23\%$.)
\end{itemize}
For those hadronic final states where the primary quarks can be
identified, the following ratios are defined:
\begin{equation}\label{eq:def_Rq}
 \Rqz~\equiv~\Gqq/\Ghad, ~ {\eg} ~ \Rbz ~=~ \Gbb/\Ghad. 
\end{equation}
Experimentally, these ratios have traditionally been treated
independently of the above set, as described in Chapter~\ref{sec:hq}
and Appendix~\ref{sec:lqappendix}.
 
The leading contribution from $\gammaZ$ interference is proportional
to the product of the vector couplings of the initial and final states
and vanishes at $\sqrt{s}=\MZ$, but becomes noticeable at off-peak
energies and therefore affects the measurement of the $\Zzero$ mass.
Because an experimental determination of all quark couplings is not
possible, the $\gammaZ$ interference term in the hadronic final state
is fixed to its predicted $\SM$ value in the analysis.  The
implications of this are discussed in Section~\ref{sec-jhad}.

The six parameters describing the leptonic and total hadronic
cross-sections around the $\Zzero$ peak are determined exclusively
from the measurements of the four LEP collaborations, due to the large
event statistics available and the precise determination of the LEP
collision energy.  In the measurement of $\Rbz$ and $\Rcz$, however,
the greater purity and significantly higher efficiency which SLD
achieved in identifying heavy quarks offset the statistical advantage
of LEP, and yield results with comparable, and in some cases better,
precision.

\subsection{Invisible Width and Number of Neutrinos}

If the $\Zzero$ had no invisible width, all partial widths could be
determined without knowledge of the absolute scale of the
cross-sections.  Not surprisingly, therefore, the measurement of
$\Ginv$ is particularly sensitive to the cross-section scale.
Assuming lepton universality, and defining $\Rinv = \Ginv/\Gll$,
Equations~\ref{eq:def_GZ} and~\ref{eq:def_shad} can be combined to
yield
\begin{equation}\label{eq:Rinv}
\Rinv ~=~ \left(\frac{12\pi \Rl}{\shad \MZ^2}\right)^{\frac{1}{2}} 
 - \Rl - (3+\delta_\tau)\,,
\end{equation}
where the dependence on the absolute cross-section scale is explicit.

\begin{figure}[thb]
\begin{center}
\mbox{\epsfig{file=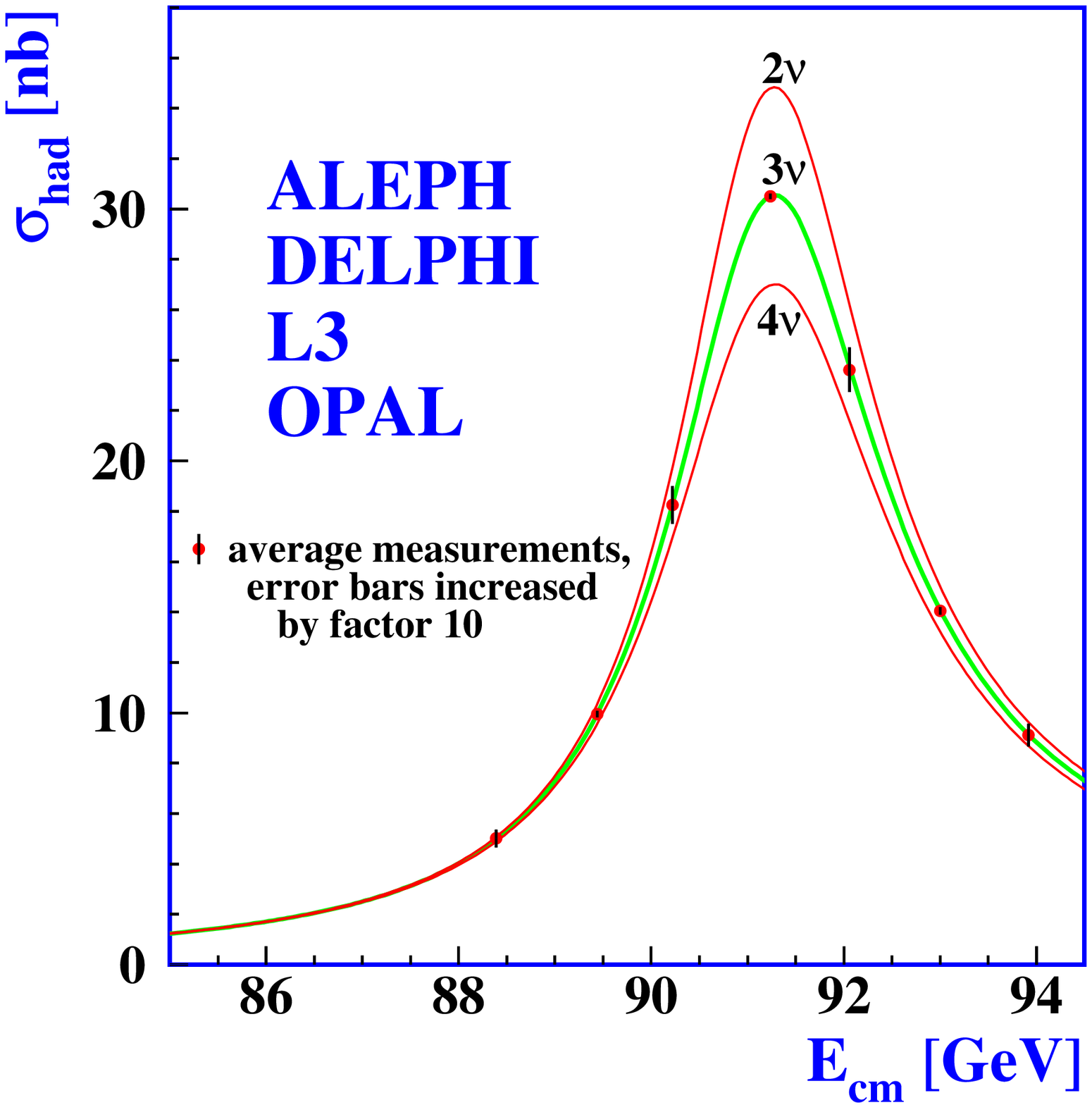,width=0.8\textwidth}} 
\end{center}
\caption[Number of neutrinos]
{\label{fig:Nnu} Measurements of the hadron production cross-section
  around the $\Zzero$ resonance. The curves indicate the predicted
  cross-section for two, three and four neutrino species
  with $\SM$ couplings and negligible mass.}
\end{figure}

Assuming that the only invisible \Zzero\ decays are to neutrinos
coupling according to $\SM$ expectations, the number of light neutrino
generations, $N_\nu$, can then be determined by comparing the measured
$\Rinv$ with the $\SM$ prediction for $\Gnn/\Gll$:
\begin{equation}
\Rinv ~ =  ~N_\nu \left(\frac{\Gnn}{\Gll}\right)_{\rm SM}\,.
\end{equation}
The strong dependence of the hadronic peak cross-section on $N_\nu$ is
illustrated in Figure~\ref{fig:Nnu}.  The precision ultimately
achieved in these measurements allows tight limits to be placed on the
possible contribution of any invisible $\Zzero$ decays originating
from sources other than the three known light neutrino species.

\subsection{Asymmetry and Polarisation}
\label{sec:intro_asymm}

Additional observables are introduced to describe the $\cos\theta$
dependent terms in Equation~\ref{eq:intro_bigugly} as well as effects
related to the helicities of the fermions in either the initial or
final state.  These observables quantify the parity violation of the
neutral current, and therefore differentiate the vector- and
axial-vector couplings of the $\Zzero$. Their measurement determines
$\swsqefff$.

Since the right- and left-handed couplings of the $\Zzero$ to fermions are
unequal, $\Zzero$ bosons can be expected to exhibit a net polarisation
along the beam axis even when the colliding electrons and positrons
which produce them are unpolarised.  Similarly, when such a polarised
$\Zzero$ decays, parity non-conservation implies not only that the
resulting fermions will have net helicity, but that their angular
distribution will also be forward-backward asymmetric.

When measuring the properties of the $\Zzero$ boson, the
energy-dependent interference between the $\Zzero$ and the purely
vector coupling of the photon must also be taken into account.  This
interference leads to an additional asymmetry component which changes
sign across the $\Zzero$-pole.

Considering the $\Zzero$ exchange diagrams and real couplings
only,\footnote{ As in the previous section, the effects of radiative
corrections, and mass effects, including the imaginary parts of
couplings, are taken into account in the analysis.  They, as well as
the small differences between helicity and chirality, are neglected
here to allow a clearer view of the helicity structure.  It is
likewise assumed that the magnitude of the beam polarisation is equal
in the two helicity states.} to simplify the discussion, the
differential cross-sections specific to each initial- and final-state
fermion helicity are:
\begin{eqnarray}
  \label{eq:z_sigll}
\frac{\mathrm{d}\sigma_{\mathrm{Ll}}}{\mathrm{d}\cos\theta}&\propto&\gle^2\glf^2(1+\cos\theta)^2 \\
  \label{eq:z_sigrr} 
\frac{\mathrm{d}\sigma_{\mathrm{Rr}}}{\mathrm{d}\cos\theta}&\propto&\gre^2\grf^2(1+\cos\theta)^2 \\
  \label{eq:z_siglr}
\frac{\mathrm{d}\sigma_{\mathrm{Lr}}}{\mathrm{d}\cos\theta}&\propto&\gle^2\grf^2(1-\cos\theta)^2 \\
  \label{eq:z_sigrl}
\frac{\mathrm{d}\sigma_{\mathrm{Rl}}}{\mathrm{d}\cos\theta}&\propto&\gre^2\glf^2(1-\cos\theta)^2. 
\end{eqnarray}
Here the upper-case subscript of the cross-section defines the
helicity of the initial-state electron, while the lower-case defines
the helicity of the final-state fermion.
Note that the designations "$+$" and "$-$" are sometimes used in place
of "r" and "l", particularly when discussing $\tau$ polarisation.
Due to the point-like nature
of the couplings and the negligible masses of the fermions involved,
the helicity of the anti-fermion is opposite that of the fermion at
each vertex.

From these basic expressions the Born level differential cross-section
for $\Zzero$ exchange only, summed over final-state helicities,
assuming an unpolarised positron beam but allowing polarisation of the
electron beam, is:
\begin{equation}
  \frac{\mathrm{d} \sigma_{\ff}}{\mathrm{d} \cos \theta}~=~\frac{3}{8}\sigma_{\ff}^{\mathrm{tot}} 
  \left[ 
     (1-\Pe \cAe)(1+\cos^2\theta) + 2 (\cAe-\Pe) \cAf  \cos\theta 
  \right] \, .
\label{eq:f_asylr_sig} 
\end{equation}
The electron beam polarisation, $\Pe$, is taken as positive for
right-handed beam helicity, negative for left.
The dependence on the fermion couplings has been incorporated into
convenient asymmetry parameters, $\cAf$:
\begin{eqnarray}
   \cAf & = &
\frac{ \glf^2 - \grf^2}{\glf^2 + \grf^2}
          =  \frac {2 \gvf \gaf} {\gvf^2 + \gaf^2}
          =  2 \frac { \gvf/\gaf} { 1 + (\gvf/\gaf)^2} \, .\label{eq:cA}
\end{eqnarray}
As the third form makes clear, the asymmetry parameters depend only on
the ratio of the couplings, and within the $\SM$ bear a one-to-one
relation with $\swsqefff$.

Although the asymmetry analyses typically utilise maximum likelihood
fits to the expected angular distributions, the simple form of
Equation~\ref{eq:f_asylr_sig} also allows the coefficients of the
$\cos\theta$ and $(1+\cos^2\theta)$ terms to be determined in terms of
the integral cross-sections over the forward or backward hemispheres.
Naturally, at SLC, the two helicity states of the polarised electron
beam also need to be distinguished.

Designating the integrals over the forward and backward hemispheres
with subscripts F and B and the cross-sections for right and left
electron helicities with subscripts R and L, three basic asymmetries
can be measured:
\begin{eqnarray}
  \Afb &=& \frac{\sigmaf - \sigmab}{\sigmaf + \sigmab}  \label{eq:intro_afb_def} \\
    \ALR &= &\frac{\sigmaL - \sigmaR}{\sigmaL + \sigmaR}%
         \frac{1}{\langle | \pole | \rangle}  \label{eq:ALR_exp}  \\
\label{eq:AFBLR_exp}
  \AFBLR  &=& \frac{   ( \sigmaf - \sigmab )_\mathrm{L} -
                     ( \sigmaf - \sigmab )_\mathrm{R}      }
                 {   ( \sigmaf + \sigmab )_\mathrm{L} +
                     ( \sigmaf + \sigmab )_\mathrm{R}      }
\frac{1}{\langle | \pole | \rangle} \, .
\end{eqnarray}
Inspection of Equation~\ref{eq:f_asylr_sig} shows that the
forward-backward asymmetry, $\Afb$, picks out the coefficient $\cAe
\cAf$ in the $\cos\theta$ term, the left-right asymmetry, $\ALR$,
picks out the coefficient $\cAe$ in the $(1+\cos^2\theta)$ term, and
the left-right forward-backward asymmetry~\cite{Blondel:1988gp},
$\AFBLR$, picks out the coefficient $\cAf$ in the $\cos\theta$ term.

The polarisation of a final-state fermion is the difference between
the cross-sections for right- and left-handed final-state helicities
divided by their sum:
\begin{equation}
  \polf~=~{\frac{\mathrm{d} (\sigmar- \sigmal)}{\mathrm{d} \cos \theta}} \left/ 
                      {\frac{\mathrm{d} (\sigmar+\sigmal)}{\mathrm{d} \cos \theta}} \right. \, .
\end{equation}
At Born level the numerator and denominator
can be derived from the helicity-specific cross-sections of
Equations~\ref{eq:z_sigll} to~\ref{eq:z_sigrl}:
\begin{eqnarray}
  \frac{\mathrm{d} (\sigmar- \sigmal)}{\mathrm{d} \cos \theta} &=&
 - \frac{3}{8}\sigma_{\ff}^{\mathrm{tot}}   \left[ 
      \cAf (1+\cos^2\theta) + 2 \cAe  \cos\theta 
  \right]  \label{eq:helicty-diff}   \\
\frac{\mathrm{d} (\sigmar+ \sigmal)}{\mathrm{d} \cos \theta} &=&
 \phantom{-} \frac{3}{8}\sigma_{\ff}^{\mathrm{tot}}   \left[ 
      (1+\cos^2\theta) + 2 \cAe \cAf  \cos\theta 
  \right] \, .
\label{eq:helicty-sum} 
\end{eqnarray}
Here we assume $\Zzero$ exchange only, and unpolarised beams.
The average final-state fermion polarisation, $\langle \polf \rangle$,
as well as the forward-backward polarisation asymmetry, $\AFBpol$, can
be found in terms of the helicity cross-sections integrated over the
forward and backward hemispheres:
\begin{eqnarray}
   \langle \polf \rangle &= &\frac{\sigmar - \sigmal}{\sigmar + \sigmal}  \label{eq:pol_f_exp}  \\
\label{eq:afbpol_exp}
   \AFBpol  &=& \frac{   ( \sigmar - \sigmal )_\mathrm{F} -
                     ( \sigmar - \sigmal )_\mathrm{B}      }
                 {   ( \sigmar + \sigmal )_\mathrm{F} +
                     ( \sigmar + \sigmal )_\mathrm{B}      }  \, .
\end{eqnarray}
Again, examination of Equations~\ref{eq:helicty-diff}
and~\ref{eq:helicty-sum} shows that $\langle \polf \rangle$ picks out
the coefficient $\cAf$ in the $(1+\cos^2\theta)$ term and $ \AFBpol$
picks out the coefficient $\cAe$ in the $\cos\theta$ term.

The net polarisation of a final-state fermion as a function of
$\cos\theta$ is simply the ratio of Equations~\ref{eq:helicty-diff}
and~\ref{eq:helicty-sum}:
\begin{equation}
\label{eq-pfcos}
\polf(\cos\theta) ~ = ~ - \frac
{\cAf(1+\cos^2\theta) + 2\cAe \cos\theta}
{(1+\cos^2\theta) + 2\cAf\cAe\cos\theta}\,.
\end{equation}
Since the polarisation of the final-state fermion can only be measured
in the case of the $\tau$-lepton, which decays in a parity violating
manner within the detectors, these quantities are measured only for
the final state $\tautau$.  As in the case of the other asymmetries, a
maximum-likelihood fit to Equation~\ref{eq-pfcos} is used in the
actual $\tau$ polarisation analyses to extract both $\langle \ptau
\rangle$ and $ \AFBpol$, rather than using the simpler integral
expressions of Equations~\ref{eq:pol_f_exp} and~\ref{eq:afbpol_exp}.

The measured asymmetries and polarisations are corrected for radiative
effects, $\gamma$ exchange and $\gammaZ$ interference to yield
``pole'' quantities designated with a superscript 0.  In the case
where the final state is $\ee$, important corrections for $t$-channel
scattering must also be taken into account.
QED corrections~\cite{Boehm:LEP1YR89VOL1} to $\Afbl$
are as large as the value of the asymmetry itself, and must be
understood precisely (see Section~\ref{sec-therr}).
Off-peak, the contributions from $\gammaZ$
interference to the forward-backward asymmetries become even larger.
The corrections to $\ALR$, $\AFBLR$, $\langle \ptau \rangle$ and
$\AFBpol$ are relatively small.

At LEP the forward-backward asymmetries, $\Afbze$, $\Afbzm$, $\Afbzt$
and $\Afbzq$ are measured for final states $\ee$, $\mumu$, $\tautau$
and $\qq$.  Tagging methods for b- and c-quarks allow $\qq$
forward-backward asymmetries for these flavours to be measured
precisely.  All four LEP experiments measure $\ptau$.

SLD measures the asymmetries involving initial-state polarisation.
The left-right asymmetry, $\ALRz$, is independent of the final state,
and the measurement is dominated by $\eeqq$.  Despite the smaller
event sample available to SLD, the measurement of $\ALRz$ provided the
single most precise determination of the initial state coupling (Z to
electron).  SLD also measures $\AFBLRz$ for each of the final states
$\ee$, $\mumu$, $\tautau$ and $\qq$, where q includes not only b- and
c-quarks, but also s-quarks.

In contrast to the partial widths, which are defined using the full
complex couplings in order to ensure that the sum over all partial
widths equals the total width, the pole asymmetries are defined purely
in terms of the real parts of the effective $\Zzero$ couplings, and
bear particularly direct relationships to the relevant asymmetry
parameters:
\begin{eqnarray}
 \Afbzf  & = & \phantom{-}{3\over 4} \cAe\cAf  \label{eq:afbzf}\\ 
  \ALRz &=&  \phantom{-\frac{3}{4}}\cAe   \label{eq:alrpole}\\
  \label{eq:afblr}
  \AFBLRz  & = &  \phantom{-}\frac{3}{4}\cAf  \\
  \label{eq:ptau}
\langle \ptau^{0} \rangle & = & -\phantom{\frac{3}{4}} \cAt  \\
  \label{eq:afbpol}
  \AFBpolz & = & - \frac{3}{4}\cAe \, .
\end{eqnarray}
The negative sign of the quantities involving the polarisation is
simply a consequence of defining the polarisation of a right-handed
fermion as positive in a world in which left-handed couplings
dominate.  It should be noted that although the pole asymmetries are
defined in terms of only the real parts of the couplings, the complex
parts are taken into account when correcting the measurements to yield
pole quantities.

Using the measurements of $\cAe$, the parameters $\cAm$, $\cAt$,
$\cAb$ and $\cAc$ can also be inferred from forward-backward asymmetry
measurements at LEP via Equation~\ref{eq:afbzf}.  Thus, the LEP and
SLC results form a complementary and practically complete set of
$\cAf$ measurements.

\begin{figure}[bt]
\begin{center}
\mbox{\epsfig{file=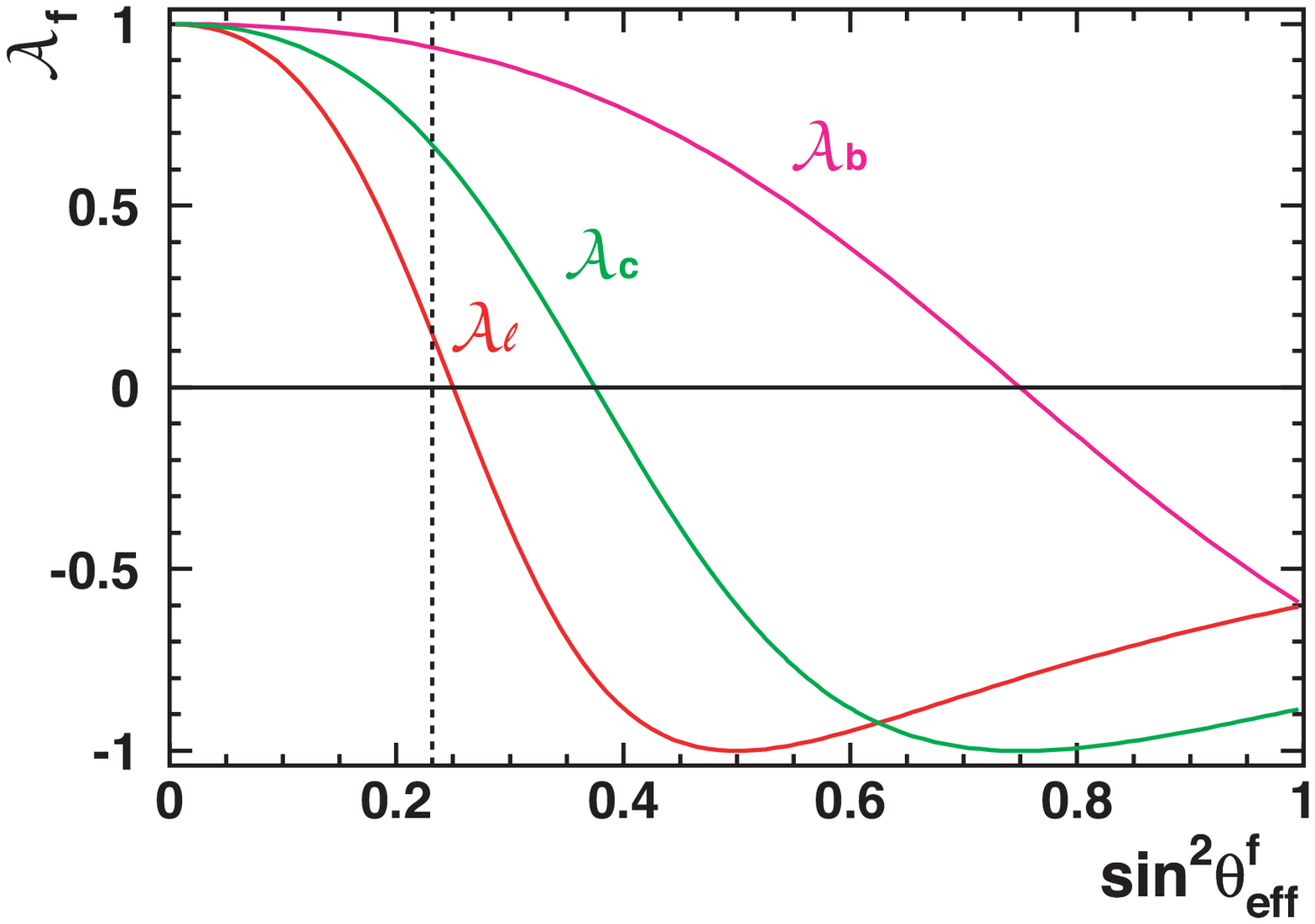,width=0.9\textwidth}} 
\end{center}
\vskip -0.75cm
\caption[$\cAf$ as a function of $\swsqefff$] {\label{fig:afvssin2} In
the $\SM$ the variation of $\cAf$ with $\swsqefff$ is controlled by
the charge and weak isospin assignment of the fermion species
concerned. The measured values of $\swsqefff$ are near the vertical
line. In this region, $\cAl$ depends strongly on $\swsqeffl$, while
$\cAb$ depends much more weakly on $\swsqeffb$.}
\end{figure}

When the couplings conform to the $\SM$ structure, then
\begin{eqnarray}
\frac {\gvf}{ \gaf} & = & 1 -  \frac {2 \Qf} {\Tf} \swsqefff
                    ~ = ~ 1 -  4|\Qf|  \swsqefff   \, , 
\label{eq:gvfovergaf}
\end{eqnarray}
and the expected variation of $\cAf$ with $\swsqefff$ is shown in
Figure~\ref{fig:afvssin2}.
Due to the proximity of $\swsqefff$ to $1/4$, $\cAl$ and the leptonic
forward-backward asymmetries at $\sqrt{s}=\MZ$ are small, but very
sensitive to $\swsqefff$.  Compared with the leptons, the coupling
parameters of the quarks in the $\SM$ are determined more by their
charge and weak isospin assignments than by the value of
$\swsqefff$. For down-type quarks, as can be seen from
Figure~\ref{fig:afvssin2}, the relative sensitivity of $\cAq$ to
changes in $\swsqeffq$ is a factor of almost 100 less than it is for
$\cAl$. It is therefore of particular interest to compare the
relatively static $\SM$ prediction for $\cAq$ with measurement.  On
the other hand, if the $\SM$ prediction for $\cAq$ is assumed to be
valid, the observed forward-backward asymmetries for quarks provide a
sensitive measurement of $\swsqeffl$ via Equation~\ref{eq:afbzf}.

\subsection{Relating Theory and Experiment
\label{sec:intro_SM_remnants}}

The parameters introduced in the preceding subsections, which describe
the main features of all measurements around the $\Zzero$ resonance,
are not ``realistic observables'' like the underlying measurements
themselves, but are defined quantities with significant theoretical
corrections.  Therefore they are commonly named pseudo-observables.
Where necessary, the pseudo-observables are denoted by a superscript
0; for example, $\sigma_{\mathrm{had}}$ is the measured hadronic
cross-section, whereas $\shad$ is the pole cross-section derived from
the measurements.  Similarly, $\Rb$ is the measured b-quark
cross-section divided by the hadronic cross-section,
$\sigma_\bb/\sigma_{\mathrm{had}}$, while $\Rbz$ is the derived ratio
of Z boson partial widths, $\Gbb/\Ghad$.

In the $\Zzero$ lineshape analysis the true realistic observables are
the experimental cross-sections and asymmetries measured in the
acceptances particular to each detector.  Before these can be further
analysed, each collaboration applies small corrections to extrapolate
them to more generic, idealized acceptances, as described in
Section~\ref{xsec:meas}.

The programs TOPAZ0 and ZFITTER are able to calculate the
cross-sections measured within these idealized acceptances, including
the effects of QED radiation, as a function of the set of nine
pseudo-parameters chosen to describe the observable features of the
$\Zzero$ resonance in a model-independent manner.  It is important to
realize that the bulk of the radiative corrections necessary to
interpret the real observables in terms of the pseudo-observables are
QED effects distinct from the deeper electroweak corrections which
modify the relations between the pseudo-parameters in the context of
any particular model, such as the $\SM$.  Further details are
discussed in Section~\ref{sec:msm:TU1}.
 
After these QED effects which depend in a model-independent manner on
the resonance properties of the $\Zzero$ have been accounted for, the
remaining differences between the pseudo-observables and the QED
deconvoluted observables at $\sqrt{s}=\MZ$ are attributable to
non-factorisable complex components, termed ``remnants'', of the
couplings $\cgaf$ and $\cgvf$ and of $\alpha(\MZ^2)$ in
Equation~\ref{eq:intro_bigugly}. These effects are found to be small
in the $\SM$.  For example, the calculated value of
$\sigma_{\ff}^{0}$, given in terms of the partial decay widths, agrees
to better than 0.05\% for both hadrons and leptons with the QED
deconvoluted cross-sections without the photon exchange contribution
at $\sqrt{s}=\MZ$. This is only a fraction of the LEP combined
experimental error. The difference between $\Afbzl$ and the QED
deconvoluted forward-backward asymmetry at the peak is dominated by a
contribution of 0.0015 from the imaginary part of $\alpha(\MZ^2)$,
which accounts, via the optical theorem, for the decay of a massive
photon to fermion pairs.  The remaining electroweak contribution in
the $\SM$ is $-0.0005$, again smaller than the LEP combined error on
$\Afbzl$.

It is therefore important to treat these complex parts correctly, but
the measurements have no sensitivity to $\SM$ parameters entering
through these components: the effects on the remnants are much
smaller than the experimental uncertainties.

Since one of the main goals of the \Zzero-pole analysis is to test
theory with experimental results, considerable effort has been
expended to make the extraction of the pseudo-observables describing
the $\Zzero$ resonance as model-in\-de\-pen\-dent as possible, so that
the meanings of ``theory'' and ``experiment'' remain distinct.  Since
the pseudo-observables do depend slightly on $\SM$ assumptions, as
explained above, a more precise definition of what we mean by
``model-in\-de\-pen\-dence'' is that our analysis is valid in any
scenario in which the predicted remnants remain small.  The very small
uncertainties arising from ambiguities in the theoretical definition
of the pseudo-observables are discussed in Section~\ref{sec-paramLS},
and quantified in Table~\ref{tab:therr}.

In the same spirit, the contribution of the 4-fermion process $\ee
\rightarrow \Zzero \rightarrow \Zzero^*\mathrm{H}
\rightarrow\ff\mathrm{H}$ entering the fermion-pair samples used for
analysis should be negligible.  The limit of $\MH >
114.4~\GeV$~\cite{LEPSMHIGGS} established by the direct search for the
Higgs boson at $\LEPII$ ensures that this is in fact the case.  Only
when hypothetical Higgs masses well below the experimental limit are
considered in the course of exploring the full parameter-space of the
$\SM$ must allowances be made for the treatment of such ZH
contributions~\cite{Kawamoto:2004pi}, both in the experimental
analyses and in the theoretical calculations.

\section{Interpretation and Impact of the Results}
\label{sec:intro_impact}

This paper aims to be an authoritative compendium of the properties of
the $\Zzero$ boson derived from precise electroweak measurements
performed at \LEPI{} and SLC.  These properties, based on $\chi^2$
combinations~\cite{\BLUE} of the results of five
experiments described in detail in this paper, are largely independent
of any model, and represent a comprehensive distillation of our
current knowledge of the $\Zzero$ pole.

Since these observed properties are found to be in good agreement with
expectations of the $\SM$, we leave theoretical speculations which go
beyond the $\SM$ context to others.  We first focus on comparing the
\Zzero-pole data with the most fundamental $\SM$ expectations (lepton
universality, consistency between the various manifestations of
$\swsq$, etc.).

We then assume the validity of the $\SM$, and perform fits which
respect all the inter-relationships among the measurable quantities
which it imposes.  These fits find optimum values of the $\SM$
parameters, and determine whether these parameters can adequately
describe the entire set of measurements simultaneously.  At first we
restrict the set of measurements to the \Zzero-pole results presented
here, and later extend the analysis to a larger set of relevant
electroweak results, including the direct measurements of the top
quark and W boson masses.  This expanded set of measurements yield the
narrowest constraints on the mass of the only particle of the $\SM$
not yet observed: the Higgs boson.

\begin{figure}[hbtp]
\begin{center}
\includegraphics[width=0.9\textwidth]{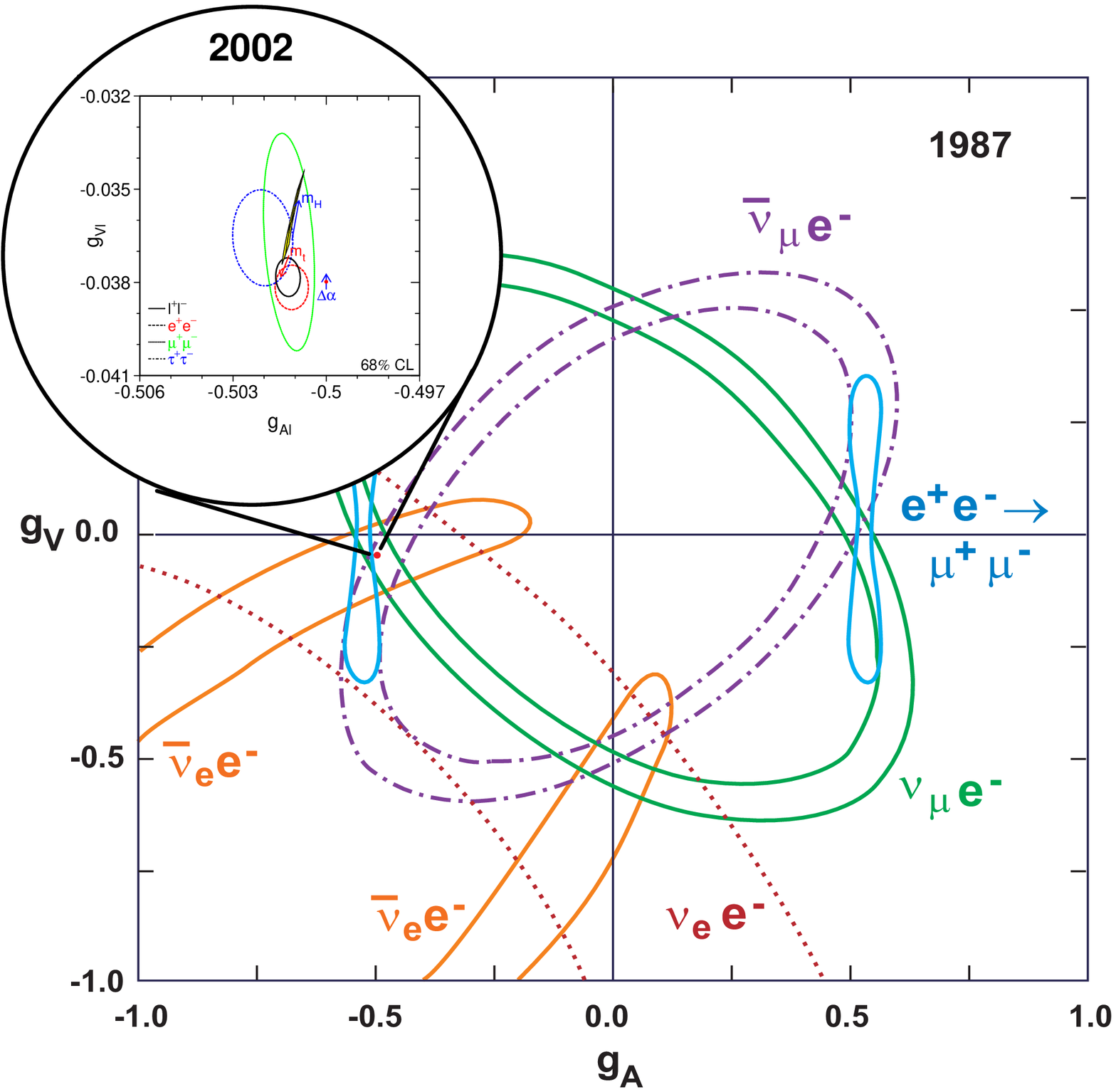}
\caption[The 1987 status of $\gvl$ and $\gal$.]{The neutrino scattering
  and $\ee$ annihilation data available in 1987 constrained the values
  of $\gvl$ and $\gal$ to lie within broad bands, whose intersections
  helped establish the validity of the $\SM$ and were consistent with
  the hypothesis of lepton universality.  The inset shows the results
  of the LEP/SLD measurements at a scale expanded by a factor of 65
  (see Figure~\ref{fig:coup:gl}).
  The flavour-specific measurements demonstrate the universal nature
  of the lepton couplings unambiguously on a scale of approximately
  0.001.}
\label{fig:gvga_preLEP}
\end{center}
\end{figure}

The LEP/SLC era represents a decade of extraordinary progress in our
experimental know\-ledge of electroweak phenomena.  It is the goal of
the remainder of this paper to demonstrate in detail how the LEP/SLD
measurements confront the theory of the $\SM$ much more precisely than
previous experiments.  The mass of the Z is now one of the most
precisely known electroweak parameters, and will long serve as an
important reference for future investigations.  The strong constraint
on the number of light neutrinos, implying that there are only three
``conventional'' generations of particles, is of particular
significance for astrophysics and cosmology.  An illustration of the
improved knowledge of the properties of the \Zzero, in addition to the
precise measurements of its mass, width and pole production
cross-section, is shown in Figure~\ref{fig:gvga_preLEP}, comparing the
$\gvl$ and $\gal$ measurements before and after the LEP and SLC
programmes.  The small dot in the 1987 plot shows the true scale of
the circle enclosing the 2002 inset.

\begin{figure}[hbtp]
\begin{center}
\includegraphics[width=0.9\textwidth]{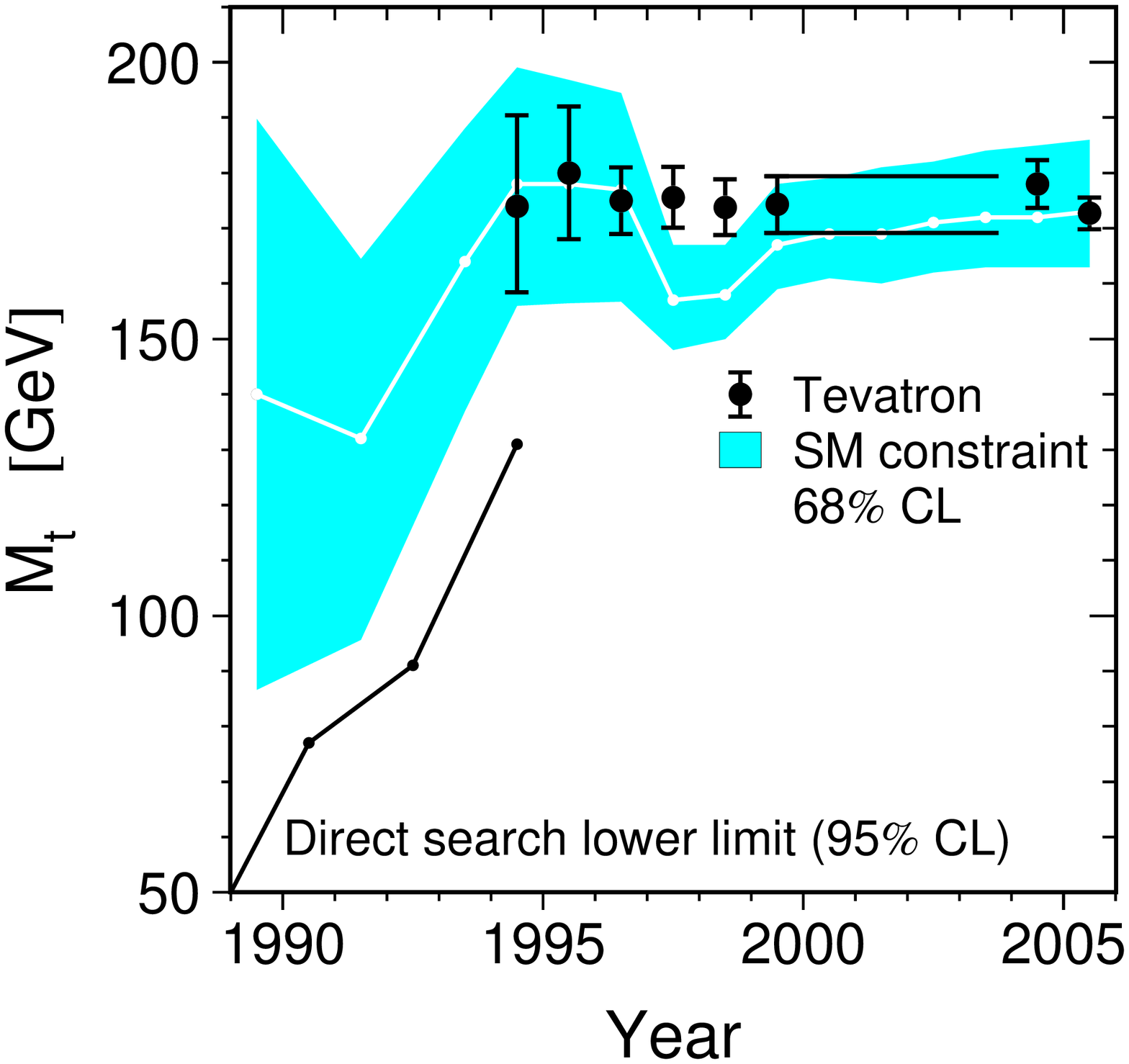}
\caption[Predictions and measurements of $\Mt$.]{Comparison of direct
and indirect determinations of the mass of the top quark, $\Mt$, as a
function of time.  The shaded area denotes the indirect determination
of $\Mt$ at 68\% confidence level derived from the analysis of
radiative corrections within the framework of the $\SM$ using
precision electroweak measurements. The dots with error bars at 68\%
confidence level denote the direct measurements of $\Mt$ performed by
the Tevatron experiments CDF and D\O.  Also shown is the 95\%
confidence level lower limit on $\Mt$ from the direct searches before
the discovery of the top quark. Predictions and measurements agree
well.}
\label{fig:history_mt}
\end{center}
\end{figure}

The good agreement between the top quark mass measured directly at the
Tevatron and the predicted mass determined indirectly within the $\SM$
framework on the basis of measurements at the \Zzero-pole, shown in
Figure~\ref{fig:history_mt}, is a convincing illustration of the
validity of $\SM$ radiative corrections and stands as a triumph of the
electroweak $\SM$.

\chapter{The Z Lineshape and the Leptonic Forward-Backward 
  Asymmetries\label{chap:lsafb}}

\section{Introduction}

The measurements described in this chapter are designed to determine
the essential parameters of the $\Zzero$ resonance, its mass, its
width, its branching fractions, and the angular distribution of its
decay products. Specifically, the nine parameters $\MZ$, $\GZ$,
$\shad$ as well as $\Rl$ and $\Afbzl$ for each of the three charged
lepton species, as defined in Section~\ref{sec:intro_xrpar}, provide a 
complete (hadron-inclusive) description of the $\Zzero$ resonance.
The mass of the $\Zzero$ is a central parameter of the Standard Model
($\SM$).  Because of the LEP programme, $\MZ$ is now measured with a
relative precision of $2.3\cdot10^{-5}$, and thus represents one of
the most precisely known parameters of the $\SM$. Together with the
Fermi constant $\GF$, known to a precision of $0.9\cdot10^{-5}$, both
$\GF$ and $\MZ$ currently act as two fixed points of the $\SM$, around
which all other quantities are forced to find their place.

The role of the total width $\GZ$ is of similar importance.  As can be
seen from Equation~\ref{eq:Gff}, the width of the $\Zzero$ to each of
its decay channels is proportional to the fundamental Z-fermion
couplings.  The total width $\GZ$ is in fact the only Z-pole
observable in the experimentally motivated nine-parameter set from
which the absolute scale of the couplings can be determined: Since
$\GZ$ is large compared to the energy spread of the colliding beams at
LEP, it does not manifest itself in terms of the apparent peak
cross-section\footnote{ The peak cross-sections would in fact remain
constant if the couplings to all final states increased or decreased
proportionally, see Equation~\ref{sig0}.}, as is the case for a narrow
resonance like the J/$\Psi$, but in terms of the measurable width of
the lineshape as the beam energy is scanned across the resonance.  In
order to determine $\GZ$, off-peak data are thus needed in addition to
peak data, as is the case for the measurement of $\MZ$.  The beam
energies of this off-peak running were carefully tuned to optimise the
precision of the measurement, and focused on a small set of
centre-of-mass energies within $\pm 3~\GeV$ around $\sqrt{s}=\MZ$.
Even after all four experiments have been combined, the dominant error
in $\GZ$ is statistical, rather than systematic.

Since the Z is expected to decay only to fermion pairs, the number of
partial decay widths to be determined is small.  The decision to treat
all Z decays to quarks as a single inclusive hadronic decay channel in
the lineshape analysis further limits the number of partial widths to
a very manageable number.  Since some of the very properties of the
hadronic decays which make the identity of the primary quarks
difficult to determine also make the experimental acceptance
quark-flavour independent, the attraction of a precise inclusive
hadronic analysis is obvious.  Separation of the primary quarks and
the determination of their couplings is therefore left to the
specialised analyses described in Chapter~\ref{sec:hq} and
Appendix~\ref{sec:lqappendix}, employing dedicated flavour and charge
tagging techniques. The expected approximate branching fractions of
the $\Zzero$ are 70\,\%, 20\,\% and 10\,\% to hadrons, neutrinos and
charged leptons, respectively. The statistical dominance of the
hadronic decays makes them decisive in determining the fundamental
parameters $\MZ$ and $\GZ$.

Due to the tight linkage between pole cross-sections, branching ratios
and partial widths implied by Equation~\ref{sig0} and the constraint
that the sum of all partial widths should equal the total width, the
parameters $\GZ$, $\shad$ and the three hadron/lepton species ratios,
$\Rl=\Ghad/\Gll$, were chosen as a less-correlated representation of
the complete set of five partial widths.  Although $\Zzero$ decays to
neutrinos escape direct detection, and are therefore referred to as
``invisible decays'', the corresponding Z decay width can be derived
from the other parameters, according to the relation described in
Equation~\ref{eq:Rinv}. Therefore the observed peak cross-sections
depend strongly on the number of existing neutrino generations, as
already shown in Figure~\ref{fig:Nnu}. The precision ultimately
achieved in the determination of the number of neutrinos thus hinges
on a precise absolute cross-section measurement, requiring a precise
determination of the integrated luminosity and an accurate calculation
of QED radiative corrections.

The spin-1 nature of the $\Zzero$ is well substantiated by the
observed $1+\cos^{2} \theta$ angular distribution of its decay
products. The $\cost$ terms of the angular decay distributions,
varying as a function of energy due to $\gammaZ$ interference,
determine the three leptonic pole forward-backward asymmetries,
$\Afbzl$.  The violation of parity conservation in $\Zzero$ production
and decay, which is most precisely quantified by the analyses of
Chapters~\ref{sec-ALR} to~\ref{sec:hq}, is evident from the non-zero
values of these three measured leptonic pole forward-backward
asymmetries.

The full LEP-I data set relevant to this analysis consists of about
200 measurements from each experiment of hadronic and leptonic cross-sections
and of leptonic forward-backward asymmetries at different
centre-of-mass energies. Although this complete set of basic
measurements carries all available experimental information on the
$\Zzero$ resonance parameters, the construction of the overall error
matrix describing all the inter-experiment correlations is too complex 
a task in practice.  Instead, each experiment has independently extracted the
agreed-upon set of nine pseudo-observables discussed above in single
multi-parameter fits to all their measurements of cross-sections and
forward-backward asymmetries.  The electroweak libraries used for this
extraction are TOPAZ0~\cite{\TOPAZref} and ZFITTER~\cite{\ZFITTERref},
which include QED and QCD corrections necessary to extract the
pseudo-parameters in a model-independent manner as well as those
electroweak corrections according to the $\SM$ which can only be
described by the imaginary parts of the $\Zzero$ couplings, as
discussed in Section~\ref{sec:intro_SM_remnants}.

The main task of the analysis undertaken here is to combine the
resulting four sets of pseudo-observables with an appropriate
treatment of common errors and especially the recognition that
re-weighting of particular datasets will occur when the balance of
statistical and systematic errors changes under the act of
combination. Much of this work involves novel techniques which were
specially developed for this analysis.

After a brief description of the key features of the experimental
analyses (Section~\ref{sec-lsafb}) and the presentation of the
individual results (Section~\ref{sec-inpdata}), the main emphasis in
the following sections is given to the hitherto unpublished aspects of
the combination procedure, namely the errors common to all experiments
(Section~\ref{sec-comerr}) and the combination procedure
(Section~\ref{sec-lsafbcombi}).  Essential cross-checks of the general
validity of the combination procedure are also discussed in this
section, which is followed by the presentation of the combined
results. Re-parametrisations in terms of partial widths and branching
fractions will be given later (see Section~\ref{chap:partrafo}).

\section{Measurements of Total Cross-Sections and Forward-Backward 
  Asymmetries\label{sec-lsafb}}

The main features of the event selection procedures for measurements
of the total hadronic and leptonic cross-sections and of the leptonic
forward-backward asymmetries are briefly described below. Detailed
descriptions of the individual experimental analyses are given in the
References~\cite{\ALEPHls,\DELPHIls,\Lls,\OPALls}.

\subsection{Event Selection \label{sec-evssel}}

The event selection for $\qq$, $\ee$, $\mumu$ and $\tautau$ final
states in each of the experiments is aimed at high selection
efficiencies within the largest possible acceptance in order to keep
corrections small.

The design of the detectors and the cleanliness of the LEP beams
allowed the experiments to trigger on hadronic and leptonic $\Zzero$
decays with high redundancy and essentially 100\% efficiency.  The
selections are as open as possible to events with initial and final
state radiation in order to benefit from cancellations between real
and virtual particle emission.  Good discrimination of $\qq$ from
$\ell^+\ell^-$ final states is mandatory for the analyses, and
excellent separation of $\ee$, $\mumu$ and $\tautau$ permits checks of
the universality of the $\Zzero$ couplings to the different lepton
species to be carried out. Machine-induced backgrounds at LEP-I were
small, and the only significant source of background from $\ee$
processes comes from two-photon reactions. The accumulated event
statistics are given in Table~\ref{tab:LSstat}, and event pictures of
each of the final states are shown in Figure~\ref{fig:intro_events} in
Chapter~\ref{sec:intro}.

\begin{figure}[thb]
\begin{center}
\mbox{\epsfig{file=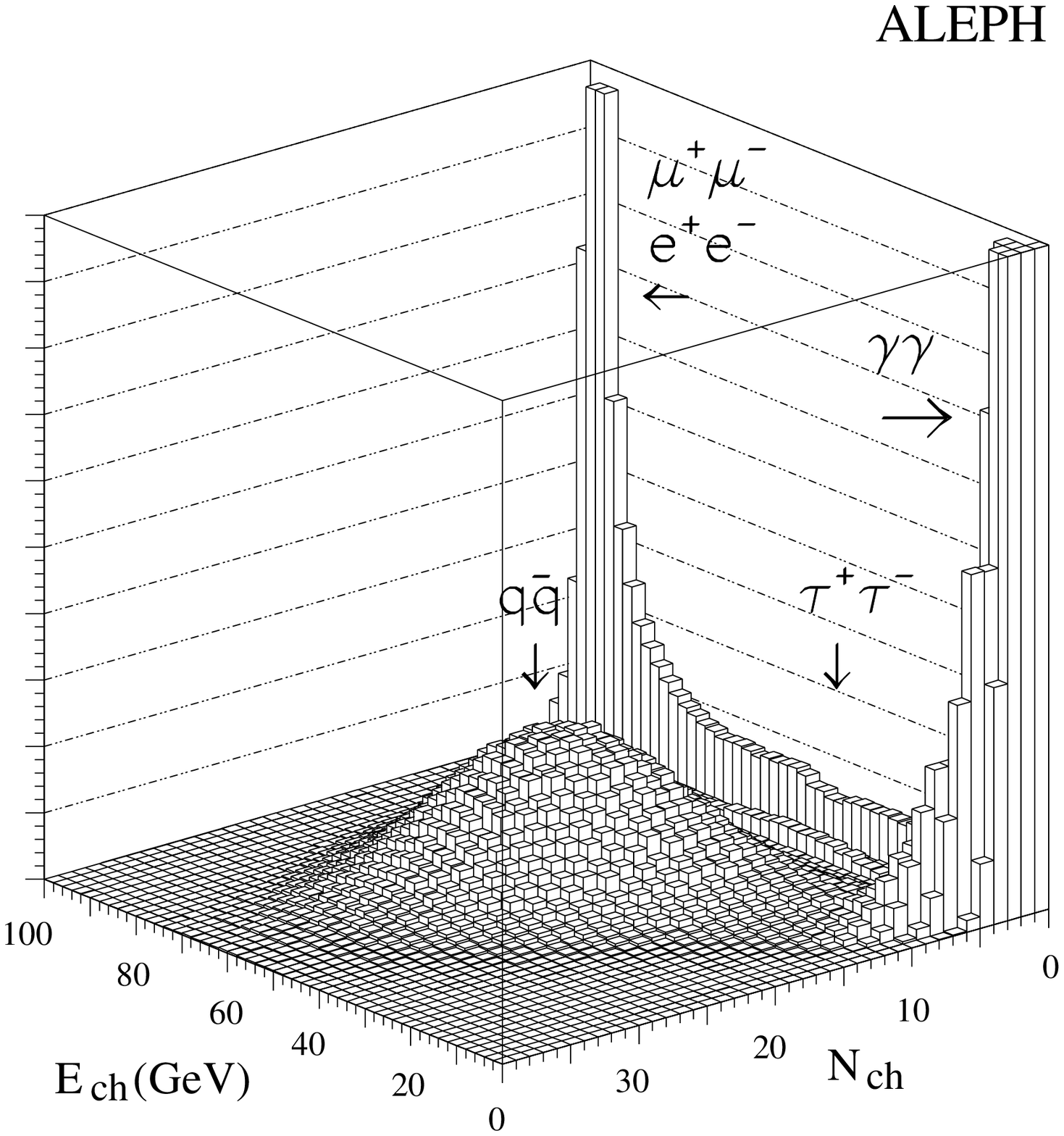,width=0.75\textwidth}} 
\end{center}
\caption[Separation of final states]
{\label{fig:EvsN} Experimental separation of the final states using
  only two variables, the sum of the track momenta, $E_{ch}$, and the
  track multiplicity, $N_{ch}$, in the central detector of the ALEPH
  experiment. }
\end{figure}

The principles used to separate leptonic and hadronic events and to
distinguish two-photon reactions are illustrated in
Figure~\ref{fig:EvsN}. A peak from $\ee$ and $\mumu$ events at high
momenta and low multiplicities is clearly separated from the
background of two-photon reactions at relatively low multiplicities
and momenta.  The intermediate momentum region at low multiplicities
is populated by $\tautau$ events.  The separation of electrons and
muons is achieved using also the information from the electromagnetic
and hadron calorimeters and from the muon chambers.  Hadronic events
populate the high multiplicity region at energies below the
centre-of-mass energy, since neutral particles in the jets are not
measured in the central detector.

In somewhat more detail, hadronic events in the detectors are
characterised by a large number of particles arising from the
hadronisation process of the originally produced quark pair.  This
leads to high track multiplicities in the central detectors and high
cluster multiplicities in the electromagnetic and hadron calorimeters.
For $\Ztoqq$ events, the deposited energy is balanced along the beam
line, which is generally not the case for hadronic events produced in
two-photon reactions.  In addition, two-photon collision events have
an almost constant production cross-section around the $\Zzero$
resonance.  It is thus possible to estimate the fraction of two-photon
reactions directly from the data by studying the energy dependence of
two event samples, one with an enriched contribution of two-photon
reactions and another with tight selection cuts for genuine $\Ztoqq$
events, which show a resonant behaviour.  Background from $\tautau$
events is subtracted using Monte Carlo simulation.

Lepton pairs are selected by requiring low track and cluster
multiplicities. Electrons are characterised by energy deposits in the
electromagnetic calorimeters that match well the measured momenta in
the tracking detectors.  Muons exhibit only minimum ionising energy
deposits in the electromagnetic and hadron calorimeters and produce
signals in the outer muon chambers.  Tau leptons decay before reaching
any detector component. Their visible decay products are either a
single electron, muon or hadron, or a collimated jet consisting of
three or five charged hadrons and a few neutral hadrons; in addition
energy is missing due to the undetectable neutrinos.  $\tautau$ events
are therefore selected by requiring the total energy and momentum sums
to be below the centre-of-mass energy to discriminate against $\Ztoee$
and $\Ztomumu$, and to be above a minimum energy to reject lepton
pairs arising from two-photon reactions. The direction of flight of
the $\tau$ is approximated by the momentum sum of the visible decay
products. Leptonic events with photons or fermion pairs radiated from
the initial- or final-state leptons are contained in the signal
definition. Initial-state pairs typically remain in the beam pipe and
are therefore experimentally indistinguishable from initial state
photon radiation. The classification of final states with radiated
fermion pairs, {\em i.e.} of four-fermion events, into one of the
three lepton categories is made by choosing the lepton pair with the
highest invariant mass.

The experiments use very detailed detector
simulations~\cite{GEANT,bib-detsimo} to understand the selection
efficiencies. Owing to the high redundancy of the detectors,
cross-checks and corrections using the actual data are possible by
comparing event samples identified with different selection criteria.
Various Monte Carlo generators are interfaced to the detector
simulations and are used to describe the kinematics of the physics
reactions of interest: $\qq$ production with gluon radiation including
phenomenological modelling of the non-perturbative hadronisation
process~\cite{JETSET,HERWIG,ARIADNE}, production of $\mumu$ and
$\tautau$ final states~\cite{KORALZ,KK}, $\ee$ final states including
the $t$-channel
contribution~\cite{BABAMC,Field:1996dk,Jadach:1997nk,UNIBAB}, and
finally $\ee$ scattering in the forward direction~\cite{BHLUMI4}, which
is dominated by $t$-channel photon exchange and serves as the
normalisation reaction in determining the luminosity of the colliding
$\ee$ beams. The effects of fermion pair radiation in the final-state
are studied using four-fermion event generators~\cite{FERMISV,grc4f}.

The Monte Carlo generators are used to apply corrections at the edges
of the experimental acceptance, and for small extrapolations of the
measured cross-sections and forward-backward asymmetries from the true
experimental cuts to sets of simple cuts that can be handled at the
fitting stage.  In the case of $\qq$ final states, this ideal
acceptance is defined by the single requirement $s'>0.01\,s$, where
$\sqrt{s'}$ is the effective centre-of-mass energy after initial-state
photon radiation.  The idealised acceptances chosen for each lepton
decay channel vary among the experiments and are specified in
Table~\ref{tab:LSeff}. The results quoted for the $\ee$ final
state either include contributions originating from $t$-channel
diagrams, or the $t$ and $s$-$t$ interference effects are explicitly
subtracted, allowing the same treatment of $\ee$ and $\mumu$ or
$\tautau$ final states in the fits for the $\Zzero$ parameters.

\subsection{Cross-Section Measurements}
\label{xsec:meas}

The total cross-section, $\sigma_{\rm tot}$, is determined from the
number of selected events in a final state, $N_{\rm sel}$, the number
of expected background events, $N_{\rm bg}$, the selection efficiency
including acceptance, $\epsilon_{\rm sel}$, and the integrated
luminosity, $\calL$, according to $\sigma_{\rm tot}=(N_{\rm
sel}-N_{\rm bg})/(\epsilon_{\rm sel} \calL )~. $

\subsubsection{Measurement of Luminosity}

The luminosity of the beams is measured~\cite{\LUMImeas} from the
process of small-angle Bhabha scattering. Further information is
available in the lineshape
publications~\cite{\ALEPHls,\DELPHIls,\Lls,\OPALls}.  Events with
forward-going electrons are recorded concurrently with all other
processes, thus ensuring that they correctly reflect any data-taking
inefficiencies arising from readout deadtimes and detector
downtimes.  Furthermore, the statistical precision of this process is
high, matching well even the high statistics of hadronic events at the
$\Zzero$ resonance. The luminosity measurement requires the detection
of back-to-back energy deposits by electrons and positrons close to
the beam direction. Their positions and energies are measured by
calorimeters placed at small angles with respect to the beam line,
typically covering a range in polar angle from 25~mrad to 60~mrad.
Depending on the experiment, the accepted cross-section in the
luminosity devices is at least twice as large as the hadronic on-peak
cross-section, and therefore the statistical errors arising from the
luminosity determination are small.
\begin{figure}[bth]
\begin{center}
\mbox{\epsfig{file=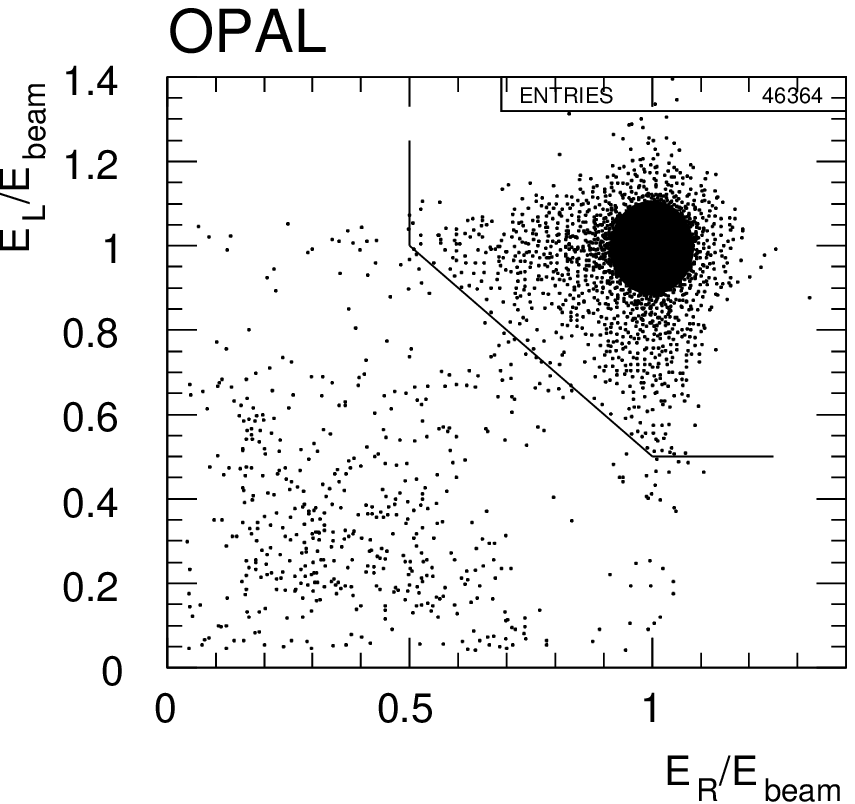,width=0.75\textwidth}} 
\end{center}
\caption[Luminosity measurement]
{\label{fig:lumi} Fraction of the beam energy observed in the left and
  right luminosity calorimeters of the OPAL experiment, after all cuts
  except the one on the deposited energies. The lines indicate the
  acceptance region for the signal events.  Initial state photon
  radiation leads to tails towards lower deposited energies.
  Background events from accidental coincidences populate the
  low-energy regions in both calorimeters.}
\end{figure}
The typical experimental signature of luminosity events is shown in
Figure~\ref{fig:lumi}. The main experimental systematic error arises
from the definition of the geometrical acceptance for this process.
Since the angular distribution is steeply falling with increasing
scattering angle ($\propto\theta^{-3}$), the precise definition of the
inner radius of the acceptance region is most critical.  Background
arises from random coincidences between the calorimeters at the two
sides and is largely beam-induced.  The integrated luminosity is given
by the ratio of the number of observed small-angle $\ee$ events
and the calculated cross-section for this process within the detector
acceptance.
The Bhabha cross-section at small scattering angles is dominated by
the well-known QED process of $t$--channel scattering, but nonetheless
calculational uncertainties give rise to an important theoretical
error of about 0.5\,per-mille affecting all experiments coherently, as
is discussed in Section~\ref{sec-lumerr}. Typical experimental
systematic errors on the luminosity are well below 1\,per-mille.

\subsubsection{Event Selection Efficiency and Background Levels}

In the hadronic channel the selection efficiencies within the
acceptance are high, typically above 99\,\%. Backgrounds are dominated
by $\Zzero\to\tautau$ and non-resonant $\qq$ production from
two-photon reactions. At the peak of the resonance these together
contribute at a level of a few per-mille.  Backgrounds in the lepton
selections are typically around 1\,\% for $\ee$ and $\mumu$ and
slightly larger for $\tautau$ final states.  The dominant background
in $\ee$ and $\mumu$ final states arises from $\tautau$ events, a
contribution which cancels when the total leptonic cross-section is
measured. Backgrounds other than $\tautau$ in the $\ee$ and $\mumu$
channels are of order $0.1\,\%$.  Backgrounds in $\tautau$ events
are larger, 2--3\,\%, and arise from low-multiplicity hadronic events,
from two-photon reactions and from $\ee$ and $\mumu$ events with small
measured lepton momenta, which may result either from undetected
radiated photons or from measurement errors.

An overview of the selection efficiencies within the acceptance and of
the background levels is presented in Table~\ref{tab:LSeff}.  The
acceptances quoted in the table are ideal ones suitable as input to
the electroweak program libraries used for fitting, while the actual
set of experimental cuts is more complicated.  Monte Carlo event
generators and detailed detector simulations in combination with
corrections derived from studies of the actual data are used to
transform the true experimental acceptances to the ideal ones. As is
shown in the table, the selection efficiencies are high, above 95\,\%
in $\ee$ and $\mumu$ and 70--90\,\% in $\tautau$ final states.

\begin{table}[bthp]
\begin{center} 
\begin{tabular}{|l||c|c|c|c|}
\hline %
~   & ALEPH & DELPHI & L3 & OPAL \\
\hline %
\hline %
\multicolumn{5}{|c|}{$\qq$ final state} \\ 
\hline %
acceptance &$s'/s>0.01$ &$s'/s>0.01$ &$s'/s>0.01$ &$s'/s>0.01$ \\
efficiency [\%] &
           99.1 & 94.8 & 99.3 & 99.5 \\
background [\%] &
           0.7 & 0.5 & 0.3 & 0.3 \\
\hline %
\hline %
\multicolumn{5}{|c|}{} \\[-0.7pc]
\multicolumn{5}{|c|}{$\ee$ final state} \\ 
\hline %
acceptance          
           & $-0.9<\cost<0.7$  & $|\cost|<0.72$
           & $|\cost|<0.72 $   & $|\cost|<0.7$ \\
 ~         & $s'>4m_\tau^2$  & $\eta<10\mydeg$ 
           & $\eta<25\mydeg$   & $\eta<10\mydeg$\\[0.3pc]
efficiency [\%] & 
           97.4 &97.0 & 98.0 & 99.0  \\
background [\%] &
           1.0 & 1.1 &   1.1 &   0.3   \\       
\hline %
\hline %
\multicolumn{5}{|c|}{} \\[-0.7pc]
\multicolumn{5}{|c|}{$\mumu$ final state} \\
\hline %
acceptance 
           &$|\cost|<0.9$ &$|\cost|<0.94$
           &$|\cost|<0.8$ & $|\cost|<0.95$           \\
   ~       &$s'>4m_\tau^2$ & $\eta<20\mydeg$ 
           &$\eta<90\mydeg$  &$m^2_\ff/s>0.01$ \\                  
efficiency [\%] &
           98.2 & 95.0& 92.8& 97.9  \\
background [\%] &
           0.2 & 1.2 &  1.5 &   1.0  \\       
\hline %
\hline %
\multicolumn{5}{|c|}{} \\[-0.7pc]
\multicolumn{5}{|c|}{$\tautau$ final state} \\
\hline %
acceptance 
           &$|\cost|<0.9$  & $0.035<|\cost|<0.94$
           &$|\cost|<0.92$ & $|\cost|<0.9$  \\
   ~       &$s'>4m_\tau^2$ & $s'>4m_\tau^2$   
           & $\eta<10\mydeg$ &$m^2_\ff/s>0.01$ \\[0.3pc]
efficiency [\%] &
           92.1 &  72.0 & 70.9 & $ 86.2 $    \\
background [\%] &
           1.7 & 3.1 &   2.3 &  2.7  \\       
\hline %
\end{tabular}
\caption[Selection efficiencies and backgrounds]{\label{tab:LSeff}
  Ideal acceptances, selection efficiencies$^*$ and background
  contribution at the peak of the resonance (1994 data). \\ $^*${The
  lepton selection efficiencies given by the experiments were in some
  cases quoted with respect to full acceptance in $\cost$; for the
  purpose of comparison, they were corrected to the fiducial cuts in
  $\cost$ actually used in the analyses, assuming a shape of the
  differential cross-section according to $(1+\cos^2\theta)$.}  }
\end{center} 
\end{table}

The idealised acceptances are defined by the scattering angle,
$\theta$, of the negatively charged lepton in the laboratory frame,
and also require a cut-off for initial-state photon radiation. The
latter may either be given by a cut on the acollinearity of the two
final-state leptons, $\eta$, or by an explicit cut on the invariant
mass of the final-state leptons, $m_\ff$; alternatively, the effective
centre-of-mass energy after initial-state photon radiation,
$\sqrt{s'}$, may be used. The experimental efficiencies for low values
of $m_\ff$ or $s'$ are small. Despite the differing definitions, the
efficiencies given in the table can nevertheless be directly compared,
because the acceptance difference between the wider definition, $s'/s
> 4m_{\tau}^2$, and a tight definition using an acollinearity cut at
$\eta<10\mydeg$ is only 2\,\%.

\subsubsection{Total Cross-Section}

\begin{figure}[p]
\begin{sideways}
\begin{minipage}{\textheight}
\vskip -1.5pc
\mbox{\epsfig{file=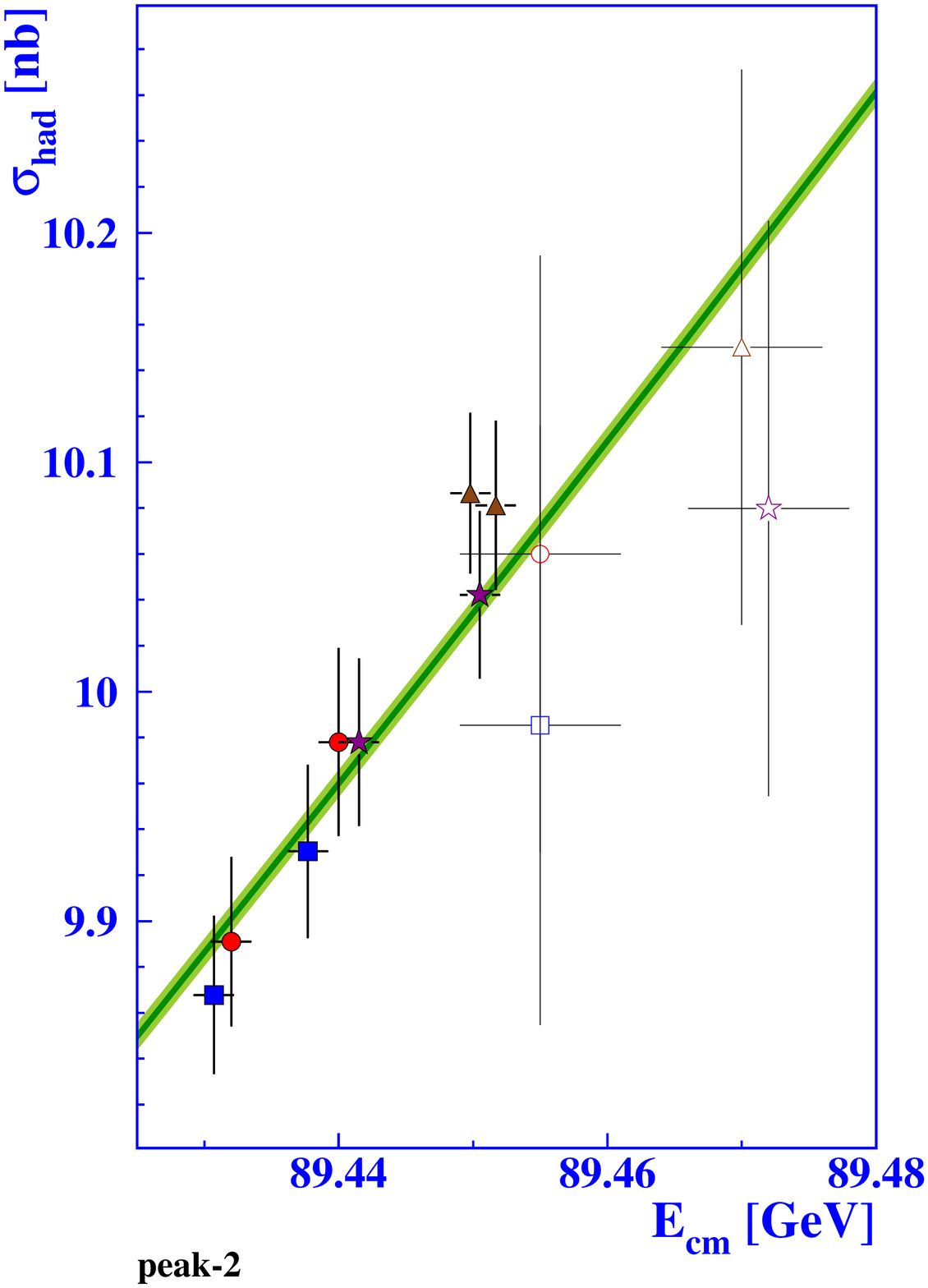,width=0.333\textheight}} 
\mbox{\epsfig{file=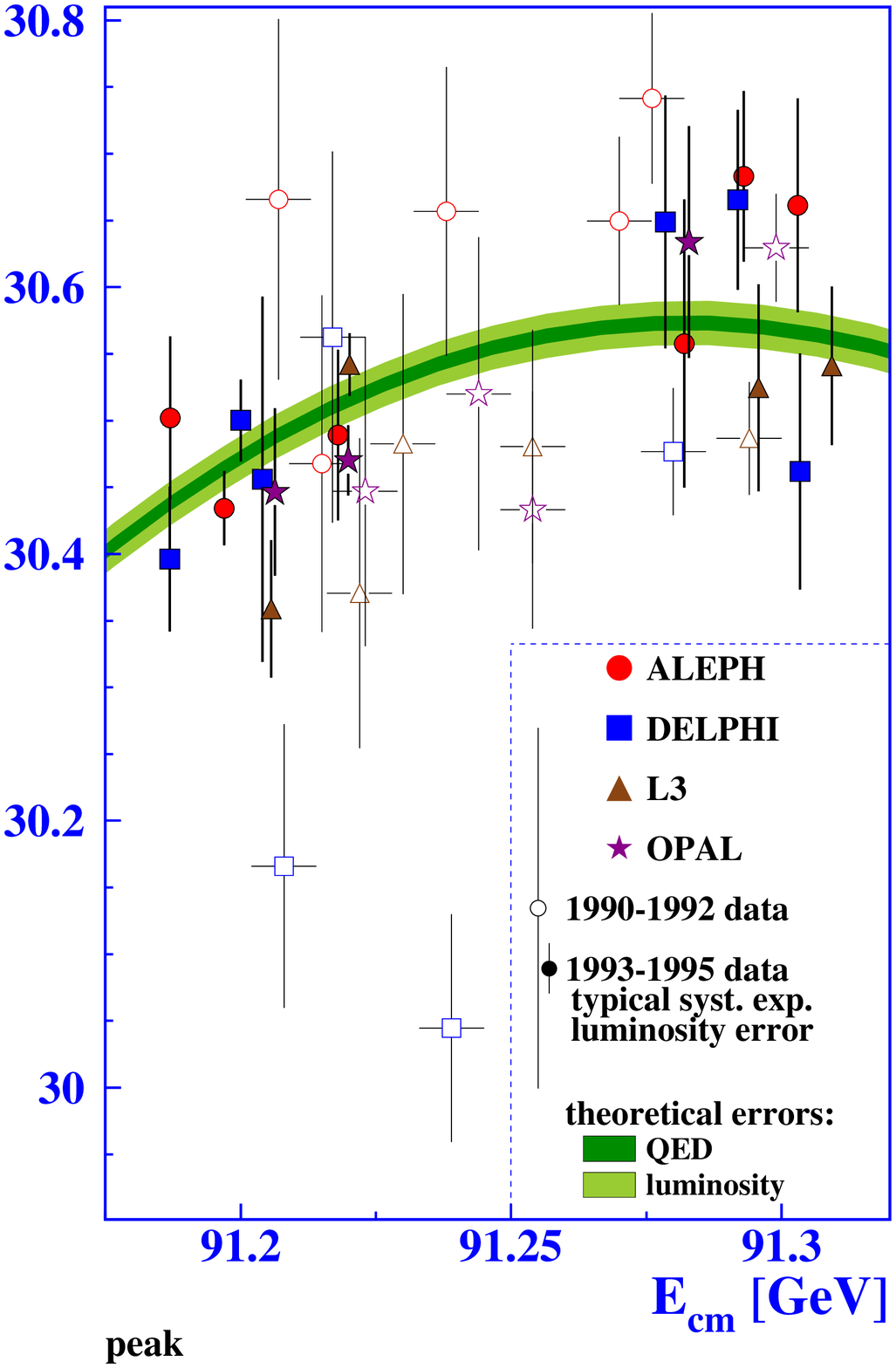,width=0.333\textheight}} 
\mbox{\epsfig{file=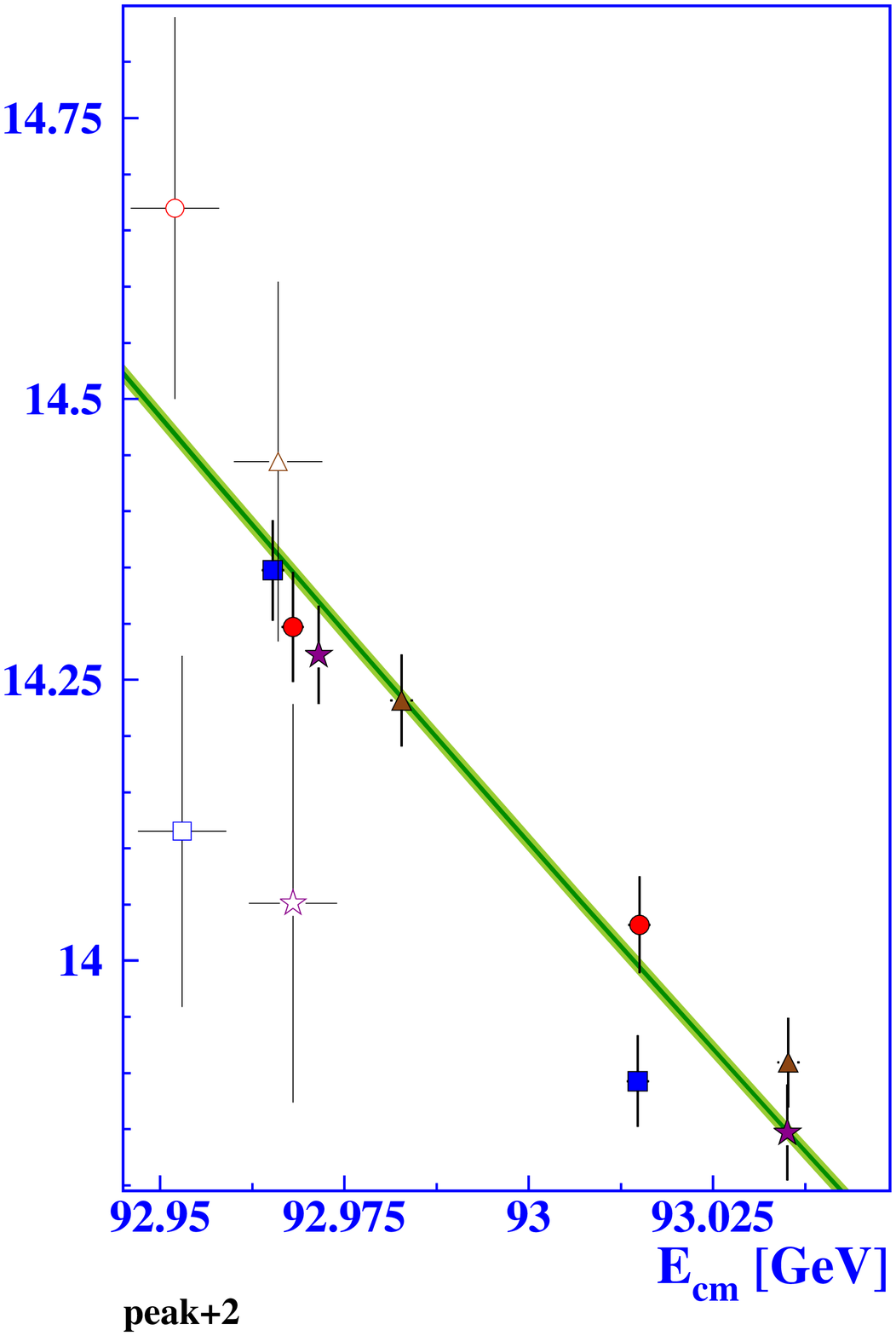,width=0.333\textheight}} \\
\vskip -1.5pc
\caption[Hadronic cross-section measurements]
{\label{fig:xsh} Measurements by the four experiments of the hadronic
  cross-sections around the three principal energies.  The vertical
  error bars show the statistical errors only. The open symbols
  represent the early measurements with typically much larger
  systematic errors than the later ones, shown as full symbols.
  Typical experimental systematic errors on the determination of the
  luminosity are indicated in the legend; these are almost fully
  correlated within each experiment, but uncorrelated among the
  experiments.  The horizontal error bars show the uncertainties in
  LEP centre-of-mass energy, where the errors for the period
  1993--1995 are smaller than the symbol size in some cases. The
  centres of the bands represent the cross-section parametrisation in
  terms of the combined pseudo-observables of the four experiments.
  The width of the bands represents the linear superposition of the
  two most important common theoretical errors from initial-state
  photon radiation and from the calculations of the small-angle Bhabha
  cross-section.}
\end{minipage}
\end{sideways}
\end{figure}

The total cross-section for the production of each final state is
obtained from the efficiency and background-corrected numbers of
selected events normalised to the luminosity. Data taken at the same
energy point and within the same year are combined into a single
cross-section measurement at the average energy.  As an example, the
measurements of the hadronic cross-section around the three principal
energies are shown in Figure~\ref{fig:xsh}. Because the hadron
statistics are almost ten times larger than the lepton statistics,
these measurements dominate the determination of the mass and the
width of the $\Zzero$.

\begin{figure}[bth]
\begin{center}
\mbox{\epsfig{file=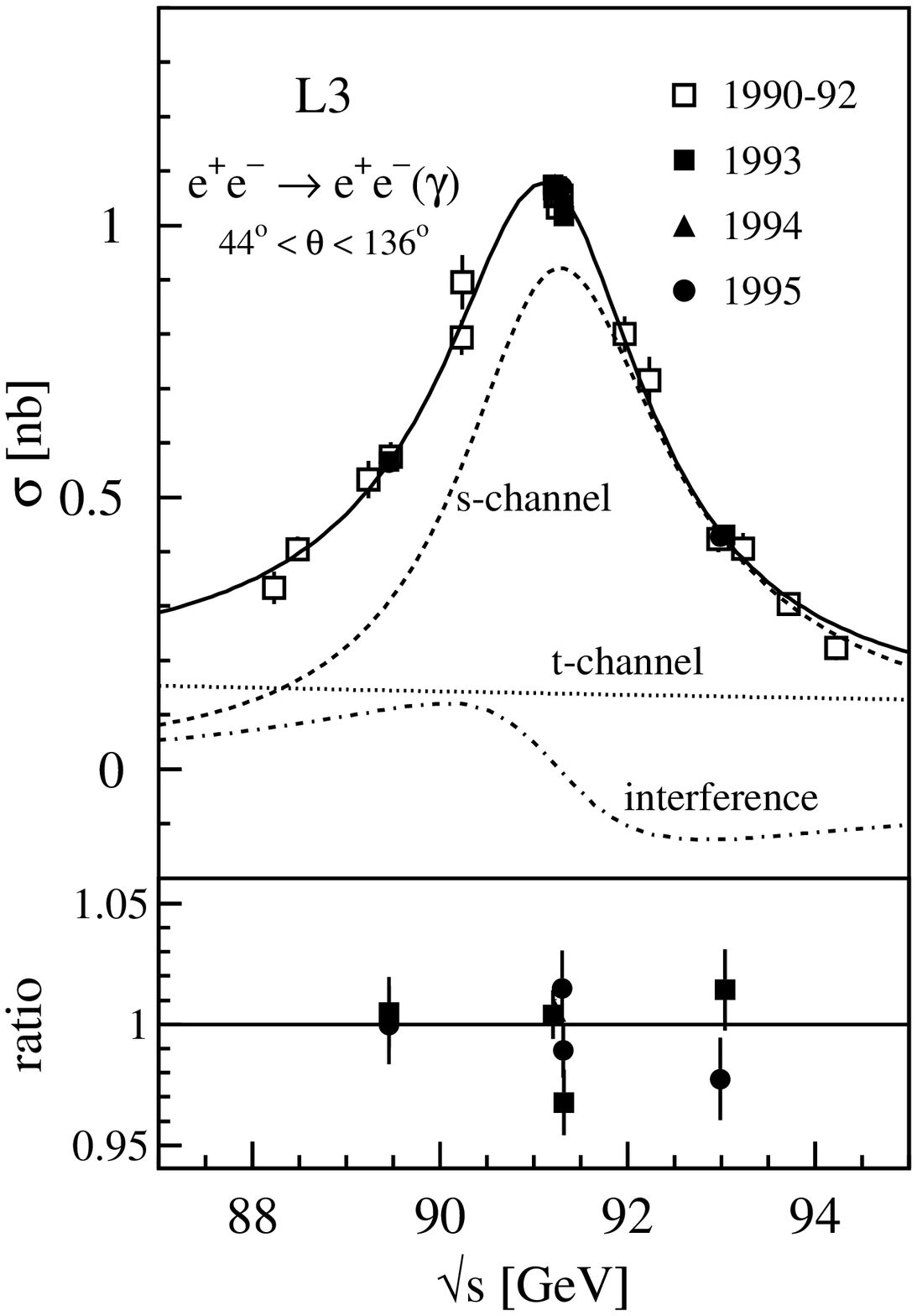,width=0.49\textwidth}} \hspace*{.1em}
\mbox{\epsfig{file=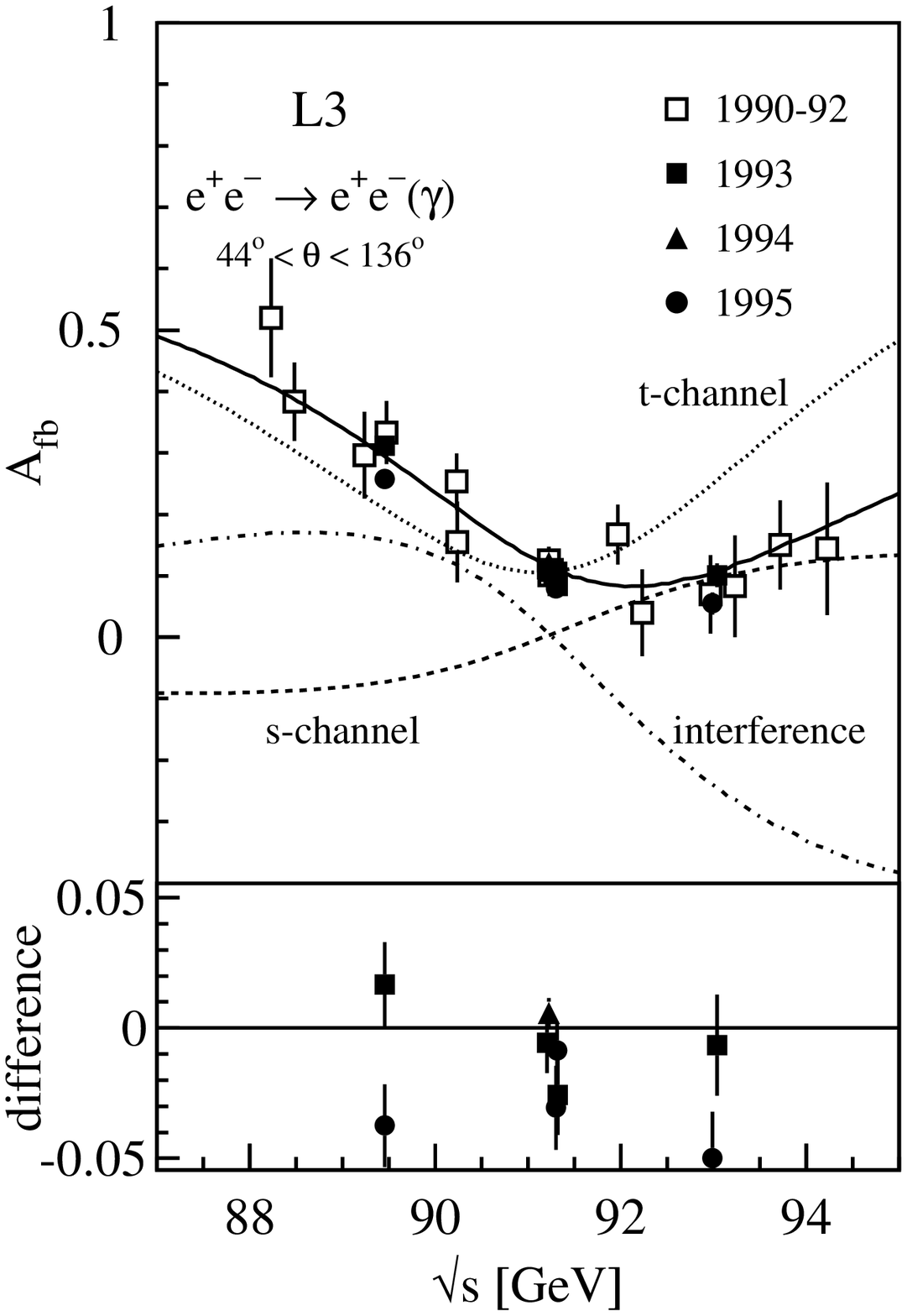,width=0.49\textwidth}}
\end{center}
\caption[$t$-channel contribution to $\ee$ final states]
{\label{fig:tchannel} The energy dependence and the contributions from
  the $s$ and $t$-channel diagrams and from the $s$-$t$ interference
  for observables in the $\ee$ channel. Shown are the total
  cross-section (left) and the difference between the forward and
  backward cross-sections after normalisation to the total
  cross-section (right). The data points measured by the L3
  collaboration refer to an angular acceptance of $|\cost|<0.72$, an
  acollinearity $\eta<25^{\circ}$ and a minimum energy of $E_{\rm
    e^\pm} \gt 1~\GeV$.  The lines represent the model-independent fit
  to all L3 data.}
\end{figure}

The energy dependence of the hadronic cross-section (the
``lineshape'') is shown in the upper plot of Figure~\ref{fig:xshafb}
in Section~\ref{sec:intro_zpar}. The energy dependence of the muon and
tau cross-section is nearly identical in shape to the hadronic one. In
$\ee$ final states however, diagrams involving photon exchange in the
$t$-channel and their interference with the $s$-channel diagrams also
contribute.  The different contributions are shown as a function of
centre-of-mass energy in the left-hand plot of
Figure~\ref{fig:tchannel}.

\subsection{Measurements of the Lepton Forward-Backward Asymmetries}

\begin{figure}[bth]
\begin{center}
\mbox{\epsfig{file=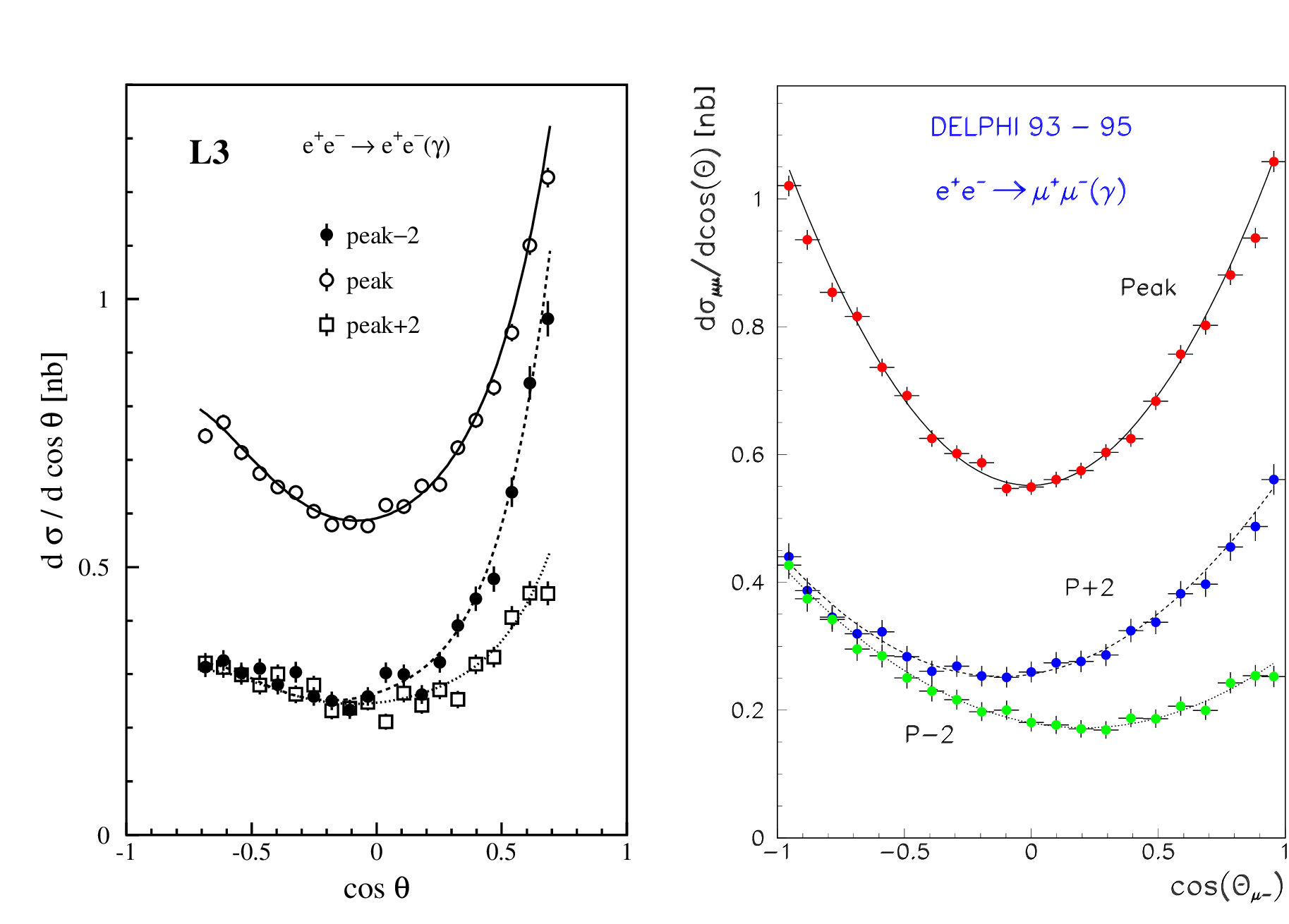,width=0.9\textwidth}}
\end{center}
\caption[$\mumu$ differential cross-section]
{\label{fig:xsdiff} Distribution of the production polar angle,
  $\cos\theta$, for $\ee$ and $\mumu$ events at the three principal
  energies during the years 1993--1995, measured in the L3 (left) and
  DELPHI (right) detectors, respectively. The curves show the $\SM$
  prediction from ALIBABA~\cite{ALIBABA} for $\ee$ and a fit to the
  data for $\mumu$ assuming the parabolic form of the differential
  cross-section given in the text. }
\end{figure}

The forward-backward asymmetry, $\Afb$, is defined by the numbers of
events, $N_{\rm F}$ and $N_{\rm B}$, in which the final state lepton
goes forward ($\cos\theta_{\ell^-}>0$) or backward
($\cos\theta_{\ell^-}<0$) with respect to the direction of the
incoming electron, \(\Afb=(N_{\rm F}-N_{\rm B})/(N_{\rm F}+N_{\rm
B}).\) This definition of $\Afb$ depends implicitly on the acceptance
cuts applied on the production polar angle, $\cost$, of the leptons.
The measurements of $\Afb(\leptlept)$ require the determination of
$\cost$ and the separation of leptons and anti-leptons based on their
electric charges, which are determined from the curvature of the
tracks in the magnetic fields of the central detectors.  For $\mumu$
and $\tautau$ final states, $\Afb$ is actually determined from
un-binned maximum-likelihood fits to the differential cross-section
distributions of the form $\mathrm{d}\sigma/\mathrm{d}\cost\propto
1+\cos^2\theta+8/3\cdot\Afb\,\cost$.  This procedure makes better use
of the available information and hence leads to slightly smaller
statistical errors. Determined this way the $\Afb$ measurements are
insensitive to any distortions of the detection efficiency as long as
these are not at the same time asymmetric in charge and asymmetric in
$\cost$.  Examples of the measured angular distributions for the $\ee$
and $\mumu$ final states are shown in Figure~\ref{fig:xsdiff}.

The shape of the differential cross-section in the electron final
state is more complex due to contributions from the $t$-channel and
the $s$-$t$-interference, which lead to a large number of events in
which the electron is scattered in the forward direction.  A
maximum-likelihood fit to obtain $\Afb(\ee)$ may be performed after
subtracting the $t$ and $s$-$t$ contributions, but usually the
asymmetry is determined from the efficiency-corrected numbers of
events with forward and backward-going electrons.

The energy dependence of the forward-backward asymmetry in the $\mumu$
final state is shown in the lower plot of Figure~\ref{fig:xshafb}
above. The forward-backward asymmetry as a function of centre-of-mass
energy in the $\ee$ final state including the $t$ and the $s$-$t$
contributions is illustrated in the right-hand plot of
Figure~\ref{fig:tchannel}.

\subsection{Experimental Systematic Errors}

In general, the systematic errors arising from the selection
procedures are small and so the accumulated statistics can be fully
exploited.  Furthermore, the purely experimental errors arising from
the limited understanding of detector acceptances are uncorrelated
among the experiments. An overview of the experimental systematic
errors is given in Table~\ref{tab:LSsyst}. Statistical errors per
experiment on the cross-sections are only around 0.5\,per-mille in the
hadronic channel and around 2.5\,per-mille in each of the three lepton
channels.  Statistical errors from the number of small-angle Bhabha
events affect all channels in a correlated way, but even on-peak they
are smaller than those in the hadronic channel by at least a factor of
$\sqrt{2}$. Experimental systematic errors on the forward-backward
asymmetries are between two and five times smaller than the
statistical errors.  Errors common to all experiments may arise from
the use of common Monte Carlo generators and detector simulation
programs. However, each experiment used its own tuning procedures for
the QCD parameters determining the simulation of the hadronisation
process; furthermore, the physical acceptances of the detectors, the
event selection procedures as well as the quantities used to define
the acceptances after all cuts vary among the experiments, and
therefore the related common errors are small and were neglected in
the combination procedure.

\begin{table}[bth]
\begin{center} \begin{tabular}{|l||ccc|ccc|}
\hline %
 &\mth{{ALEPH}}&\multicolumn {3}{c|}{{DELPHI}}\\ 
\cline{2-7}
 &1993&1994&1995   &1993  &1994&1995  \\ 
\hline
\hline
$\calL^{\rm exp}$
 &0.067\%&0.073\%&0.080\%&0.24\%&0.09\% &0.09\% \\
\hline
$\sigma_{\rm {had}}$
&0.069\% &0.072\%& 0.073\% &0.10\%& 0.11\% &0.10\% \\ 
$\sigma_{\rm e}$
&0.15\% &0.13\%& 0.15\%& 0.46\% &0.52\% &0.52\% \\
$\sigma_{\mu}$
&0.11\% &0.09\%& 0.11\%& 0.28\% &0.26\% &0.28\% \\
$\sigma_{\tau}$
&0.26\% &0.18\%& 0.25\%&0.60\% &0.60\%&0.60\%  \\
\hline %
\hline %
$\Afb^{\rm e}$
   &0.0006 &0.0006 &0.0006 &0.0026&0.0021 &0.0020 \\
$\Afb^{\mu}$
   &0.0005 &0.0005 &0.0005 &0.0009&0.0005 &0.0010 \\
$\Afb^{\tau}$
   &0.0009 &0.0007 &0.0009 &0.0020&0.0020& 0.0020 \\
\hline %
\multicolumn{7}{c}{~} \\[-0.5pc]
\hline %
&\mth{{L3}}&\multicolumn{3}{c|}{{OPAL}}\\
\cline{2-7}
&1993  &1994 &1995  &1993  &1994 &1995\\
\hline
\hline
$\calL^{\rm exp}$
&0.086\%& 0.064\%&0.068\%&0.033\%&0.033\%&0.034\%\\
\hline
$\sigma_{\rm {had}}$
&0.042\%& 0.041\%& 0.042\%& 0.073\%& 0.073\%&0.085\%\\
$\sigma_{\rm e}$
&0.24\%& 0.17\%&0.28\% &0.17\%&0.14\%& 0.16\%\\
$\sigma_{\mu}$
&0.32\%&0.31\%&0.40\%&0.16\%&0.10\%& 0.12\%\\
$\sigma_{\tau}$
&0.68\%&0.65\%&0.76\%&0.49\%&0.42\%&0.48\%\\
\hline %
\hline %
$\Afb^{\rm e}$
&0.0025&0.0025   & 0.0025&0.001&0.001& 0.001 \\
$\Afb^{\mu}$
&0.0008&0.0008& 0.0015& 0.0007&0.0004& 0.0009\\
$\Afb^{\tau}$
&0.0032&0.0032& 0.0032& 0.0012&0.0012& 0.0012\\
\hline %
\end{tabular}
\caption[Experimental systematic errors]{\label{tab:LSsyst}
  Experimental systematic errors for the analyses at the $\Zzero$
  peak. The errors are relative for the cross-sections and absolute
  for the forward-backward asymmetries. None of the common errors
  discussed in Section~\ref{sec-comerr} are included here.}
\end{center}
\end{table}

Errors arising from limitations in theoretical precision, such as the
calculation of the small-angle Bhabha cross-section, the $t$-channel
contribution in the $\ee$ final state or pure QED corrections to the
cross-section, are common to all experiments. They are discussed in
detail in Section~\ref{sec-comerr}.

\subsection{Energy Calibration \label{sec-ecal}}

Precise knowledge of the centre-of-mass energy is essential for the
determination of the mass and width of the $\Zzero$ resonance.  The
uncertainty in the absolute energy scale, {\em i.\,e.}  uncertainties
correlated between the energy points, directly affect the
determination of the $\Zzero$ mass, whereas the $\Zzero$ width is only
influenced by the error in the difference in energy between energy
points.  The determination of the mass and width are completely
dominated by the high-statistics scans taken at the off-peak points
approximately $\pm2~\GeV$ away from the resonance in 1993 and 1995,
and the errors due to energy calibration are therefore given by
\begin{equation} 
\begin{array}{lcl}
\Delta \MZ & \approx & \frac{1}{2}\cdot\Delta(E_{+2} + E_{-2}) {\rm ~and}\\
\Delta \GZ &\approx  & \frac{\GZ}{E_{+2}-E_{-2}}\Delta(E_{+2} - E_{-2})\,. \\
\end{array} 
\end{equation}

\begin{figure}[t]
\begin{center}
\mbox{\epsfig{file=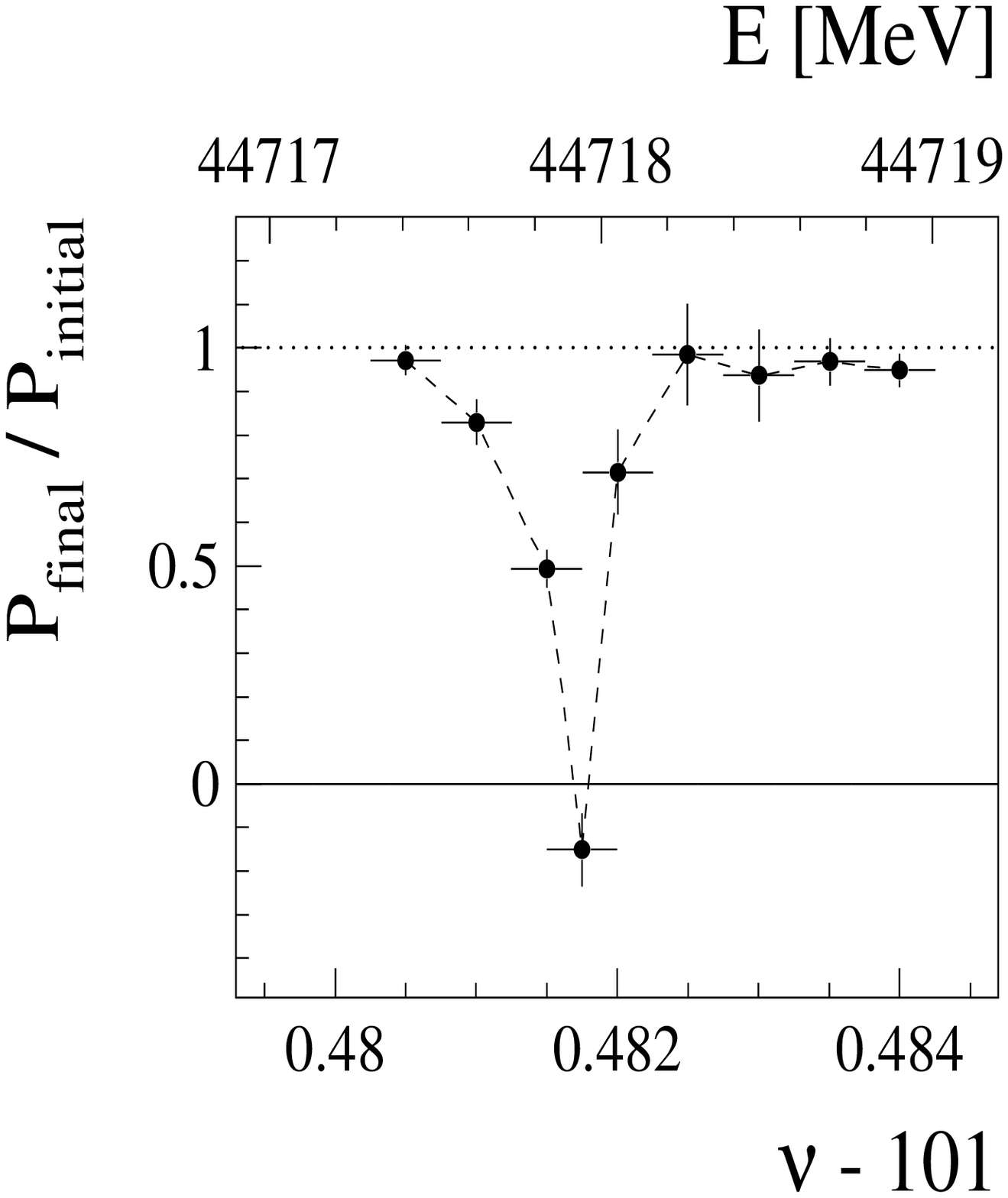,width=0.6\textwidth}}
\end{center}
\caption[Width of the depolarising spin resonance]
{\label{fig:resdepol} Measurement of the width of the artificially
  excited spin resonance which is used for the energy calibration of
  LEP (from Reference~\citen{bib-EPOL2}).  The drop in the observed
  polarisation level is shown as a function of the ``fractional spin
  tune'', {\em i.\,e.} the spin tune $\nu$ minus its integer part of
  101.}
\end{figure}

The average momentum of particles circulating in a storage ring is
proportional to the magnetic bending field integrated over the path of
the particles. The very accurate determination of the average energy
of the beams in LEP was based on the technique of resonant spin
depolarisation~\cite{bib-EPOL,bib-EPOL2}, which became available in
1991, after transverse polarisation of the electron beam in LEP had
first been observed in 1990~\cite{bib-firstPOL} with a Compton
polarimeter~\cite{bib-polarimeter}. Transverse polarisation of single
or separated beams due to the Sokolov-Ternov
mechanism~\cite{bib-transpol} was observed in LEP after careful
adjustment of the beam orbit in order to avoid any static depolarising
resonances.  The same magnetic bending field seen by the particles
along their path leads to precession of the average spin vector of the
polarised bunches. The beam energy is therefore proportional to the
number of spin precessions per turn, the ``spin tune'', $\nu$. The
spin precession frequency is measured by observing the depolarisation
which occurs when an artificial spin resonance is excited with the
help of a weak oscillating radial magnetic field.  This method offers
a very high precision, as good as $\pm$0.2~$\MeV$, on the beam energy
at the time of the measurement.  The resolution of the method is
illustrated in Figure~\ref{fig:resdepol}, which shows the observed
drop in polarisation as a function of the oscillations per turn of the
depolarising magnetic field, corresponding to the fractional spin tune
of the beam particles.

Measurements with resonant depolarisation were only possible outside
normal data taking, typically at the end of fills.  About 40\% of the
recorded off-peak luminosity in the 1993 scan and about 70\% in the
1995 scan was taken during fills with at least one such precise
calibration of the beam energy.  Other techniques had to be employed
to extrapolate these calibrations back to earlier times in a fill and
to those fills where no calibrations by resonant depolarisation could
be made. This required precise knowledge of the values and time
evolution of numerous parameters and careful modelling of their impact
on the beam energy~\cite{\ECAL,bib-ECAL95}.

For particles on central orbit the magnetic bending field is given by
the field produced by the bending dipoles and corrector magnets and by
small contributions from the Earth's magnetic field and from remnant
fields in the beam pipe. In addition, magnetic fields originating from
leakage DC currents produced by trains in the Geneva area had to be
taken into account.  The magnetic field of the dipoles was initially
measured with a single nuclear magnetic resonance probe (``NMR'')
installed only in a reference dipole on the surface. In 1995, two NMR
probes were installed in two of the tunnel dipoles, which measured the
magnetic field directly above the beam pipe.

\begin{figure}[th] \begin{center}
\mbox{\epsfig{file=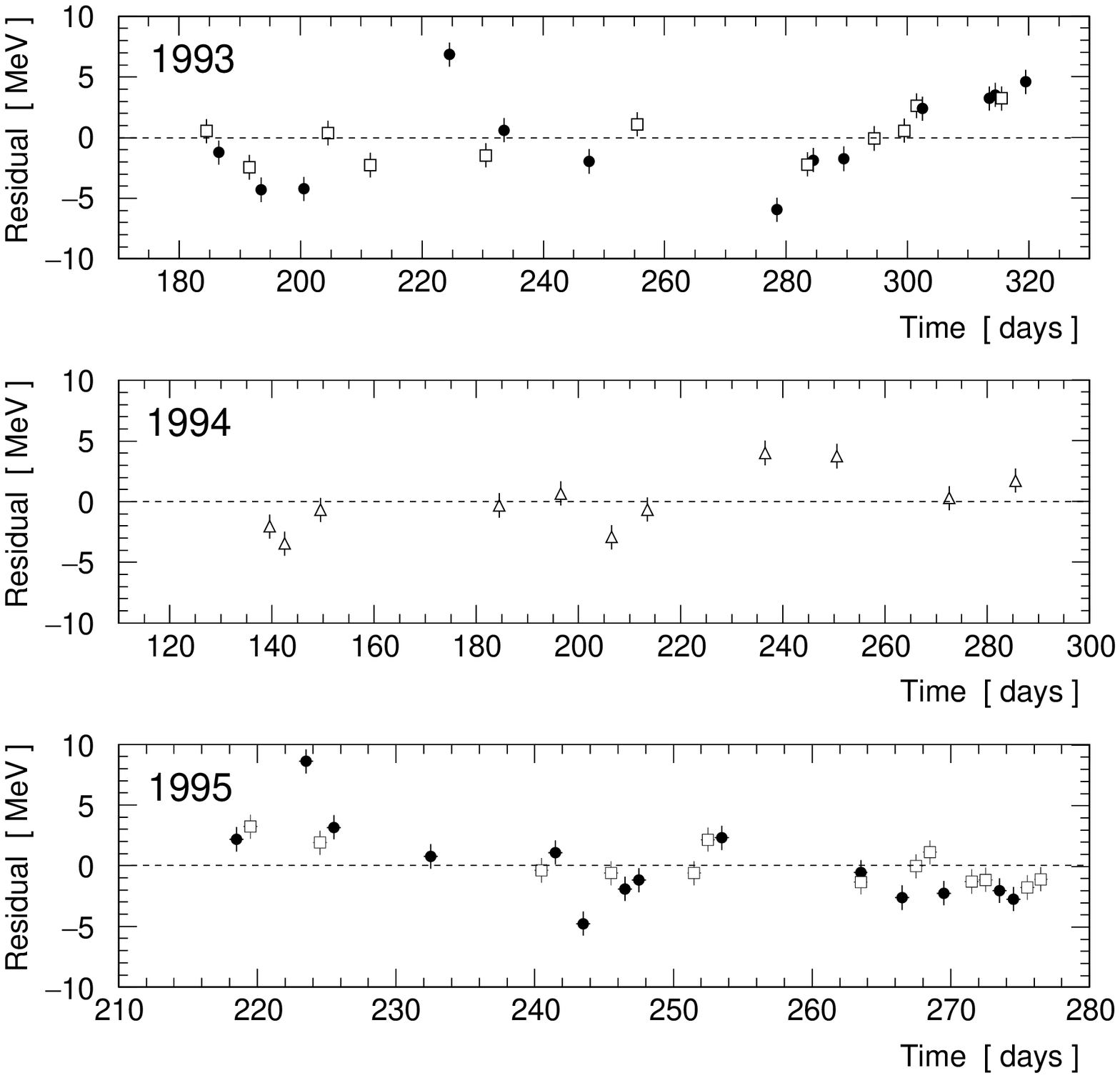,width=0.8\columnwidth}}
\end{center}
\caption[Energy from model \emph{vs.} depolarisation calibrations] 
{\label{fig:ECALresid} Difference between centre-of-mass energies
  measured by resonant depolarisation and from the energy model (from
  Reference~\citen{bib-ECAL95}).  The black circles are for
  \mbox{peak-2}, the open triangles for peak energies and the open
  squares are for \mbox{peak+2} energy points.  The error bars have a
  size of $\pm1~\MeV$. }
\end{figure}

Contributions from the quadrupoles and sextupoles must also be
considered if the beam particles do not pass, on average, through the
centres of these magnets, i.e. if the particles oscillate around
non-central orbits. Because the ultra-relativistic electrons and
positrons circulate synchronously with the frequency of the
accelerating radio frequency cavities with a speed which is constant
to a very high level of precision, the path length per revolution
remains constant. Movements of the LEP equipment, caused by geological
deformations of the LEP tunnel, therefore brought the beam orbit away
from the central position, where the beam particles now sensed the
extra magnetic fields of the quadrupoles. As a consequence, the
bending field became different, and the particle energies changed
accordingly through changes of their phases relative to the radio
frequency clock. Among the identified origins of such movements of the
LEP equipment relative to the beam orbit were tidal effects from the
Sun and the Moon, the water level in Lake Geneva and rainfall in the
Jura Mountains. These could all be tracked by frequent and precise
measurements of the beam orbit position inside the LEP beam pipe. An
energy model was developed that was able to predict the beam energy at
any given time. The quality of this model and remaining uncertainties
can be estimated by comparing the energy predicted by the model with
the precise energy determinations by resonant depolarisation, as is
shown in Figure~\ref{fig:ECALresid}.

\begin{figure}[bt] \begin{center}
\mbox{\epsfig{file=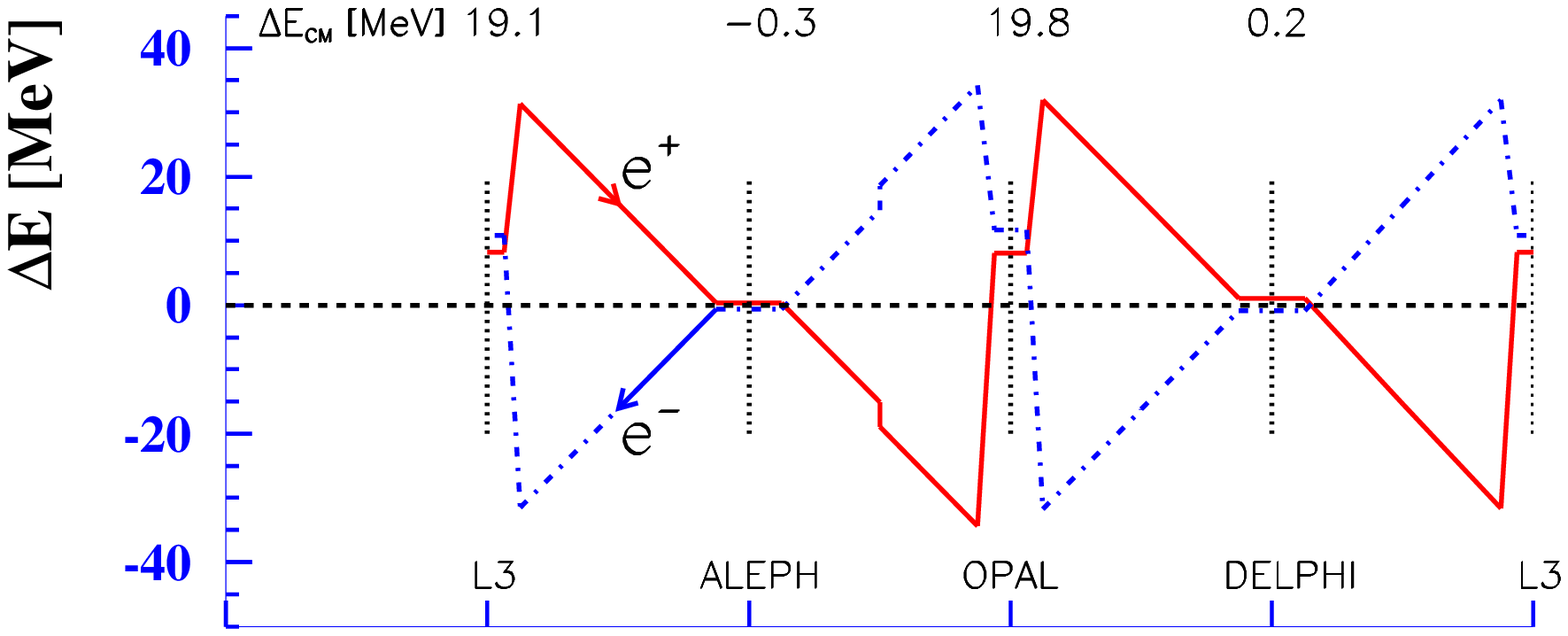,width=\columnwidth}}
\end{center}
\vskip -2pc \caption[Beam energy variations around the LEP ring]
{\label{fig:ECALsawtooth} Typical variations of the beam energy around
  the LEP ring during the 1993 run. Energy losses from synchrotron
  radiation in the arcs, and in wiggler magnets between the ALEPH and
  OPAL IPs, are compensated by acceleration in the radio frequency
  cavities mounted in the straight sections on both sides of L3 and
  OPAL. The detailed modelling results in significant corrections on
  the centre-of-mass energy at the IPs between acceleration sections,
  as indicated by the numbers on the top.}
\end{figure}

In order to obtain the energy of the particles colliding at an
interaction point (``IP''), additional effects have to be considered.
Figure~\ref{fig:ECALsawtooth} shows the variations of the beam energy
of electrons and positrons as they travel round the ring and the large
energy corrections at the interaction points.  Precise knowledge of
all relevant parameters of the radio frequency system at any time is
mandatory for the reliable calculation of these corrections in a
detailed ``RF model''. Frequent measurements of the synchrotron tune and
of beam orbit positions as well as measurements of the position of the
collision vertex performed by the experiments and comparisons of these
measurements with predictions from the RF model were essential to
ensure the internal consistency of all the input parameters and to
keep systematic errors small.  If the bunches in a collider do not
precisely collide head-on at an IP, a possible energy-dependence of
the distribution of particle positions in a bunch, so-called
``dispersion effects'', may lead to shifts in the average collision
energy. Due to the operation of LEP in bunch train mode in 1995,
unlike-sign dispersion of the colliding electron and positron bunches
in the vertical direction was present, which would have led to
significant energy displacements of about $2~\MeV$ for collision
offsets of one $\mu$m between the bunches.  Such collision offsets
therefore had to be minimised during data-taking, which was achieved
by small vertical movements of the beams and adjusting them such that
the luminosity was maximised.

For each experiment a value of the beam energy was provided every 15
minutes. Errors on the centre-of-mass energy are largely dominated by
the uncertainties in the energy model mentioned above.  A summary of
the typical size of the main effects and of their contributions to the
error is shown in Table~\ref{tab:ECALsys}.

\begin{table}[tb] \begin{center}
\begin{tabular}{|l||rr|rr|}
\hline %
    ~     & \multicolumn{2}{c|}{Correction to E$_{\rm CM}$} 
          & \multicolumn{2}{c|}{Error on} \\ 

  Origin of correction & Size & Error & $\MZ$& $\GZ$ \\
                       & [\MeV]~& [\MeV]~& [\MeV]& [\MeV]\\
\hline
\hline
Energy measurement by resonant depolarisation  & & 0.5 & 0.4 & 0.5\\
Mean fill energy, from
uncalibrated fills       & & [0.5--5.0] & 0.5 & 0.8 \\
\hline
Dipole field changes        & up to 20 & [1.3--3.3] & 1.7 & 0.6\\ 
Tidal deformations      & $\pm$10 & [0.0--0.3] & 0.0 & 0.1 \\ 
$\mathrm{e}^+$ energy difference  & $<$0.3 & 0.3 & 0.2 & 0.1\\
Bending field from horizontal correctors  &[0--2] & [0.0--0.5] & 0.2 &0.1\\
\hline
IP dependent RF corrections &[0--20]  & [0.5--0.7] & 0.4 & 0.2\\
Dispersion at IPs           & 0.5   & [0.4--0.7] & 0.2 & 0.1 \\
\hline %
\end{tabular}
\caption[Summary of errors on the centre-of-mass energy]
{\label{tab:ECALsys} Breakdown of effects on the centre-of-mass
  energy, for illustrative purposes only.  The last two columns give
  the approximate contribution of each effect to the error on $\MZ$
  and $\GZ$. The full evaluation of the energy errors used values
  specific to each year and energy, and also took into account their
  correlations. (See Reference~\citen{bib-ECAL95} for a complete
  discussion.)  }
\end{center}
\end{table}

The energy errors vary slightly among the interaction points, mainly
due to different configurations of the radio frequency cavities. The
energy errors for different experiments and data taking periods have
large common parts, and therefore the use of a full correlation matrix
is necessary.  Assuming that all experiments contribute with the same
weight allows all the {LEP} energy errors to be conveniently
summarised in a single error matrix, common to all interaction points,
as given in Reference~\citen{bib-ECAL95}.

The energy of individual beam particles is usually not at the mean
value considered above, but oscillates around the mean energy.
Therefore observables are not measured at a sharp energy, $E^0_{cm}$,
but instead their values are averaged over a range in energies
$E^0_{cm} \pm \delta E_{cm}$.  With the assumption of a Gaussian shape
of the energy distribution, the total cross-sections,
$\sigma(E_{cm})$, receive a correction proportional to $\delta
E_{cm}^2$ and the second derivative of $\sigma(E_{cm})$ with respect
to $E_{cm}$.  At LEP-I, typical values of the centre-of-mass energy
spread were around 50~$\MeV$. The effects of the correction lead to an
increase of the cross-section at the peak of the $\Zzero$ resonance by
0.16\% and a decrease of the width by about 5~$\MeV$. The beam energy
spread is affected by the operation of wiggler magnets used to
optimise the luminosity. It is also related through some machine
parameters to the length of the luminous region at the interaction
points, which was precisely measured by the experiments and thus
allowed a permanent monitoring of the beam energy spread. Bunch length
measurements also served as a cross-check in evaluating the
uncertainties on the energy spread from the uncertainties in numerous
machine and beam parameters.  Uncertainties on the centre-of-mass
energy spread were around $\pm 1~\MeV$ in 1993--1995, and constitute
an almost negligible source of error common to all experiments.

Changes in the mean beam energy due to changes of machine parameters
have an effect similar to the natural beam energy spread.  Data taking
periods with a very similar centre-of-mass energy were combined into a
single energy point in the experimental analyses by performing a
luminosity-weighted average. The additional energy spread resulting
from this grouping was only around $10~\MeV$, which is added in
quadrature to the natural beam energy spread of the accelerator.

Uncertainties from the energy calibration as described in this
subsection and corrections for the beam energy spread were taken into
account by all experiments in the fits from which the $\Zzero$
parameters were extracted; the related common uncertainties are
discussed in Section~\ref{sec-comenerr}.

\section{Experimental Results \label{sec-inpdata}}

\begin{table}[p] \begin{center}{%
\begin {tabular} {|lr||@{\,}r@{\,}r@{\,}r@{\,}r@{\,}r@{\,}r@{\,}r@{\,}r@{\,}r|}
\hline %
\multicolumn{2}{|c||}{~}& \multicolumn{9}{c|}{Correlations} \\
\multicolumn{2}{|c||}{~} & $\MZ$ & $\GZ$ & $\shad$ &
     $\Ree$ &$\Rmu$ & $\Rtau$ & $\Afbze$ & $\Afbzm$ & $\Afbzt$ \\
\hline %
\hline %
\multicolumn{2}{|l}{ $\pzz \chidf\,=\, 169/176$}& 
                                      \multicolumn{9}{c|}{ALEPH} \\
\hline %
 $\MZ$\,[\GeV{}]\hspace*{-.5pc} & 91.1891 $\pm$ 0.0031     &   
  1.000 & \multicolumn{8}{c|}{}\\
 $\GZ$\,[\GeV]\hspace*{-2pc}  &  2.4959 $\pm$ 0.0043     & 
  0.038 & 1.000 & \multicolumn{7}{c|}{}\\ 
 $\shad$\,[nb]\hspace*{-2pc}  &  41.558 $\pm$ 0.057$\pz$ &   
 $-$0.091 & $-$0.383 & 1.000 & \multicolumn{6}{c|}{}\\
 $\Ree$        &  20.690 $\pm$ 0.075$\pz$ &   
  0.102  &~0.004 & ~0.134 & 1.000 & \multicolumn{5}{c|}{}\\
 $\Rmu$        &  20.801 $\pm$ 0.056$\pz$ &   
 $-$0.003 & ~0.012 & ~0.167 & ~0.083 & 1.000 & \multicolumn{4}{c|}{}\\ 
 $\Rtau$       &  20.708 $\pm$ 0.062$\pz$ &   
 $-$0.003 & ~0.004 & ~0.152 & ~0.067 & ~0.093 & 1.000 & \multicolumn{3}{c|}{}\\ 
 $\Afbze$      &  0.0184 $\pm$ 0.0034     &   
 $-$0.047 & ~0.000 & $-$0.003 & $-$0.388 & ~0.000 & ~0.000 & 1.000 & \multicolumn{2}{c|}{}\\ 
 $\Afbzm$      &  0.0172 $\pm$ 0.0024     &   
 0.072 & ~0.002 & ~0.002 & ~0.019 & ~0.013 & ~0.000 & $-$0.008 & 1.000 & \multicolumn{1}{c|}{}\\ 
 $\Afbzt$      &  0.0170 $\pm$ 0.0028     &   
 0.061 & ~0.002 & ~0.002 & ~0.017 & ~0.000 & ~0.011 & $-$0.007 & ~0.016 & ~1.000\\
\hline %
\hline %
\multicolumn{2}{|l}{$\pzz \chidf\,=\, 177/168$} & 
                                        \multicolumn{9}{c|}{DELPHI} \\
\hline %
 $\MZ$\,[\GeV{}]\hspace*{-.5pc}   &  91.1864 $\pm$ 0.0028    &
 1.000 & \multicolumn{8}{c|}{}\\ 
 $\GZ$\,[\GeV]\hspace*{-2pc}   &  2.4876 $\pm$ 0.0041     &
 ~0.047 & 1.000 & \multicolumn{7}{c|}{}\\ 
 $\shad$\,[nb]\hspace*{-2pc}   &  41.578 $\pm$ 0.069$\pz$ &
 $-$0.070 & $-$0.270 & 1.000 & \multicolumn{6}{c|}{}\\ 
 $\Ree$        &  20.88  $\pm$ 0.12$\pzz$ &
 ~0.063 & ~0.000 & ~0.120 & 1.000 & \multicolumn{5}{c|}{}\\ 
 $\Rmu$        &  20.650 $\pm$ 0.076$\pz$ &
 $-$0.003 & $-$0.007 & ~0.191 & ~0.054 & 1.000 & \multicolumn{4}{c|}{}\\ 
 $\Rtau$       &  20.84  $\pm$ 0.13$\pzz$ &
 ~0.001 & $-$0.001 & ~0.113 & ~0.033 & ~0.051 & 1.000 & \multicolumn{3}{c|}{}\\ 
 $\Afbze$      &  0.0171 $\pm$ 0.0049     &
 ~0.057 & ~0.001 & $-$0.006 & $-$0.106 & ~0.000 & $-$0.001 & 1.000 & \multicolumn{2}{c|}{}\\ 
 $\Afbzm$      &  0.0165 $\pm$ 0.0025    &
 ~0.064 & ~0.006 & $-$0.002 & ~0.025 & ~0.008 & ~0.000 & $-$0.016 & 1.000 & \multicolumn{1}{c|}{}\\ 
 $\Afbzt$      &  0.0241 $\pm$ 0.0037     & 
 ~0.043 & ~0.003 & $-$0.002 & ~0.015 & ~0.000 & ~0.012 & $-$0.015 & ~0.014 & 1.000 \\
\hline %
\hline %
\multicolumn{2}{|l}{$\pzz \chidf\,=\, 158/166 $}  & 
                                    \multicolumn{9}{c|}{L3} \\
\hline %
 $\MZ$\,[\GeV{}]\hspace*{-.5pc}   &  91.1897 $\pm$ 0.0030      & 
 1.000 & \multicolumn{8}{c|}{}\\ 
 $\GZ$\,[\GeV]\hspace*{-2pc}   &   2.5025 $\pm$ 0.0041      & 
 ~0.065 & 1.000 & \multicolumn{7}{c|}{}\\ 
 $\shad$\,[nb]\hspace*{-2pc}   &   41.535 $\pm$ 0.054$\pz$  & 
 ~0.009 & $-$0.343 & 1.000 & \multicolumn{6}{c|}{}\\ 
 $\Ree$        &   20.815  $\pm$ 0.089$\pz$ & 
 ~0.108 & $-$0.007 & ~0.075 & 1.000 & \multicolumn{5}{c|}{}\\ 
 $\Rmu$        &   20.861  $\pm$ 0.097$\pz$ & 
 $-$0.001 & ~0.002 & ~0.077 & ~0.030 & 1.000 & \multicolumn{4}{c|}{}\\ 
 $\Rtau$       &   20.79 $\pz\pm$ 0.13$\pzz$& 
 ~0.002 & ~0.005 & ~0.053 & ~0.024 & ~0.020 & 1.000 & \multicolumn{3}{c|}{} \\ 
 $\Afbze$      &   0.0107 $\pm$ 0.0058      & 
 $-$0.045 & ~0.055 & $-$0.006 & $-$0.146 & $-$0.001 & $-$0.003 & 1.000 & \multicolumn{2}{c|}{} \\ 
 $\Afbzm$      &   0.0188 $\pm$ 0.0033      & 
 ~0.052 & ~0.004 & ~0.005 & ~0.017 & ~0.005 & ~0.000 & ~0.011 & 1.000 & \multicolumn{1}{c|}{} \\ 
 $\Afbzt$      &   0.0260 $\pm$ 0.0047      & 
 ~0.034 & ~0.004 & ~0.003 & ~0.012 & ~0.000 & ~0.007 & $-$0.008 & ~0.006 & 1.000 \\
\hline %
\hline %
\multicolumn{2}{|l}{$\pzz \chidf\,=\, 155/194 $} & 
                                      \multicolumn{9}{c|}{OPAL} \\
\hline %
 $\MZ$\,[\GeV{}]\hspace*{-.5pc}   & 91.1858 $\pm$ 0.0030 &
 1.000 & \multicolumn{8}{c|}{}\\ 
 $\GZ$\,[\GeV]\hspace*{-2pc}   & 2.4948  $\pm$ 0.0041 & 
 ~0.049 & 1.000 & \multicolumn{7}{c|}{}\\ 
 $\shad$\,[nb]\hspace*{-2pc}   & 41.501 $\pm$ 0.055$\pz$ & 
 ~0.031 & $-$0.352 & 1.000 & \multicolumn{6}{c|}{}\\ 
 $\Ree$        & 20.901 $\pm$ 0.084$\pz$ & 
 ~0.108 & ~0.011 & ~0.155 & 1.000 & \multicolumn{5}{c|}{}\\ 
 $\Rmu$        & 20.811 $\pm$ 0.058$\pz$ &
 ~0.001 & ~0.020 & ~0.222 & ~0.093 & 1.000 &  \multicolumn{4}{c|}{}\\ 
 $\Rtau$       & 20.832 $\pm$ 0.091$\pz$ &
 ~0.001 & ~0.013 & ~0.137 & ~0.039 & ~0.051 & 1.000 & \multicolumn{3}{c|}{}\\ 
 $\Afbze$      & 0.0089 $\pm$ 0.0045 &
 $-$0.053 & $-$0.005 & ~0.011 & $-$0.222 & $-$0.001 & ~0.005 & 1.000 & \multicolumn{2}{c|}{}\\ 
 $\Afbzm$     & 0.0159 $\pm$ 0.0023 &
 ~0.077 & $-$0.002 & ~0.011 & ~0.031 & ~0.018 & ~0.004 & $-$0.012 & 1.000 & \multicolumn{1}{c|}{}\\ 
 $\Afbzt$      & 0.0145 $\pm$ 0.0030 & 
 ~0.059 & $-$0.003 & ~0.003 & ~0.015 & $-$0.010 & ~0.007 & $-$0.010 & ~0.013 & 1.000 \\
\hline %
\end{tabular}}%
\end{center}
\vskip-1.5pc\caption[Nine parameter results] {\label{tab:Zparinput}
  Individual results on Z parameters and their correlation
  coefficients from the four experiments. Systematic errors are
  included here except those summarised in Table~\ref{tab:QEDtherr}.}
\end{table} 

The common set of pseudo-observables used for the parametrisation of
the differential cross-section, as described in the introductory
chapter, was extracted by each experiment independently from the
largely model-independent fits to their measured cross-sections and
forward-backward asymmetries~\cite{\ALEPHls,\DELPHIls,\Lls,\OPALls}.
The results presented here deviate slightly from those published by
the experiments in order to facilitate the combination procedure. The
four dedicated sets of fit results for the combination are summarised
in Table~\ref{tab:Zparinput}.

All fits are based on versions 6.23 of ZFITTER and 4.4 of TOPAZ0.  The
published ALEPH results were derived using version 6.10 of ZFITTER,
which did not yet contain the improved treatment of fermion pairs
radiated from the initial state~\cite{Arbuzov}. For the combination
presented here, the ALEPH measurements were re-analysed using version
6.23 of ZFITTER, leading to small changes at the level of a few tenths
of $\MeV$ in $\MZ$ and $\GZ$.

While the individual publications were based on the
experiment-specific energy error matrices, the combined energy error
matrix described above~\cite{bib-ECAL95} was used in the fits for the
input to the combination.  This makes a small difference at the level
of $0.1~\MeV$ on $\MZ$ and $\GZ$ and their errors for L3 only, where
uncertainties arising from the modelling of the radio frequency
cavities are largest.

The calculated $s$-$t$ interference in the Bhabha final state has a
small dependence on the assumed value of the $\Zzero$ mass. Although
this is practically negligible for a single experiment, a consistent
treatment becomes important for the combination. Despite some
different choices in the publications of the individual analyses, all
experiments evaluated the $t$,$s$-$t$ channel correction at their own
value of $\MZ$ for the results presented here. The resulting
interdependencies between the $\Zzero$ mass and the parameters from
the Bhabha final state are explicitly included in the error
correlation coefficients between $\MZ$ and $\Ree$ or $\Afbze$.

The {LEP} experiments agreed to use a standard set of parameters for
the calculation of the $\SM$ remnants (see
Section~\ref{sec:intro_SM_remnants}) in the theory programs. The
important parameters are the $\Zzero$ mass, $\MZ=91.187~\GeV$, the
Fermi constant, $\GF=1.16637\cdot10^{-5}~\GeV^{-2}$, the
electromagnetic coupling constant,
$\alpha(\MZ^2)=1/128.886$,\footnote{ This corresponds to a value of
the correction due to hadronic vacuum polarisation of
$\Delta\alpha^{(5)}_{\rm had}=0.02804\pm0.00065$~\cite{bib-JEG2}.
Note that a more precise value of $\Delta\alpha^{(5)}_{\rm
had}=0.02758\pm0.00035$~\cite{bib-BP05} became available after these
analyses had been finalised.} the strong coupling constant,
$\alpha_s(\MZ^2)=0.119$, the top quark mass, $\Mt=175~\GeV$, and
finally the Higgs mass, $\MH=150~\GeV$.  The dependence of the fit
results arising from uncertainties in these parameters is negligible
except for $\MH$, as discussed in Section~\ref{sec-parerr}.

All experiments also provided fits to their measured cross-sections
and asymmetries with lepton universality imposed, {\em i.\,e.} $\Ree$,
$\Rmu$ and $\Rtau$ are replaced by $\Rl$, and $\Afbze$, $\Afbzm$ and
$\Afbzt$ are replaced by $\Afbzl$ in the model-independent
parametrisation of the differential cross-section.  Here $\Rl$ is not
a simple average over the three lepton species, but refers to $\Zzero$
decays into pairs of a single massless charged lepton species. The
individual experimental results and the correlation matrices are given
in Table~\ref{tab:Zpar5input}.  A graphical overview of the results is
given in Figure~\ref{fig:lsafb}; the averages are those discussed in
Section~\ref{sec-lsafbcombi} below.

\begin{table}[tb] \begin{center}{%
\begin {tabular} {|lr||r@{\,}r@{\,}r@{\,}r@{\,}r@{\,}|}
\hline %
\multicolumn{2}{|c||}{~}& \multicolumn{5}{c|}{Correlations} \\
\multicolumn{2}{|c||}{~} & $\MZ$ & $\GZ$ & $\shad$ & $\Rl$ & $\Afbzl$ \\
\hline %
\hline %
\multicolumn{2}{|l}{$\pzz \chidf\,=\,172/180$}     & 
                            \multicolumn{5}{c|}{ALEPH} \\
\hline %
 $\MZ$\,[\GeV]\hspace*{-3em} & 91.1893 $\pm$ 0.0031     &   
 1.000 & \multicolumn{4}{c|}{}\\
 $\GZ$\,[\GeV]\hspace*{-1em}  &  2.4959 $\pm$ 0.0043     & 
 0.038 & 1.000 & \multicolumn{3}{c|}{}\\ 
 $\shad$\,[nb]\hspace*{-1em}  &  41.559 $\pm$ 0.057$\pz$ &   
 $-$0.092 & $-$0.383 & 1.000 & \multicolumn{2}{c|}{}\\ 
 $\Rl$      &  20.729 $\pm$ 0.039$\pz$ &   
 0.033 & 0.011 & 0.246 & 1.000 & \multicolumn{1}{c|}{}\\
 $\Afbzl$     &  0.0173 $\pm$ 0.0016     &   
 0.071  & 0.002 & 0.001 & $-$0.076 & ~1.000 \\
\hline %
\hline %
\multicolumn{2}{|l}{$\pzz \chidf\,=\,183/172 $} & 
                             \multicolumn{5}{c|}{DELPHI} \\
\hline %
 $\MZ$\,[\GeV]\hspace*{-0.3em}   &  91.1863 $\pm$ 0.0028    &
 1.000 & \multicolumn{4}{c|}{}\\
 $\GZ$\,[\GeV]\hspace*{-1em}   &  2.4876 $\pm$ 0.0041     &
 0.046 & 1.000 & \multicolumn{3}{c|}{}\\
 $\shad$\,[nb]\hspace*{-1em}   &  41.578 $\pm$ 0.069$\pz$ &
 $-$0.070 & $-$0.270  &1.000 & \multicolumn{2}{c|}{}\\
 $\Rl$      &  20.730  $\pm$ 0.060$\pz$ &
 0.028 & $-$0.006 & 0.242 & 1.000 & \multicolumn{1}{c|}{}\\
 $\Afbzl$     &  0.0187 $\pm$ 0.0019     &
 0.095 & 0.006 & $-$0.005 & 0.000 & 1.000 \\
\hline %
\hline %
\multicolumn{2}{|l}{$\pzz \chidf\,=\,163/170$}  & 
                                     \multicolumn{5}{c|}{L3} \\
\hline %
 $\MZ$\,[\GeV]\hspace*{-.3em}   &  91.1894 $\pm$ 0.0030      & 
 1.000 & \multicolumn{4}{c|}{}\\
 $\GZ$\,[\GeV]\hspace*{-1em}   &   2.5025 $\pm$ 0.0041      & 
 0.068 & 1.000 & \multicolumn{3}{c|}{}\\
 $\shad$\,[nb]\hspace*{-1em}   &   41.536 $\pm$ 0.055$\pz$  & 
 0.014 & $-$0.348 & 1.000 & \multicolumn{2}{c|}{}\\
 $\Rl$      &   20.809  $\pm$ 0.060$\pz$ & 
 0.067 & 0.020 & 0.111 & 1.000 & \multicolumn{1}{c|}{}\\
 $\Afbzl$     &   0.0192 $\pm$ 0.0024      & 
 0.041 & 0.020 & 0.005 & $-$0.024 & 1.000 \\
\hline %
\hline %
\multicolumn{2}{|l}{$\pzz \chidf\,=\,158/198$} 
                                      & \multicolumn{5}{c|}{OPAL} \\
\hline %
 $\MZ$\,[\GeV]\hspace*{-.3em}   & 91.1853 $\pm$ 0.0029       &
 1.000 & \multicolumn{4}{c|}{}\\
 $\GZ$\,[\GeV]\hspace*{-1em}   & 2.4947 $\pm$ 0.0041        &
 0.051 & 1.000 & \multicolumn{3}{c|}{}\\
 $\shad$\,[nb]\hspace*{-1em}   & 41.502 $\pm$ 0.055$\pz$    &
 0.030 & $-$0.352 & 1.000 & \multicolumn{2}{c|}{}\\
 $\Rl$      & 20.822 $\pm$ 0.044$\pz$    &
 0.043 & 0.024 & 0.290 & 1.000 & \multicolumn{1}{c|}{}\\
 $\Afbzl$     & 0.0145 $\pm$ 0.0017        &
 0.075 & $-$0.005 & 0.013 & $-$0.017 & 1.000 \\ 
\hline %
\end{tabular}}%
\end{center}
\caption[Five parameter results]{\label{tab:Zpar5input} 
  Results on Z parameters and error correlation matrices from the four
  experiments, with lepton universality imposed. Systematic errors are
  included here except those summarised in Table~\ref{tab:QEDtherr}.}
\end{table} 

Compared with the nine-parameter results of Table~\ref{tab:Zparinput},
there is a noticeable change in $\MZ$ of a few tenths of $\MeV$ in all
experiments. This is a consequence of the dependence of the
$t$-channel correction on $\MZ$, as discussed in
Section~\ref{sec-tcherr}. When $\Ree$ and $\Afbze$ are replaced by the
leptonic quantities $\Rl$ and $\Afbzl$, their correlation with the
$\Zzero$ mass leads to a shift, which is driven by the (statistical)
difference between $\Ree$ and $\Rl$ and $\Afbze$ and $\Afbzl$.
Similarly, replacing $\Ree$ and $\Afbze$ from the values of a single
experiment by the {LEP} average introduces a shift in $\MZ$ in the
presence of these particular correlation coefficients. Such a shift
should be smaller when averaged over the four experiments, and indeed
this is observed with the average of the shifts being only
$-0.2~\MeV$.

\begin{figure}[p]
\begin{center}
\begin{tabular}{ll}
\mbox{\epsfig{file=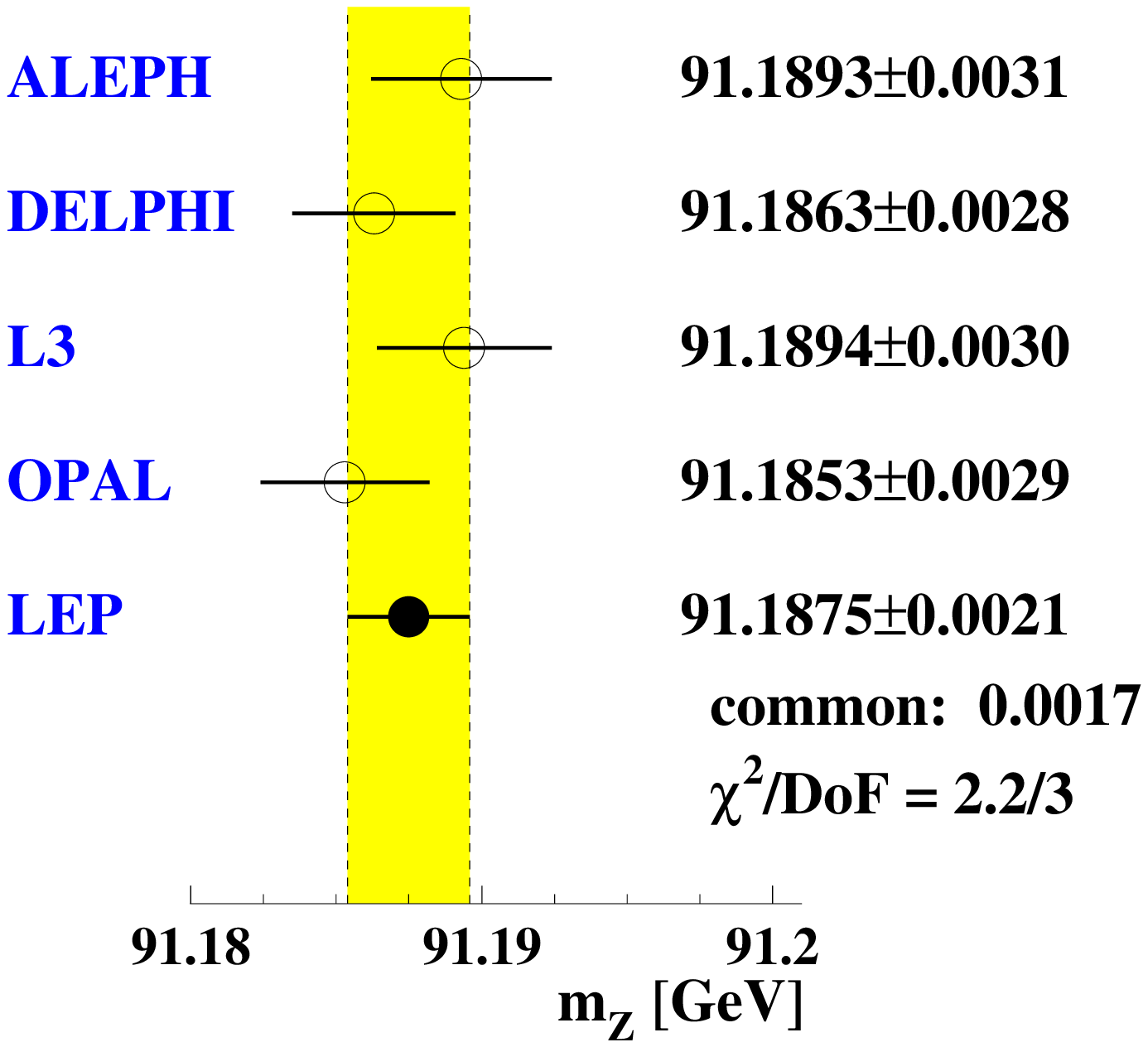,width=0.49\textwidth}}  &
\mbox{\epsfig{file=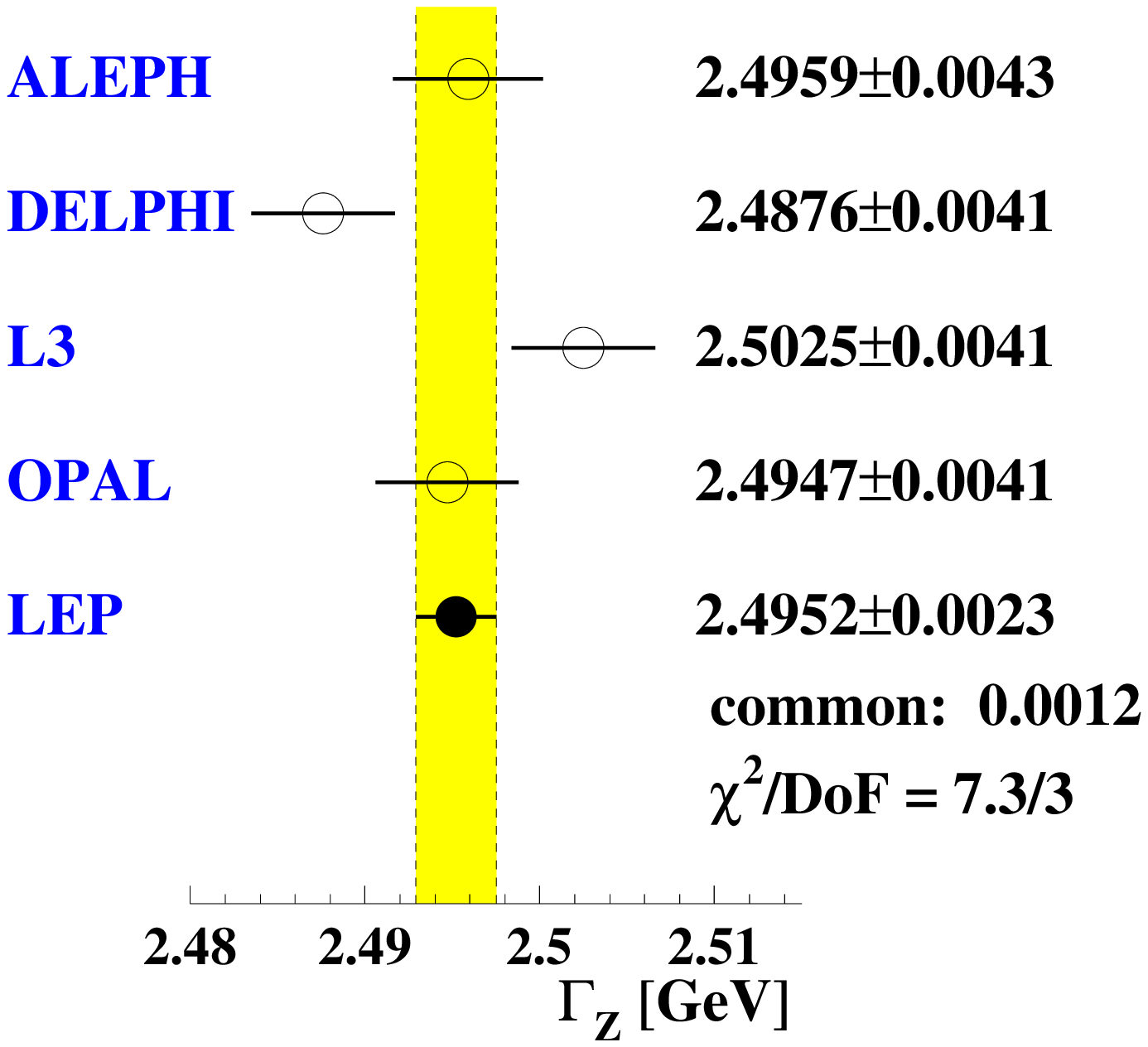,width=0.49\textwidth}} \\
\mbox{\epsfig{file=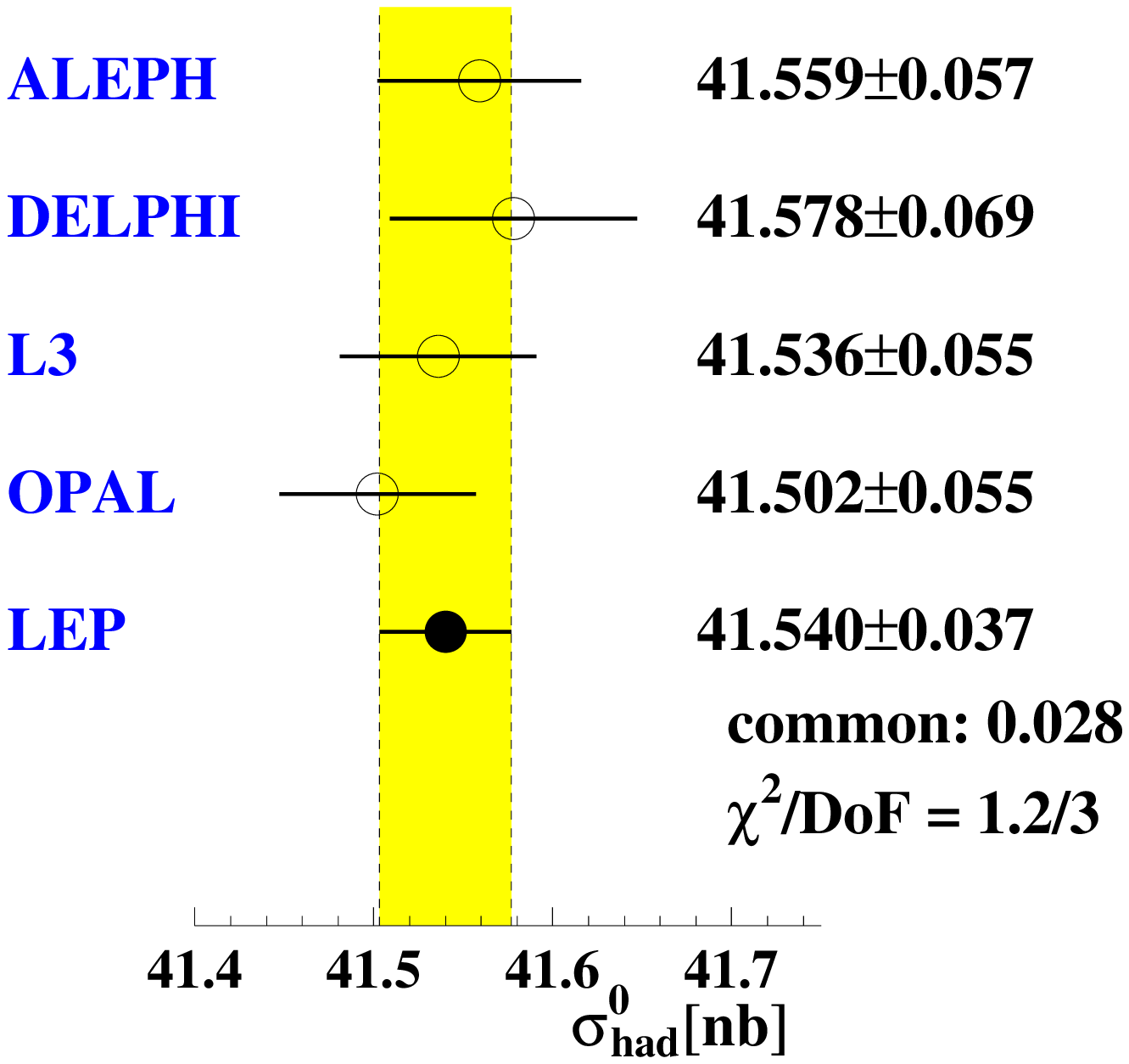,width=0.49\textwidth}} &
\mbox{\epsfig{file=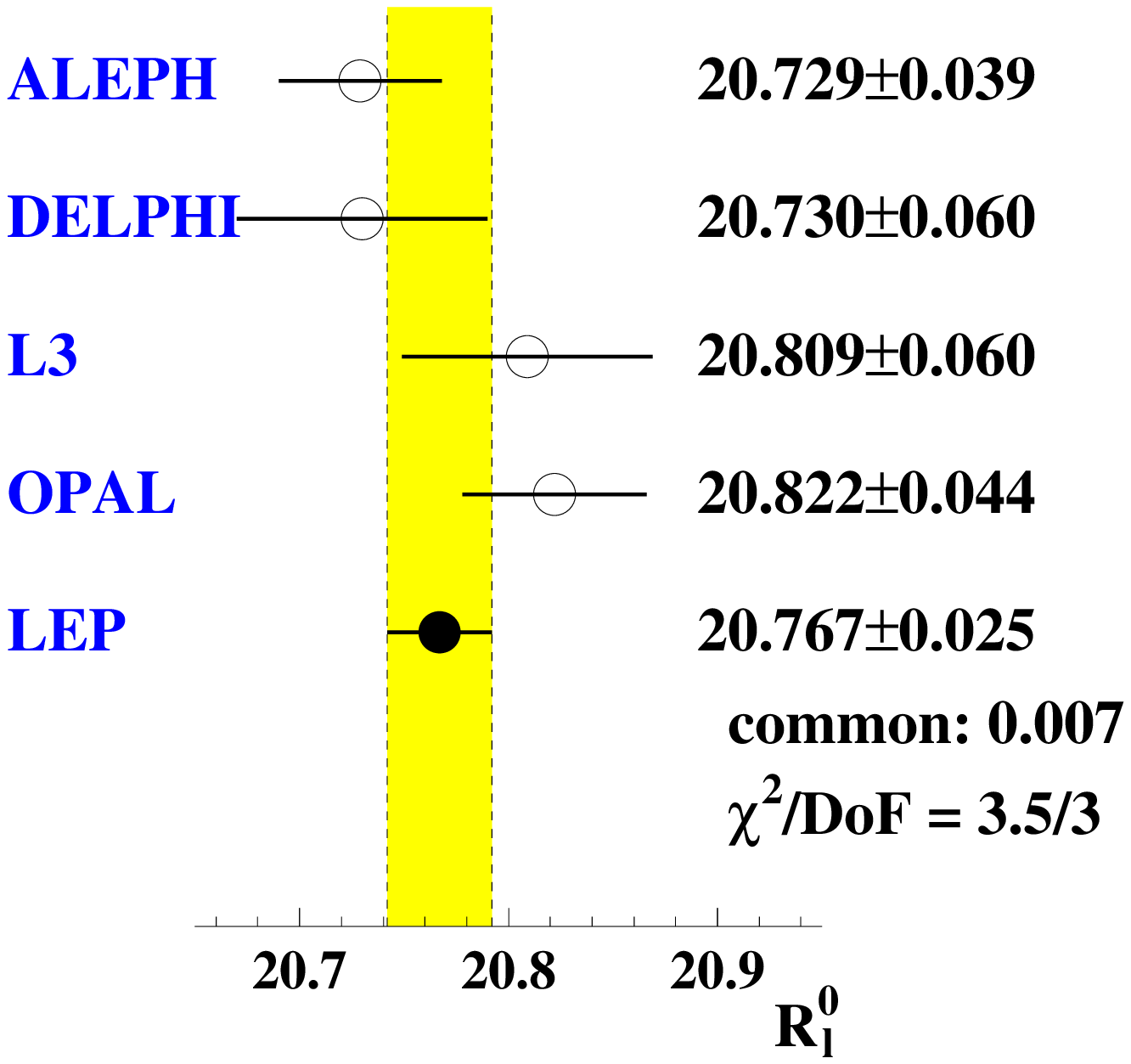,width=0.49\textwidth}} \\
\mbox{\epsfig{file=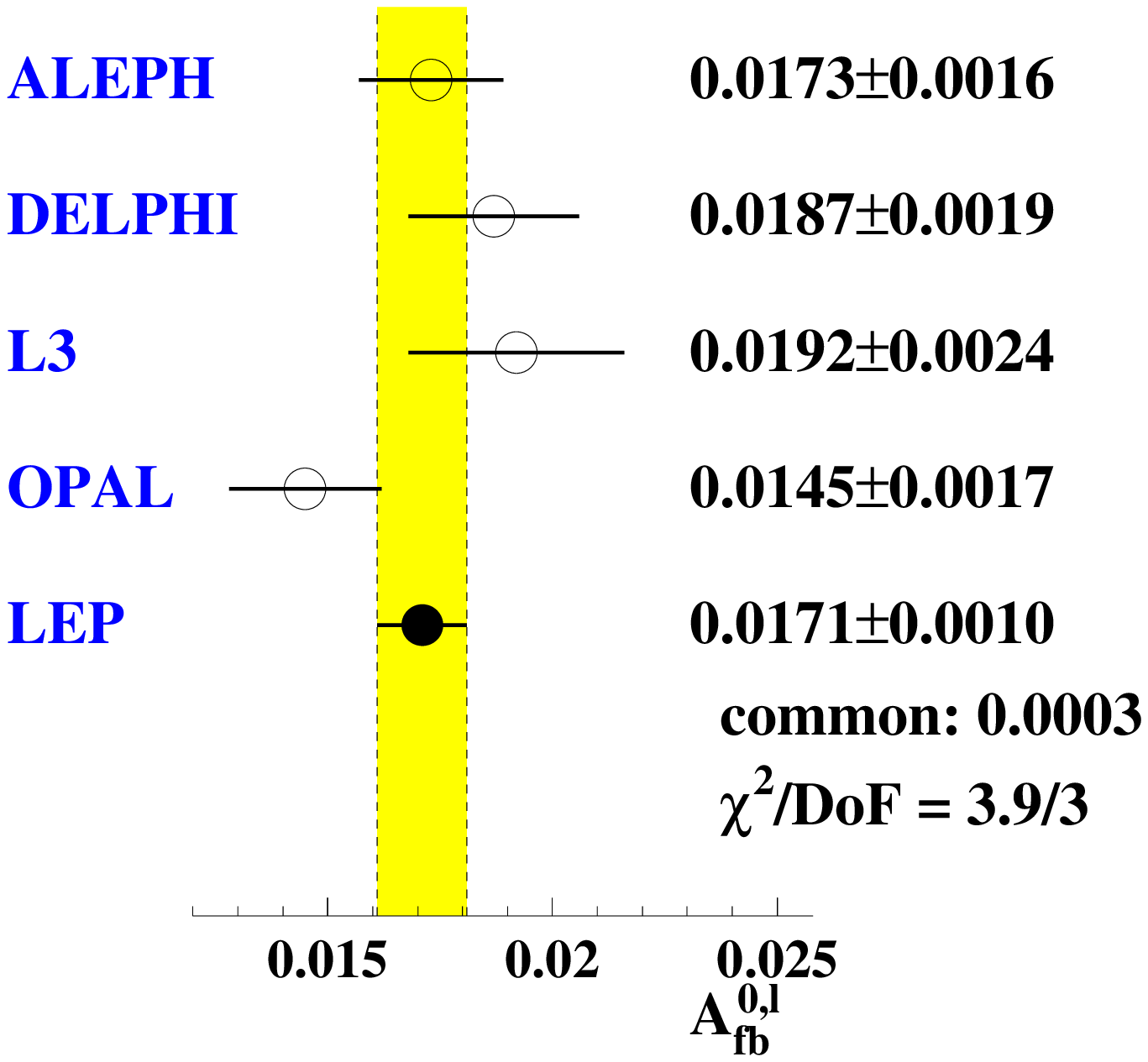,width=0.49\textwidth}} &
\begin{minipage}[b]{0.45\textwidth}
\caption[Measurements of of $\MZ$,$\GZ$, $\shad$, $\Rl$ and $\Afbzl$]
{\label{fig:lsafb} Measurements of $\MZ$, $\GZ$, $\shad$, $\Rl$ and
  $\Afbzl$. The averages indicated were obtained using the common
  errors and combination method discussed in the text.  The values of
  $\chi^2$ per degree of freedom were calculated considering error
  correlations between measurements of the same parameter, but not
  error correlations between different parameters.  \\}
\end{minipage}
\end{tabular}
\end{center}
\end{figure}

\section{Common Uncertainties \label{sec-comerr}}

Important common errors among the results from all {LEP} experiments
arise from several sources. These include the calibration of the beam
energy, the theoretical error on the calculation of the small-angle
Bhabha cross-section used as the normalisation reaction, the
theoretical uncertainties in the $t$-channel and $s$-$t$ interference
contribution to the differential large-angle Bhabha cross-section, the
theoretical uncertainties in the calculations of QED radiative effects
and, finally, from small uncertainties in the parametrisation of the
electroweak cross-section near the $\Zzero$ resonance in terms of the
standard set of pseudo-observables. These common errors are quantified
below and are used in the combination.

For the purpose of combining the experimental results at the parameter
level, the common sources of error on each individual cross-section or
asymmetry measurement need to be transformed into errors on the
extracted pseudo-observables.  A popular method to measure the
contribution of an error component of an input quantity on a fit
parameter is to set the input error to zero and repeat the fit.
However, this will also lead to shifts in central values if the
omitted error component is large. Such shifts indicate that the
internal weighting of inputs has changed, and the estimated error
contribution may be wrong. A better method therefore is to examine the
changes in the error matrices resulting from special fits with only
slightly modified input errors.  The contribution of each input error
component to the full covariance matrix of the fit parameters can then
be determined under the constraint of maintaining the weights at their
actual values. This method will be used and illustrated more clearly
in the following sections.

\subsection{Energy Calibration Uncertainties \label{sec-comenerr}}

The first step in the determination of common energy related
uncertainties on the pseudo-observables is for each experiment to
scale the energy errors by factors of $1\pm\epsilon$, while
maintaining the experimental errors fixed. Typical values of
$\epsilon$ used are between 5\,\% and 20\,\%. Performing the standard
fits to the cross-section and asymmetry measurements with these scaled
errors generates two pseudo-observable covariance matrices, $V_\pm$,
from which the covariance matrix due to energy errors, $V_E$, can be
separated from the other errors, $V_{\rm exp}$, using the relation
$(V_\pm) = (1 \pm \epsilon)^2 (V_E) + (V_{\rm exp})$. The validity of
this procedure was verified using a data set restricted to the
hadronic cross-section measurements of the years 1993--1995, which
were combined both at the cross-section level and at the parameter
level.

The estimated energy errors differ only slightly depending on which
experimental data set is used to derive them. Combinations may be
attempted based on each of them, or on the average.  For all such
choices the central values and errors of each of the averaged
parameters agree to well within 5\,\% of the error.  It is therefore
most appropriate to take the average of the error estimates over the
experiments as the common energy errors, which are shown in
Table~\ref{tab:cove9}.

\begin{table}[ht] \begin{center} \begin{tabular}{lr}
\begin{minipage}[c]{0.57\textwidth}
\begin{tabular} {|l||rrrr|}
\hline %
  ~     &   $\MZ$     & $\GZ$      & $\shad$ & $\Ree$ \\
\hline %
\hline %
$\MZ$\,[\GeV]\hspace*{-.3em}   &   0.0017 & & &  \\
$\GZ$\,[\GeV]\hspace*{-1em}     &$-$0.0006 &   0.0012 &  & \\
$\shad$\,[nb]\hspace*{-1em}     &$-$0.0018 &$-$0.0027 &   0.011 & \\
$\Ree$  &   0.0017 &$-$0.0014   &   0.0073 & 0.013 \\
\hline %
\end{tabular}
\end{minipage}  
      &
\begin{minipage}[c]{0.4\textwidth}
\begin{tabular} {|l||rrr|}
\hline %
    ~    & $\Afbze$ & $\Afbzm$ & $\Afbzt$ \\
\hline %
\hline %
$\Afbze$ &   0.0004 &  &  \\
$\Afbzm$ &$-$0.0003 &   0.0003 &  \\
$\Afbzt$ &$-$0.0003 &   0.0003 & 0.0003 \\
\hline %
\end{tabular}
\end{minipage}  \\
\end{tabular}
\caption[Common energy errors for nine-parameter fits]
{\label{tab:cove9} Common energy errors for nine-parameter fits.
  Values are given as the signed square root of the covariance matrix
  elements; elements above the diagonal have been omitted for
  simplicity.  The anti-correlation between electron and muon or tau
  asymmetries arises from the different energy dependence of the
  electron asymmetry due to the $t$-channel contribution.}
\end{center}
\end{table}

\subsection{Uncertainties Related to the $t$-Channel \label{sec-tcherr}}

The $t$ channel and $s$-$t$ interference contributions are calculated
in the $\SM$ using the programs ALIBABA~\cite{ALIBABA} and
TOPAZ0~\cite{\TOPAZref}.  The theoretical uncertainty on the
$t$-channel correction is discussed in detail in
Reference~\citen{bib-tierror1}.  The size of the uncertainty is
typically 1.1~pb for the forward cross-section and 0.3~pb for the
backward cross-section and depends slightly on the acceptance
cuts~\cite{bib-tierror2}.  All collaborations incorporate the theory
uncertainty as an additional error on the electron pair cross-section
and asymmetry. In order to evaluate the common error due to the
$t$,\,$s$-$t$ theory error, each collaboration performed two fits,
with and without the theory error, and the quadratic differences of
the covariance matrix elements for $\Ree$ and $\Afbze$ are taken as an
estimate of the common error. The unknown error correlation between
energy points below and above the peak is included in the error
estimates by adding in quadrature the observed shifts in mean values
of $\Ree$ and $\Afbze$ when varying this correlation between $-1$ and
$+1$. The $t$,\,$s$-$t$ related errors estimated by individual
experiments are all very similar, and therefore the average is taken
as the common error matrix, as shown in Table~\ref{tab:tch_syst}.

\begin{table}[ht] \begin{center}
\begin{tabular} {|l||rr|}
\hline %
     ~    & $\Ree$ & $\Afbze$ \\
\hline %
\hline %
 $\Ree$   &   0.024$\pz$ &            \\
 $\Afbze$ &$-$0.0054  & 0.0014     \\
\hline %
\end{tabular}
\caption [Common $t$,\,$s-t$ uncertainties]
{\label{tab:tch_syst} Common uncertainties arising from the $t$
  channel and $s$-$t$ interference contribution to the $\ee$ final
  states, given as the signed square root of the covariance matrix
  elements.}
\end{center}
\end{table} 

The $s$-$t$ interference contribution to the $t$-channel correction in
Bhabha final states depends on the value of the $\Zzero$ mass. For the
purpose of this combination, all experiments parametrise the $t$ and
$s$-$t$ contributions as a function of $\MZ$. This allows the
$t$,~$s$-$t$ correction to follow the determination of $\MZ$ in the
fits, which results in correlations between $\MZ$ and $\Ree$ or
$\Afbze$. Typical changes of the correlation coefficients amount to
about +10\,\% for the correlation $\MZ$--$\Ree$ and $-$10\,\% for
$\MZ$--$\Afbze$. The presence of these correlations induces changes in
$\Ree$ and $\Afbze$ when $\MZ$ takes its average value in the
combination of the four experiments.

\subsection{Luminosity Uncertainties \label{sec-lumerr}}

The four collaborations use similar techniques to measure the
luminosity of their data samples by counting the number of small-angle
Bhabha-scattered electrons.  The experimental details of the four
measurements are sufficiently different that no correlations are
considered to exist in the experimental component of the luminosity
errors.  All four collaborations, however, use
BHLUMI~4.04~\cite{BHLUMI4}, the best available Monte Carlo generator
for small-angle Bhabha scattering, to calculate the accepted
cross-section of their luminosity counters. Therefore significant
correlations exist in the errors assigned to the scale of the measured
cross-sections due to the uncertainty in this common theoretical
calculation.

The total theoretical uncertainty, including an estimate large enough
to cover the entire contribution from light fermion pair production,
which is not included in BHLUMI, is 0.061\,\%~\cite{BHLUMI061}. The
contribution of light pairs has been calculated~\cite{\Lpairs}, and 
explicit inclusion of this effect allowed
OPAL to reduce the theoretical uncertainty to 0.054\,\%.  This
0.054\,\% error is taken to be fully correlated with the errors of the
other three experiments, which among themselves share a mutual
correlated error of 0.061\,\%.

These errors affect almost exclusively the hadronic pole
cross-section, and contribute about half its total error after
combination.  The common luminosity error also introduces a small
contribution to the covariance matrix element between $\GZ$ and
$\shad$.  This correlation was neglected in the common error tables
given above, as it had no noticeable effect on the combined result.

\subsection{Theory Uncertainties \label{sec-therr}}

An additional class of common theoretical errors arises from the
approximations and special choices made in the fitting codes.  These
comprise contributions from QED radiative corrections, including
initial-state pair radiation, and the parametrisation of the
differential cross-section around the $\Zzero$ resonance in terms of
pseudo-observables defined precisely at the peak and for pure $\Zzero$
exchange only.  In order to estimate the uncertainties from the
parametrisation of the electroweak cross-sections near the $\Zzero$
resonance the two most advanced calculational tools,
TOPAZ0~\cite{\TOPAZref} and ZFITTER~\cite{\ZFITTERref} were compared. In
addition, there are ``parametric uncertainties'' arising from
parameters of the $\SM$ that are needed to fix the $\SM$ remnants.

\subsubsection{QED Uncertainties \label{sec-QEDerr}}

The effects of initial state radiation (ISR) are more than two orders
of magnitude larger than the experimental precision, which is below
the per-mille level in the case of the hadronic cross-section. The
radiation of fermion pairs (ISPP), although much smaller than ISR in
absolute effect, exhibits a larger uncertainty. Therefore these
corrections play a central role in the extraction of the
pseudo-observables from the measured cross-sections and asymmetries.

The most up-to-date evaluations of photonic corrections to the
measurements are complete in $\calO(\alpha^2)$ and for the total
cross-sections also include the leading contributions up to
$\calO(\alpha^3)$. Two different schemes are available to estimate the
remaining uncertainties:
\begin{enumerate}
{\parsep=0pt \itemsep=0pt \topsep=0pt \parskip=0pt \partopsep=0pt}
\item KF: $\calO(\alpha^2)$ calculations~\cite{Berends88} including
  the exponentiation scheme of Kuraev-Fadin~\cite{KF} with
  $\calO(\alpha^3)$~\cite{Montagna}.\footnote{ Third-order terms for
  the KF scheme had also been calculated
  previously~\cite{Skrzypek92}.}
\item YFS: the 2$^{\rm nd}$ order inclusive exponentiation scheme of
  References~\citen{JSW} and~\citen{Skrzypek92}, based on the YFS
  approach~\cite{YFS}.  Third order terms are also known and have only
  a small effect~\cite{JPS}.
\end{enumerate}
Differences between these schemes, which are both implemented in
TOPAZ0, ZFITTER and MIZA~\cite{\MIZA}, and uncertainties due to missing
higher order corrections~\cite{JPS}, amount to at most $\pm 0.1~\MeV$
on $\MZ$ and $\GZ$, and $\pm 0.01$\,\% on $\shad$.

The influence of the interference between initial and final state
radiation on the extracted parameters has also been studied~\cite{bib-ifi}, 
and uncertainties on $\MZ$ of at most $\pm
0.1~\MeV$ from this source are expected for experimental measurements
which accept events down to small values of $s'$, the effective
squared centre-of-mass energy after photon radiation from the initial
state.  The methods for the extrapolation of the leptonic $s$-channel
cross-sections to full angular acceptance and from large to small $s'$
differ among the experiments and therefore the resulting uncertainties
are believed to be largely uncorrelated.

Although contributing only 1\,\% of the ISR correction, the radiation
of fermion pairs from the initial state dominates the QED related
uncertainties. Starting from the full second order pair
radiator~\cite{Berends88,ISPP88}, a simultaneous exponentiation scheme
for radiated photons and pairs was proposed in
Reference~\citen{JSMpairs}.  A third-order pair radiator was
calculated~\cite{Arbuzov} and compared with the other existing
schemes, which are all available in ZFITTER since version 6.23.
Independent implementations of some schemes exist in TOPAZ0 and in
MIZA. The largest uncertainty arises from the sub-sub-leading terms of
the third order and from the approximate treatment of hadronic pairs.
The maximum differences are $0.3~\MeV$ on $\MZ$, $0.2~\MeV$ on $\GZ$
and $0.015$\,\% on $\shad$.
 
In summary, comparing the different options for photonic and
fermion-pair radiation leads to error estimates of $\pm 0.3~\MeV$ on
$\MZ$ and $\pm 0.2~\MeV$ on $\GZ$. The observed differences in
$\shad$ are slightly smaller than the error estimate of $\pm 0.02\,$\%
in Reference~\citen{JPS}, which is therefore taken as the error for
the QED-related uncertainties.

\subsubsection{Choice of Parametrisation of Lineshape and Asymmetries
\label{sec-paramLS}}

In a very detailed comparison~\cite{PCP99} of TOPAZ0 and ZFITTER,
cross-sections and asymmetries from $\SM$ calculations and from
differing choices in the model-independent parametrisation were
considered. Uncertainties on the fitted pseudo-observables may be
expected to arise from these choices in parametrisation of the
electroweak cross-sections near the $\Zzero$ resonance.  To evaluate
such differences, cross-sections and forward-backward asymmetries were
calculated with TOPAZ0 and these results fitted with ZFITTER. Errors
were assigned to the calculated cross-sections and forward-backward
asymmetries which reflect the integrated luminosity taken at each
energy, thus ensuring that each energy point entered with the
appropriate weight.

The dominant part of the small differences between the two codes
results from details of the implementation of the cross-section
parametrisation in terms of the pseudo-observables. This is
particularly visible for the off-peak points, where the assignment of
higher-order corrections to the $\Zzero$ resonance or to the $\SM$
remnants is not in all cases unambiguous. The size of the differences
also depends on the particular values of the pseudo-observables, since
these do not necessarily respect the exact $\SM$ relations.  Slightly
different choices are made in the two codes if the $\SM$ relations
between the pseudo-observables are not fulfilled.  Finally, variations
of factorisation schemes and other options in the electroweak
calculations may affect the fit results through the $\SM$ remnants,
but were found to have a negligible effect.

In Table~\ref{tab:therr} differences between TOPAZ0 and ZFITTER are
shown, which are taken as systematic uncertainties. They were
evaluated around the set of pseudo-observables representing the
average of the four experiments; cross-sections and asymmetries were
calculated for full acceptance with only a cut on $s'\,>\,0.01\,s$.
The only non-negligible systematic error of this kind is that on
$\Rl$, which amounts to 15\,\% of the combined error.

\begin{table}[ht] \begin{center}
\begin{tabular}{|c|c|c|c|c|}
\hline %
$\Delta\MZ$  &$\Delta\GZ$  &$\Delta\shad$ &$\Delta\Rl$ &$\Delta\Afbzl$ \\
{}[\GeV]   & [\GeV]   & [nb]     &   ~      &   ~       \\
\hline %
\hline %
 0.0001  &  0.0001  & 0.001    &  0.004   &  0.0001 \\
\hline %
\end{tabular} 
\caption[TOPAZ0-ZFITTER differences]{\label{tab:therr}
  Differences in fit results obtained with TOPAZ0 and ZFITTER, taken
  as part of the common systematic errors.}
\end{center} 
\end{table} 

Putting all sources together, the overall theoretical errors as listed
in Table~\ref{tab:QEDtherr} are obtained, and these are used as
common errors in the combination.

\begin{table}[bt] \begin{center}
\begin{tabular} {|l||rrrrrrrrr|}
\hline %
  ~ &$\MZ$&$\GZ$&$\shad$&$\Ree$&$\Rmu$&$\Rtau$&$\Afbze$&$\Afbzm$&$\Afbzt$\\
\hline %
\hline %
$\MZ$[\GeV]&0.0003& & & & & & & & \\
$\GZ$[\GeV]& ~   & 0.0002 & & & & & & & \\
$\shad$[nb] & ~   &   ~   & 0.008 & & & & & &  \\
$\Ree$  & ~   &   ~   &    ~  & 0.004 & & & & & \\
$\Rmu$  & ~   &   ~   &    ~  & 0.004 & 0.004 & & & & \\     
$\Rtau$ & ~   &   ~   &    ~  & 0.004 & 0.004 & 0.004 & & & \\   
$\Afbze$ & ~  &   ~   &    ~  &  ~    &       &  ~ &0.0001& & \\
$\Afbzm$ & ~  &   ~   &    ~  &  ~    &       &  ~ &0.0001&0.0001& \\
$\Afbzt$ & ~  &   ~   &    ~  &  ~    &       &  ~ &0.0001&0.0001&0.0001\\
\hline %
\end{tabular} 
\caption[QED-related common errors]{\label{tab:QEDtherr}
  Common theoretical errors due to photon and fermion-pair radiation
  and the choice of model-independent parametrisation, given as the
  signed square root of the covariance matrix elements.}
\end{center} 
\end{table} 

\subsubsection{Parametric Uncertainties \label{sec-parerr}}

Through the $\SM$ remnants the fit results depend slightly on the
values of some $\SM$ parameters. Varying these within their present
experimental errors, or between $100~\GeV$ and $1000~\GeV$ in case
of the Higgs boson mass, leads to observable effects only on the
$\Zzero$ mass, which is affected through the $\gammaZ$ interference
term. The dominant dependence is on $\MH$, followed by
$\alpha^{(5)}_{\rm had}(\MZ^2)$.

The effect on $\MZ$ from a variation of $\Delta\alpha^{(5)}_{\rm
had}(\MZ^2)$ by its error of $\pm$0.00065 is $\pm0.05~\MeV$, which is
negligibly small compared to the systematic error on $\MZ$ arising
from other QED-related uncertainties (see Table~\ref{tab:QEDtherr}).
The change in $\MZ$ due to $\MH$ amounts to $+0.23~\MeV$ per unit
change in $\log_{10}(\MH/\GeV)$.  Note that this is small compared to
the total error on $\MZ$ of $\pm2.1~\MeV$ and is not considered as an
error, but rather as a correction to be applied if and when a $\SM$
Higgs boson is found and its mass measured. The consequences of a
completely model-independent treatment of the $\gammaZ$ interference
in the hadronic channel are discussed in Section~\ref{sec-jhad}.

\section{Combination of Results \label{sec-lsafbcombi}}

The combination of results on the $\Zzero$ parameters is based on the
four sets of nine parameters $\MZ$, $\GZ$, $\shad$, $\Ree$, $\Rmu$,
$\Rtau$, $\Afbze$, $\Afbzm$ and $\Afbzt$ and the common errors given
in the previous chapter.

For this purpose it is necessary to construct the full $(4 \times 9
)\, \times\,(4 \times 9)$ covariance matrix of the errors. The four
on-diagonal $9 \times 9$ matrices consist of the four error matrices
specified by each experiment (Table~\ref{tab:Zparinput}). The $9
\times 9$ common error matrices build the off-diagonal elements.

A symbolic representation of the full error matrix is shown in
Table~\ref{tab:covlsafb}. Each table element represents a $9 \times 9$
matrix; $(\calC_{exp})$ for $exp$~=~A, D, L and O are the covariance
matrices of the experiments (see Table~\ref{tab:Zparinput}), and
$(\calC_c) = (\calC_E) + (\calC_\calL) + (\calC_t) + (\calC_{\rm
QED,th})$ is the matrix of common errors.  $(\calC_E)$
(Table~\ref{tab:cove9}) is the error matrix due to {LEP} energy
uncertainties, $(\calC_\calL)$ (Section~\ref{sec-lumerr}) arises
from the theoretical error on the small-angle Bhabha cross-section
calculations, $\calC_t$ (Table~\ref{tab:tch_syst}) contains the
errors from the $t$-channel treatment in the $\ee$ final state, and
$(\calC_{\rm QED,th})$ contains the errors from initial state photon
and fermion pair radiation and from the model-independent
parametrisation (Table~\ref{tab:QEDtherr}).  Since the latter errors
were not included in the experimental error matrices, they were also
added to the block matrices in the diagonal of
Table~\ref{tab:covlsafb}.

\begin{table} [tb] \begin{center} \begin{tabular}{|l||cccc|}
\hline %
 $(\calC)$ &  {ALEPH}        & {DELPHI}        &  {L3}          &  {OPAL}       \\ 
\hline
\hline
A &$(\calC_A)+(\calC_{\rm QED,th})$ & & & \\
D &$(C_c)$  & $(\calC_D)+(\calC_{\rm QED,th})$ & & \\
L &$(C_c)$  &$(C_c)$   & $(\calC_L)+(\calC_{\rm QED,th})$ & \\ 
O &$(C_c)$  &$(C_c)$   & $(C_c)$   &$(\calC_O)+(\calC_{\rm QED,th})$ \\ 
\hline %
\end{tabular}
\caption[Covariance matrix of combined lineshape and asymmetry measurements]
{\label{tab:covlsafb} Symbolic representation of the covariance
  matrix, $(\calC)$, used to combine the lineshape and asymmetry
  results of the four experiments. The components of the matrix are
  explained in the text.}
\end{center}
\end{table}

The combined parameter set and its covariance matrix are obtained from
a $\chi^2$ minimisation, with
\begin{equation}\label{equ-chi2comb}
 \chi^2 ~=~ ({\bf X} - {\bf X_m})^T (\calC)^{-1} ({\bf X}-{\bf X_m});
\end{equation}
$({\bf X} - {\bf X_m})$ is the vector of residuals of the combined
parameter set to the individual results.

Some checks of the combination procedure outlined above are described
in the following subsections, and the combined results are given in
the tables of Section~\ref{sec-lsafbcombres}.

\subsection{Multiple Z-Mass Fits\label{sec-elevenpar}}

In 1993 and 1995, the two years when {LEP} performed precision energy
scans to measure the $\Zzero$ lineshape, the experimental errors are
very comparable, but the {LEP} energy was appreciably better
understood in 1995 than in 1993. For a single experiment the errors
are not dominated by those from the energy. This changes if the
combined data set is considered, since then energy errors are
comparable in size to the combined experimental errors.  In
determining the optimum value of $\MZ$ in a statistical sense,
therefore, more weight should be given to the 1995 data for four
experiments combined than is given to the 1995 data in the independent
determinations. To quantify this issue the measurements of each
experiment were fit to determine independent values of $\MZ$ for the
three periods 1990--1992, 1993--1994 and 1995.  In this
``eleven-parameter fit'', each of the three mass values $\MZA$, $\MZB$
and $\MZC$ has its specific energy error reflecting the different
systematic errors on the absolute energy scale of {LEP}. The
combination of these four sets of 11 parameters was carried out and
thus the relative importance of energy-related and independent
experimental errors on the mass values is properly treated.

When the three values of $\MZ$ are condensed into one, the effects of
the time dependence of the precision in the energy calibration are
taken into account.  The difference of $-$0.2~$\MeV$ from the $\MZ$
value from the nine-parameter fits corresponds to 10\,\% of the
combined error.  All other parameters are identical to their values
from the nine-parameter fit to within less than 5\,\% of the combined
error. This result justifies using the standard combination based on
the nine parameters.

\begin{figure}[t]
\begin{center}
\vskip -1.5cm
\mbox{\epsfig{file=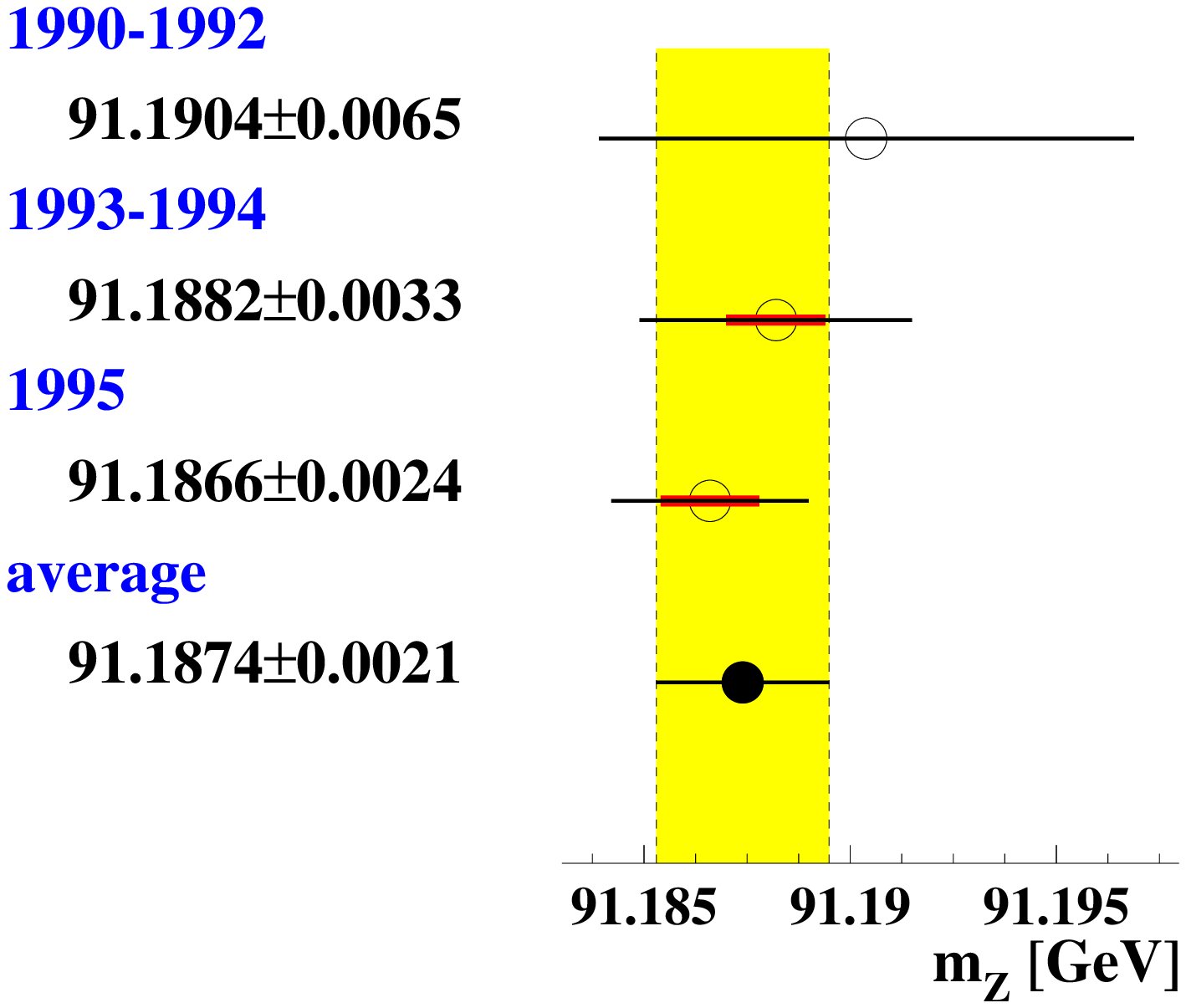,width=0.9\columnwidth}}
\end{center}
\caption[Multiple $\MZ$ fits]
{\label{fig:MZcheck} $\MZ$ in $\GeV$ for three different periods of
  data taking, before 1993, 1993--1994 and 1995.  The second, smaller
  error bar represents the correlated error component of $1.2~\MeV$
  between $\MZB$ and $\MZC$. The value of $\MZA$ is essentially
  uncorrelated with the other two.}
\end{figure}

The averages over the four experiments of the three values $\MZA$,
$\MZB$ and $\MZC$ also provide a cross-check on the consistency of the
energy calibration, which dominates the errors on $\MZ$ in each of the
periods considered. The mass values for the three different periods
and the correlated and uncorrelated parts of their errors are shown in
Figure~\ref{fig:MZcheck}.  The differences amount to $ | \MZA - \MZB
| = 31\,\%$, $ | \MZA - \MZC | = 56\,\%$ and $ | \MZB - \MZC | =
43\,\%$ of the uncorrelated errors, {\em i.\,e.} the three $\Zzero$
mass values are consistent.

\subsection{Shifts for Halved Experimental Errors \label{sec-halvederr}}

When the average over the experiments is performed at the level of the
pseudo-observables, information on the individual contribution of
particular data points to the average is lost. Thus, if the average
were to be performed over the data points themselves, the relative
importance of independent experimental errors and the common errors
would be expected to change.  The examples of $\MZ$ and the importance
of the $t$-channel errors for $\Rl$, as discussed in the previous
subsections, provide good illustrations of such effects.

While $\MZ$ is properly treated by the eleven-parameter fits, other
pseudo-observables may suffer from similar shifts due to weight
changes. A generic ``combined'' experiment can be simulated by halving
the independent experimental errors in each experiment.  The observed
shifts in central values resulting from fits to the measurements with
errors modified in such a way can be used as a monitoring tool. The
average of these shifts over the four {LEP} experiments serves to
estimate the differences between an average at the parameter level
compared to an average at the level of the raw measurements. Of
course, this simple procedure assumes that all measurements from
individual experiments enter into the average with the same weight. As
a test of the validity of this technique, it was demonstrated that
averaging the shifts in $\MZ$ which each experiment observed when
halving its experimental errors also reproduced the shift seen in a
full combined fit to the four sets of hadronic cross-section
measurements.  The observed shifts are small for each of the nine
pseudo-observables, as is summarised in Table~\ref{tab:halvederr}.
The shift downwards in $\MZ$ of $0.3~\MeV$ corresponds to the
slightly smaller shift of $0.2~\MeV$ already seen in the
multiple-$\MZ$ fits.

\begin{table}\begin{center}
\begin{tabular}{|l||rrrr|r|r|}
\hline %
       &    A~~   &   D~~ &  L~~  &     O~~         & Average & \% of error\\
\hline %
\hline %
$\MZ$\,[\GeV]\hspace*{-0.3em} 
   &$-$0.0006  &   0.0000  &$-$0.0004  &$-$0.0001  &$-$0.00028  &13\\
$\GZ$\,[\GeV]\hspace*{-1em} 
   &$-$0.0002  &$+$0.0001  &$-$0.0004  &   0.0000  &$-$0.00013  & 5\\
$\shad$\,[nb]\hspace*{-1em} 
   &$+$0.006\pz&   0.000\pz&$+$0.008\pz&$+$0.0036  &$+$0.0037\pz&10\\
$\Ree$ &$+$0.004\pz&$+$0.017\pz&   0.000\pz&$+$0.004\pz&$+$0.0063\pz&13\\
$\Rmu$ &   0.000\pz&   0.000\pz&   0.000\pz&$+$0.001\pz&  ~0.0000\pz& 0\\
$\Rtau$&   0.000\pz&   0.000\pz&$-$0.001\pz&$+$0.002\pz&$+$0.0003\pz& 1\\
$\Afbze$&$-$0.0001 &$-$0.0003  &   0.0000  &$-$0.0000  &$-$0.00011  & 5\\
$\Afbzm$&$+$0.0002 &$+$0.0003  &   0.0000  &$+$0.0001  &$+$0.00014  &11\\
$\Afbzt$&$+$0.0002 &$+$0.0003  &   0.0000  &$+$0.0001  &$+$0.00015  & 9\\
\hline %
\end{tabular}
\caption[Shifts in central values with halved experimental errors]
{\label{tab:halvederr} Shifts in central values of the fitted
  pseudo-observables seen when halving the independent experimental
  errors, for individual experiments and the average. }
\end{center}
\end{table}

The average changes in $\MZ$, $\shad$, $\Ree$, $\Afbzm$ and $\Afbzt$
amount to about 10\,\% of the combined errors, in all other cases they
are even smaller.  This is an estimate of the magnitude of the changes
in the final results that would arise from a combination of the four
experiments at the cross-section level rather than the averaging at
the parameter level. Given the smallness of the observed effects it is
obvious that the parameter-level average is adequate.

\subsection{Influence of the $\gammaZ$ Interference Term
  \label{sec-jhad}}

In the nine-parameter analyses discussed here, the $\gammaZ$
interference terms in the differential cross-sections for leptons are
expressed using the effective coupling constants and the electric
charges of the electron and the final state fermion (see
Equation~\ref{eq:intro_bigugly}). This dependence can be exploited
in the fits to fix the interference terms in leptonic final states,
although the actual experimental procedures are slightly different in
detail, as described in the original
publications~\cite{\ALEPHls,\DELPHIls,\Lls,\OPALls}.  For the inclusive
hadronic final state, however, the $\gammaZ$ interference term must be
fixed to the $\SM$ value.  Fits with a free interference term are
possible in the S-matrix scheme~\cite{\bibSMAT}.  The OPAL
collaboration also studied a similar approach based on an extension of
the standard parameter set~\cite{\OPALls}.  In the S-matrix approach
the interference terms are considered as free and independent
parameters. The hadronic interference term is described by the
parameter $j_{\rm tot}^{\rm had}$, given in the $\SM$ by
\begin{equation}
j_{\rm tot}^{\rm had}  =
 \frac{\GF \MZ^2}{\sqrt{2}\pi\alpha(\MZ^2)}\, 
\Qe\,\gve 
 \cdot 3\sum_{\rm q \ne t} \Qq\,\gvq \,. 
\end{equation}
Note that the running of $\alpha$ as well as final state QED and QCD
corrections are also included in the definition of the S-matrix
parameters. The $\SM$ value of $j_{\rm tot}^{\rm had}$ is
$0.21\pm0.01$.

The dependence of the nine parameters on possible variations of the
hadronic $\gammaZ$ interference term away from the $\SM$ value is
studied by considering a set of ten parameters consisting of the
standard nine parameters extended by the parameter $j_{\rm tot}^{\rm
had}$ from the S-matrix approach.  The extra free parameter $j_{\rm
tot}^{\rm had}$ is strongly anti-correlated with $\MZ$, resulting in
errors on $\MZ$ enlarged by a factor of almost three, as is observed
in the existing S-matrix analyses of LEP-I data~\cite{\Smatrix}.

The dependence of $\MZ$ on $j_{\rm tot}^{\rm had}$ is given by:
\begin{eqnarray}
\frac{{\rm d}\MZ}{{\rm d}j_{\rm tot}^{\rm had}} & = & -1.6~\MeV/0.1\,.
\end{eqnarray}
The changes in all other parameters are below 20\,\% of their combined
error for a change in $j_{\rm tot}^{\rm had}$ of 0.1\,.

Improved experimental constraints on the hadronic interference term
are obtained by including measurements of the hadronic total
cross-section at centre-of-mass energies further away from the Z pole
than just the off-peak energies at LEP-I.  Including the measurements
of the TRISTAN collaborations at KEK, TOPAZ~\cite{TOPAZ} and
VENUS~\cite{VENUS}, at $\sqrt(s)=58~\GeV$, the error on $j_{\rm
tot}^{\rm had}$ is about $\pm0.1$, while its central value is in good
agreement with the $\SM$ expectation.  Measurements at centre-of-mass
energies above the Z resonance at
LEP-II~\cite{bib-LEP2smata,bib-LEP2smatd,\LEPIIsmatl,bib-LEP2smato}
also provide constraints on $j_{\rm tot}^{\rm had}$, and in addition
test modifications to the interference terms arising from the possible
existence of a heavy Z$^\prime$ boson.

The available experimental constraints on $j_{\rm tot}^{\rm had}$ thus
lead to uncertainties on $\MZ$, independent of $\SM$ assumptions in
the hadronic channel, which are already smaller than its error.  No
additional error is assigned to the standard nine-parameter results
from effects which might arise from a non-$\SM$ behaviour of the
$\gammaZ$ interference.

\subsection{Direct Standard Model Fits to Cross-Sections and 
Asymmetries}
\label{sec-sm}

Since an important use of the combined results presented here is to
test the validity of the $\SM$ and to determine its parameters, it is
crucial to verify that the parameter set chosen for the combination
represents the four sets of experimental measurements with no
significant loss in precision.  When the set of pseudo-observables is
used in the framework of the $\SM$, the role of $\MZ$ changes from an
independent parameter to that of a Lagrangian parameter of the theory,
intimately linked with other quantities. This imposes additional
constraints which can be expected to shift the value of $\MZ$.

To check whether the nine parameters adequately describe the reaction
to these constraints, each collaboration provided results from direct
$\SM$ fits to their cross-section and asymmetry data.  The comparison
of these results with those obtained from $\SM$ fits using the set of
pseudo-observables as input is shown in Table~\ref{tab:MZsm}. $\MH$
and $\alpha_s$ were free parameters in these fits, while the
additional inputs $\Mt=174.3\pm5.1~\GeV$~\cite{PDG2004} and
$\Delta\alpha^{(5)}_{\rm had}=0.02804\pm0.00065$~\cite{bib-JEG2}
(corresponding to $1/\alpha(\MZ^2)=128.886\pm0.090$) provided external
constraints.

\begin{table}[ht]
\begin{center}
\begin{tabular}{|l||rrrr|r|r|}
\hline %
                    &  A~~  &  D~~   &  L~~   &  O~~ & Average & \% of error \\
\hline %
\hline %
$\pzz \chidf$ & $174/180$ & 184/172 & $168/170$ & $161/198$ & & \\
\hline %
$\Delta\MZ$ [\MeV] &$-$0.7   &$+$0.5   &0.0   &$+$0.1   &$-$0.03 & 1  \\
$\Delta\Mt$ [\GeV] &   0.0   & 0.0     & 0.0  &   0.0   &   0.0  & $<$2 \\
$\Delta\log_{10}(\MH/\GeV)$ 
                 &$-$0.01  &$+$0.04  &$+$0.02  &$+$0.04   &$+$0.02  &  4 \\
$\Delta\alpha_s$    &   0.0000&$-$0.0002&$+$0.0002&$+$0.0002 &$+$0.0001 & 4 \\
$\Delta(\Delta\alpha^{(5)}_{\rm had})$
                      &$+$0.00002&$-$0.00004&   0.00000&$-$0.00004 &$-$0.00002 & 2 \\
\hline %
  fit value           &        &        &        &         &     &    \\ 
  of $\MH$ [$\GeV$]   &   40.  &   10.  &    35. &    390. &     &    \\  
\hline %
$\Delta\MZ$ [$\MeV$]     &        &        &        &         &      &   \\
corr. to              &        &        &        &        &      &   \\
 150\,$\GeV$ $\MH$    &$-$0.6  &$+$0.7  &$+$0.1  &  0.0   &$+$0.05 & 2 \\
\hline %
\end{tabular} 
\caption[Direct SM fits] {\label{tab:MZsm}
  Shifts in $\SM$ parameters, when fit directly to the cross-sections
  and forward-backward asymmetries compared to when fit to the
  nine-parameter results.  The numbers in the last line of the table
  give the shifts in $\MZ$ if the results from the first line are
  corrected to a common value of the Higgs mass of $150~\GeV$.}
\end{center}
\end{table}

Significant shifts in $\MZ$ of up to 20\,\% of its error are observed
in some experiments, which however cancel out to almost zero in the
average over the four experiments. One anticipated source of these
shifts has already been mentioned: the $\Zzero$ couplings defining the
$\gammaZ$ interference term depend on $\MH$, which is allowed to move
freely in the first fit, but is fixed to 150~$\GeV$ for the extraction
of the pseudo-observables.  The approximate values of $\MH$ preferred
by the $\SM$ fit to the cross-sections and asymmetries are indicated
in the second part of the table. Using the dependence of $\MZ$ on the
value of $\MH$ given in Section~\ref{sec-parerr}, the differences in
$\MZ$ can be corrected to a common value of the Higgs mass of
$\MZ=150~\GeV$, as is shown in the last line of
Table~\ref{tab:MZsm}. The results indicate that the expected $\MZ$
dependence on $\MH$ is not the dominant mechanism responsible for the
differences. Since the two procedures compared here represent
different estimators for $\MZ$, such differences may be expected due
to fluctuations of the measurements around the exact $\SM$
expectations.

It was verified that the two procedures lead to identical results if
applied to pseudo-data calculated according to the $\SM$.  If the
origin of the shifts is due to fluctuations of the measurements within
errors, a reduction of the shifts with increased statistical precision
is expected to occur, which is indeed what is observed when averaging
over the four experiments.  The net average difference in $\MZ$
directly from the realistic observables or through the intermediary of
the pseudo-observables is less than $0.1~\MeV$. Shifts in the other
$\SM$ parameters, in the individual data sets as well as in the
average, are all well under 5\,\% of the errors, and therefore also
negligible.

The conclusion of this study is that $\SM$ parameters extracted from
the pseudo-ob\-ser\-vab\-les are almost identical to the ones that
would be obtained from the combined cross-sections and asymmetries.
Within the $\SM$ the combined set of pseudo-observables provides a
description of the measurements of the $\Zzero$ parameters that is
equivalent to the full set of cross-sections and asymmetries. This is
also true for any theory beyond the $\SM$ which leads to corrections
that are absorbed in the pseudo-observables. An exception to this are
those theories which lead to significant modifications of the
$\gammaZ$ interference term, like theories with 
additional $\Zzero'$-bosons.  (See the discussion in Section~\ref{sec-jhad}.)

\section{Combined Results \label{sec-lsafbcombres}}

The full result of the combination of the four sets of nine
pseudo-observables including the experimental and common error
matrices
is given in Table~\ref{tab:lsafbresult}. The central values and errors
on the combined results are presented graphically and compared with
the corresponding input values of the four experiments in
Figure~\ref{fig:lsafb}. The parametric uncertainties due to the
residual dependence on the choice of $\SM$ parameters used to
calculate the remnants are not included. The only significant such
uncertainty concerns the value of the Higgs boson mass, which is taken
to be $150~\GeV$ and is relevant only for the value of $\MZ$. The
value of $\MZ$ changes by $+0.23~\MeV$ per unit change in
$\log_{10}(\MH/\GeV)$, as was discussed in Section~\ref{sec-parerr}.

\begin{table}[tb]\begin{center}
\begin {tabular} {|lr||r@{\,}r@{\,}r@{\,}r@{\,}r@{\,}r@{\,}r@{\,}r@{\,}r|}
\hline %
\multicolumn{2}{|c||} {Without lepton universality} & 
                                    \multicolumn{9}{l|}{~~~Correlations} \\
\hline %
\hline %
\multicolumn{2}{|c||}{$\pzz \chidf\,=\,32.6/27 $} &
   $\MZ$ & $\GZ$ & $\shad$ &
     $\Ree$ &$\Rmu$ & $\Rtau$ & $\Afbze$ & $\Afbzm$ & $\Afbzt$ \\
\hline %
 $\MZ$ [\GeV{}]  & 91.1876$\pm$ 0.0021 &
 ~1.000 & \multicolumn{8}{c|}{} \\
 $\GZ$ [\GeV]  & 2.4952 $\pm$ 0.0023 &
 $-$0.024 & ~1.000 & \multicolumn{7}{c|}{}\\ 
 $\shad$ [nb]  & 41.541 $\pm$ 0.037$\pz$ &
 $-$0.044 & $-$0.297 & ~1.000 & \multicolumn{6}{c|}{}\\ 
 $\Ree$        & 20.804 $\pm$ 0.050$\pz$ &
 ~0.078 & $-$0.011 & ~0.105 & ~1.000 & \multicolumn{5}{c|}{}\\ 
 $\Rmu$        & 20.785 $\pm$ 0.033$\pz$ & 
 ~0.000 & ~0.008 & ~0.131 & ~0.069 & ~1.000 & \multicolumn{4}{c|}{}\\ 
 $\Rtau$       & 20.764 $\pm$ 0.045$\pz$ &  
 ~0.002 & ~0.006 & ~0.092 & ~0.046 & ~0.069 & ~1.000 & \multicolumn{3}{c|}{}\\ 
 $\Afbze$      & 0.0145 $\pm$ 0.0025 &
 $-$0.014 & ~0.007 & ~0.001 & $-$0.371 & ~0.001 & ~0.003 & ~1.000 & \multicolumn{2}{c|}{}\\ 
 $\Afbzm$      & 0.0169 $\pm$ 0.0013 &
 ~0.046 & ~0.002 & ~0.003 & ~0.020 & ~0.012 & ~0.001 & $-$0.024 & ~1.000 & \multicolumn{1}{c|}{}\\ 
 $\Afbzt$      & 0.0188 $\pm$ 0.0017 &
 ~0.035 & ~0.001 & ~0.002 & ~0.013 & $-$0.003 & ~0.009 & $-$0.020 & ~0.046 & ~1.000 \\ 
\hline %
\multicolumn{3}{c}{~}\\
\end{tabular} \\
\begin {tabular} {|lr||r@{\,}r@{\,}r@{\,}r@{\,}r|}
\hline %
\multicolumn{2}{|c||} {With lepton universality} & \multicolumn{5}{c|}{Correlations} \\
\hline %
\hline %
\multicolumn{2}{|c||}{$\pzz \chidf\,=\,36.5/31 $}  & 
   $\MZ$ & $\GZ$ & $\shad$ & $\Rl$ &$\Afbzl$ \\
\hline %
 $\MZ$ [\GeV{}]  & 91.1875$\pm$ 0.0021$\pz$    &
 ~1.000 & \multicolumn{4}{c|}{}\\ 
 $\GZ$ [\GeV]  & 2.4952 $\pm$ 0.0023$\pz$    &
 $-$0.023  & ~1.000 & \multicolumn{3}{c|}{}\\ 
 $\shad$ [nb]  & 41.540 $\pm$ 0.037$\pzz$ &
 $-$0.045 & $-$0.297 &  ~1.000 & \multicolumn{2}{c|}{}\\ 
 $\Rl$         & 20.767 $\pm$ 0.025$\pzz$ &
 ~0.033 & ~0.004 & ~0.183 & ~1.000 & \multicolumn{1}{c|}{}\\ 
 $\Afbzl$     & 0.0171 $\pm$ 0.0010   & 
 ~0.055 & ~0.003 & ~0.006 & $-$0.056 &  ~1.000 \\ 
\hline %
\end{tabular} \end{center}
\vskip-1pc\caption[Combined results] {\label{tab:lsafbresult} Combined
  results for the $\Zzero$ parameters of the four sets of nine
  pseudo-observables from Table~\ref{tab:Zparinput}. The errors
  include all common errors except the parametric uncertainty on $\MZ$
  due to the choice of $\MH$.}
\end{table}

The value of $\chi^2$ per degree of freedom of the combination of the
nine-parameter results is 32.6/27 and corresponds to a probability of
21\,\% to find a value of $\chi^2$ which is larger than the one
actually observed. The correlation matrix of the combined result shows
significant correlations of $\shad$ with $\GZ$ and $\Ree$, $\Rmu$ and
$\Rtau$, and between $\Ree$ and $\Afbze$.

\begin{figure}[htb]\begin{center}
\mbox{\epsfig{file=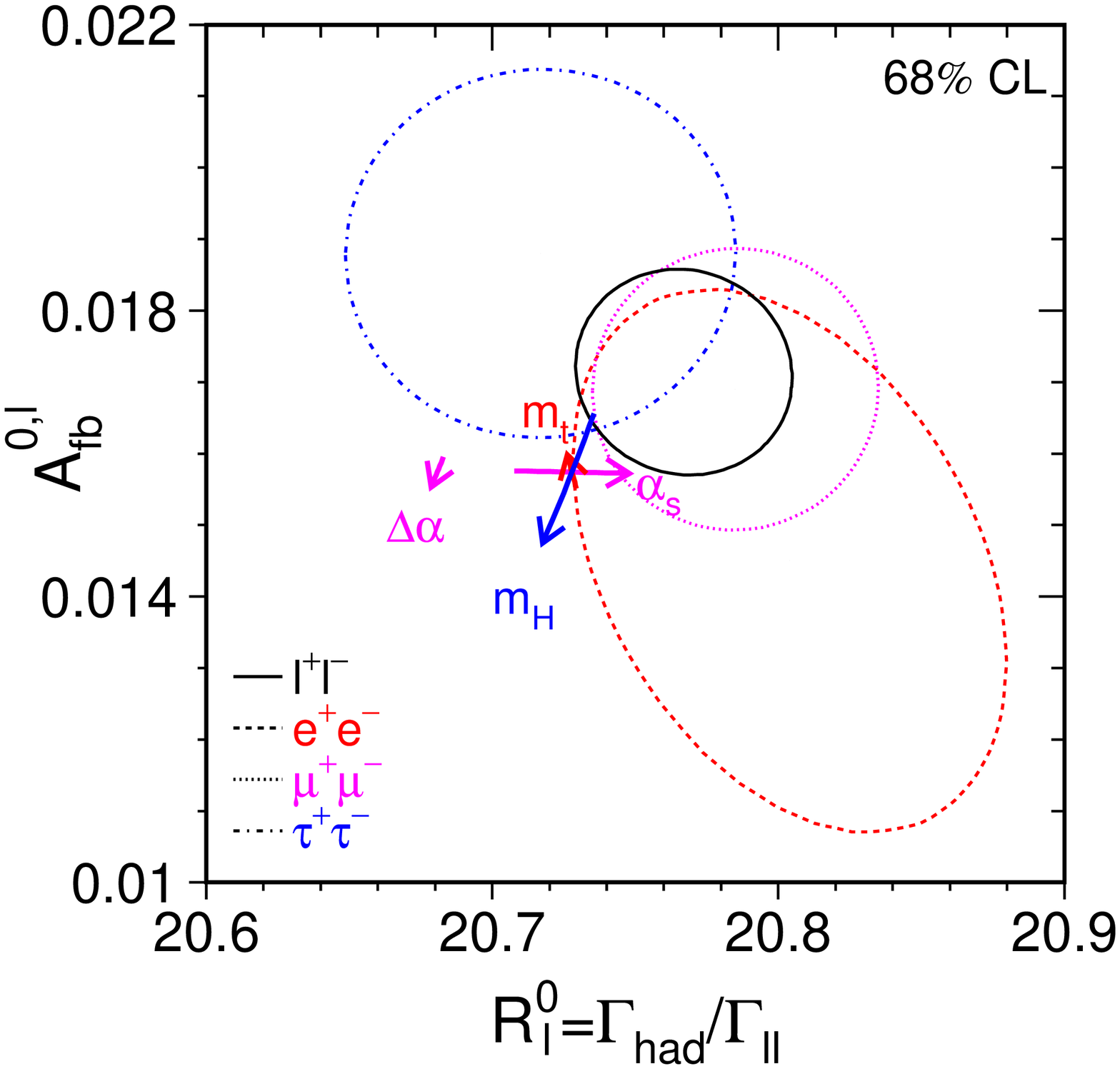,width=0.9\linewidth}}
\end{center} \vskip-1pc \caption[$\Rl$ vs. $\Afbzl$]
{\label{fig:rlafb} Contour lines (68\,\% CL) in the $\Rl$--$\Afbzl$
  plane for $\ee$, $\mumu$ and $\tautau$ final states and for all
  leptons combined. For better comparison the results for the $\tau$
  lepton are corrected to correspond to the massless case.  The $\SM$
  prediction for $\MZ=91.1875$~\GeV{}, $\Mt=178.0$~\GeV{},
  $\MH=300$~\GeV{}, and $\alfmz=0.118$ is also shown as the
  intersection of the lines with arrows, which correspond to the
  variation of the $\SM$ prediction when $\Mt$, $\MH$ and $\alfmz$ are
  varied in the intervals $\Mt=178.0\pm4.3$~\GeV{},
  $\MH=300^{+700}_{-186}~\GeV$, and $\alfmz=0.118\pm0.003$,
  respectively. The arrow showing the small dependence on the hadronic
  vacuum polarisation $\dalhad=0.02758\pm0.00035$ is displaced for
  clarity. The arrows point in the direction of increasing values of
  these parameters. }
\end{figure}

A comparison of the leptonic quantities $\Ree$, $\Rmu$ and $\Rtau$,
and of $\Afbze$, $\Afbzm$ and $\Afbzt$ shows that they agree within
errors. Note that $\Rtau$ is expected to be larger by 0.23\,\% because
of $\tau$ mass effects.  Figure~\ref{fig:rlafb} shows the
corresponding 68\,\% confidence level contours in the $\Rl$--$\Afbzl$
plane.

Imposing the additional requirement of lepton universality in the
nine-parameter combination leads to the results shown in the second
part of Table~\ref{tab:lsafbresult}.\footnote{ Performing the average
at the level of the five-parameter results leads to slightly different
values for $\Rl$ due to weight shifts originating from the common
$t$-channel error on $\Ree$, which is not treated properly in this
case.} Note that $\Rl$ is defined for massless leptons. The value of
$\chidf$ of $36.5/31$ for the combination of the four sets of nine
pseudo-observables into the five parameters of
Table~\ref{tab:lsafbresult} corresponds to a $\chi^2$ probability of
23\,\%. The central ellipse in Figure~\ref{fig:rlafb} shows the
68\,\%-CL contour for the combined values of $\Rl$ and $\Afbzl$
determined from all three lepton species.

While the errors on most of the pseudo-observables are dominated by
independent experimental or statistical errors, the combined errors on
$\MZ$ and $\shad$ have large contributions from a single, common
systematic error. The dominant contribution to the error in $\MZ$
arises from the uncertainty in the calibration of the energy of the
beams in LEP, and amounts to $\pm 1.7~\MeV$. The uncertainty on
$\shad$ arising from the theoretical error on the small-angle Bhabha
cross-section amounts to $\pm 0.025$~\nb, the total contribution of
common systematic errors is $\pm 0.028$~\nb. The systematic error on
$\GZ$, $\pm 1.2~\MeV$, is dominated by the uncertainty of the beam
energy.  Common systematics on $\Rl$ amount to $\pm 0.007$ and
contribute $\pm 0.0003$ to $\Afbzl$.

\boldmath
\chapter{Measurement of Left-Right and Lepton Asymmetries at the SLC}
\unboldmath
\label{sec-ALR}

\section{Left-Right Asymmetry Measurements}

The measurement of the left-right cross-section asymmetry, $\ALR$, by
SLD~\cite{\SLDalr} at the SLC provides a determination of the
asymmetry parameter $\cAe$, which is presently the most precise single
measurement of this quantity, with the smallest systematic error.  In
addition, $\ALR$, along with the tau polarisation measurements, is the
observable most sensitive to the effective weak mixing angle among the
asymmetries, with $\delta\ALR\approx8\delta\swsqeffl$.

In principle the analysis is straightforward: one counts the numbers
of Z bosons produced by left and right longitudinally polarised
electron bunches, $N_{\mathrm{L}}$ and $N_{\mathrm{R}}$, forms the
asymmetry, and then divides by the luminosity-weighted e$^-$ beam
polarisation magnitude $\apolel$ - the e$^+$ beam is not polarised
(see Equation~\ref{eq:ALR_exp}) :
\begin{equation}
  \label{eq:ALR}
  \ALR = \frac{N_{\mathrm{L}} - N_{\mathrm{R}}}%
              {N_{\mathrm{L}} + N_{\mathrm{R}}}%
         \frac{1}{\apolel}.
\end{equation}
The measurement requires no detailed final-state event identification:
$\ee$ final-state events are removed since they contain non-resonant
t-channel contributions, as are all other backgrounds not due to Z
decay.  It is also insensitive to all acceptance and efficiency
effects.  In order to relate $\ALR$ at a particular value of \Ecm\ to
a determination of the effective weak mixing angle, the result is
converted into a ``Z-pole'' value by the application of an
approximately $2.0\%$ relative correction for initial-state radiation
and $\gammaZ$ interference, $\Delta\ALR$~\cite{\ZFITTERref},
\begin{equation}
  \label{eq:ALR0}
  \ALR + \Delta\ALR = \ALRz \equiv \cAe.
\end{equation}
The calculation of this correction requires a good measurement of the
luminosity-weighted average centre-of-mass collision energy \Ecm.  The
$\Delta\ALR$ correction is small compared to the analogous QED
corrections for the leptonic forward-backward asymmetry measurements,
while similar in size to those required for the tau polarization and
bottom quark asymmetries.  There is no need for QCD corrections for
the measurement of $\ALR$.

For the data of 1997 and 1998, the small total relative
systematic error of 0.65\% is dominated by the 0.50\% relative
systematic error in the determination of the luminosity-weighted
average e$^-$ polarisation, with the second largest error, 0.39\%,
arising from uncertainties in the determination of the
luminosity-weighted average centre-of-mass energy.  A number of much
smaller contributions to the systematic error is discussed below.  The
relative statistical error on $\ALR$ from all data is about 1.3\%.  In
what follows, some of the details of the $\ALR$ measurement are
described and some historical context for the $\ALR$ programme at
SLC/SLD from 1992 until 1998 is provided.

\subsection{Electron Polarisation at the SLC}
\label{subsec:SLCepol}

In Section~\ref{sec:intro_SLC-data}, the operation of the SLC is
briefly outlined, and Figure~\ref{fig:SLC} provides a schematic of the
machine.  The SLC produced longitudinally polarised electrons by
illuminating a GaAs photocathode with circularly polarised light
produced by a Ti-Sapphire laser.  Following the advent of high
polarisation ``strained lattice'' GaAs photocathodes in
1994~\cite{\SLDstrain}, where mechanical strain induced in a 0.1$\mu$m
GaAs layer lifts an angular momentum degeneracy in the valence band of
the material, the average electron polarisation at the $\ee$
interaction point (IP) was in the range 73\% to 77\%, only slightly
lower than the value produced at the source, see
Figure~\ref{fig:slc_polar}.  The corresponding polarisation results were about
22\% in 1992 using an unstrained ``bulk'' GaAs cathode, and 63\% in 1993 using
a 0.3$\mu$m strained-layer cathode design.  The electron helicity was
chosen randomly pulse-to-pulse at the machine repetition rate of 120
Hz by controlling the circular polarisation of the source laser.

The electron spin orientation was longitudinal at the source and
remained longitudinal until it was transported to the damping ring (DR).
In this linac-to-ring (LTR) transport line, the electron spins precessed
horizontally due to the dipole bending magnets, where the spin precession
angle is given in terms of the anomalous magnetic moment: $
\theta_{\mathrm{precession}} = ({g-2 \over {2} }) {E \over m}
\theta_{\mathrm{bend}}$.  By design, the bend angle
$\theta_{\mathrm{bend}}$ resulted in transverse spin orientation at
the entrance to the LTR spin rotator magnet.  This superconducting
solenoid magnet was used to rotate the polarisation about the beam
direction into the vertical orientation for storage in the DR.  This
was necessary as any horizontal spin components precessed rapidly and
were completely dissipated during the 8.3 msec (1/120 seconds) storage time
due to energy spread in the bunch.  The polarised electron bunches could
be stored in one of two possible configurations by the reversal of the
LTR spin rotator solenoid magnet.  These reversals, typically done at
three month intervals, were useful for identifying and minimising the
small ($\calO(10^{-4})$) polarisation asymmetries produced at the
source.

The electron spin was vertical in the linac and had to be
reoriented for maximal longitudinal polarisation at the IP.  Spin
manipulation was possible during transport through the electron arc by
employing two large vertical betatron oscillations in the beam orbit
(``spin bumps'').  As the betatron phase advance closely matched the
spin precession, 1080 and 1085 degrees, respectively, in each of the
23 bending-magnet assemblies (``achromats'') used in the arc, the
electron arc operated close to a spin-tune
resonance, and hence an iterative spin bump
procedure was effective in optimising IP polarization~\cite{\SLDspin}.
As a result,
excepting for the 1992 running, the two additional SLC spin rotator
solenoids located downstream of the damping ring were not necessary 
for spin orientation and were used only
in a series of specialised polarisation experiments.

\subsection{Polarimetry at the SLC}

The SLD collaboration monitored the longitudinal polarization of the
electron beam with a Compton scattering polarimeter.  The Compton
polarimeter detected beam electrons that had been scattered by photons
from a circularly polarised laser.  The scattered electrons were
momentum analysed by magnets and swept into a multi-channel detector.
The choice of a Compton-scattering polarimeter was dictated by the
requirements that the device be operated continually while beams were
in collision and that uncertainties in the physics of the scattering
process not be a limiting factor in the systematic error. Both of
these requirements are troublesome issues for M\o ller scattering
instruments due to their magnetic alloy targets.  In addition, the
pulse-to-pulse controllability of the laser polarisation sign, as well
as its high polarisation value of 99.9\%, are additional advantages of
a Compton polarimeter relative to other options.

In Compton scattering of longitudinally polarised electrons from
circularly polarised photons, the differential cross-section in terms
of the normalised scattered photon energy fraction $x$ is given by:
\begin{equation}
  {{\mathrm{d} \sigma}\over{\mathrm{d}x}} ~ = ~ 
  {{\mathrm{d} \sigma_0}\over{\mathrm{d}x}} [1 - \polg \pole A(x)] \,, 
\label{eq:comptoncross}
\end{equation}
where ${{\mathrm{d} \sigma_0}/{\mathrm{d}x}}$ is the unpolarised
differential cross-section, $\polge$ are the photon and electron
polarisations, and $A(x)$ is the Compton asymmetry function.  The
asymmetry arises due to the difference between spin parallel and spin
anti-parallel cross-sections, $\sigma(j=3/2) > \sigma(j=1/2)$.  Both
the asymmetry function and differential cross-section are well known
theoretically~\cite{Lipps:1954}.  The unpolarized cross-section is a
relatively slowly varying function of the energy of the scattered
electron or photon.  At the SLC, typical laser/electron-beam
luminosities led to about 1000 Compton scatters per laser pulse.  The
asymmetry function changes sign (corresponding to going from forward
to backward photon scattering in the electron rest frame), and reaches
extreme values at the kinematic endpoints, corresponding to full
forward or back scattering.  In the SLD polarimeter, scattered
electrons of minimum energy and maximum deflection in the spectrometer
exhibited the maximum asymmetry.

In a polarimeter, the Compton-scattered photons or electrons are
detected, and the requisite instrumental effects are incorporated into
an energy dependent detector response function.  The normalised
weighting of $A(x)$ with ${{\mathrm{d} \sigma_0}/{\mathrm{d}x}}$ and
the response function $R(x)$, all functions of the fractional energy
$x$, is known as the ``analysing power'' $a$:
\begin{equation}
  a~ = ~
{ {\int A(x) R(x) {{\mathrm{d} \sigma_0}\over{\mathrm{d}x}} \mathrm{d}x}\over
         {\int R(x) {{\mathrm{d} \sigma_0}\over{\mathrm{d}x}} \mathrm{d}x} }\,, 
\label{eq:analypower}
\end{equation}
where the integration is over the relevant acceptance in $x$.  For a
multichannel detector, as was used by the SLD, $a_i$ and $R_{i}(x)$
are defined for the $i$th channel.  For example, in the case of the
Cherenkov detector discussed below, the response function for a given
detector channel quantified the Cherenkov light produced by an
incident electron as a function of the range of $x$ or equivalently,
the transverse position in the spectrometer bending plane,
corresponding to the acceptance of that channel.

As mentioned above, the laser/electron-beam luminosities for the SLD
polarimeter led to a large number of Compton scatters per laser pulse.
All of the channels of the polarimeter detector were hit on virtually
every laser pulse, and for every pulse, each channel integrated the
response to many Compton scattered electrons as well as backgrounds.
Linearity of response was therefore an essential detector requirement.
The Compton scattering asymmetry in the SLD polarimeter was formed
from the time averaged detector channel responses, typically taken
over a few minutes, for each of the four possible electron-photon
helicity combinations.

For the $i^{th}$ detector channel, the two spin aligned configurations
were combined to give $\langle N \rangle^i_{3/2}$, while the two spin
opposed configurations yielded $\langle N \rangle^i_{1/2}$.  The SLC
operated at 120 Hz and the polarimeter laser fired every 7th beam
crossing,\footnote{ The laser firing sequence was automatically
shifted by one beam crossing at regular intervals to avoid undesirable
synchronisation with periodic effects in the SLC.} so that the six
intervening ``laser-off'' beam crossings were used to monitor the
polarimeter background responses $\langle N \rangle^i_{off}$.  The
measured asymmetry is
\begin{equation}
  \label{eq:rawasym}
  A^i ~ = ~ \frac{\langle N \rangle^i_{3/2} - \langle N \rangle^i_{1/2}}%
              {\langle N \rangle^i_{3/2} + \langle N \rangle^i_{1/2} 
            - 2\langle N \rangle^i_{off}}.
\end{equation}
The set of $A^i$ are corrected for small effects due to electronics
noise and detector non-linearity, as described below, and the result
can then be related to the known analysing powers $a^i$ and laser
polarization as:
\begin{equation}
 A^i_C ~ = ~ \polg \pole a^i, 
\label{eq:correctasym}
\end{equation}
which can be solved for $\pole$.

\begin{figure}[p]
\begin{center}
\mbox{\epsfig{file=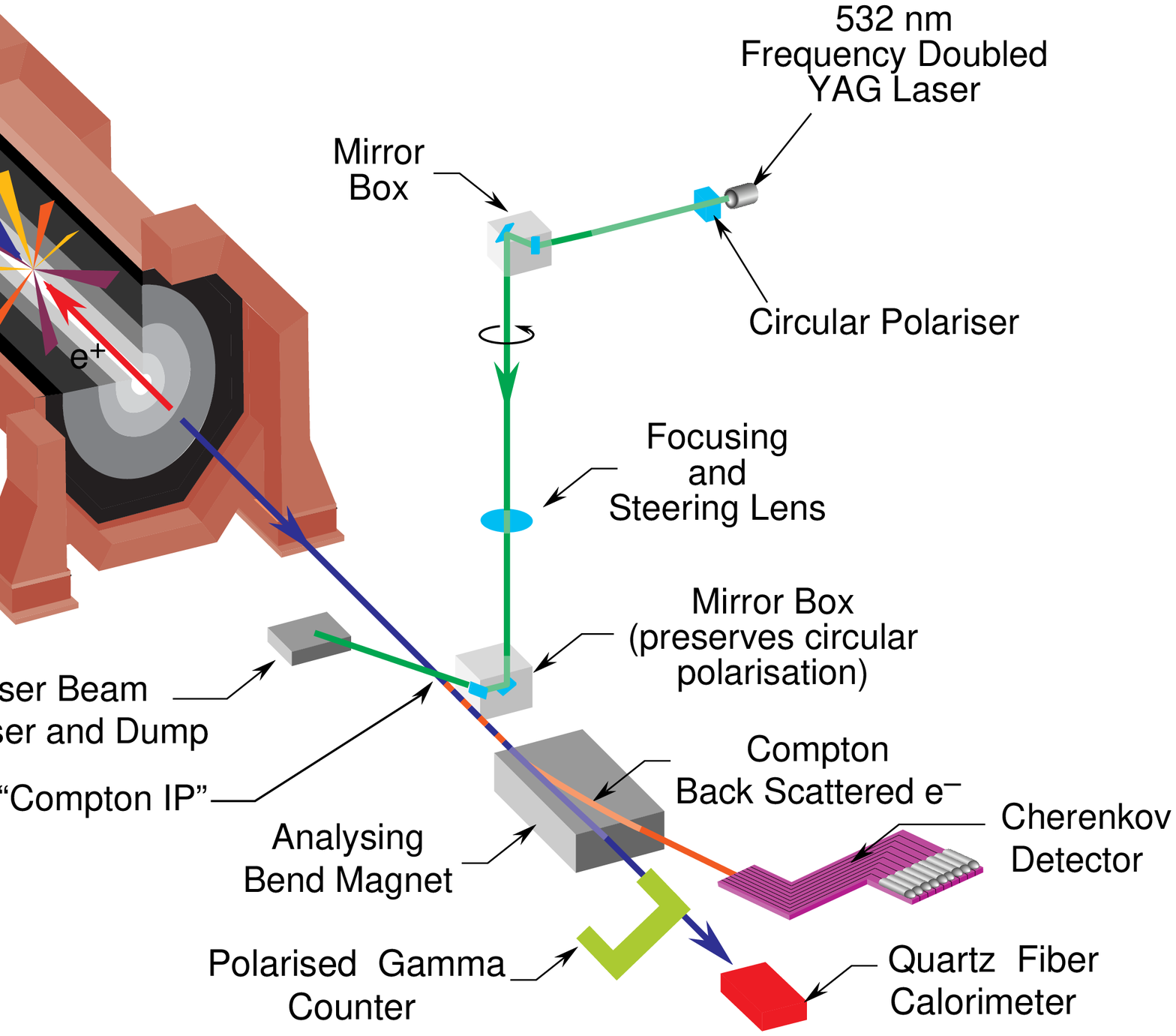,width=\linewidth}}
\vskip 1cm
\caption[The SLC Compton polarimeter setup]
{A conceptual diagram of the SLD Compton Polarimeter.  The laser beam,
  consisting of 532~nm wavelength 8~ns pulses produced at 17
  Hz and a peak power of typically 25~MW, were 
  circularly polarised and transported into collision with
  the electron beam at a crossing angle of 10~mrad
  approximately 30 meters from the IP.  Following
  the laser/electron-beam collision, the electrons and
  Compton-scattered photons, which are strongly boosted along the
  electron beam direction, continue downstream until analysing bend
  magnets deflect the Compton-scattered electrons into a
  transversely-segmented Cherenkov detector.  The photons continue
  undeflected and are detected by a gamma counter (PGC) and a
  calorimeter (QFC) which are used to cross-check the polarimeter
  calibration.  }
\label{fig:alr:compton}
\end{center}
\end{figure}

Figure~\ref{fig:alr:compton} illustrates the essential features of the
polarimeter setup: Frequency doubled Nd:YAG laser pulses were
circularly polarised by a linear polariser and a Pockels cell pair.
The laser beam was transported to the SLC beamline by four sets of
phase-compensating mirror pairs and into the vacuum chamber through a
reduced-strain quartz window.  About 30 meters downstream from the IP,
the laser beam was brought into nearly head-on collisions with the
outgoing electron beam at the Compton Interaction Point (CIP), and
then left the beampipe through a second window to an analysis station.
The pair of Pockels cells on the optical bench allowed for full
control of elliptical polarisation and was used to automatically scan
the laser beam polarisation at regular intervals in order to monitor,
and maximise, laser polarisation at the CIP.  This procedure
significantly improved the magnitude of the laser circular
polarisation, and the precision of its
determination~\cite{ref:sld-pockel}.  In colliding a $\sim 45~\GeV$
electron beam with visible light, the scattered photons are very
strongly boosted along the electron beam direction and are essentially
collinear with the Compton-scattered electrons.\footnote{
Compton-scattered photons with energies in the range from the
kinematically allowed maximum of $28~\GeV$ down to $1~\GeV$ are
contained within an angle of about 100~$\mu$rad with respect to the
electron beam direction.}  Downstream from the CIP, a pair of bend
magnets swept out the off-energy Compton-scattered electrons, which
passed through a thin window and out of the beamline vacuum into a
nine-channel transversely segmented gas Cherenkov detector, each
channel covering 1~cm.  By detecting the Compton-scattered electrons
with a Cherenkov device whose threshold was about $11~\MeV$, the
copious soft backgrounds originating from the beam-beam interactions
and from synchrotron radiation, were dramatically reduced.

The minimum energy $17.4~\GeV$ electrons, corresponding to full
backscattering, generally fell into the 7th channel (see
Figure~\ref{fig:comptondata}).  At this point in the electron
spectrum, known as the ``Compton edge'', the polarisation asymmetry
function reached its maximum value of 0.748.  A number of effects,
including electron scattering and showering in the detector and in the
detector vicinity, signal response in the detector, and beam steering
and focusing, tend to slightly smear the asymmetry function.  Small
deviations from the theoretical Compton energy dependent asymmetry
function, of order 1\% near the Compton edge, were determined by
modelling the detector response functions for each of the nine
channels.  An EGS4 simulation was used for this calculation, which
included a detailed Monte Carlo of the detector geometry and relevant
beamline elements, the Cherenkov light generation and transport, and
the magnetic spectrometer~\cite{ref:sld-EGS4, Torrence:1997bdMC}.  The
detector was mounted on a movable platform and the Compton edge was
scanned across several channels at regular intervals in order to
constrain the individual channel gains, monitor the location of the
Compton edge and to experimentally constrain the detector/spectrometer
simulation.  For each edge scan, a multi-parameter fit to the channel
scan data for the 16 scan positions was performed to determine the
beam position, the relative channel gains of a set of outer channels
(usually, channels 5-8), a normalization (luminosity) at each platform
position, and the polarisation product $\polg \pole$.  From these
fits, the reliability of the simulation was tested.  For example, the
main cause of the small deviations from ideal response functions arose
due to showering in a lead pre-radiator in front of the detector that
had been installed to optimize signal to noise.  The resulting
smearing of the acceptance of each channel was shown to be well
modeled when compared with the edge-scan data.  Additional
cross-checks tracked the stability of the analysing powers during time
periods between edge scans (for example, the ratios between selected
channel asymmetries were monitored).  Representative data showing the
corrected Compton asymmetry as well as the magnitude of the
correction, as a function of position and scattered electron energy,
is shown in Figure~\ref{fig:comptondata}. There is good agreement
between the corrected asymmetry and the data in each channel.

\begin{figure}[h]
\begin{center}
\mbox{\epsfig{file=compt_asym3.eps,width=0.9\linewidth}}
\caption[Compton scattering asymmetry as a function of channel position]
{Compton scattering asymmetry as a function of channel position.  The
  deflection of the electron beam in the spectrometer is shown on the
  horizontal axis as the distance in mm from the centre of the
  detector channels, each 1~cm wide, to the path of a hypothetical
  infinite momentum electron beam.  The inset shows the seven inner
  detector channels, sized to match the horizontal scale.  The per
  channel asymmetry data is plotted as open circles, and the corrected
  asymmetry function is the solid curve.  The relative size of the
  correction to the theoretical QED calculation is indicated by the
  dashed curve and the right-side vertical scale.}
\label{fig:comptondata}
\end{center}
\end{figure}

\begin{figure}[h]
\begin{center}
\mbox{\epsfig{file=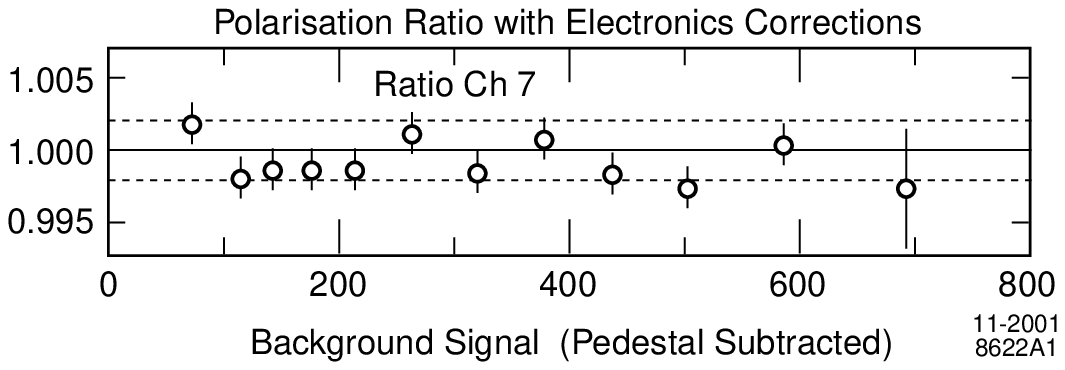,width=0.9\linewidth}}
\caption[The linearity of a polarimeter channel]
{The linearity of channel 7 is shown for a wide range of detector
  background levels (expressed here in pedestal subtracted ADC
  counts).  Plotted on the vertical axis is the fully corrected
  polarisation result, normalised to the ``zero background'' case.
  The result here is seen to be constant to within 0.2\% over the full
  range. Running conditions varied widely, but on average were a total
  response of 500 counts, including 300 counts of signal.  }
\label{fig:comptonlinear}
\end{center}
\end{figure}

Detector effects such as non-linearity in the electronics and/or the
photomultiplier tubes, and electronic noise, mainly due to pickup from
the laser Q-switch used to produce the short high peak power laser
pulse, are measured from the data. Firstly, the highly variable $\ee$
collision-related backgrounds in the polarimeter, as well as the
varying CIP luminosity, produced signals over a wide dynamic range in
each channel, which allowed for an effective linearity measurement, as
shown in Figure~\ref{fig:comptonlinear}.

The ``zero-backgrounds'' condition was determined from polarisation
measurements taken when the positron beam was absent, as backgrounds
were dominated by beam-beam interaction effects.  Secondly, the
electronic pickup effects were conveniently studied using the
occasional machine cycles without either the electron or positron
beams.  A number of offline electronics tests and specialised test
procedures during running, for example, photomultiplier tube voltage
scans, were also useful in establishing the size of systematic
uncertainties.

\begin{figure}[t]
\begin{center}
\mbox{\epsfig{file=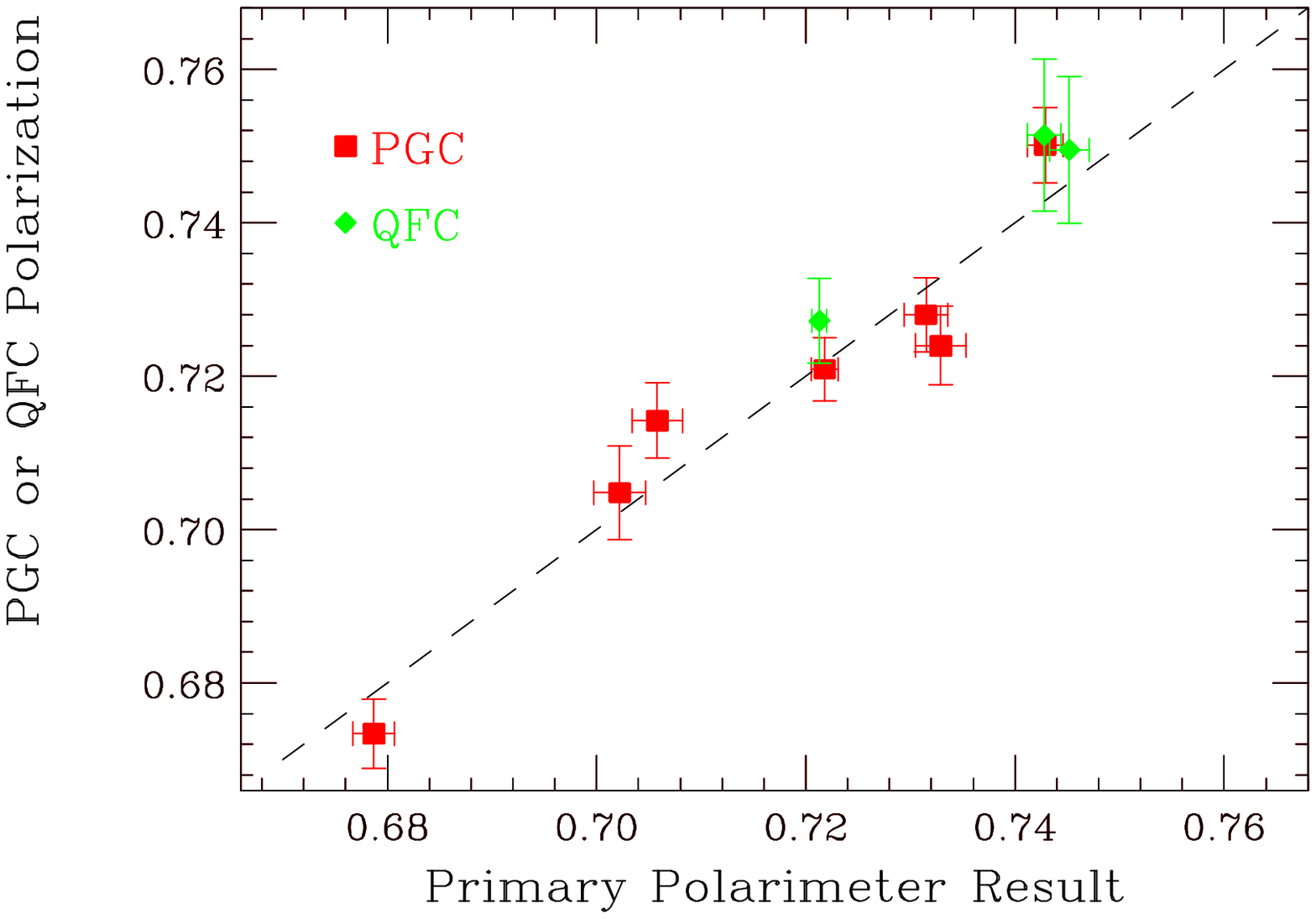,width=0.9\linewidth}}
\caption[Comparison between the SLD photon and electron polarimeter results]
{The polarised gamma counter (PGC) and quartz fiber calorimeter (QFC)
  photon detector polarisation results (vertical axis) compared to the
  primary electron detector polarimeter measurements (horizontal
  axis).}
\label{fig:polcrosscheck}
\end{center}
\end{figure}

Starting in 1996, two additional polarimeter
detectors~\cite{\SLDpgc,\SLDqfc} that were sensitive to the
Compton-scattered photons and which were operated in the absence of
positron beam, were used to verify the precision polarimeter
calibration.  These two devices were of different design, one was a
threshold-gas Cherenkov detector and the other was a quartz-fiber
calorimeter, with different systematic errors, and had in common with
the primary electron polarimeter only the instrumental errors due to
the polarised laser.  The cross-check provided by these photon
detectors was used to establish a calibration uncertainty of 0.4\%, as
shown in Figure~\ref{fig:polcrosscheck}.  The systematic errors due to
polarimetry are summarised in Table~\ref{tab:alr:polsys}.  During the
period 1992-1998, this total fractional systematic error decreased
from 2.7\% to its final value of 0.50\%, with the most significant
reductions coming from greatly improved understanding of the laser
polarisation and Cherenkov detector nonlinearities.  The final
systematic error is dominated by the analysing power calibration
uncertainty discussed above.

The polarimeter result was corrected for higher order QED and
accelerator-related effects by a total of $(-0.22\pm0.15)$\% for
1997-1998 data. The higher order QED offset was small and determined
to be $-0.1$\%~\cite{ref:sld-morrisQED}.  The primary
accelerator-related effect arose from energy-to-polarisation
correlations and energy-to-luminosity correlations that, together with
the finite energy spread in the beam, caused the average beam
polarisation measured by the Compton Polarimeter to differ slightly
from the luminosity-weighted average beam polarisation at the IP.
When first observed in 1993, this {\it chromaticity} correction and
its associated error was $(1.1\pm1.7)\%$.  In 1994-1998 a number of
changes in the operation of the SLC and in monitoring procedures, such
as smaller and better determined beam energy spread and polarisation
energy dependence, reduced the size of this effect and its uncertainty
to below 0.2\%.  An effect of comparable magnitude arose due to the
small precession of the electron spin in the final focusing elements
between the IP and the polarimeter.  The contribution of collisional
depolarisation was determined to be negligible, as expected, by
comparing polarimeter data taken with and without beams in collision.
All effects combined yielded a correction with the uncertainty given
in Table~\ref{tab:alr:polsys}.

The luminosity-weighted average polarisation $\apolel$ for each run
was estimated from measurements of $\pole$ made when Z events were
recorded:
\begin{equation}
\apolel ~ = ~ 
(1+\xi)\cdot\frac{1}{N_{\mathrm{Z}}}\sum_{i=1}^{N_{\mathrm{Z}}}{\poll}_i\,,
\label{eq:poldef}
\end{equation}
where $N_{\mathrm{Z}}$ is the total number of Z events, ${\poll}_i$ is
the polarisation measurement associated in time with the $i^{th}$
event, and $\xi$ is the small total correction described in the
previous paragraph.  The polarimeter was operated continually, where
for typical background conditions about three minutes were required to
achieve a relative statistical precision of order 1\% for each
polarisation measurement.

The fully corrected luminosity weighted average polarisations
corresponding to each of the SLD runs are given in
Table~\ref{tab:sld-alrresults}.  The evolution of GaAs photo-cathode
performance is evident in 1993 and again in 1994-1995.  Changes in the
achieved polarisation in later years mainly reflect variations in
photo-cathode manufacture.

\subsection{Energy Spectrometry}
\label{subsec:alr-ecm-cal}

\begin{figure}[h]
\begin{center}
\mbox{\epsfig{file=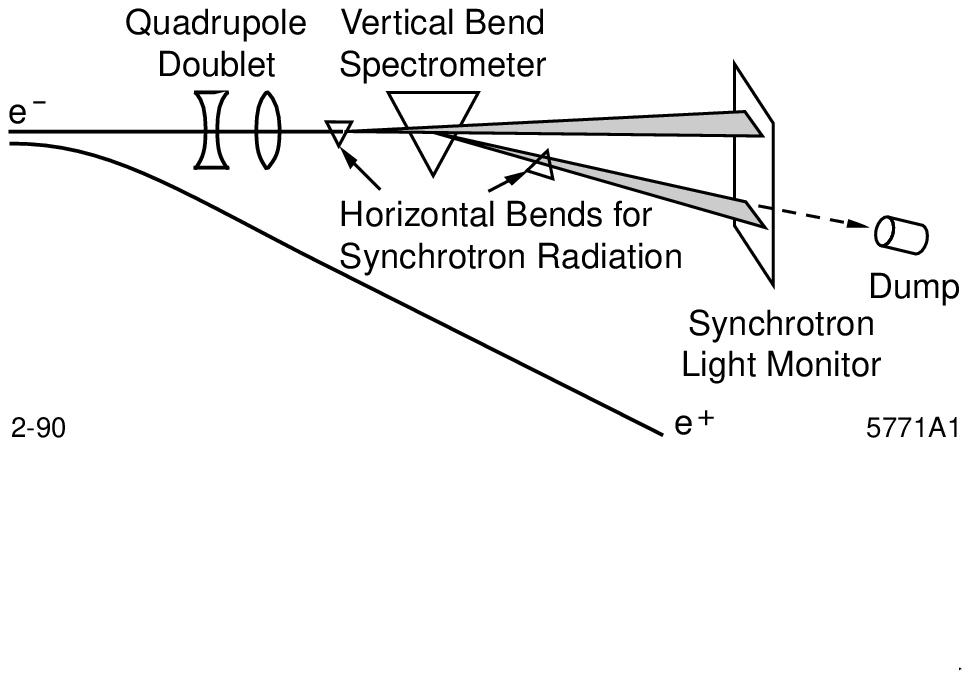,width=0.9\linewidth}}
\vskip -3cm
\caption[The SLC extraction-line energy spectrometer]
{ The energy spectrometer for electrons (a similar device is used on
  the positron side) uses a precision bend magnet and
  synchrotron-radiation-producing deflector magnets before and after
  the bend, in order to determine the beam bend angle.}
\label{fig:alr:espec}
\end{center}
\end{figure}

The SLC employed a pair of energy spectrometers located in the
electron and positron extraction lines (Figure~\ref{fig:SLC}).  The
beam deflection by a precision dipole magnet was detected and measured
using the separation between synchrotron radiation swathes emitted by
the beam in deflector magnets, oriented perpendicular to the bending
plane and located before and after the bend, see
Figure~\ref{fig:alr:espec}.

These devices were first operated in their final configuration in 1989
by the Mark II experiment, and the calibration of the two precision
spectrometer magnets was performed in 1988~\cite{\SLDespec}.  Their
expected precision was about $\pm 20~\MeV$ on the measured
centre-of-mass collision energy $\Ecm$.  The importance of these
devices to the $\ALR$ measurement is quantified by the approximate
rule of thumb that an $80~\MeV$ uncertainty in $\Ecm$ corresponds to a
$1\%$ error on the Z-pole asymmetry $\ALRz$.  For this reason, in 1998
a Z peak scan was performed in order to calibrate the spectrometers to
the LEP measurement of the Z mass.  The scan used two optimised
off-peak points at $+0.88~\GeV$ and $-0.93~\GeV$, and approximately
$9\,000$ on-peak Z equivalents of luminosity ($~300$ $\rm{nb}^{-1}$)
to reach a statistical precision on the peak position of $20~\MeV$.
The results of a complete analysis of systematic effects determined an
offset of $-46~\MeV$ and a total $\Ecm$ uncertainty of $29~\MeV$, the
latter corresponding to a 0.39\% uncertainty on $\ALRz$, as shown in
Table~\ref{tab:alr:polsys}~\cite{ref:sld-zpeak}.  The measured offset
appeared to be correlated with energy spectrometer backgrounds during
the high luminosity operation of the SLC in 1997-1998, and a
correction was applied to only this run (which constituted about
$60\%$ of the data).

\begin{table}[h]
\begin{center}
\renewcommand{\arraystretch}{1.25}
\begin{tabular}{|l||c|c|c|}
\hline
Uncertainty & $\delta\pole/\pole$      & $\delta\ALR/\ALR$      &
$\delta\ALRz/\ALRz$     \\
            &                     [\%] &                   [\%] &
                    [\%]\\
\hline
\hline
Laser polarisation & 0.10 & & \\
Detector linearity & 0.20 & & \\
Analysing power calibration & 0.40 & & \\
Electronic noise & 0.20 & & \\ \hline
Total polarimeter uncertainty  & 0.50 & 0.50 & \\
Chromaticity and IP corrections &  & 0.15 & \\
Corrections in Equation~\ref{eq:ALRsys} &  & 0.07 & \\ \hline
$\ALR$ Systematic uncertainty &    & 0.52 & 0.52\\
Electroweak interference correction &    &  & 0.39\\ \hline
$\ALRz$ Systematic uncertainty  &    &  & 0.64\\
\hline
\end{tabular}
\caption[Systematic uncertainties in the $\ALR$ measurement for
1997/98] {Systematic uncertainties that affect the $\ALR$ measurement
  for 1997/98.  The uncertainty on the electroweak interference
  correction is caused by the uncertainty on the SLC energy scale. }
\label{tab:alr:polsys}
\end{center}
\end{table}

\subsection{Event Selection}

A simple calorimetric event selection in the Liquid Argon Calorimeter
(LAC), supplemented by track multiplicity and topology requirements in
the Central Drift Chamber (CDC), were used to select hadronic Z
decays.  For each event candidate, energy clusters were reconstructed
in the LAC.  Selected events were required to contain at least
$22~\GeV$ of energy observed in the clusters and to have a normalised
energy imbalance of less than 0.6.\footnote{ The energy imbalance is
defined as a normalised vector sum of the energy clusters as follows,
$E_{\rm imb}=|\sum \vec E_{\rm cluster}|/\sum |E_{\rm cluster}|$.}
The left-right asymmetry associated with final-state $\ee$ events is
expected to be diluted by the $t$-channel photon exchange subprocess.
Therefore, we excluded $\ee$ final states by requiring that each event
candidate contain at least 4 selected CDC tracks, with at least 2
tracks in each hemisphere, defined with respect to the beam axis, or
at least 4 tracks in either hemisphere.  This track topology
requirement excludes Bhabha events which contain a reconstructed gamma
conversion.

Aside from $t$-channel effects in $\ee$ production, $\ALR$ is
independent of the final-state fermion flavour, and hence tau and muon
pairs are not a background.  However, the event selection was
optimised for hadronic events.  Tau and muon pairs were almost
completely removed,\footnote{ Tau pairs constituted (0.3 $\pm $ 0.1)\%
of the sample, while muon pair events deposited little energy in the
calorimeter and were completely removed by the cuts.}  but were
instead included in the complementary analysis described in section
3.2.  Small backgrounds in the $\ALR$ data sample were due to residual
$\ee$ final-state events, and to two-photon events, beam-related
noise, and cosmic rays.  For the data collected from 1996 to 1998, the
total background contamination was estimated to be $<0.05\%$ for a
hadronic event selection efficiency of $(91\pm1)\%$.  For a discussion
of event selection used in the earlier $\ALR$ analyses, see
reference~\cite{Ben-David:1994es} and reference~\cite{Park:1993es} for
the 1992 and 1993 datasets, respectively.

\subsection{Control of Systematic Effects}

The $\ALR$ measurement is remarkably resistant to detector dependent
systematic effects and Monte Carlo modelling uncertainties.  By far the
dominant systematic effects arise from polarimetry and from the
determination of the collision energy, rather than from any details of
the analysis or the operation of the SLD.  The simple expression given
in Equation~\ref{eq:ALR} applies to the ideal case in the absence of
systematic effects, and as such it is a good approximation to better
than a relative 0.2\%.

Nevertheless, systematic left-right asymmetries in luminosity,
polarisation, beam energy, and acceptance, as well as background and
positron polarisation effects, can be incorporated into an extended
expression for the cross-section asymmetry.  Note that while the
random helicity of the delivered electron bunches is exactly 50\%
right-handed, it is in principle possible that the magnitude of the
luminosity is not equal for the two helicities.  In addition, the
individual polarisation measurements of Equation~\ref{eq:poldef}
average over many beam crossings and over any systematic left-right
polarisation difference, and hence additional information is needed to
make the required correction.  One finds the measured asymmetry
$\ameas$ is related to $\ALR$ by the following expression which
incorporates a number of small correction terms in lowest-order
approximation,
\begin{eqnarray}
\label{eq:ALRsys}
\ALR & = & \frac{\ameas}{\avPe} +\frac{1}
{\avPe}\biggl[
f_{\mathrm{bkg}}(\ameas-A_{\mathrm{bkg}})-A_\calL+\ameas^2\apol \nonumber \\
 & & ~~~~~~~~~~~~~~~~
-\Ecm\frac{\sigma^\prime(\Ecm)}{\sigma(\Ecm)}\aengy
-\aeff + \avPe\polp \biggr]\,,
\end{eqnarray}
where \avPe\ is the mean luminosity-weighted polarisation;
$f_{\mathrm{bkg}}$ is the background fraction; $\sigma(E)$ is the
unpolarised Z boson cross-section at energy $E$; $\sigma^\prime(E)$ is
the derivative of the cross-section with respect to $E$;
$A_{\mathrm{bkg}}$, $A_\calL$, $\apol$, $\aengy$, and $\aeff$ are the
left-right asymmetries\footnote{ The left-right asymmetry for a
quantity $Q$ is defined as
$A_Q\equiv(Q_{\mathrm{L}}-Q_{\mathrm{R}})/(Q_{\mathrm{L}}+Q_{\mathrm{R}})$
where the subscripts ${\mathrm{L}}$ and ${\mathrm{R}}$ refer to the
left- and right-handed beams, respectively.}  of the residual
background, the integrated luminosity, the beam polarisation, the
centre-of-mass energy, and the product of detector acceptance and
efficiency, respectively; and $\polp$ is any longitudinal positron
polarisation of constant helicity. Since the colliding electron and
positron bunches were produced on different machine cycles and since
the electron helicity of each cycle was chosen randomly, any positron
helicity arising from the polarisation of their parent electrons was
uncorrelated with electron helicity at the IP.  The net effect of
positron polarisation from this process vanishes rigorously.  However,
positron polarisation of constant helicity would affect the
measurement.

The close ties between this measurement and the SLC accelerator
complex are evident from numerous accelerator-based experiments
dedicated to the SLD physics programme, for which the
energy-calibrating Z-peak scan is one example. Other examples include:

\begin{itemize}
  
\item {\it Communication of the $\mathrm{e}^-$ bunch helicity from the
    polarised source was verified (1992-1993).}  Although the electron
  bunch polarisation state was transmitted via reliable and redundant
  paths to the SLD detector/polarimeter complex, the SLD electroweak
  group proposed a series of independent tests of the synchronisation
  of this data and the SLD event data.  In one such test, the laser
  optics at the SLC polarised source were temporarily modified by the
  addition of a polariser and quarter-wave plate so that photo-cathode
  illumination was nulled for one of the two circular polarisation
  states.  The positron beam was turned off, and the electron beam was
  delivered to the IP.  Beam-related background in the SLD
  liquid-argon calorimeter (LAC) was detected, but only for the
  non-extinct pulses.  By this means, the expected correlation between
  helicity and the presence of beam, and hence the LAC data stream,
  was verified~\cite{ref:sld-extinct}.  In addition, the helicity
  sequence generated at the source was pseudo-random and
  deterministic, and pulse patterns received at the SLD could be
  verified.
  
\item {\it Moderate precision M\o ller and Mott polarimeters confirmed
    the high precision Compton polarimeter result to $\sim 3\%$
    (1993-1995).}  M\o ller polarimeters located at the end of the
    SLAC linac and in the SLC electron extraction line were used to
    cross-check the Compton polarimeter.  The perils of employing a
    less reliable method to test a precision device were apparent when
    large corrections for atomic electron momentum effects in the M\o
    ller target were discovered~\cite{ref:sld-levchuk}, after which,
    good agreement was obtained.  In addition, a less direct
    comparison was provided by Mott polarimeter bench tests of the
    GaAs photo-cathodes~\cite{ref:sld-Mott}.
  
\item {\it SLC arc spin transport was extensively studied (1993-1998),
    and was frequently monitored and adjusted}.  A series of
  experiments was done that studied the beam polarisation reported by
  the Compton polarimeter as a function of beam energy, beam energy
  spread and beam trajectory in the SLC arcs.  Two spin rotators (in
  the linac, and in the ring-to-linac return line) were scanned in
  order to determine the IP polarisation maximum.  An important result
  of these experiments was the discovery that the SLC arcs operate
  near a spin tune resonance, leading to the advent of spin
  manipulation via ``spin bumps" in the SLC arcs mentioned earlier.
  This procedure eliminated the need for the two spin rotators and
  allowed the spin chromaticity (${\mathrm{d {\mathcal P}/d}}E$) to be
  minimised, reducing the resulting polarisation correction from
  $>1\%$ in 1993 to $<0.2\%$ by 1995.  In subsequent years the spin
  transport properties of the SLC arcs were monitored at regular
  intervals.
  
\item{\it Positron polarisation was experimentally constrained.}  In
  1998, a dedicated experiment was performed in order to directly test
  the expectation that unintended polarisation of the positron beam
  was negligible ; the positron beam was delivered to a M\o ller
  polarimeter in the SLAC End Station A (ESA).  Experimental control
  was assured by first delivering the polarised electron beam, and
  then an unpolarised electron beam (sourced from SLAC's thermionic
  electron gun), to the ESA, confirming polarimeter operation.  In
  addition, the spin rotator magnet located in the Linac was reversed
  halfway through the positron beam running, reversing the sense of
  polarisation at the M\o ller target and reducing systematic error.
  The final result verified that positron polarisation was consistent
  with zero $(-0.02\pm 0.07)\%$~\cite{ref:sld-posipol}.

\end{itemize}

The asymmetries in luminosity, polarisation, and beam energy,
approximately $10^{-4}$, $10^{-3}$ and $10^{-7}$, respectively, were
all continually monitored using a small-angle radiative Bhabha counter
located $\approx 40\rm{m}$ from the IP, beamstrahlung monitors, beam
current monitors, the Compton polarimeter, and energy spectrometer
data.  The long-term average values of all asymmetries of this type
were reduced by the roughly tri-monthly reversal of the transverse
polarisation sense in the electron damping ring referred to in
Section~\ref{subsec:SLCepol}.  The dominant cause of the observed
asymmetries was the small current asymmetry produced at the SLC
polarised source.  This effect arose because of the source
photo-cathode sensitivity to linearly polarised light, together with
residual linear polarisation in the source laser light that was
correlated with the light helicity.  This effect was minimised by a
polarisation control and intensity feedback system starting in 1993,
and was generally maintained at below $10^{-4}$.

The value of $\ALR$ is unaffected by decay-mode-dependent variations
in detector acceptance and efficiency provided, for the simple case of
Z decay to a fermion pair, that the efficiency for detecting a fermion
at some polar angle is equal to the efficiency for detecting an
antifermion at the same polar angle.  In hadronic Z decays, the
fermions in question are the initial quark-antiquark pair, which
materialise as multi-particle jets.  These facts, and the high degree
of polar symmetry in the SLD detector, render $\aeff$ completely
negligible.  Finally, $\polp$ was experimentally demonstrated to be
consistent with zero to a precision of $7 \times 10^{-4}$ as described
above. Calculations based on polarisation buildup in the positron
damping ring suggested a much smaller number, $\polp <
\calO(10^{-5})$. Hence, no correction for $\polp$ was applied to
the data.

The systematic effects discussed in this section are summarised in
Table~\ref{tab:alr:alrcorr}.
\begin{table}[tbp]
\begin{center}
\renewcommand{\arraystretch}{1.25}
\begin{tabular}{|c||c|c|c|c|c|}
\hline 
Source
 & {\bf 1992} & {\bf 1993} & {\bf 1994-95} & {\bf 1996} & {\bf 1997-98} \\
\hline
\hline
$N_{\mathrm{L}}$  &   5,226 & 27,225 & 52,179 & 29,016 & 183,335 \\
\hline
$N_{\mathrm{R}}$  &   4,998 & 22,167 & 41,465 & 22,857 & 148,259 \\
\hline
$\ameas$  & $0.0223$ & $0.1024$ & $0.1144$ & $0.1187$ & $0.1058$  \\
         &$\pm 0.0099$ & $\pm 0.0045$ & $\pm 0.0032$ & $\pm 0.0044$ & $\pm 0.0017$ \\
\hline
\hline
$f_{\rm bkg}$ (\%) & $1.4$  & $0.25$ & $0.11$ & $0.029$ & $0.042$     \\
                   & $\pm 1.4$ & $\pm 0.10$ & $\pm 0.08$ & $\pm 0.021$ & $\pm 0.032$ \\
\hline
$A_{\rm bkg}$ & & $0.031$     & $0.055$     & $0.033$     & $0.023$      \\
              & & $\pm 0.010$ & $\pm 0.021$ & $\pm 0.026$ & $\pm 0.022$ \\
\hline
$A_\calL$ $(10^{-4})$ & $1.8$ & $0.38$  & $-1.9$  & $+0.03$ & $-1.3$  \\
                      & $\pm\: 4.2$ & $\pm\: 0.50$ & $\pm\: 0.3$ 
                       & $\pm\: 0.50$ & $\pm\: 0.7$ \\
\hline  
$A_\calP$ $(10^{-4})$ & $-29$ & $-33$ & $+24$  & $+29$ & $+28$      \\
                       & & $\pm\: 1$ & $\pm\: 10$ & $\pm\: 43$ & $\pm\: 69$ \\
\hline
$A_{E}$ $(10^{-4})$ & & $0.0044$ & $0.0092$ & $-0.0001$ & $+0.0028$      \\
                    & & $\pm 0.0001$ & $\pm 0.0002$ 
                                & $\pm 0.0035$ & $\pm 0.0014$ \\
\hline
$E_{\rm cm} \frac{\sigma '(E_{\rm cm})}{\sigma (E_{\rm cm})}$
                  & & $-1.9$ & $0.0$       & $2.0$       & $4.3$  \\
                  & &        & $\pm\: 2.5$ & $\pm\: 3.0$ & $\pm\: 2.9$ \\
\hline
$A_{\varepsilon}$ & 0 & 0 & 0 & 0 & 0 \\
                  & & & & & \\
\hline
$\polp$ $(10^{-4})$  & $<0.16$ & $<0.16$ & $<0.16$ & $<0.16$ & $-2$ \\
                       &     &         &         &         & $\pm\: 7$ \\
\hline
\hline
Total correction,   & $\it + \: 2.2$ & $+ \: 0.10$  & $+ \: 0.2$   & $+0.02$ & $+0.16$   \\
$\Delta \ALR / \ALR$, (\%) & $\it \pm\: 2.3$ & $\pm\: 0.08$ & $\pm\: 0.06$ 
                           & $\pm\: 0.05$ & $\pm\: 0.07$ \\
\hline
\hline
$\delta\Pe/\Pe$ (\%) & 2.7 & 1.7 & 0.67 & 0.52 & 0.52 \\
\hline
\hline
Electroweak interference &$\it - \: 2.4$ &$+ \: 1.7$ &$+ \: 1.8$ &$+ \: 2.2$ &$+ \: 2.5$ \\
correction [relative (\%)] & $\it \pm\: 1.4$ &$\pm\: 0.3$&$\pm\: 0.3$&$\pm\: 0.4$&$\pm\: 0.39$ \\
\hline
\hline
Total systematic & 3.9 & 1.7 & 0.75 & 0.63 & 0.64 \\
error [relative (\%)] & & & & & \\
\hline
\end{tabular}
\caption[SLD Z event counts and corrections]
{Z event counts and corrections (see Equation~\ref{eq:ALRsys}) for all
  SLD run periods.  Also shown are the total polarimetry errors
  (including chromaticity and IP effects), the relative errors due to
  the electroweak interference correction needed for the conversion of
  $\ALR$ to $\ALRz$, and the total systematic errors.  Note that due
  to low statistics a number of effects were ignored for the 1992 data
  and no corrections were applied (given here in italics).  Also, the
  total systematic error reported in 1992 (3.6\%) ignored the uncertainty
  due to the electroweak correction.  }
\label{tab:alr:alrcorr}
\end{center}
\end{table}
The corrections for backgrounds and accelerator asymmetries, and the
associated uncertainties, were much smaller than the leading
systematic errors due to polarimetry and energy uncertainties, as can
be seen by comparing the penultimate three rows of
Table~\ref{tab:alr:alrcorr}.

\subsection{Results}

\begin{figure}[btp]
\begin{center}
\mbox{\epsfig{file=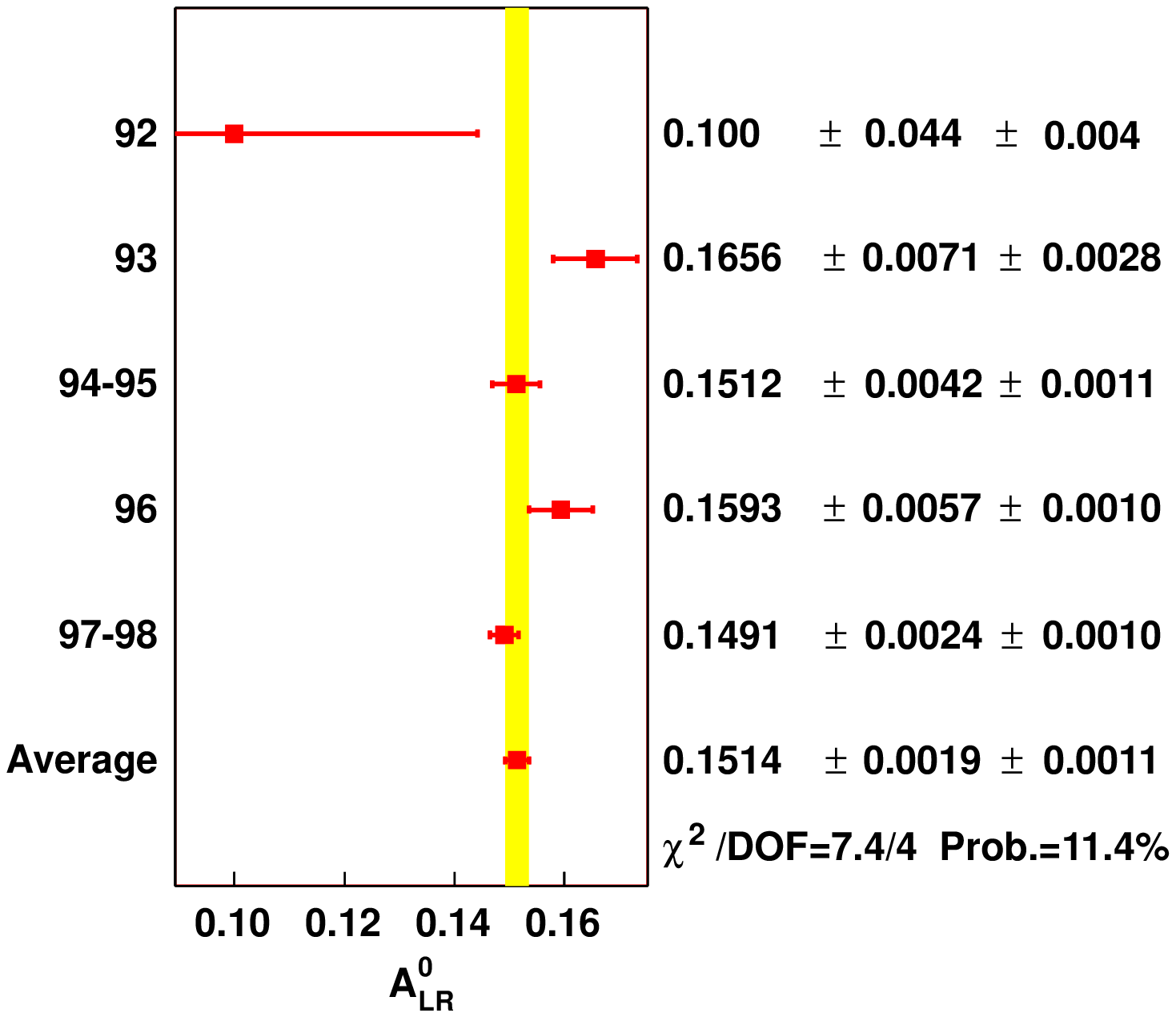,width=\linewidth}}
\caption[History of the SLD $\ALRz$ measurements.]
{A compilation of the published SLD $\ALRz$ results, ordered by year.
  The final average is formed including correlations in systematic
  errors.}
\label{fig:alrhistory}
\end{center}
\end{figure}

The run-by-run $\ALR$ results are shown in
Table~\ref{tab:sld-alrresults}.  The $\Ecm$ dependent radiative
correction, and its uncertainty, is evident in the difference between
$\ALR$ and $\ALRz$.  These five results show a $\chi^2$ of 7.44 for 4
degrees of freedom, corresponding to a probability of 11.4\%
(Figure~\ref{fig:alrhistory}).  The $\swsqeffl$ results derive from
the equivalence $\ALRz \equiv \cAe$, which along with
Equations~\ref{eq:cA} and~\ref{eq:gvfovergaf} provide that
\begin{equation}
\ALRz ~=~ \frac{2(1 - 4 \swsqeffl )}
{1 + (1 - 4 \swsqeffl)^{2}}.
\label{eq:alrsin}
\end{equation}
The average for the complete SLD data sample is:
\begin{eqnarray}
\ALRz     & = & 0.1514  \pm 0.0022
\end{eqnarray}
or equivalently:
\begin{eqnarray}
\swsqeffl & = & 0.23097 \pm 0.00027 \,.
\end{eqnarray}
Small correlated systematic effects due to polarimetry are accounted
for in forming this average.  The estimated systematic uncertainties for
these results are $\pm 0.0011$ and $\pm 0.00013$, respectively.

\begin{table}[p]
\begin{center}
\renewcommand{\arraystretch}{1.5}
\begin{tabular}{|c||c|c|c|c|}
\hline
 & $\apolel$ & $\ALR$ & $\ALRz$ & $\swsqeffl$ \\
\hline
\hline
1992 & $0.244$ & $0.100$ & $0.100$ & $0.2378$ \\
 & $\pm 0.006$ &  $\pm 0.044 \pm 0.004$ & $\pm 0.044 \pm 0.004$ & 
   $\pm 0.0056 \pm 0.0005$ \\
\hline
1993 & $0.630$ & $0.1628$ & $0.1656$ & $0.2292$ \\
 & $\pm 0.011$ &  $\pm 0.0071 \pm 0.0028$ &  $\pm 0.0071 \pm 0.0028$ & 
   $\pm 0.0009 \pm 0.0004$ \\
\hline
1994/95 & $0.7723$  & $0.1485$ & $0.1512$ & $0.23100$ \\
 & $\pm 0.0052$ &  $\pm 0.0042 \pm 0.0010$ &  $\pm 0.0042 \pm 0.0011$ &
   $\pm 0.00054 \pm 0.00014$ \\
\hline
1996 & $0.7616$  & $0.1559$ & $0.1593$ & $0.22996$ \\
 & $\pm 0.0040$ &  $\pm 0.0057 \pm 0.0008$ &  $\pm 0.0057 \pm 0.0010$ & 
   $\pm 0.00073 \pm 0.00013$ \\
\hline
1997/98 & $0.7292$ & $0.1454$ & $0.1491$ & $0.23126$ \\
 & $\pm 0.0038$ &  $\pm 0.0024 \pm 0.0008$ &  $\pm 0.0024 \pm 0.0010$ &
   $\pm 0.00030 \pm 0.00012$ \\
\hline
\hline
All    & & & $0.1514 $ & $0.23097 $ \\
combined & & & $\pm 0.0019 \pm 0.0011$ & $\pm 0.00024 \pm 0.00013$ \\
\hline
\end{tabular}
\caption[Summary of the SLD $\ALR$ measurements]
{Summary of the SLD $\ALR$ measurements for all runs.  Listed are the
  luminosity-weighted mean electron polarisation $(\apolel)$, the
  measured $\ALR$, its value corrected to the Z-pole $(\ALRz)$ and
  $\swsqeffl$.  For $\apolel$ the total error shown is dominantly
  systematic.  For the other quantities, the errors are the
  statistical and systematic components respectively. The final
  combined result accounts for correlated uncertainties. }
\label{tab:sld-alrresults}
\end{center}
\end{table}

\clearpage

\section{Lepton Asymmetry Measurements}

The SLD collaboration determined the individual lepton asymmetry
parameters using lepton final-state
events~\cite{ref:sld-al2000,ref:sld-al1997}.  Electron polarisation
allows one to determine the final-state asymmetry parameter $\cAl$ for
lepton $\ell$ in a single measurement using the left-right
forward-backward asymmetry, $\AFBLRzl=\frac{3}{4}|\pole|\cAl$.  An
advantage of polarisation is that with $\pole = 75\%$, the left-right
forward-backward asymmetries yield a statistical precision equivalent
to measurements of the unpolarised forward-backward asymmetry using a
25 times larger event sample.

If lepton universality is assumed, the results for all three lepton
flavours can be combined to yield a determination of $\swsqeffl$,
which in turn can be combined with the more precise result from
$\ALR$.  The event sample used for $\ALR$ is almost purely hadronic as
there is only a very small, $(0.3\pm0.1)\%$, admixture of tau pair
events - hence the left-right asymmetry of the lepton events was a
statistically independent measurement.  While the lepton final-state
analysis described in what follows is more sophisticated than an
$\ALR$-style counting measurement, essentially all the information on
$\swsqeffl$ is obtained from the left-right asymmetry of these events.
The inclusion of the distributions in polar angle that are essential
for the extraction of the final-state asymmetries improves the
resulting precision on $\swsqeffl$, but only to $\pm 0.00076$ compared
to about $\pm0.00078$ obtained from a simple left-right event count.

The differential cross-section for the pure Z amplitude $\ee
\rightarrow {\Zzero} \rightarrow \ff$ is factorized as follows:
\begin{eqnarray}
\frac{d}{dx}\sigma_{\Zzero}(x,s,\pole ;\cAe,\cAl)  
& \equiv & 
  f_{\Zzero}(s)\Omega_{\Zzero}(x,\pole ;\cAe,\cAl)  \nonumber \\
&   =    & 
  f_{\Zzero}(s)\left[ (1-\pole \cAe)(1+x^2)+(\cAe-\pole)\cAl 2x \right] \,,
  \label{eq:alepdifcs}
\end{eqnarray}
where $f_{\Zzero}$ isolates dependence on $s$, the squared
centre-of-mass energy, and $\Omega_{\Zzero}$ contains the dependence
on $x=\cos\theta$, which gives the direction of the outgoing lepton
$\ell^-$ with respect to the electron-beam direction.  For a complete
description of lepton pair production, photon exchange terms and, if
the final-state leptons are electrons, $t$-channel contributions have
to be taken into account, as we describe below.

\subsection{Analysis Method}

\begin{figure}[p]
\begin{center}
\mbox{\epsfig{file=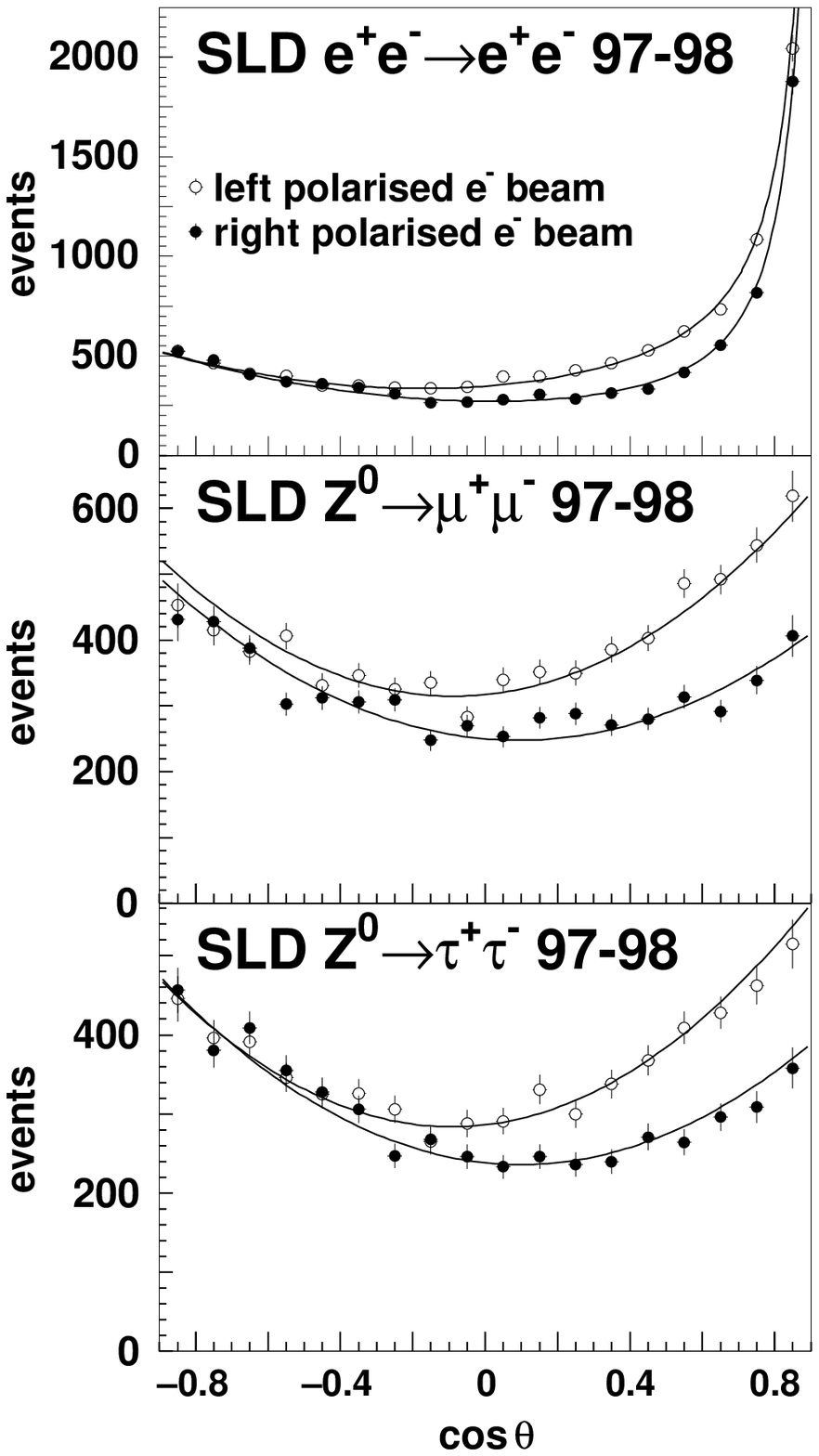,width=0.7\linewidth}}
\caption[Polarisation dependent angular distributions for leptons]
{Polar-angle distributions for Z decays to e, $\mu$ and $\tau$ pairs
  for the 1997-1998 SLD run.  The solid line represents the fit, while
  the points with error bars show the data in bins of 0.1 in
  $\cos\theta_{\mathrm{thrust}}$.  For
  $|\cos\theta_{\mathrm{thrust}}|>0.7$, the data are corrected for a
  decrease in the detection efficiency with increasing
  $|\cos\theta_{\mathrm{thrust}}|$.  Note that the polarization
  independence at $\cos\theta = -1$ implied by
  Equation~\ref{eq:alepdifcs}, for the case of lepton universality, is
  apparent. }
\label{fig:alr:distr}
\end{center}
\end{figure}

Figure~\ref{fig:alr:distr} shows the $\cos\theta$ distributions for
$\ee$, $\mu^+\mu^-$, and $\tau^+\tau^-$ candidates for the 1997-1998
data.  Leptonic final-state events are identified, and
Table~\ref{tab:sld:alevts} summarises the selection efficiencies,
backgrounds and numbers of selected candidates for $\ee$,
$\mu^+\mu^-$, and $\tau^+\tau^-$ final states.  The pre-1997 results
are similar but have smaller acceptance $|\cos\theta| \le 0.8$,
reflecting the improved acceptance of an upgraded vertex detector used
for the newer data, which allowed for efficient track finding up to
$|\cos\theta|=0.9$.  The SLD event totals, including all data, are
$22\,254$, $16\,844$ and $16\,084$ for the electron-, muon- and
tau-pair final states respectively.

An event-by-event maximum likelihood fit is used to incorporate the
contributions of all the terms in the cross-section and to include the
effect of initial-state radiation.  There are three likelihood
functions for individual lepton final states.  All three lepton final
states contribute to the measurement of $\cAe$, while $\mu^+\mu^-$ and
$\tau^+\tau^-$ final states are used to determine $\cAm$ and $\cAt$
respectively.

The likelihood function for muon- and tau-pair final states is defined
as follows:
\begin{eqnarray}
\calL(x,s,\pole ;\cAe,\cAl) & = & 
\int \mathrm{d}s^\prime
H(s,s^\prime)
\left\{
 \frac{\mathrm{d}}{\mathrm{d}x}\sigma_{\Zzero}(x,s^\prime,\pole ;\cAe,\cAl) \right. \nonumber \\
 &  & \left. + \frac{\mathrm{d}}{\mathrm{d}x}\sigma_{Z\gamma}(x,s^\prime,\pole ;\cAe,\cAl)
 + \frac{\mathrm{d}}{\mathrm{d}x}\sigma_{\gamma}(x,s^\prime) 
\right\} \,,
\label{Eq:likelihood function(MIZA1)}
\end{eqnarray}
where $\cAe$ and $\cAl$ (=$\cAm$ or $\cAt$) are free parameters and
$H(s,s^\prime)$ is a radiator function.  The integration over
$s^\prime$ was done with the program MIZA~\cite{\MIZA} to take into
account the initial-state radiation.  The spread in the beam energy
had a negligible effect.
$(\mathrm{d}\sigma_{\Zzero}/\mathrm{d}x)(...)$,
$(\mathrm{d}\sigma_{\gamma}/\mathrm{d}x)(...)$, and
$(\mathrm{d}\sigma_{\Zzero\gamma}/\mathrm{d}x)(...)$ are the
tree-level differential cross-sections for Z exchange, photon
exchange, and their interference.  The integration was performed
before the fit to obtain the coefficients $\bar{f}_{\Zzero}$,
$\bar{f}_{\Zzero\gamma}$, and $\bar{f}_\gamma$, resulting in the
likelihood function for muon- and tau-pair final states :
\begin{equation}
\calL(x,s,\pole ;\cAe,\cAl) ~=~
  \bar{f}_{\Zzero}(s) \Omega_{\Zzero}(x,\pole ;\cAe,\cAl)
 + \bar{f}_\Zzero\gamma(s) \Omega_\Zzero\gamma(x,\pole ;\cAe,\cAl)
 + \bar{f}_{\gamma} (s) \Omega_{\gamma}(x)\,,
\label{Eq:likelihood function(MIZA2)}
\end{equation}
where the differential cross-sections have been factorized in analogy
with Equation~\ref{eq:alepdifcs}.  These coefficients gave the
relative sizes of the three terms at the SLC centre-of-mass energy,
e.g., $\sqrt{s}=91.237 \pm 0.029~\GeV$ for the 1997-1998 run.

\begin{table}[t]
\begin{center}
\renewcommand{\arraystretch}{1.2}
\begin{tabular}{|c||c|c|c|}
\hline
Event  & Background    & Efficiency in      & Selected \\
Sample & Fraction [\%] & $|\cos\theta|<0.9$ [\%] & Events\\ 
\hline
\hline
$\ee\rightarrow\ee$     &
   $\tau^+\tau^-$: 0.7 &              75 & 15675 \\ \hline
$\ee\rightarrow\mumu$   &
   $\tau^+\tau^-$: 0.2 &              77 & 11431 \\ \hline
$\ee\rightarrow\tautau$ &
   $\ee$: $\mumu$: ~$\gamma\gamma$: had: & 70 & 10841    \cr
 &~0.9~~~~~2.9~~~~~0.9~~~~0.6~ &    &  \\\hline
\end{tabular}
\caption[Properties of the SLD $\ee \to \ell^+\ell^-$ event selections]
{Summary of event selections, efficiency, and purity for $\ee \to
  \ell^+\ell^-$ for the 1997-1998 SLD data}
\label{tab:sld:alevts}
\end{center}
\end{table}

The $\ee$ final state includes both $s$-channel and $t$-channel Z and
photon exchanges which yields four amplitudes and ten cross-section
terms.  All ten terms are energy-dependent.  A maximum likelihood
function for $\ee$ final states was defined by modifying
Equations~\ref{Eq:likelihood function(MIZA1)} and~\ref{Eq:likelihood
function(MIZA2)} to include all ten terms.  The integration over
$s^{\prime}$ was performed with DMIBA~\cite{ref:sld-dmiba} to obtain
the coefficients for the relative size of the ten terms.

\subsection{Systematic Errors}

Systematic uncertainties are summarised in Table~\ref{tab:sld:alsys},
from which it is clear that this measurement is entirely statistics
dominated.  The errors for the 1997-98 dataset, which dominates the
sample, are shown.

\begin{table}[ht]
\begin{center}
\renewcommand{\arraystretch}{1.1}
\begin{tabular}{|l||r|r|r|r|r|}
\hline
Observable       & $\cAe     $ & $\cAe     $ & $\cAe     $ & $\cAm     $ & $\cAt     $   \\ 
\hline
Channel $\ee\to$ & $     \ee $ & $   \mumu $ & $ \tautau $ & $   \mumu $ & $ \tautau $   \\ 
\hline
Uncertainty      & $[10^{-4}]$ & $[10^{-4}]$ & $[10^{-4}]$ & $[10^{-4}]$ & $[10^{-4}]$ \\ 
\hline
\hline
Statistics              & 110 & 130 & 130 & 180 & 180 \\
\hline
Polarisation            &   8 &   8 &   8 &   8 &   8 \\
Backgrounds             &   5 &  -- &  13 &  -- &  14 \\
Radiative Correction    &  23 &   2 &   2 &   3 &   2 \\
V-A                     &  -- &  -- &  -- &  -- &  18 \\
Charge Confusion        &  -- &  -- &  -- &   7 &  11 \\
Detector asymmetry      &  -- &  -- &  -- &  -- &   4 \\
Nonuniform efficiency   &   2 &  -- &  -- &  -- &  -- \\ 
\hline
\end{tabular}
\caption[Summary of uncertainties
for the 1997-1998 SLD $\ee \to \ell^+\ell^-$ data]
{\label{tab:sld:alsys} Summary of statistical and systematic
uncertainties, in units of $10^{-4}$, for the 1997-1998 SLD $\ee \to
\ell^+\ell^-$ data.  }
\label{Table:systematics}
\end{center}
\end{table}

The uncertainty on the beam polarisation is correlated among all the
measurements and corresponds to an uncertainty on $\cAl$ of $\pm
0.0008$.  The uncertainty in the amount of background and its effect
on the fitted parameters are taken into account.  The background
contaminations have been derived from detailed Monte Carlo simulations
as well as from studying the effect of cuts in background-rich samples
of real data.

The radiative corrections and their systematic errors are estimated
using MIZA~\cite{\MIZA} and DMIBA~\cite{ref:sld-dmiba}.
The uncertainty in the asymmetry parameters due to a $\pm1\sigma$
variation of $\sqrt{s}$, the dominant systematic effect for radiative
corrections, is of the order $10^{-4}$, except for the $\cAe$
determination from $\ee$ final states for which it is of order
$10^{-3}$.

The dominant systematic error in the tau analysis results from the V-A
structure of tau decay, which introduces a selection bias in the
analysis.  For example, if both taus decay to $\pi\nu$, helicity
conservation requires that both pions generally have lower momentum
for a left-handed $\tau^-$ and right-handed $\tau^+$ and higher
momentum otherwise.  This effect, which biases the reconstructed event
mass, is large at the SLD because the high beam polarisation induces a
very high and asymmetric tau polarisation as a function of polar
angle.  The value of $\cAe$ extracted from $\tau^+\tau^-$ final states
is not affected since the overall relative efficiencies for
left-handed beam and right-handed beam events are not changed
significantly, only the polar angle dependence of the efficiencies is
changed.

\subsection{Results}

Results for all SLD runs are combined while accounting for small
effects due to correlations entering through the systematic
uncertainties in polarisation and average SLD centre-of-mass energy.
From purely leptonic final states, one obtains $\cAe=0.1544 \pm
0.0060$.  This $\cAe$ result is combined with the left-right asymmetry
measurement in the final tabulation of SLD leptonic asymmetry results
which is reported in Table~\ref{tab:alr:result}.

\begin{table}[htbp]
\begin{center}
\renewcommand{\arraystretch}{1.25}
\begin{tabular}{|c||r@{$\pm$}l||rrr|}
\hline
Parameter & \multicolumn{2}{|c||}{Average} 
          & \multicolumn{3}{|c| }{Correlations}    \\
          & \multicolumn{2}{|c||}{ }
          & {$\cAe$} & {$\cAm$}& {$\cAt$}          \\
\hline
\hline
$\cAe$ &$0.1516$&$0.0021$ & $ 1.000$&$      $&$      $ \\
$\cAm$ &$0.142 $&$0.015 $ & $ 0.038$&$ 1.000$&$      $ \\
$\cAt$ &$0.136 $&$0.015 $ & $ 0.033$&$ 0.007$&$ 1.000$ \\
\hline
\end{tabular}
\caption[Results on the leptonic asymmetry parameters $\cAl$ from SLD]
{Results on the leptonic asymmetry parameters $\cAl$ not assuming
  neutral-current lepton universality obtained at SLD. The result on
  $\cAe$ includes the result on $\ALRz$. }
\label{tab:alr:result}
\end{center}
\end{table}

\section{Combined Results}

These results are consistent with lepton universality and hence can be
combined into $\cAl$, which in the context of the Standard Model is
simply related to the electroweak mixing angle.  Assuming lepton
universality and accounting for correlated uncertainties, the combined
result is:
\begin{equation}
 \cAl = 0.1513 \pm 0.0021,
\label{eq:al:result}
\end{equation}
where the total error includes the systematic error of $\pm 0.0011$.
This measurement is equivalent to a determination of:
\begin{equation}
 \swsqeffl = 0.23098 \pm 0.00026,
\end{equation}
where the total error includes the systematic error of $\pm 0.00013$.

\boldmath
\chapter{The Tau Polarisation Measurements}
\label{sec-TP}
\unboldmath

\section{Introduction}
Parity violation in the weak neutral current results in a non-zero
longitudinal polarisation of fermion pairs produced in the reaction
$\eeff$, with the $\tau$ lepton being the only fundamental final-state
fermion whose polarisation is experimentally accessible at
LEP~\cite{Eberhard:1989ve}.  The $\tau$ polarisation, \ptau, is given
by
\begin{eqnarray}
\ptau & \equiv  & (\sigma_+ - \sigma_-)/(\sigma_+ + \sigma_-)\,,
\end{eqnarray}
where $\sigma_{+}$ represents the cross-section for producing positive
helicity $\tau^-$ leptons and $\sigma_{-}$ those of negative helicity.
The $\gl$ and $\gr$ neutral current couplings, introduced in
Equations~\ref{eq:gl} and~\ref{eq:gr}, quantify the strength of the
interaction between the $\Zzero$ and the chiral states of the
fermions.  A subtle, but conceptually important, point is that the polarisation
measurements involve the fermion helicity states, as opposed to their
chiral states.  The ($1 \pm \gamma_5$)/2 operators project out states
of a definite chirality: ($1 - \gamma_5$)/2 projects out the
left-handed chiral fermion (and right-handed anti-fermion) states and
($1 + \gamma_5$)/2 the right-handed chiral fermion (and left-handed
anti-fermion) states. In contrast, helicity is the projection of the
spin onto the direction of the fermion momentum: if the spin and
momentum are oppositely aligned, the helicity is negative whereas if
the spin and momentum are aligned, the helicity is positive. In the
extreme relativistic limit, ($1 - \gamma_5$)/2 projects out negative
helicity states and ($1 + \gamma_5$)/2 positive helicity states. The
left-handed chiral fermion (and right-handed anti-fermion) states
become indistinguishable from the measured negative helicity states
and the right-handed chiral fermion (and left-handed anti-fermion)
states from the positive helicity states.  Consequently, at LEP, where
the $\tau$ leptons are produced with highly relativistic energies,
$\ptau$ provides a direct measurement of the chiral asymmetries of the
neutral current.  By convention, $\ptau \equiv \ptaum$ and since, to a
very good approximation, the $\tau^-$ and $\tau^+$ have opposite
helicities at LEP, $\ptaum=-\ptaup$.

For pure $\Zzero$ exchange in the interaction of the unpolarised $\ee$
beams at LEP, the dependence of \ptau ~on $\theta_{\tau^-}$, the angle
between the $\tau^-$ momentum and e$^-$ beam, can be described by a
simple relation expressed in terms of the two neutral current
asymmetry parameters, $\cAt$ and $\cAe$, and the forward-backward
asymmetry of the $\tau$, $\Afb^{\tau}$:
\begin{equation}
\label{eq-ptcos}
\ptau(\cos\theta_{\tau^-})= - \frac
{\cAt(1+\cos^2\theta_{\tau^-}) + 2\cAe \cos\theta_{\tau^-}}
{(1+\cos^2\theta_{\tau^-}) + \frac{8}{3}\Afb^{\tau}\cos\theta_{\tau^-}}.
\end{equation}
The $\tau$ polarisation measurements allow for the determination of
$\cAt$ and $\cAe$ and are largely insensitive to $\Afb^{\tau}$.

The four LEP experiments use kinematic distributions of the observable
$\tau$ decay products, and the V$-$A nature of the charged weak
current decays, to measure the polarisation as a function of
$\cos\theta_{\tau^-}$ in data collected during the 1990-95 $\Zzero$
running period.  Because the actual reaction does not only contain the
pure $\Zzero$ propagator but also includes contributions from the
photon propagator, $\gammaZ$ interference, and other photonic
radiative corrections, the parameters obtained using
Equation~\ref{eq-ptcos} are approximations to $\cAt$ and $\cAe$.  In
order to distinguish between these pure $\Zzero$ parameters and those
which include the small non-$\Zzero$ effects, the measured parameters
are denoted as $\pta$ and $\AFBpol$ in the literature.  \pta ~is the
average $\tau$ polarisation over all production angles and \AFBpol ~is
the forward-backward polarisation asymmetry.  If one had only pure
$\Zzero$ exchange, these would be trivially related to the neutral
current asymmetry parameters: $\pta=-\cAt$ and
$\AFBpol=-\frac{3}{4}\cAe$.  ZFITTER~\cite{\ZFITTERref} is used to
convert from $\pta$ and $\AFBpol$ to $\cAt$ and $\cAe$, respectively,
by correcting for the contributions of the photon propagator,
$\gammaZ$ interference and electromagnetic radiative corrections for
initial state and final state radiation. These corrections have a
$\roots$ dependence which arises from the non-$\Zzero$ contributions
to \pta ~and \AFBpol.  This latter feature is important since the
off-peak data are included in the event samples for all experiments.
Ultimately, all LEP collaborations express their $\tau$ polarisation
measurements in terms of $\cAt$ and $\cAe$.

It is important to remark that this method of measuring
$\ptau(\cos\theta_{\tau^-})$ yields nearly independent determinations
of $\cAt$ and $\cAe$. Consequently, the $\tau$ polarisation
measurements provide not only a determination of $\swsqeffl$ but also
test the hypothesis of the universality of the couplings of the
$\Zzero$ to the electron and $\tau$ lepton.

A general overview describing the experimental methods for measuring
the $\tau$ polarisation at LEP is contained in
Section~\ref{TauPolExperimentalMethods}.  This is followed in
Section~\ref{TauPolSystematicErrors} by a discussion of the dominant
systematic uncertainties relevant to these measurements. The results
for $\cAt$ and $\cAe$ from each of the four LEP experiments are
presented in Section~\ref{TauPolResults} as well as the combined
results with and without the assumption of lepton universality.  The
treatment of correlations between the measurements in the combined
results is also discussed in that section.

\section{Experimental Methods}
\label{TauPolExperimentalMethods}

The polarisation measurements rely on the dependence of kinematic
distributions of the observed $\tau$ decay products on the helicity of
the parent $\tau$ lepton.  Because the helicity of the parent cannot
be determined on an event-by-event basis, the polarisation measurement
is performed by fitting the observed kinematic spectrum of a
particular decay mode to a linear combination of the positive and
negative helicity spectra associated with that mode.

\begin{figure}
\begin{center}
\includegraphics[width=0.9\linewidth]{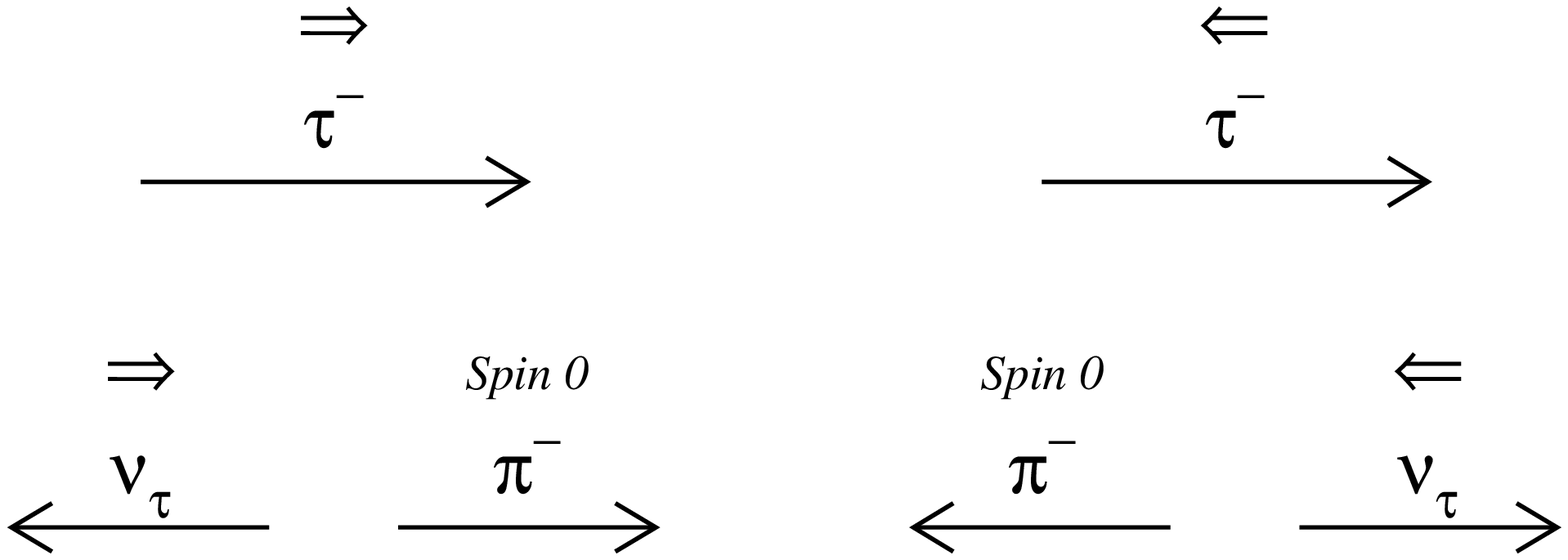}
\caption[Decay configurations for two $\tau$ helicity states]
{Decay configurations for two $\tau$ helicity states for the decay
  $\tau^-\rightarrow\pi^-\nu_{\tau}$. The positive helicity configuration
  is on the left and the negative configuration is on the right. For
  each particle, the long arrow depicts the momentum direction while
  the short double arrow that of the spin. The lower pair of figures is
  depicted in the helicity rest frame of the parent $\tau$. }
\label{fig:taupol_1}
\end{center}
\end{figure}

\begin{figure}[hbtp]
\begin{center}
\includegraphics[width=\linewidth]{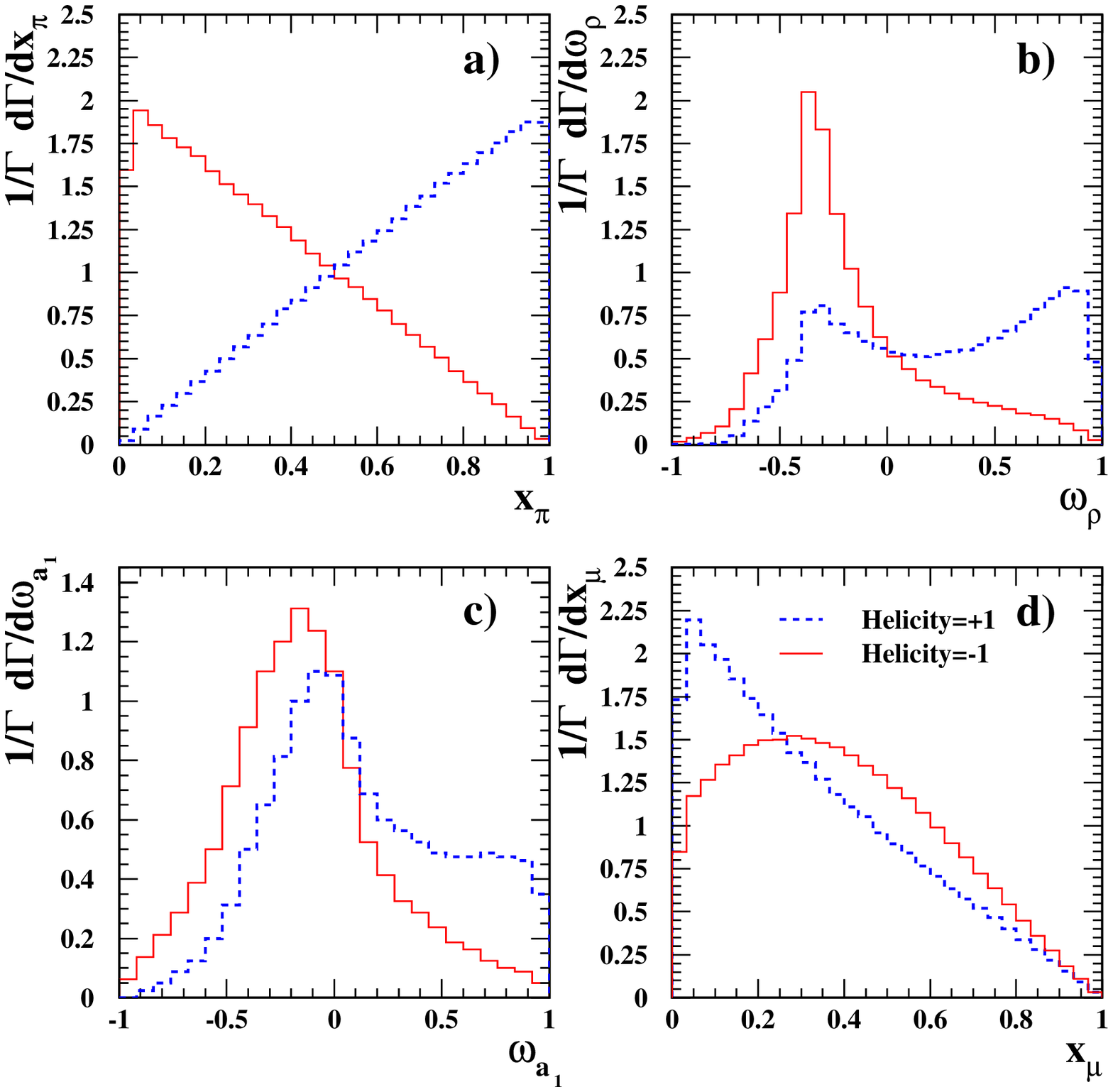}
\end{center}
\caption[ Distributions of polarisation sensitive kinematic variables]
{Monte Carlo simulated distributions of polarisation sensitive
  kinematic variables defined in the text for (a) $\tpinu$, (b)
  $\trhonu$, (c)$\tanu$ and (d) $\tmununu$ decays for positive and
  negative helicity $\tau$ leptons excluding the effects of selection
  and detector response.}
\label{fig:taupol_2} 
\end{figure}

For the simplest case, that of the two-body decay of a $\tau$ lepton
to a spin-zero $\pi$ meson and $\nu_{\tau}$, $\tau \rightarrow \pi
\nu_{\tau}$, the maximum sensitivity is provided by the energy
spectrum of the $\pi$ in the laboratory frame.  The pure V$-$A charged
weak current decay of the $\tau$ together with angular momentum
conservation produces a $\pi$ with momentum preferentially aligned
with the helicity of the $\tau$ as depicted in
Figure~\ref{fig:taupol_1}. In the laboratory frame this means that a
$\pi^-$ produced from the decay of a positive-helicity $\tau^-$ will,
on average, be more energetic than a $\pi^-$ produced from the decay
of a negative-helicity $\tau^-$.\footnote{ For $\tau^+$ decays the
current is V$+$A and the opposite kinematic relations hold. However,
because the $\tau^-$ and $\tau^+$ are produced with opposite
helicities, for a given \ptau ~the decay distributions are the same.}
In the helicity rest frame of the $\tau$,\footnote{ The $\tau$ rest
frame whose z-axis is aligned with the $\tau$ momentum as measured in
the laboratory frame.}
 the differential decay width is
\begin{equation}
\frac{1}{\Gamma}\frac{d\Gamma}{d \cos\theta_{\pi}} ~ = ~ 
 \frac{1}{2}  \left( 1 +  \ptau \cos \theta_{\pi} \right) \,,
\end{equation}
where $\theta_{\pi}$ is the polar angle of $\pi$ momentum 
in the $\tau$ helicity rest frame and $\ptau$ is the net polarisation
for an ensemble of $\tau$ leptons. This expression, when
boosted into the lab frame, gives a differential decay width of
\begin{equation}
 \frac{1}{\Gamma}\frac{d\Gamma}{d x_{\pi}} ~ = ~ 
 1 + \ptau ( 2 x_{\pi} -1 ) \,,
\end{equation}
where $x_{\pi}=E_{\pi}/E_{\tau}$ is the pion energy in the lab frame
scaled by the maximum energy available and terms of order
$(m_{\pi}/m_{\tau})^2$ have been neglected.  This is depicted in
Figure~\ref{fig:taupol_2}a for both helicity states.

More complex is the $\trhonu$ decay. The charged $\rho$ is a vector
meson with a $770~\MeV$ mass which decays promptly via $\rho
\rightarrow \pi \pi^0$. Having spin-1, the $\rho$ itself is polarised
with either helicity $\lambda_{\rho}$=0 or $\lambda_{\rho}$=$\pm$1
 for each $\tau$ helicity
configuration.  The cases where the $\rho$ is polarised with 
$\lambda_{\rho}$=0
 are equivalent to the $\tpinu$ configurations, but the
$\lambda_{\rho}$=$\pm$1 polarised cases produce the opposite angular
distribution.

The differential widths for $\trhonu$ are given
by~\cite{Hagiwara:1990fn}
\begin{equation}
\frac{1}{\Gamma} \frac{d\Gamma^{\lambda_{\rho}=0}}{d\cos\theta^*} ~ = ~
 \frac{m_{\tau}^2/2}{m^2_{\tau} +2 m^2_{\rho}} 
  \left( 1 + \ptau \cos\theta^* \right) 
\end{equation}
\begin{equation}
\frac{1}{\Gamma}
\frac{d\Gamma^{\lambda_{\rho}=\pm1}}{d\cos\theta^*} ~ = ~
 \frac{m_{\rho}^2}{m^2_{\tau} +2 m^2_{\rho}} 
  \left( 1 - \ptau \cos\theta^* \right) 
\end{equation}
where $\theta^*$ is the
angle between the $\rho$ momentum in the $\tau$ rest frame
 and the direction of the $\tau$ 
in the laboratory frame. The latter case effectively diminishes the
sensitivity to \ptau ~when only the $\theta^*$ angle is used, or,
equivalently, in the laboratory frame when only the $\rho$ energy is
used.

Much of this sensitivity, however, may be recovered by using
information from the $\rho$ decay products by, in effect,
spin-analysing the $\rho$.  The kinematic variable that provides this
information is the angle, $\psi$, between the charged pion momentum
in the $\rho$ rest frame and the $\rho$ flight direction in the
laboratory frame.  The two variables can be combined to form a single
variable without loss of polarisation
sensitivity~\cite{Davier:1993nw}.  This `optimal variable',
$\omega_{\rho}$, is given by
\begin{equation}
\omega_{\rho} ~ = ~ \frac{ 
\mbox{W}_{\mbox{\scriptsize +}}(\theta^*,\psi) - 
\mbox{W}_{\mbox{\scriptsize $-$}}(\theta^*,\psi)} 
{\mbox{W}_{\mbox{\scriptsize +}}(\theta^*,\psi) +
\mbox{W}_{\mbox{\scriptsize $-$}}(\theta^*,\psi)}\,, 
\end{equation}
where $\mbox{W}_{\mbox{\scriptsize +($-$)}}$ is proportional to the
differential decay width for positive (negative) helicity $\tau$ leptons,
as a function of $\theta^*$ and $\psi$.  The distributions of
$\omega_{\rho}$, for both positive and negative helicity $\tau$
decays, are shown in Figure~\ref{fig:taupol_2}b.

As with the $\trhonu$ decay, the $\tanu$ channel exhibits
significantly reduced polarisation sensitivity when only the a$_1$
energy is measured in the laboratory frame.  The a$_1$ is an
axial-vector meson with mass and width of approximately $1230~\MeV$
and $500~\MeV$, respectively, and decays to $\pi^-\pi^-\pi^+$ or
$\pi^-\pi^0\pi^0$ with nearly equal probability. There are again two
possible spin configurations where much of the sensitivity can be
regained through a spin analysis of the a$_1$ decay.  In this case six
variables are used which include: the angle $\theta ^{*}$ between the
a$_1$ momentum in the $\tau$ rest frame and the $\tau$ laboratory
flight direction; the angle $\psi$ in the rest frame of the a$_1$
between the vector perpendicular to the a$_1$ decay plane and the
a$_1$ laboratory flight direction; the angle $\gamma$ in the a$_1$
rest frame between the unlike-sign pion momentum and the a$_1$
laboratory flight direction projected into the a$_1$ decay
plane;\footnote{ The `unlike-sign pion' is defined as the $\pi^+$ in
the $\pi^-\pi^-\pi^+$ decay and the $\pi^-$ in the $\pi^-\pi^0\pi^0$
decay.} the $3\pi$-invariant mass; and the two unlike-sign pion mass
combinations present in the decays. In complete analogy with the
$\trhonu$, the polarisation information from these six variables is
fully contained in a single optimal variable, $\omega_{\mathrm
a_1}$~\cite{Davier:1993nw}.  The $\omega_{\mathrm a_1}$ distributions
for both positive and negative helicity $\tau$ decays are plotted in
Figure~\ref{fig:taupol_2}c.

For the leptonic channels, $\tenunu$ and $\tmununu$, the situation is
less favourable: all three final state particles carry off angular
momentum, but only one of the particles is measured. This causes a
substantial unrecoverable reduction in sensitivity relative to the
$\tpinu$ channel.  For these decays the variable is the scaled energy
of the charged decay product: $x_\ell = E_\ell/E_{\tau}$ for
$\ell=\mathrm{e},\mu$.  The decay distributions of the two leptonic
channels are almost identical.  Ignoring the masses of the daughter
leptons, the differential decay width is~\cite{Tsai:1971vv}:
\begin{equation}
 \frac{1}{\Gamma}\frac{d\Gamma}{d x_{\ell}} ~ = ~
\frac{1}{3}\left[ (5 - 9 x_{\ell}^2 + 4 x_{\ell}^3) + 
  \ptau (1 - 9 x_{\ell}^2 + 8 x_{\ell}^3 )\right].
\end{equation}
Shown in Figure~\ref{fig:taupol_2}d are the distributions for both
positive and negative helicity $\tmununu$ decays where the decrease in
sensitivity is apparent.  It should also be noted that, in contrast to
the $\tpinu$ channel, the positive helicity case now produces a
charged particle with lower energy on average than the negative
helicity case.

Each LEP experiment measures $\ptau$ using the five $\tau$ decay modes
e$\nu \overline{\nu}$, $\mu\nu \overline{\nu}$, $\pi\nu$, $\rho\nu$
and $a_{1}\nu$~\cite{\ALEPHTAU,\DELPHITAU,\LTAU,\OPALTAU} comprising
approximately 80\% of $\tau$ decays.\footnote{ As no experiment
discriminates between charged pions and kaons, the $\tpinu$ channel
also includes $\tau\rightarrow$K$\nu$ decays and the $\trhonu$ channel
also includes $\tau\rightarrow$K$\pi^0\nu$ decays.  Negligible
sensitivity is lost by combining these modes.}  As just demonstrated,
the five decay modes do not all have the same sensitivity to the
$\tau$ polarisation.  The maximum sensitivity for each decay mode,
defined as $\frac{1}{\sqrt{N} \sigma}$ where $\sigma$ is the
statistical error on the polarisation measurement using $N$ events for
$\ptau$=0, is given in Table~\ref{table-channelsens}. It assumes that
all the available information in the decay is used with full
efficiency both for the case when the three-dimensional $\tau$
direction information is not used and for the case when it is used.
The additional information provided by the $\tau$ direction is an
azimuthal angle of the decay of the hadronic system in the $\tau$ rest
frame~\cite{Davier:1993nw}.  When included in the decay distributions
of spin-1 hadronic channels with even modest precision an improvement
in the sensitivity is achieved.  A measure of the weight with which a
given decay mode ideally contributes to the overall measurement of the
polarisation is given by that decay mode's sensitivity squared
multiplied by its branching fraction. Normalised ideal weights, which
are calculated assuming maximum sensitivity and perfect identification
efficiency and purity, for each decay mode, are also given in
Table~\ref{table-channelsens}.  As can be seen, the $\trhonu$ and
$\tpinu$ channels are expected to dominate the combined polarisation
measurement, especially if information from the $\tau$ direction is
not used. The actual sensitivity achieved by each experiment for its
selected event sample is degraded because of inefficiencies in the
process of selecting a sample of decays, by the presence of background
in the sample and, to a lesser extent, by resolution effects. Much of
the background from cross-contamination from other $\tau$ decay
channels, however, retains some polarisation information which is
exploited by the fitting procedure.

\begin {table}[t]
\begin{center}
\renewcommand{\arraystretch}{1.2}
\begin{tabular}{|l||c|c|c|c|c|} \hline
                       & $\trhonu$   & $\tpinu$ & $\tenunu$ & $\tmununu$  & $\tanu$ \\
  &        &       &      &       &
   a$_1\ra\pi^{\pm}\pi^+\pi^-$ \\ \hline \hline
 Branching fraction        & 0.25   & 0.12  & 0.18 & 0.17 & 0.09 \\ \hline
 Maximum sensitivity:      &        &       &      &      &      \\
 ~~~~no 3D $\tau$ direction    & 0.49   & 0.58  & 0.22 & 0.22 & 0.45 \\
 ~~~~with 3D $\tau$ direction  & 0.58   & 0.58  & 0.27 & 0.27 & 0.58 \\ \hline
 Normalised ideal weight:  &        &       &      &      &      \\
 ~~~~no 3D $\tau$ direction    & 0.44   & 0.30  & 0.06 & 0.06 & 0.13 \\
 ~~~~with 3D $\tau$ direction  & 0.47   & 0.22  & 0.07 & 0.07 & 0.17 \\
\hline 
\end {tabular}
\caption[Tau polarisation decay channel sensitivity]
 {The branching fractions, maximum sensitivity~\cite{Davier:1993nw}
  and normalised ideal weight for the five decay modes listed.  The
  ideal weight is calculated as the product of the branching fraction
  and the square of the maximum sensitivity.  Presented in the last
  two lines of the table is the ideal weight for each channel divided
  by the sum of the ideal weights of the five channels.}
\label{table-channelsens}
\end{center}
\end{table}

In all analyses, a value of $\ptau$ is extracted from the data by
fitting linear combinations of positive and negative helicity
distributions in kinematic variables appropriate to each $\tau$ decay
channel to the data, where the two distributions are obtained from
Monte Carlo simulation. As discussed above, in the $\tmununu$,
$\tenunu$ and $\tpinu$ channels, the energy of the charged particle
from the $\tau$
decay divided by the beam energy is the appropriate kinematic variable
while for the $\trhonu$ and $\tanu$ channels, the appropriate optimal
variable, $\omega$, is employed.  Using Monte Carlo distributions in
the fitting procedure allows for simple inclusion of detector effects
and their correlations, efficiencies and backgrounds.  Any
polarisation dependence in the backgrounds from other $\tau$ decays
are automatically incorporated into these analyses. The systematic
errors associated with the detector then amount to uncertainties in
how well the detector response is modelled by the Monte Carlo
simulation, whereas the errors associated with
uncertainties in the underlying physics content in the distributions
arise from uncertainties in the Monte Carlo generators of the signal
and backgrounds.  The spin correlations between the two $\tau$-leptons
produced in a $\Zzero$ decay are treated differently in the different
experiments and are discussed below.

\begin{figure}[hbtp]
\begin{center}
\includegraphics[width=0.9\linewidth]{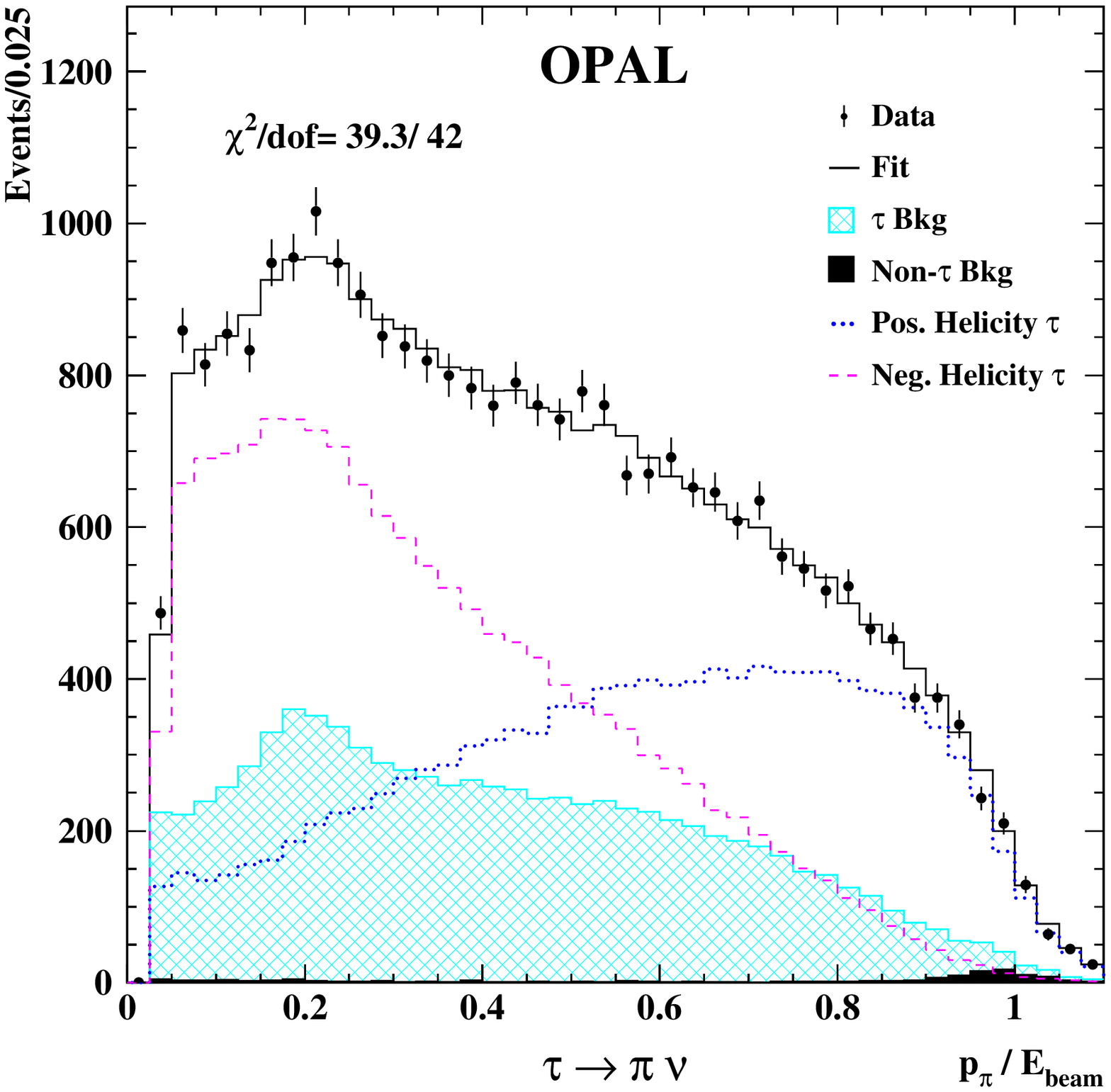}
\end{center}
\caption[OPAL $\tau$ polarisation: $\tpinu$ momentum spectrum]
{ \label{fig:taupol_3} The measured distributions in the
  polarisation-sensitive variable for the $\tpinu$ decays in the OPAL
  experiment.  The variable is the ratio of the measured charged
  hadron momentum to the beam energy, which is an approximation of
  $x_{\pi} = E_{\pi}/E_{\tau}$.  The data, shown by points with error
  bars, are integrated over the whole $\cos\theta_{\tau^-}$ ~range.
  Overlaying this distribution are Monte Carlo distributions for the
  positive (dotted line) and negative (dashed line) helicity $\tau$
  leptons and for their sum including background, assuming a value for
  $\pta$ equal to the fitted polarisation.  The hatched histogram
  represents the Monte Carlo expectations of contributions from
  cross-contamination from other $\tau$ decays and the dark shaded
  histogram the background from non-$\tau$ sources.  The level of
  agreement between the data and Monte Carlo distributions is
  quantified by quoting the $\chi^2$ and the number of degrees of
  freedom.  }
\end{figure}

\begin{figure}[hbtp]
\begin{center}
\includegraphics[width=0.9\linewidth]{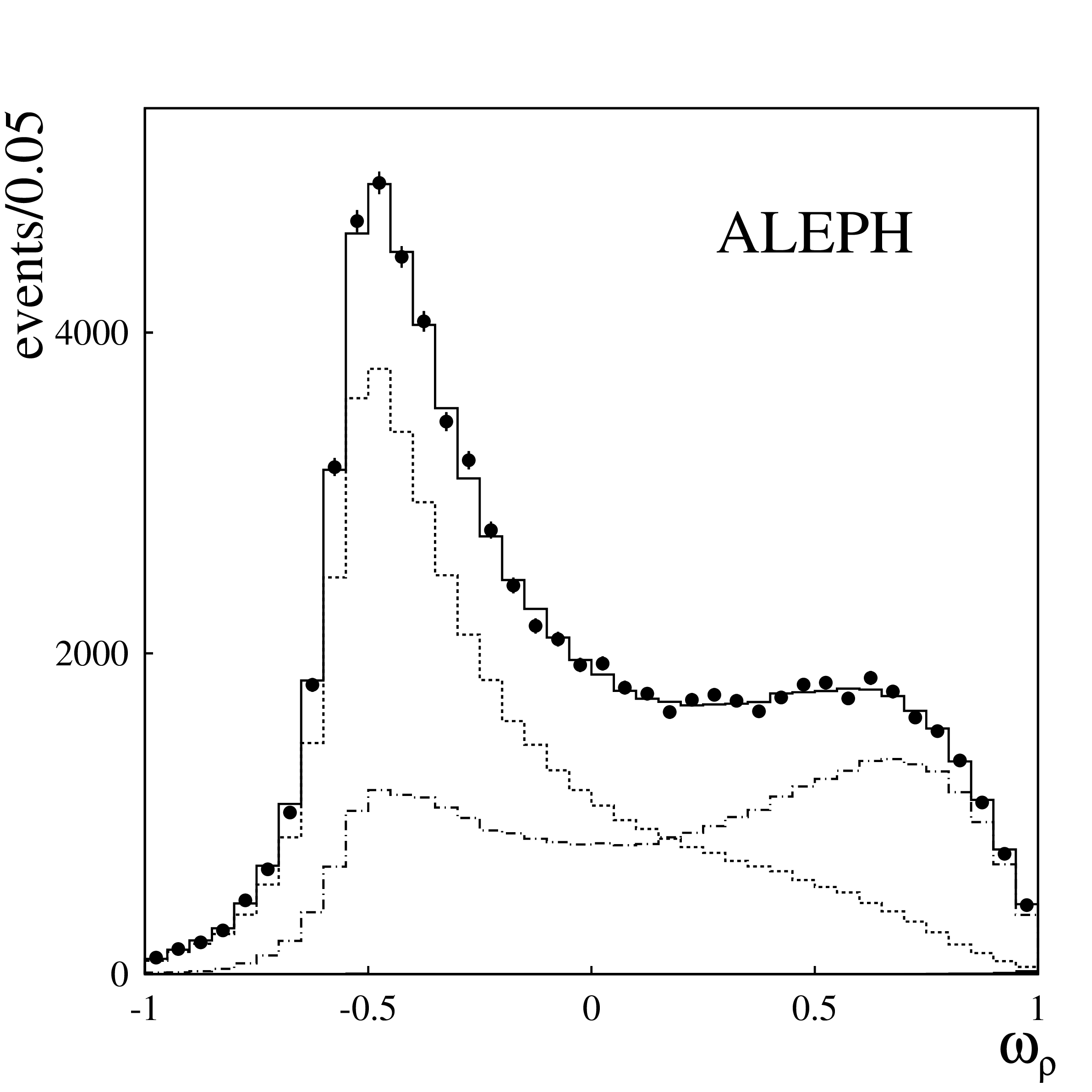}
\end{center}
\caption[ ALEPH $\tau$ polarisation:  $\tau\rightarrow \rho \nu$ optimal
variable spectrum] { \label{fig:taupol_4} 
 The measured spectrum of the
  polarisation-sensitive variable  $\omega_{\rho}$, described in the text,
 for the $\tau\rightarrow \rho \nu$  decays in the ALEPH experiment.
   The dashed and dashed-dotted lines correspond to the
  contributions of negative and positive helicity $\tau$'s,
  respectively. The small shaded area near $\omega$=1 is the
  non-$\tau$ background contribution.}
\end{figure}

\begin{figure}[hbtp]
\begin{center}
\includegraphics[width=0.9\linewidth]{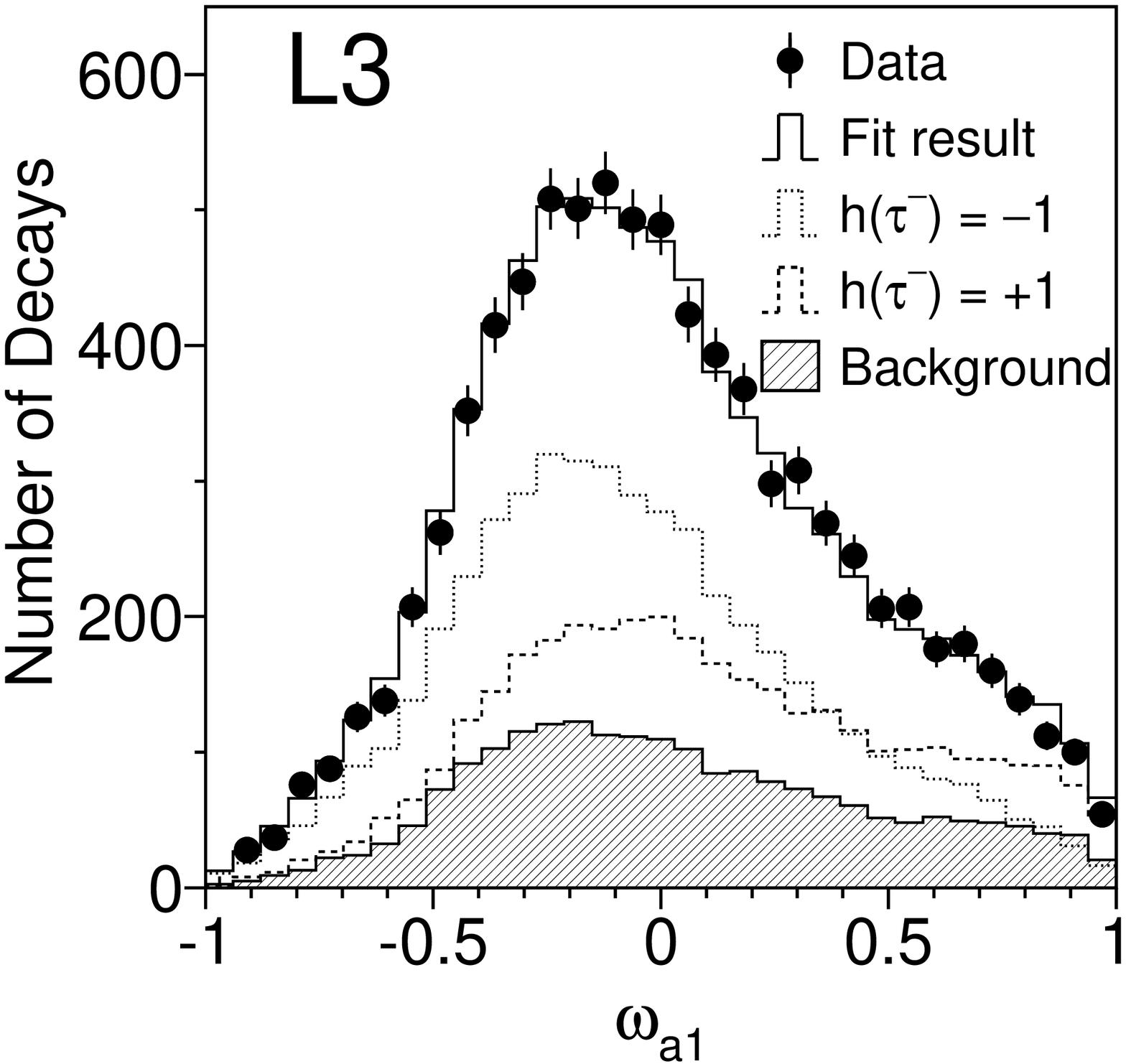}
\end{center}
\caption[L3 $\tau$ polarisation: $\tau\rightarrow$ e $\nu \bar{\nu}$
energy spectrum ] {\label{fig:taupol_5} 
 The measured spectrum of the
  polarisation-sensitive variable  $\omega_{\mathrm a_1}$, described in the text,
 for the $\tanu$  decays in the L3 experiment.
  The distributions for both
  a$^-_1 \rightarrow \pi^- \pi^+ \pi^-$ and a$^-_1 \rightarrow \pi^-
  \pi^0 \pi^0$ decays are combined in this figure.  The two helicity
  components and the background are shown separately.}
\end{figure}

\begin{figure}[hbtp]
\begin{center}
\includegraphics[width=0.9\linewidth]{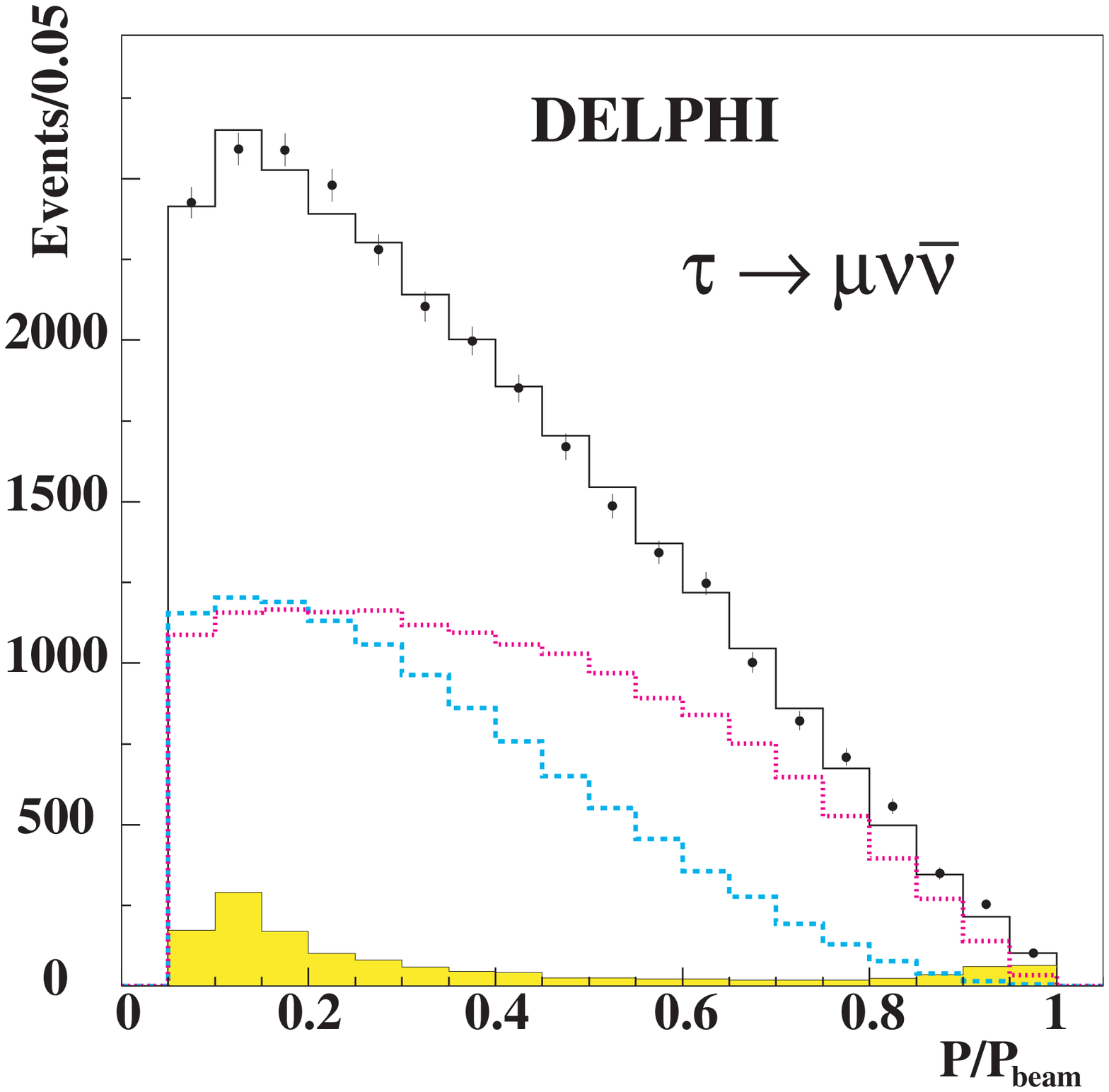}
\end{center}
\caption[DELPHI $\tau$ polarisation: $\tau\rightarrow \mu\nu \bar{\nu}$ 
momentum spectrum] { \label{fig:taupol_6} The measured spectrum of the
  polarisation-sensitive variable for the $\tau\rightarrow\mu\nu
  \bar{\nu}$ decays in the DELPHI experiment. The variable is the
  ratio of the measured muon momentum to the beam energy, which is an
  approximation of $x_{\mu} = E_{\mu}/E_{\tau}$.  The data are
  compared to the results of the polarisation fit.  The points with
 error bars are data and the solid line is simulated data for
  the fitted values of
  $\cAt$ and $\cAe$. The shaded area is background and the dashed and
  dotted lines correspond to the positive and negative helicity
  contributions respectively.}
\end{figure}

All four LEP experiments analyse the five exclusive channels listed in
Table~\ref{table-channelsens}~\cite{\ALEPHTAU,\LTAU,\DELPHITAU,\OPALTAU}.
In addition to those, ALEPH, DELPHI and L3 include the
$\tau\rightarrow \pi 2\pi^0 \nu$ mode in their exclusive channel
$\tanu$ analyses.  ALEPH also uses information from the $\tau$
direction for the hadronic decays, as discussed
in~\cite{Davier:1993nw}.  The addition of the $\tau$ direction
information ideally increases the sensitivity of the $\tanu$ and
$\trhonu$ channels by the amounts indicated in
Table~\ref{table-channelsens}.  Examples of the different kinematic
distributions from the different experiments are shown in
Figures~\ref{fig:taupol_3} to~\ref{fig:taupol_6}.

To their exclusive channel analyses, ALEPH~\cite{\ALEPHTAU},
DELPHI~\cite{\DELPHITAU} and L3~\cite{\LTAU} add an inclusive hadronic
decay analysis in which the single charged track (one-prong) hadronic
decay modes are collectively analysed. This approach yields a high
overall efficiency for these modes by sacrificing the optimal
sensitivity characterising the analysis of high purity channels.  For
DELPHI and L3 the correlations between the polarisation measurements
from their inclusive hadronic analysis and measurements using
separately identified single-track hadronic channels are small enough
that significant improvements are achieved when both results are
combined.  In the case of ALEPH, however, the exclusive reconstruction
efficiencies are high enough to produce strong correlations between
the exclusive and inclusive measurements, and little is gained from
the inclusive analysis.

The OPAL~\cite{\OPALTAU} $\cAt$ and $\cAe$ results are based entirely
on an analysis in which all five exclusive channels listed in
Table~\ref{table-channelsens} are combined in a global binned maximum
likelihood analysis. A single fit to all distributions in the
kinematic observables of all decay modes and $\cos\theta_{\tau^-}$
yields $\pta$ and $\AFBpol$. When both $\tau^+$ and $\tau^-$ decays of
a given event are identified, the event is analysed as a whole. This
global analysis approach fully accounts for the intrinsic correlation
between the helicities of the $\tau^+$ and $\tau^-$ produced in the
same $\Zzero$ decay, an effect which is accounted for by the other
experiments by applying a correction to the statistical errors of the
fit results.  In such a global analysis, the evaluation of the
systematic errors automatically incorporates all correlations between
the systematic uncertainties in the different channels.  For the
channel-by-channel analyses of ALEPH, L3 and DELPHI, the correlation
in the systematic errors between channels are taken into account in
the combination.

DELPHI~\cite{\DELPHITAU} augments their exclusive five channel and
inclusive one-prong analysis with a separate neural network analysis
of its 1993-1995 one-prong data set.  The neural network is used to
classify all one-prong decays as either $\trhonu$, $\tpinu$,
$\tenunu$, $\tmununu$ or $\tau \rightarrow \pi 2\pi^0 \nu$.  A
simultaneous fit for $\ptau$ as a function of $\cos\theta_{\tau^-}$ is
performed with $\cAt$ and $\cAe$ determined from a separate fit to the
$\ptau$($\cos\theta_{\tau^-}$) functional form as described below. As
with OPAL's global analysis, the channel-to-channel correlated
systematic errors are automatically evaluated in this analysis.

ALEPH~\cite{\ALEPHTAU} and L3~\cite{\LTAU} complement their analyses
of the kinematic distributions of the different decay modes, by
including information from event acollinearity to measure the $\tau$
polarisation.  Although of modest polarisation sensitivity, this
information has the advantage of being sensitive to detector-related
systematic errors that are different from those associated with the
measurements of spectra.

In order to extract $\cAt$ and $\cAe$ from their data, ALEPH, DELPHI
and L3 measure the polarisation as a function of $\cos\theta_{\tau^-}$
and then perform a separate fit for the two parameters using the
theoretical expectation of the dependence.  The results quoted by
OPAL~\cite{\OPALTAU}, which depend on a single maximum likelihood fit,
do not explicitly use measurements of the polarisation as a function
of $\cos\theta_{\tau^-}$, although such fits are performed as
cross-checks and for graphical presentation. ALEPH, L3 and OPAL use
Equation~\ref{eq-ptcos} in their fits but treat $\Afb^{\tau}$
differently as discussed in Section~\ref{TauPolSystematicErrors}.
Small corrections for the effects of initial state radiation, the
photon propagator and $\gammaZ$ interference, and the fact that not
all data are collected at the peak of the $\Zzero$ resonance are
incorporated into the quoted values of $\cAt$ and $\cAe$.  These
corrections, of $\calO(0.005)$, are calculated using
ZFITTER~\cite{\ZFITTERref}.  DELPHI incorporates these corrections
directly into the fit they perform by using ZFITTER to predict $\ptau
(\cos\theta_{\tau^-})$, averaged over the luminosity-weighted
centre-of-mass energies, as a function of $\cAt$ and $\cAe$.  This
automatically includes the QED and weak effects as a function of
$\cos\theta_{\tau^-}$, rather than as a separate correction as in the
approach taken by the other three LEP experiments.

Although the size of the event samples used by the four experiments
are roughly equal, smaller errors on the asymmetries are obtained by
ALEPH (see Table~\ref{tab-taupol}).
This is largely associated with the higher angular granularity of the
ALEPH electromagnetic calorimeter.  The $\tau$ `jets' produced in
hadronic $\tau$ decays are tightly collimated at LEP energies which
results in a substantial overlap of the energy deposited in the
calorimeter by different particles. A calorimeter with a higher
granularity is better able to identify the individual photons from
$\pi^0$ decay and therefore provides greater discrimination between
hadronic decay channels of the $\tau$. This results in improved
signal-to-noise thereby providing greater polarisation sensitivity and
smaller systematic errors.

The LEP combination is made using the overall results of $\cAt$ and
$\cAe$ from each experiment, rather than by first combining the
results for each decay mode.  Correlations between decay channels are
dominated by detector-specific systematic errors which are most
reliably taken into account by the individual experiments as discussed
in Section~\ref{TauPolSystematicErrors}.  The combinations of the
eight measurements, four each of $\cAt$ and $\cAe$, take into account
all other correlations and are presented in the following sections.

\section{Systematic Errors}
\label{TauPolSystematicErrors}

As will be shown in Section~\ref{TauPolResults},
the combined statistical errors on $\cAt$ and $\cAe$ are 0.0035 and
0.0048, respectively.  Systematic errors on these parameters which are
less than 0.0003 will not alter the combined errors when two
significant figures are quoted. Therefore, such systematic errors are
considered to be negligible. The one exception is the systematic error
associated with ZFITTER, which contributes $\pm$0.0002 to all
measurements of $\cAt$ and $\cAe$.

The systematic errors on $\cAe$ are considerably smaller than those on
$\cAt$ because, for the most part, the systematic effects are
symmetric in $q\times\cos\theta$ and consequently cancel in $\cAe$ but
not in $\cAt$.\footnote{ Here, $q$ is the charge of a $\tau$ lepton
whose decay is analysed and contributes to the measurements of $\cAt$
and $\cAe$. Systematic differences in the responses of different
$\cos\theta$ regions of a detector generate systematic errors in
$\cAt$, but to contribute to $\cAe$ there must be uncontrolled
differences in the response of the detector to positively and
negatively charged particles in the same $\cos\theta$ region of a
detector.  As detector responses are approximately charge-symmetric
for particles passing through the same region of a detector, there are
smaller systematic uncertainties associated with quantities that are
symmetric in $q\times\cos\theta$.}  This includes large cancellations
of the Monte Carlo statistical errors which arise by using the same
Monte Carlo samples in reflected $\cos\theta$ bins.  Different
approaches to evaluating the degree of cancellation of the $\cAe$
systematic errors are adopted by the four experiments and are detailed
in References~\citen{ALEPHTAU,DELPHITAU,L3TAU,OPALTAU}.

There are two broad categories of systematic error in these
measurements: those associated with the uncertainty of the underlying
physics assumptions and their treatment, and those associated with the
modelling of the detector. The systematic errors in the latter category
depend on the details of each of the individual detectors.  Together
with Monte Carlo statistical errors, these detector modelling errors
tend to dominate the systematic uncertainties. Although three of the
four experiments depend on the same detector simulation software,
GEANT~\cite{GEANT}, the designs of the four detectors are sufficiently
different that these detector related errors are uncorrelated between
experiments. However, these uncertainties can be strongly correlated
between measurements from different decay channels performed with the
same detector. For example, the uncertainty in the momentum scale for
one of the detectors is independent of that in the other three
detectors, but the momentum scale error is correlated between the
$\ptau$ measurements from different decay modes made with the same
detector.  Because each of the experiments takes these correlations
into account when quoting a systematic error on the measurements of
$\cAt$ and $\cAe$ using all channels, only the global results from
each of the four experiments can be reliably combined to give a LEP
average.

Turning now to the uncertainty of the treatment of the physics of
$\tau$ production and decay, there are a number of systematic
uncertainties in this category that are common to all four
experiments. One set of these uncertainties affects all decay modes in
the same way while others are different for each $\tau$ decay mode.
The origins of some of the common uncertainties are the common
software tools that are used to describe the production and decay of
the $\tau$~\cite{KORALZ} and the major
backgrounds~\cite{KORALZ,Jadach:1997nk,JETSET,Smith:1977rr,
Berends:1984sd, Berends:1985gf}; and the tools~\cite{\ZFITTERref} used
to interpret the data in terms of the Standard Model.  Another source
of common errors arises from a reliance on the same physics input used
in the analyses of the four experiments, such as the branching
fractions of $\tau$ decay modes.

\subsection{Decay-Independent Systematic Uncertainties}

In the category of systematic uncertainty that affects all $\tau$
decay modes, the following have been identified as potential sources
of error common to all experiments:

\subsubsection{Electromagnetic radiative corrections}

Initial state radiation from the e$^+$ and e$^-$ and final state
radiation from the $\tau^+$ and $\tau^-$ influence the measurement in
two ways.  The first relates to the fact that the experiments measure
$\pta$ and $\AFBpol$ integrated over ${\sqrt \spr}$, ${\sqrt \spr}$
being the centre-of-mass energy of the $\tau$-pair system excluding
initial state radiation.  This effect is included in the ZFITTER
correction discussed below.  The second influence relates to changes
to the kinematic distributions caused by initial and final state
radiation and potential ${\sqrt \spr}$ biases introduced in the
selection procedure.  In this case, the four experiments rely on the
KORALZ Monte Carlo event generator to take these effects into
account. This radiation is calculated to $\calO(\alpha^{2})$ and
includes exclusive exponentiation in both initial and final state
radiation. Although interference between the initial and final state
radiation is not included in the generator when producing the
simulated events, such effects have negligible impact on the $\ptau$
measurements. Because of its precision, the treatment of initial and
final state radiation, although common to all experiments, introduces
no significant contribution to the systematic error.

\subsubsection[Energy dependence of the \protect$\tau$ polarisation]
              {Energy dependence of the \protect\boldmath$\tau$ polarisation}

The ZFITTER treatment of $\roots$ dependence of $\ptau$, including the
effects of initial state radiation, and of photon propagator and
$\gammaZ$ interference amounts to the application of the \SM{}
interpretation of $\pta$ and $\AFBpol$ in terms of $\cAt$ and $\cAe$.
Although the experiments introduce this interpretation at different
stages of their analyses, it effectively involves applying corrections
of $\calO(0.005)$ to both $\pta$ and $\AFBpol$.  For data at the
peak of the $\Zzero$ resonance, the photon propagator and $\gammaZ$
interference introduce the dominant component of the correction, having
a value of approximately +0.005 for both $|\pta|$ and
$\frac{4}{3}|\AFBpol|$.  Because relatively little data is collected
off the peak and because the corrections below the peak are of
opposite sign to those above the peak, the impact of the $\roots$
dependence is small, contributing $\calO(+0.0003)$ to the
corrections.  Initial state radiation changes the relative
contribution of the pure $\Zzero$ exchange and introduces a small
distortion to the \ptau($\cos\theta_{\tau^-})$ relationship of
Equation~\ref{eq-ptcos}. The actual value of this component of the
correction depends on the details of the individual experiment.
However, the uncertainty on the total correction is significantly
smaller than the correction itself as given by variations of the
unknown parameters in the model. The variation of the Higgs mass
alters this correction by $\pm 0.0002$ and is used to estimate this
uncertainty.  Since all experiments rely on ZFITTER for this
treatment, the error is common across experiments as well as to $\cAt$
and $\cAe$. 

\subsubsection{Mass effects}

Born level mass terms lead to helicity flip configurations.  At the
$\calO(10^{-3})$ level, the $\tau^-$ and $\tau^+$ will have the
same instead of opposite helicities. Although this effect cannot be
seen in the quoted measurements at this level of precision, it is
included in the KORALZ treatment nonetheless.

\subsubsection[The value of \protect$\Afb^{\tau}$ used in the fit]
              {The value of \protect\boldmath$\Afb^{\tau}$ used in the fit}

The different experiments treat this differently.  ALEPH and DELPHI
use the \SM{} values of $\Afb^{\tau}$ with appropriate $\roots$
dependence.  OPAL uses its measured values of $\Afb^{\tau}$ for
$\tau$-pairs at the different values of $\roots$.  L3 assumes the
relation $\Afb^{\tau}=\frac{3}{4}\cAe\cAt$ in the denominator of
Equation~\ref{eq-ptcos}.  Since $\Afb^{\tau}$ enters into the analysis
as a small number in the denominator, its uncertainty introduces a
correspondingly small systematic error for each experiment. Although
the \SM{} assumptions regarding $\Afb^{\tau}$ by ALEPH and DELPHI
imply that some correlation exists from this source between the
measurements of these two experiments, it is negligible and
consequently ignored in the combined LEP results.  The OPAL treatment
introduces a small correlation between the $\tau$-polarisation
measurement and the OPAL $\Afb^{\tau}$ measurement. Varying the value
of $\Afb^{\tau}$ by 0.001, however, introduces negligible changes to
the $\cAt$ and $\cAe$ measurements.

\subsubsection{Summary}

In conclusion, all of these effects are theoretically well defined and
have been calculated to more than adequate precision for the
measurements at hand.  Of these, only the ZFITTER error of $\pm$0.0002
is included as a common error in the LEP combination, see 
Table~\ref{table-systematics}.

\subsection{Decay-Dependent Systematic Uncertainties}

Concerning the category of uncertainty that affects each $\tau$ decay
mode separately, the following sources of potentially common
systematic error have been identified:

\subsubsection[Branching fractions of the \protect$\tau$ decay modes]
              {Branching fractions of the \protect\boldmath$\tau$ decay modes}

These arise since the purity for selecting any particular decay mode
for polarisation analysis is not unity.  All experiments use the world
average values of the branching fractions as determined by the
Particle Data Group~\cite{PDG98,PDG2000}, along with the quoted
errors.  Consequently, the components of the systematic error which
are associated with uncertainties in the branching fractions are
correlated between experiments.  These errors are taken into account
in the combined error, and are shown in Table~\ref{table-systematics}
for the combined error on $\cAt$ and $\cAe$ for each of the
experiments.

\begin{table}[t]
\begin{center}
\renewcommand{\arraystretch}{1.2}
\begin{tabular}{|l||cc|cc|cc|cc|} \hline
                      & \multicolumn{2}{|c|}{ALEPH} 
                      & \multicolumn{2}{|c|}{DELPHI} 
                      & \multicolumn{2}{|c|}{L3} 
                      & \multicolumn{2}{|c|}{OPAL} \\ 
                      & $\delta\cAt$ & $\delta\cAe$ 
                      & $\delta\cAt$ & $\delta\cAe$ 
                      & $\delta\cAt$ & $\delta\cAe$ 
                      & $\delta\cAt$ & $\delta\cAe$ \\ \hline \hline
ZFITTER  &0.0002&0.0002&0.0002&0.0002&0.0002&0.0002&0.0002&0.0002 \\
$\tau$ branching fractions &0.0003&0.0000&0.0016&0.0000&0.0007&0.0012&    0.0011    & 0.0003   \\
two-photon bg              &0.0000&0.0000&0.0005&0.0000&0.0007&0.0000&    0.0000    & 0.0000   \\
had. decay model           &0.0012&0.0008&0.0010&0.0000&0.0010&0.0001&    0.0025    & 0.0005   \\
\hline 
\end{tabular}
\caption[Common systematic errors in $\tau$ polarisation measurements]
{The magnitude of the major common systematic errors on $\cAt$ and
  $\cAe$ by category for each of the LEP experiments.}
\label{table-systematics}
\end{center}
\end{table}

\subsubsection[Radiative corrections for \protect$\tau$ leptonic decays]
              {Radiative corrections for \protect\boldmath$\tau$ leptonic decays}

The radiation in the decays $\tenunu$ and $\tmununu$ are treated
exactly to $\calO(\alpha)$ in KORALZ and negligible contributions
to the systematic error are introduced by this treatment.

\subsubsection{Bhabha background}
 
OPAL uses the BHWIDE Monte Carlo generator~\cite{Jadach:1997nk} to
describe this background while DELPHI uses the BABAMC~\cite{BABAMC}
and UNIBAB~\cite{UNIBAB} in addition to BHWIDE.  ALEPH primarily uses
UNIBAB but also uses BABAMC as an auxiliary generator.  L3 uses the
BHAGENE3~\cite{Field:1996dk} generator.  The use of common generators
by some of the experiments potentially introduces a common systematic
error.  However, in the case of experiments where there is very little
$\eeee$ background, the errors are negligible.  It should be noted
that much of the uncertainty associated with this is
detector-specific, and in fact has been found to constitute a
negligible common systematic error.

\subsubsection{Two-photon background}

The background from two-photon collision processes can be problematic
since the two-photon Monte Carlo generators used by the experiments do
not include initial state radiation, although these QED radiative effects
are expected to be small. The potential danger is that
the measured event transverse momentum ($p_T$), a quantity which
discriminates between $\tau$-pair events, which have large $p_T$, and
two-photon events which have small $p_T$, is sensitive to initial
state radiation. Consequently,  low energy events, which can have a high $\ptau$
analysing power, do not have perfectly modelled backgrounds. This is
common to all experiments, but the sensitivity of a given experiment
to the effect depends on the effectiveness with which two-photon
events are identified and removed from the sample.  These errors are
taken into account in the combined error with the contributions from
each experiment shown in Table~\ref{table-systematics} but do not
represent a significant correlation because some experiments make
corrections to this background based on control samples in their own
data.

\subsubsection{Modelling of hadronic decays}

Model dependent uncertainties in the a$_1$ decay mode have been
evaluated by all experiments. These uncertainties arise both in the
analysis of the $\tanu$ channel itself and in the analysis of channels
where backgrounds from the a$_1$ can be significant, such as the
$\trhonu$.  These errors can be common to all experiments, but will
vary in sensitivity depending on the purity of the samples and details
of the analysis.  The TAUOLA~\cite{\KORALZ} Monte Carlo simulation of
the $\tau\rightarrow \pi \ge 3\pi^0 \nu_{\tau}$ and $\tau\rightarrow
3\pi^{\pm} \ge 2\pi^0 \nu_{\tau}$ decays, which are backgrounds to
some of the $\ptau$ analysing channels, also have model dependencies
with a corresponding uncertainty.  Consequently, each experiment
estimates how much these deficiencies affect their $\ptau$
measurements and, because they depend on the channel selection purity,
there is variation in the magnitude of these effects between
experiments.
  
Another aspect of hadronic decay modelling is the treatment of
radiative corrections for $\tau$ hadronic final states. Unlike
radiation from leptons, there is no precise formalism for handling
these corrections. The KORALZ generator uses an $\calO(\alpha)$
correction in the leading logarithmic approximation as implemented in
the PHOTOS software package~\cite{PHOTOS}. In the $\tpinu$ channel,
this radiation affects the polarisation at the 0.01 level absolute,
while for $\trhonu$ the effects are less than half that.  Theoretical
work~\cite{Decker,Finkemeier} indicates that the treatment of
radiation in the decay $\tpinu\gamma$ is valid to the 5\% level of the
decay rate. Consequently, the uncertainties in the decay radiation
treatment contribute at the 0.0005 level to the systematic error of
the $\tpinu$ measurement of $\cAt$, and much less than that to the
error on the combined measurements.  Unfortunately, no analogous
theoretical studies have been performed for the $\trhonu\gamma$ decay.
Following reference~\cite{PHOTOS}, the error on the treatment of the
radiation is approximately $1/\ln(m_{\tau}/m_{\rho})$ of the magnitude
of the effect of the radiation on the measurement of $\ptau$. This
results in an error of no more than 0.001 on $\cAt$ and a negligible
error on $\cAe$.  The equivalent radiation effects for the other
hadronic decay modes introduce a negligible contribution to the
combined systematic error.  These hadronic modelling errors are
summarised in Table~\ref{table-systematics} and are found to
contribute a small effect to the measurements over all channels.

\subsubsection{Modelling of multihadronic background}

The modelling uncertainty of the multihadronic background introduces
negligible errors in all channels but the $\tanu$.  However, because
the background itself is small and the weight of the $\tanu$
measurement is not high, this is a negligible contribution to the
error on $\ptau$ from all channels.

\subsubsection{Modelling of muon-pair background}

The modelling of $\mu$-pair background has a negligible error.  Any
uncertainty arising from $\mu$-pair events is evaluated as a
detector-related systematic error.

\section{Results}
\label{TauPolResults}

\begin{figure}[p]
\begin{center}
\includegraphics[width=0.95\linewidth]{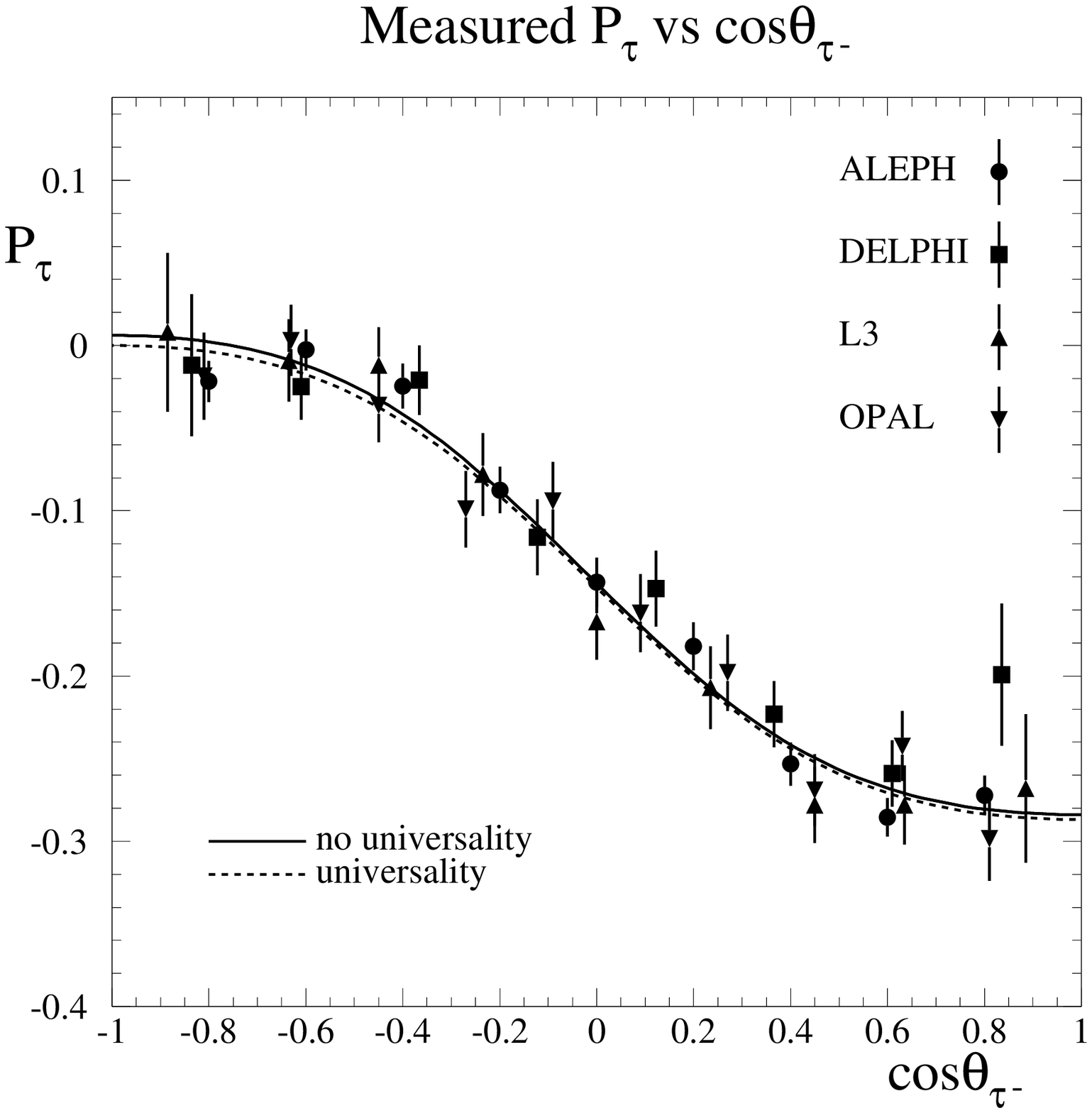}
\end{center}
\caption[Measured $\tau$ polarisation {\em vs.}
$\cos\theta_{\tau}$ for all LEP experiments] {\label{fig:taupol_7} The
  values of \ptau ~as a function of $\cos\theta_{\tau^-}$ as measured
  by each of the LEP experiments.  Only the statistical errors are
  shown.  The values are not corrected for radiation, interference or
  pure photon exchange. The solid curve overlays
  Equation~\ref{eq-ptcos} for the LEP values of $\cAt$ and $\cAe$. The
  dashed curve overlays Equation~\ref{eq-ptcos} under the assumption
  of lepton universality for the LEP value of $\cAl$.}
\end{figure}

Figure~\ref{fig:taupol_7} shows the measured values of \ptau ~as a
function of $\cos\theta_{\tau^-}$ for all four LEP experiments. The
curves overlaying the figure depict Equation~\ref{eq-ptcos} for the
combined results with and without assuming lepton universality.  It is
interesting to remark that if lepton universality is assumed, \ptau
~is forced to be zero at $\cos\theta_{\tau^-}=-1$, regardless of the
actual values of the \SM{} couplings.  From Figure~\ref{fig:taupol_7}
it is evident that the data are indeed consistent with \ptau=0 at
$\cos\theta_{\tau^-}=-1$.

\begin{figure}[p]
\begin{center}
\includegraphics[width=0.9\linewidth]{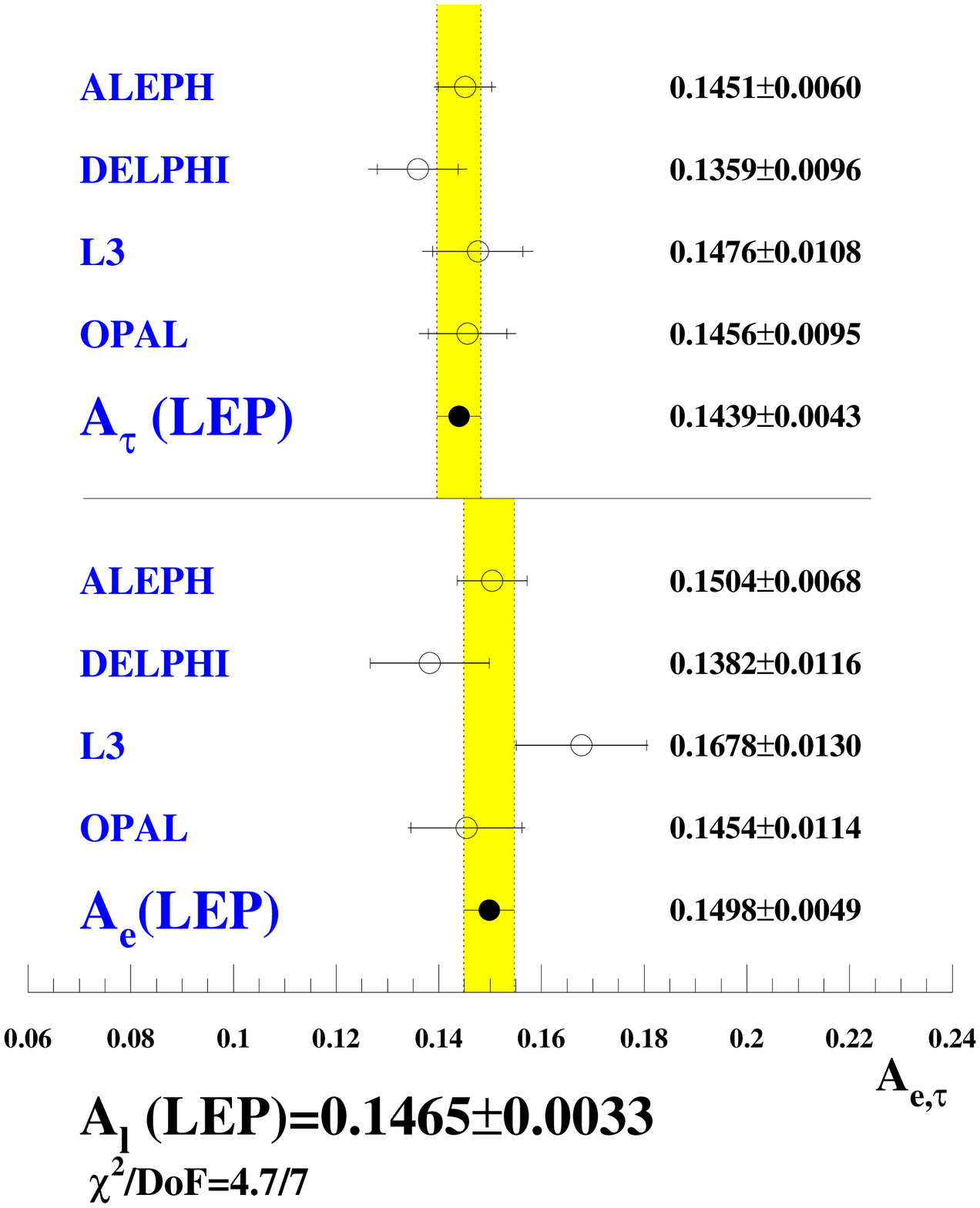}
\end{center}
\caption[Comparison of $\tau$ polarisation measurements of $\cAt$ and $\cAe$]
{\label{fig:taupol_8} Measurements of $\cAt$ and $\cAe$ from the four
  LEP experiments. The error bars indicate the quadrature sum of the
  statistical and systematic errors.  The magnitude of the statistical
  error alone is indicated by the small tick marks on each error bar.
  The value of $\cAl$ and the $\chi^2$ of the fit assuming lepton
  universality are also quoted.  }
\end{figure}

The results for $\cAt$ and $\cAe$ obtained by the four LEP
collaborations~\cite{\ALEPHTAU,\DELPHITAU,\LTAU,\OPALTAU} are shown in
Table~\ref{tab-taupol}. The measurements from all experiments are
consistent with each other and are combined to give values of $\cAt$
and $\cAe$ from a fit which includes the effects of correlated errors.
The combined results are included in Table~\ref{tab-taupol} and are
also summarised in Figure~\ref{fig:taupol_8}.

There are small ($\le$ 5\%) statistical and, in some cases, systematic
correlations between $\cAt$ and $\cAe$ performed by a single
experiment. There are also systematic correlations between the
different experimental values as discussed in the previous section.
Therefore a single fit to all of the data using the complete
8$\times$8 error correlation matrix, given in Table~\ref{tab-tau8by8},
is used to obtain the LEP combined values of these two parameters.

We take the $\pm 0.0002$ ZFITTER errors to be fully correlated between
$\cAt$ and $\cAe$.  Other systematic errors listed in
Table~\ref{table-systematics} are taken to be fully correlated between
either $\cAt$ or $\cAe$ measurements. These are used to calculate the
inter-experiment off-diagonal elements of the error correlation
matrix.  The correlated errors between $\cAt$ and $\cAe$ for a given
experiment as quoted by the experiment are also included in the error
correlation matrix.

The fitted values for $\cAt$ and $\cAe$ with no assumption of lepton
universality are:
\begin{eqnarray} 
  \cAt & = & 0.1439 \pm 0.0043 \label{eq:ptau:At}\\
  \cAe & = & 0.1498 \pm 0.0049 \label{eq:ptau:Ae}\,,
\end{eqnarray}
where the $\chi^2$ is 3.9 for six degrees of freedom and the
correlation is +0.012.  These asymmetries are consistent with each
other, in agreement with lepton universality. Assuming
$\mathrm{e}$-$\tau$ universality, the values for $\cAt$ and $\cAe$ can
be combined in a fit with a single lepton asymmetry parameter which
yields a result of:
\begin{eqnarray}
  \cAl & = & 0.1465 \pm 0.0033 \label{eq:ptau:Al} \,,
\end{eqnarray}
where the total error contains the systematic error of 0.0015.  The
$\chi^2$ is 0.8 for one degree of freedom, considering this to be a
combination of $\cAt$ and $\cAe$.  If one considers the eight
measurements contributing to $\cAl$, the $\chi^2$ is 4.7 for seven
degrees of freedom.  This value of $\cAl$ corresponds to a value of:
\begin{eqnarray}
\swsqeffl = 0.23159 \pm   0.00041 \,.
\end{eqnarray}

\begin{table}[p]
\renewcommand{\arraystretch}{1.15}
\begin{center}
\begin{tabular}{|l||c|c|}
\hline
Experiment  & $\cAt$ & $\cAe$ \\
\hline
\hline
ALEPH   & $0.1451\pm0.0052\pm0.0029$  & $0.1504\pm0.0068\pm0.0008$  \\
DELPHI  & $0.1359\pm0.0079\pm0.0055$  & $0.1382\pm0.0116\pm0.0005$  \\
L3      & $0.1476\pm0.0088\pm0.0062$  & $0.1678\pm0.0127\pm0.0030$  \\
OPAL    & $0.1456\pm0.0076\pm0.0057$  & $0.1454\pm0.0108\pm0.0036$  \\
\hline
LEP     & $0.1439\pm0.0035\pm0.0026$  & $0.1498\pm0.0048\pm0.0009$  \\
\hline
\end{tabular}
\caption[LEP results for $\cAt$ and $\cAe$]{  
  LEP results for $\cAt$ and $\cAe$.  The first error is statistical
  and the second systematic.  }
\label{tab-taupol}
\end{center}
\end{table}
\begin{table}[p]
\renewcommand{\arraystretch}{1.15}
\begin{center}
\begin{tabular}{|r||rrrrrrrr|}
\hline
&$\cAt$(A)&$\cAt$(D)&$\cAt$(L)&$\cAt$(O)
&$\cAe$(A)&$\cAe$(D)&$\cAe$(L)&$\cAe$(O)\\
\hline
\hline
$\cAt$(A) &   1.000 &       &       &       &       &        &      &      \\ 
$\cAt$(D) &   0.029 & 1.000 &       &       &       &        &      &      \\ 
$\cAt$(L) &   0.022 & 0.024 & 1.000 &       &       &        &      &      \\ 
$\cAt$(O) &   0.059 & 0.047 & 0.032 & 1.000 &       &        &      &      \\ 
$\cAe$(A) &$-$0.002 & 0.000 & 0.000 & 0.000 & 1.000 &        &      &      \\
$\cAe$(D) &   0.000 & 0.025 & 0.000 & 0.000 & 0.000 & 1.000  &      &      \\
$\cAe$(L) &   0.000 & 0.000 & 0.032 & 0.000 & 0.001 & 0.000  &1.000 &      \\
$\cAe$(O) &   0.000 & 0.000 & 0.000 & 0.025 & 0.005 & 0.000  &0.002 &1.000 \\
\hline
\end{tabular}
\caption[Tau polarisation error correlation matrix for $\cAt$ and $\cAe$]
{ Error correlation matrix for the total error of the eight
  measurements, used for the combination of the LEP results for $\cAt$
  and $\cAe$. The order is: $\cAt$ for ALEPH, DELPHI, L3 and OPAL;
  followed by $\cAe$ for ALEPH, DELPHI, L3 and OPAL.}
\label{tab-tau8by8}
\end{center}
\end{table}

\chapter{Results from b and c Quarks}
\label{sec:hq}

\section{Introduction}
\label{sec:hqintro}

Heavy flavours can be identified with high efficiency and purity at
LEP and SLD, allowing precise electroweak measurements.  As already
explained in Chapter~\ref{sec:intro}, the b and c partial widths,
normalised to the total hadronic width of the Z: $\Rbz$, $\Rcz$, the
forward-backward asymmetries with unpolarised beams: $\Abb$, $\Acc$,
and, with polarised beams, the left-right-forward-backward
asymmetries: $\Abblr$, $\Acclr$, can be measured.  These measurements
probe the fundamental charge and weak-isospin structure of the
Standard Model ($\SM$) couplings for quarks and, in case of $\Abb$ and
$\Acc$ for the initial state leptons.  The ratio $\Rbz$ is of special
interest since it probes corrections to the ${\rm Z}\bb$ vertex which
are sensitive to new physics, for example from a super-symmetric Higgs
sector.  As illustrated in Figure~\ref{fig:afvssin2}, the $\SM$
predictions for the quark asymmetry parameters, $\cAq$ are essentially
fixed points in the model, insensitive even to variations in
$\swsqeffq$.  If $\cAq$ agrees with the $\SM$ prediction this fact
makes the $\Aqq$ measurements sensitive probes of the initial-state
${\rm Z}\ee$-couplings, resulting in one of the most precise
measurements of $\swsqeffl$.  Even considering departures from the
$\SM$, predictions for $\cAq$ are basically invariant in any model
where new physics appears only in loops, making $\Aqqlr = 3/4 \cAq$ a
good experimental test for new Born-level physics like Z-Z$'$ mixing.
Due to the inferior tagging possibilities for light quarks, as
discussed in Appendix~\ref{sec:lqappendix}, electroweak tests of
similar precision are not possible for u, d or s quarks.

The LEP experiments and SLD measure these quantities with a variety of
methods.  Since all the measurements make some assumptions about the
fragmentation of b- and c-quarks and decays of hadrons containing
these heavy quarks, there are many sources of systematic correlations
between them.  In addition, different observables are sometimes
measured simultaneously, giving rise to statistical correlations
between the results.  For these reasons a simple average of the
different results is not sufficient. A more sophisticated procedure is
needed as described below.

To derive consistent averages the experiments have agreed on a common
set of input parameters and associated uncertainties.
These parameters are described in Section~\ref{sec:hqinputs}.
They consist of the electroweak parameters of interest plus some auxiliary
parameters that are included in the combination for technical reasons.
These auxiliary parameters from LEP and SLD are either
measured together with some of the electroweak quantities
or they share systematic
uncertainties with them. To treat the dependences of these
parameters on the electroweak parameters correctly they are included in the 
electroweak heavy flavour fit.  
The fit parameters in the electroweak heavy flavour fit thus are:
\begin{itemize}
\item the Z partial decay widths to b- and c-quarks normalised to the Z
  hadronic width: $\Rbz,\, \Rcz$,
\item the b- and c-quark forward-backward asymmetries:
  $\Afb^{\rm{b,c}}(\sqrt{s})$, either at three different centre-of-mass
  energies around the Z-peak or with all asymmetries transported to
  the peak,
\item the b- and c-quark asymmetry parameters: $\cAb,\,\cAc$ measured
  from the left-right-forward-backward asymmetries at SLD,
\item the $\BB$ effective mixing parameter $\chiM$, which is the
  probability that a semileptonically decaying b-quark has been
  produced as an anti-b-quark,
\item the prompt and cascade semileptonic branching fraction of the
  b-hadrons $\Brbl$\footnote{ Unless otherwise stated, charge
  conjugate modes are always included.} and $\Brbclp$ and the prompt
  semileptonic branching fraction of the c-hadrons $\Brcl$.
\item the fraction of charm hemispheres fragmenting into a specified
  weakly-decaying charmed hadron: $\fDp$, $\fDs$, $\fcb$,\footnote{ 
  The quantity $\fDz$ is calculated from the constraint $\fDz + \fDp +
  \fDs + \fcb = 1.$ }
\item the probability that a c-quark fragments into a $\Dstarp$ that
  decays into $\Dzero \pi^+$: $P\mathrm{( c \rightarrow D^{*+})}
  \times \BR\mathrm{( D^{*+} \rightarrow \pi^+ D^0 )}$, denoted
  $\PcDst$ in the following.
\end{itemize}
The input parameters used in the combination are either the fit
parameters themselves or simple combinations of them that make a
correct error treatment easier.

The methods of tagging heavy flavours at LEP and SLD are described in
Section~\ref{sec:hqtag}. The different measurements of the electroweak
and auxiliary parameters used in the heavy flavour combinations are
outlined in Sections~\ref{sec:hqrbc} to~\ref{sec:hqaux}.
Section~\ref{sec:hqinputs} describes the agreed common external
parameters.  In Section~\ref{sec:hqcorr} the corrections to the
electroweak parameters due to physics effects such as QED and QCD
corrections are described, and the combination procedure is explained
in Section~\ref{sec:hqcomb}. Finally the results are summarised in
Section~\ref{sec:hqresults}.

\section{Heavy Flavour Tagging Methods}
\label{sec:hqtag}

In principle, the rate measurements $R_{\rm q} = \sigma_{\rm
q}/\sigma_{\rm had}$ only require a selection of the quark flavour q
from hadronic events with an identification algorithm, usually
referred to as a tag, that has efficiencies and purities that are known
to high precision. The asymmetry measurements require in addition that
a distinction between quark and antiquark is made, with a known
charge-tagging efficiency. Cancellations in the asymmetry definition
make these measurements largely independent of the flavour tagging
efficiency, apart from background corrections.

At LEP and SLD three basic methods are used for flavour tagging.
In the first method the finite path traversed by the hadron containing
the heavy quark during its long lifetime is utilised.  Due to the
somewhat longer lifetime and the larger mass these methods are
especially efficient for b-quarks. They tag only the flavour of the
quarks. To obtain the quark charge additional methods have to be used.

The second and historically oldest method is to tag prompt leptons.
b- and c-quarks can decay semileptonically and, because of the higher
b-mass, the two quark species can be separated by the transverse
momentum of the lepton with respect to the jet axis.  For direct
decays the sign of the quark charge is equal to that of the lepton, so
that leptons tag simultaneously the quark flavour and charge.

The third method is the reconstruction of charmed hadrons. Most
charmed hadrons have low multiplicity decay modes with relatively high
branching fractions so that they can be used for flavour
tagging. Since most charmed hadron decays are not charge symmetric
they can also be used for quark charge tagging.  Charmed hadrons tag
charm quarks and, via the decay ${\rm b} \rightarrow {\rm c}$,
b-quarks. Properties of the fragmentation or lifetime tags have thus
to be used to separate the two.

\subsection{Lifetime Tagging}
\label{sec:btag}

Lifetime tagging represents the most efficient and pure way of
selecting b-hadrons from hadronic Z decays.  The two principal
techniques are based on the reconstruction of secondary vertices and
on the measurement of the large impact parameter of the b-hadron decay
products.  Since the average b lifetime is about 1.6 ps and the
b-hadrons are produced with a mean energy of $32~\GeV$ at the Z peak,
they travel for about 3 mm before decaying. Their mean charged
multiplicity is $\sim$5 (see Section~\ref{sec:hq_bmult}).  The silicon
vertex detectors of the LEP experiments and SLD have a resolution for
the secondary vertex position about one order of magnitude smaller
than the mean decay length.

Since the b-hadron decay vertex is separated from the
$\ee$-interaction point, some of the tracks originating from the decay
will appear to miss the reconstructed primary vertex.  The impact
parameter is defined as the distance of closest approach of the
reconstructed track to the interaction point.  It is given by
\begin{equation}
\delta= \gamma \beta  c \tau \sin\psi \,,
\end{equation}
where $\tau$ is the particle proper decay time and $\psi$ is the angle
between the secondary particle and the b-hadron flight direction in
the lab frame.

For a high momentum track, $\sin \psi$ is proportional to $1
/\beta\gamma$, and the average impact parameter is then proportional
to the average lifetime $\tau$: $\delta \propto c \tau$, independent
of the b-hadron energy.  Since at LEP the b-hadron momentum is high,
the uncertainty on the b-hadron momentum distribution, i.e. the 
b-fragmentation function, has only a small effect on the impact
parameter distribution.  The impact parameter of the b-hadrons is
about 300$\,\mu$m, to be compared with the experimental resolution of
20 to 70 $\mu$m, depending on the track momentum.  ALEPH, L3, and SLD
compute the impact parameter in 3D space, while DELPHI and OPAL
compute the impact parameter separately in the two projections $R\phi$
and $Rz$.  The two projections are then treated as two separate
variables.

The precise determination of the Z decay point, the so called primary
vertex, is required in lifetime b-tagging techniques.  It is
determined separately for each hadronic event using the location of
the $\ee$ interaction region (the beam spot) as a constraint.
At LEP the width along the horizontal $x$-axis varies with time but is
typically 100 to 150\,$\mu$m.  The width along the vertical $y$-axis
is around 5\,$\mu$m, which is below the detector resolution, and the
longitudinal length along the z-axis is about 1\,cm.  Since the beam
spot width in $z$ is much larger than the detector resolution, the
exact position and width in this direction does not influence the
tagging efficiencies.  At SLC the beam spot is only a few microns wide
in the transverse ($R\phi$) plane, giving an almost point-like primary
vertex resolution. Only the vertex position along the $z$-axis needs
to be reconstructed event by event.

The event primary vertex is determined by a fit to all tracks after
having excluded tracks classified to originate from decays of long
lived particles or hadronic interaction products.  The precision of
the reconstructed primary vertex position depends on the algorithm
used, on the geometry of the silicon vertex detectors and on the size
of the beam spot.  The parameters of the various vertex detectors and
the relevant resolution for a lifetime b-tag are summarised for the
LEP and SLC experiments in Table~\ref{tab:hq_vd}.

{%
\begin{table} 
\begin{center}
\renewcommand{\arraystretch}{1.1}
\begin{tabular} {| l || c | c  | c | c | c |} \hline
                 & ALEPH    & DELPHI  &    L3   & OPAL   & SLD \\ 
\hline
\hline
Number of layers   &   2      &   3     &    2    &  2     &  3      \\\hline
Radius of layers (cm) & 6.5/11.3 & 6.3/9/11  & 6.2/7.7 &6.1/7.5 & 2.7-4.8 \\
\hline
$R\phi$ imp. par.
res. ($\mu$m)&        & 20      &30       & 16     & 8      \\
\cline{1-1} \cline{3-6}
$z$ imp. par.
res. ($\mu$m)&  \raisebox{1.5ex}[-1.5ex]{25$^*$}
                     & 30      &100      & 35     & 10       \\\hline   
Primary vertex res.  & $58\times 10$ &  $60 \times 10$ & 
$77\times 10$ & $80 \times 12$ & $4\times 4$ \\
$x \times y \times z$ ($\mu$m)
& $\times 60$   &  $ \times 70$  & $\times 100$  & $\times 85$ & $\times 17$
  \\\hline
\end{tabular}
\caption[Vertex detector characteristics and experimental
resolutions.]  {Vertex detector characteristics and experimental
resolutions: the impact parameter resolution is given for $45~\GeV$
muons and the vertex resolution is given for $\bb$-events when
including the beam spot information.\\ $^*$ for ALEPH the 3D impact
parameter resolution is given.  }
\label{tab:hq_vd}
\end{center}
\end{table}
}

\begin{figure}[htb]
\begin{center} 
\includegraphics[width=0.6\linewidth]{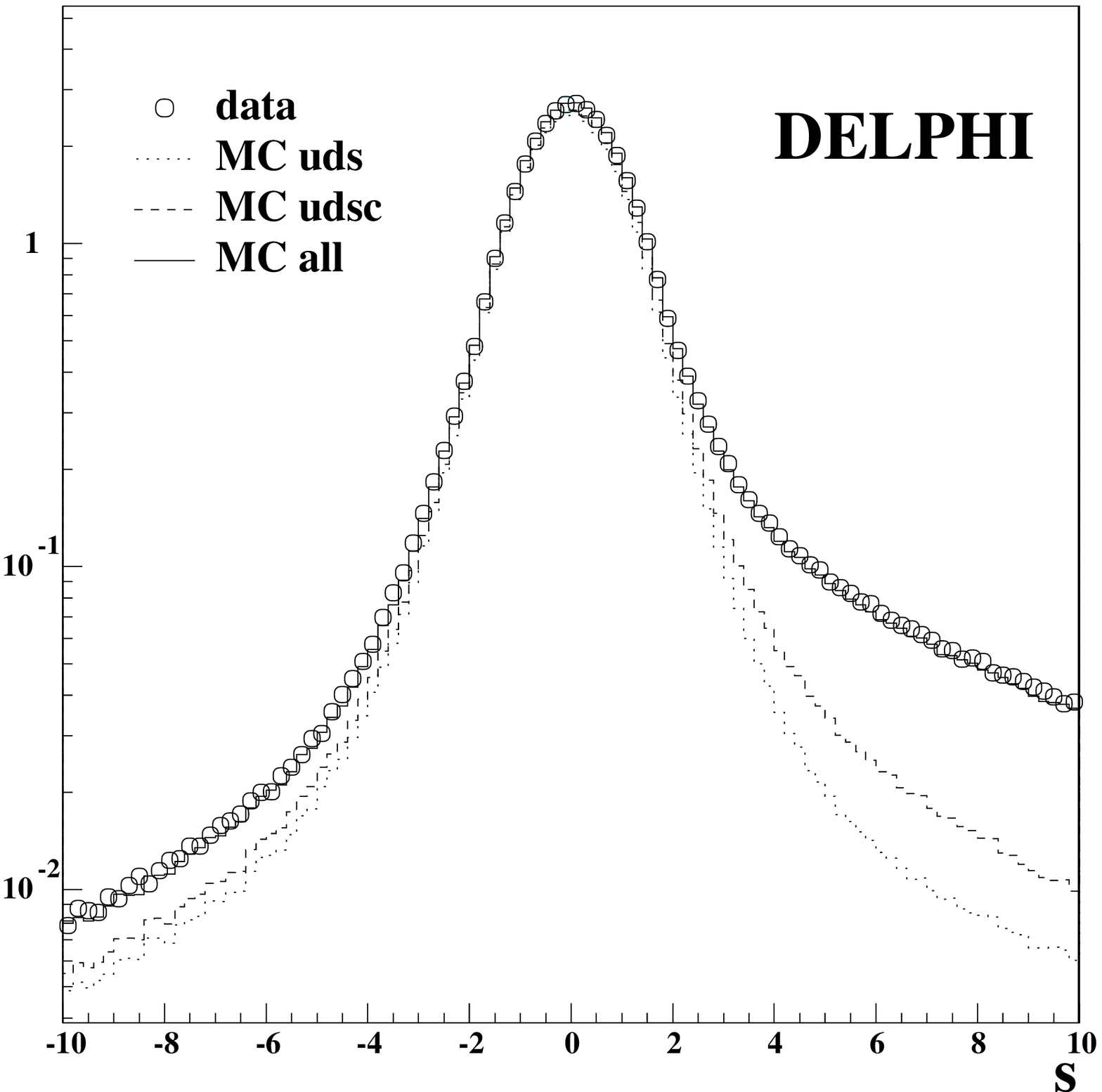}
\end{center}
\caption[Impact parameter significance from DELPHI for data and simulation.]
{Impact parameter significance from DELPHI for data and simulation.
 The contributions of the different quark flavours are shown separately.
 The normalisation is arbitrary.}
\label{fig:hq_ip}
\end{figure}

A lifetime sign is assigned to each track impact parameter.  This is
positive if the extra\-po\-lated track is consistent with a secondary
vertex which lies on the same side of the primary vertex as the track
itself, otherwise it is negative.
Due to the finite resolution of the detector, the relevant quantity
for the identification of the b-quark is the impact parameter
significance $S$, defined as the lifetime-signed impact parameter
divided by its error.  In Figure~\ref{fig:hq_ip} the projection in the
$R\phi$ plane of the lifetime-signed impact parameter significance
distribution is shown for tracks coming from the different quark
flavours. Decay tracks of a $\rm K^0_s$ and $\Lambda$
are removed, so that the distribution of the light quark reflects the
resolution of the apparatus (DELPHI in this case).

A good description of $S$ in the simulation is crucial for a reliable
estimate of the tagging efficiencies.  Negative significance values
arise mainly from primary-vertex-tracks, which have no
lifetime information and show the effects of finite resolution. This
allows a calibration of the tag from the negative side of the
significance distribution.
Even for tracks coming from the primary vertex the distribution of $S$
is expected to be non-Gaussian.  This is caused by pattern recognition
mistakes, non-Gaussian tails of multiple scattering and elastic
hadronic interactions. It has been verified by simulation that these
tails are symmetric for primary tracks.

A simple b-tag can use the number of tracks with a large positive
significance.  A better estimator is constructed by combining all the
positive track significances: first the negative part of the
significance distribution is fitted to a functional form that defines
the resolution of the detector, then for each track the integral of
this function from negative infinity to the $S$ of the track is
computed giving the probability that the track originates from the primary
vertex, which by construction is flat from zero to one.  The
probability that all tracks in a jet, hemisphere or event, come from
the primary vertex is calculated by combining the probabilities for
all tracks in that jet, hemisphere or event \cite{ref:atag}.  By
construction it is flat if all tracks originate from the primary
vertex.  The probability for a group of tracks from an u, d or
s event is then flat between zero and one. The probability for a group
of tracks from a b-quark event, however, is peaked at zero.

\begin{figure}[htb]
\begin{center} 
\includegraphics[width=0.6\linewidth,bb=0 30 514 514]{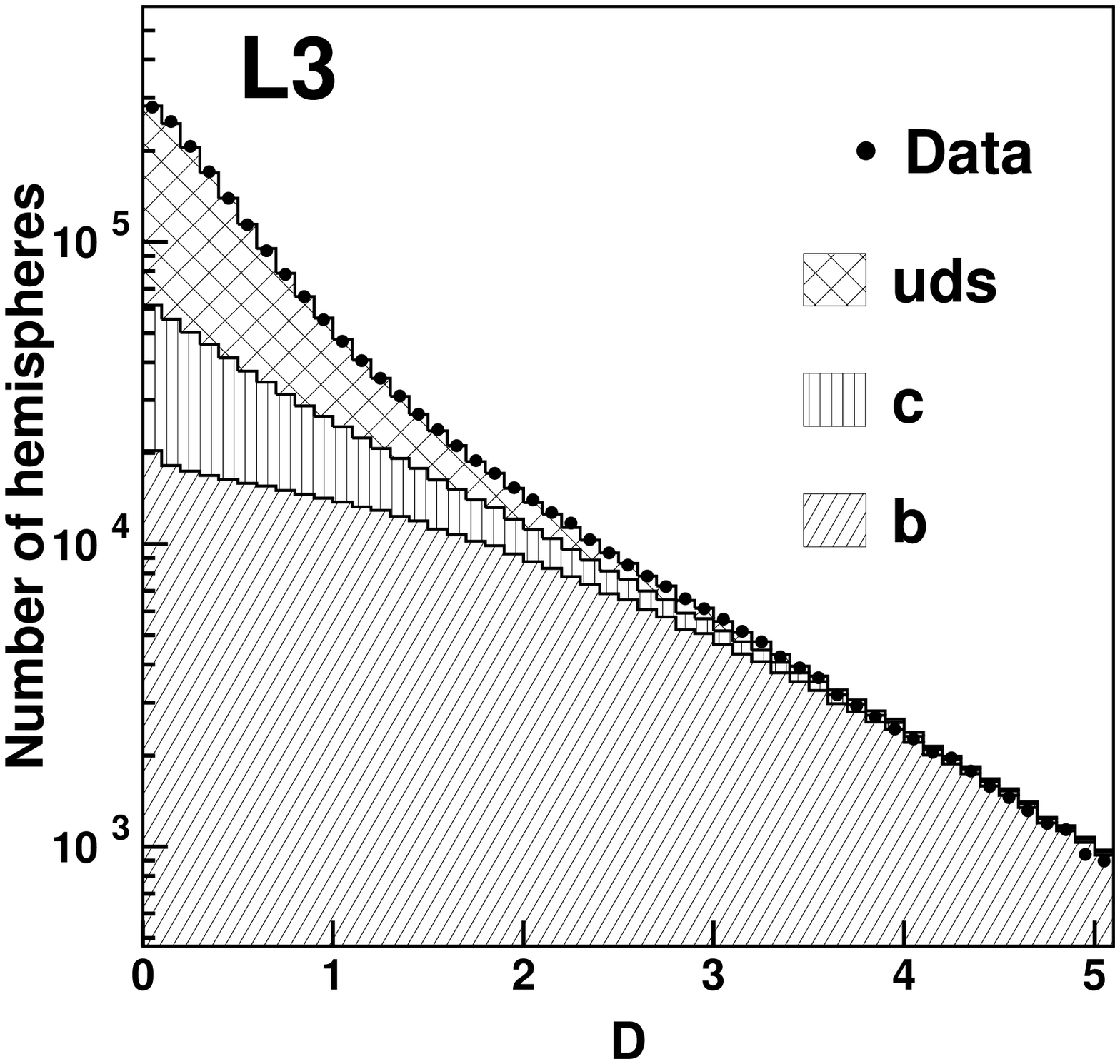}
\end{center}
\caption[Impact parameter b-tag from L3.]  {Impact parameter b-tag
from L3~\cite{ref:lrbmixed}.  D is the negative logarithm of the
hemisphere impact parameter probability.}
\label{fig:lthreebtag} 
\end{figure}

Figure~\ref{fig:lthreebtag} shows the distribution of the L3 b-tagging
variable $D$ which is the negative logarithm of the hemisphere impact
parameter probability.  It can be seen that at large values of $D$
high tag purities can be achieved with impact parameters only.

\begin{figure}[htb]
\begin{center} 
\includegraphics[width=0.7\linewidth,bb=0 20 514 514]{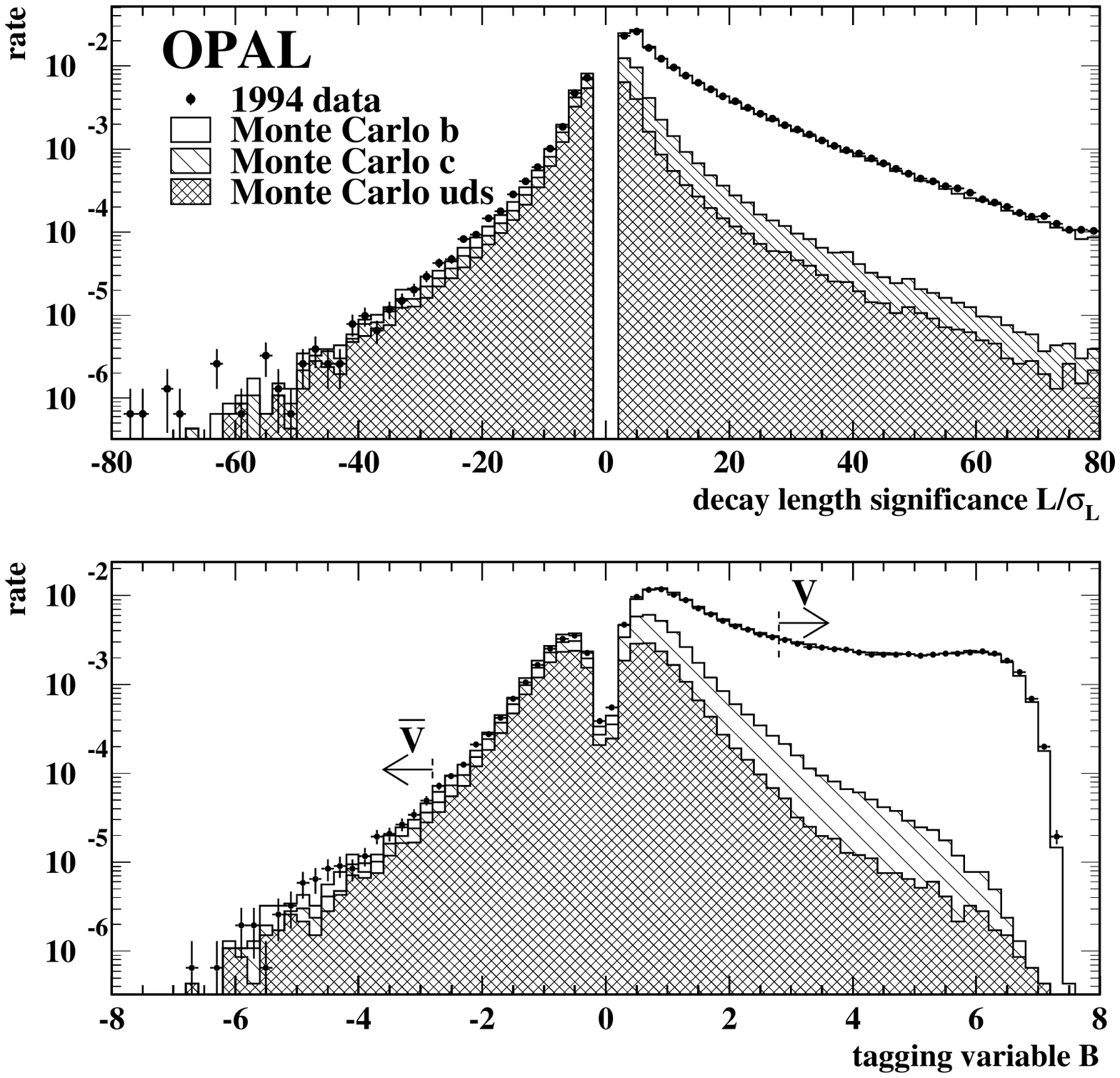}
\end{center}
\caption[Decay length significance and neural network tagging variable
for OPAL.]  {Decay length significance $L/\sigma_L$ (top) and neural
network tagging variable (bottom) for the OPAL secondary vertex based
b-tag~\cite{ref:omixed}.  The gaps around zero significance are due to
neural network preselection cuts removing jets with no significant
secondary vertex. $v$ and $\bar{v}$ are the cut values used in the
\Rb{} analysis.}
\label{fig:hqobtag}
\end{figure}

An alternative lifetime-based tag uses the reconstruction of secondary
vertices. OPAL fits all well-reconstructed high momentum tracks in a
jet to a single secondary vertex, then progressively removes those
which do not fit well. The decay length significance $L/\sigma_L$ (the
reconstructed distance between the primary and secondary vertices
divided by its error) is used as the b-tagging variable, signed
depending on whether the secondary vertex is reconstructed in front of
or behind the primary vertex (see Figure~\ref{fig:hqobtag}).  This
allows the background from light quark events with $L/\sigma_L>0$ to
be estimated using the number of events with $L/\sigma_L<0$.

The extremely precise SLD vertex detector and small stable SLC beam
spot allow a different approach to secondary vertex finding, based on
representing tracks as Gaussian `probability tubes'~\cite{SLD_ZVTOP}.
Spatial overlaps between the probability tubes give regions of high
probability density corresponding to candidate vertices, to which
tracks are finally attached.  This algorithm finds at least one
secondary vertex in 73\% (29\%) of the hemispheres in \bb\ (\cc)
events. Among the b hemispheres that have at least one secondary
vertex, two or more secondary vertices are found in 30\% of them
mostly coming from the decay of the secondary charmed hadron.

\subsection{Combined Lifetime Tag}
\label{sec:hf_ctag}

The pure lifetime tags have an intrinsic limitation because D-mesons
have a lifetime comparable to B-mesons. However this can be overcome
if additional information is used.  Since B-mesons are much heavier
than D-mesons, the most obvious variable is the invariant mass of the
particles fitted to the secondary vertex.
In SLD this mass is used as a b-tag with an additional correction for
the neutral decay products of the B.  From the flight direction of the
B, calculated from the primary and the B-decay vertex, and the
momentum vector of the charged decay products of the B, fitted to the
secondary vertex, the transverse momentum, $p_t$, of the sum of the
neutral decay products can be calculated.  Adding a massless
pseudo-particle with momentum $p_t$ to the secondary vertex gives an
improved lower limit for the mass of the decaying particle.

\begin{figure}[htb]
\begin{center} 
\includegraphics[width=0.6\linewidth]{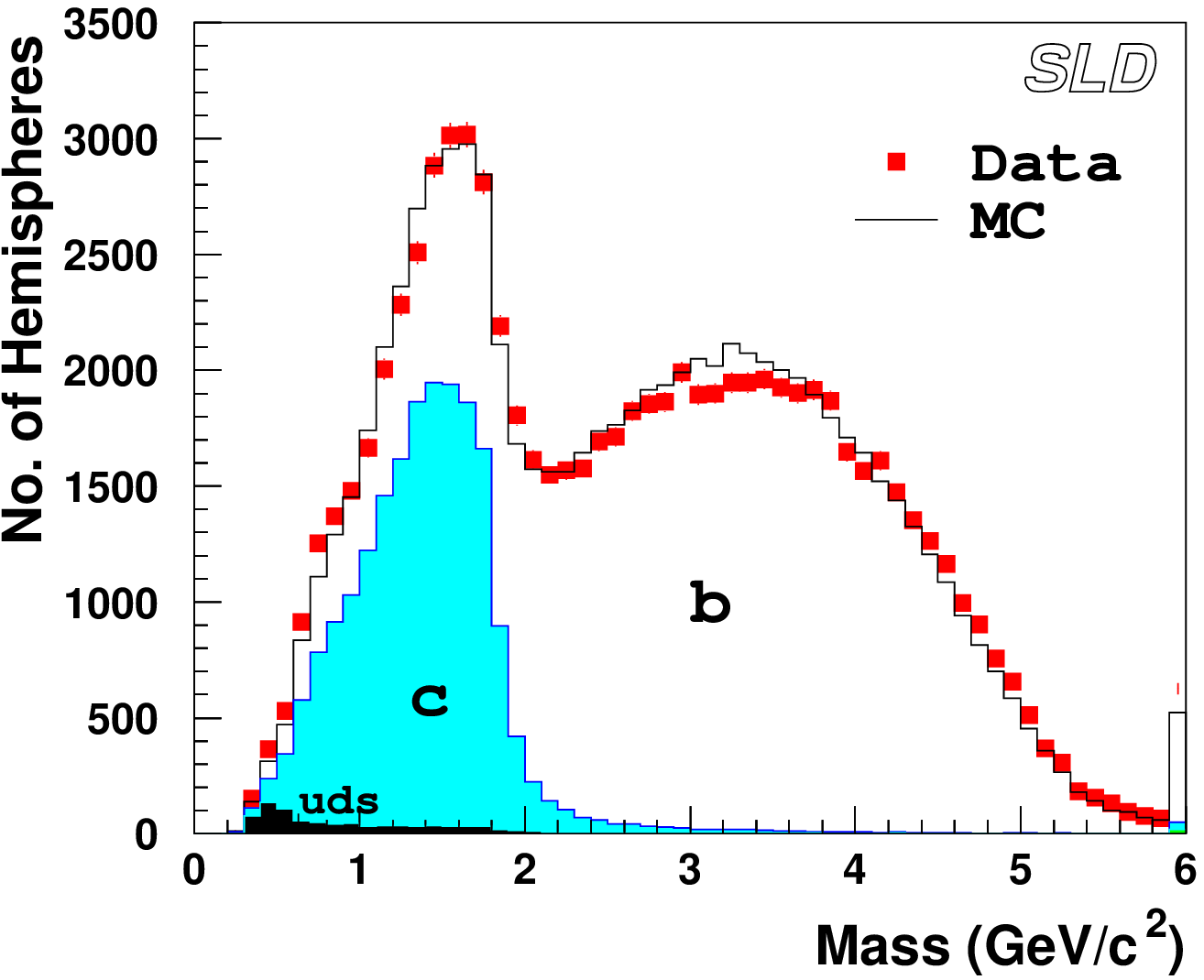}
\end{center}
\caption{Reconstructed vertex mass from SLD for data and simulation.}
\label{fig:sld_vmass} 
\end{figure}

In Figure~\ref{fig:sld_vmass} the $p_t$-corrected mass of the
secondary vertex is shown for b events and for the uds and c
background.  The high efficiency for assigning the correct tracks to
the decay vertex results in a very high b-tagging purity of 98\% for
53\% efficiency, simply by requiring the $p_t$-corrected mass to be
above the D-meson mass.  A further improvement of the performance is
obtained with the introduction of a neural network to optimise the
track to vertex association and a second neural network to improve the
c-b separation by using the vertex decay length, multiplicity and
momentum in addition to the $p_t$-corrected vertex mass. With this
improved tag the b-tagging efficiency increases to 62\% with the same
purity~\cite{SLD_ZVTOP}.

The LEP beam spot is much larger in the $x$ and $y$ directions than
that of SLC, the resolution of the SLD CCD-detectors are about a
factor two better than the ones of the microstrips used at LEP, and
the innermost silicon layers of the LEP vertex detectors have to be at
approximately twice the innermost radius of the SLD vertex detector,
as indicated in Table~\ref{tab:hq_vd}.  This limits the b-tagging
performance of the LEP detectors and motivates development of tags
that combine additional information together with the impact parameter
or decay length information.

DELPHI utilises a likelihood technique combining 4 variables: the
probability that the tracks in the jet come from the primary vertex
(see Section~\ref{sec:btag}), the mass of the reconstructed secondary
vertex, the energy of the charged tracks belonging to the secondary
vertex and their rapidity~\cite{ref:drb,ref:dbtag}.  Combining track
properties with the information from the reconstructed secondary
vertices makes the tag more robust against detector resolution
effects.  A considerable improvement is obtained if the direction
defined by the primary and secondary vertex is used as the b-hadron
direction, instead of the jet axis.

ALEPH uses a linear combination of two lifetime-related
variables~\cite{ref:aimp}.  The first is the probability that the
tracks from each hemisphere come from the primary vertex (as defined
in Section~\ref{sec:btag}). The second variable is correlated with the
mass of the hadron produced. In each jet the tracks are combined in
order of decreasing inconsistency with the primary vertex until their
mass exceeds $1.8\, \GeV$.  The mass-sensitive variable is defined as
the impact parameter probability of the last track added.

L3 identifies b-hemispheres using the impact parameter tag
only~\cite{ref:lrbmixed}.

OPAL uses a vertex tag based on a neural network combining five
variables~\cite{ref:omixed}.  The first four are derived from the
reconstructed secondary vertex: the decay length significance
$L/\sigma_L$, the decay length $L$, the number of tracks in the
secondary vertex and a variable that measures the stability of the
vertex against mismeasured tracks. The fifth variable exploits the
high mass of b-hadrons. For each track in the jet, the relative
probabilities that it came from the primary and secondary vertex are
calculated, using impact parameter and kinematic information. As in
the ALEPH tag, these tracks are then combined in decreasing order of
secondary vertex probability until the charm-hadron mass is exceeded,
and the secondary vertex probability of the last track added is used
as input to the main neural network. The neural network output is
signed according to the sign of $L$, preserving the `folding' symmetry
of the simple $L/\sigma_L$ tag and allowing the light quark background
to be subtracted (see Figure~\ref{fig:hqobtag}).

The b-tag performance of SLD and the LEP experiments at the
purity/efficiency working point used for the $\Rb$ analysis are shown
in Table~\ref{tab:hq_iptag}.

\begin{table}
\begin{center}
\renewcommand{\arraystretch}{1.1}
\begin{tabular} {| l || c | c  | c | c | c |} \hline
                  & ALEPH    & DELPHI  &    L3   & OPAL   & SLD \\
\hline
\hline
b Purity     [\%] &  97.8    & 98.6    &    84.3 & 96.7   & 98.3 \\
b Efficiency [\%] &  22.7    & 29.6    &    23.7 & 25.5   & 61.8 \\
\hline
\end{tabular}
\caption[b-Tagging performance of the different experiments.]
{b-Tagging performance of the different experiments at the cut
where the $\Rb$ analyses are performed.
The lifetime tagging is combined with other information (see text).
The OPAL tag is an OR of a secondary vertex and a lepton tag.}
\label{tab:hq_iptag}
\end{center}
\end{table}

\subsection{Lepton Tagging}
\label{sec:hq_ltag}

\begin{figure}[tbp]
\begin{center}
 \mbox{\includegraphics[width=0.495\linewidth,bb=20 175 535 655]{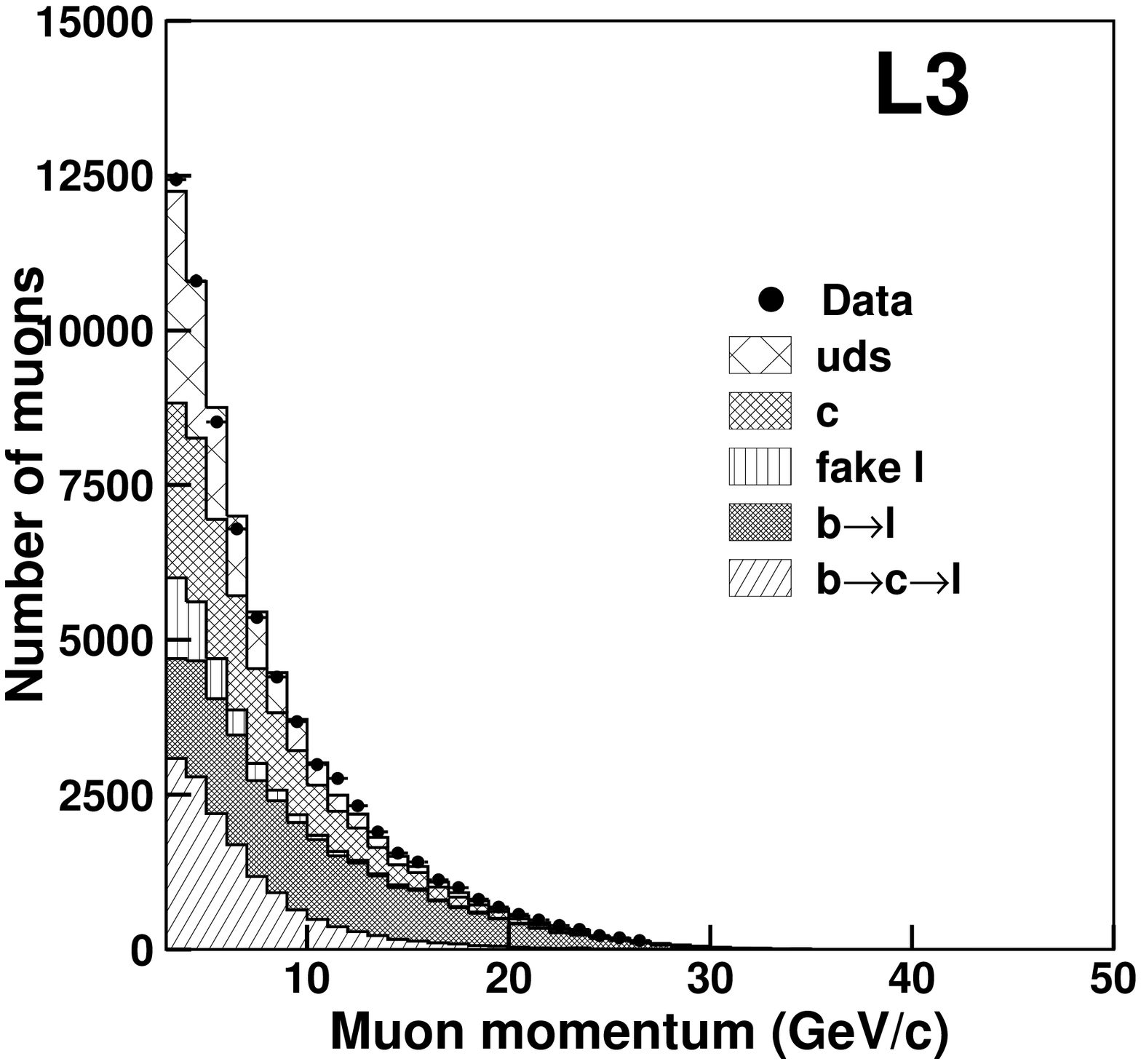}}
\hfill
 \mbox{\includegraphics[width=0.495\linewidth,bb=20 175 535 655]{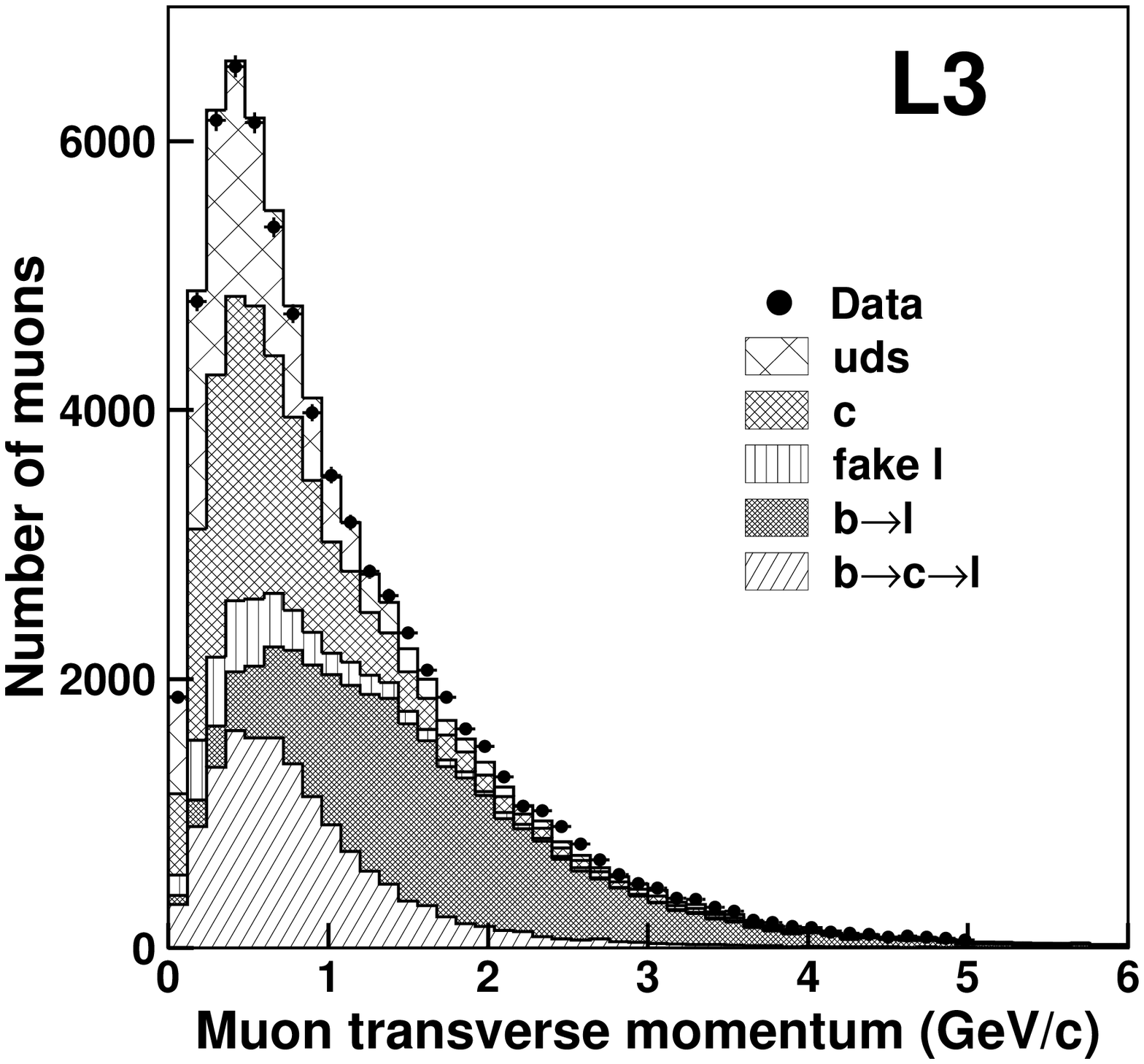}}
\end{center}
\caption[Muon momentum and transverse momentum spectra obtained by L3]{%
Muon momentum and transverse momentum spectra obtained by L3,
together with expectations from simulation for the contributions from
the various sources.  }
\label{fig:hq_leptspec} 
\end{figure}

The semileptonic decays of heavy quarks provide a clean signature
that was the basis of the first methods used to identify the flavour
composition of jets.  Due to the hard fragmentation and the large mass
of b hadrons leptons from b-decays are characterised by large total
and transverse momenta.  Leptons from c-decays also have high
momentum, but a significantly smaller transverse momentum.  The
dominant semileptonic decay modes are $\bl$, $\cl$ and the cascade
decay $\bcl$.  The transverse momentum, $p_t$, of the decay lepton
with respect to the decaying hadron direction is limited to half the
hadron mass.  The direction of the jet containing the lepton, which
experimentally serves as the reference for measuring $p_t$ provides a
good approximation of the hadron direction.  Since b-quarks have a
harder fragmentation spectrum than c-quarks, additional separation
power is given by the lepton momentum.  Figure~\ref{fig:hq_leptspec}
shows the muon $p$ and $p_t$ spectrum from L3 compared to the
simulation of the different sources.  ${\rm b} \rightarrow \ell$ can
be separated cleanly with a simple cut on $p_t$. However the other
sources overlap strongly and can only be separated from each other on
a statistical basis.  At SLD the good resolution of their vertex
detector can also be used to separate $\bl$ and $\bcl$.

The charge of the lepton from a b- or c-decay is correlated to the
charge of the decaying quark. Therefore in the asymmetry measurements
the lepton tag can be used simultaneously to tag the quark flavour and
to distinguish between the quark and the antiquark.  b- and c-quarks
decay semileptonically into either electrons or muons with
approximately equal branching fractions of about 10\%.  While the
lepton always carries the sign of the parent quark charge, the
possibility exists to confuse $\rm c \rightarrow \ell^+$ and $\rm
\overline{b} \rightarrow \ell^+$. Due to the fermion / anti-fermion
flip in the case of c- but not b-quarks, and because the sign of the
two quark asymmetries is the same, this leads to a large sensitivity
of the asymmetry measurements with leptons to the sample composition.
Apart from these three main sources, there are also some other sources
with different charge correlations, mainly $\bcbl$, $\btaul$ and
$\bpsill$. In addition there are misidentified hadrons and electrons
from photon conversion.

As a b flavour tag the lepton tag is not competitive with the lifetime
tag.  As one can see from Figure~\ref{fig:hq_leptspec} only the $\bl$
decay allows a tag with sufficient purity and efficiency about 20\%.
Even from this efficiency roughly half is lost due
to the lepton tag efficiency and a necessary $p_t$ cut. However due to
the simultaneous b-charge tag the lepton tag provides precise
asymmetry measurements.

\subsection{D-Meson Tags}
\label{sec:hq_dtag}

\begin{figure}[tb]
\begin{center}
\includegraphics[width=0.6\linewidth]{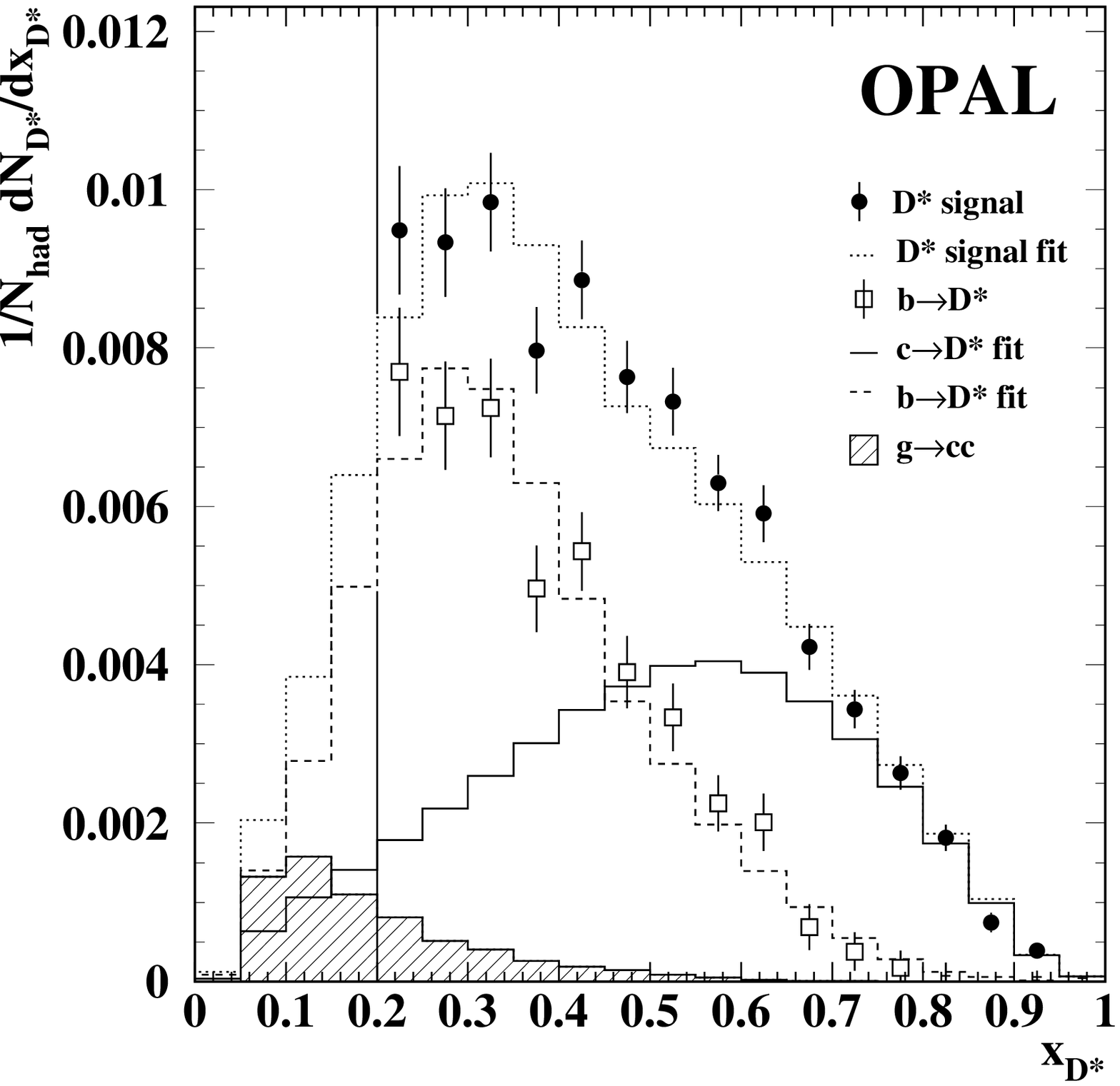}
\vspace{-0.5cm}
\end{center}
\caption[$\Dstarpm$ momentum spectra from OPAL]{ $\Dstarpm$ momentum
spectrum for all events and for $\bb$ and $\cc$ events from OPAL
normalised to the beam energy after subtraction of combinatorial
background~\cite{ref:orcd}.  }
\label{fig:hq_cmom}
\end{figure}

Since charmed hadrons are only rarely produced during light quark
fragmentation, their presence tags c-quarks coming either from the
primary Z-decay or from decay products of a b-quark.  Charmed hadrons
from a primary c-quark have on average a higher momentum than those
from a b-decay (see Figure~\ref{fig:hq_cmom}).  In addition, the decay
length of the reconstructed hadron or lifetime tagging on the whole
event can be used to separate the two sources.

\begin{figure}[htbp]
\begin{center}
\includegraphics[width=0.495\linewidth]{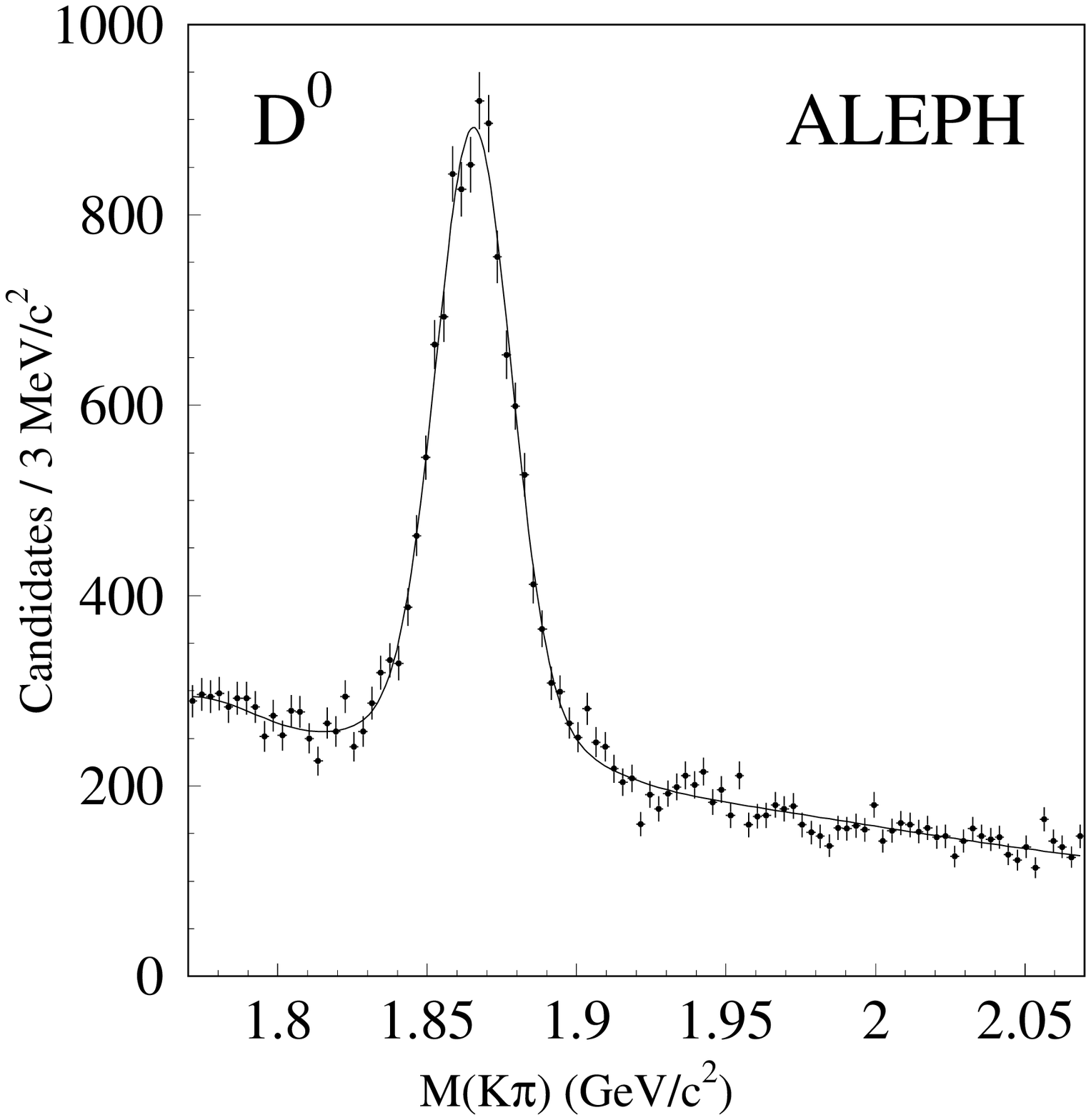}\hfill
\includegraphics[width=0.495\linewidth]{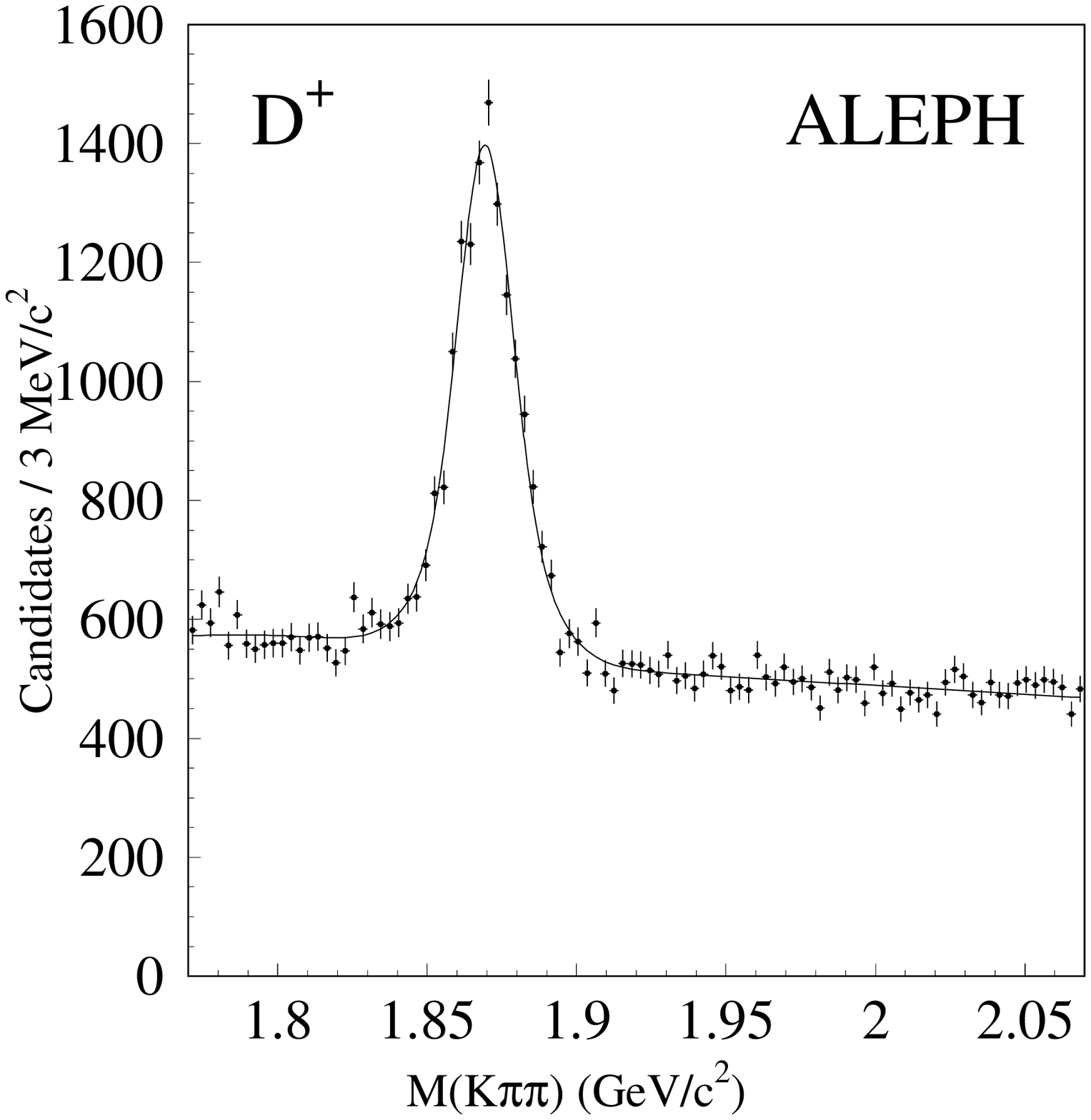}\vskip 1cm
\includegraphics[width=0.495\linewidth]{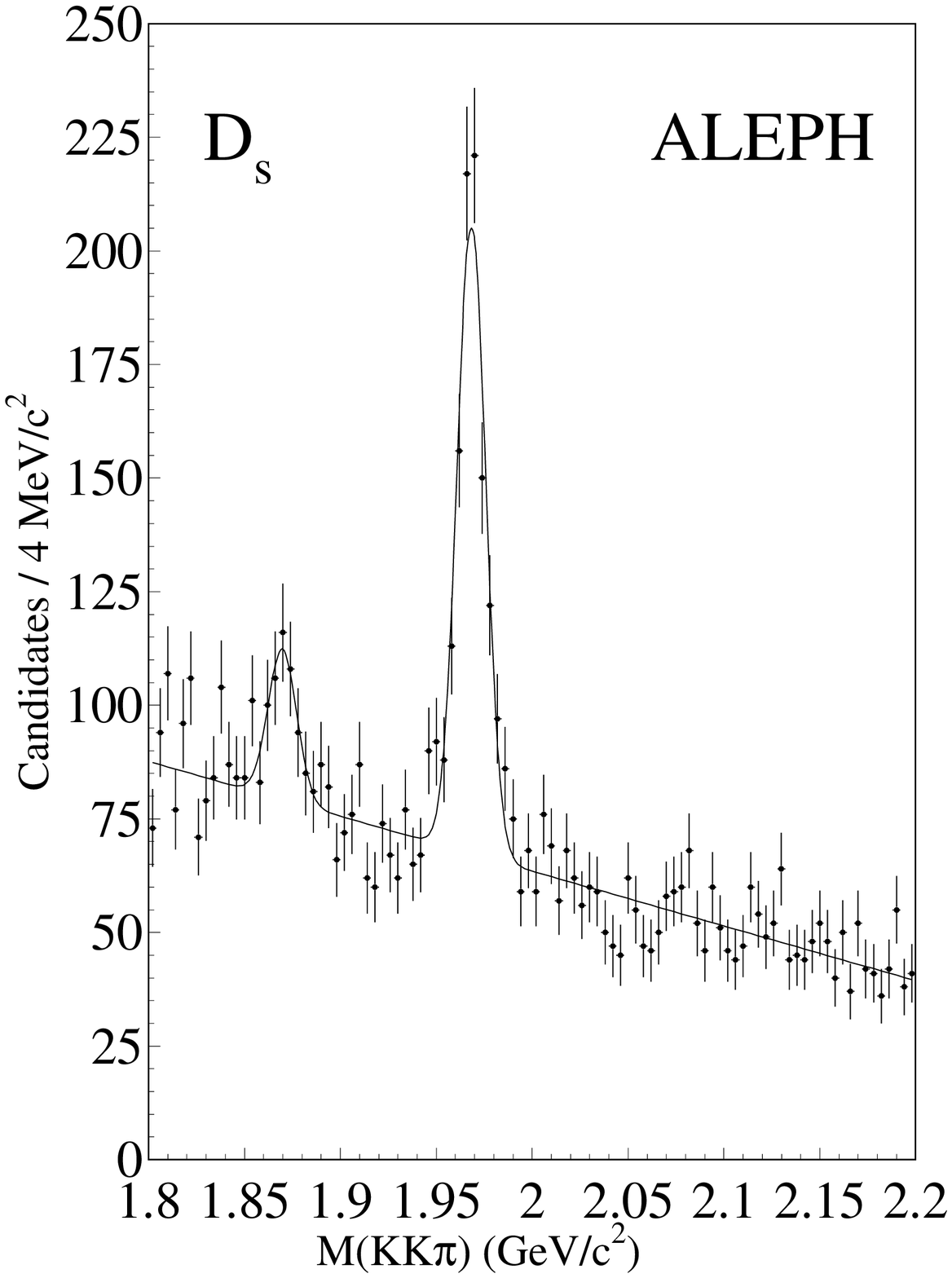}\hfill
\includegraphics[width=0.495\linewidth]{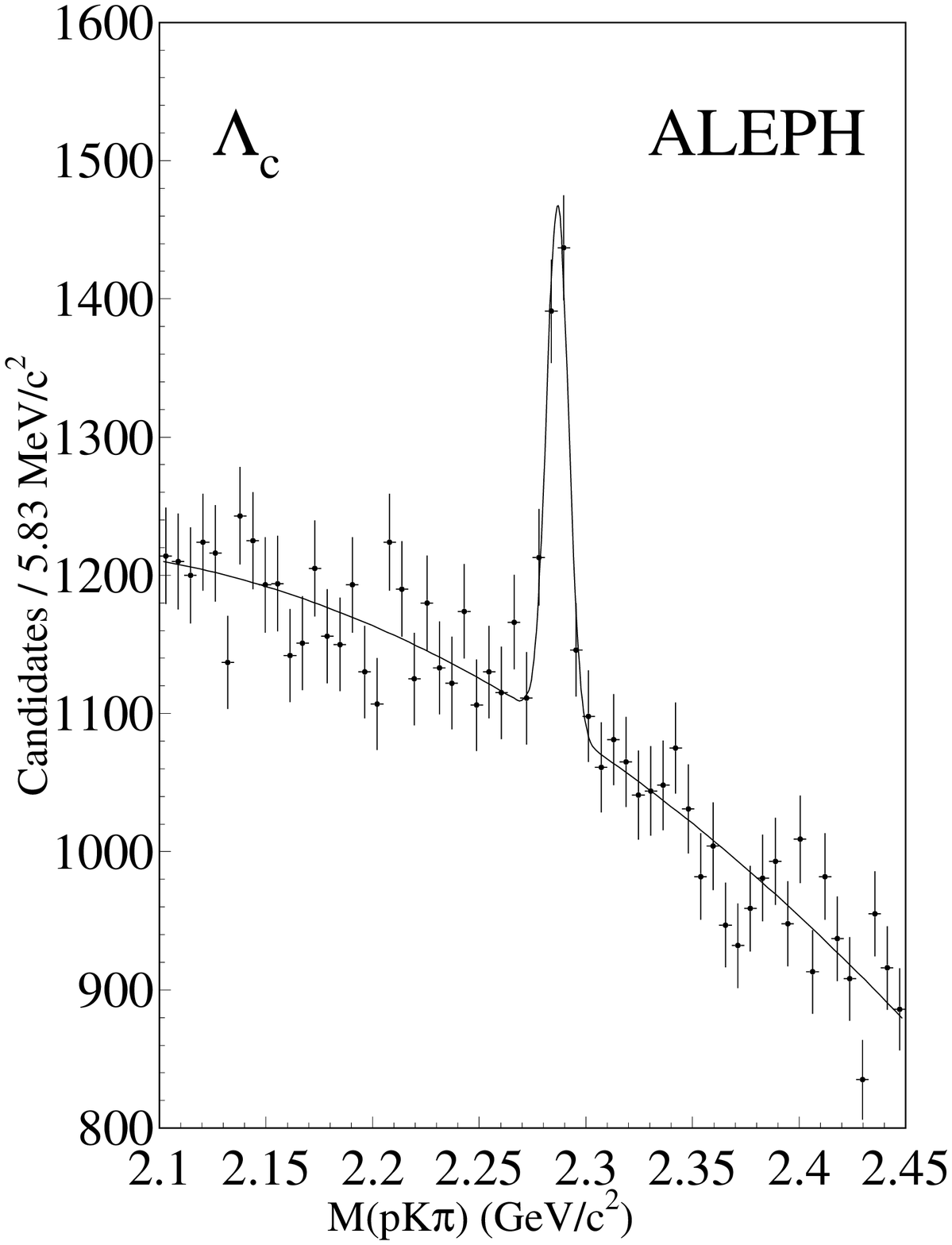}
\end{center}
\caption[$\Dzero \rightarrow {\rm K}^-\pi^+,\, 
\Dplus \rightarrow {\rm K}^-\pi^+\pi^+,\, 
\Ds \rightarrow {\rm K}^+{\rm K}^-\pi^+$ and 
$\Lc \rightarrow {\rm p K}^-\pi^+$ mass spectra]{
Mass spectra for $\Dzero \rightarrow {\rm K}^-\pi^+,\, 
\Dplus \rightarrow {\rm K}^-\pi^+\pi^+,\, 
\Ds \rightarrow {\rm K}^+{\rm K}^-\pi^+$ and 
$\Lc \rightarrow {\rm p K}^-\pi^+$ obtained by ALEPH~\cite{ref:arcc}.
}
\label{fig:hq_dmes}
\end{figure}

\begin{figure}[htbp]
\begin{center}
\includegraphics[width=\linewidth,bb=5 15 520 515]{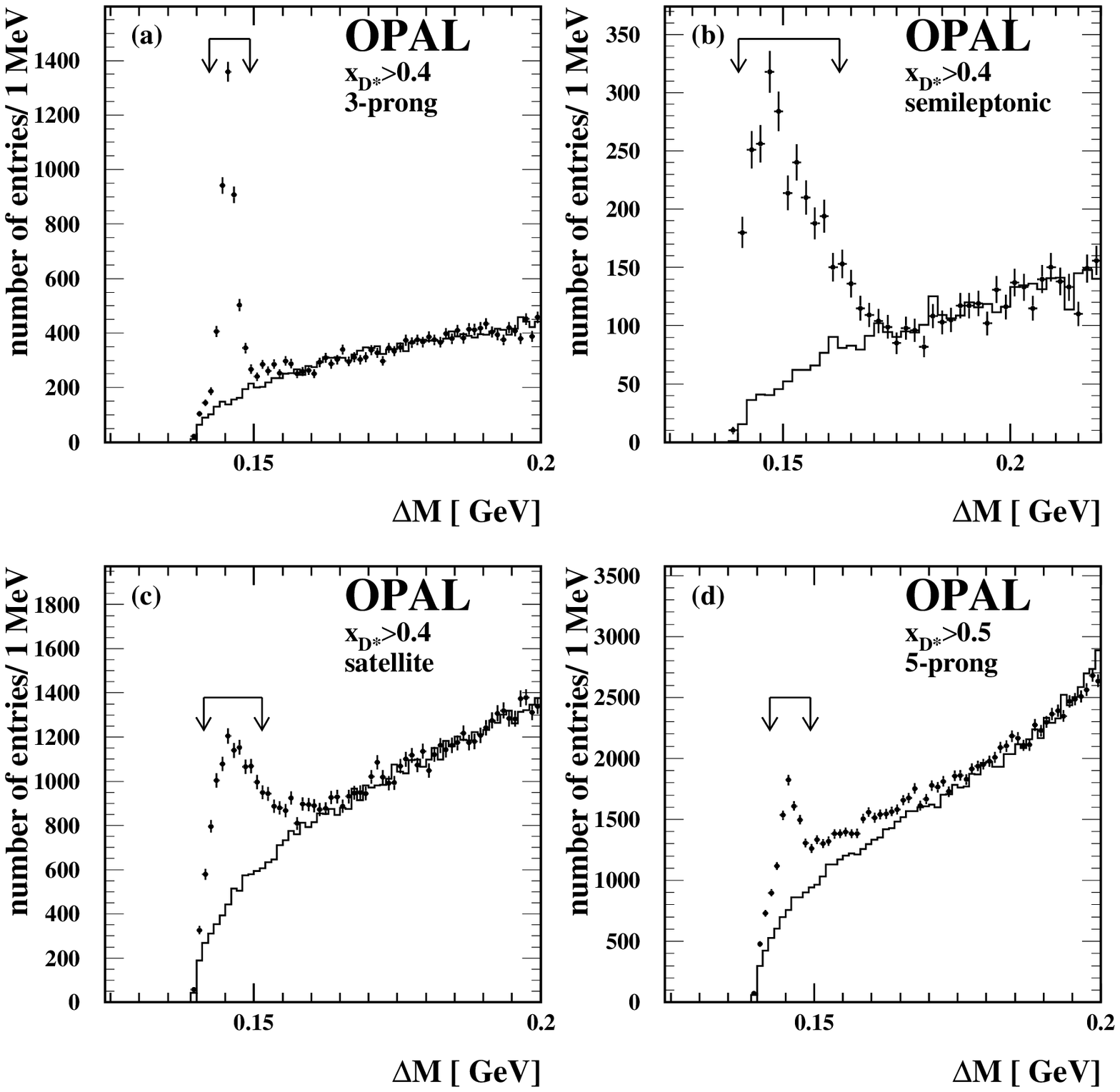}
\end{center}
\caption[Mass difference ${m}(\pi^+ \rm D^0)-{m}(\rm D^0) $ spectrum
from OPAL in different channels.]{ Mass difference spectrum ${m}(\pi^+
D^0)-{m}(D^0)$ from OPAL in different channels~\cite{ref:orcd}.  In a)
and d) the $\Dzero$ is fully reconstructed in the decay modes $\Dzero
\rightarrow {\rm K^-} \pi^+$ and $\Dzero \rightarrow {\rm K^-} \pi^+
\pi^+ \pi^-$ In c) the decay mode $\Dzero \rightarrow \rm K^- \pi^+
\pi^0$ is used, where the $\pi^0$ is not reconstructed. This ${\rm
K}^- \pi^+$ mass peak is enhanced due to the large polarisation of the
intermediate $\rho^+$ produced in the $\Dzero\rightarrow\rm
K^-\rho^+\rightarrow K^-\pi^+\pi^0$ decay.  In b) the $\Dzero$ is
partially reconstructed in the semileptonic decay mode.  The points
with the error bars are the measured data. The solid histogram is the
background estimated from measured data by a hemisphere mixing
technique.
}
\label{fig:hq_dstar}
\end{figure}

At LEP and SLD the weakly-decaying charmed hadrons
$\Dzero,\,\Dplus,\,\Ds$ and $\Lc$ can be reconstructed in particular
exclusive final states (see Figure~\ref{fig:hq_dmes}). The charm
tagging efficiency is limited by the low branching fractions for these
decay modes, which are typically only a few percent.  The decay
$\Dstarp \rightarrow \pi^+ \Dzero$ can be reconstructed particularly
cleanly, due to the small mass difference $\Delta m = m_{\Dstarp} -
m_{\Dzero}$, which leads to a characteristic narrow peak with little
background, as shown in
Figure~\ref{fig:hq_dstar}. Because of the good resolution, even
$\Dzero$ decays which are not fully reconstructed, such as $\Dzero
\rightarrow \ell \nu X$ or $\Dzero \rightarrow \rm K^- \pi^+ \pi^0$,
where the $\pi^0$ is not seen, can be used.

\begin{figure}[htb]
\begin{center}
\includegraphics[width=0.7\linewidth,bb=0 20 520 520]{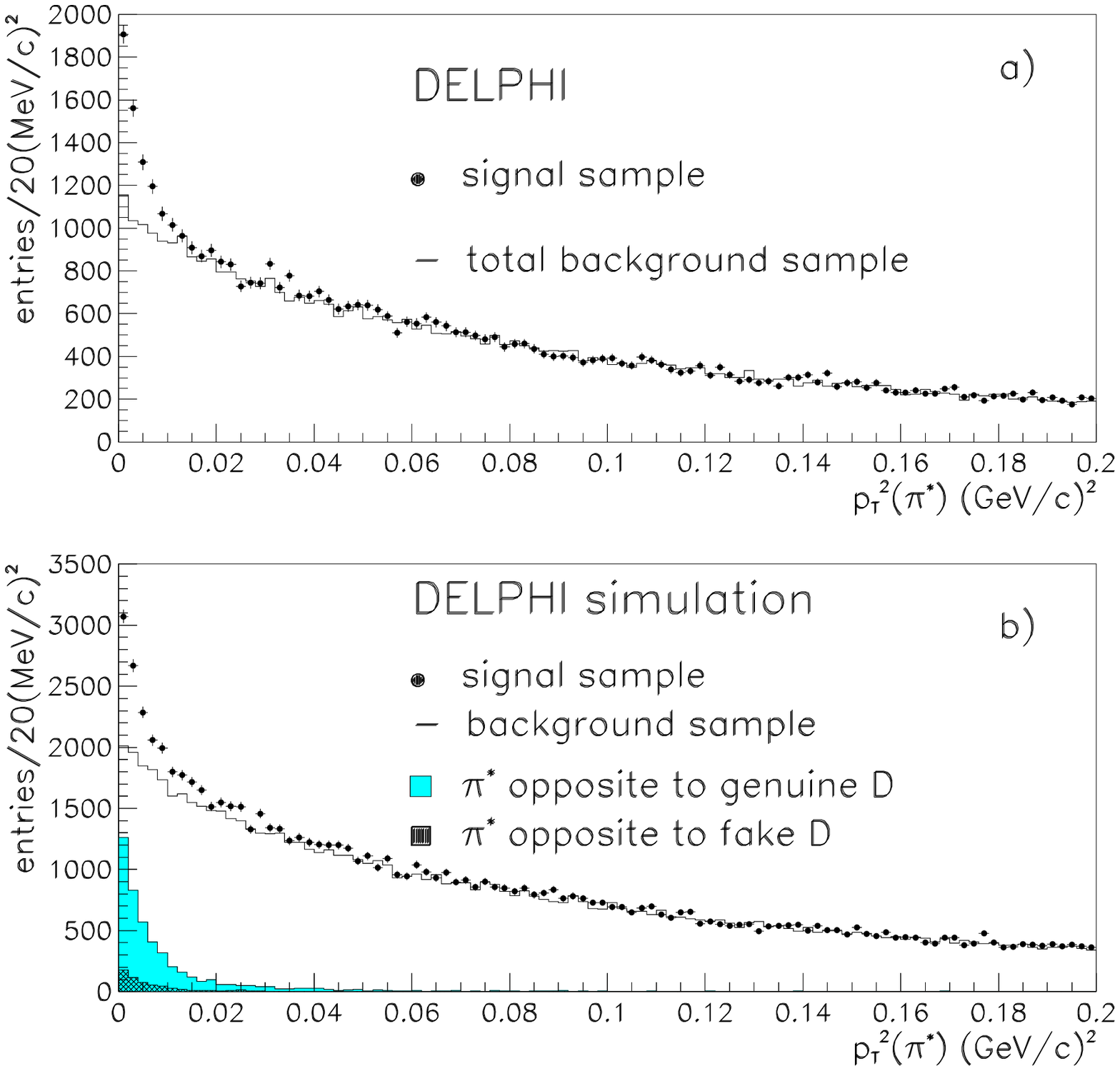}
\end{center}
\caption[$p_t^2$ spectrum of the slow pion opposite a high energy
D-meson.]  { $p_t^2$ spectrum of the slow pion opposite a high energy D-meson
  candidate\cite{ref:drcd}.  The points show pions with charge opposite that
  of the charm quark in the D-meson, while the histogram shows those with the
  same charge.  The data is shown in a) and the simulation in b), with the
  signal component indicated separately.  }
\label{fig:hq_lowpt} 
\end{figure}

The decay $\Dstarp \rightarrow \pi^+ \Dzero$ can also be tagged
inclusively without specifically recognising any of the decay products
of the $\Dzero$.  The small mass difference between the $\Dstarp$ and
$\Dzero$ and the low mass of the pion result in a very low pion
momentum in the $\Dstarp$ rest-frame.  Therefore in the laboratory
frame the pion closely follows the $\Dstarp$ direction and has a very
low transverse momentum, $p_t$, with respect to the jet direction.  As
shown in Figure~\ref{fig:hq_lowpt}, the number of $\Dstarp$ in a
sample can thus be measured from the excess in the $p_t^2$ spectrum at
very low values.
Because of the large background, this tag is typically used to count
c-quarks on a statistical basis in conjunction with other tags.

The flavour of D-mesons also measures the flavour of the original
quark.  In $\cc$-events the primary quark is directly contained in the
D-meson, while in $\bb$-events the c-quark comes from the decay chain
${\rm b} \rightarrow {\rm c}$.  The decay ${\rm b} \rightarrow
\bar{\rm c}$ (via ${\rm b} \rightarrow {\rm c W}^-, {\rm W}^-
\rightarrow \bar{\rm c}{\rm s}$) is suppressed, so that the quark
flavour tag using D-mesons is almost always correct, in contrast to
the lepton tag, where the $\bl$ and $\bcl$ decays have to be
separated.  Another advantage of a D-meson tag is that it separates
directly the quark from the antiquark and not positively from
negatively charged quarks as in the lepton case, so that the
sensitivity of the asymmetry measurements to the sample composition is
significantly reduced.  Since the absolute efficiency cancels in the
asymmetries many different states can be used. Because of the low
background, however, the most sensitive decay is $\Dstarp \rightarrow
\pi^+ \Dzero$ with $\Dzero \rightarrow\rm K^- \pi^+$.  Asymmetry
measurements with D-mesons contribute with a significant weight to
$\Acc$ while they contribute only little to $\Abb$.

\section{Partial Width Measurements}
\label{sec:hqrbc}

In principle, the partial width measurements only require counting the
fraction of hadronic events tagged as a particular flavour, and
knowing the efficiency and purity of the tag to a high precision.
This ``single-tag'' approach has been adopted in some of the $\Rc$
measurements.
  
The single-tag approach is highly sensitive to knowledge of the
tagging efficiency, which in any case is best extracted from the data
itself.  So-called double-tag methods have been developed which
provide a simultaneous determination of the tagging efficiency and
quark rate through comparison of the probabilities that one or both of
the two hemispheres in each event is tagged. Since the statistical
precision of these methods varies as the tagging efficiency squared,
they are most useful when this efficiency is high.

A summary of all individual measurements of $\Rb$ and $\Rc$, used in
the combination is given in Tables~\ref{tab:Rbinp} and~\ref{tab:Rcinp}
of Appendix~\ref{sec:hqappendix}.

\boldmath
\subsection{$\Rb$ Measurements}
\unboldmath
\label{sec:hq_rbmeas}

\subsubsection{The Double-tag Method}
\label{sec:hq_rbdble}

All precise measurements of \Rb\ are primarily based on counting
events with either one or both hemispheres tagged.  The fraction of
hemispheres which are b-tagged, $f_s$, and the fraction of events
where both hemispheres are b-tagged, $f_d$, are given by:
\begin{eqnarray}
f_s & = & \epsb \Rb + \epsc \Rc + \epsl(1-\Rb-\Rc) \label{eq:rb}\\
f_d & = & \epsb^{(d)} \Rb + \epsc^{(d)} \Rc + \epsl^{(d)} (1-\Rb-\Rc) \,,
\nonumber
\end{eqnarray}
where $\epsf$ is the hemisphere tagging efficiency for flavour f.  The
double-tagging efficiency $\epsf^{(d)} $ can be written as
\begin{equation}
\epsf^{(d)} = (1+\calCf) \epsf^2 
\label{eq:hq_hcor}
\end{equation}
where the correction factor $\calCf \not= 0$ accounts for the fact
that the two hemispheres in an event are slightly correlated.  For the
high purity b-tags used in the analyses $\calCc$ and $\calCuds$ can be
safely neglected.
Neglecting all hemisphere correlations and background one has $\Rb =
f_s^2/f_d$, independent of the b-tagging efficiency $\epsb$ which then
does not need to be determined from simulation.
In reality, corrections dependent on the background efficiencies
$\epsc,\,\epsl$ and hemisphere correlations $\calCb$ must be applied
and these have to be determined from Monte Carlo.  The uncertainties
on these parameters are included in the systematic errors. The effect
of an uncertainty $\Delta \varepsilon_x$ from a background source $x$
is approximately given by $\Delta \Rb = 2 {\Delta
\varepsilon_x}/{\epsb} R_x$ and for an uncertainty on the correlation
by $\Delta \Rb = \Delta \calCb \Rb$.
The statistical uncertainty is dominated by the double tag statistics
so that the number of events needed to achieve the same statistical
precision is proportional to ${1}/{\epsb^2}$.  A large b-tag
efficiency also reduces the sensitivity to the uncertainty on
hemisphere correlations since it usually results in less dependence on
parameters like the b-hadron momentum.
Therefore it is essential to develop a high efficiency and high purity
b-tag to enable the double tag scheme to achieve the necessary
statistical and systematic precision.  Details about the hemisphere
correlations are explained in Section~\ref{sec:hq_rbhemcol}.

OPAL~\cite{ref:omixed} and SLD~\cite{Abe:2005nq} measure $\Rb$ with the
double-tag technique, %
OPAL with a logical OR of secondary vertex and lepton tags and SLD
with only their neural network improved vertex mass tag explained in
Section~\ref{sec:hf_ctag}.

\subsubsection{The Multi-tag Method}

In the double-tag method, hemispheres are tagged simply as b or non-b.
This leads to two equations and six unknowns, $\Rb,\ \Rc, \ \epsb,\
\epsc, \ \epsl$ and $\calCb$. Three of them ($\epsc, \ \epsl$ and
$\calCb$) are taken from simulation and \Rc\ is fixed.  The
\Rc-dependence is then accounted for in the systematic error of the
experimental publications and in the combination procedure described
in Section~\ref{sec:hqcomb}.  The method can be extended by adding
more tags, {\em e.g.}, additional b-tags with lower purity, or charm
and light flavour tags~\cite{fernando}.  The tags are made exclusive,
such that each hemisphere is counted as tagged by only one tag method,
and the untagged hemispheres are counted as an extra `null' tag.

With $T$ separate hemisphere tags, there are then
$T(T+1)/2$ double tag fractions $f_d^{ij}$ ($i,j=1 \ldots T$), given
(analogously with Equations~\ref{eq:rb}) by:

\begin{equation}
f_d^{ij}  =  \epsb^i\epsb^j (1+\calCb^{ij}) \Rb + 
\epsc^i\epsc^j (1+\calCc^{ij}) \Rc + 
\epsl^i\epsl^j (1+\calCuds^{ij}) (1-\Rb-\Rc) \,,
\label{eq:rbmult}
\end{equation}
where $\epsf^i$ is the hemisphere tagging efficiency for flavour f
with tag $i$, and $\calCf^{ij}$ is the hemisphere correlation
coefficient for tagging an event of flavour f with tag $i$ in one
hemisphere and $j$ in the other.  The single tag rates do not give
additional information in this case, since they can be written as sums
over the appropriate double tag fractions.

With $T$ tags and $F$ event types, there are $F(T-1)$ unknown
efficiencies $\epsq^j$ (since the $T$ efficiencies for each flavour
must add up to one) and $F-1$ unknown partial width ratios $R_f$. If
all the correlation coefficients $\calCf^{ij}$ are taken from
simulation, that leaves $F(T-1)+(F-1)=TF-1$ unknowns to be determined
from
$T(T+1)/2-1$ independent double tag rates, $f_d^{ij}$. With $F=3$
event types (b, c, uds), the minimum number of tags for an
over-constrained system is six.

ALEPH~\cite{ref:aimp,ref:alife} and DELPHI~\cite{ref:drb} both use
this multi-tag method for measuring $\Rb$. The six tags used are:
three b-tags with different purities, a charm tag, a light quark tag
and the ``untagged'' hemispheres. However, even with these six tags,
the solution for all efficiencies and partial widths is still not well
determined.  This problem is solved by exploiting the very high purity
of the high-purity b-tag, taking the small background efficiencies for
charm and light quark events from Monte Carlo, as in the simple
double-tag analysis. $\Rc$ is also fixed in the analysis to its $\SM$
value and the dependence on the assumed $\Rc$ is taken into account in
the combination.

Since the auxiliary b-tags contribute to the measurement, the
statistical error of a multi-tag analysis is smaller than a double-tag
analysis using the same high purity b-tag alone. The charm and light
quark tags also allow the data to constrain the backgrounds in the
additional b-tags.  The systematic error due to the backgrounds in the
high purity b-tag stays the same as in the double tag method.  It can
be reduced by changing the working point of the high purity b-tag
towards higher purity, thus sacrificing some of the gain in
statistical error.  Many additional hemisphere correlations have to be
estimated from Monte Carlo, but the impact of the most important,
between two hemispheres tagged with the high purity b-tag, is
reduced. The total systematic uncertainty from hemisphere correlations
is therefore almost unchanged.

L3~\cite{ref:lrbmixed} also use a multi-tag analysis for $\Rb$, but
with only two tags, based on lifetime and leptons, and determine the
background efficiencies for both tags from simulation. The b-tagging
efficiency for the lepton tag is used to provide a measurement of the
semileptonic branching fraction $\Brbl$.

\boldmath
\subsection{$\Rc$ Measurements}
\unboldmath
\label{sec:hq_rcmeas}

For $\Rc$ the situation is more complicated than for $\Rb$. Especially
at LEP, the c-tags are less efficient and less pure than the
b-tags. To obtain the optimal $\Rc$ precision under these
circumstances a variety of methods are employed.

\subsubsection{Double Tag Measurements}

In the normal double tag analyses the statistical error is determined
by the size of the double tagged sample, which is proportional to the
square of the tagging efficiency.  Thus only SLD is able to present a
high-precision $\Rc$ measurement with the normal double tagging
technique~\cite{Abe:2005nq}. The charm tag is based on the same neural
network used for the b tag. An output value of the network greater
than 0.75 is considered a b tag and a value below 0.3 a charm tag. In
addition, two intermediate tags are introduced covering intervals from
0.3 to 0.5 and 0.5 to 0.75. The charm tag has an efficiency of 18\% at
a purity of 85\%.  The tag has a very low uds-background, which can be
estimated with sufficient precision from simulation.  The b background
is relatively high, but can be measured accurately in hemispheres
opposite a high-purity b-tag. $R_c$ is extracted from a simultaneous
fit to the count rates of the 4 different tags. The b and charm
efficiencies are fitted from data.

ALEPH also presents a double tag measurement of $\Rc$ using fully
reconstructed D-mesons.  Due to the small branching fractions,
however, the efficiency is low and the statistical error relatively
large~\cite{ref:arcd}.

\subsubsection{Inclusive/Exclusive Double Tags}

At LEP more precise results can be obtained with the
inclusive/exclusive double-tag method.  In the first step $\RcPcDst$
is measured from a sample of exclusively reconstructed $\Dstarp$ (the
`exclusive tag').  In the second step $\PcDst$ is obtained using an
inclusive $\Dstarp$ tag where only the charged pion from the $\Dstarp$
decay is identified (see Section~\ref{sec:hq_dtag}).  A fit is made to
the $\pi^-$ $p_t$ spectrum in hemispheres tagged as containing a charm
quark using a high energy $\Dstarp$ reconstructed in the other
hemisphere of the event.  The uds background in this tagged charm
sample is estimated from the sidebands in the mass spectra of the high
energy $\Dstarp$, and the b-background is measured using lifetime tags
and the $\Dstarp$ momentum distribution.  The fragmentation background
under the low $p_t$ pion $\Dstarp$ signal can be estimated by
exploiting the charge correlation between the pion and the $\Dstar$ in
the opposite hemisphere. Genuine signal pions have the opposite charge
to that of the $\Dstar$, while background pions can have either charge
(see Figure~\ref{fig:hq_lowpt}).

In this method the reconstruction efficiency for the $\Dzero$ and the
relevant decay branching fraction (normally $\Dzero \rightarrow {\rm
K}^- \pi^+$) still need to be known from simulation or external
measurements. However the probability that a c-quark fragments into a
$\Dstarp$, which is hard to calculate, is measured from the data.
ALEPH~\cite{ref:arcd}, DELPHI~\cite{ref:drcd,ref:drcc} and
OPAL~\cite{ref:orcd} present such inclusive/exclusive double tag
measurements.  DELPHI and OPAL give both $\Rc$ and $\PcDst$ as results
while ALEPH does the unfolding internally and presents only $\Rc$.

\subsubsection{Charm Counting}

Another method for measuring $\Rc$ is known as charm counting.  All
charm quarks finally end up in a weakly-decaying charmed hadron. The
production rate of a single charmed hadron ${\rm D_i}$ is proportional
to $\Rc \fDi$, where $\fDi$ is the fraction of charm quarks that
eventually produce a ${\rm D_i}$. However if all weakly-decaying
charmed hadrons can be reconstructed, the constraint $\sum_{\rm i}
{\fDi} = 1$ can be exploited and $\Rc$ can be measured without the
unknown fragmentation probabilities $\fDi$.  In practice
$\Dzero,\,\Dplus,\,\Ds$ and $\Lc$ are reconstructed and small
corrections for unmeasured strange charmed baryons have to be 
applied~\cite{ref:arcc}:
\begin{equation}
\fcb ~ = ~ (1.15 \pm 0.05) \fLc \,.
\end{equation}
$\Rc$ is then obtained using the constraint
\begin{equation}
\fDz + \fDp + \fDs + \fcb ~ = ~ 1\,.
\label{eq:fdi}
\end{equation}
A priori there is the same amount of charmed hadrons coming from
primary c-quarks and from b-decays. The b component can, however be
efficiently separated using lifetime tags and the momentum of the
reconstructed charmed hadrons. The efficiency to reconstruct a given
decay channel has to be taken from simulation.  As a by-product these
measurements obtain the production rates of the weakly-decaying
charmed hadrons $\fDi$, which are needed to calculate the charm
tagging efficiency of the lifetime b-tags.  ALEPH~\cite{ref:arcc},
DELPHI~\cite{ref:drcc} and OPAL~\cite{ref:orcc} present charm counting
$\Rc$ analyses. The method is however limited by the knowledge of
branching fractions to the decay modes used in calculating the
reconstruction efficiency, especially for the $\Ds$ and the $\Lc$.

\subsubsection{Lepton Tag}

ALEPH also measures $\Rc$ with leptons~\cite{ref:arcd}. They measure
the lepton total and transverse momentum spectrum and subtract the
contribution from b decays. This is determined from the lepton spectra
measured in b events tagged in the opposite hemisphere by a
lifetime-based b-tag.  The result is proportional to $\Rc \Brcl$,
where $\Brcl$ is measured by DELPHI~\cite{ref:drcd} and
OPAL~\cite{ref:ocl} in charm events tagged in the opposite hemisphere
by a high energy $\Dstarp$.

\section{Asymmetry Measurements}
\label{sec:hqasy}

The forward backward asymmetry for a quark flavour q is defined as
\begin{equation}
  \Aqq ~ = ~ \frac{\sigma^{\rm q}_{\rm{F}}-\sigma^{\rm q}_{\rm{B}}}
                {\sigma^{\rm q}_{\rm{F}}+\sigma^{\rm q}_{\rm{B}}}\, ,
  \label{eq:hq_asydef}
\end{equation}
where the cross-sections are integrated over the full forward (F) and
backward (B) hemisphere.  "Forward" means that the quark, rather than
the antiquark, is produced at positive $\cos\theta$. The differential
cross-section with respect to the scattering angle is, on Born level,
given by
\begin{equation}
  \frac{d \sigma^{\rm q}}{d \cos \theta} ~ = ~
  \sigma_{\rm tot}^{\rm q}\left[ \frac{3}{8} \left (1 +\cos^2 \theta \right)
    + \Aqq \cos \theta \right]\,.
  \label{eq:hq_asysig}
\end{equation}
This dependence can be used to correct for a non-uniform efficiency or
can be fitted directly to the data.  The asymmetry at a quark
production angle $\theta$ can be written as
\begin{equation}
  \Aqq(\cos \theta) ~=~ \frac{8}{3} \Aqq \frac{\cos \theta}
  {1 + \cos^2 \theta} ~=~ \cAe\cAq \frac{2\cos \theta}{1 + \cos^2 \theta}\,.
  \label{eq:hq_difasy}
\end{equation}
Most experimental analyses measure $\Aqq(\cos \theta)$ and then use
Equation~\ref{eq:hq_difasy} to fit $\Aqq$. This is statistically
slightly more powerful than simple event counting.
The exact angular form is slightly modified by QCD and mass effects. 
This is corrected for by simulation. A more detailed description of these
effects can be found in \cite{ourpap}.
The quark asymmetries share a
similar freedom from systematic detector effects as is enjoyed by the
lepton asymmetries. Neither a detector asymmetry in $\cos \theta$ nor
charge alone is sufficient to disturb the measurement. Both must
simultaneously be present, and be correlated.

Mass effects are formally of order $m_q^2/s$ as a relative correction to the
asymmetry. Especially for b-quarks they are additionally suppressed by an
accidental cancellation of the mass effect in the numerator and denominator
\cite{Boehm:LEP1YR89VOL1}. The very small residual mass effects are included
in the asymmetry corrections explained in section \ref{sec:hqocor}.

With the availability of beam polarisation, as in the case of the SLD
experiment, the forward-backward-left-right asymmetry can be formed as
\begin{equation}
  \Aqqlr ~=~ \frac{1}{|\Pe|}
            \frac{   ( \sigmaf - \sigmab )_\mathrm{L} -
                     ( \sigmaf - \sigmab )_\mathrm{R}      }
                 {   ( \sigmaf + \sigmab )_\mathrm{L} +
                     ( \sigmaf + \sigmab )_\mathrm{R}      }\, ,
  \label{eq:hq_asylrdef}
\end{equation}
where L,R denote the cross-sections with left- and right-handed
electron beams and $\Pe$ is the beam polarisation.  The more general
Born level differential cross-section with polarised electron beam is
given by
\begin{equation}
  \frac{d \sigma^{\rm q}}{d \cos \theta} ~=~ \frac{3}{8}\sigma^q_{tot} 
  \left[ 
     (1-\Pe \cAe)(1+\cos^2\theta) + 2 (\cAe-\Pe) \cAq  \cos\theta 
  \right]
\label{eq:hq_asylr_sig}
\end{equation}
where the electron beam polarisation $\Pe$ is positive for
right-handed beam. The asymmetry $\Aqqlr$ as a function of polar angle
can therefore be expressed as
\begin{equation}
  \Aqqlr(\cos \theta) ~=~ |\Pe|\cAq \frac{2\cos \theta}{1 + \cos^2 \theta}\,.
\label{eq:hq_asylr_difasy}
\end{equation}  
Comparing Equations~\ref{eq:hq_asylr_difasy} and~\ref{eq:hq_difasy},
it can be seen that $\Aqq$ measures the product of $\cAe\cAq$ while
$\Aqqlr$ measures $\cAq$ directly. $\Aqqlr$ also gives a significant
statistical advantage for sensitivity to $\cAq$ compared to $\Aqq$ by
a factor of $(|\Pe|/\cAe)^2\sim 25$, given a highly polarised electron
beam with $|\Pe|\sim 75\%$. The analysis procedure for $\Aqqlr$ is
otherwise similar to $\Aqq$. As for $\Aqq$ the total tagging
efficiencies and the luminosity cancel in the calculation of $\Aqqlr$,
although one needs to ensure there is no luminosity asymmetry between
the two beam polarisation states by monitoring low angle Bhabhas.
The actual analyses at SLD use a maximum likelihood fit to the
differential cross sections (Equation~\ref{eq:hq_asylr_sig}) in order
to extract $\cAq$. This procedure is equivalent to $\Aqqlr$
(Equation~\ref{eq:hq_asylr_difasy}) at first order, although with
slightly improved statistical precision on $\cAq$ and a very small
dependence on $\cAe$.

As a first step of both asymmetry analyses, the thrust axis of the
event is used to define the quark direction $\theta$, signed by the
charge tagging methods described in the following.  The thrust axis is
stable against infrared and collinear divergences, so that it can be
calculated in perturbative QCD and it is relatively insensitive to
fragmentation effects.

In order to measure a quark asymmetry two ingredients are needed. The
quark flavour needs to be tagged and the quark has to be distinguished
from the antiquark.  For the flavour tagging the methods described in
Sections~\ref{sec:btag} to~\ref{sec:hq_dtag} can be used.  For the
charge tagging essentially five methods have been used, relying on
leptons, D-mesons, jet-charge, vertex-charge and kaons.  Some analyses
also combine the information from the different methods.

In every $\Aqq$ analysis the measured asymmetry is given by
\begin{equation}
A_{\rm FB}^{\rm meas} ~=~ 
\sum_{\rm q} (2 \omega_{\rm q} -1) \eta_{\rm q} \Aqq\,,
\label{eq:afbsum}
\end{equation}
where $\eta_{\rm q}$ is the fraction of $\qq$ events in the sample,
$\omega_{\rm q}$ is the probability to tag the quark charge correctly
and the sum is taken over all quark flavours.  It should be noted that
the tagging methods often tag the quark charge and not the flavour, so
that in these cases $(2 \omega_{\rm q} -1)$ is close to $-1$ for charm
if it is constructed to be positive for b-quarks.  Similar flavour
composition and quark charge tag factors also apply to corresponding
equation for $\Aqqlr$ analyses.

\begin{figure}[p]
\begin{center} 
\includegraphics[width=\linewidth,bb=8 47 555 285]{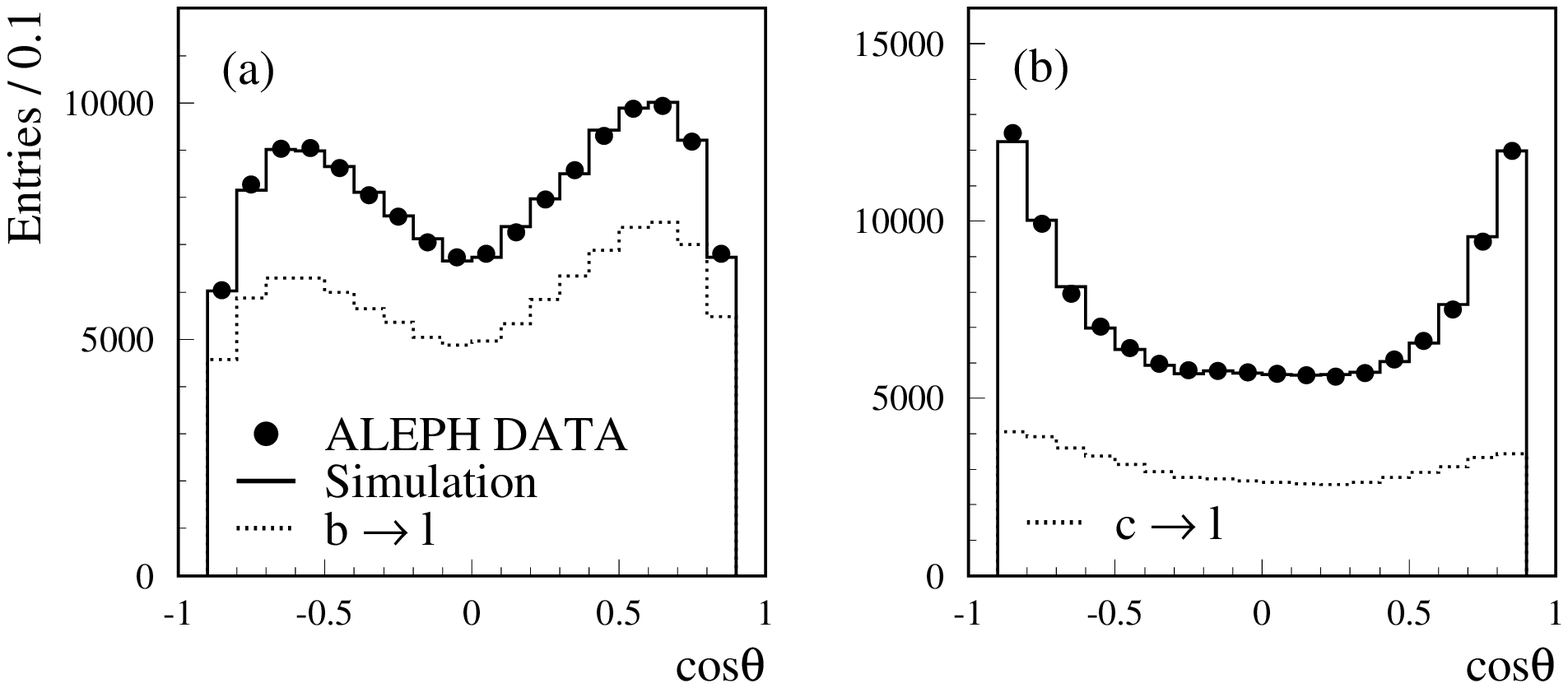}
\end{center}
\caption[$\cos \theta$ distribution from the ALEPH b-asymmetry
measurement with leptons] {Reconstructed $\cos \theta$ distribution
from the ALEPH asymmetry measurements with leptons for a) the
b-enriched and b) the c-enriched sample~\cite{\alasy}.  The full
histogram shows the expected raw angular distribution in the
simulation. The dashed histogram show the signal component.}
\label{fig:hq_asyplot}
\end{figure}

\begin{figure}[p]
\begin{center} 
\includegraphics[width=\linewidth]{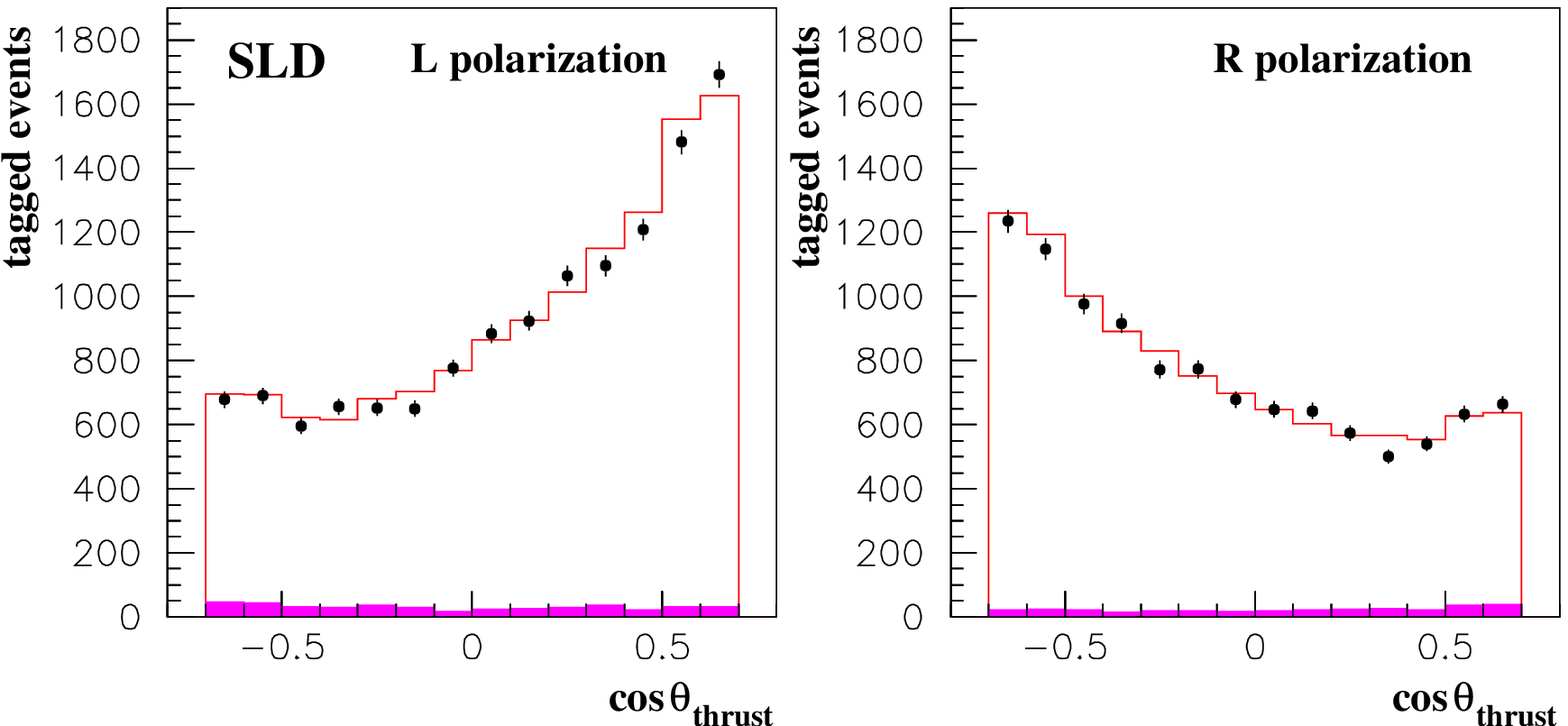}
\end{center}
\caption[Reconstructed $\cos\theta$ distributions from the SLD vertex
charge $\cAb$ analysis.]  {%
Reconstructed $\cos\theta$ distributions
from the SLD vertex charge $\cAb$ analysis for events with left-handed
and right-handed electron beam polarisations. The shaded region
corresponds to udsc background in the sample estimated from Monte
Carlo.  }
\label{fig:hq_asyplot_pol}
\end{figure}

As an example, Figure~\ref{fig:hq_asyplot} shows the reconstructed
$\cos \theta$ distribution from the ALEPH $\Abb$ and $\Acc$
measurement with leptons. The asymmetry of about 10\% for $\Abb$ and
6\% for $\Acc$ can clearly be seen.  An example of the event angular
distributions for the SLD vertex charge $\Aqqlr$ analysis is shown in
Figure~\ref{fig:hq_asyplot_pol}. The much larger forward-backward
asymmetry is a result of the highly polarised electron beam.  The
slightly larger number of events in the left-handed sample is due to
the cross-section asymmetry $\ALR$. The change of asymmetry sign
between the left-handed and right-handed samples, and the slightly
steeper asymmetry in the left-handed sample can be understood from the
proportionality to $(\cAe-\Pe)$ in Equation~\ref{eq:hq_asylr_sig},
dominated by the large $\Pe$.

\subsection{Lepton and D-Meson Measurements}
\label{sec:hq_ldasy}

As described in Sections~\ref{sec:hq_ltag} and~\ref{sec:hq_dtag} the
identification of leptons and D-mesons simultaneously provide flavour
and charge tagging. A simple cut on the lepton transverse momentum
provides good enhancement of ${\rm b} \rightarrow \ell$, as seen in
Figure~\ref{fig:hq_leptspec}. Table~\ref{tab:hq_lsamp} provides an
example of sample compositions for the ALEPH lepton sample with a
transverse momentum cut of $p_t > 1.25\, \GeV$ together with the
correlation between the lepton charge and decaying quark charge.  The
quark charge at production is however the relevant quantity for the
asymmetry determination, requiring correction for the effects of $\BB$
mixing via the integrated mixing parameter $\chiM$.

\begin{table}[t]
\begin{center}
\renewcommand{\arraystretch}{1.1}
\begin{tabular}{|l||c|c|}
\hline
Lepton source & charge correlation & fraction for $p_t > 1.25 \GeV$ \\
\hline
\hline
$\bl,\,\bclm$ & $\phantom{-} 1$ & $0.795$ \\
$\bclp$ & $-1$ & $0.046$ \\
$\cl$ & $\phantom{-} $1 & $0.048$ \\
background & weak & $0.111$ \\
\hline
\end{tabular}
\caption[Correlation between the lepton charge and the quark charge at
decay time.]  {Correlation between the lepton charge and the quark
charge at decay time. The sample composition for $p_t > 1.25 \, \GeV$
in the ALEPH lepton sample is also shown.  }
\label{tab:hq_lsamp} 
\end{center}
\end{table}

To enhance the sensitivity of lepton-based analyses to $\Acc$, the
experiments use additional information like lifetime tagging, jet
charge in the opposite hemisphere or hadronic information from the
lepton jet.  Tagging D-mesons also provides a relatively pure charm
sample after a momentum cut and additional b-tagging requirements are
used to enhance the sensitivity to $\Abb$.

In both cases the sample composition is usually taken from simulation.
For the lepton tag analyses the uncertainties on the sample
composition due to the modelling of the semileptonic decays are
generally rather large.  Therefore, in addition to the asymmetries,
the experiments measure the $\BB$ effective mixing parameter $\chiM$,
the prompt and cascade semileptonic branching fraction of b-hadrons
$\Brbl$ and $\Brbclp$ and the prompt semileptonic branching fraction
of c-hadrons $\Brcl$. If the same analysis cuts are used in both
cases, these auxiliary measurements serve as an effective
parametrisation of the lepton spectrum, greatly reducing the modelling
errors.

In the case of the D-meson analyses the fragmentation function for
D-mesons from b- and c-quarks is measured from data.
However, there is only one important source of D-mesons per quark
flavour, and the correlation between the quark flavour and the D-meson
flavour is the same for b- and c-quarks, so that the sign of the
D-meson asymmetry for the two quark species is the same.  For these
reasons the sensitivity to the sample composition is much smaller than
in the lepton case.

\subsection{Jet and Vertex Charge}

The average charge of all particles in a jet, or jet charge, retains
some information on the original quark charge.  Usually the jet charge
is defined as:
\begin{equation}
  Q_h ~ = ~ \frac{ \sum_i q_i p_{\| i}^\kappa }{ \sum_i p_{\| i}^\kappa }\,,
\label{eq:jetch}
\end{equation}
where the sum runs over all charged particles in a hemisphere with
charge $q_i$ and longitudinal momentum with respect to the thrust axis
$p_{\| i}$, and $\kappa$ is a tunable parameter with typical values
between 0.3 and 1.

For B- and D-mesons the meson charge is correlated with the flavour of
the b- or c-quark. If all charged particles of a jet can be uniquely
assigned to the primary or the decay vertex, the charge sum of the
decay vertex, if non-zero, uniquely tags the quark charge.  At SLD the
$\cAb$ measurement with vertex charge is the most precise measurement
of this quantity.  At LEP the vertex charge has also been used in
conjunction with other tags, however the impact parameter resolutions
at LEP limit the efficiencies in comparison with SLD.

For both charge tagging methods, it is difficult to estimate the
charge tagging efficiency from simulation due to uncertainties from
fragmentation and B-decays. However, the efficiency can be obtained
reliably from data using double tags.  In a cut based analysis,
defining $\omega_q$ as the efficiency to tag the quark charge
correctly in a pure sample of q-quarks, the fraction of same sign
double tags in the sample of all double tags is given by
\begin{equation}
f_{SS} ~=~ 2 \omega_q (1-\omega_q)\,,
\label{eq:hq_sstag}
\end{equation}
apart from small corrections due to hemisphere correlations.
Equation~\ref{eq:hq_sstag} can then be used to obtain $\omega_q$.
Corrections for background and hemisphere correlations are obtained
from simulation.

Since the charge tagging efficiency for the jet charge is rather
modest, a statistical method to extract the asymmetry is usually used.
With $Q_{\rm{F/B}}$ being the jet charge of the forward/backward
hemisphere and $Q_{{\rm q}/\overline{\rm q}}$ the jet charge of the
quark/antiquark hemisphere, one has
\begin{eqnarray}
  \avQfb & = & \langle Q_{{\rm F}} - Q_{{\rm B}} \rangle \label{eq:hf_qfb}\\
         & = & \delta_{\rm q} \Aqq \nonumber \\
  \delta_{\rm q} & = & \langle Q_{\rm q} - Q_{\bar{{\rm q}}} \rangle\,, 
  \nonumber
\end{eqnarray}
for a pure sample of $\qq$-events.  The ``charge separation''
$\delta_{\rm q}$ can be measured from data using:\footnote{ The exact
formulae used by the experiments vary slightly, however the general
formalism is identical.  }
\begin{equation}
\left( \frac{\delta_{\rm q}}{2} \right)^2 ~ = ~
\frac {\langle Q_{\rm{F}} \cdot Q_{\rm{B}} \rangle + 
\rho_{\qq} \sigma(Q)^2 + \mu(Q)^2}{1+\rho_{\qq}},
\label{eq:hf:deltaq}
\end{equation}
where $\mu(Q)$ is the mean value of $Q$ for all hemispheres and
$\sigma(Q)$ is its variance.  $\mu(Q)$ is slightly positive due to an
excess of positive particles in secondary hadronic interactions.  The
hemisphere correlations, $\rho_{\qq}$, arise from charge conservation,
hard gluon radiation and some other small effects and have to be taken
from simulation.

The analyses select a relatively pure sample of $\bb$ events using
lifetime tagging techniques.  Light quark background is always
subtracted using Monte Carlo simulation. The charge separation for
charm is either taken from Monte Carlo or determined by performing the
analysis in bins of different b-purities and fitting $\delta_{\rm b}$
and $\delta_{\rm c}$ from the data.  It should be noted that dilution
due to $\BB$-mixing is completely absorbed into the measured
$\delta_{\rm b}$. Effects from gluon radiation are also included to a
large extent, so that only small QCD corrections have to be applied.

\begin{figure}[htb]
\begin{center} 
\includegraphics[width=0.7\linewidth,bb=30 12 515 345]{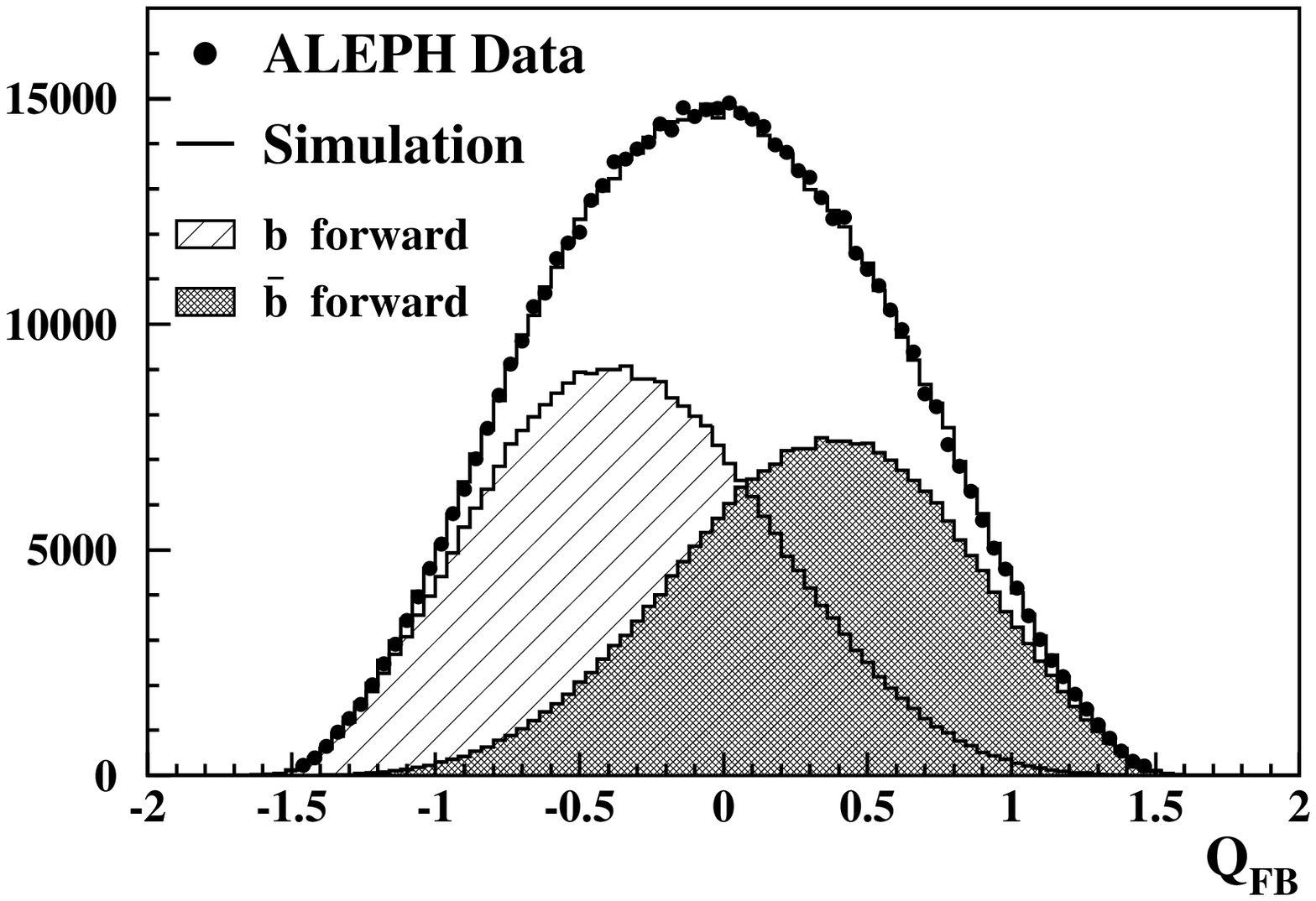}
\end{center}
\caption[Charge separation of the ALEPH neural net tag] {Charge
separation of the ALEPH neural net tag using jet charge, vertex charge
and charged kaons~\cite{ref:ajet}.  The asymmetry reflects $\Abb$
diluted by the non-perfect charge tagging.  }
\label{fig:hq_vtxnn}
\end{figure}

The above formalism can be generalised to any variable sensitive to
the quark charge, including the combination of several different
charge tagging techniques.
As an example Figure~\ref{fig:hq_vtxnn} shows the charge tagging from
ALEPH, which combines jet charge, vertex charge and charged kaon
information using a neural net to reach almost perfect tagging at high
$Q_{\rm FB}$ values.

\subsection{Kaons}

Charged kaons from b- and c-decays are also sensitive to the quark
flavour, via the decay chains ${\rm b} \rightarrow {\rm c} \rightarrow
{\rm s}$ and ${\rm c} \rightarrow {\rm s}$.  Only kaons with large
impact parameters are used, to suppress those produced in the
fragmentation process. As with other methods, the charge tagging
efficiency is measured using the double tag technique.

In the SLD measurements of the asymmetries using kaons, only
identified kaons coming from a secondary vertex are used. The kaon tag
is in fact not used in the $\cAb$ measurement for the bulk of the data
from 1996--1998, since the vertex charge tag dominates hemispheres with a
charged b hadron, while the neutral B mixing significantly limits the
additional contribution from the kaon tag in the remaining
hemispheres. On the other hand, the kaon tag has a good correct quark
charge tag probability of 86\% for a charm hemisphere which is
comparable to the 91\% achieved by the vertex charge tag in the $\cAc$
analysis. They are therefore combined as a joint tag and the joint
correct tag efficiency is calibrated from the data for the SLD $\cAc$
measurement.

\subsection{Asymmetry Measurements used in the Combination}

The forward-backward asymmetry measurements included in the average
are:
\begin{itemize}
\item Measurements of $\Abb$ and $\Acc$ using leptons from
  ALEPH~\cite{\alasy}, DELPHI~\cite{\dlasy}, L3~\cite{\llasy} and
  OPAL~\cite{ref:olasy}: L3 measures $\Abb$ only from a sample of high
  $p_t$ leptons.  ALEPH, DELPHI and OPAL measure $\Abb$ and $\Acc$
  using leptons combined with lifetime tagging and some additional
  information.  ALEPH adds properties of hadrons in the events and
  information from the missing energy due to escaping neutrinos. OPAL
  also uses hadronic properties while DELPHI includes the jet charge
  of the hemisphere opposite the lepton.
\item Measurements of \Abb{} based on lifetime tagged events with a
  jet charge measurement using the weight method (see
  Equation~\ref{eq:hf_qfb}) from ALEPH~\cite{ref:ajet},
  L3~\cite{ref:ljet} and OPAL~\cite{ref:ojet}. ALEPH and OPAL combine
  their jet charge with additional information like vertex charge and
  kaons.  The DELPHI analysis of $\Abb$, also combining jet charge,
  vertex charge, kaons and some other variables sensitive to the
  b-quark charge in a neural net~\cite{ref:dnnasy}, is based on a cut
  on the charge estimator.
\item Analyses with D-mesons from ALEPH~\cite{ref:adsac},
  DELPHI~\cite{ref:ddasy} and OPAL~\cite{ref:odsac}: ALEPH measures
  $\Acc$ only from a sample of high momentum D-mesons.  DELPHI and
  OPAL measure $\Abb$ and $\Acc$ by fitting the momentum spectrum of
  the D-mesons and including lifetime information.
\end{itemize}
The left-right-forward-backward asymmetry measurements from SLD are
directly quoted in terms of $\cAb$ and $\cAc$. The following results
are included:
\begin{itemize}
\item Measurements of \cAb{} and \cAc{} using
  leptons~\cite{\SLDacbl};
\item A measurement of $\cAc$ using D-mesons~\cite{ref:SLD_ACD};
\item A measurement of $\cAb$ using jet charge~\cite{\SLDabj};
\item A measurement of $\cAb$ using vertex
  charge~\cite{ref:SLD_vtxasy};
\item A measurement of $\cAb$ using kaons~\cite{\SLDabk};
\item A measurement of $\cAc$ using vertex charge and
  kaons~\cite{ref:SLD_vtxasy}.
\end{itemize}
All these measurements are listed in detail in
Appendix~\ref{sec:hqappendix}.

\section{Auxiliary Measurements}
\label{sec:hqaux}

The measurements of the charmed hadron fractions $\PcDst$, $\fDp$,
$\fDs$ and $\fcb$ are included in the \Rc{} analyses and are described
there.

ALEPH~\cite{ref:abl,\alasy}, DELPHI~\cite{ref:dbl},
L3~\cite{ref:lbl,ref:lrbmixed,\llasy} and OPAL~\cite{ref:obl,ref:olasy}
measure $\Brbl$, $\Brbclp$ and $\chiM$ or a subset from a sample of
leptons opposite a b-tagged hemisphere and from a double lepton
sample.  DELPHI~\cite{ref:drcd} and OPAL~\cite{ref:ocl} measure $\Brcl$
from a sample opposite a high energy $\Dstarpm$.  All the auxiliary
measurements used in the combination are listed in
Appendix~\ref{sec:hqappendix}.

\section{External Inputs to the Heavy Flavour Combination}
\label{sec:hqinputs}

All the measurements contributing to the heavy flavour combination
require some input from simulated events.  Quantities derived from the
simulation are affected by uncertainties related to the modelling of
the detector response, as well as by the limited knowledge of the
physics processes that are simulated. These latter sources are common
to all experiments, and they have to be treated as correlated when
averaging individual results. Furthermore, in order to produce
consistent averages, the external physics parameters or models used in
the simulation must be the same for all analyses in all experiments.

The choice of the external physics parameters and models relevant for
electroweak heavy flavour analyses is discussed below.  Whenever
possible measurements at LEP/SLD or at lower energies are used to
constrain the models used in the simulations.  The uncertainties on
the fitted partial widths and asymmetries due to the knowledge of the
external parameters can be seen from Table~\ref{tab:hferrbk}. If a
parameter does not appear in Table~\ref{tab:hferrbk} the error is
negligible either because the parameter is relatively unimportant or
because it is known very precisely.

In many cases the world averages of the Particle Data Group are used.
They are consistently taken from the 1998 edition of the
RPP~\cite{PDG98}.  It has been checked that updates published in the
2004 edition~\cite{PDG2004} do not change any of the results.
Table~\ref{tab:hfextpar} summarises the important external parameters
used.  Details of their choice are explained in the remainder of this
section

\begin{table}[htbp]
\begin{center}
\renewcommand{\arraystretch}{1.2}
\begin{tabular}{|l||l|}
\hline
Error Source & Used Range  \\ 
\hline
\hline
$\meanxb$    &  $ 0.702 \pm 0.008$  \\
$\meanxc$    &  $0.484 \pm 0.008$ \\
Choice of b fragmentation function & See sec.~\ref{sec:hqbfrag} \\
Choice of c fragmentation function & See sec.~\ref{sec:hqbfrag} \\
\hline
$\BR(\bclm)$    &  $(1.62 \, ^{+ 0.44} _{- 0.36})$\%  \\
$\BR(\btaul)$   &  $(0.419\pm0.055)$\% \\
$\BR(\bpsill)$  &  $(0.072\pm 0.006)$\%  \\
Semilept.  model $\bl$  
              & ACCMM ($\mathrm{^{+ISGW}_{-ISGW**}}$) 
              (sec.~\ref{sec:hqldmod}) \\
Semilept.  model $\cl$  
              & ACCMM1 ($\mathrm{^{+ACCMM2}_{-ACCMM3}}$)
              (sec.~\ref{sec:hqldcmod}) \\
$\bD$ model & Peterson $\epsilon = 0.42\pm0.07$  \\
\hline
$\Dzero$ lifetime & $0.415\pm0.004$ ps \\
$\Dplus$ lifetime & $1.057\pm0.015$ ps \\
$\Ds$ lifetime    & $0.467\pm0.017$ ps \\
$\Lc$ lifetime    & $0.206\pm0.012$ ps \\
B lifetime        & $1.576 \pm 0.016$ ps \\
\hline
$  \BR({\rm D^0\to K^- \pi^+}) $&$ 0.0385 \pm 0.0009$\\
$  \BR({\rm D^+ \to  K^- \pi^+ \pi^+}) $&$ 0.090 \pm 0.006$\\
$  \BR({\rm D_s^+ \to \phi \pi^+}) $&$ 0.036 \pm 0.009 $\\
$  \frac{{\BR\rm(D_s^+ \to {\overline K}^{\star 0} K^+)}}
   {\BR({\rm D_s^+ \to \phi \pi^+)}} $&$ 0.92 \pm 0.09 $\\
$  \BR({\rm \Lambda_c \to p K^- \pi^+ }) $&$ 0.050 \pm 0.013$\\
\hline
B charged decay multiplicity & $4.955\pm0.062$ \\
D charged decay multiplicity & See sec.~\ref{sec:hq_cmult} \\
D neutral decay multiplicity & See sec.~\ref{sec:hq_cnmult} \\
\hline
$\mathrm{g} \ra \cc$  per multi-hadron 
                  & $(2.96 \pm  0.38 ) \% $ \\
$\mathrm{g} \ra \bb$  per multi-hadron 
                  & $(0.254 \pm  0.051 ) \% $ \\
\hline
Rate of long-lived light hadrons & Tuned JETSET$\pm10\%$ 
(sec.~\ref{sec:hqlighth}) \\
Light quark fragmentation & See sec.~\ref{sec:hqfrag}\\
QCD hemisphere correlations & See sec.~\ref{sec:hq_rbhemcol} \\
\hline
\end{tabular}
\end{center}
\caption{The most important external parameters used in the heavy
flavour analyses}
\label{tab:hfextpar}
\end{table}

\boldmath
\subsection{Fragmentation of Heavy Quarks}
\label{sec:hqbfrag}
\unboldmath

The process of hadron production is modelled as the convolution of a
perturbative part (hard gluon radiation), and a non-perturbative part,
called fragmentation, described with phenomenological models.

In the JETSET~\cite{JETSET} simulation the fragmentation model by
Peterson~\etal~\cite{pet} is used, which describes the process in terms
of the variable $z={(E+p_\parallel)}_{\mathrm{hadron}} /
{(E+p_\parallel)}_{\mathrm{quark}}$, where $p_\parallel$ is the
momentum component in the direction of the fragmenting quark.  The
model contains one free parameter, $\epsQ$, which is tuned to
reproduce a given value of the mean energy of the heavy hadrons
produced. Such tuning depends on the cut-off used for the transition
between the perturbative and the non-perturbative part, therefore
$\epsQ$ can not be given an absolute meaning.  The energy spectrum is
more conveniently described in terms of the variable $x_Q$, defined as
the energy of the weakly-decaying hadron containing the heavy quark
$Q$ normalised to the beam energy.

The analyses quoted in Reference
\mcite{epsb_results_1,epsb_results_2,epsb_results_3,
epsb_results_4,epsb_results_5}, provide values for the mean energy of
weakly-decaying b hadrons, which are averaged to obtain:
\begin{equation}
\meanxb ~=~ 0.702 \pm 0.008  \ \, 
\end{equation}
where the error includes the uncertainty due to the modelling of the
fragmentation function. This uncertainty is estimated by using the
functional forms proposed by Collins and Spiller~\cite{collins}, and
by Kartvelishvili~\etal~\cite{kart} as alternatives to
Peterson~\etal~\cite{pet} when extracting $\meanxb$. Only analyses which
are close to the heavy flavour analyses, especially with leptons, are
used in the average. This ensures a consistent treatment of
B-fragmentation leading to some error cancellations.

The energy of charmed hadrons is measured in analyses which make use
of lepton tags or inclusive reconstruction of
$\Dzero/\Dplus$-mesons~\cite{epsb_results_2,ref:orcc}, and in analyses
with full reconstruction of
$\Dstarp$-mesons~\cite{epsc_dstar_1,epsc_dstar_2,epsc_dstar_3}.  The
former have a larger dependence on the modelling of the spectrum,
while the latter need an additional correction to obtain the energy of
the weakly-decaying hadron. In all cases the contribution from charmed
hadrons produced by hard gluons splitting to heavy quarks is removed.
The average energy of weakly-decaying charmed hadrons is found to be
\begin{equation}
\meanxc ~=~ 0.484 \pm 0.008  \ \, 
\end{equation}
which again includes the estimated uncertainty from the modelling of
the spectrum.

\boldmath
\subsection{Heavy Quarks from Gluon Splitting}
\unboldmath

Gluons can occasionally split to heavy quark pairs.  In several
analyses these contributions need to be subtracted. In particular the
uncertainty on the rate of gluons splitting to $\bb$ pairs is the
single largest contribution to the systematic error on the $\Rb$ world
average.

The rates $\gcc$ and $\gbb$ are defined as the number of hadronic Z
decays containing a gluon splitting to a $\cc$ or $\bb$ pair,
normalised to the total number of hadronic Z decays.

Measurements of $\gbb$~\cite{\GBBmeas} rely on an inclusive
lifetime-based tag applied to the jets reconstructed in the event,
while measurements of $\gcc$~\cite{ref:arcc,gcc_l,gcc_o} make use of
exclusive $\Dstar$ reconstruction, final states containing leptons, or
are based on the combination of event shape variables.

Averaging published results yields:
\begin{eqnarray}
  \gcc  &=&  0.0296 \pm 0.0038 \, , \\
  \gbb & = & 0.00254 \pm 0.00051 \, , \nonumber
\end{eqnarray}
with only a very small correlation between the two values \cite{gsplit}.

\boldmath
\subsection{Multiplicities in Heavy Flavour Decays}
\unboldmath
\label{sec:hq_hqmult}

Many analyses make use of inclusive b tagging methods which exploit
the long lifetimes of b hadrons.  The discrimination is based on the
presence, in a jet, or a hemisphere, or the whole event, of charged
tracks with significant impact parameter from the primary vertex of
the events. Therefore the tagging efficiency is directly affected by
the number of charged tracks produced in the long-lived hadron decay.

In $\Rb$ measurements the tag is applied to hemispheres, and the b
efficiency is measured directly in the data from the fraction of
events with both hemispheres tagged.  The b charged multiplicity only
enters, as a simulation uncertainty, through the hemisphere
correlation.  Measurements of the average b charged multiplicity
performed at LEP are used. Results from lower energy experiments
cannot be used because of the different b-hadron mixture.

However, the charm selection efficiency is taken from the simulation,
at least for the samples with highest purity. It is therefore crucial
to propagate correctly the uncertainty due to the decay charged
multiplicities of the various charmed hadrons. This is done separately
for each hadron species due to the significant differences in
lifetimes.

The charm selection efficiency also depends on the number of neutral
particles accompanying the charged particles in a given topological
decay channel. The size of this effect depends on how invariant mass
cuts are implemented and might vary substantially in different
analyses. The uncertainty is evaluated varying the ${\rm K}^0$ and
$\pi^0$ production rates in charmed hadron decays.

\boldmath
\subsubsection{Average Charged Multiplicity in b Hadron Decays}
\unboldmath
\label{sec:hq_bmult}

Inclusive measurements of the mean b-hadron charged multiplicity at
LEP~\cite{\bmulti} are combined to obtain:
\begin{equation}
\langle n_b^{\mathrm ch} \rangle ~=~ 4.955 \pm 0.062\ . 
\end{equation}
Particles coming from the decay of ${\rm B}^{\star\star}$ or other
possible excited b states are excluded; the result is also corrected
to exclude charged particles originating from the decay of ${ \rm
K}^0$ and $\Lambda$.

\boldmath
\subsubsection{Charged Multiplicities of c Hadron Decays}
\unboldmath
\label{sec:hq_cmult}

Inclusive topological branching fractions have been measured for
$\Dzero$, $\Dplus$ and $\Ds$~\cite{ctopo}.  For each species, each
channel is varied within its uncertainty, except for the channel with
the highest rate, which is used to compensate the variation.  The
resulting errors are combined using the corresponding correlation
coefficients.  The values $f_i$ of the branching fractions for the
decays into $i$ charged particles, the corresponding errors $\sigma_i$
and correlation coefficients $C_{ij}$ are given in
Table~\ref{tab.topo}.  For charm baryons, for which no measurements
are available, an uncertainty of $\pm 0.5$ in the overall charged
decay multiplicity was used.

\begin{table}[htbp]
\begin{center}
\renewcommand{\arraystretch}{1.2}
\begin{tabular}{|l||l|l|l|l|}
\hline
D meson &  \multicolumn{4}{|c|}{Topological Decays}\\
\hline
\hline
        &$f_0 = 0.054$   &$f_2 =0.634$  &  $f_4 =0.293$  &$f_6 = 0.019$    \\
$\Dzero$&$\sigma_0=0.011$&              &$\sigma_4=0.023$&$\sigma_6=0.009$ \\
        &$C_{04} = 0.07$ &              &$C_{46}=-0.46$  &$C_{06} = 0$ \\
\hline
        &$f_1 = 0.384$   &$f_3 = 0.541$ &$f_5 =0.075$     & \\
$\Dplus$&$\sigma_1=0.023$&              &$\sigma_5 =0.015$& \\
        &$C_{15} = -0.33$&              &                 & \\
\hline
        &$f_1 = 0.37$    &$f_3 = 0.42$  &$f_5 =0.21$      &\\
 $\Ds$  &$\sigma_1=0.10$ &              &$\sigma_5 =0.11$ &\\
        &$C_{15}=-0.02$  &              &                 &\\
\hline
\end{tabular}
\caption[Topological rates for the different charm-meson species.]
{Topological rates for the different charm-meson species, with
estimated errors and correlation coefficients. The subscripts indicate
the number of charged particles produced.}
\label{tab.topo}
\end{center}
\end{table}

\boldmath
\subsubsection{Neutral Particle Production in c Hadron Decays}
\label{sec:hq_cnmult}
\unboldmath

The procedure to estimate the residual dependence of the lifetime tag
efficiency on the average rate of neutral particles produced in charm
decays is tailored, case by case, on the specific properties of the
tag and based on the measurements available~\cite{PDG98}.
Although the procedures differ somewhat between experiments, the
resulting estimated uncertainties are taken as fully correlated.

\boldmath
\subsection{Heavy Flavour Lifetimes}
\unboldmath

The lifetimes of heavy hadrons are relevant to many analyses, in
particular all those which make use of lifetime-based b tagging
methods.  As for the charged multiplicity, in the case of the $\Rb$
analyses charm lifetimes enter directly in the estimate of the charm
contamination in high purity samples, whereas b hadron lifetimes only
affect the estimate of the hemisphere correlations.

\boldmath
\subsubsection{Average b Hadron Lifetime}
\unboldmath

The average lifetime of b hadrons is taken~\cite{PDG98} to be
\begin{equation}
\tau_b ~=~ 1.576 \pm 0.016\ {\mathrm{ps}}\ , 
\end{equation}
which is obtained from analyses of fully inclusive b final states.
The lifetime difference between b species has in general little impact
in all analyses.  It is considered as a source of uncertainty in the
$\Rb$ analyses either by using the individual lifetimes~\cite{PDG98} or
by enlarging the error to $0.05\ {\mathrm{ps}}$.

\boldmath
\subsubsection{Lifetimes of c Hadrons}
\unboldmath

The lifetimes of the different c hadron species are considered as
individual sources of uncertainties.  The values and
errors~\cite{PDG98} are:
\begin{eqnarray}
  \tau({\rm D}^0) &=& 0.415 \pm 0.004 \ {\mathrm{ps}} \ , \\
  \tau({\rm D}^+) &=& 1.057 \pm 0.015 \ {\mathrm{ps}} \ , \nonumber \\
  \tau({\rm D_s}) &=& 0.467 \pm 0.017 \ {\mathrm{ps}} \ , \nonumber \\ 
  \tau(\Lambda_c^+) &=& 0.206 \pm 0.012 \ {\mathrm{ps}} \ \nonumber .
\end{eqnarray}

\boldmath
\subsection{Charmed Hadron Decays to Exclusive Final States}
\unboldmath

Charm counting measurements determine the production rates of
individual c-hadron species by tagging exclusive final states, using
the branching fraction for the appropriate decay mode as input.  The
values and errors used are~\cite{PDG98,cleo_lc}:
\begin{eqnarray}
  \BR({\rm D^0\to K^- \pi^+}) &=& 0.0385 \pm 0.0009 \ , \\
  \BR({\rm D^+ \to  K^- \pi^+ \pi^+}) &=& 0.090 \pm 0.006 \ , \nonumber \\
  \BR({\rm D_s^+ \to \phi \pi^+}) &=& 0.036 \pm 0.009 \ , \nonumber \\ 
  \frac{{\BR\rm(D_s^+ \to {\overline K}^{\star 0} K^+)}}
  {\BR({\rm D_s^+ \to \phi \pi^+})} &=& 0.92 \pm 0.09 \ , \nonumber \\ 
  \BR({\rm \Lambda_c \to p K^- \pi^+ }) &=& 0.050 \pm 0.013 \  .\nonumber 
\end{eqnarray}

\boldmath
\subsection{Heavy Flavour Leptonic Decays}
\unboldmath

Many analyses rely on semileptonic final states in order to tag the
presence of heavy hadrons and possibly their charge.  Assessing the
performance of such tags involves estimating the rates of the
different sources of lepton candidates in hadronic events, and
modelling the kinematics of the leptons produced in the decay of heavy
hadrons.

The rates for the major sources (direct decays, $\bl$ and $\cl$,
cascade b decays, $\bcl$) are measured at LEP, and included as fitted
parameters. The modelling of the decay kinematics is a common source
of systematic uncertainty. The rates for the other sources are taken
from external measurements.

\boldmath
\subsubsection{Modelling of Direct Semileptonic b Decays}
\label{sec:hqldmod}
\unboldmath
 
For the semileptonic decays of $\Bzero$ and $\Bplus$ mesons the CLEO
collaboration has compared decay models to their data and measured the
free parameters of the models.  Based on the CLEO fits~\cite{cleospe},
the LEP experiments quote results for three different models.
\begin{itemize}
\item The model proposed by Altarelli~\etal~\cite{accmm} is an
  extension of the free quark model which attempts to account for
  non-perturbative effects kinematically. The two free parameters of
  the model, the Fermi momentum of the constituent quarks inside the
  heavy meson and the mass of the final quark, are determined from
  CLEO data to be $\pF= 298\ \MeV$, $\mc = 1673\ \MeV$.
\item The form-factor model proposed by Isgur~\etal~\cite{isgw}, with
  the model prediction that $11 \%$ of semileptonic B meson decays
  result in an L=1 charm meson, ${\rm D}^{\star\star}$.
\item The same model with the rate of ${\rm D}^{\star\star}$ mesons
  increased to $32 \%$, as preferred by the CLEO
  data~\cite{cleospe,isgw}.
\end{itemize}
The model of Altarelli~\etal\ is used to derive the central values of
the analyses, while the two others, which give respectively harder and
softer lepton spectra, are used to give an estimate of the associated
uncertainty.

Reweighting functions are constructed to adjust the lepton spectrum of
semileptonic $\Bzero$ and $\Bplus$ decays in the LEP Monte Carlo
samples to the three models based on CLEO data.  For use in Z decays,
the same reweighting functions have been assumed to be valid for the
$\Bs$ meson and b baryons.  This would be correct in the simplest
spectator model, and is thought more generally to be adequate for the
$\Bs$. The baryon contribution is only about 10\%, and no additional
systematic error is assigned.

\boldmath
\subsubsection{Modelling of Direct Semileptonic c Decays}
\label{sec:hqldcmod}
\unboldmath
  
The measurements of DELCO~\cite{delcocl} and MARK III~\cite{mark3cl} for
$\Dzero$ and $\Dplus$ semileptonic decays have been combined and
parametrised using the model of Altarelli~\etal\ as a convenient
functional form.  The D boost and the experimental resolution are
taken into account in the fit to the data.  Based on this
fit~\cite{ref:lephf}, the model parameters are fixed to $\pF = 467\
\MeV$, $\ms = 1\ \MeV$ and they are varied to $\pF = 353\ \MeV$, $\ms
= 1\ \MeV$ and $\pF = 467\ \MeV$, $\ms = 153\ \MeV$ to derive an
estimate of the associated uncertainty.  The reweighting functions
derived from $\Dzero$ and $\Dplus$ decays are assumed to be valid for
all charm hadrons.

\boldmath
\subsubsection{Modelling of Cascade Semileptonic b Decays}
\unboldmath
 
For the cascade decays, $\bcl$, the three models used for $\cl$ decays
are combined with the measured $\bD$ spectrum from CLEO~\cite{cleobd}
to generate three models for the lepton momentum spectrum in the rest
frame of the b hadron.  The CLEO $\bD$ decay spectrum can be conveniently
modelled by a Peterson function~\cite{pet} with free parameter
$\varepsilon = 0.42 \pm 0.07$.  The effect of this $\bD$ model
uncertainty on the $\bcl$ spectrum is negligible compared to the
uncertainty from the $\cl$ models.

\boldmath
\subsubsection{Rate of $\bcbarl$ Transitions}
\unboldmath

Several quantities related to the rate of leptons from c hadrons
produced from the ``upper vertex'' in b-hadron decays\footnote{ The
term ``upper vertex'' is used in the literature for the decay of the
virtual W from the b-quark decay.}  have been measured. An estimate of
this rate is therefore possible, based upon experimental results.

The inclusive and flavour-specific ${\rm B}\to {\rm D,X}$ and ${\rm
  B}\to \lamc,{\rm X}$ rates measured at CLEO~\cite{\cleoupver}, which
  are sensitive to the sum $({\rm B} \to {\rm c}) + ({\rm B} \to
  \overline{{\rm c}})$, are combined with the ${\rm B}\to {\rm D
  \overline{D} (X)}$ rates measured in ALEPH~\cite{aleph2d} to extract
  the probabilities of producing the different c-hadrons from the
  upper vertex in b decays.  These are combined with the c-hadron
  semileptonic branching fractions to obtain a value for the
  $\BR(\bcbarl)$.

The estimate obtained is
\begin{equation}
  \BR(\bcbarl) ~=~   0.0162 \, ^{+ 0.0044} _{- 0.0036}  \, .
\end{equation}

\subsubsection{Other Semileptonic Decays}

The rate for $\btaul$ decays is derived from existing measurements of
$\BR(\btau)$~\cite{\btotau} combined with the $\tau$ leptonic
branching fraction~\cite{PDG98}. The procedure yields:
\begin{equation}
\BR(\btaul) ~=~ 0.00419 \pm 0.00055 \ .
\end{equation}

The rate for $\bpsill$ decays is calculated from the production rate
of $\jpsi$ and $\psipr$ in $\Ztobb$ decays, and the $\jpsi$ and
$\psipr$ leptonic branching fractions~\cite{PDG98}, yielding
\begin{equation}
\BR(\bpsill) ~=~ 0.00072 \pm 0.00006 \ . 
\end{equation}

\boldmath
\subsection{Hemisphere Correlations in Double-Tag Methods}
\label{sec:hq_rbhemcol}
\unboldmath

In analyses where a b-tagging algorithm is applied in one hemisphere,
the tagging efficiency can be measured from the data by comparing the
fraction of hemispheres that are tagged and the fraction of events
with both hemispheres tagged. However, the correlation between the
tagging efficiencies in the two hemispheres, defined in
Equation~\ref{eq:hq_hcor}, must then be estimated from simulation.
This is particularly crucial for the precise $\Rb$ double tag
measurements.

There are basically three physics sources for such a correlation:
\begin{itemize}
\item detector inhomogeneities,
\item the use of a common primary vertex,
\item kinematic correlations, mainly due to gluon radiation.
\end{itemize}
Detector effects are easily controllable from the data by measuring
the tagging rate as a function of the jet direction and then
calculating the correlation from this rate assuming that the quarks in
an event are back-to-back.  This error source is of statistical nature
and uncorrelated between the experiments.

The second of these sources is relatively small for algorithms based
on the reconstruction of the b decay length, since this is dominated
by the uncertainty on the position of the secondary vertex. However,
it is a major issue for tags based on track impact parameters, and it
is particularly difficult to control since it heavily influences the
other sources. Therefore in the $\Rb$ analyses
the primary vertex is generally reconstructed independently in the two
hemispheres, rendering this source of correlation negligible.

The kinematic correlations are correlated between the experiments.
They mainly arise from the fact that the tagging efficiency depends on
the b hadron momentum and that a gluon emitted at a large angle
reduces the energy of both quarks.

If the efficiency is proportional to the b hadron momentum, the
efficiency correlation is directly given by the momentum correlation.
Analytic $\calO(\alfas)$ QCD calculations predict effects of
about $1.4~\%$~\cite{paolo} for the correlation between the two
b-quark momenta.  At the parton level, fragmentation models agree at
the $0.2~\%$ level with this number. At the hadron level
HERWIG~\cite{HERWIG} gives a correlation up to 0.8~\% larger than
JETSET or ARIADNE.

Since the proportionality between the B momentum and the tagging
efficiency is only approximate, in practice the experiments have
derived test quantities that are sensitive to the kinematical
correlations and the systematic uncertainties are derived from
data/Monte Carlo comparisons. These methods are described in detail in
the experimental papers.  As an example the momentum of the fastest
jet, assuming a three-jet topology, can be calculated and the tagging
rate for the hemisphere containing this jet and for the opposite
hemisphere are measured.  Although these errors have a large
statistical component, they are conservatively taken as fully
correlated between the experiments.

Events where the radiated gluon is so hard that the two b hadrons are
in the same hemisphere are particularly relevant for the estimate of
the correlation. The rate of such events (about 1~\% of all $\rm\Ztobb$
events) is varied by $30-40~\%$, motivated by a comparison of matrix
element and parton shower models, and by studies of the modelling of
events with two b-tags in the same hemisphere.

Furthermore, the hemisphere correlation also depends on b hadron
production and decay properties. Such a dependence is a small second
order effect for analyses which reconstruct the primary vertex
independently in the two hemispheres, but can be substantial if a
common primary vertex is used, due to the inclusion of tracks which
actually come from b hadrons in the primary vertex determination. The
sources of uncertainty considered are:
\begin{itemize}
\item average charged track multiplicity in b-hadron decay,
\item b fragmentation,
\item b hadron lifetimes,
\end{itemize}
and the errors are evaluated according to the prescription in this
section.

\boldmath
\subsection{Light Quark Background in Lifetime Tagged Samples}
\unboldmath

The amount of light quark background in lifetime-tagged samples is
mainly determined by the rate of long-lived light hadrons, namely
${\rm K}^0_s$ and $\Lambda$, produced in the fragmentation. This is
only a significant source of uncertainty for the precise $\Rb$
measurements.  In the case of forward-backward asymmetry measurements,
details of light quark fragmentation are relevant in the extraction of
the asymmetry from the measured charge flow.

\subsubsection{Rate of Long-Lived Light Hadrons}
\label{sec:hqlighth}
All experiments have measured the rates of long-lived light hadrons
and tuned their fragmentation model accordingly. Variations of $10~\%$
around the central value are used to estimate the uncertainty.

\subsubsection{Fragmentation of Light Quarks}
\label{sec:hqfrag}

The JETSET model contains many free parameters, several of which
influence the charge flow predictions. These parameters have been
tuned individually by the experiments and it is not possible to define
a common procedure to evaluate the errors due to light quark
fragmentation. Fortunately these errors turn out to be relatively
small, and they are assumed to be fully correlated even if the
procedures to evaluate them vary between the experiments.

\section{Corrections to the Electroweak Observables}
\label{sec:hqcorr}

\boldmath
\subsection{Corrections to \Rb\ and \Rc}
\unboldmath

Small corrections have to be applied to the raw experimental
measurements.  The \Rb{} and \Rc{} analyses measure the ratio of
production cross-sections $R_{\rm{q}} =
\sigma_{\rm{q\bar{q}}}/\sigma_{\rm{had}}$. To obtain the ratios of
partial widths $R_{\rm{q}}^0 =
\Gamma_{\rm{q\bar{q}}}/\Gamma_{\rm{had}}$, small corrections for
photon exchange and $\gammaZ$ interference have to be applied.  These
corrections are typically $+0.0002$ for \Rb{} and $-0.0002$ for \Rc,
and are applied by the experiments before the combination as their
size depends slightly on the invariant mass cutoff of the
$\rm{q\bar{q}}$-system imposed in the analysis.

\boldmath
\subsection{QCD Corrections to the Forward-Backward Asymmetries}
\label{sec:hf_qcdcor}
\unboldmath

Due to QCD effects the measured forward-backward asymmetries do not
correspond to the underlying quark asymmetries on the electroweak
level.  The dominant corrections are due to radiation of gluons from
the final state quarks. The QCD corrections do not depend on the beam
polarisation and are thus identical for the unpolarised
forward-backward asymmetry and the left-right-forward-backward
asymmetries.  All statements on $\Aqq$ in this section equally apply
to $\Aqqlr$.

Theoretical calculations use either the quark direction or the thrust
direction to compute the asymmetry.  In case the thrust direction is
used, the thrust axis is signed by the projection of the quark
direction on this axis.  Since the reconstructed thrust axis is
generally used as the heavy quark direction estimator in experimental
measurements, calculations based on the thrust axis are considered.

The effect on the asymmetry at the scale $\mu^2 = \MZ^2$ is
parametrised as~\cite{ourpap}:
\begin{eqnarray}
  \left(\Aqq \right)_{\rm meas} & = &
(1-C_{\mathrm{QCD}}) \left(\Aqq \right)_{\rm no \, QCD} \\
 & = &
  \left( 1 - \frac{\alpha_s(\MZ^2)}{\pi} c_1 -   
    \left( \frac{\alpha_s(\MZ^2)}{\pi} \right)^2 c_2 \right)
  \left(\Aqq \right)_{\rm no \, QCD}  \,. \nonumber 
\end{eqnarray}
The first-order corrections are known including mass
effects~\cite{lampe}.  Taking the thrust axis as the direction and
using the pole mass, they are $c_1=0.77$ for $\Abb$ and $c_1=0.86$ for
$\Acc$.

The second-order corrections have been recalculated in~\cite{neerv}
and~\cite{sey} and both calculations agree well if the quark direction
is used. However only the latter contains also the case where the
thrust axis is used as a reference so that this one is used to correct
the LEP and SLD measurements. The two calculations disagree with
previous results~\cite{lampe}, however there is a general consensus
that the newer ones, which are in agreement amongst each other, should
be trusted.  The calculation of~\cite{sey} is strictly massless and
also neglects the corrections from triangle diagrams involving top
quarks, given in~\cite{lampe}. Corrections arising from diagrams which
lead to two-parton final states are the largest, and they can be added
to the results of~\cite{sey}, as they apply in the same way to
calculations based either on the thrust or the quark direction.

The second order coefficients used are $c_2 = 5.93$ for $\Abb$ and
$c_2 = 8.5$ for $\Acc$. The final QCD correction coefficients,
including further corrections due to fragmentation effects
and using the thrust axis as reference
direction ($C_{\mathrm{QCD}}^{\mathrm{had,T}}$), are
$C_{\mathrm{QCD}}^{\mathrm{had,T}} = 0.0354 \pm 0.0063$ for $\Abb$ and
$C_{\mathrm{QCD}}^{\mathrm{had,T}} = 0.0413 \pm 0.0063$ for $\Acc$.
The breakdown of the errors is given in Table~\ref{tab:qcderr}.

\begin{table}[htbp]
\begin{center}
\renewcommand{\arraystretch}{1.2}
\begin{tabular}{|lc||l|l|}
\hline
\multicolumn{2}{|l||}{Error on $C_{\mathrm{QCD}}^{\mathrm{had,T}}$
 \rule{0pt}{12pt} }
& \multicolumn{1}{c|}{$\bb$} & \multicolumn{1}{c|}{$\cc$} \\ 
\hline
\hline
Higher orders              &\cite{sey}    & 0.0025& 0.0046\\
Mass effects               &\cite{ourpap} & 0.0015& 0.0008\\
Higher order mass          &\cite{sey}    & 0.005 & 0.002\\
$\alpha_s = 0.119\pm 0.003$&              & 0.0012& 0.0015\\
Hadronisation              &\cite{ourpap} & 0.0023& 0.0035\\
\hline Total               &              & 0.0063& 0.0063\\
\hline
\end{tabular}
\caption{Error sources for the QCD corrections to the forward-backward
  asymmetries.}
\label{tab:qcderr}
\end{center}
\end{table}

The procedure to implement QCD corrections in the experimental
analyses is non-trivial. It is described in detail in~\cite{ourpap} and
briefly summarised in the following.

The corrections provided by theoretical calculations are not directly
applicable to experimental measurements for two main reasons.  First,
the thrust axis used in theoretical calculations is defined using
partons in second order QCD, where the axis is signed by the
projection of the b-quark on the thrust axis; a further smearing is
caused by the hadronisation of partons into hadrons. This effect,
about ten times smaller than the correction itself, is taken from the
simulation using the JETSET model, and its full size is taken as an
additional uncertainty.  Second, and much more important, the
experimental selection and analysis method can introduce a bias in the
topology of the events used, or intrinsically correct for the effects.
This analysis bias is calculated using the full detector simulation
with JETSET for event-generation, where it has been verified that
JETSET reproduces the analytical calculations very well for full
acceptance.  It turns out that analyses based on semileptonic decays
typically need half of the full correction.  In the jet charge
analyses the QCD corrections are partly included in the measured
charge separation and partly in the hemisphere correlations which are
corrected for internally. The remaining corrections are very small.
The experimental asymmetries are then corrected by a factor 
$1/(1-C_{\mathrm{QCD}}^{\mathrm{had,T}} \cdot b)$ where $b$ is the bias factor
calculated with the simulation.

Because of the analysis dependence of the QCD corrections all
asymmetries quoted in this chapter are already corrected for QCD
effects.

The uncertainty on the theoretical calculation of the corrections, as
well as on the additional effect due to hadronisation, are taken as
fully correlated between the different measurements. The ``scaling
factor'' applied for each individual analysis to account for the
experimental bias is instead evaluated case by case together with its
associated uncertainty, and these errors are taken as uncorrelated.
For the jet charge measurement, the part of the QCD correction that is
included in the hemisphere correlations is also accounted for in the
error estimate. This part is estimated from the dependence of the
hemisphere correlations on the thrust value.

\subsection{Other Corrections to the Asymmetries}
\label{sec:hqocor}

The forward-backward asymmetries at LEP vary strongly as a function of
the centre-of-mass energy because of $\gammaZ$ interference.  Since
the mean energies at the different points vary slightly with time (see
e.g. Figure~\ref{fig:xsh}), the mean energies of the different
analyses are also not completely identical. The experiments quote the
mean centre-of-mass energy for each asymmetry measurement. In a first
fit the asymmetries are corrected to the closest of the three energies
$\sqrt{s}=89.55 \, \GeV{}(-2),\ 91.26 \, \GeV{}(\rm{pk}),\ 92.94 \,
\GeV(+2)$ assuming the $\SM$ energy dependence.

The slope of the asymmetries depends only on the well known fermion
charges and axial couplings while the asymmetry value on the Z-pole is
sensitive to the effective weak mixing angle. The first fit verifies
that the energy dependence is indeed consistent with the one expected
in the $\SM$. In a second fit all asymmetries are then corrected to
the peak energy (91.26 \GeV) before fitting.

To obtain the pole asymmetry, $\Afbzq$, which is defined by the real
parts of the Z-fermion couplings,
the fitted asymmetries at the peak energy, denoted as
$A_{\mathrm{FB}}^{\qq}({\rm pk})$ need to be corrected further as
summarised in Table~\ref{tab:aqqcor}.  These corrections are due to
the energy shift from 91.26 \GeV{} to $\MZ$, initial state radiation,
$\gamma$ exchange, $\gammaZ$ interference and imaginary parts of the
couplings.  A very small correction due to the nonzero value of the b
quark mass is also included.
All corrections are calculated using
ZFITTER~6.42~\cite{\ZFITTERref}. Further details can be found
in~\cite{hfasycor}.  The uncertainties on these corrections have been
estimated to be $\Delta (\delta \Abb) = 0.0002$ and $\Delta (\delta
\Acc) = 0.0001$~\cite{hfasycor}.  Compared to the experimental errors
on the quark asymmetries they can be safely neglected.  Similar
corrections have been applied to the left-right-forward-backward
asymmetries. The corrections are only about one tenth of the
experimental error and the asymmetries are directly presented in terms
of $\cAb$ and $\cAc$ by SLD.

\begin{table}[h]
\begin{center}
\renewcommand{\arraystretch}{1.1}
\begin{tabular}{|l||l|l|}
\hline
Source   & $\delta \Abb$
         & $\delta \Acc$ \\
\hline
\hline
$\sqrt{s} = \MZ $       & $ -0.0014 $  & $ -0.0035$  \\
QED corrections         & $ +0.0039 $  & $ +0.0107$  \\
other                   & $ -0.0006 $  & $ -0.0008$  \\
\hline
Total                   & $ +0.0019 $  & $ +0.0064$  \\
\hline
\end{tabular}
\end{center}
\caption[Corrections to be applied to the quark asymmetries.]
{Corrections to be applied to the quark asymmetries as $\Afbzq =
A_{\mathrm{FB}}^{\qq}({\rm pk}) + \delta A_{\mathrm{FB}}$.  The row
labelled ``other'' denotes corrections due to $\gamma$ exchange,
$\gammaZ$ interference, quark-mass effects and imaginary parts of the
couplings.  The uncertainties of the corrections are negligible. }
\label{tab:aqqcor}
\end{table}

\section{Combination Procedure}
\label{sec:hqcomb}

The heavy flavour results are combined~\cite{ref:lephf} using a
$\chi^2$ minimisation technique. In the case of the lineshape, each
experiment measures the same 5 or 9 parameters. Here, the set of
measurements is different for each experiment.  Nonetheless, a
$\chi^2$ minimisation can be used to find the best estimate of each of
the electroweak parameters. The formulation must be sufficiently
flexible to allow any number of measurements of each electroweak
parameter by each experiment. The measured values of closely related
auxiliary parameters, detailed in Appendix~\ref{sec:hqappendix} are included
in the averaging procedure. Their treatment will be explained more
fully below.

In order to write down an expression for this $\chi^2$, the average
value, i.e. the best estimate of the set of electroweak parameters is
denoted $x^{\mu}$, where the index ${\mu}$ refers to the different fit
parameters.
\begin{eqnarray}
\label{eq:hf_fitpar}
x^{{\mu}} &=& \Rb, \, \Rc    , \,  \\ \nonumber
&&\Abl   , \, \Acl, \,  \Abp, \, \Acp, \,  \Abh, \,  \Ach , \, \\ \nonumber
&&\cAb   , \,  \cAc   , \,   \\ \nonumber
&&\Brbl  , \,  \Brbclp, \,  \Brcl  , \,  \chiM  , \,   \\ \nonumber
&&\fDp   , \,  \fDs   , \,  \fcb   , \,  \PcDst .
\end{eqnarray}
Note that the forward-backward asymmetries can either be averaged at
three different centre-of-mass energies or be interpreted as
measurements of the asymmetry at the Z-peak, $\Abp$ and $\Acp$, as
described in Section~\ref{sec:hqocor}.

Each experimental result is referred to as $r_i$ and is a measurement of any
of the parameters $\mu(i)$ introduced in Equation~\ref{eq:hf_fitpar}: $\Rb$
corresponds to ${\mu}(i)=1$, $\Rc$ corresponds to ${\mu}(i)=2$ and so on.  A
group of $k$ results can be measured simultaneously in the same analysis to
give: $r_i$, $r_{i+1}$ ... $r_{i+k-1}$.

The averages are given by minimising the $\chi^2$:
\begin{equation}
\chi^2 ~=~ \sum_{ij} ( r_i - x^{\mu(i)} ) \calC_{ij}^{-1} 
                     ( r_j - x^{\mu(j)} ) \,.
\end{equation}
Since the uncertainties on the branching fractions of some of the
decay modes used in the charm counting $\Rc$ analyses are rather
large, two refinements are added to the fit to correct for non-linear
effects. The products $\Rc\PcDst$, $\Rc\fDp$, $\Rc\fDz$, $\Rc\fDs$ and
$\Rc\fcb$ are given as experimental results $r_i$ and are compared
to the product of the relevant fit parameters in the $\chi^2$
calculation.  $\fDz$ is calculated in this case from the other charmed
hadron fractions using Equation~\ref{eq:fdi}.  In addition the errors
on these parameters, again mainly the branching fraction errors, are
more Gaussian if they are treated as relative errors. For this reason
the logarithm of the products is fitted instead of the products
themselves.  It has been found that only in the case of $\Rc\fDs$ and
$\Rc\fcb$ do the fit results depend on whether the logarithms or the
values themselves are used.  However these two measurements are
completely dominated by the branching fraction error for which it is
clear that the logarithmic treatment is the better one.

Almost all the complications in building the $\chi^2$ are in
calculating the $n\times n$ covariance matrix, $\calC$, relating
the $i=1,n$ measurements. This matrix must take into account
statistical and systematic correlations.  Statistical correlations
arise from overlap of samples within an experiment, and for groups of
measurements of closely related parameters in the same fit.  Some
systematic errors lead only to correlations between measurements made
by the same experiment, for example errors due to the modelling of
track resolutions in a particular detector. Others are potentially
common to all the measurements.  The experiments provide their
measurements in the form of input tables, which list the central
values, the statistical errors, any correlations between statistical
errors and a detailed breakdown of the systematic errors.  This
breakdown is used to calculate the systematic error contribution to
the covariance matrix by assuming that any particular systematic
uncertainty, for example the uncertainty due to the lifetime of the
$\Bzero$ meson, is fully correlated for all
measurements~\cite{ref:lephf}.  This assumption is legitimate since
common values and uncertainties are used for those quantities taken
from external experimental measurements.  All results are corrected,
if necessary, to use the agreed set of external parameters. The input
parameters are discussed in Section~\ref{sec:hqinputs}.  In summary,
the covariance matrix has the form:
\begin{equation}
\calC_{ij} ~=~ \calC^{\mathrm{stat}}_{ij} + 
               \sum_k \sigma_i^k \sigma_j^k \,,
\end{equation}
where $\calC^{\mathrm{stat}}_{ij}$ is the covariance matrix of
statistical errors and $\sigma_i^k$ is the systematic error in
measurement $i$, due to the source of systematic uncertainty $k$.
Some errors, such as the error from Monte Carlo statistics, are
uncorrelated for all results and therefore contribute only to the
diagonal elements of $\calC$. Others, such as those connected with
lepton identification or tracking efficiency, are correlated for any
measurements made by the same experiment. The remaining errors,
arising from the physics sources discussed in
Section~\ref{sec:hqinputs}, are assumed to be fully correlated for all
measurements.

It is also important to take into account that even when two
electroweak parameters are not measured in the same fit, the measured
value of one will depend on the value assumed for the other. For
example, a measurement of \Rb\ often depends on the fraction of charm
contamination in the sample, and therefore on the value of \Rc\ that
was assumed.  
Let $r_i$ be a measurement of \Rb.
The explicit first order dependence of the value of 
$r_i$, on the assumed value of \Rc, $x^{\Rc}$, is then included as
follows:
\begin{equation}
r_i ~=~ \Rb^{\mathrm{meas}} + 
            a_i^{\Rc} \frac{ ( x^{\Rc} - \Rc^{\mathrm{used}} ) } 
                           {  x^{\Rc} }\,.
\end{equation}
Here $\Rb^{\mathrm{meas}}$ is the central value of $\Rb$ measured by
the experiment, assuming a value for $\Rc = \Rc^{\mathrm{used}}$.  The
constant $a_i^{\Rc}$ is given by
\begin{equation}
\frac {a_i^{\Rc}} {x^{\Rc}} ~=~ 
\frac { \mathrm{d} r_i^{\Rb} } { \mathrm{d} x^{\Rc} }
\left( { x^{\Rc} = \Rc^{\mathrm{used}} } \right) \,.
\label{eq:hfslope}
\end{equation}
The dependence of any measurement on any of the other fitted
parameters can be expressed in the same way.

The system of including measurements by input tables has proved to be
very flexible. Different subsets of results can be combined together
in cross-checks, to verify that the results are robust.

As an example, Table~\ref{tab:Rbinp_exa} shows the measurements of
\Rbz{} used in the fit.  The line labelled ``\Rbz(published)'' shows
the value published by the collaborations while in the line
``\Rbz(input)'' the values corrected for the agreed external
parameters are given. The errors labelled ``uncorrelated'' are either
internal to the analysis or to the experiment while the ones labelled
``correlated'' are potentially in common with other experiments.  Also
the dependences of the \Rbz{} measurements on the other input
parameters are given.

\section{Results}
\label{sec:hqresults}

\begin{figure}[p]
\begin{center}
 \includegraphics[width=0.7\linewidth,bb=24 76  472 485]{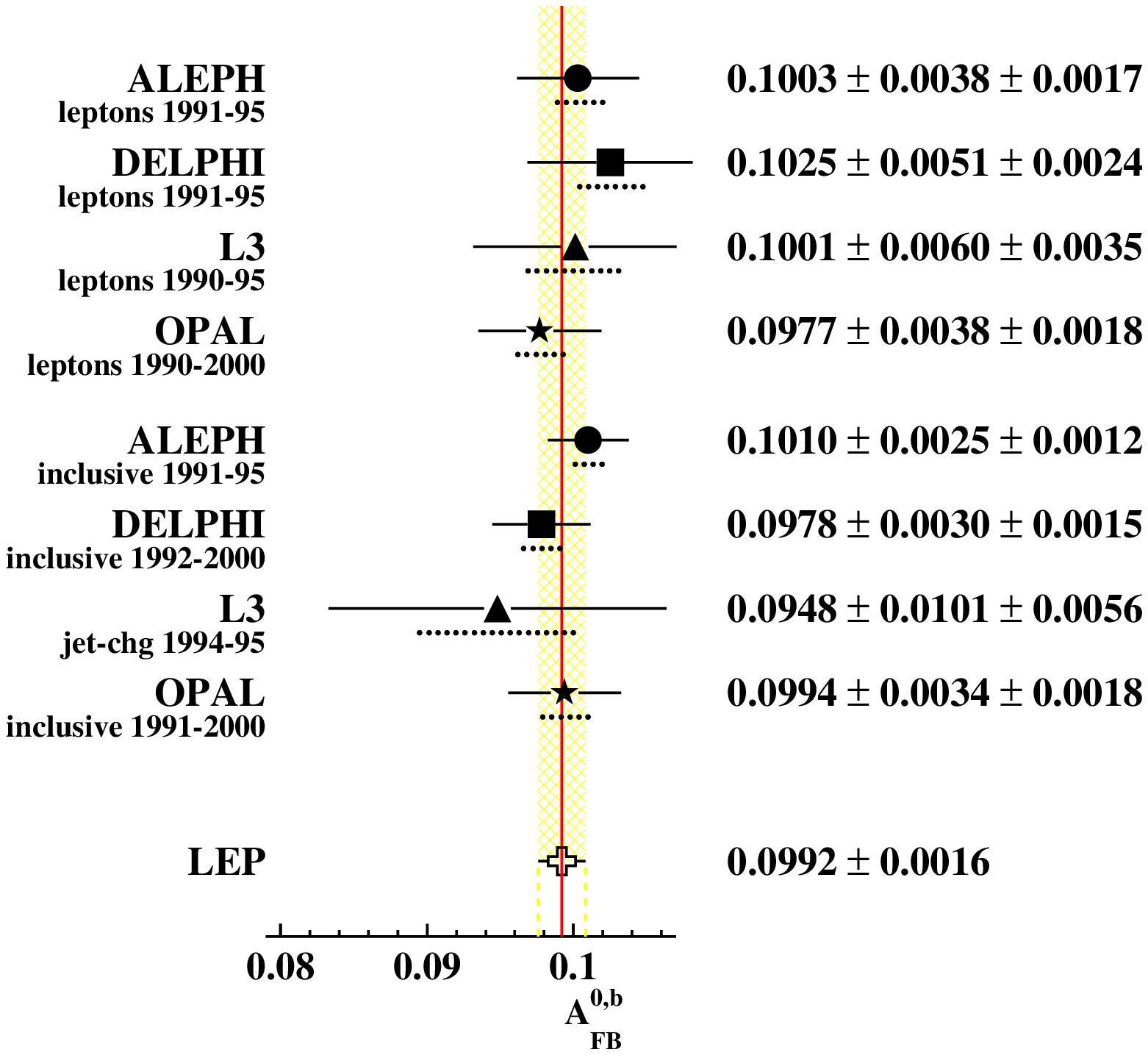}
 \includegraphics[width=0.7\linewidth,bb=24 76  472 485]{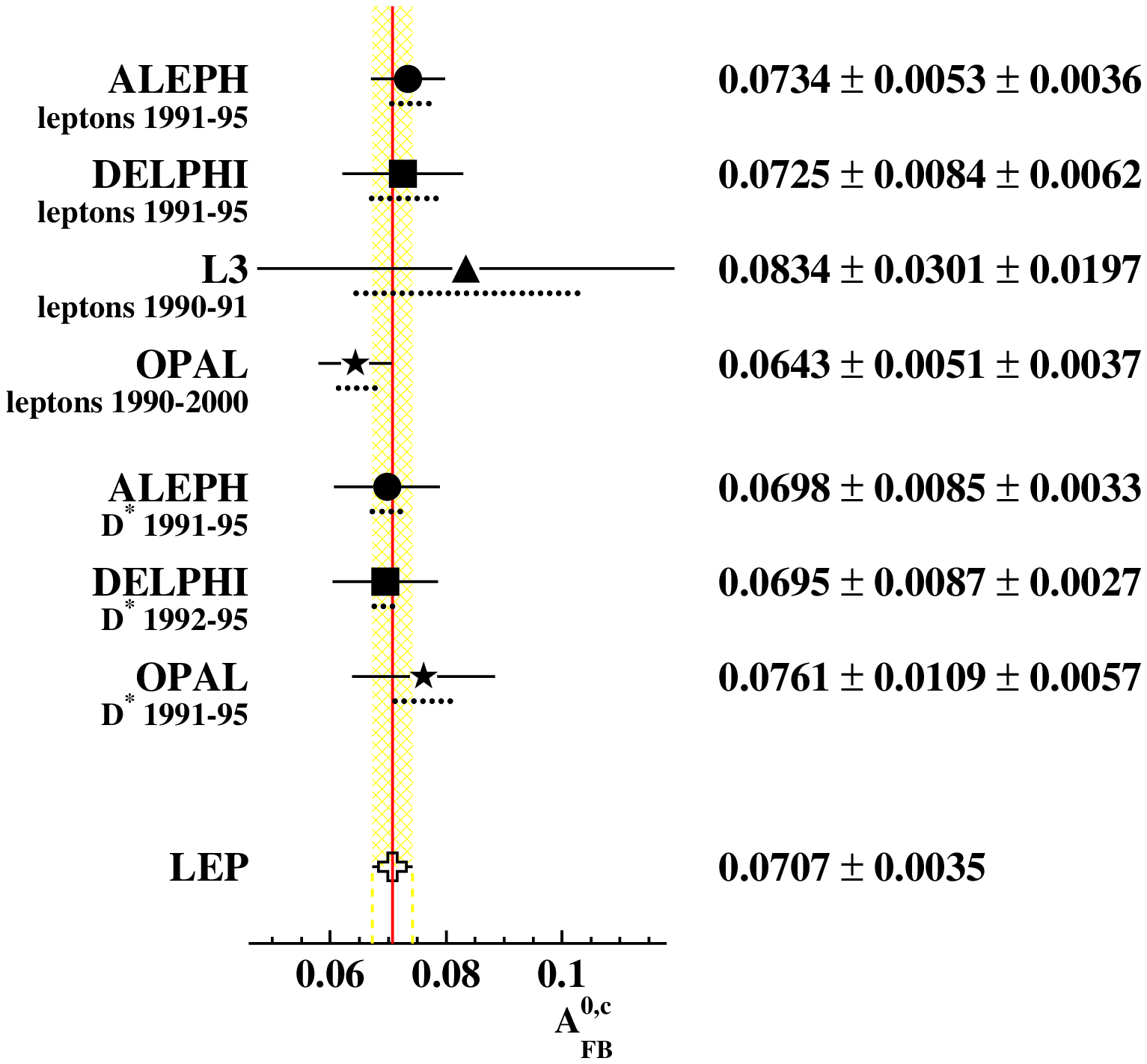}
\end{center}
\caption[$\Afbzb$ and $\Afbzc$ measurements used in the heavy flavour
  combination.]{$\Afbzb$ and $\Afbzc$ measurements used in the heavy
  flavour combination, corrected for their dependence on parameters
  evaluated in the multi-parameter fit described in the text. The
  $\Afbzb$ measurements with D-mesons do not contribute significantly
  to the average and are not shown in the plots.  The experimental
  results are derived from the ones shown in Tables~\ref{tab:Ablinp}
  to~\ref{tab:Achinp} combining the different centre of mass
  energies. The dotted lines indicate the size of the systematic
  error. }
\label{fig:hqbar_asy}
\end{figure}

\begin{figure}[p]
\begin{center}
 \includegraphics[width=0.9\linewidth,bb=33 60 511 313]{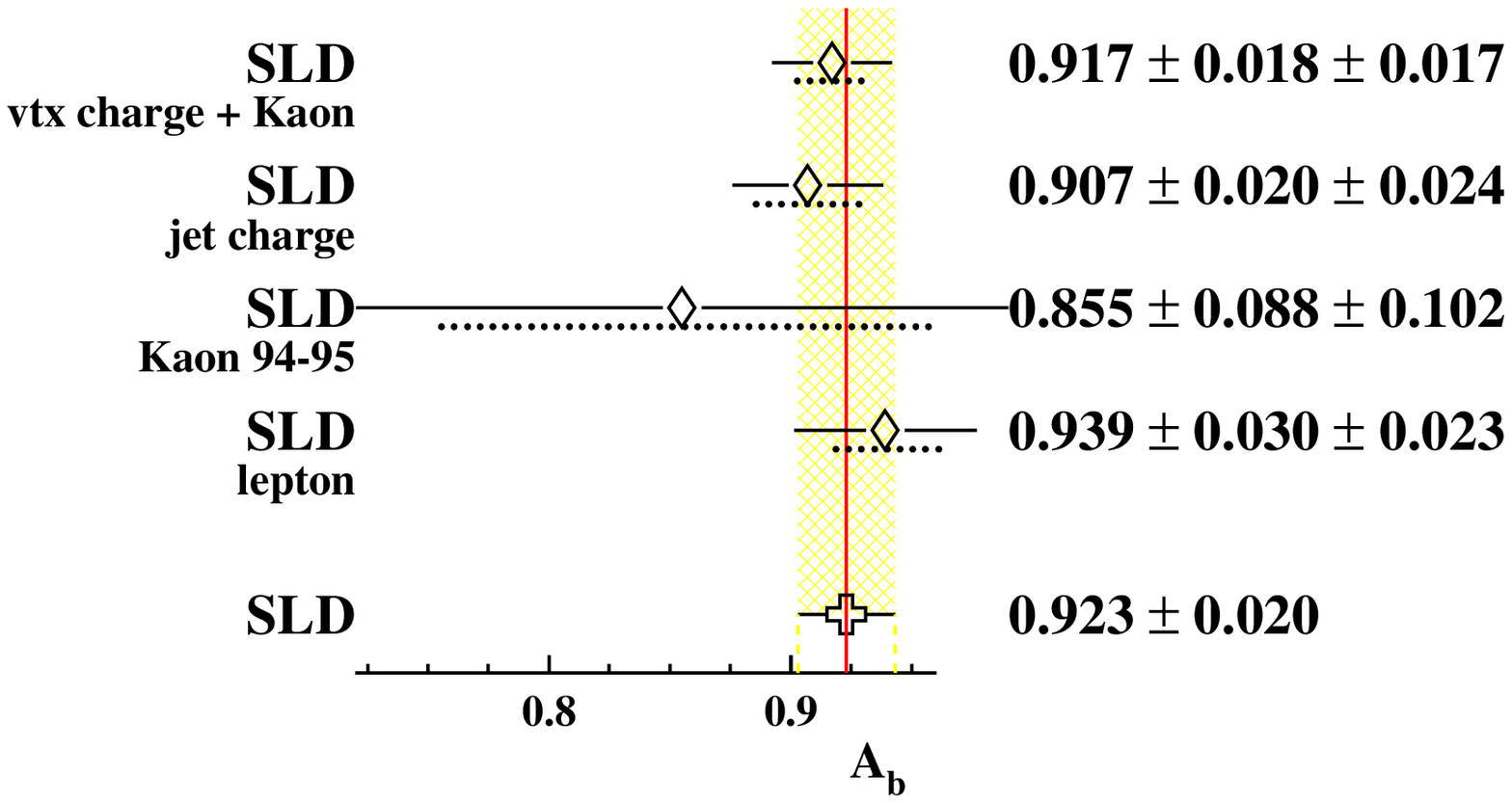}
\vskip 1cm
 \includegraphics[width=0.9\linewidth,bb=33 60 511 313]{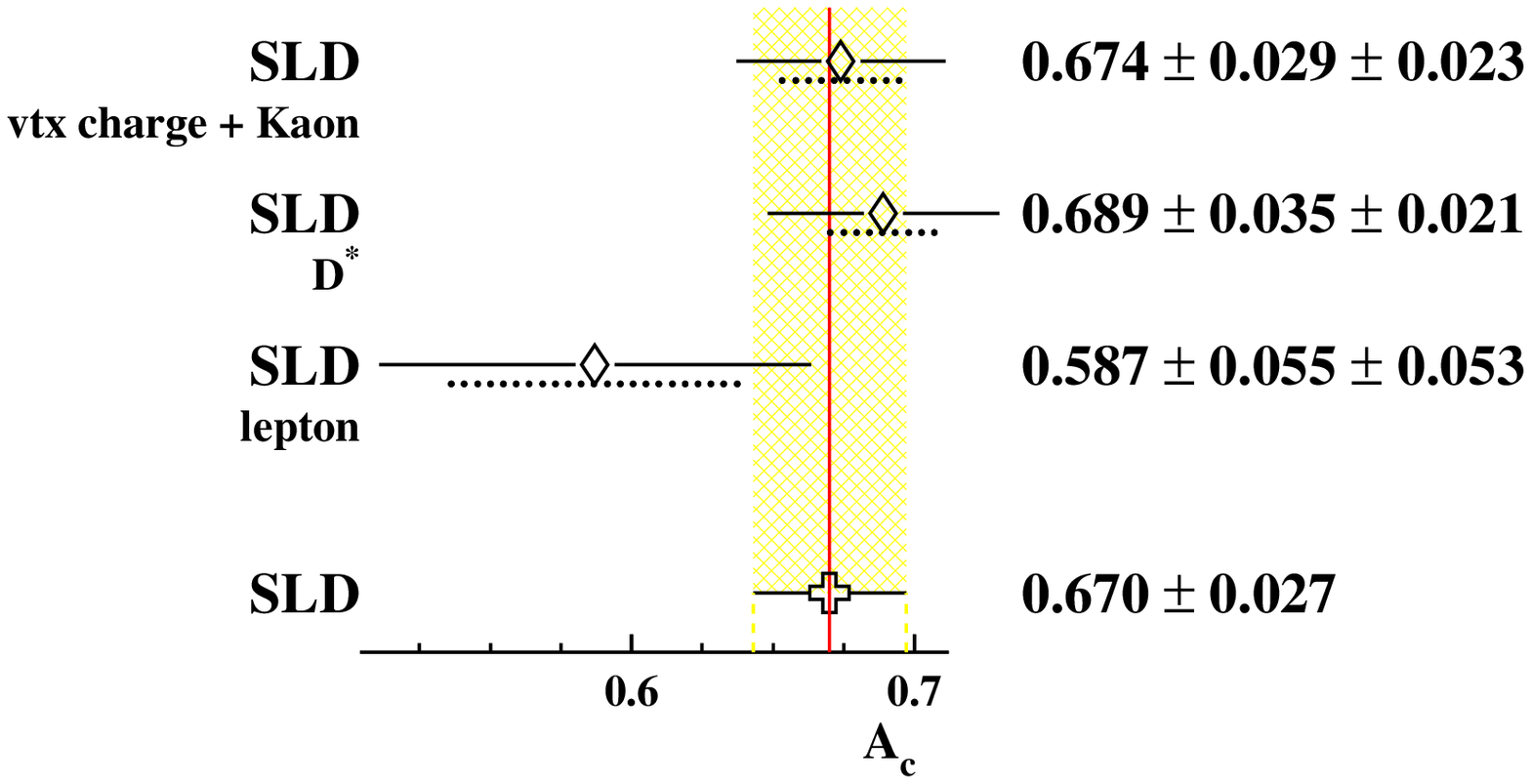}
\end{center}
\caption[ $\cAb$ and $\cAc$ measurements used in the heavy flavour
  combination.]{$\cAb$ and $\cAc$ measurements used in the heavy
  flavour combination, corrected for their dependence on parameters
  evaluated in the multi-parameter fit described in the text. The
  dotted lines indicate the size of the systematic error. }
\label{fig:hqbar_abc}
\end{figure}

The results used in this combination have been described in
Sections~\ref{sec:hqrbc},~\ref{sec:hqasy} and~\ref{sec:hqaux} and are
summarised in Tables~\ref{tab:Rbinp} to~\ref{tab:RcPcDstinp} in
Appendix~\ref{sec:hqappendix}.  Figures~\ref{fig:hqbar_rbc}
to~\ref{fig:hqbar_abc} compare the main electroweak results of the
different experiments.

\begin{figure}[p]
\begin{center}
 \includegraphics[width=0.6\linewidth,bb=10 20 483 455]{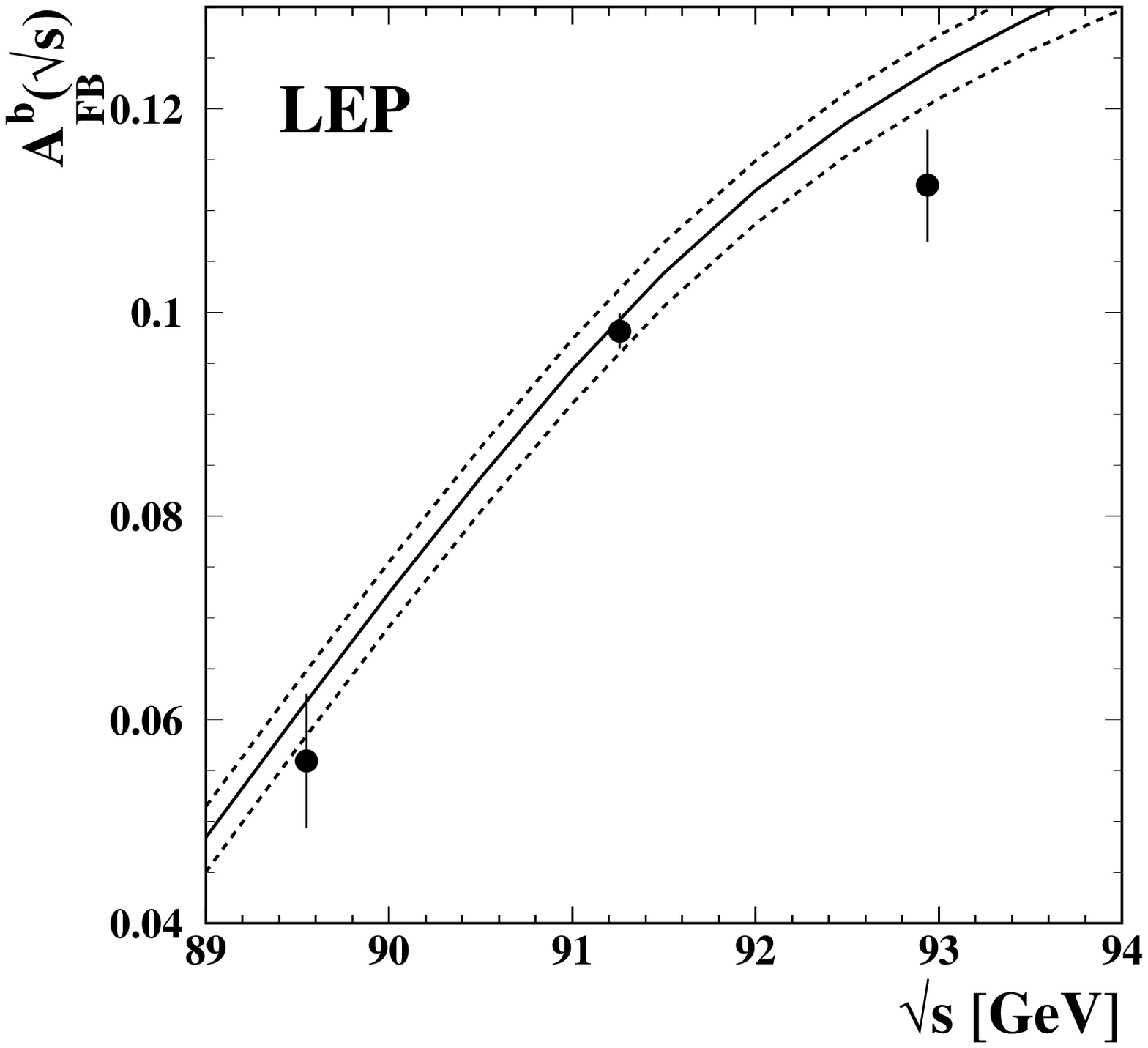}
\vskip 1cm
 \includegraphics[width=0.6\linewidth,bb=10 20 483 455]{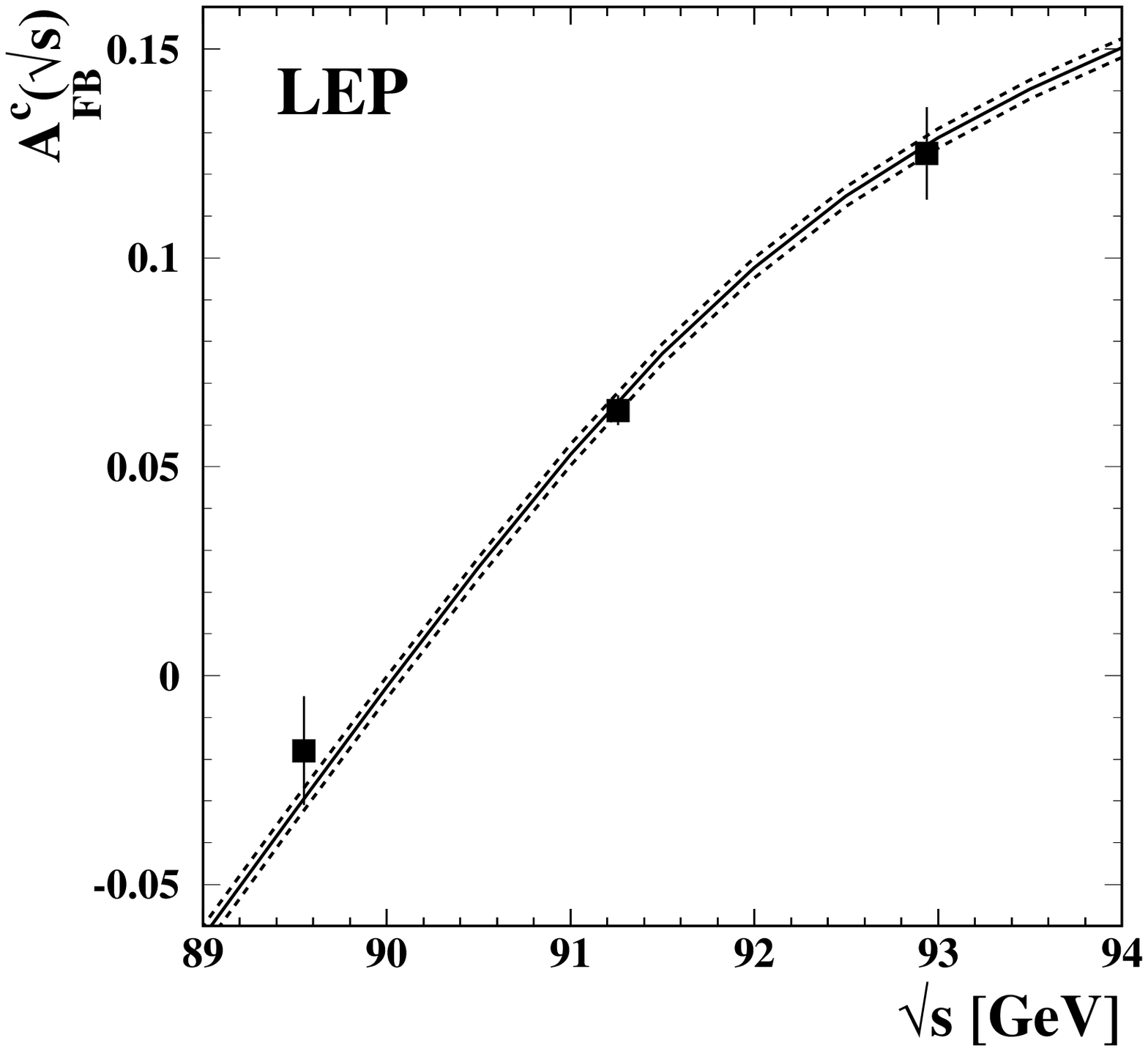}
\end{center}
\caption[Energy dependence of $\Afb^b$ and $\Afb^c$. ]{Energy
  dependence of $\Afb^b$ and $\Afb^c$. The solid line represents the
  $\SM$ prediction for $\Mt=178 \, \GeV,\, \MH = 300 \,\GeV$, the
  upper (lower) dashed line is the prediction for $\MH = 100 \, (1000)
  \, \GeV$.  }
\label{fig:hfafbvsene} 
\end{figure}

In the first fit the different analyses have been combined with the
asymmetries kept at the three different energies, yielding in total 18
free parameters.  The results of this fit for the asymmetries are
listed in Table~\ref{tab:18parasy} including their correlations.
These asymmetries are only corrected for QCD effects.  The full fit
results including the correlation matrix is shown in
Appendix~\ref{sec:hqappfit}.  The $\chidf$ of the fit is
$48/(105-18)$.
Applying the corrections explained in Section~\ref{sec:hqocor} to the
peak asymmetry only one obtains for the pole asymmetries:\footnote{ To
correct the peak asymmetries to the pole asymmetries only a number
with negligible additional uncertainty is added, see
Table~\ref{tab:aqqcor}.  All errors and correlations thus remain
unchanged.}
\begin{eqnarray}
\Afbzb & = & 0.1000 \pm 0.0017 \\
\Afbzc & = & 0.0699 \pm 0.0036 \,,
\end{eqnarray}
with a correlation of $+0.15$.  Figure~\ref{fig:hfafbvsene} shows the
energy dependence of $\Afb^{\rm{b}}$ and $\Afb^{\rm{c}}$ compared to
the $\SM$ prediction.

\begin{table}[htb]
\begin{center}
\renewcommand{\arraystretch}{1.1}
\begin{tabular}{|l||c|c|c|c|c|}
\hline
  & \mca{1}{ALEPH} & \mca{1}{DELPHI} & \mca{1}{L3} 
  & \mca{1}{OPAL} & \mca{1}{SLD} \\
\hline
           &92-95 &92-95 &94-95 &92-95 &93-98\\
           &\tmcite{ref:alife} &\tmcite{ref:drb} &\tmcite{ref:lrbmixed} &\tmcite{ref:omixed} 
           &\tmcite{Abe:2005nq}\\
 \hline\hline
 \Rbz(published)      &    0.2159 &   0.2163 &   0.2174 &   0.2178 &   0.2159\\
 \hline
 \Rbz(input)          &    0.2158 &   0.2163 &   0.2173 &   0.2174 &   0.2159\\
 \hline
 Statistical          &    0.0009 &   0.0007 &   0.0015 &   0.0011 &   0.0009\\
 \hline
 Uncorrelated         &    0.0007 &   0.0004 &   0.0015 &   0.0009 &   0.0005\\
 Correlated           &    0.0006 &   0.0004 &   0.0018 &   0.0008 &   0.0005\\
 \hline
 Total Systematic     &    0.0009 &   0.0005 &   0.0023 &   0.0012 &   0.0007\\
 \hline
 \hline
 $a( \Rc      )$ &   -0.0033 &  -0.0041 &  -0.0376 &  -0.0122 &  -0.0056\\
 $\Rc     ^{\mathrm{used}}$ &    0.1720 &   0.1720 &   0.1734 &   0.1720 &   0.17123\\
 \hline
 $a( \Brcl    )$ &           &          &  -0.0133 &  -0.0067 &         \\
 $\Brcl   ^{\mathrm{used}}$ &           &          &     9.80 &     9.80 &         \\
 \hline
 $a( \fDp     )$ &   -0.0010 &  -0.0010 &  -0.0086 &  -0.0029 &  -0.0008\\
 $\fDp    ^{\mathrm{used}}$ &    0.2330 &   0.2330 &   0.2330 &   0.2380 &   0.2330\\
 \hline
 $a( \fDs     )$ &   -0.0001 &   0.0001 &  -0.0005 &  -0.0001 &  -0.0003\\
 $\fDs    ^{\mathrm{used}}$ &    0.1020 &   0.1030 &   0.1030 &   0.1020 &   0.1020\\
 \hline
 $a( \fLc     )$ &    0.0002 &   0.0003 &   0.0008 &   0.0003 &  -0.0002\\
 $\fLc    ^{\mathrm{used}}$ &    0.0650 &   0.0630 &   0.0630 &   0.0650 &   0.0650\\
\hline
\end{tabular}
\caption[The measurements of \Rbz.]  {The measurements of \Rbz.  All
measurements use a lifetime tag enhanced by other features like
invariant mass cuts or high $p_T$ leptons.  The lines $a(X)$ and
$x^{\mathrm{used}}$ refer to the dependences defined in
Equation~\ref{eq:hfslope}. The dependence on $\Brcl$ is only present
for the measurements that use leptons in their primary b-tag.}
\label{tab:Rbinp_exa}
\end{center}
\end{table}

\begin{table}[htb]
\begin{center}
\renewcommand{\arraystretch}{1.1}
\begin{tabular}{|l||c|cccccc|}
\hline
Observable & Result & \multicolumn{6}{|c|}{Correlations} \\
 & & \makebox[1.3cm]{$\Abl$} & \makebox[1.3cm]{$\Acl$} & 
 \makebox[1.3cm]{$\Abp$} & \makebox[1.3cm]{$\Acp$} & 
 \makebox[1.3cm]{$\Abh$} & \makebox[1.3cm]{$\Ach$} \\
\hline
\hline
 $ \Abl $&$\phantom{-}    0.0560    \pm  0.0066 $
 & $ 1.00$ &         &         &         &         &        \\
 $ \Acl $&$              -0.018     \pm  0.013  $
 & $ 0.13$ & $ 1.00$ &         &         &         &        \\
 $ \Abp $&$\phantom{-}    0.0982    \pm  0.0017 $
 & $ 0.03$ & $ 0.01$ & $ 1.00$ &         &         &        \\
 $ \Acp $&$\phantom{-}    0.0635    \pm  0.0036 $
 & $ 0.00$ & $ 0.02$ & $ 0.15$ & $ 1.00$ &         &        \\
 $ \Abh $&$\phantom{-}    0.1125    \pm  0.0055 $
 & $ 0.01$ & $ 0.01$ & $ 0.08$ & $ 0.02$ & $ 1.00$ &        \\
 $ \Ach $&$\phantom{-}    0.125     \pm  0.011  $
 & $ 0.00$ & $ 0.01$ & $ 0.02$ & $ 0.15$ & $ 0.13$ & $ 1.00$\\
\hline
\end{tabular}
\end{center}
\caption[The forward-backward asymmetry results from the 18-parameter
fit]{The forward-backward asymmetry results from the 18-parameter fit,
including their correlations.}
\label{tab:18parasy}
\end{table}

\clearpage

\begin{figure}[p]
\begin{center}
 \includegraphics[width=0.8\linewidth,bb=30 190 495 505]{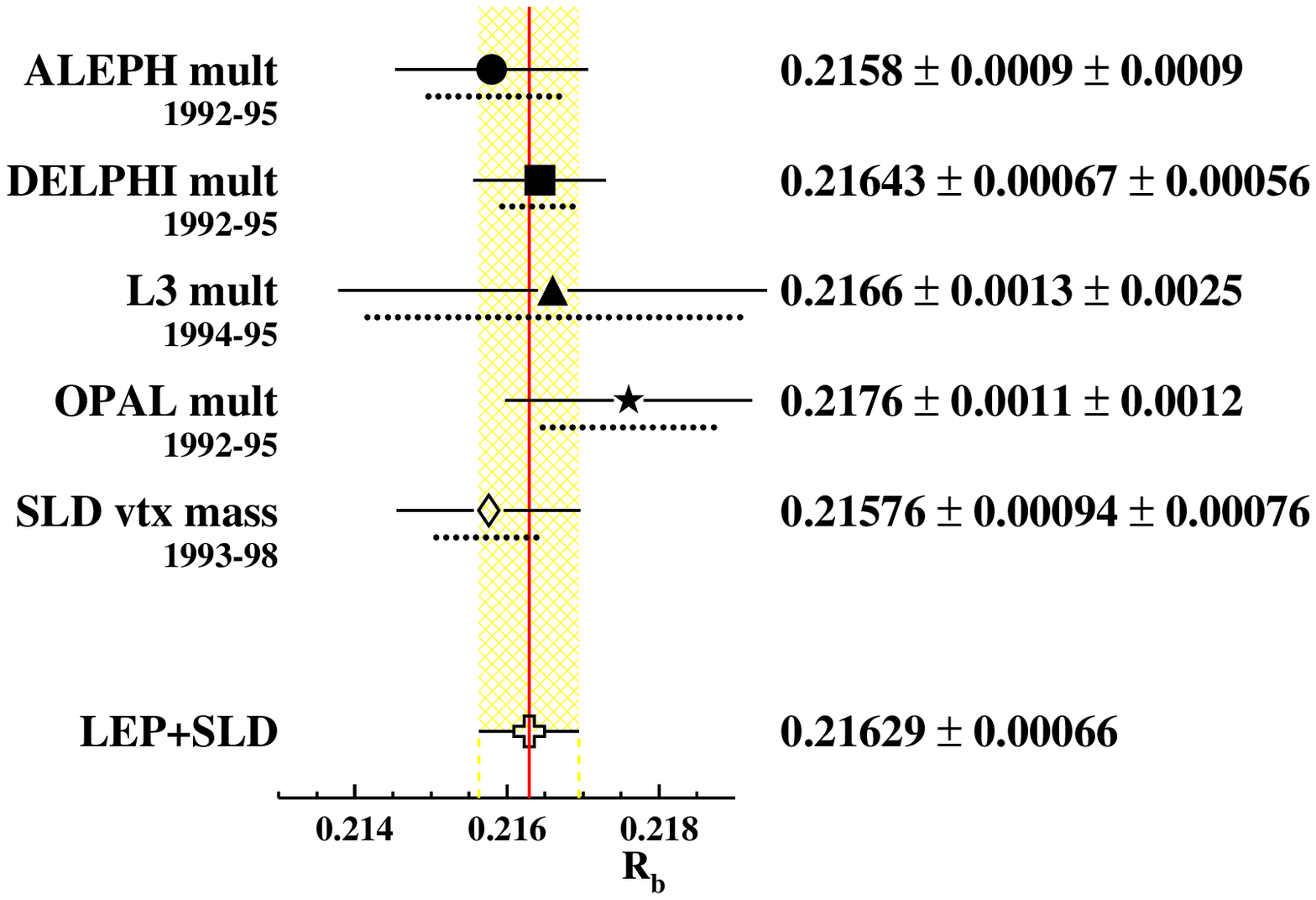}
\vskip 1cm
 \includegraphics[width=0.8\linewidth,bb=10  80 475 480]{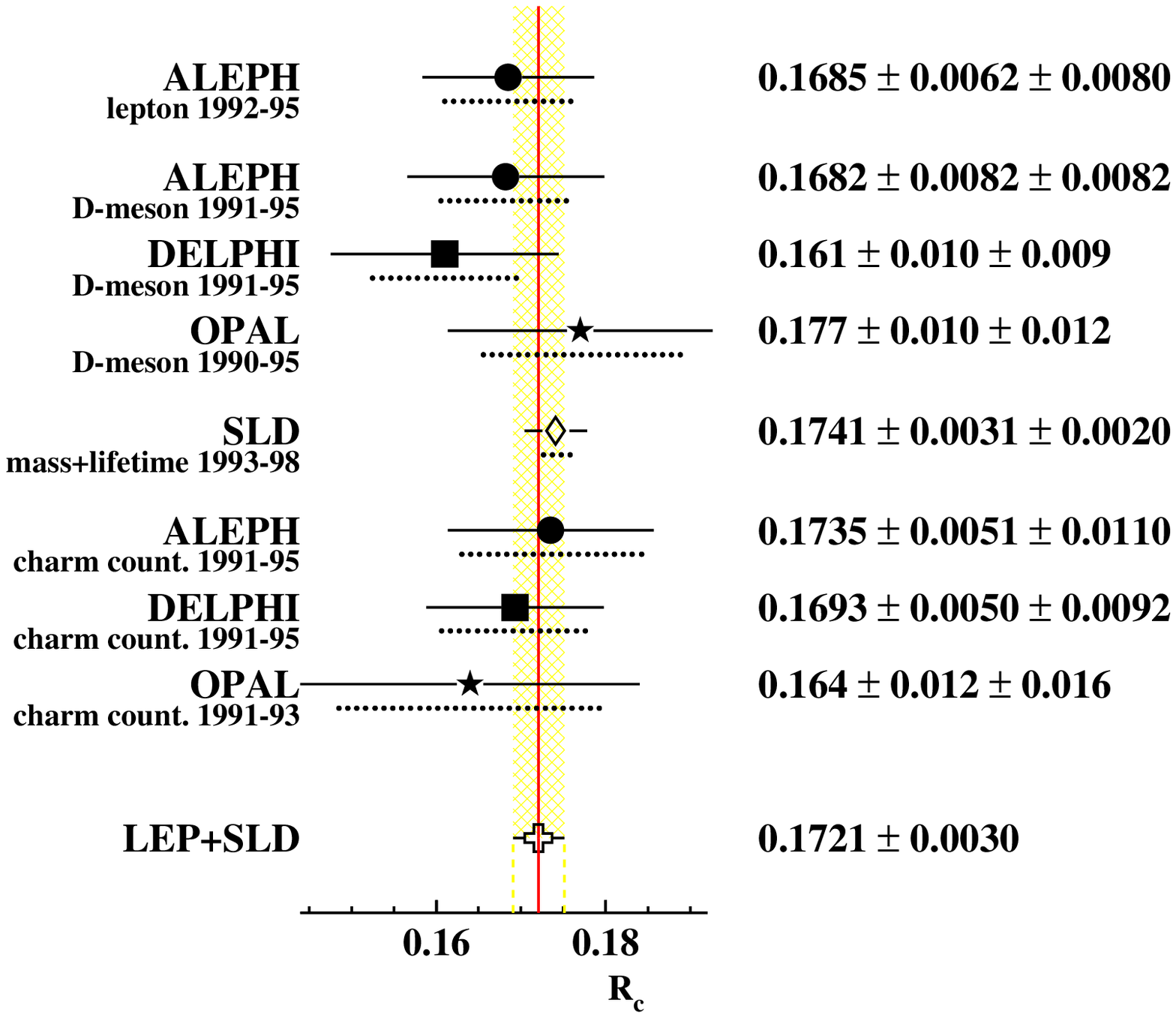}
\end{center}
\caption[$\Rbz$ and $\Rcz$ measurements used in the heavy flavour
  combination.]{$\Rbz$ and $\Rcz$ measurements used in the heavy
  flavour combination, corrected for their dependence on parameters
  evaluated in the multi-parameter fit described in the text.  The
  dotted lines indicate the size of the systematic error.  }
\label{fig:hqbar_rbc}
\end{figure}

\clearpage

Since the energy dependence of the asymmetries is described by the
$\SM$ prediction, in a second fit all asymmetries are corrected to the
peak energy before fitting, resulting in 14 free parameters. The
results of this fit are shown in Table~\ref{tab:14parres}.  The
$\chidf$ of the fit is $53/(105-14)$.  The corresponding correlation
matrix is given in Table~\ref{tab:14parcor}.  Note that here the
values of $A^{\qq}_{\rm FB}({\rm pk})$ actually found in the fit have
already been corrected to pole asymmetries, as described in
Section~\ref{sec:hqocor}.\footnote{ To correct the peak asymmetries to
the pole asymmetries only a number with negligible additional
uncertainty is added, see Table~\ref{tab:aqqcor}.  All errors and
correlations thus remain unchanged.}
If the off-peak asymmetries are included in the fit the pole asymmetry
$\Afbzb$ is about half a sigma below the values without these
asymmetries.  This is due to the somewhat low b-asymmetry at
$92.94\,\GeV$.

\begin{table}[htb]
\begin{center}
\renewcommand{\arraystretch}{1.1}
\begin{tabular}{|l||c|}
\hline
Observable & Result \\
\hline
\hline
 $ \Rbz    $&$ 0.21629  \pm 0.00066 $\\
 $ \Rcz    $&$ 0.1721   \pm 0.0030  $\\
 $ \Afbzb  $&$ 0.0992   \pm 0.0016  $\\
 $ \Afbzc  $&$ 0.0707   \pm 0.0035  $\\
 $ \cAb    $&$ 0.923    \pm 0.020   $\\
 $ \cAc    $&$ 0.670    \pm 0.027   $\\
 $ \Brbl   $&$ 0.1071    \pm 0.0022 $\\
 $ \Brbclp $&$ 0.0801    \pm 0.0018 $\\
 $ \Brcl   $&$ 0.0969    \pm 0.0031 $\\
 $ \chiM   $&$ 0.1250    \pm 0.0039 $\\
 $ \fDp    $&$ 0.235     \pm 0.016  $\\
 $ \fDs    $&$ 0.126     \pm 0.026  $\\
 $ \fcb    $&$ 0.093     \pm 0.022  $\\
 $ \PcDst  $&$ 0.1622    \pm 0.0048 $\\
\hline
\end{tabular}
\end{center}
\caption[The results of the 14-parameter fit to the LEP/SLD heavy
flavour data]{The results of the 14-parameter fit to the LEP/SLD heavy
flavour data.  The correlations are listed in
Table~\ref{tab:14parcor}.}
\label{tab:14parres}
\end{table}

\begin{table}[p]
\begin{center}
\renewcommand{\arraystretch}{1.1}
\begin{sideways}
\begin{minipage}[b]{\textheight}
\begin{center}
\begin{tabular}{|l||rrrrrrrrrrrrrr|}
\hline
&\makebox[0.45cm]{\Rb}
&\makebox[0.45cm]{\Rc}
&\makebox[0.45cm]{$\Afbzb$}
&\makebox[0.45cm]{$\Afbzc$}
&\makebox[0.45cm]{\cAb}
&\makebox[0.45cm]{\cAc}
&\makebox[0.45cm]{$\BR(1)$}
&\makebox[0.45cm]{$\BR(2)$}
&\makebox[0.45cm]{$\BR(3)$}
&\makebox[0.45cm]{\chiM}
&\makebox[0.45cm]{$\fDp$}
&\makebox[0.45cm]{$\fDs$}
&\makebox[0.45cm]{$f(c_{bar.})$}
&\makebox[0.55cm]{$P$}\\
\hline
\hline
 {\Rb}          & $ 1.00$ &         &         &         &         &         
&         &         &         &         &         &         &         &        \\ 
 {\Rc}          & $-0.18$ & $ 1.00$ &         &         &         &         
&         &         &         &         &         &         &         &        \\ 
 {$\Afbzb$}     & $-0.10$ & $ 0.04$ & $ 1.00$ &         &         &         
&         &         &         &         &         &         &         &        \\ 
 {$\Afbzc$}     & $ 0.07$ & $-0.06$ & $ 0.15$ & $ 1.00$ &         &         
&         &         &         &         &         &         &         &        \\ 
 {\cAb}         & $-0.08$ & $ 0.04$ & $ 0.06$ & $-0.02$ & $ 1.00$ &         
&         &         &         &         &         &         &         &        \\ 
 {\cAc}         & $ 0.04$ & $-0.06$ & $ 0.01$ & $ 0.04$ & $ 0.11$ & $ 1.00$ 
&         &         &         &         &         &         &         &        \\ 
 {$\BR(1)$}     & $-0.08$ & $ 0.05$ & $-0.01$ & $ 0.18$ & $-0.02$ & $ 0.02$
& $ 1.00$ &         &         &         &         &         &         &        \\ 
 {$\BR(2)$}     & $-0.03$ & $-0.01$ & $-0.06$ & $-0.23$ & $ 0.02$ & $-0.04$ 
& $-0.24$ & $ 1.00$ &         &         &         &         &         &        \\ 
 {$\BR(3)$}     & $ 0.00$ & $-0.30$ & $ 0.00$ & $-0.21$ & $ 0.03$ & $-0.02$ 
& $ 0.01$ & $ 0.10$ & $ 1.00$ &         &         &         &         &        \\ 
 {\chiM}        & $ 0.00$ & $ 0.02$ & $ 0.11$ & $ 0.08$ & $ 0.06$ & $ 0.00$ 
& $ 0.29$ & $-0.23$ & $ 0.16$ & $ 1.00$ &         &         &         &        \\ 
 {$\fDp$}       & $-0.15$ & $-0.10$ & $ 0.01$ & $-0.04$ & $ 0.00$ & $ 0.00$ 
& $ 0.04$ & $ 0.02$ & $ 0.00$ & $ 0.02$ & $ 1.00$ &         &         &        \\ 
 {$\fDs$}       & $-0.03$ & $ 0.13$ & $ 0.00$ & $-0.02$ & $ 0.00$ & $ 0.00$ 
& $ 0.01$ & $ 0.00$ & $-0.01$ & $-0.01$ & $-0.40$ & $ 1.00$ &         &        \\ 
 {$f(c_{bar.})$}& $ 0.11$ & $ 0.18$ & $-0.01$ & $ 0.04$ & $ 0.00$ & $ 0.00$ 
& $-0.02$ & $-0.01$ & $-0.02$ & $ 0.00$ & $-0.24$ & $-0.49$ & $ 1.00$ &        \\ 
 {$P$}          & $ 0.13$ & $-0.43$ & $-0.02$ & $ 0.04$ & $-0.02$ & $ 0.02$ 
& $-0.01$ & $ 0.01$ & $ 0.13$ & $ 0.00$ & $ 0.08$ & $-0.06$ & $-0.14$ & $ 1.00$\\ 
 \hline
\end{tabular}
\normalsize
\end{center}
\caption[The correlation matrix for the set of the 14 heavy flavour
parameters.]{ The correlation matrix for the set of the 14 heavy
flavour parameters. $\BR(1)$, $\BR(2)$ and $\BR(3)$ denote $\Brbl$,
$\Brbclp$ and $\Brcl$ respectively, $P$ denotes $\PcDst$.  }
\label{tab:14parcor}
\end{minipage}
\end{sideways}
\end{center}
\end{table}

In all cases, the fit $\chi^2$ is smaller than expected. As a cross
check the fit has been repeated using statistical errors only,
resulting in consistent central values and a $\chidf$ of $92
/(105-14)$. In this case a large contribution to the $\chi^2$ comes
from $\Brbl$ measurements, which is sharply reduced when detector
systematics are included. Subtracting the $\chi^2$ contribution from
$\Brbl$ measurements one gets $\chidf = 65/(99-13)$.  This shows that
the low $\chi^2$ largely comes from a statistical fluctuation.  In
addition many systematic errors are estimated very conservatively.
Several error sources are evaluated by comparing test quantities
between data and simulation. The statistical errors of these tests are
taken as systematic uncertainties but no explicit correction is applied
because of this test.
Also in some cases fairly conservative assumptions are used for the error
evaluation. Especially for the $\bl$ model only fairly old publications exist
where the central spectrum describes the data well, but the two alternatives
that are used for the error evaluation are no longer really compatible with
the data.
However it should be noted that especially for the
quark forward backward asymmetries the systematic errors are much
smaller than the statistical ones so that a possible overestimate of
these errors cannot hide disagreements with other electroweak
measurements.

Table~\ref{tab:hferrbk} summarises the dominant errors for the
electroweak parameters. In all cases the two largest error sources are
statistics and systematics internal to the experiments.  The internal
systematics consist mainly of errors due to Monte Carlo statistics,
data statistics for cross-checks and the knowledge of detector
resolutions and efficiencies.  The error labelled ``QCD effects'' is
due to hemisphere correlation for \Rbz{} and \Rcz{}
(Section~\ref{sec:hq_rbhemcol}) and due to the theoretical uncertainty
in the QCD corrections for the asymmetries
(Section~\ref{sec:hf_qcdcor}).  For the asymmetries on average about
50~\% of the QCD corrections are seen.  The uncertainties due to the
knowledge of the beam energy are negligible in all cases.

\begin{table}[t]
\begin{center}
\renewcommand{\arraystretch}{1.1}
\begin{tabular}{|c||c|c|c|c|c|c|}
\hline
Source
&\makebox[1.2cm]{\Rbz}
&\makebox[1.2cm]{\Rcz}
&\makebox[1.2cm]{$\Afbzb$}
&\makebox[1.2cm]{$\Afbzc$}
&\makebox[0.9cm]{\cAb}
&\makebox[0.9cm]{\cAc}\\
 & $[10^{-3}]$ & $[10^{-3}]$ & $[10^{-3}]$ & $[10^{-3}]$ 
 & $[10^{-2}]$ & $[10^{-2}]$ \\
\hline
\hline
statistics & 
$0.44$ & $2.4$ & $1.5$ & $3.0$ & $1.5$ & $2.2$ \\
\hline
internal systematics &
$0.28$ & $1.2$ & $0.6$ & $1.4$ & $1.2$ & $1.5$ \\
\hline
QCD effects &
$0.18$ & $0  $ & $0.4$ & $0.1$ & $0.3$ & $0.2$ \\
$\BR$(D $\rightarrow$ neut.)&
$0.14$ & $0.3$ & $0$   & $0$ & $0$ & $0$ \\
D decay multiplicity &
$0.13$ & $0.6$ & $0  $ & $0.2$ & $0$ & $0$ \\
B decay multiplicity &
$0.11$ & $0.1$ & $0  $ & $0.2$ & $0$ & $0  $ \\
$\BR$(D$^+ \rightarrow$ K$^- \pi^+ \pi^+) $&
$0.09$ & $0.2$ & $0  $ & $0.1$ & $0$ & $0  $ \\
$\BR$($\Ds \rightarrow \phi \pi^+) $&
$0.02$ & $0.5$ & $0  $ & $0.1$ & $0$ & $0$ \\
$\BR$($\Lambda_{\mathrm{c}} \rightarrow $p K$^- \pi^+) $&
$0.05$ & $0.5$ & $0  $ & $0.1$ & $0$ & $0  $ \\
D lifetimes&
$0.07$ & $0.6$ & $0  $ & $0.2$ & $0$ & $0$ \\
B decays&
$0$ & $0$ & $0.1$ & $0.4$ & $0$ & $0.1$  \\
decay models&
$0$ & $0.1$ & $0.1$ & $0.5$ & $0.1$ & $0.1$ \\
non incl. mixing&
$0$ & $0.1$ & $0.1$ & $0.4$ & $0$ & $0$ \\
gluon splitting &
$0.23$ & $0.9$ &$0.1$& $0.2$ & $0.1$ & $0.1$ \\
c fragmentation &
$0.11$ & $0.3$ & $0.1$ & $0.1$ & $0.1$ & $0.1$ \\
light quarks&
$0.07$ & $0.1$ & $0  $ & $0  $ & $0$ & $0  $ \\
beam polarisation&
$0$ & $0$ & $0$ & $0$ & $0.5$ & $0.3$ \\
\hline
total correlated&
$0.42$ & $1.5$ & $0.4$ & $0.9$ & $0.6$ & $0.4$ \\
\hline
total error&
$0.66$ & $3.0$ & $1.6$ & $3.5$ & $2.0$ & $2.7$ \\
\hline
\end{tabular}
\caption[Dominant error sources for the heavy-flavour electroweak
parameters]{ The dominant error sources for the heavy-flavour
electroweak parameters from the 14-parameter fit, see text for
details.  }
\label{tab:hferrbk}
\end{center}
\end{table}

Amongst the non-electroweak observables the B semileptonic branching
fraction is of special interest ($\Brbl \, = \, 0.1071 \pm 0.0022$).
The largest error source for this quantity is the dependence on the
semileptonic decay model $\bl$ with
\begin{equation}
\Delta \Brbl (\bl\,\rm{modelling}) ~=~ 0.0012.
\end{equation}
Extensive studies have been made to understand the size of this error.
Amongst the electroweak quantities, the quark asymmetries measured
with leptons depend on the assumptions of the decay model while the
asymmetries using other methods usually do not. The fit implicitly
requires that the different methods give consistent results. This
effectively constrains the decay model and thus reduces the error 
in $\Brbl$ from this source in the fit result.

To get a conservative estimate of the modelling error in $\Brbl$ the
fit has been repeated removing all asymmetry measurements. The result
of this fit is
\begin{equation}
\Brbl ~ = ~ 0.1069 \pm 0.0022
\end{equation}
with
\begin{equation}
\Delta \Brbl (\bl\,\rm{modelling}) ~=~ 0.0013.
\end{equation}

The other B-decay related observables from this fit are
\begin{eqnarray}
  \Brbclp & = &  0.0802    \pm 0.0019 \\
  \chiM   & = &  0.1259    \pm 0.0042. \nonumber 
\end{eqnarray}

\begin{figure}[hb]
\begin{center}
  \includegraphics[width=0.6\linewidth,bb=16 16 488 457]{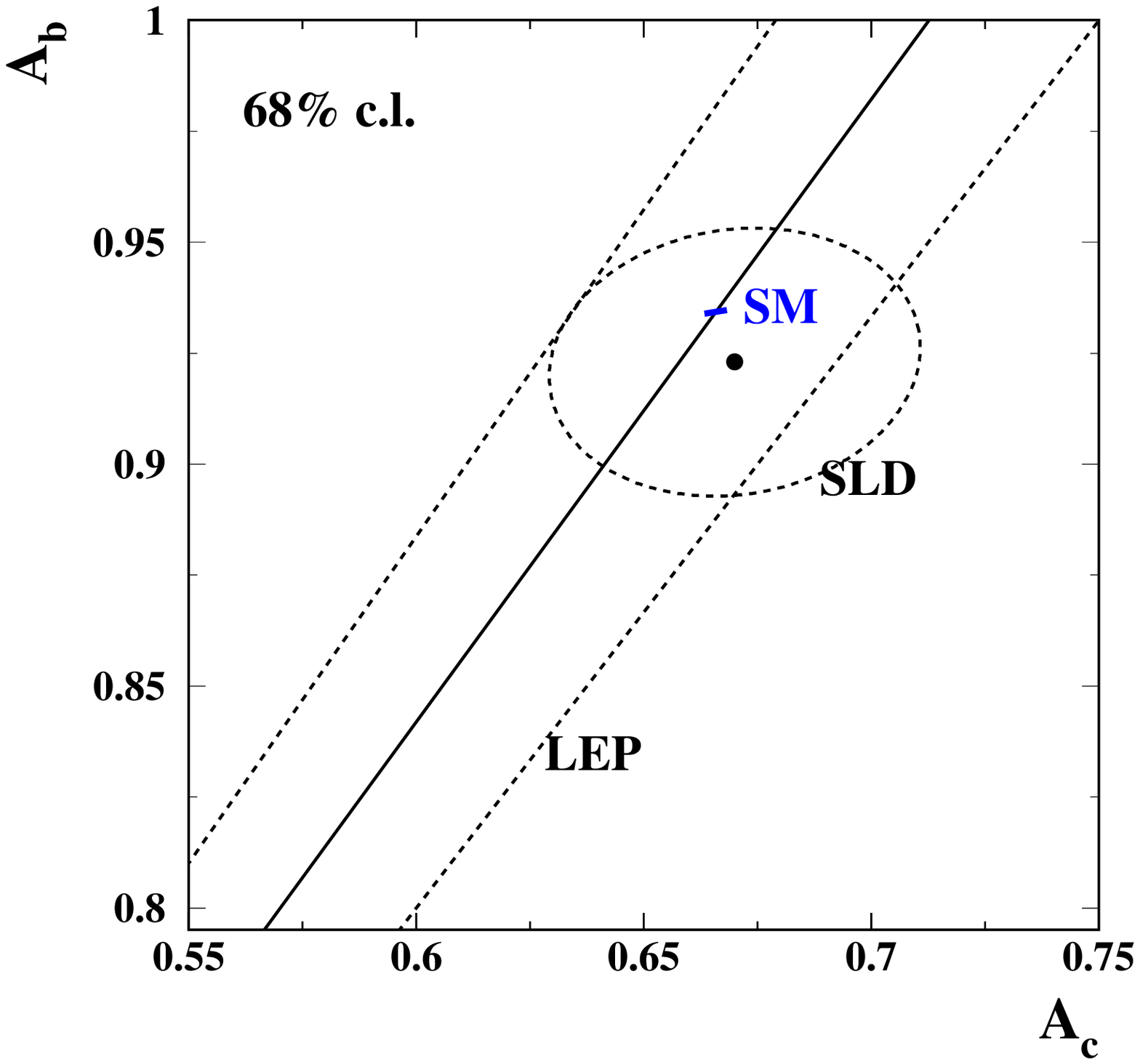}
\vskip -0.5cm
\end{center}
\caption[Contours in the $\cAc$-$\cAb$ plane from the Z-pole data] {
Contours in the $\cAc$-$\cAb$ plane and ratios of forward-backward
asymmetries from the SLD and LEP, corresponding to 68~\% confidence
levels assuming Gaussian systematic errors. The $\SM$ prediction for
$\Mt=178.0 \pm 4.3\,\GeV$, $\MH = 300^{+700}_{-186}\,\GeV$ and
$\dalhad=0.02758\pm0.00035$ is also shown.  } \label{fig:hfsmcomp1}
\end{figure}

\begin{figure}[p]
\begin{center}
\vskip -1cm
\includegraphics[width=0.6\linewidth]{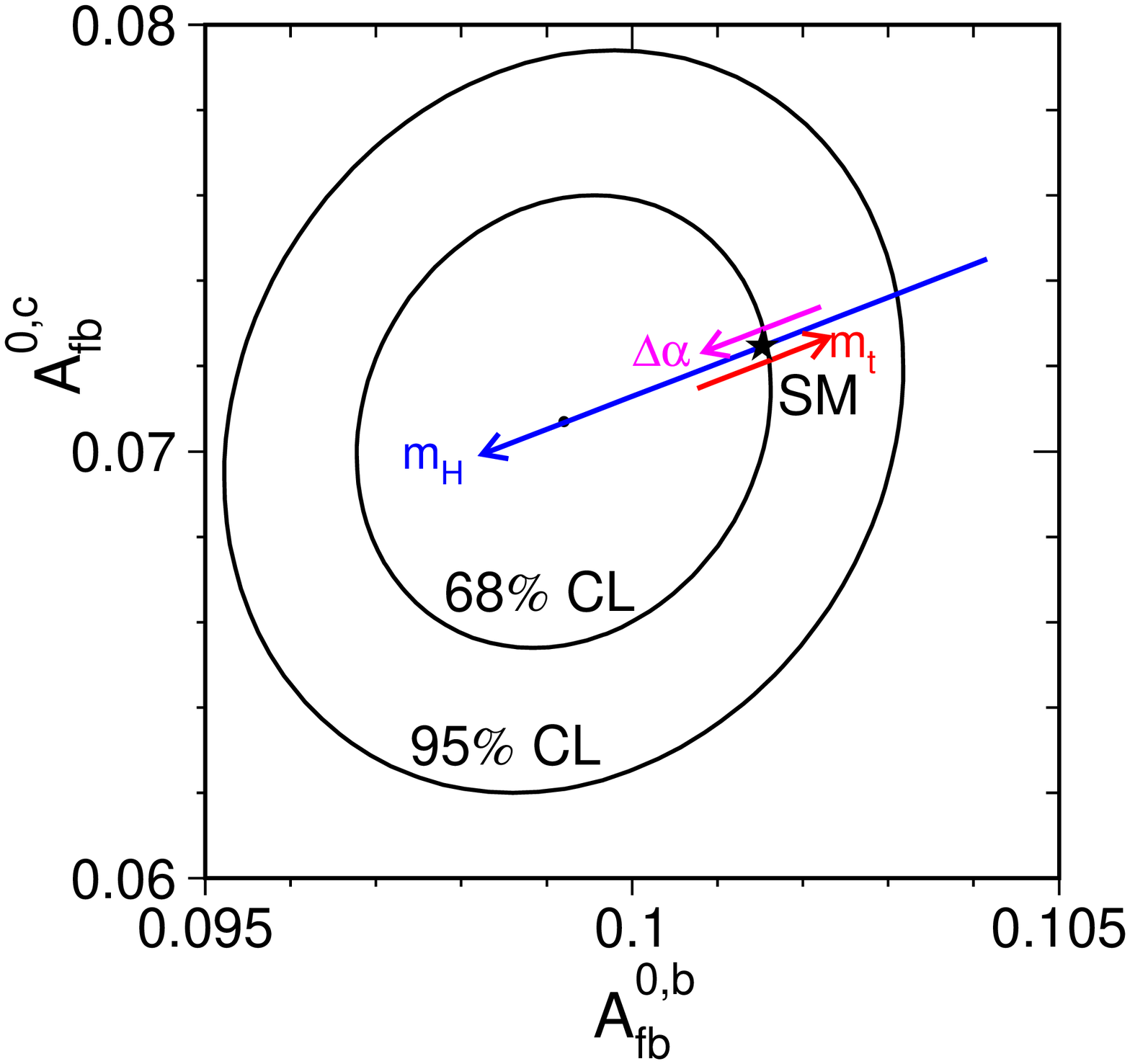}
\vskip -1cm
\includegraphics[width=0.6\linewidth]{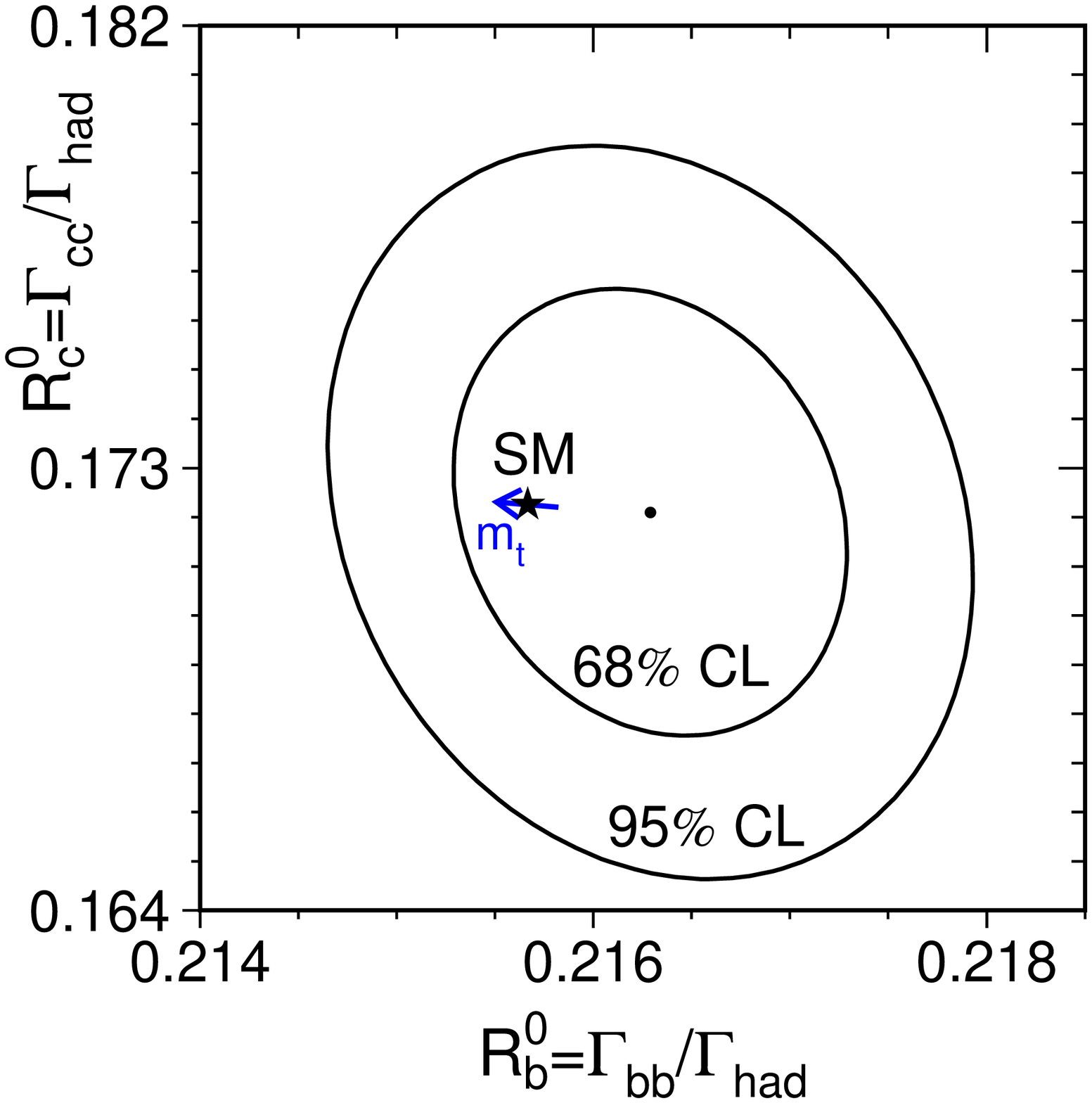}
\vskip -0.5cm
\end{center}
\caption[Contours in the $\Afbzc$-$\Afbzb$ and $\Rcz$-$\Rbz$ planes
from the Z-pole data]{ Contours in the $\Afbzc$-$\Afbzb$ and
$\Rcz$-$\Rbz$ planes from the LEP and SLD data, corresponding to 68~\%
confidence levels assuming Gaussian systematic errors. The $\SM$
prediction for $\Mt=178.0 \pm 4.3\,\GeV$, $\MH =
300^{+700}_{-186}\,\GeV$ and $\dalhad=0.02758\pm0.00035$ is also
shown.  } \label{fig:hfsmcomp2}
\end{figure}

Figures~\ref{fig:hfsmcomp1} and~\ref{fig:hfsmcomp2} compare
$(\cAb,\cAc)$, $(\Acc,\Abb)$ and $(\Rbz,\Rcz)$ with the $\SM$
prediction.  Good agreement is found everywhere.  However, unlike the
asymmetries in lepton pair production, the quark asymmetries favour a
Higgs mass of a few hundred $\GeV$.  In case of $\cAb$-$\cAc$ the
ratio $\Afbzb/\Afbzc$ from LEP is also shown in
Figure~\ref{fig:hfsmcomp1}. This ratio is equal to $\cAb/\cAc$ and
thus, unlike $\Acc$ and $\Abb$ themselves, is free from assumptions
about the leptonic couplings of the Z.  The data are interpreted
further, together with the leptonic observables, in
Chapters~\ref{chap:Z+coup} and~\ref{chap:MSM}.

\vfill

\chapter{Inclusive Hadronic Charge Asymmetry}
\label{sec:lq}

The measurement of the total hadronic partial width of the $\Zzero$
described in Chapter~\ref{chap:lsafb} makes no attempt to distinguish
different quark flavours. Similarly, an inclusive forward-backward
asymmetry measurement can be made using the samples of all hadronic
events, taking advantage of the high statistics. This measurement is
generically referred to as the $\Qfbhad$ measurement, since all the
methods use some kind of forward-backward charge asymmetry in
inclusive hadronic events.  The up-type (charge $2/3$) and down-type
(charge $-1/3$) quarks contribute to the average forward-backward
charge asymmetry with opposite sign.  The average asymmetry is
therefore particularly sensitive to the flavour ratios in the sample.
To interpret the measurement, these ratios are usually taken from the
predictions of the Standard Model ($\SM$).  Indeed, the result of the
measurement is often quoted directly as a value of $\swsqeffl$, in the
context of the $\SM$.  The systematic errors are much more significant
than for the high-efficiency and high-purity heavy flavour samples
already discussed in Chapter~\ref{sec:hq}.

Tagging methods to enhance the fraction of specific light flavour
quarks (up, down or strange) have also been developed, and used to
measure forward-backward asymmetries and partial widths.  Further
information on the partial widths of the $\Zzero$ to up-type and
down-type quarks in hadronic $\Zzero$ decays has been inferred from
the observed rate of direct photon production, by exploiting the fact
that the probability of photon radiation from final-state quarks is
proportional to the square of the
quark charge.  The tagged light quark and direct
photon results are summarised in Appendix~\ref{sec:lqappendix}, where
the limited tests of the light quark couplings to the $\Zzero$ that
they allow are also presented.

\section{Asymmetry of Flavour-Inclusive Hadronic Events}
\label{sec:lq:qfb}

Many of the ideas developed in Chapter~\ref{sec:hq} have been extended
and applied to an inclusive sample of $\Ztoqq$ decays by the four LEP
experiments~\cite{ALEPHcharge1996,DELPHIcharge,ref:ljet,OPALcharge}.
However, the DELPHI and OPAL publications only include data from 1990
and 1991, and the collaborations did not update the measurements with
more data due to the implicit SM dependence of the technique.  The
details of the methods vary, but all use some variant of the jet
charge, as defined in Equation~\ref{eq:jetch}.  The event is divided
into two hemispheres by the plane perpendicular to the thrust axis.
The electron beam points into the forward hemisphere, and the jet
charges are evaluated in the forward and backward hemispheres, giving
$\QF$ and $\QB$.  ALEPH and DELPHI then consider the observable
$\avQfb \equiv \langle \QF - \QB \rangle$, the average value of the
difference between the hemisphere charges.  This quantity is referred
to as the forward-backward charge flow.  The observable $\avQfb$ is
given by:
\begin{equation}
\avQfb ~ = ~ \sum_{\mathrm{q}} \Rq \Aqq 
\delta_{\mathrm{q}} C_{\mathrm{q}}\, ,
\end{equation}
where the sum runs over the 5 primary quark flavours, and the
coefficients $C_{\mathrm{q}}$ account for the acceptance of each
flavour subsample.  The charge separation, $\delta_{\mathrm{q}} =
\langle Q_{\mathrm{q}} - Q_{\mathrm{\overline{q}}} \rangle$, is the
mean jet charge difference between the hemispheres containing quark
and the anti-quark, which can equivalently be expressed in terms of
the jet charges in the hemispheres containing the negatively charged
parton, $Q_{-}$, and the positively charged parton, $Q_{+}$:
\begin{equation}
\delta_{\mathrm{q}} ~ = ~ s_{\mathrm{q}} \langle Q_{-} - Q_{+} \rangle \, ,
\end{equation}
where $s_{\mathrm{q}} = +1$ for down-type quarks and $-1$ for up-type
quarks.  This choice of notation makes explicit the fact that the
contributions to $\avQfb$ from the different quark types are of
opposite sign. The main benefit of the method is that the charge
separation can be evaluated from the data, as shown by
Equation~\ref{eq:hf:deltaq}.  The evaluation of the charge separation
is discussed further below.  The parameters $\Rq$ and $\Aqq$ can be
expressed in the $\SM$ as a function of the effective weak mixing
angle, $\swsqeffl$.  Once the charge separations $\delta_{\mathrm{q}}$
are known, the measurement of $\avQfb$ can then be interpreted as a
measurement of $\swsqeffl$.

\begin{figure}[htb]
\begin{center} 
\includegraphics[width=0.9\linewidth]{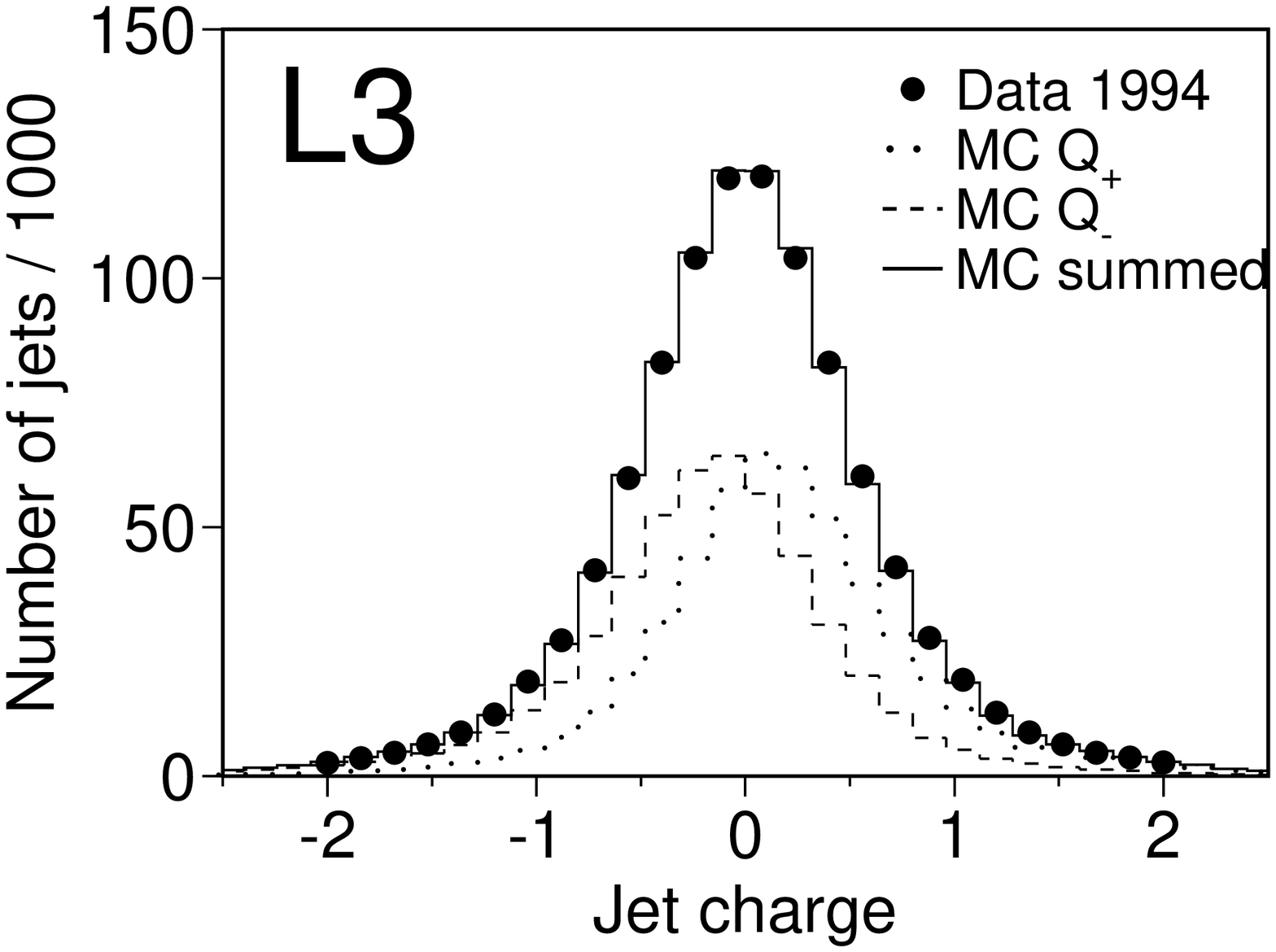}
\caption[The $Q_{+}$ and $Q_{-}$ distributions obtained from 
Monte Carlo simulation by L3] {The $Q_{+}$ and $Q_{-}$ distributions
  obtained from Monte Carlo simulation by L3. Their sum is compared to
  the sum of the $ \QF + \QB \equiv Q_+ + Q_- $ distributions for 1994
  data.}
\label{fig:Qfbpic} 
\end{center}
\end{figure}

L3 use a very closely related approach, calling an event forward if
$\QF$ is larger than $\QB$. The probability that an event is forward
simply depends on the charge separation and the width of the
distributions of $Q_{-}$ and $Q_{+}$, with a correction for hemisphere
correlations, and can be derived from data in a very similar manner to
$\avQfb$.  The degree of charge separation between $Q_{+}$ and $Q_{-}$
is illustrated in Figure~\ref{fig:Qfbpic}. The width of the
distribution of $\QF + \QB \equiv Q_+ + Q_-$ agrees well between data
and the Monte Carlo simulation.

The OPAL analysis calculates overall event weights using the three
highest weight tracks per hemisphere. The overall
event weight is 
the probability that the event is forwards.  An observable average
forward-backward charge asymmetry is derived in an iterative
procedure, adjusting the value of $\swsqeffl$ in the Monte Carlo
simulation.  The Monte Carlo modelling of the weights is controlled by
comparisons with data. DELPHI also present an alternative measurement,
where the value of $\QF-\QB$ is used event-by-event to decide if it is
forward or backward, and an effective observable average charge
asymmetry is derived.

Experimentally, the crux of the measurement is to determine the mean
charge separations for each flavour.  As described in
Chapter~\ref{sec:hq} when discussing measurements of the
forward-backward asymmetry in $\Ztobb$ events using jet charges, the
mean charge separation for $\Ztobb$ events can be determined directly
from the data, in a sample of b-tagged events (see
Equation~\ref{eq:hf:deltaq}).  In a similar way, charm tagging may be
used to determine the mean charge separation in $\Ztocc$ events.
However, in each case a correction must be made to account for any
difference between the charge separation for tagged and untagged
events of the same flavour.  The reduction in systematic errors from
assessing the charge separation for heavy flavours from data rather
than taking it from the Monte Carlo simulation outweighs the
uncertainties introduced by the correction.  The charge separation in
light-quark events can only be determined from Monte Carlo models.
This is the largest source of systematic uncertainty in the analyses.
The mean charge separation for the inclusive sample may also be
determined from the data.  It can be used as an additional constraint
on the light-quark mean charge separations, although it is not
directly applicable to the charge flow.

The only practical way to combine these analyses is at the level of
the derived $\swsqeffl$ values.  The observed values of $\avQfb$ or
hadronic charge asymmetries reflect the experimental acceptance and
resolution, and cannot be combined directly.  It must be emphasised
that because the measurements in this section are interpreted as
$\swsqeffl$ measurements entirely within the context of the $\SM$,
they must be used with care when comparing with alternative
models. This is in contrast with results such as $\Abb$ and $\Rb$.
For example, the value of $\swsqeffl$ discussed here can only
legitimately be used to test a model that does not change the relative
fractions of each flavour.

\section{Systematic Uncertainties}

Due to the lack of high-purity and high-efficiency tags for specific
light flavours, by far the dominant systematic uncertainties in these
results arise from the model input required to describe the light
quark properties.  All experiments use the JETSET Monte Carlo as a
reference fragmentation and hadronisation model, while the HERWIG
model is used for systematic comparisons. The parameter set within
JETSET is also often varied as part of the assessment of the
fragmentation/hadronisation model uncertainties. However, neither the
parameter set used for the central values nor the method for parameter
variation is common to the experiments, with different experimental
measurements being used by the experiments to constrain the model
parameters.  In addition, there are typically code changes made to the
Monte Carlo programs to improve the overall description in each
experiment. Thus, there is far from 100\% correlation between the
quoted uncertainties due to fragmentation and hadronisation modelling.

The remaining significant uncertainties are all specific to a given
experiment, for example due to the modelling of detector resolution,
or due to the evaluation of the charge biases such as differences in
the reconstruction of the tracks of positive and negative particles,
or the charge-dependence of hadronic interactions in the material of
the detector.

The theoretical QCD corrections applied to the forward-backward
asymmetries for each flavour are potentially another common
uncertainty (see Section~\ref{sec:hf_qcdcor}).  In practice, the
corrections for QCD effects such as hard gluon radiation are all
derived from JETSET, and are not distinguished from the overall
correction for fragmentation and hadronisation effects.  The
theoretical QCD correction uncertainties are all much smaller than the
quoted fragmentation/hadronisation uncertainties and other
experimental errors, and treating them as an additional common error
would have no impact on the result.

\section{Combination Procedure}

The derived values of the effective weak mixing angle, $\swsqeffl$,
are combined by first forming a full covariance matrix for the
uncertainties, assuming that the errors associated with quark
fragmentation and hadronisation are the only source of correlation.
As explained above, these dominant systematic uncertainties are not
fully correlated because they are not evaluated in the same way for
each experiment. The off-diagonal terms are therefore taken to be the
smaller of the two quoted fragmentation/hadronisation uncertainties
for each pair of measurements (so-called ``minimum-overlap''
estimate).  A $\chi^2$ minimisation is then performed for the single
free parameter, $\swsqeffl$.  The fit has a $\chidf$ of 0.43/3. In
order to assess the sensitivity of the combined result to the
assumptions made in calculating the covariance matrix, different
approaches have also been considered.  The resulting weights for each
input result and the final combined $\swsqeffl$ change very little
when the assumptions are changed.  For example, taking as the
off-diagonal elements the smallest error common to all the inputs
only changes the central value by 0.00007, and the uncertainty on
the average by 0.00009.  However, if the common systematic errors are
incorrectly assumed to be fully correlated, the system is badly
behaved, with some measurements getting a negative weight. This is
symptomatic of an unphysical over-correlation in a set of
measurements.

\section{Combined Results and Discussion}
\label{sec:lq:res}

\begin{figure}[htb]
\begin{center} 
\includegraphics[width=0.8\linewidth,bb=120 25 440 500]{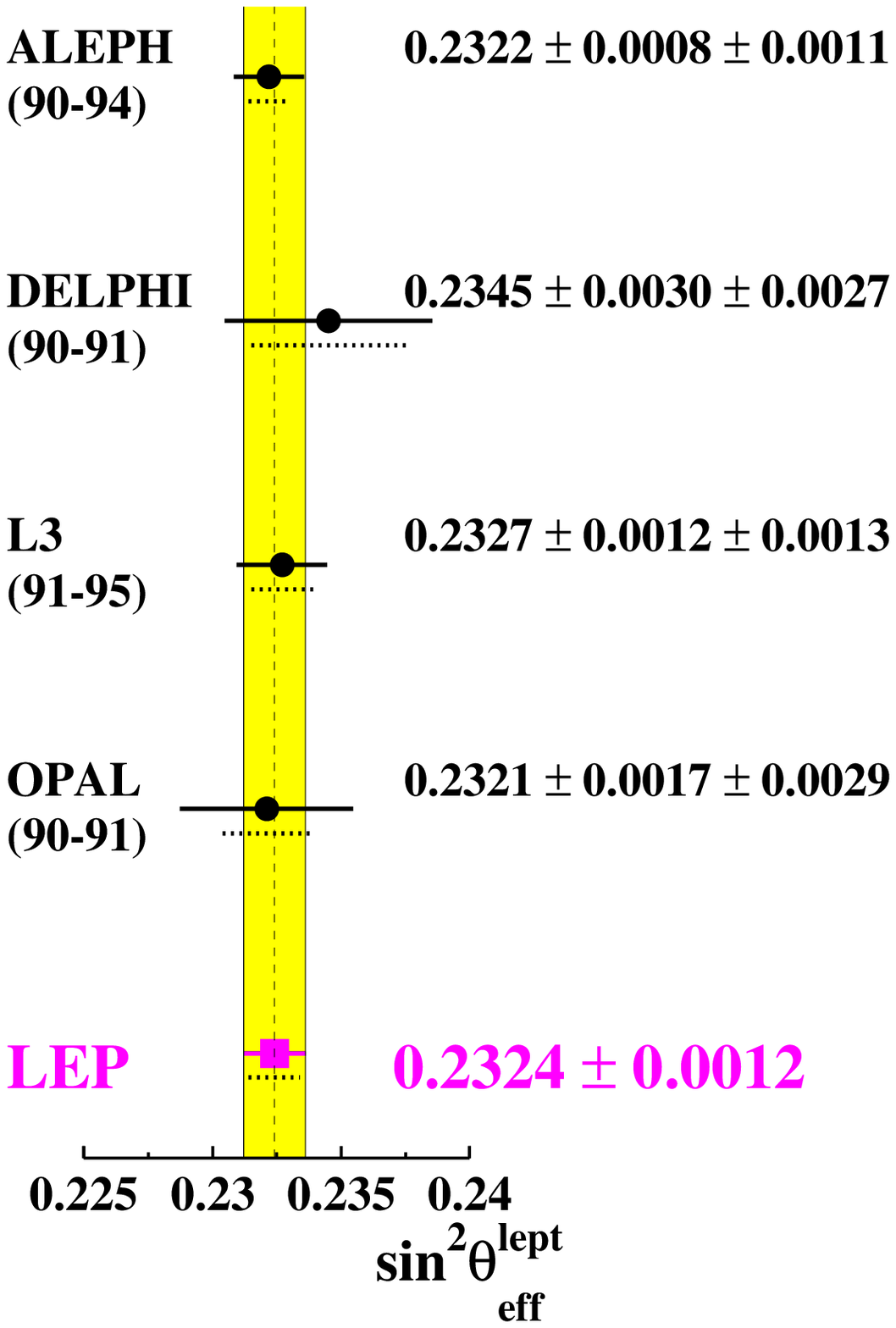}
\caption[Results on $\swsqeffl$ from $\Qfbhad$ measurements]{The input
  values and derived average of $\swsqeffl$ from $\Qfbhad$
  measurements. The total uncertainties are indicated by the solid
  lines, and the systematic contribution to the uncertainties by the
  dotted lines. }
\label{fig:Qfbhadbar} 
\end{center}
\end{figure}

The results from the four LEP experiments have been combined using the
procedure described above. The inputs and the correlation matrix for
the total errors are given in Table~\ref{tab:avQfb}.  The combined
result is:
\begin{eqnarray}
\swsqeffl & = & 0.2324 \pm 0.0012\,,
\label{eq:avQfb}
\end{eqnarray}
where the total error includes a systematic component of 0.0010.  The
experimental results and the average are presented graphically in
Figure~\ref{fig:Qfbhadbar}.

The values of $\swsqeffl$ given here for a particular experiment can
be correlated with the measurement of \Abb\ using the jet charge
method in the same experiment and the same years' data. The
correlation coefficient can be up to 25\% for one experiment.
However, the overall correlation between the average value of
$\swsqeffl$ given here and the average value of \Abb\ has been
estimated to be less than 4\%, taking into account the additional
significant contribution of lepton tag measurements to \Abb, and the
fact that the DELPHI and OPAL inclusive hadronic charge asymmetry
measurements only use 1990--91 data.  A 4\% correlation has a
negligible effect when determining a global combined $\swsqeffl$ value
and in the $\SM$ fits.

\begin{table}[tb]
\begin{center}
\renewcommand{\arraystretch}{1.2}
\begin{tabular}{|ll||c|cccc|}
\hline
Experiment & & $\swsqeffl$ & \multicolumn{4}{|c|}{Correlations} \\
\hline
\hline
ALEPH &(1990-94)&$0.2322\pm0.0008\pm0.0011$& $1.00$ & $    $ & $    $ & $    $\\
DELPHI&(1990-91)&$0.2345\pm0.0030\pm0.0027$& $0.12$ & $1.00$ & $    $ & $    $\\
L3    &(1991-95)&$0.2327\pm0.0012\pm0.0013$& $0.27$ & $0.13$ & $1.00$ & $    $\\
OPAL  &(1990-91)&$0.2321\pm0.0017\pm0.0029$& $0.14$ & $0.37$ & $0.15$ & $1.00$\\
\hline
LEP Average & &$0.2324\pm0.0007\pm0.0010$& \multicolumn{4}{|c|}{ } \\
\hline
\end{tabular}
\caption[$\swsqeffl$ from inclusive hadronic charge asymmetry]{
  Summary of the determination of $\swsqeffl$ from inclusive hadronic
  charge asymmetries at LEP. For each experiment, the first error is
  statistical and the second systematic. The latter is dominated by
  fragmentation and hadronisation uncertainties.  Also listed is the
  `minimum overlap' correlation matrix for the total errors, summing
  statistical and systematic uncertainties in quadrature, used in the
  final average of $\Qfbhad$ results.}
\label{tab:avQfb}
\end{center}
\end{table}

\chapter{Z Boson Properties and Effective Couplings}
\label{chap:Z+coup}

The final combined Z-pole results as derived from the SLD and LEP
measurements, including their correlations, constitute the main result
of this report.  The definitions of the pseudo-observables describing
the resonance properties of the heavy Z boson have been introduced in
Chapter~\ref{sec:intro}. The experimental measurements have been
discussed in Chapters~\ref{chap:lsafb} to~\ref{sec:lq} and are briefly
summarised again here, including an assessment of global correlated
uncertainties.  Based on this input, physics analyses are presented
showing clearly the high precision obtained in the measurement of Z
resonance parameters, and the resulting predictive power of the Z-pole
measurements.

The partial decay widths and the decay branching fractions of the Z
boson are presented in Section~\ref{chap:partrafo}, obtained from the
results of Chapters~\ref{chap:lsafb} and~\ref{sec:hq} using simple
parameter transformations.  An important aspect is the determination
of the number of light neutrino species, a crucial result also in
astrophysics and cosmology.  The effective neutral weak current
couplings, such as the asymmetry parameters and the effective coupling
constants, are derived in Section~\ref{chap:coupling} in largely
model-independent analyses.  Tests of lepton universality and
comparisons with expectations in the framework of the Standard Model
($\SM$) are included.  In a first step towards the $\SM$, the
effective $\rhof$ parameters and the effective electroweak mixing
angles $\swsqefff$ are derived.  As described in
Section~\ref{sec:coup:radcor}, the high precision of the Z-pole data
allows stringent tests of radiative corrections, which are now
unambiguously demonstrated to exist beyond QED.  Analyses and tests
within the constraining framework of the $\SM$, such as the indirect
determination of the mass of the top quark and the mass of the $\SM$
Higgs boson, are deferred to Chapter~\ref{chap:MSM}.  Predictions of
many observables within the $\SM$ framework are reported in
Appendix~\ref{app:SM:preds}.

\section{Summary of Z-Pole Results}
\label{chap:corr}

\subsection{Overview}

The final combined Z-pole results are presented at the following
locations:
\begin{itemize}
\item Chapter~\ref{chap:lsafb}: Z lineshape and leptonic
  forward-backward asymmetries from LEP in
  Table~\ref{tab:lsafbresult};
\item Chapter~\ref{sec-ALR}: Left-right and leptonic left-right
  forward-backward asymmetries from SLD in Table~\ref{tab:alr:result}
  and Equation~\ref{eq:al:result};
\item Chapter~\ref{sec-TP}: Tau polarisation from LEP in
  Equations~\ref{eq:ptau:At}, \ref{eq:ptau:Ae} and~\ref{eq:ptau:Al};
\item Chapter~\ref{sec:hq}: Heavy quark flavour electroweak results
  from SLD and LEP in Table~\ref{tab:14parres} and
  Table~\ref{tab:14parcor};
\item Chapter~\ref{sec:lq}: Inclusive hadronic charge asymmetry from
  LEP in Equation~\ref{eq:avQfb}.
\end{itemize}
The interpretation of these pseudo observables describing the
properties of the Z resonance is largely model independent.  Based on
the discussions presented in Chapter~\ref{sec:intro_SM_remnants}, the
few underlying assumptions concerning the event samples selected for
the measurements and the interpretation of the results are:
\begin{itemize}
\item Associated ZH production is negligible (implying Higgs-boson
masses in excess of about $50~\GeV$)~\cite{LEPSMHIGGS};
\item Contributions from non-resonant processes such as 4-fermion
production are described by the $\SM$~\cite{\fourferm} - or are at
least centre-of-mass energy independent close to the Z-pole, as
discussed in Chapter~\ref{chap:lsafb};
\item Effects of a second heavy neutral boson (Z$'$) are
negligible~\cite{\Zprime};
\item Effects of the strong interaction in heavy flavour production,
namely for asymmetries needed to extract the pole asymmetries
(Section~\ref{sec:hf_qcdcor}) and for partial widths needed to extract
effective coupling constants (Section~\ref{sec:intro_xrpar}) are
correctly described by QCD;
\item Electromagnetic radiative effects are described by QED to the
required level of precision.
\end{itemize}
All these points are either well supported by the cited experimental
results, or are believed to be well-understood theoretically.

All correlations within each group of measurements have been discussed
in the previous chapters and included in the correlation matrices.
The majority of these correlations concern the heavy quark flavour
measurements, where most quantities have correlations exceeding 10\%
with four or more others, as described in Table~\ref{tab:14parcor}.
In addition, important correlations exist between the lineshape
parameters ($\GZ$, $\shad$) and ($\Afbze$,$\Ree$), as described in
Table~\ref{tab:lsafbresult}.

Considering possible correlations between results extracted from
different groups of measurements, including the SLC beam polarisation,
the QCD correction for quark-pair final-state asymmetries, and the
correlation between inclusive and tagged heavy-flavour asymmetries,
only the uncertainty in the SLC beam polarisation creates a
non-negligible effect.  Thus the following additional correlation
coefficients $\calC(\cAl,\cAq)$ between the results on $\cAl$
(Chapter~\ref{sec-ALR}) and on $\cAq$, (Chapter~\ref{sec:hq}) arise
and are taken into account:
\begin{eqnarray}
\calC(\cAl,\cAb) & = & +0.09 \label{eq:coup:c-lb}\\
\calC(\cAl,\cAc) & = & +0.05 \label{eq:coup:c-lc}\,.
\end{eqnarray}
These correlations modify values of quantities derived from the
combined averages at the level of several \% of the respective total
uncertainty.

Even though the various sets of parameters representing the Z-pole
measurements are constructed in such a way as to minimise correlations
between sets and inside sets of parameters, the correlations exceed
10\% in a few sets and thus need to be taken into account for any
precision analysis using these final Z-pole results.

\section{Z-Boson Decay Widths and Branching Fractions}
\label{chap:partrafo}

As discussed in Chapter~\ref{chap:lsafb}, the electroweak measurements
are quoted in terms of experimentally motivated pseudo-observables
defined such that correlations between them are reduced.  Other, more
familiar pseudo-observables describing Z-boson production and decays,
such as leptonic pole cross-sections, Z-boson partial decay widths and
branching fractions, are obtained through simple parameter
transformations.

Assuming lepton universality, the leptonic pole cross-section
$\slept$, defined in analogy to the hadronic pole cross-section, is
measured to be:
\begin{eqnarray}
\slept & \equiv & {12\pi\over\MZ^2}{\Gll^2\over\GZ^2} 
       ~    =   ~ \frac{\shad}{\Rl}
       ~    =   ~ 2.0003\pm0.0027~\mathrm{nb}\,,
\end{eqnarray}
in very good agreement with the $\SM$ expectation.  Note that this
purely leptonic quantity has a higher sensitivity to $\alfmz$ than any
of the hadronic Z-pole observables, as discussed in
Section~\ref{sec:msm:msm}.

\subsection{Z-Boson Decay Parameters}

The partial $\Zzero$ decay widths are summarised in
Table~\ref{tab:width}.  Note that they have larger correlations than
the original set of results reported in Table~\ref{tab:lsafbresult}.
If lepton universality is imposed, a more precise value of $\Ghad$ is
obtained, because $\Gee$ in the relation between the hadronic pole
cross-section and the partial widths is replaced by the more precise
value of $\Gll$.  The $\Zzero$ branching fractions, \ie, the ratios
between each partial decay width and the total width of the Z, are
shown in Table~\ref{tab:brfrac}.

In order to test lepton universality in Z decays quantitatively, the
ratios of the leptonic partial widths or equivalently the ratios of
the leptonic branching fractions are calculated. The results are:
\begin{eqnarray}
\frac{\Gmumu  }{\Gee} & = & \frac{B(\Zzero\to\mumu  )}{B(\Zzero\to\ee)}
                    ~ = ~ 1.0009\pm0.0028 \\
\frac{\Gtautau}{\Gee} & = & \frac{B(\Zzero\to\tautau)}{B(\Zzero\to\ee)}
                    ~ = ~ 1.0019\pm0.0032
\end{eqnarray}
with a correlation of $+0.63$.  In both cases, good agreement with
lepton universality is observed.  Assuming lepton universality, $\tau$
mass effects are expected to decrease $\Gtautau$ and
B$(\Zzero\to\tautau)$ as quoted here by 0.23\% relative to the light
lepton species e and $\mu$.

\begin{table}[ht] \begin{center} 
\renewcommand{\arraystretch}{1.15}
\begin{tabular} {|l||r||rrrrrrr|}
\hline
Parameter & Average &  \multicolumn{7}{|c|}{Correlations}\\
$\Gff$    & [\MeV]~ &  \multicolumn{7}{|c|}{            }\\
\hline
\hline
\multicolumn{9}{|c|}{Without Lepton Universality} \\
\hline %
 & & $\Ghad$ & $\Gee$ & $\Gmumu$ & $\Gtautau$ & $\Gbb$ & $\Gcc$ & $\Ginv$ \\
\hline %
$\Ghad$    & 1745.8 $\pzz\pm$ 2.7$\pzz$ &    1.00 & \multicolumn{6}{c|}{} \\
$\Gee$     &   83.92 $\pz\pm$ 0.12$\pz$ & $-$0.29&1.00& \multicolumn{5}{c|}{} \\ 
$\Gmumu$   &   83.99 $\pz\pm$ 0.18$\pz$ &   ~0.66& $-$0.20&1.00&\multicolumn{4}{c|}{} \\
$\Gtautau$ &   84.08 $\pz\pm$ 0.22$\pz$ &    0.54& $-$0.17&  ~0.39&1.00& \multicolumn{3}{c|}{}\\  
$\Gbb$     &  377.6 $\pzz\pm$ 1.3$\pzz$ &    0.45& $-$0.13&   0.29&  ~0.24&1.00&\multicolumn{2}{c|}{}\\
$\Gcc$     &  300.5 $\pzz\pm$ 5.3$\pzz$ &    0.09& $-$0.02&   0.06&   0.05&$-$0.12&1.00&\multicolumn{1}{c|}{}\\
$\Ginv$    &  497.4 $\pzz\pm$ 2.5$\pzz$ & $-$0.67&    0.78&$-$0.45&$-$0.40&$-$0.30&$-$0.06&1.00\\
\hline %
\hline %
\multicolumn{9}{|c|}{With Lepton Universality} \\
\hline %
 & & $\Ghad$ & $\Gll$ & $\Gbb$ & $\Gcc$ & $\Ginv$ & & \\
\hline %
$\Ghad$ & 1744.4 $\pzz\pm$ 2.0$\pzz$   &  1.00&\multicolumn{6}{c|}{} \\
$\Gll$  &   83.985 $\pm$   0.086       &   ~0.39&1.00&\multicolumn{5}{c|}{} \\ 
$\Gbb$  &  377.3 $\pzz\pm$ 1.2$\pzz$   &   ~0.35&~0.13&1.00&\multicolumn{4}{c|}{} \\
$\Gcc$  &  300.2 $\pzz\pm$ 5.2$\pzz$   &   ~0.06&~0.03&$-$0.15&1.00& \multicolumn{3}{c|}{}\\  
$\Ginv$ &  499.0 $\pzz\pm$ 1.5$\pzz$   & $-$0.29&~0.49&$-$0.10&$-$0.02&1.00&\multicolumn{2}{c|}{}\\
\hline %
\end{tabular} 
\caption[Partial $\Zzero$ widths] {\label{tab:width} Partial $\Zzero$
decay widths, derived from the results of
Tables~\ref{tab:lsafbresult}, \ref{tab:14parres}
and~\ref{tab:14parcor}. The width denoted as $\leptlept$ is that of a
single charged massless lepton species.  The width to invisible
particles is calculated as the difference of total and all other
partial widths.  }
\end{center} 
\end{table}

\subsection{Invisible Width and Number of Light Neutrino Species}

\begin{figure}[ht]
\begin{center}
\includegraphics[width=0.9\linewidth,clip=true,origin=br,bb=0 265 540 800]{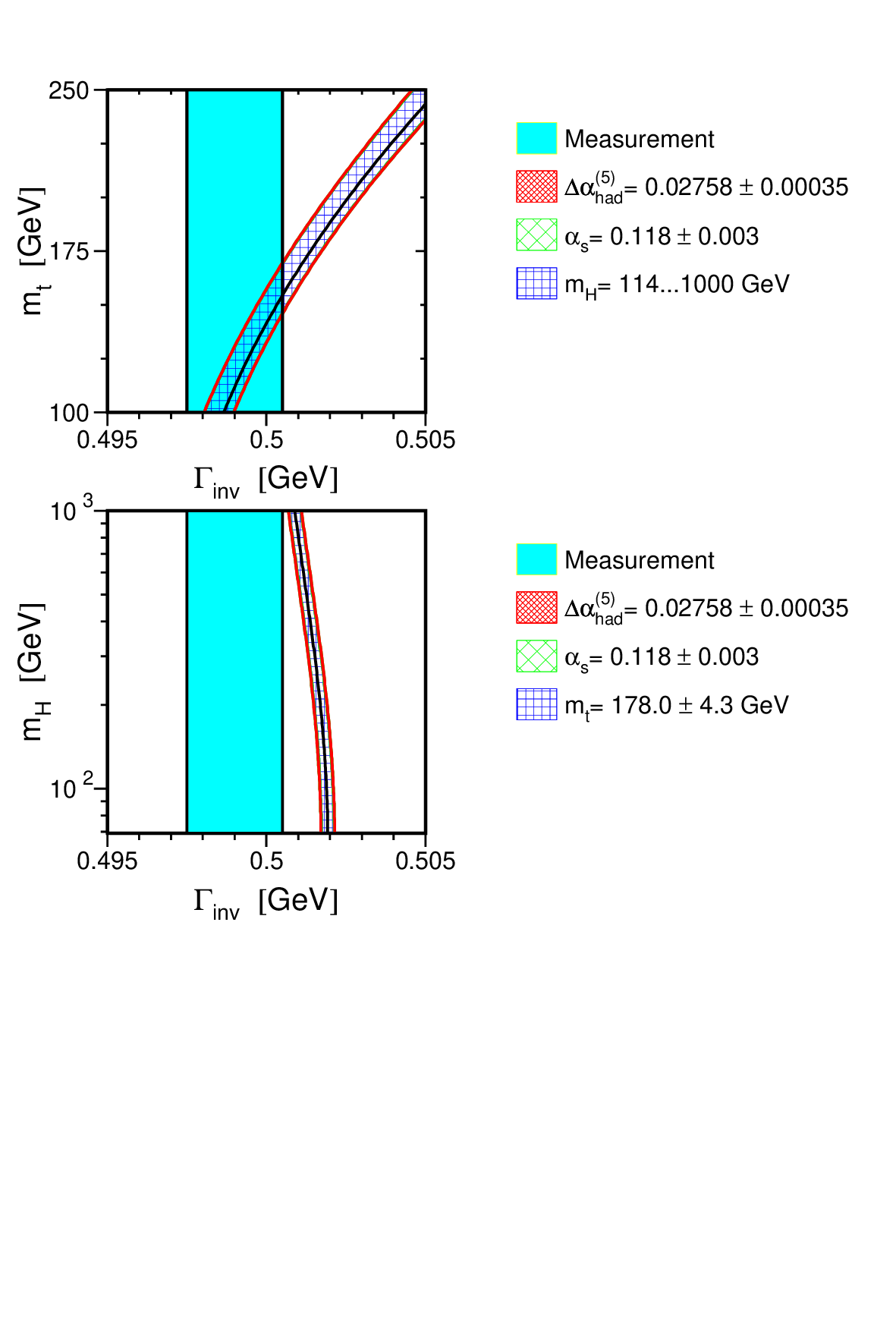}
\caption[Top and Higgs sensitivity of $\Ginv$] {Comparison of the LEP
combined result on $\Ginv$ with the $\SM$ prediction as a function of
(top) the top quark mass, and (bottom) the Higgs boson mass.  The
measurement with its uncertainty is shown as the vertical band.  The
width of the $\SM$ band arises due to the uncertainties in $\dalhad$,
$\alfmz$, $\MH$ and $\Mt$ in the ranges indicated.  The total width of
the band is the linear sum of these uncertainties.  }
\label{fig:coup:inv} 
\end{center}
\end{figure}

The invisible width, $\Ginv=\GZ-(\Ghad+\Gee+\Gmumu+\Gtautau)$, is also
shown in Table~\ref{tab:width}.  The branching fraction to invisible
particles, reported in Table~\ref{tab:brfrac}, is derived by
constraining the sum of the inclusive hadronic, leptonic and invisible
branching fractions to unity, and therefore does not constitute an
independent result.  The result on $\Ginv$ is compared to the $\SM$
expectation calculated as a function of $\Mt$ and $\MH$ in
Figure~\ref{fig:coup:inv}. It shows a small deficit of about
$2.7~\MeV$ or 1.8 standard deviations compared to the $\SM$
expectation calculated for $\Mt=178~\GeV$, mainly reflecting the
observation that the hadronic pole cross-section is slightly larger
than expected.

The limit on extra, non-standard contributions to the invisible width,
\ie, not originating from $\Zzero\to\nu\overline{\nu}$, is calculated
by taking the difference between the value given in
Table~\ref{tab:width} and the $\SM$ expectation of
$\left({\Ginv}\right)_{\SM}=501.7\pm0.2^{+0.1}_{-0.9}~\MeV$, where the
first error is due to the uncertainties in the $\SM$ input parameters
and the second one is due to the unknown mass of the Higgs boson,
taken to be between $114~\GeV$ and $1000~\GeV$ with a central value of
$150~\GeV$.  This gives $\Ginvx\,=\,-2.7^{+1.8}_{-1.5}~\MeV$, or
expressed as a limit, $\Delta\Ginvx < 2.0~\MeV$ at 95\% CL.  This
limit is conservatively calculated allowing only values of $\Ginv$
above the minimal value of the $\SM$ prediction for $\MH=1000~{\GeV}$.
In the same way, upper limits on non-standard contributions to other
$\Zzero$ decays can be calculated and are summarized in
Appendix~\ref{sec:NP-Z-decays}.

\begin{table}[ht] \begin{center}
\renewcommand{\arraystretch}{1.15}
\begin{tabular} {|l||r||rrrrrrr|}
\hline
Parameter & Average &  \multicolumn{7}{|c|}{Correlations}\\
$B(\Zzero\to\ff)$   & [\%]~~~ &  \multicolumn{7}{|c|}{            }\\
\hline
\hline
\multicolumn{9}{|c|}{Without Lepton Universality} \\
\hline %
 & & $\qq$ & $\ee$ & $\mumu$ & $\tautau$ & $\bb$ & $\cc$ & $\inv$ \\
\hline %
$\qq$     & 69.967 $\pz\pm$ 0.093$\pz$ & 1.00&\multicolumn{6}{c|}{} \\
$\ee$     &  3.3632 $\pm$   0.0042     & $-$0.76&1.00&\multicolumn{5}{c|}{} \\ 
$\mumu$   &  3.3662 $\pm$   0.0066     &   ~0.59& $-$0.50&1.00&\multicolumn{4}{c|}{} \\
$\tautau$ &  3.3696 $\pm$   0.0083     &    0.48& $-$0.40&  ~0.33&1.00& \multicolumn{3}{c|}{}\\  
$\bb$     & 15.133 $\pz\pm$ 0.050$\pz$ &    0.40& $-$0.30&   0.24&  ~0.19&1.00&\multicolumn{2}{c|}{}\\
$\cc$     & 12.04 $\pzz\pm$ 0.21$\pzz$ &    0.08& $-$0.06&   0.05&   0.04&$-$0.13&1.00&\multicolumn{1}{c|}{}\\
\hline
$\inv$    & 19.934 $\pz\pm$ 0.098$\pz$ & $-$0.99&    0.75&$-$0.63&$-$0.54&$-$0.40&$-$0.08&1.00\\
\hline %
\hline %
\multicolumn{9}{|c|}{With Lepton Universality} \\
\hline %
 & & $\qq$ & $\leptlept$ & $\bb$ & $\cc$ & $\inv$ & & \\
\hline %
$\qq$     & 69.911 $\pz\pm$ 0.057$\pz$ & 1.00&\multicolumn{6}{c|}{} \\
$\leptlept$
          &  3.3658 $\pm$   0.0023     & $-$0.29&1.00&\multicolumn{5}{c|}{} \\
$\ee,\mumu,\tautau$ 
          & 10.0899 $\pm$   0.0068     & $-$0.29&1.00&\multicolumn{5}{c|}{} \\
$\bb$     & 15.121 $\pz\pm$ 0.048$\pz$ &   ~0.26&$-$0.08&1.00&\multicolumn{4}{c|}{} \\
$\cc$     & 12.03 $\pzz\pm$ 0.21$\pzz$ &   ~0.05&$-$0.01&$-$0.16&1.00& \multicolumn{3}{c|}{}\\   
\hline
$\inv$    & 20.000 $\pz\pm$ 0.055$\pz$ & $-$0.99&  ~0.18&$-$0.25&$-$0.05&1.00&\multicolumn{2}{c|}{}\\ 
\hline %
\end{tabular} 
\caption[$\Zzero$ branching fractions] {\label{tab:brfrac} $\Zzero$
branching fractions, derived from the results of
Tables~\ref{tab:lsafbresult}, \ref{tab:14parres}
and~\ref{tab:14parcor}.  The branching fraction denoted as $\leptlept$
is that of a single charged massless lepton species.  The branching
fraction to invisible particles is fully correlated with the sum of
the branching fractions of leptonic and inclusive hadronic decays.  }
\end{center}
\end{table}

Assuming only $\SM$ particles as Z decay products, the invisible
Z-decay width determines the number $N_\nu$ of light neutrinos
species: $\Ginv\,=\,N_\nu\,\Gnn$.  Since the ratio $\Ginv/\Gll$ is
experimentally determined with higher precision than $\Ginv$, and the
$\SM$ prediction of $\Gnn/\Gll$ shows a reduced dependence on the
unknown $\SM$ parameters, the number of neutrinos is derived from:
\begin{eqnarray} 
\Rinv & \equiv &  \frac{\Ginv}{\Gll}
      ~ = ~  N_\nu \left(\frac{\Gnn}{\Gll}\right)_{\SM}\,.
\end{eqnarray}
Recall that in case of lepton universality, $\Gll$ is defined as the
partial decay width for massless leptons, and the correction for the
$\tau$ mass is applied explicitly in the analysis.

The $\SM$ value for the ratio of the partial widths to neutrinos and
to charged leptons is 1.99125$\pm$0.00083, where the uncertainty
arises from variations of the top quark mass within its experimental
error, $\Mt=178.0\pm4.3~\GeV$, and of the Higgs mass within
$100~\GeV\,<\,\MH\,<\,1000~\GeV$.  Assuming lepton universality, the
measured value of $\Rinv$ is:
\begin{eqnarray} 
\Rinv & = & 5.943\pm0.016\,,
\end{eqnarray}
and the corresponding number of light neutrino species is therefore
determined to be:
\begin{eqnarray}
N_{\nu} & = & 2.9840 \pm 0.0082\,,
\label{eq:n_nu}
\end{eqnarray}
about 2.0 standard deviations smaller than three, driven by the
observed value of $\Ginv$.  This result fixes the number of fermion
families with light neutrinos to the observed three.  The
decomposition of the error on $N_{\nu}$ is given by:
\begin{eqnarray}
\delta N_{\nu} 
          & \simeq & 10.5 \frac{\delta n_{\mathrm{had}}}{n_{\mathrm{had}}} 
              \oplus  3.0 \frac{\delta n_{\mathrm{lep}}}{n_{\mathrm{lep}}} 
              \oplus  7.5 \frac{\delta \calL}{\calL} \,,
\end{eqnarray}
where ${\delta n_{\mathrm{had}}}/{n_{\mathrm{had}}}$, ${\delta
  n_{\mathrm{lep}}}/{n_{\mathrm{lep}}}$ and $\delta\calL/\calL$ denote
  respectively the total errors on the number $n$ of selected hadronic
  and leptonic events, and cross-section scale uncertainties from the
  luminosity determination, while $\oplus$ denotes addition in
  quadrature. The luminosity theory error of 0.061\% is one of the
  largest contributions to the total error on the number of neutrinos,
  causing an error of 0.0046 on $N_{\nu}$.

\clearpage 

\section{Effective Couplings of the Neutral Weak Current}
\label{chap:coupling}

The experimental measurements and results on electroweak Z-pole
observables are now used to derive values for the effective couplings
of the neutral weak current at the Z pole, namely: the asymmetry
parameters $\cAf$ in Section~\ref{sec:coup:asym}, the effective
coupling constants $(\gvf,\gaf)$ and $(\grf,\glf)$ in
Section~\ref{sec:coup:ga-gv}, the $\rhof$ parameters and the effective
electroweak mixing angles $\swsqefff$ in
Section~\ref{sec:coup:rho-sef2}, and the leptonic effective
electroweak mixing angle $\swsqeffl$ in
Section~\ref{sec:coup:sef2lept}.  The partial width results determine
the overall scale of the effective coupling constants, while the
asymmetry results determine their ratio.  The results of these largely
model-independent analyses are compared to the expectations within the
framework of the $\SM$, thereby testing its validity.  A concluding
discussion of these analyses is given in Section~\ref{sec:coup:disc},
with special emphasis on the observation of non-QED electroweak
radiative corrections in Section~\ref{sec:coup:radcor}.

The inputs consist of the results presented in
Chapter~\ref{chap:lsafb} to~\ref{sec:lq} and summarised in
Section~\ref{chap:corr}.  The derived couplings are determined in fits
to these input results, based on the simple expressions, listed in
Section~\ref{sec:intro_ew}, of the input observables in terms of the
couplings to be determined.  Input observables such as $\MZ$, $\GZ$,
$\shad$ or $\Rfz$, which cannot be expressed by the asymmetry
parameters or the couplings to be determined, are allowed to vary in
the fits as well.

For the determination of the leptonic couplings, including tests of
lepton universality, the results of
Chapters~\ref{chap:lsafb}, \ref{sec-ALR} and~\ref{sec-TP} are used.
For the determination of quark couplings, the results presented in
Chapter~\ref{sec:hq} are included as well and lepton universality is
assumed in the analysis.  In the analysis for the leptonic effective
electroweak mixing angle, its determination based on the hadronic
charge asymmetry, Chapter~\ref{sec:lq}, is also added.

In general, the results which have been obtained without imposing lepton
universality are used as inputs. However, when quarks and leptons are
considered in a joint analysis, the issue is no longer one of testing
lepton universality, hence for leptonic pseudo-observables, the lepton
universality results are used, and the correlations listed in
Equations~\ref{eq:coup:c-lb} and~\ref{eq:coup:c-lc} are included.

\subsection[The Asymmetry Parameters \protect$\cAf$]%
{The Asymmetry Parameters \boldmath{$\cAf$}}
\label{sec:coup:asym}

The polarised electron beams at SLC allow the SLD collaboration to
measure the asymmetry parameters $\cAf$ directly by analysing the
left-right and left-right forward-backward asymmetry, $\ALRz=\cAe$ and
$\AFBLRf=(3/4)\cAf$.  The analyses of the tau polarisation and its
forward-backward asymmetry at LEP determine $\cAt$ and $\cAe$
separately.  The forward-backward pole asymmetries,
$\Afbzf=(3/4)\cAe\cAf$, constrain the product of two asymmetry
parameters.  The measurements are performed separately for all three
charged lepton species and the heavy-quark flavours b and c.

The results on the leptonic asymmetry parameters derived from various
measurements which do not involve quark asymmetry parameters are
reported in Table~\ref{tab:coup:al:separate}, with combined values
including correlations reported in Table~\ref{tab:coup:al}.  The
values of the asymmetry parameters obtained for the three lepton
species agree well.  Under the assumption of neutral-current lepton
universality, the combined result is:
\begin{eqnarray}
\cAl & = & 0.1501\pm0.0016\,.
\label{eq:coup:al}
\end{eqnarray}
This average has a $\chidf$ of 7.8/9, corresponding to a
probability of 56\%.

The analysis is now expanded to include the results on quark-pair
production.  The values of $\cAq$ and $\Afbzq$, which have been
extracted from realistic observables in Chapter~\ref{sec:hq}, have
been corrected for the QCD and QED effects expected in the $\SM$ and
calculated with ZFITTER~\cite{\ZFITTERref} (see
Section~\ref{sec:hf_qcdcor}). They therefore rest on the same footing
as the corresponding pole quantities for leptons.  As already
discussed in Section~\ref{sec:hqresults} and shown in
Figure~\ref{fig:hfsmcomp1}, the ratio of the forward-backward pole
asymmetries $\Afbzb/\Afbzc=\cAb/\cAc$ agrees well with the ratio of
the direct measurements of the asymmetry parameters $\cAb$ and $\cAc$.

\begin{table}[p]
\begin{center}
\renewcommand{\arraystretch}{1.25}
\begin{tabular}{|c||r@{$\pm$}l|r@{$\pm$}l|r@{$\pm$}l|}
\hline
Parameter & \multicolumn{2}{|c|}{$\Afbzl$} 
          & \multicolumn{2}{|c|}{$\ALRz,\AFBLRl$}
          & \multicolumn{2}{|c|}{$\ptau$ } \\
\hline
\hline
$\cAe$ & $0.139$&$0.012$ & $0.1516$&$0.0021$ &$0.1498$&$0.0049$ \\
$\cAm$ & $0.162$&$0.019$ & $0.142 $&$0.015 $ &\multicolumn{2}{|c|}{---} \\
$\cAt$ & $0.180$&$0.023$ & $0.136 $&$0.015 $ &$0.1439$&$0.0043$ \\
\hline
\end{tabular}
\caption[Comparison of the leptonic asymmetry parameters $\cAl$]
{Comparison of the leptonic asymmetry parameters $\cAl$ using the
electroweak measurements of Tables~\ref{tab:lsafbresult}
and~\ref{tab:alr:result}, and Equations~\ref{eq:ptau:At}
and~\ref{eq:ptau:Ae}. The results derived from $\Afbzl$ are strongly
correlated, with correlation coefficients of $-0.75$, $-0.70$ and
$+0.55$ between e$\mu$, e$\tau$ and $\mu\tau$, respectively. }
\label{tab:coup:al:separate}
\end{center}
\end{table}
\begin{table}[p]
\begin{center}
\renewcommand{\arraystretch}{1.25}
\begin{tabular}{|c||r@{$\pm$}l||rrr|}
\hline
Parameter & \multicolumn{2}{|c||}{Average} 
          & \multicolumn{3}{|c| }{Correlations}    \\
          & \multicolumn{2}{|c||}{ }
          & {$\cAe$} & {$\cAm$}& {$\cAt$}          \\
\hline
\hline
$\cAe$ &$0.1514$&$0.0019$ & $ 1.00$&$     $&$     $ \\
$\cAm$ &$0.1456$&$0.0091$ & $-0.10$&$ 1.00$&$     $ \\
$\cAt$ &$0.1449$&$0.0040$ & $-0.02$&$ 0.01$&$ 1.00$ \\
\hline
\end{tabular}
\caption[Results on the leptonic asymmetry parameters $\cAl$] {Results
on the leptonic asymmetry parameters $\cAl$ using the 14 electroweak
measurements of Tables~\ref{tab:lsafbresult} and~\ref{tab:alr:result},
and Equations~\ref{eq:ptau:At} and~\ref{eq:ptau:Ae}.  The combination
has a $\chidf$ of 3.6/5, corresponding to a probability of 61\%.}
\label{tab:coup:al}
\end{center}
\end{table}
\begin{table}[p]
\begin{center}
\renewcommand{\arraystretch}{1.25}
\begin{tabular}{|c||r@{$\pm$}l|r@{$\pm$}l||r@{$\pm$}l|}
\hline
Flavour $q$  & \multicolumn{2}{|c|}{$\cAq=\frac{4}{3}\frac{\Afbzq}{\cAl}$} 
             & \multicolumn{2}{|c||}{Direct $\cAq$}
             & \multicolumn{2}{|c|}{SM} \\
\hline
\hline
b  &$0.881 $&$0.017 $& $0.923$&$0.020$ & $0.935$&$0.001$ \\
c  &$0.628 $&$0.032 $& $0.670$&$0.027$ & $0.668$&$0.002$ \\
\hline
\end{tabular}
\caption[Determination of the quark asymmetry parameters $\cAq$]
{Determination of the quark asymmetry parameters $\cAq$, based on the
ratio $\Afbzq/\cAl$ and the direct measurement $\AFBLRq$.  Lepton
universality for $\cAl$ is assumed. The correlation between
$4\Afbzb/3\cAl$ and $4\Afbzc/3\cAl$ is $+0.24$, while it is $+0.11$
between the direct measurements $\cAb$ and $\cAc$.  The expectation of
$\cAq$ in the $\SM$ is listed in the last column.}
\label{tab:coup:test}
\end{center}
\end{table}
\begin{table}[p]
\begin{center}
\renewcommand{\arraystretch}{1.25}
\begin{tabular}{|c||r@{$\pm$}l||rrr|}
\hline
Parameter & \multicolumn{2}{|c||}{Average} 
          & \multicolumn{3}{|c|}{Correlations} \\
          & \multicolumn{2}{|c||}{ }
          & {$\cAl$} & {$\cAb$}& {$\cAc$} \\
\hline
\hline
$\cAl$ &$0.1489$&$0.0015$& $ 1.00$& $     $&$     $ \\
$\cAb$ &$0.899 $&$0.013 $& $-0.42$& $ 1.00$&$     $ \\
$\cAc$ &$0.654 $&$0.021 $& $-0.10$& $ 0.15$&$ 1.00$ \\
\hline
\end{tabular}
\caption[Results on the asymmetry parameters $\cAf$]
{Results on the quark asymmetry parameters $\cAq$ and the leptonic
  asymmetry parameter $\cAl$ assuming neutral-current lepton
  universality using the 13 electroweak measurements of
  Tables~\ref{tab:lsafbresult}, \ref{tab:14parres}
  and~\ref{tab:14parcor}, and Equations~\ref{eq:al:result}
  and~\ref{eq:ptau:Al}.  The combination has a $\chidf$ of 4.5/4,
  corresponding to a probability of 34\%. }
\label{tab:coup:aq}
\end{center}
\end{table}

\begin{figure}[p]
\begin{center}
\vskip -1cm
\mbox{\epsfig{file=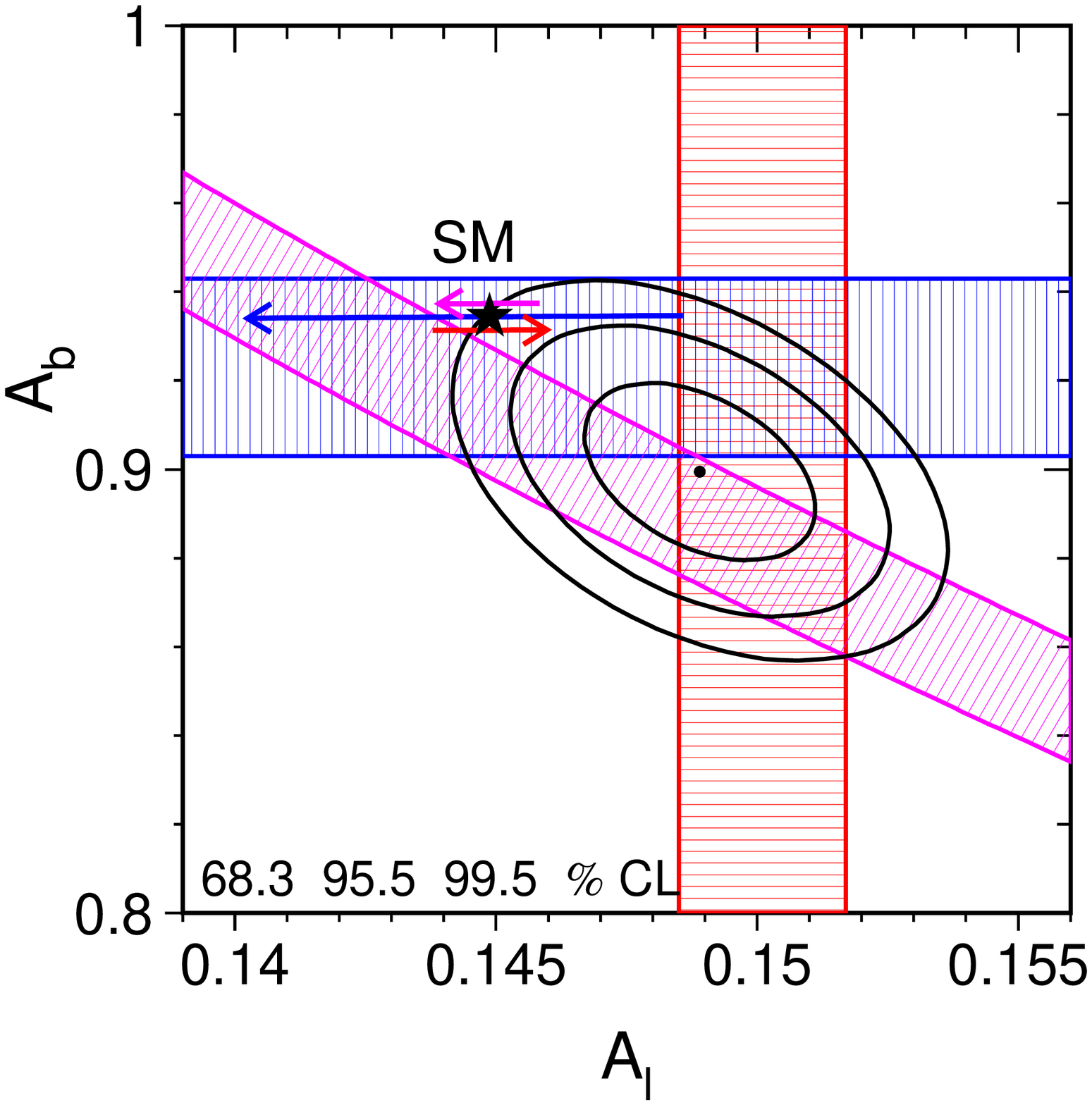,width=0.6\linewidth}} 
\vskip -1.5cm
\mbox{\epsfig{file=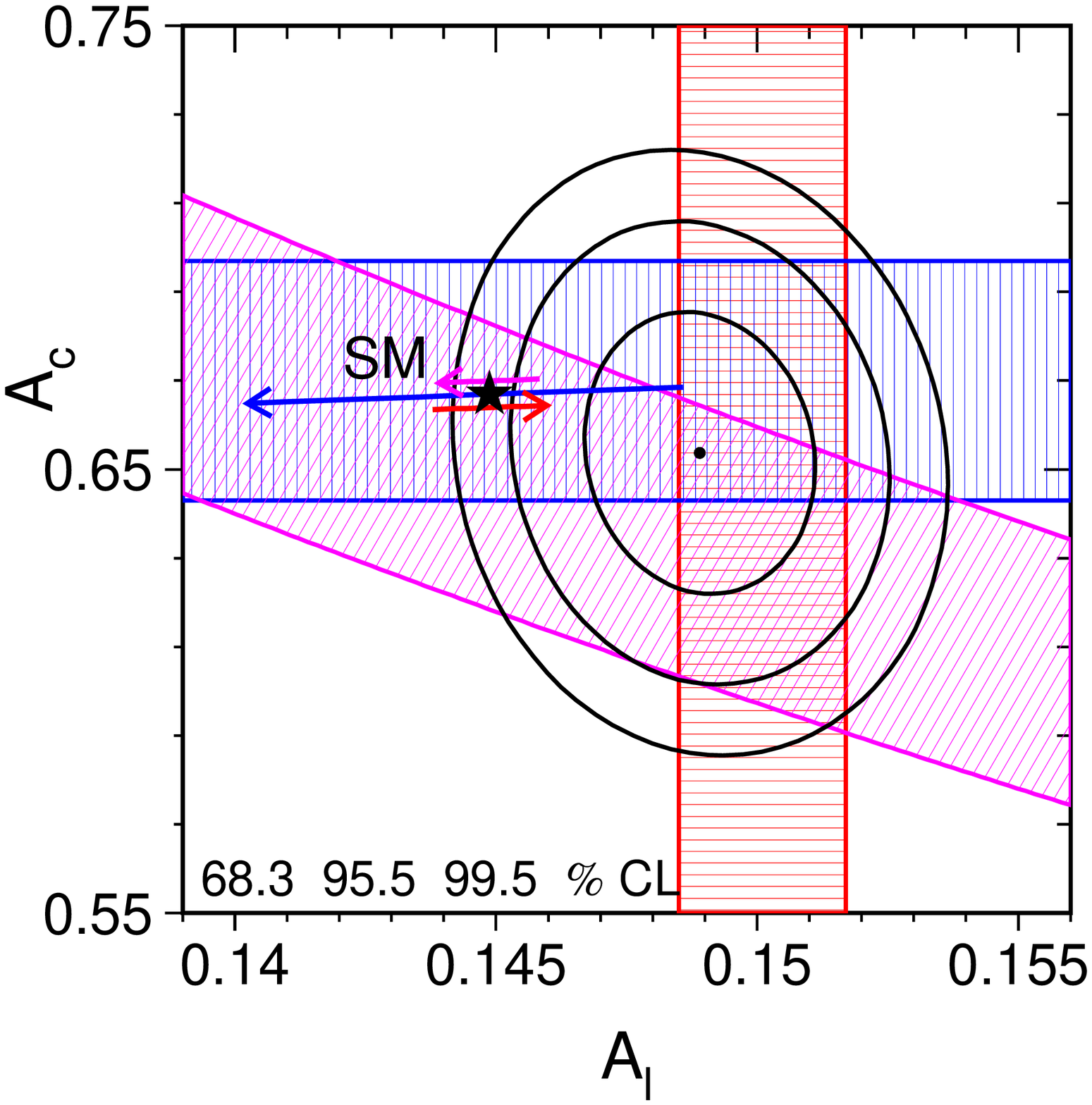,width=0.6\linewidth}} 
\vskip -1cm
\caption[Comparison of the asymmetry parameters $\cAf$] { Comparison
of the measurements of $\cAl$, $\cAq$ and $\Afbzq$ for (top) b-quarks,
and (bottom) c-quarks, assuming lepton universality.  Bands of $\pm1$
standard deviation width in the $(\cAl,\cAq)$ plane are shown for the
measurements of $\cAl$ (vertical band), $\cAq$ (horizontal band), and
$\Afbzq=(3/4)\cAe\cAq$ (diagonal band).  Also shown are the 68\%, 95\%
and 99.5\% confidence level contours for the two asymmetry parameters
resulting from the joint analysis (Table~\ref{tab:coup:aq}).  The
arrows pointing to the right and to the left show the variation in the
$\SM$ prediction for varying $\dalhad$ in the range
$0.02758\pm0.00035$ (arrow displaced vertically upwards), $\MH$ in the
range of $300^{+700}_{-186}~\GeV$, and $\Mt$ in the range
$178.0\pm4.3~\GeV$ (arrow displaced vertically downwards).  All arrows
point in the direction of increasing values of these parameters.  }
\label{fig:coup:aq} 
\end{center}
\end{figure}

Numerical results on the asymmetry parameters $\cAb$ and $\cAc$ are
compared in Table~\ref{tab:coup:test}.  The direct measurements of
both $\cAb$ and $\cAc$ are seen to agree well with $\SM$ expectations.
Each ratio $(4/3)\Afbzq/\cAl$ also determines $\cAq$ indirectly, with
a precision comparable to that of the direct measurements.  Reasonable
agreement between the direct measurement and the ratio is observed for
c quarks.  In the case of b quarks, the ratio $(4/3)\Afbzb/\cAl$ is
lower than the direct measurement of $\cAb$ by 1.6 standard
deviations, and lower than the $\SM$ expectation for $\cAb$ by 3.2
standard deviations.  

The mutual consistency of the measurements of $\cAq$,
$\Afbzq=(3/4)\cAe\cAq$ and $\cAl$ assuming lepton universality is
shown in Figure~\ref{fig:coup:aq}.  The results of the joint analysis
of the leptonic and heavy-flavour measurements in terms of the
asymmetry parameters $\cAf$ are reported in Table~\ref{tab:coup:aq}
and shown as the error ellipses in Figure~\ref{fig:coup:aq}, where the
constraint of lepton universality is also imposed.  Since $\cAl$ and
$\Afbzb$ are already determined with relatively small errors, the
joint analysis primarily improves the determination of the b-quark
asymmetry parameter $\cAb$ and pulls $\cAl$ towards lower values.

As explained in connection with Figure~\ref{fig:afvssin2}, the
hadronic asymmetry parameters, $\cAq$, are very much less sensitive to
$\SM$ parameters than is the case for $\cAl$.  This is a consequence
of the $\SM$ structure of the couplings in terms of the electric
charge $\Qf$ and of the third component of the weak isospin $\Tf$,
see Equations~\ref{eq:gl} to~\ref{eq:ga}.  Particularly compared to
the larger experimental uncertainties of the hadronic measurements,
the $\SM$ predictions for $\cAq$ have negligible dependence on $\SM$
parameters such as $\Mt$, $\MH$ and $\alqed$.  The measured quark
asymmetry parameters $\cAq$ allow the $\SM$ to be tested in a manner
which is insensitive to electroweak radiative corrections or the
knowledge of other $\SM$ parameters.

\subsection{The Effective Coupling Constants}
\label{sec:coup:ga-gv}

The asymmetry parameters $\cAf$ depend only on the ratio $\gvf/\gaf$
of the effective vector and axial-vector coupling constants as shown
in Equation~\ref{eq:cA}.  In contrast, the partial decay widths of the
Z boson determine the sum of the squares of these two coupling
constants (Equation~\ref{eq:Gff}).  The expressions for both
observables are invariant under the exchange
$\gvf\leftrightarrow\gaf$, and only the relative sign between $\gvf$
and $\gaf$ is determined by $\cAf$. The energy dependence of the
forward backward asymmetries measured at LEP resolves the
$\gvf\leftrightarrow\gaf$ ambiguity, and the absolute sign of all
couplings is established by the convention $\gae<0$.  It is thus
possible to disentangle the effective coupling constants $\gvf$ and
$\gaf$ by analysing both the asymmetry measurements as well as the
partial Z decay widths.

\begin{figure}[p]
\begin{center}
\mbox{\epsfig{file=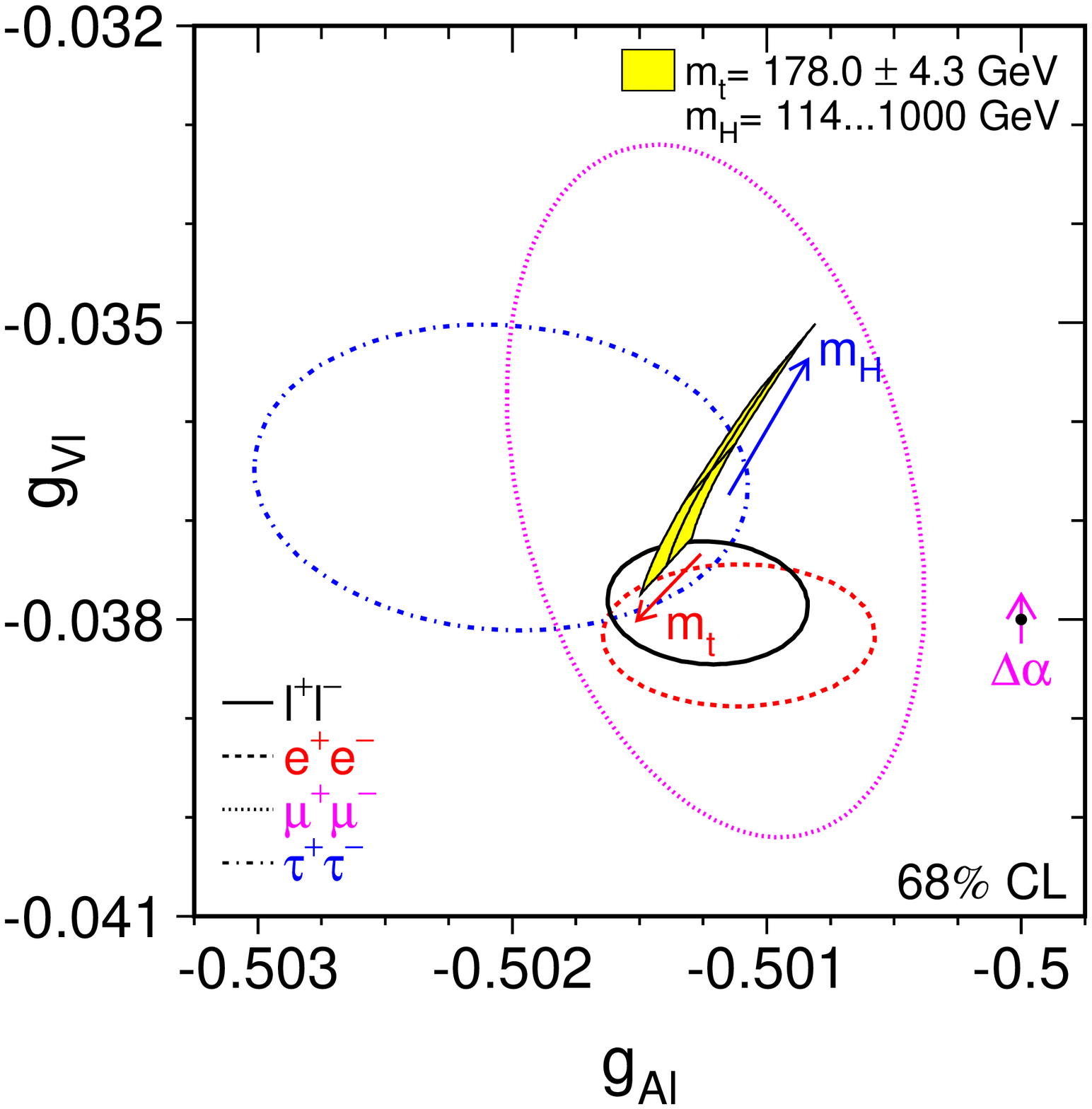,width=0.9\linewidth}}
\caption[Comparison of the effective coupling constants for leptons] {
Comparison of the effective vector and axial-vector coupling constants
for leptons (Tables~\ref{tab:coup:gemt} and~\ref{tab:coup:gl}).  The
shaded region in the lepton plot shows the predictions within the
$\SM$ for $\Mt=178.0\pm4.3~\GeV$ and $\MH=300^{+700}_{-186}~\GeV$;
varying the hadronic vacuum polarisation by
$\dalhad=0.02758\pm0.00035$ yields an additional uncertainty on the
$\SM$ prediction shown by the arrow labeled $\Delta\alpha$. }
\label{fig:coup:gl} 
\end{center}
\end{figure}

For charged leptons and neutrinos, the results on $\gvf$ and $\gaf$
are reported in Table~\ref{tab:coup:gemt}.  The factors
$R_{\mathrm{Af}}$ and $R_{\mathrm{Vf}}$ of Equation~\ref{eq:Gff} which
are used to extract the couplings for charged leptons contain only
small final-state QED corrections, $R_{\mathrm{QED}}=1+3\alqed/4\pi$,
while for neutrinos $R_{\mathrm{QED}}=1$.  The term
$\Delta_{\mathrm{ew/QCD}}$ vanishes for both.  By attributing the
entire invisible decay width of the Z to the production of neutrino
pairs, the magnitude of the effective coupling of the Z boson to
neutrinos can be determined.  Three light neutrino families with equal
effective coupling constants and $\gvn\equiv\gan$ are assumed.  The
comparison of different charged lepton species in the $(\gvl,\gal)$
plane is also shown in Figure~\ref{fig:coup:gl}.  Good agreement is
observed.

\begin{table}[p]
\begin{center}
\renewcommand{\arraystretch}{1.25}
\begin{tabular}{|c||r@{$\pm$}l||rrrrrrr|}
\hline
Parameter & \multicolumn{2}{|c||}{Average} 
          & \multicolumn{7}{|c| }{Correlations} \\
          & \multicolumn{2}{|c||}{ }
          & {$\gan$}
          & {$\gae$} & {$\gamu$} & {$\gatau$} 
          & {$\gve$} & {$\gvmu$} & {$\gvtau$}    \\
\hline
\hline
$\gan\equiv\gvn$ 
        &$+0.5003 $&$0.0012 $&$ 1.00$&$     $&$     $&$     $&$     $&$     $&$     $\\
$\gae$  &$-0.50111$&$0.00035$&$-0.75$&$ 1.00$&$     $&$     $&$     $&$     $&$     $\\
$\gamu$ &$-0.50120$&$0.00054$&$ 0.39$&$-0.13$&$ 1.00$&$     $&$     $&$     $&$     $\\
$\gatau$&$-0.50204$&$0.00064$&$ 0.37$&$-0.12$&$ 0.35$&$ 1.00$&$     $&$     $&$     $\\
$\gve$  &$-0.03816$&$0.00047$&$-0.10$&$ 0.01$&$-0.01$&$-0.03$&$ 1.00$&$     $&$     $\\
$\gvmu$ &$-0.0367 $&$0.0023 $&$ 0.02$&$ 0.00$&$-0.30$&$ 0.01$&$-0.10$&$ 1.00$&$     $\\
$\gvtau$&$-0.0366 $&$0.0010 $&$ 0.02$&$-0.01$&$ 0.01$&$-0.07$&$-0.02$&$ 0.01$&$ 1.00$\\
\hline
\hline
Parameter & \multicolumn{2}{|c||}{Average} 
          & \multicolumn{7}{|c| }{Correlations} \\
          & \multicolumn{2}{|c||}{ }
          & {$\gln$}
          & {$\gle$} & {$\glmu$} & {$\gltau$} 
          & {$\gre$} & {$\grmu$} & {$\grtau$}    \\
\hline
\hline
$\gln$  &$+0.5003 $&$0.0012 $&$ 1.00$&$     $&$     $&$     $&$     $&$     $&$     $\\
$\gle$  &$-0.26963$&$0.00030$&$-0.52$&$ 1.00$&$     $&$     $&$     $&$     $&$     $\\
$\glmu$ &$-0.2689 $&$0.0011 $&$ 0.12$&$-0.11$&$ 1.00$&$     $&$     $&$     $&$     $\\
$\gltau$&$-0.26930$&$0.00058$&$ 0.22$&$-0.07$&$ 0.07$&$ 1.00$&$     $&$     $&$     $\\
$\gre$  &$+0.23148$&$0.00029$&$ 0.37$&$ 0.29$&$-0.07$&$ 0.01$&$ 1.00$&$     $&$     $\\
$\grmu$ &$+0.2323 $&$0.0013 $&$-0.06$&$-0.06$&$ 0.90$&$-0.03$&$-0.09$&$ 1.00$&$     $\\
$\grtau$&$+0.23274$&$0.00062$&$-0.17$&$ 0.04$&$-0.04$&$ 0.44$&$-0.03$&$ 0.04$&$ 1.00$\\
\hline
\end{tabular}
\caption[Results on the effective coupling constants for leptons]
{Results on the effective coupling constants for leptons, using the 14
electroweak measurements of Tables~\ref{tab:lsafbresult}
and~\ref{tab:alr:result}, and Equations~\ref{eq:ptau:At}
and~\ref{eq:ptau:Ae}.  The combination has a $\chidf$ of 3.6/5,
corresponding to a probability of 61\%.}
\label{tab:coup:gemt}
\end{center}
\end{table}
\begin{table}[p]
\begin{center}
\renewcommand{\arraystretch}{1.25}
\begin{tabular}{|c||r@{$\pm$}l||rrr|}
\hline
Parameter & \multicolumn{2}{|c||}{Average} 
          & \multicolumn{3}{|c|}{Correlations} \\
          & \multicolumn{2}{|c||}{ }
          & {$\gn $} & {$\gal$}& {$\gvl$} \\
\hline
\hline
$\gan\equiv
 \gvn$ &$+0.50076$&$0.00076$& $ 1.00$& $     $&$     $ \\
$\gal$ &$-0.50123$&$0.00026$& $-0.48$& $ 1.00$&$     $ \\
$\gvl$ &$-0.03783$&$0.00041$& $-0.03$& $-0.06$&$ 1.00$ \\
\hline
\hline
Parameter & \multicolumn{2}{|c||}{Average} 
          & \multicolumn{3}{|c|}{Correlations} \\
          & \multicolumn{2}{|c||}{ }
          & {$\gln$} & {$\gll$}& {$\grl$} \\
\hline
$\gln$ &$+0.50076$&$0.00076$& $ 1.00$& $     $&$     $ \\
$\gll$ &$-0.26953$&$0.00024$& $-0.29$& $ 1.00$&$     $ \\
$\grl$ &$+0.23170$&$0.00025$& $ 0.22$& $ 0.43$&$ 1.00$ \\
\hline
\end{tabular}
\caption[Results on the effective coupling constants for leptons]
{Results on the effective coupling constants for leptons, using the 14
electroweak measurements of Tables~\ref{tab:lsafbresult}
and~\ref{tab:alr:result}, and Equations~\ref{eq:ptau:At}
and~\ref{eq:ptau:Ae}.  Lepton universality is imposed. The combination
has a $\chidf$ of 7.8/9, corresponding to a probability of 56\%.}
\label{tab:coup:gl}
\end{center}
\end{table}

The combined result under the assumption of neutral-current lepton
universality is reported in Table~\ref{tab:coup:gl}.  The neutrino
coupling is smaller by about 1.8 standard deviations than the $\SM$
expectation listed in Appendix~\ref{app:SM:preds}; this is the same
effect as observed above for $\Ginv$.  The value of $\gal$ is
different from the corresponding Born-level value of $\Tl=-1/2$
by 4.7 standard deviations, indicating the presence of non-QED
electroweak radiative corrections.

Including the heavy-quark measurements and assuming lepton
universality, the couplings of quarks and charged leptons are reported
in Table~\ref{tab:coup:gq}.  As in the case of the leptonic couplings,
final-state corrections affect the partial widths used for determining
the scale of the quark couplings.  Here $R_{\mathrm{Af}} $,
$R_{\mathrm{Vf}}$ and $\Delta_{\mathrm{ew/QCD}}$ of
Equation~\ref{eq:Gff} include also QCD corrections, which are
calculated according to the $\SM$ with ZFITTER~\cite{\ZFITTERref} when
extracting effective quark couplings from the partial widths.
Similarly, the quoted heavy-quark asymmetries have already been
corrected for final-state QCD and QED effects as expected in the
$\SM$, see Section~\ref{sec:hf_qcdcor}.  The uncertainties in the
extracted quark couplings due to the uncertainty in the strong
coupling constant are negligible.

\begin{figure}[p]
\begin{center}
\vskip -1cm
\mbox{\epsfig{file=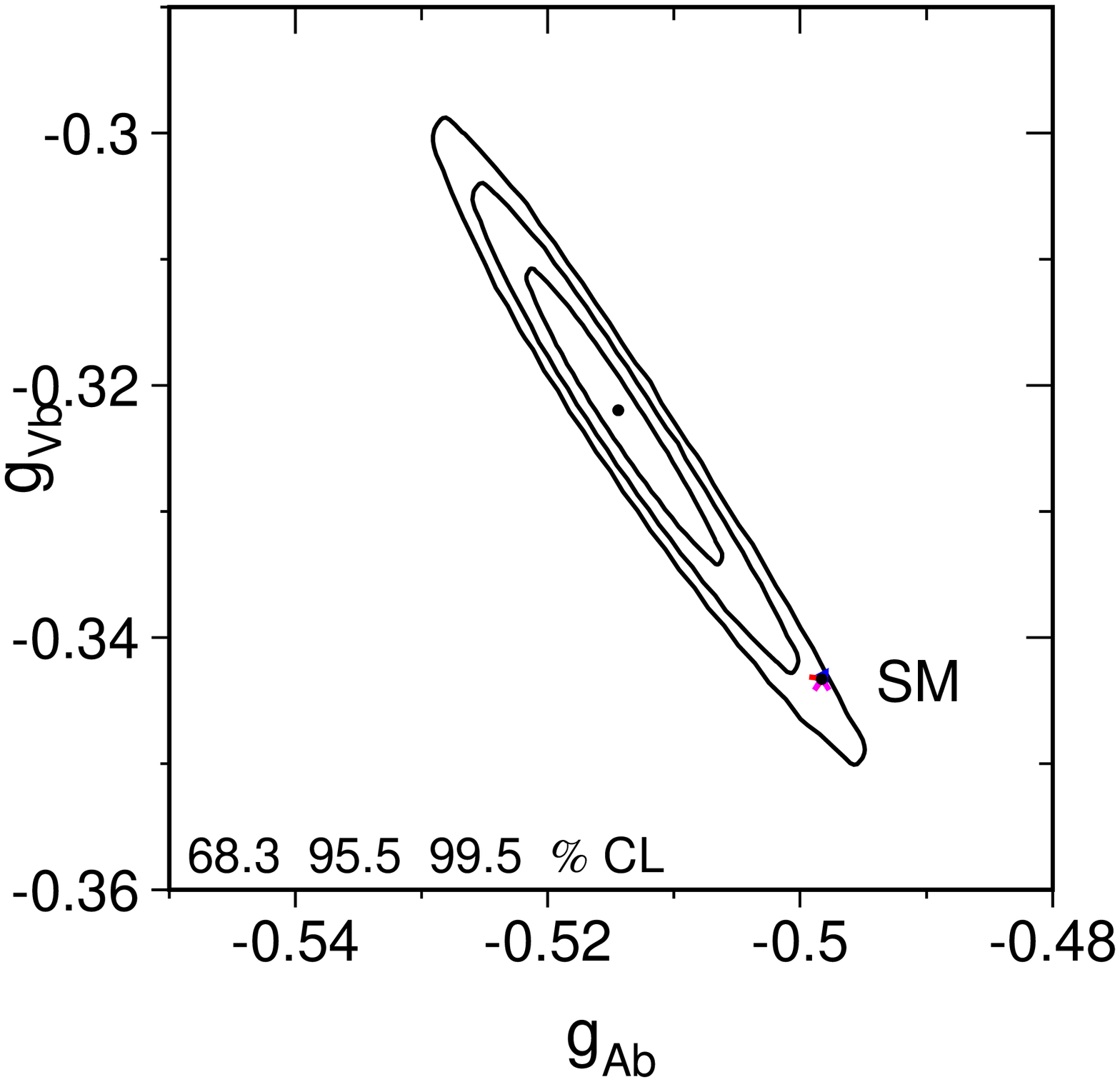,width=0.6\linewidth}} 
\vskip -1cm
\mbox{\epsfig{file=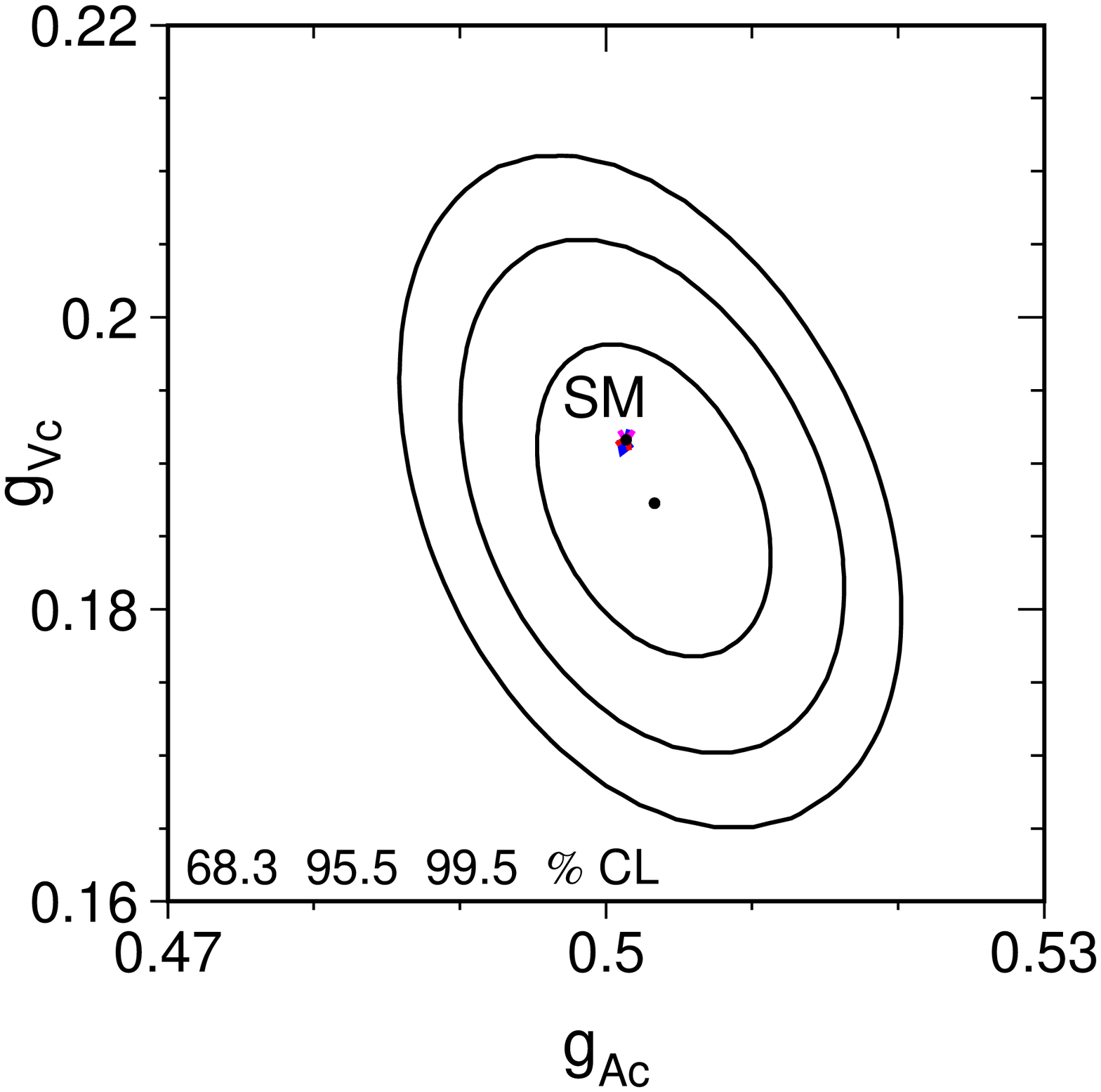,width=0.6\linewidth}} 
\vskip -0.5cm
\caption[Comparison of the effective coupling constants for heavy
quarks] { Comparison of the effective vector and axial-vector coupling
constants for heavy quarks, using results on leptons and assuming
lepton universality (Table~\ref{tab:coup:gq}). Top: b quarks; bottom:
c quarks.  Compared to the experimental uncertainties, the $\SM$
predictions for the heavy quarks b and c have negligible dependence on
the $\SM$ input parameters.}
\label{fig:coup:gq} 
\end{center}
\end{figure}

The vector coupling constant for charged leptons is decreased in
magnitude compared to Table~\ref{tab:coup:gl} as already observed
for the asymmetry parameter $\cAl$ in the previous section.  For the
quark flavours b and c, the results are also shown in
Figure~\ref{fig:coup:gq}.  The strong anti-correlation between the
b-quark couplings arises from the tight constraint on the sum of their
squares due to the measurement of $\Rbz$, which agrees with the $\SM$
prediction.  The apparent deviation of the measured b-quark coupling
constants from the $\SM$ expectation is a direct consequence of the
combined result on $\cAb$ being lower than the $\SM$ expectation as
discussed in the previous section.

\begin{table}[t]
\begin{center}
\renewcommand{\arraystretch}{1.25}
\begin{tabular}{|c||r@{$\pm$}l||rrrrrrr|}
\hline
Parameter & \multicolumn{2}{|c||}{Average} 
          & \multicolumn{7}{|c| }{Correlations} \\
          & \multicolumn{2}{|c||}{ }
          & {$\gan$} 
          & {$\gal$} & {$\gab$} & {$\gac$} 
          & {$\gvl$} & {$\gvb$} & {$\gvc$}          \\
\hline
\hline
$\gan\equiv\gvn$ 
      &$+0.50075$&$0.00077$&$ 1.00$&$     $&$     $&$     $&$     $&$     $&$     $\\
$\gal$&$-0.50125$&$0.00026$&$-0.49$&$ 1.00$&$     $&$     $&$     $&$     $&$     $\\
$\gab$&$-0.5144 $&$0.0051 $&$ 0.01$&$-0.02$&$ 1.00$&$     $&$     $&$     $&$     $\\
$\gac$&$+0.5034 $&$0.0053 $&$-0.02$&$-0.02$&$ 0.00$&$ 1.00$&$     $&$     $&$     $\\
$\gvl$&$-0.03753$&$0.00037$&$-0.04$&$-0.04$&$ 0.41$&$-0.05$&$ 1.00$&$     $&$     $\\
$\gvb$&$-0.3220 $&$0.0077 $&$ 0.01$&$ 0.05$&$-0.97$&$ 0.04$&$-0.42$&$ 1.00$&$     $\\
$\gvc$&$+0.1873 $&$0.0070 $&$-0.01$&$-0.02$&$ 0.15$&$-0.35$&$ 0.10$&$-0.17$&$ 1.00$\\
\hline
\hline
Parameter & \multicolumn{2}{|c||}{Average} 
          & \multicolumn{7}{|c| }{Correlations} \\
          & \multicolumn{2}{|c||}{ }
          & {$\gln$} 
          & {$\gll$} & {$\glb$} & {$\glc$} 
          & {$\grl$} & {$\grb$} & {$\grc$}          \\
\hline
\hline
$\gln$&$+0.50075$&$0.00077$&$ 1.00$&$     $&$     $&$     $&$     $&$     $&$     $\\
$\gll$&$-0.26939$&$0.00022$&$-0.32$&$ 1.00$&$     $&$     $&$     $&$     $&$     $\\
$\glb$&$-0.4182 $&$0.0015 $&$ 0.05$&$-0.27$&$ 1.00$&$     $&$     $&$     $&$     $\\
$\glc$&$+0.3453 $&$0.0036 $&$-0.02$&$ 0.04$&$-0.09$&$ 1.00$&$     $&$     $&$     $\\
$\grl$&$+0.23186$&$0.00023$&$ 0.25$&$ 0.34$&$-0.37$&$ 0.07$&$ 1.00$&$     $&$     $\\
$\grb$&$+0.0962 $&$0.0063 $&$ 0.00$&$-0.33$&$ 0.88$&$-0.14$&$-0.35$&$ 1.00$&$     $\\
$\grc$&$-0.1580 $&$0.0051 $&$ 0.00$&$ 0.08$&$-0.17$&$ 0.30$&$ 0.08$&$-0.13$&$ 1.00$\\
\hline
\end{tabular}
\caption[Results on the effective coupling constants] {Results on the
effective coupling constants for leptons and quarks assuming
neutral-current lepton universality, using the 13 electroweak
measurements of Tables~\ref{tab:lsafbresult}, \ref{tab:14parres}
and~\ref{tab:14parcor}, and Equations~\ref{eq:al:result}
and~\ref{eq:ptau:Al}.  The combination has a $\chidf$ of 4.5/4,
corresponding to a probability of 34\%.}
\label{tab:coup:gq}
\end{center}
\end{table}

\subsection[The \protect$\rhof$ Parameters and the Effective
  Electroweak Mixing Angles]%
{The \boldmath{$\rhof$} Parameters and the Effective Electroweak
Mixing Angles}
\label{sec:coup:rho-sef2}

The effective vector and axial-vector coupling constants obey simple
relations with the $\rho$ parameter and the effective electroweak
mixing angle, given by Equations~\ref{eq:gveff} and~\ref{eq:gaeff}.
For the following analyses, the electric charge $\Qf$ and the third
component of the weak isospin $\Tf$ are assumed to be given by the
$\SM$ assignments as listed in Table~\ref{tab:intro_SM}.  Tests of
fermion universality, \ie, a comparison between leptons and quarks in
terms of $\rhof$ and $\swsqefff$, now become possible.

Considering the leptonic measurements alone and assuming lepton
universality, the combined results on $\rhof$ and $\swsqeffl$ are
reported in Table~\ref{tab:coup:rsl}.  As noted earlier, the neutrino
coupling is smaller by about 1.8 standard deviations than the $\SM$
expectation listed in Appendix~\ref{app:SM:preds}, while for charged
leptons the results are in good agreement with the $\SM$ prediction.

The results on $\rhof$ and the effective electroweak mixing angle for
leptons and quarks are reported in Table~\ref{tab:coup:rsq}.  As
before, neutral-current lepton universality is assumed.  The
measurement of $\swsqeffl$ based on the hadronic charge asymmetry,
Equation~\ref{eq:avQfb}, is not included here as that result is
derived under the assumption of quark universality.  The value of
$\rhol$ is different from the corresponding Born-level value of unity
by 5.0 standard deviations, again indicating the presence of non-QED
electroweak radiative corrections.  The strong correlation between
$\rhob$ and $\swsqeffb$ arises, as the anti-correlation between $\gvb$
and $\gab$ above, from the tight constraint given by the measurement
of $\Rbz\propto\gvb^2+\gab^2$.

\begin{figure}[p]
\begin{center}
\mbox{\epsfig{file=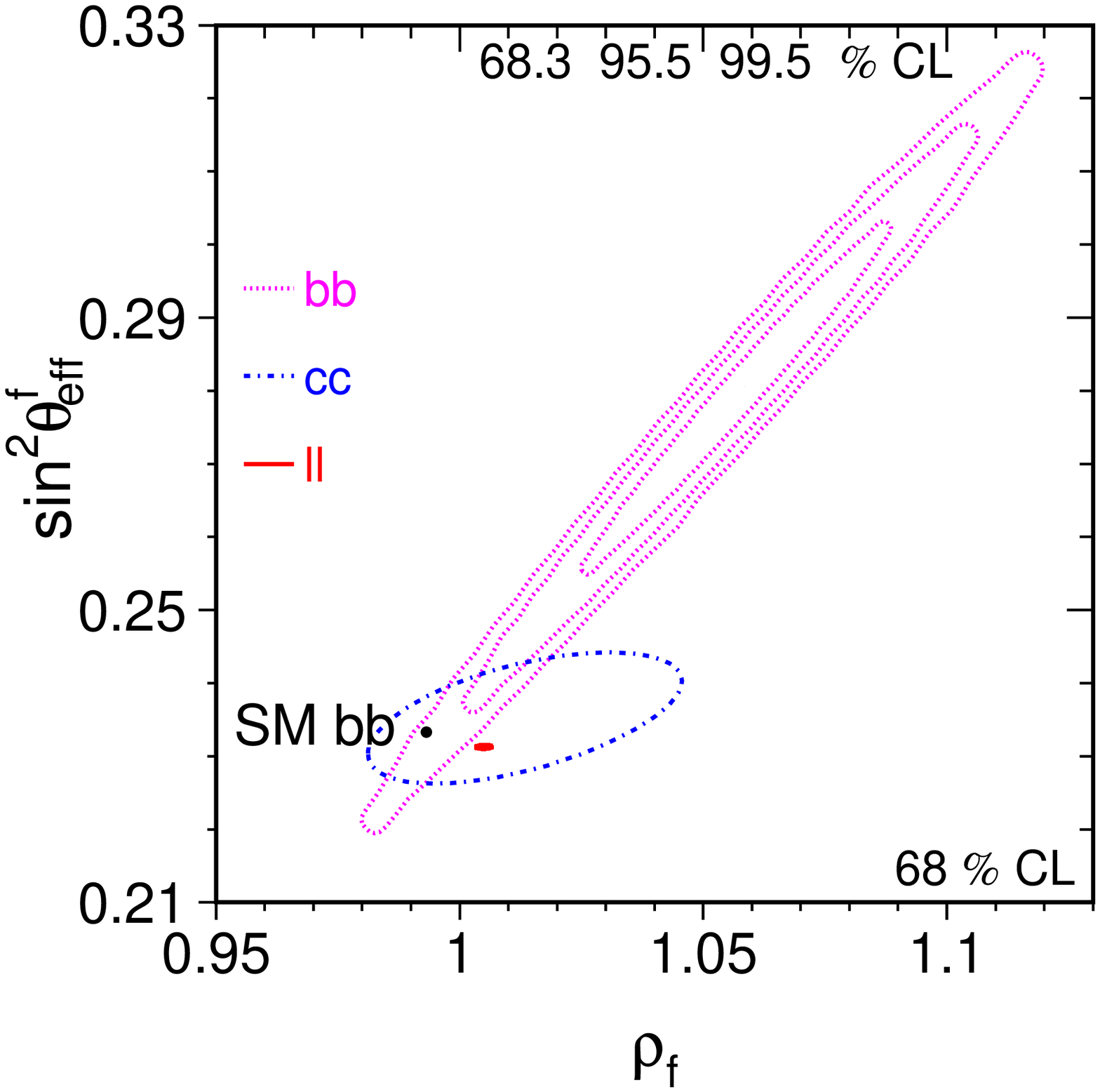,width=0.9\linewidth}} 
\caption[Comparison of $\rhof$ and the effective electroweak mixing
angle $\swsqefff$] { Comparison of $\rhof$ and the effective
electroweak mixing angle $\swsqefff$ for leptons, b and c quarks.  The
$\SM$ expectation for b quarks is shown as the dot ($\rhob<1$); those
of c quarks and of leptons are not drawn as they lie at the same area
as the experimental contour curve for leptons ($\rhol>1$).  Only the
68\% CL contour is shown for c quarks and leptons.  On the scale of
this plot, variations of the $\SM$ prediction with $\Mt$, $\MH$ and
$\dalhad$ are negligible.  }
\label{fig:coup:rho-sef} 
\end{center}
\end{figure}

The comparison between the different fermion species is shown
graphically in Figure~\ref{fig:coup:rho-sef}.  Within the $\SM$,
slightly different values for both $\rhof$ and $\swsqefff$ are
expected for different fermions due to non-universal flavour-specific
electroweak radiative corrections.  These specific corrections are
largest for b quarks, $\rhob-\rhol\approx-0.011$ and
$\swsqeffb-\swsqeffl\approx0.0014$, and more than a factor of five
smaller for the other quark flavours, as visible in
Figure~\ref{fig:coup:rho-sef}.  Except for b-quarks, the non-universal
flavour-specific corrections expected in the $\SM$ are small compared
to the experimental errors.

Increasing the measured value of $\Rbz$ while keeping the measured
value of $\Afbzb$ fixed moves the b-quark contour parallel to the
$\rho$-axis in the direction of increasing $\rhob$ values, since if
$\swsqeffb$ is fixed, $\Rbz$ is simply proportional to $\rhob$.
Changing the measured values of $\Afbzb$, $\cAb$ or $\cAl$ while
keeping the measured value of $\Rbz$ fixed moves the b-quark contour
along its major axis; this is because changing $\swsqeffb$ moves both
the b asymmetries and the b width, therefore $\rhob$ also changes in
order to keep $\Rbz$ fixed.  Increasing $\Afbzb$ or $\cAb$ moves the
contour towards the $\SM$ expectation, with roughly equal sensitivity
to a one standard deviation shift of either parameter.  Decreasing
$\cAl$ moves the contour in the same direction, but a one standard
deviation shift in $\cAl$ has a smaller effect.

\begin{table}[htb]
\begin{center}
\renewcommand{\arraystretch}{1.25}
\begin{tabular}{|c||r@{$\pm$}l||rrr|}
\hline
Parameter & \multicolumn{2}{|c||}{Average} 
          & \multicolumn{3}{|c|}{Correlations} \\
          & \multicolumn{2}{|c||}{ }
          & {$\rhon$} & {$\rhol$}& {$\swsqeffl$} \\
\hline
$\rhon$     & $1.0030$&$0.0031$ & $ 1.00$& $     $&$     $ \\
$\rhol$     & $1.0049$&$0.0010$ & $ 0.48$& $ 1.00$&$     $ \\
$\swsqeffl$ &$0.23113$&$0.00021$& $-0.01$& $ 0.11$&$ 1.00$ \\
\hline
\end{tabular}
\caption[Results on $\rhof$ and $\swsqefff$ for leptons]
{Results on $\rhof$ and $\swsqefff$ for leptons, using the 14
electroweak measurements of Tables~\ref{tab:lsafbresult}
and~\ref{tab:alr:result}, and Equations~\ref{eq:ptau:At}
and~\ref{eq:ptau:Ae}.  Lepton universality is imposed. The combination
has a $\chidf$ of 7.8/9, corresponding to a probability of 56\%.}
\label{tab:coup:rsl}
\end{center}
\end{table}

\begin{table}[htb]
\begin{center}
\renewcommand{\arraystretch}{1.25}
\begin{tabular}{|c||r@{$\pm$}l||rrrrrrr|}
\hline
Parameter & \multicolumn{2}{|c||}{Average} 
          & \multicolumn{7}{|c| }{Correlations} \\
          & \multicolumn{2}{|c||}{ }
          & {$\rhon$} 
          & {$\rhol$} & {$\rhob$} & {$\rhoc$} 
          & {$\swsqeffl$} & {$\swsqeffb$} & {$\swsqeffc$}      \\
\hline
\hline
$\rhon$    &$1.0030 $&$0.0031 $&$ 1.00$&$     $&$     $&$     $&$     $&$     $&$     $\\
$\rhol$    &$1.0050 $&$0.0010 $&$ 0.49$&$ 1.00$&$     $&$     $&$     $&$     $&$     $\\
$\rhob$    &$1.059  $&$0.021  $&$-0.01$&$-0.02$&$ 1.00$&$     $&$     $&$     $&$     $\\
$\rhoc$    &$1.013  $&$0.021  $&$-0.02$&$ 0.02$&$ 0.00$&$ 1.00$&$     $&$     $&$     $\\
$\swsqeffl$&$0.23128$&$0.00019$&$-0.01$&$ 0.09$&$-0.41$&$-0.05$&$ 1.00$&$     $&$     $\\
$\swsqeffb$&$0.281  $&$0.016  $&$ 0.00$&$-0.04$&$ 0.99$&$ 0.03$&$-0.42$&$ 1.00$&$     $\\
$\swsqeffc$&$0.2355 $&$0.0059 $&$ 0.00$&$-0.01$&$ 0.14$&$ 0.56$&$-0.10$&$ 0.15$&$ 1.00$\\
\hline
\end{tabular}
\caption[Results on $\rhof$ and $\swsqefff$] {Results on the $\rhof$
parameter and the effective electroweak mixing angle $\swsqefff$
assuming neutral-current lepton universality, using the 13 electroweak
measurements of Tables~\ref{tab:lsafbresult}, \ref{tab:14parres}
and~\ref{tab:14parcor}, and Equations~\ref{eq:al:result}
and~\ref{eq:ptau:Al}.  The combination has a $\chidf$ of 4.5/4,
corresponding to a probability of 34\%.}
\label{tab:coup:rsq}
\end{center}
\end{table}

\subsection{The Leptonic Effective Electroweak Mixing Angle}
\label{sec:coup:sef2lept}

The measurements of the various asymmetries determine the effective
electroweak mixing angle $\swsqefff$ independently of $\rhof$, because
they depend only on the ratio $\gvf/\gaf$ of the effective coupling
constants.  As illustrated in Figure~\ref{fig:afvssin2} the charge and
weak isospin assignments of the quarks imply that the relative
sensitivity of $\cAq$ to $\swsqeffq$ is much smaller than is the case
for leptons.  In particular for b-quarks this sensitivity is almost a
factor 100 less than it is for leptons.  This is also visible in
Figures~\ref{fig:coup:aq} and~\ref{fig:coup:gq}, showing that for
up-type quarks as well as down-type quarks both the asymmetry
parameters $\cAq$ and the effective coupling constants $\gaq$ and
$\gvq$ are, on the scale of the experimental uncertainties, nearly
independent of $\SM$ parameters.  Therefore, the heavy quark
forward-backward asymmetries $\Afbzq=(3/4)\cAe\cAq$ as well as the
hadronic charge asymmetry $\Qfbhad$, are sensitive to $\swsqeffl$
through the factor $\cAe$ and rather insensitive to $\swsqeffq$.  The
latter fact is also evident from the $\swsqefff$ results reported in
Table~\ref{tab:coup:rsq}, showing that the direct measurements of
$\cAq$ do not impose stringent constraints on $\swsqeffq$.

\begin{figure}[p]
\begin{center}
\mbox{\epsfig{file=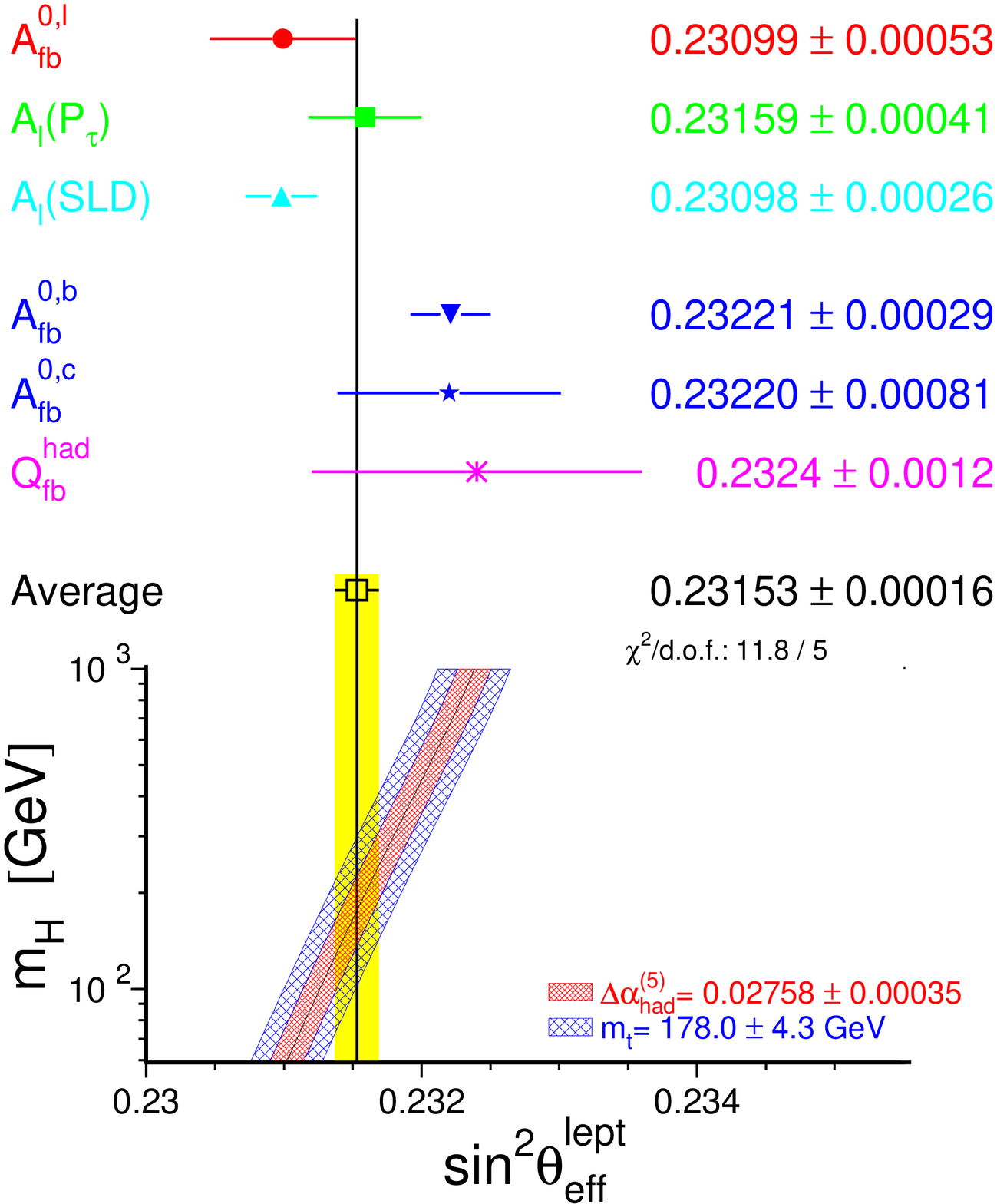,width=0.9\linewidth}} 
\caption[Comparison of the effective electroweak mixing angle
$\swsqeffl$] { Comparison of the effective electroweak mixing angle
$\swsqeffl$ derived from measurements depending on lepton couplings
only (top) and also quark couplings (bottom).  Also shown is the $\SM$
prediction for $\swsqeffl$ as a function of $\MH$.  The additional
uncertainty of the $\SM$ prediction is parametric and dominated by the
uncertainties in $\dalhad$ and $\Mt$, shown as the bands.  The total
width of the band is the linear sum of these effects. }
\label{fig:coup:sef2} 
\end{center}
\end{figure}

The resulting determinations of $\swsqeffl$ derived from each of the
six asymmetry measurements sensitive to $\swsqeffl$ are compared in
Figure~\ref{fig:coup:sef2}.  The measurements fall into two sets of
three results each.  In the first set, the results on $\swsqeffl$ are
derived from measurements depending on leptonic couplings only,
$\Afbzl$, $\cAl(P_\tau)$ and $\cAl$(SLD).  In this case, only lepton
universality is assumed, and no further corrections to interpret the
results in terms of $\swsqeffl$ are necessary.  In the second set,
consisting of $\Afbzb$, $\Afbzc$ and $\Qfbhad$, quark couplings are
involved.  In this case, the small non-universal flavour-specific
electroweak corrections, making $\swsqeffl$ different from
$\swsqeffq$, are taken from the $\SM$.  The size of the applied $\SM$
flavour-specific corrections can be seen in
Figure~\ref{fig:coup:rho-sef}.  Only the correction for b-quarks is
large enough to be visible.  The effect of these corrections and their
uncertainties on the extracted value of $\swsqeffl$ is, as discussed
above, negligible.

The average of all six $\swsqeffl$ determinations is:
\begin{eqnarray}
\swsqeffl & = & 0.23153\pm0.00016\,,
\label{eq:coup:swsqeffl}
\end{eqnarray}
with a $\chidf$ of 11.8/5, corresponding to a probability of only
3.7\%.  This enlarged $\chidf$ is solely driven by the two most
precise determinations of $\swsqeffl$, namely those derived from the
measurements of $\cAl$ by SLD, dominated by the $\ALRz$ result, and of
$\Afbzb$ at LEP, which yield the largest pulls and fall on opposite
sides of the $\swsqeffl$ average.  These two $\swsqeffl$ measurements
differ by 3.2 standard deviations.  Thus, the sets of leptonic and
hadronic measurements, yielding average values for $\swsqeffl$ of
$0.23113\pm0.00021$ ($\chidf=1.6/2$) and $0.23222\pm0.00027$
($\chidf=0.02/2$), respectively, also differ, by 3.2 standard
deviations.  This is a consequence of the same effect as discussed in
the previous sections: the deviation in $\cAb$ as extracted from
$\Afbzb$ discussed in Section~\ref{sec:coup:asym} is reflected in the
value of $\swsqeffl$ extracted from $\Afbzb$.

\subsection{Discussion}
\label{sec:coup:disc}

The unexpectedly large shifts and differences observed in the various
analyses for asymmetry parameters, effective coupling constants,
$\rhof$ and $\swsqeffl$ all show the consequences of the same effect.
It is most clearly visible in the effective couplings and $\swsqeffl$
averages and stems from the measurements of $\ALRz$ and $\Afbzb$.

The results as shown in Figure~\ref{fig:coup:gq} suggest that the
effective couplings for b-quarks cause the main effect; both $\gvb$
and $\gab$ deviate from the $\SM$ expectation at the level of two
standard deviations.  In terms of the left- and right-handed couplings
$\glb$ and $\grb$, which are much better aligned with the axes of the
error ellipse, only $\grb$ shows a noticeable deviation from the
expectation.  The value of $\glb$, which is essentially equivalent to
$\Rbz\propto\grb^2+\glb^2$ due to the smallness of $\grb$, shows no
discrepancy.  The data therefore invite an economical explanation in
terms of a possible deviation of the right-handed
b quark coupling alone, even at Born level (see Equation~\ref{eq:gr}),
from the $\SM$ prediction.
This would affect $\cAb$ and $\Afbzb$, which both depend only on the
ratio $\grb/\glb$, more strongly than $\Rbz$.

From the experimental point of view, no systematic effect potentially
explaining such shifts in the measurement of $\Afbzb$ has been
identified.  While the QCD corrections are significant, their
uncertainties are small compared to the total errors and are taken
into account, see Section~\ref{sec:hf_qcdcor}.  Within the $\SM$,
flavour specific electroweak radiative corrections as listed above and
their uncertainties are much too small to explain the difference in
the extracted $\swsqeffl$ values.  All known uncertainties are
investigated and are taken into account in the analyses. The same
holds for the $\ALRz$ measurement, where the most important source of
systematic uncertainty, namely the determination of the beam
polarization, is small and well-controlled.

Thus the shift is either a sign for new physics which invalidates the
simple relations between the effective parameters assumed in this
chapter, or a fluctuation in one or more of the input measurements.
In the following we assume that measurement fluctuations are
responsible.  Furthermore, we largely continue to assume a Gaussian
model for the experimental errors, despite the fact that this results
in a value for $\swsqeffl$, with small errors, which is in poor
agreement with both $\ALRz$ and $\Afbzb$.  As a direct consequence,
the $\chidf$ in all analyses including these measurements will be
inflated due to the contribution of at least $11.8$ units from the six
asymmetry measurements.  To acknowledge the possibility that a
Gaussian model may in fact poorly represent the tails of the
experimental uncertainties, we also consider how subsequent analyses
are affected if one or the other of the high-pull measurements in the
$\swsqeffl$ sector is excluded from consideration.

\section{Sensitivity to Radiative Corrections Beyond QED}
\label{sec:coup:radcor}

\begin{figure}[p]
\begin{center}
\mbox{\epsfig{file=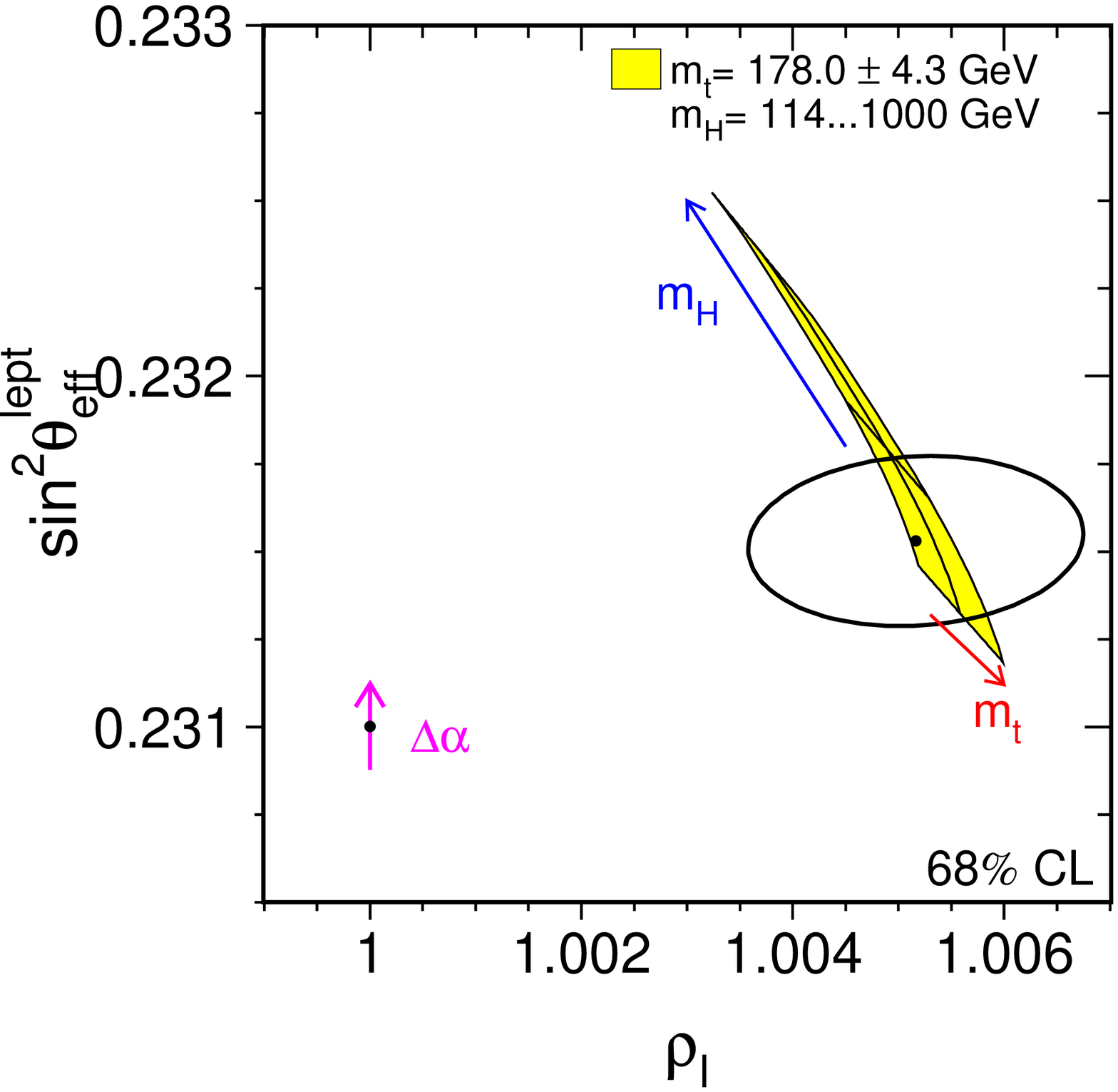,width=0.9\linewidth}} 
\caption[Contour curve in the ($\rhol,\swsqeffl$) plane] {Contour
curve of 68\% probability in the ($\rhol,\swsqeffl$) plane.  The
prediction of a theory based on electroweak Born-level formulae and
QED with running $\alpha$ is shown as the dot, with the arrow
representing the uncertainty due to the hadronic vacuum polarisation
$\dalhad$.  The same uncertainty also affects the $\SM$ prediction,
shown as the shaded region drawn for fixed $\dalhad$ while $\Mt$ and
$\MH$ are varied in the ranges indicated. }
\label{fig:coup:qed-ew-cor} 
\end{center}
\end{figure}

A fundamental question is whether the experimental Z-pole results
indeed confirm the existence of electroweak radiative corrections
beyond those predicted by the well known and tested theory of
QED.  Including only the running of $\alpha$, the expectations based
on Born-term expressions for the $\rho$ parameter and the electroweak
mixing angle are obtained from the equations given in
Section~\ref{sec:intro_ew}, by setting
$\Delta\rho=\Delta\kappa=\Drw=0$.  The results are:
\begin{eqnarray}
\rho_0 & = & 1 \label{eq:sm:ro0}\\
\stzsq & = & \frac{1}{2}
\left(1-\sqrt{1-4\frac{\pi\alqed}{\sqrt{2}\GF\MZ^2}}\,\right)
       ~ = ~ 0.23098\pm0.00012 \label{eq:sm:s20} \,,
\end{eqnarray}
where the uncertainty on $\stzsq$ arises due to the uncertainty on
$\alqed$ mainly caused by the hadronic vacuum polarisation, see
Equations~\ref{eq:Dalpha}, \ref{eq:alpharun}, \ref{eq:sm:alpha}
and~\ref{eq:dalhad:exp:new}.  The measured values of $\rhol$
(Table~\ref{tab:coup:rsq}) and the effective electroweak mixing angle
(Equation~\ref{eq:coup:swsqeffl}):
\begin{eqnarray}
\rhol      & = & 1.0050\phantom{0}\pm0.0010 \label{eq:sm:rol}\\
\swsqeffl  & = & 0.23153\pm0.00016 \label{eq:sm:s2l}\,,
\end{eqnarray}
differ significantly, particularly in the case of the $\rho$
parameter, from these expectations, indicating that electroweak
radiative corrections beyond QED are needed to describe the Z-pole
measurements.  This is also shown in Figure~\ref{fig:coup:qed-ew-cor}.
Further tests of electroweak radiative corrections based on dedicated
sets of parameters are presented in Appendix~\ref{sec:msm:eps-stu}.

In the case of the effective electroweak mixing angle, the uncertainty
on the prediction of $\swsqeffl$ within the $\SM$ due to the
uncertainty on $\dalhad$ is nearly as large as the accuracy of the
experimental measurement of $\swsqeffl$.  This observation underlines
the importance of a precise cross-section measurement of
electron-positron annihilation into hadrons at low centre-of-mass
energies.  In contrast to $\swsqeffl$, the $\SM$ prediction for the
$\rho$ parameter is not affected by the uncertainty in $\dalhad$.

\begin{figure}[tb]
\begin{center}
\includegraphics[width=0.9\linewidth,clip=true,origin=br,bb=0 530 540 800]{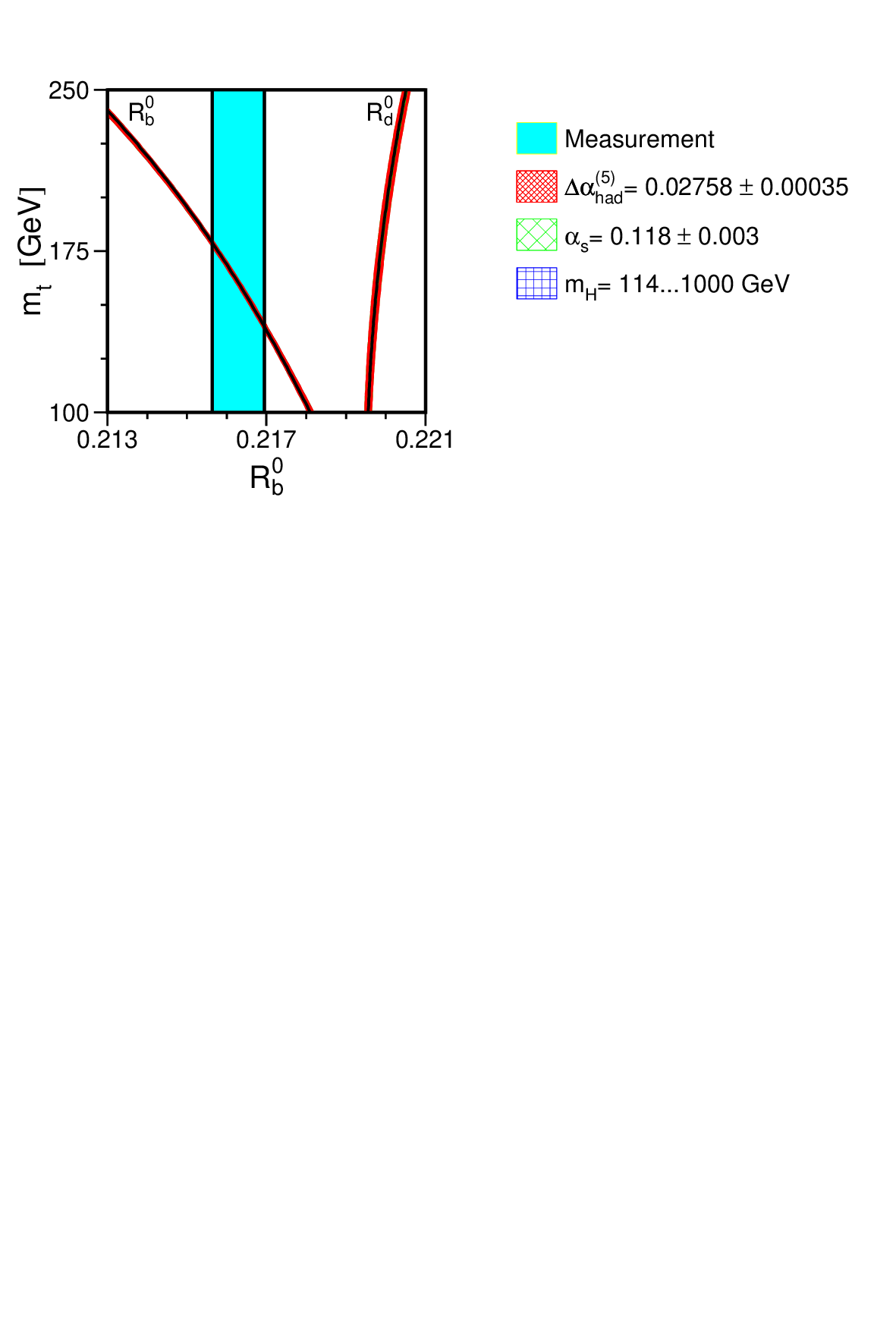}
\caption[Top sensitivity of $\Rbz$] {Comparison of the LEP combined
measurement of $\Rbz$ with the $\SM$ prediction as a function of the
mass of the top quark.  The measurement with its uncertainty is shown
as the vertical band. The two curves show the expectations for $\Rbz$
and $\Rdz$ as a function of $\Mt$. The $\SM$ prediction for $\Rdz$
displays a markedly smaller $\Mt$ dependence which is also of opposite
sign.  The parametric uncertainties on the $\SM$ expectations due to
the uncertainties in $\dalhad$, $\alfmz$ and $\MH$ are shown as the
width of the curves, they are negligible compared to the uncertainty
of the measurement.  }
\label{fig:coup:b-vertex} 
\end{center}
\end{figure}

As discussed in Section~\ref{sec:intro_ew} in connection with
Figure~\ref{fig:b_vertex}, in the $\SM$ the $\bb$ final state is
subject to additional large vertex corrections which depend on the
top-quark mass.  These flavour-specific vertex corrections are
particularly significant for the measurement of $\Rbz$, shown in
Figure~\ref{fig:coup:b-vertex} and compared to theory predictions.
The measurement of $\Rbz$ is able to discriminate between the
different predictions for b-quarks versus light down-type quarks,
showing that also b-specific vertex corrections are observed with high
significance.  The much weaker $\Mt$ dependence of $\Rdz$, which is
also shown, results mainly from b-quark contributions in the
denominator of $\Rdz=\Gdd/\Ghad$.  Due to the fact that other
radiative corrections affect all quark species about equally, $\Rbz$,
as a ratio, benefits from small parametric uncertainties arising from
other $\SM$ parameters, and therefore imposes a particularly direct
constraint on the top-quark mass in the $\SM$.

As will be shown in Chapter~\ref{chap:MSM}, also the mass of the W
boson, measured at the Tevatron and at \LEPII, implies the existence
of genuine electroweak radiative corrections through $\Dr$ and $\Drw$,
with even higher significance.  It is interesting to note that in
1987, before the advent of SLC and LEP, electroweak radiative
corrections - including the large QED contributions - had been seen at
the level of three standard deviations based on a variant of
$\Dr$~\cite{Amaldi:1987fu}, while the pure electroweak components of
the corrections could not be separated.  Today, it is the pure
electroweak correction $\rhol$ which is demonstrated above to deviate
from unity with a significance of five standard deviations.
Furthermore, using the current Z-pole results alone, the error in
$\Dr$ has been reduced by a factor of about 20 compared to
1987~\cite{Amaldi:1987fu}, see the next chapter.

\chapter{Constraints on the Standard Model}
\label{chap:MSM}

In the previous sections, several figures have already shown
comparisons between the experimental results and the expectations from
the Standard Model ($\SM$)~\cite{\GWS}.  In this chapter, the
experimental results are used to constrain the input parameters of the
$\SM$.  As discussed in Section~\ref{sec:intro_ew}, the $\SM$
prediction for each Z-pole observable depends on free parameters which
are not predicted by the theory, such as the coupling constants of the
various interactions and the masses of the fundamental fermions
(quarks and leptons) and bosons (Z, W, and H).  Consistency of the
$\SM$ framework requires that all measurements are accommodated by the
same values of these free $\SM$ parameters.  Owing to this dependence,
directly at Born level or through electroweak radiative corrections,
the experimental measurements of Z-pole and other observables allow us
to constrain these free parameters. Most importantly it is possible to
determine the mass of the top quark precisely and also the mass of the
Higgs boson, albeit with less precision.

The input parameter set chosen for $\SM$ calculations is discussed in
Section~\ref{sec:msm:ips}.  An important ingredient, the hadronic
vacuum polarisation, is discussed in Section~\ref{sec:msm:vacpol}.
Additional measurements from other experiments, used for comparisons,
or to increase the precision of the $\SM$ constraints, are reported in
Section~\ref{sec:msm:add}.  Parameter dependencies and theoretical
uncertainties in the calculation of $\SM$ predictions for measured
observables are discussed in Section~\ref{sec:msm:TU}.  The analysis
and fitting procedure used in this chapter is described in
Section~\ref{sec:msm:proc}.  The remaining sections present the
results of the $\SM$ analyses: Constraints on the input parameters of
the $\SM$, in particular on the masses of the heavy particles top
quark, W boson and Higgs boson are reported in
Section~\ref{sec:msm:msm}.  A concluding discussion is presented in
Section~\ref{sec:msm:disc}.  Predictions of many observables within
the $\SM$ framework are reported in Appendix~\ref{app:SM:preds}.

\section{Parameters of the Standard Model}
\label{sec:msm:ips}

For the electromagnetic, weak and strong interactions described by the
Minimal $\SM$, the corresponding coupling constants are not predicted,
but must be inferred from measurements.  Because the $\SM$ gives an
integrated description of the electromagnetic and weak interactions in
the form of the electroweak theory, the weak coupling is related to
the electromagnetic coupling and the masses of the charged and neutral
heavy gauge bosons W and Z.  Therefore, just two coupling constants,
those of the electromagnetic and the strong interaction, $\alpha$ and
$\alfas$, remain to be determined, together with the masses of the
heavy gauge bosons W and Z.  The mass of the electromagnetic gauge
boson, the photon, is fixed at zero as required by the theory of QED.

The masses of all known fundamental fermions with the exception of the
top quark are small compared to $\MZ$, and precisely enough measured
so that their influence on Z-pole observables through kinematic
effects is both rather small and calculable to more than adequate
precision. In the following analyses the masses of the light fermions
are therefore considered fixed.  In particular, the results are
insensitive to small neutrino masses corresponding to current
experimental limits~\cite{PDG2004}.

The mass of the Z boson is precisely measured as described in
Chapter~\ref{chap:lsafb}. Although it is treated formally as a free
parameter of the theory, the precision of its experimental
determination is sufficient to ensure that no $\SM$ constraints can
pull it appreciably, and it could just as well be taken as a fixed
quantity in our analysis.

Within the $\SM$, the mass of the W, measured directly at the Tevatron
and \LEPII\, is related to $\MZ$ and the Fermi constant $\GF$ through
Equation~\ref{eq:MW_gone}. A very precise value for the latter,
$\GF=1.16637(1)\cdot10^{-5}~\GeV^{-2}$~\cite{PDG2004}, is derived from
measurements of the muon lifetime using two-loop
corrections~\cite{\bibGmu}.  This 9 ppm precision on $\GF$ greatly
exceeds the relative precision with which $\MW$ can be measured in the
foreseeable future. Indeed, the current precision of $\GF$ is so great
that it is treated as a constant in the following analyses.  This also
motivates our substitution of $\GF$ for $\MW$ as an input parameter
for $\SM$ calculations.\footnote{ Note, however, that this replacement
is purely technical: none of the results would change if $\MW$ rather
than $\GF$ were taken as an input parameter for $\SM$ calculations.}
In addition, radiative corrections are smaller when calculated with
respect to lowest-order expressions formulated in terms of $\GF$.  The
mass of the W boson, $\MW$, is then predicted within the $\SM$ in
terms of $\GF$, $\MZ$, and the radiative correction $\Dr$, which is a
function of the other $\SM$ input parameters. Comparing this
prediction with the direct measurements of $\MW$ performed at the
Tevatron and \LEPII\ yields a stringent test of the $\SM$.

Electroweak radiative corrections such as those shown in
Figure~\ref{fig:intro_ewcor} modify the calculation of Z-pole
observables. Comparisons with the measurements thus allow constraints
to be placed on $\SM$ input parameters beyond those accessible
directly.  The top quark, with a mass of about $175~\GeV$, and
apparently the Higgs boson, are too heavy to be produced directly in
$\ee$ collisions at \LEPI/SLC centre-of-mass energies close to the Z
pole, \ie, $88~\GeV<\sqrt{s}<95~\GeV$.  Loop corrections in $\ee$
interactions involving virtual top quarks and Higgs bosons depend,
however, on the masses of these two particles.  To leading one-loop
order these corrections do not depend on the species of fermion to
which the $\Zzero$ decays, and show a dependence quadratic in the
top-quark mass and logarithmic in the Higgs-boson mass, as illustrated
in Equations~\ref{eq:deltarho} and~\ref{eq:deltakappa}. Non-leading,
higher-order and fermion-specific corrections (see
Equation~\ref{eq:b_vertex}) allow the effects from the top quark and
the Higgs boson to be disentangled.  The resulting indirect
determination of the top-quark mass $\Mt$ is precise, and its
comparison with the direct measurement obtained from $\TT$ production
in proton-antiproton collisions at the Tevatron constitutes another
important test of the $\SM$.  Our determination of the Higgs mass is
consistent with moderate values on the electroweak scale, and
establishes a useful upper limit to guide future searches.

Loop corrections also induce a running of the electromagnetic coupling
constant $\alpha$ with momentum transfer (or $s$), as described in
Equation~\ref{eq:alpharun}. The running of the strong coupling,
$\alfas(s)$, is even larger. The $\Zzero$ resonance is sufficiently
dominant for Z-pole observables, however, that the Z-pole
approximation can be taken, and the relevant coupling constants become
simply $\alqed$ and $\alfmz$.

The five input parameters of the $\SM$ relevant for the calculation of
Z-pole observables are therefore the coupling constants of QED and QCD
at the Z pole, $\alqed$ and $\alfmz$, and the masses of the Z boson,
the top quark and the Higgs boson.  The measurements of electroweak
observables presented in the previous chapters are used to find
optimal values for these five $\SM$ input parameters, and to test
whether all the measurements can be simultaneously accommodated by
this single set.  Besides the mass of the Z boson, the interesting
input parameters of the $\SM$ are the mass of the top quark and of
course the mass of the Higgs boson.  Since the electroweak sector of
the $\SM$ is well understood, the hadronic Z-pole observables will
give rise to one of the most precise determinations of $\alfmz$.  The
treatment of $\alqed$ is discussed in the following section.  The
programs TOPAZ0~\cite{\TOPAZref} and ZFITTER~\cite{\ZFITTERref} are
used to calculate all Z-pole observables including radiative
corrections in the framework of the $\SM$ and as a function of these
five $\SM$ input parameters. They include the equations shown in
Section~\ref{sec:intro_ew}, supplemented by more complicated
high-order expressions for improved theoretical accuracy.

\section{Hadronic Vacuum Polarisation}
\label{sec:msm:vacpol}

The running of the electromagnetic coupling with momentum transfer,
$\alpha(0)\rightarrow\alpha(s)$, caused by fermion-pair loop
insertions in the photon propagator, is customarily written as given
in Equations~\ref{eq:Dalpha} and~\ref{eq:alpharun}:
\begin{eqnarray}
\alpha(s) & = & \frac{\alpha(0)}{1-\Delta\alpha(s)}
          ~ = ~ 
         \frac{\alpha(0)}{1
                            -\Delta\alpha_{\mathrm{e\mu\tau}}(s)
                            -\Delta\alpha_{\mathrm{top     }}(s)
                            -\Delta\alpha_{\mathrm{had     }}^{(5)}(s) 
                          }\,,
\label{eq:sm:alpha}
\end{eqnarray}
with $\alpha(0)=1/137.036$~\cite{PDG2004}.  The contribution of
leptons is calculated diagrammatically up to third order:
$\Delta\alpha_{\mathrm{e\mu\tau}}(\MZ^2)=0.03150$ with negligible
uncertainty~\cite{bib-alphalept}.  Since heavy particles decouple in
QED, the top-quark contribution is small:
$\Delta\alpha_{\mathrm{top}}(\MZ^2)=-0.00007(1)$; it is calculated by
TOPAZ0 and ZFITTER as a function of the pole mass of the top quark,
$\Mt$.  The running electromagnetic coupling is insensitive to new
particles with high masses.  For light-quark loops the diagrammatic
calculations are not viable as at such low energy scales perturbative
QCD is not applicable.  Therefore, the total contribution of the five
light quark flavours to the hadronic vacuum polarisation, $\dalhad$,
is more accurately obtained through a dispersion integral over the
measured hadronic cross-section in electron-positron annihilations at
low centre-of-mass energies.  In this case the uncertainty on
$\dalhad$ is given by the experimental uncertainties in the measured
hadronic cross-section at low centre-of-mass energies, leading
to~\cite{bib-JEG2,bib-Burk}:
\begin{eqnarray}
\dalhad & = & 0.02804\pm0.00065\,,
\label{eq:dalhad:exp:old}
\end{eqnarray}
as used in Chapter~\ref{chap:lsafb} for the extraction of the Z
resonance parameters.  Based on the same analysis technique but
including new measurements of the hadronic cross-section at low
energies, in particular the precise measurements of the BES
collaboration in the range $2~\GeV<\sqrt{s}<5~\GeV$~\cite{BES_01} as
well as measurements by the CMD-2 and KLOE experiments below that
energy in $\pi^+\pi^-$ production~\cite{CMD_03,KLOE_04}, the
uncertainty is much reduced~\cite{bib-BP05}:
\begin{eqnarray}
\dalhad & = & 0.02758\pm0.00035\,,
\label{eq:dalhad:exp:new}
\end{eqnarray}
leading to $\Delta\alpha(\MZ^2)=0.05901\pm0.00035$.  During the course
of the last few years, more theory-driven determinations of $\dalhad$
have appeared~\cite{\dalphaQCD,bib-Troconiz-Yndurain-2004}, which
employ perturbative QCD to calculate the hadronic cross-section in the
continuum region at low $\sqrt{s}$, outside the region populated by
the hadronic resonances.  Since the theoretical uncertainty on the
predicted cross-section is assumed to be smaller than that of the
experimental measurements, a reduced error on $\dalhad$ is achieved,
for example~\cite{bib-Troconiz-Yndurain-2004}:
\begin{eqnarray}
\dalhad & = & 0.02749\pm0.00012\,,
\label{eq:dalhad:qcd}
\end{eqnarray}
which also takes the new results from BES into account.  All updated
evaluations of $\dalhad$ are consistent with, but lower than, the
previous evaluation of Equation~\ref{eq:dalhad:exp:old}.  In the
following analyses, the experiment-driven value of $\dalhad$ as given
in Equation~\ref{eq:dalhad:exp:new} will be used, on the same footing
as any other experimental measurement with its associated uncertainty.

\section{Additional Measurements}
\label{sec:msm:add}

Obviously, a wealth of measurements are performed in particle physics
experiments elsewhere, using various particle beams and targets.  The
results of these experiments are crucial to explore the predictive
power of the $\SM$ in as large a breadth as possible.  Of all these
measurements, those which have a high sensitivity to the five $\SM$
input parameters introduced above are particularly interesting here.

The additional results considered in some of the $\SM$ analyses
presented in the following are the mass of the top quark and the mass
and the total width of the W boson.  In addition, the $\SM$ analyses
are used to obtain predictions for electroweak observables measured in
reactions at low momentum transfer, $Q^2\ll\MZ^2$, namely those
measuring parity violation effects in atomic transitions, in polarised
M\o ller scattering and in neutrino-nucleon scattering.  These results
are sensitive to different types of new-physics effects than the
Z-pole observables.  However, since the precision of these results is
insufficient to provide additional power in determining the five $\SM$
input parameters, they are not included in our fits, but used to test
their compatibility with the SM predictions based on the high-$Q^2$
fits.  Predictions of the observables within the $\SM$ framework are
reported in Appendix~\ref{app:SM:preds}.

\subsection{Mass of the Top Quark}
\label{sec:msm:add:top}

In 1995, the top quark was discovered in proton-antiproton
interactions recorded at the Tevatron collider by the experiments
CDF~\cite{\CDFtop} and D\O~\cite{D0-top:1995}.  Both experiments
measure its mass directly, exploiting various decay chains.  The
published results from CDF~\cite{\CDFtopm} and D\O~\cite{\Dztopm}
obtained from data collected in Run-I (1992-1996) are combined taking
correlated uncertainties into account. The Tevatron Run-I world
average value for the pole mass of the top quark is:
$\Mt=178.0\pm4.3~\GeV$~\cite{PP-MT:combination}.  Improved direct
measurements of $\Mt$ are expected from the currently ongoing Run-II
of the Tevatron.

\subsection{Mass and Width of the W Boson}
\label{sec:msm:add:W}

Initially, the mass of the W boson was measured in proton-antiproton
collisions, first by the experiments UA1~\cite{UA1-MW} and
UA2~\cite{UA2-MW} at the SPS collider, which discovered the W and Z
bosons, and subsequently with much higher precision by the experiments
CDF~\cite{\CDFWm} and D\O~\cite{\DzWm} at the Tevatron.  Also the
total width of the W boson, $\GW$, is measured by the Tevatron
experiments CDF~\cite{CDF-GW} and D\O~\cite{D0-GW}.  The results based
on the data collected during Run-I of the Tevatron are final and are
combined taking correlated systematic uncertainties into account. The
combined results are~\cite{PP-MW-GW:combination}:
$\MW=80.452\pm0.059~\GeV$ and $\GW=2.102\pm0.106~\GeV$ with an overall
correlated error of $-0.033~\GeV$ or a correlation coefficient of
$-0.174$ between mass and width.  Improved direct measurements of
$\MW$ and $\GW$ are expected from the currently ongoing Run-II of the
Tevatron.

The LEP experiments ALEPH, DELPHI, L3 and OPAL also measure the
W-boson mass and width directly in the process $\mathrm{\ee\rightarrow
W^+W^-}$, after the centre-of-mass energy of the LEP accelerator was
more than doubled (\LEPII).  Combining all published~\cite{\LEPMW} and
{\em preliminary} \LEPII\ measurements, the \LEPII\ results
are~\cite{LEP-MW:combination}: $\MW=80.412\pm0.042~\GeV$ and
$\GW=2.150\pm0.091~\GeV$.

The results obtained at hadron and lepton colliders are in very good
agreement with each other.  Combining the independent sets of results
leads to {\em preliminary} direct determinations of the W-boson mass
and width with high accuracy: $\MW=80.425\pm0.034~\GeV$ and
$\GW=2.133\pm0.069~\GeV$ with a correlation coefficient of $-0.067$
between mass and width.  As for the Z boson, the mass and width of the
W boson quoted here are defined according to a Breit-Wigner
denominator with $s$-dependent width, $|s-\MW^2 + i s \GW/\MW|$.

\subsection{Measurements at Low Momentum Transfer}
\label{sec:msm:add:lowQ2}

Combinations of effective coupling constants are also measured in
low-$Q^2$ processes, $Q^2\ll\MZ^2$.  However, owing to the running of
effective coupling constants with $Q^2$, the couplings measured in
low-$Q^2$ reactions are different from those measured at the Z pole,
$Q^2=\MZ^2$. This running has to be accounted for before comparisons
can be made.

\subsubsection{Parity Violation in Atoms}
\label{sec:msm:add:apv}

The measurement of parity violation in atomic transitions determines
the weak charge of the atomic nucleus as probed by the shell electron,
$\QW(Z,N) = -2[(2Z+N)C_{1\mathrm{u}}+(Z+2N)C_{1\mathrm{d}}]$ for a
nucleus with $Z$ protons and $N$ neutrons.  The weak charges
$C_{1\mathrm{q}}$ of up and down quarks as seen by the electron
through the parity-violating $t$-channel $\gamma$/Z exchange are
expressed in terms of effective vector and axial-vector coupling
constants, $C_{1\mathrm{q}}=2\gae\gvq$ for $Q^2\to0$.

Precise measurements of $\QW$ are performed for
cesium~\cite{QWCs:exp:1,QWCs:exp:2}, while less precise results are
available for thallium~\cite{QWTl:exp:1,QWTl:exp:2}.  In recent years,
certain aspects in nuclear many-body perturbation theory and QED
radiative corrections needed in the experimental analyses have been
investigated, see Reference~\citen{QWCs:theo:2003:new} for a review.
The newly corrected experimental results for cesium is:
$\QWCs=-72.74\pm0.46$~\cite{QWCs:theo:2003:new}.  This result is now
in good agreement with the $\SM$ calculation~\cite{QW:MSM} included in
TOPAZ0 and ZFITTER.

\subsubsection{Parity Violation in M\o ller Scattering}
\label{sec:msm:add:Moller}

The measurement of parity violation in fixed-target M\o ller
scattering, $\emem$, with beam polarisation, determines the weak
charge of the electron, $\QW(\mathrm{e})=-4\gae\gve$.  The experiment
E-158 at SLAC has published its final
measurement~\cite{E158RunI,E158RunI+II+III}, performed at an average
momentum transfer of $Q^2=0.026~\GeV^2$.  In terms of the weak mixing
angle the result is: $\swsqeff(Q^2)=0.2397\pm0.0013$ or $\swsqMSb =
0.2330\pm0.0015$ using the $\SM$ running of the electroweak mixing
angle with $Q^2$.  Adding 0.00029~\cite{PDG2004} to $\swsqMSb$ yields
the effective electroweak mixing angle, $\swsqeffl$.

\subsubsection{Parity Violation in Neutrino-Nucleon Scattering}
\label{sec:msm:add:R-}

The measurement of the neutrino-nucleon neutral-to-charged current
cross-section ratio also determines a combination of effective
coupling constants. In the ideal case of an iso-scalar target and
using both a $\nu_\mu$ and $\bar\nu_\mu$ beam, the Paschos-Wolfenstein
relations hold~\cite{PaschosWolfenstein}: $R_\pm \equiv
(\sigma_{NC}(\nu)\pm\sigma_{NC}(\bar\nu))/
(\sigma_{CC}(\nu)\pm\sigma_{CC}(\bar\nu)) = \gnlq^2\pm\gnrq^2$, where
$\gnlq^2=4\gln^2(\glu^2+\gld^2) =
[1/2-\swsqsqeff+(5/9)\swsqsqeff]\rhon\rho_{\mathrm{ud}}$ and
$\gnrq^2=4\gln^2(\gru^2+\grd^2) =
(5/9)\swsqsqeff\rhon\rho_{\mathrm{ud}}$.  Historically, the result
is often quoted in terms of the on-shell electroweak mixing angle,
adding small electroweak radiative corrections and assuming the $\SM$
values of the $\rhof$ parameters for light quarks and neutrinos.

Using both muon neutrino and muon anti-neutrino beams, the NuTeV
collaboration has published by far the most precise result in
neutrino-nucleon scattering~\cite{bib-NuTeV-final}, obtained at an
average $Q^2\simeq20~\GeV^2$.  Based on an analysis mainly exploiting
$R_-$, the results for the effective couplings defined above are:
$\gnlq^2=0.30005\pm0.00137$ and $\gnrq^2=0.03076\pm0.00110$, with a
correlation of $-0.017$.  While the result on $\gnrq$ agrees with the
$\SM$ expectation, the result on $\gnlq$, measured nearly eight times
more precisely, shows a deficit with respect to the expectation at the
level of 3.0 standard deviations.  Possible large theoretical
uncertainties in the area of radiative corrections and QCD effects
affecting this measurement are discussed in
References~\citen{Zeller:2002du, McFarland:2003jw, Olness:2003wz,
Kretzer:2003wy, Martin:2004dh}.

Assuming the $\SM$ value of $\swsq$, the result corresponds to a
deficit of ($1.2\pm0.4$)\% in either $\rhon$ or
$\rho_{\mathrm{ud}}$~\cite{KMcF:NuInt01}.  Recall that the neutrino
coupling $\rhon$ as derived from $\Ginv$ measured at LEP and discussed
in Chapter~\ref{chap:Z+coup} shows a deficit of $(0.5\pm0.3)\%$.

Assuming the $\SM$ value of the $\rhof$ parameters for light quarks
and neutrinos, the result converts to: $\swsq \equiv 1-\MW^2/\MZ^2 =
0.2277\pm0.0016 -0.00022\frac{\Mt^2-(175~\GeV)^2}{(50~\GeV)^2}
+0.00032\ln\frac{\MH}{150~\GeV}$~\cite{bib-NuTeV-final}, where the
residual dependence of the result on the $\SM$ electroweak radiative
corrections is explicitly parametrised.  Using $\MZ$ from \LEPI,
Table~\ref{tab:lsafbresult}, and ignoring the small $\Mt$ and $\MH$
dependence, the $\swsq$ result corresponds to a W-boson mass of
$\MW=80.136\pm0.084~\GeV$.  This value differs from the direct
measurement of $\MW$ discussed above by 3.2 standard deviations.

\section{Parametric and Theoretical Uncertainties}
\label{sec:msm:TU}

Since the interesting electroweak radiative corrections involving
top-quark and Higgs-boson masses are typically on the order of 1\% or
less at the Z pole, all other effects must be controlled at the
per-mille level in order to extract quantitatively these interesting
$\SM$ effects and the parameters governing them.  The precision with
which the pseudo-observables can be calculated within the framework of
the $\SM$ is determined by both theoretical uncertainties and by the
manner in which the pseudo-observables depend on the five $\SM$ input
parameters.  When a pseudo-observable depends on several $\SM$
parameters, some of which are only poorly determined, the resulting
parametric uncertainty can become significant.  Due to their
importance in determining the precision with which the five $\SM$
input parameters can be measured, these parameter dependencies and
theoretical uncertainties are discussed in the following.

\subsection{Parameter Dependence}

The fact that the pseudo-observables depend on the five $\SM$ input
parameters is of course essential for allowing these parameters to be
extracted from the data. For determining the interesting $\SM$ input
parameters, namely the mass of the top quark and the mass of the Higgs
boson, a large parametric dependence, or sensitivity, is advantageous,
while dependence on the other $\SM$ input parameters, in particular
the hadronic vacuum polarisation, should be small, in order to limit
the resulting parametric uncertainties.  Since all five $\SM$ input
parameters are determined in parallel, these intertwined dependencies
are properly accounted for automatically by the analysis procedure
discussed in Section~\ref{sec:msm:proc}.

In general, the pseudo-observables fall into three groups.  First,
there are the pseudo-observables which are also $\SM$ input
parameters, namely the mass of the Z boson, the hadronic vacuum
polarisation, and the mass of the top quark.  Second, there are the
pseudo-observables which have, compared to their experimental
uncertainties, very little dependence on the five $\SM$ input
parameters, such as $\shad$ or the quark left-right forward-backward
asymmetries determining the quark asymmetry parameter $\cAq$.
Nevertheless they test the $\SM$ independent of radiative corrections
in terms of its static properties, such as the number of fermion
generations or the quantum numbers for weak isospin and electric
charge assigned to the fundamental fermions.  Third, there are the
pseudo-observables which are highly sensitive to electroweak radiative
corrections, such as partial widths and the various asymmetries.  They
determine the $\rho$ parameter and the effective electroweak mixing
angle as discussed in the previous chapter.

\begin{figure}[p]
\begin{center}
\mbox{\epsfig{file=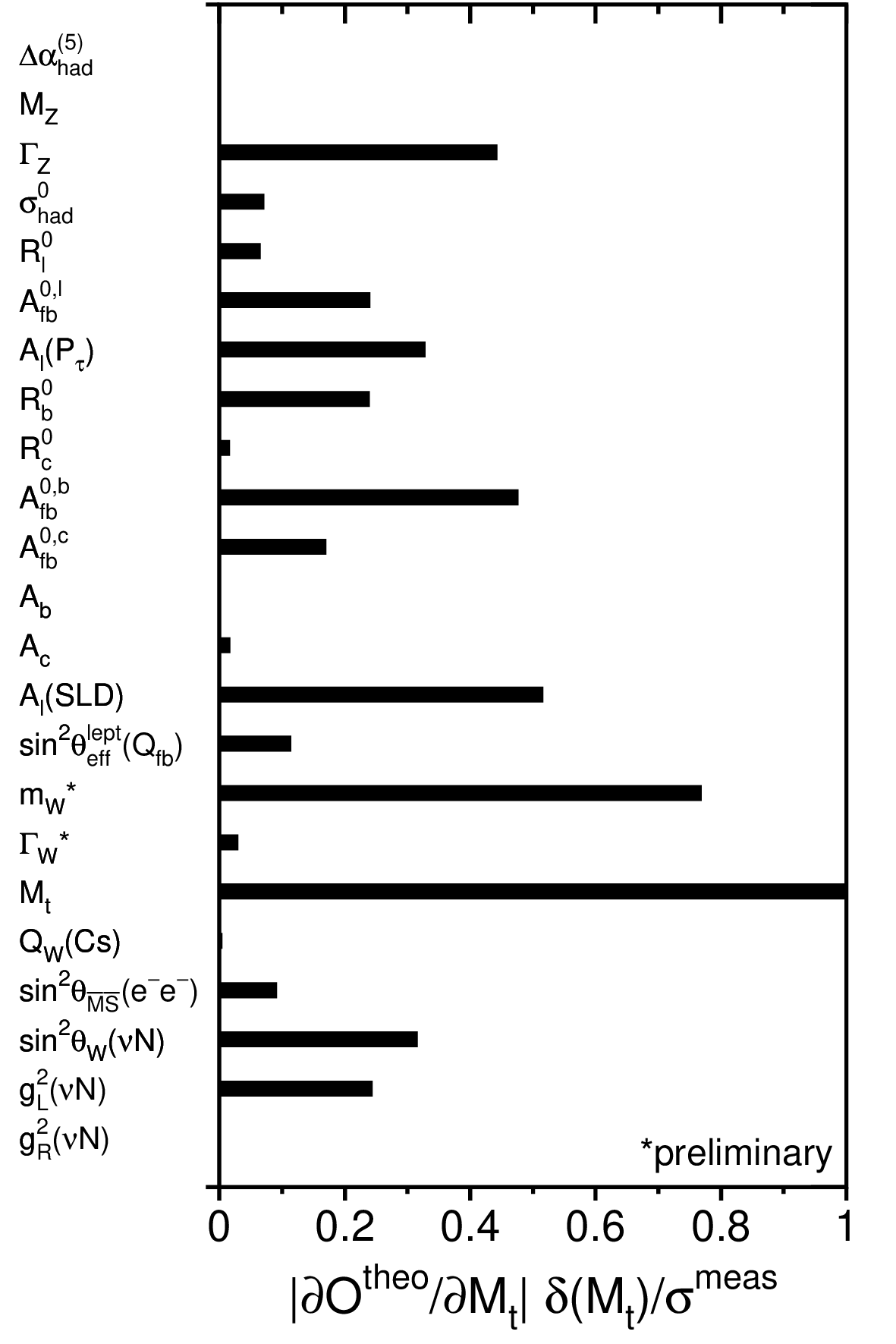,width=0.8\linewidth}} 
\caption[Top-quark mass sensitivity] {Sensitivity of each
pseudo-observable to the mass of the top quark, defined as the partial
derivative of the $\SM$ calculation of the observable with respect to
$\Mt$, relative to the total measurement error $\sigma$ on the
pseudo-observable, and multiplied by the $\pm4.3~\GeV$ uncertainty
$\delta$ in the Tevatron Run-I measurement of $\Mt$.  The other $\SM$
input parameters are kept fixed at values $\dalhad=0.02758$,
$\alfmz=0.118$, $\MZ=91.1875~\GeV$, and $\MH=150~\GeV$.  The direct
measurements of $\MW$ and $\GW$ used here are preliminary. }
\label{fig:msm:mt:sens} 
\end{center}
\end{figure}

\begin{figure}[p]
\begin{center}
\mbox{\epsfig{file=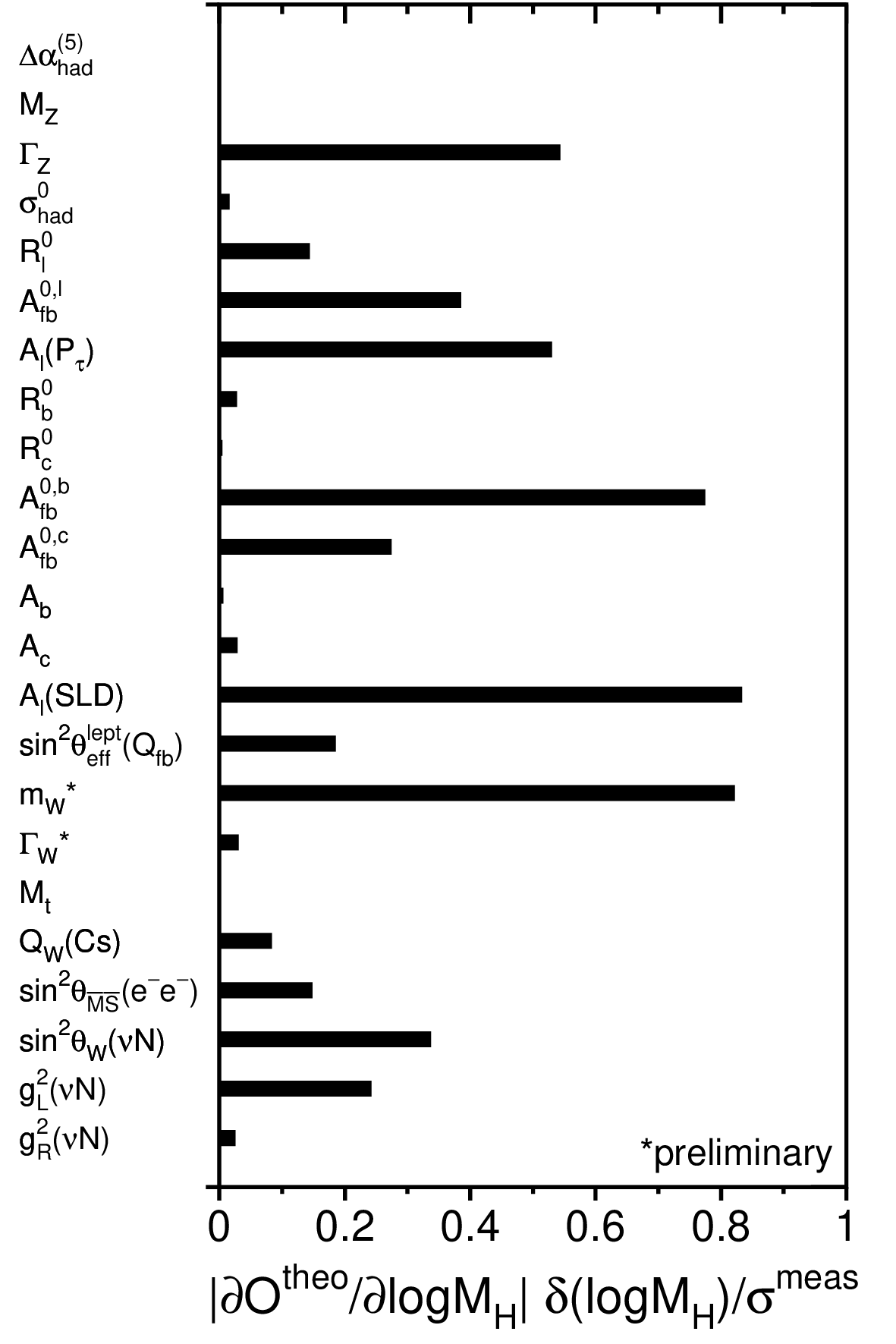,width=0.8\linewidth}} 
\caption[Higgs-boson mass sensitivity] {Sensitivity of each
pseudo-observable to the mass of the Higgs boson, defined as the
partial derivative of the $\SM$ calculation of the observable with
respect to $\LOGMH$, relative to the total measurement error $\sigma$
on the pseudo-observable, and multiplied by the $\pm0.2$ uncertainty
$\delta$ in $\LOGMH$ (see Tables~\ref{tab:TU}
and~\ref{tab:msmfit-all}).  The other $\SM$ input parameters are kept
fixed at values $\dalhad=0.02758$, $\alfmz=0.118$, $\MZ=91.1875~\GeV$,
and $\Mt=178~\GeV$.  The direct measurements of $\MW$ and $\GW$ used
here are preliminary. }
\label{fig:msm:mh:sens} 
\end{center}
\end{figure}

Numerical results for parametric uncertainties of several selected
pseudo-observables are reported in Table~\ref{tab:TU}.  Direct
quantitative comparisons of the interesting $\SM$ top-quark and
Higgs-boson mass sensitivities of the observables are shown in
Figures~\ref{fig:msm:mt:sens} and~\ref{fig:msm:mh:sens}; where the
sensitivities are quantified as the partial derivative of the $\SM$
calculation of the observable with respect to $\Mt$ or $\LOGMH$,
relative to the total measurement error of the observable and
multiplied by the uncertainty $\delta$ in $\Mt$ or $\LOGMH$ as listed
in Table~\ref{tab:TU}, so that they are dimensionless and thus
comparable in terms of the ratios of the standard deviations of
observable and $\SM$ input parameter.  For measured observables which
are also $\SM$ input parameters, their scaled sensitivity is unity
with respect to themselves, and vanishes with respect to the other
$\SM$ input parameters.  However, through correlations in
multi-parameter fits, measurements of $\SM$ input parameters do
influence the values and errors of all $\SM$ input parameters
extracted from fits to the data set, including the mass of the Higgs
boson.

\begin{table}[t]
\begin{center}
\renewcommand{\arraystretch}{1.25}
\begin{tabular}{|c|l||rrrrrrr|}
\hline
Source     & $\delta$           
&$\GZ$   & $\shad$ & $\Rl$   &  $\Rbz$  & $\rhol$  &$\swsqeffl$& $\MW$   \\
           & 
&$[\MeV]$ & $[$nb$]$&        &          &          &           &$[\MeV]$ \\
\hline
\hline
$\dalhad $ & $0.00035$ 
& $0.3$ & $0.001$ & $0.002$ & $0.00001$ &    ---   & $0.00012$ & $\phantom{0}6$ \\
$\alfmz  $ & $0.003$    
& $1.6$ & $0.015$ & $0.020$ &   ---     &    ---   & $0.00001$ & $\phantom{0}2$ \\
$\MZ$ & $2.1~\MeV$         
& $0.2$ & $0.002$ &    ---  &   ---     &    ---   & $0.00002$ & $\phantom{0}3$ \\
$\Mt$ & $4.3~\GeV$       
& $1.0$ & $0.003$ & $0.002$ & $0.00016$ & $0.0004$ & $0.00014$ & $26$ \\
$\LOGMH  $ & $0.2$
& $1.3$ & $0.001$ & $0.004$ & $0.00002$ & $0.0003$ & $0.00022$ & $28$ \\
\hline
\hline
Theory  & 
& $0.1$ & $0.001$ & $0.001$ & $0.00002$ &    ---   & $0.00005$ & $\phantom{0}4$ \\
\hline
\hline
Experiment & 
& $2.3$ & $0.037$ & $0.025$ & $0.00065$ & $0.0010$ & $0.00016$ & $34$ \\
\hline
\end{tabular}
\caption[Theoretical Uncertainties] {Uncertainties on the theoretical
calculations of selected Z-pole observables and $\MW$.  Top:
parametric uncertainties caused by the five $\SM$ input parameters.
For each observable, the change is shown when varying the $\SM$ input
parameter listed in the first column by the amount $\delta$ listed in
the second column, around the following central values:
$\dalhad=0.02758$, $\alfmz=0.118$, $\MZ=91.1875~\GeV$, $\Mt=178~\GeV$,
$\MH=150~\GeV$.  Where no number is listed, the effect is smaller than
half a unit in the number of digits quoted.  Bottom: theoretical
uncertainties due to missing higher-order corrections estimated
through variation of calculational schemes implemented in ZFITTER
(half of full range of values).  For comparison, the uncertainties on
the experimental measurements are shown in the last row.  }
\label{tab:TU}
\end{center}
\end{table}

\clearpage

\begin{figure}[p]
\begin{center}
\mbox{\epsfig{file=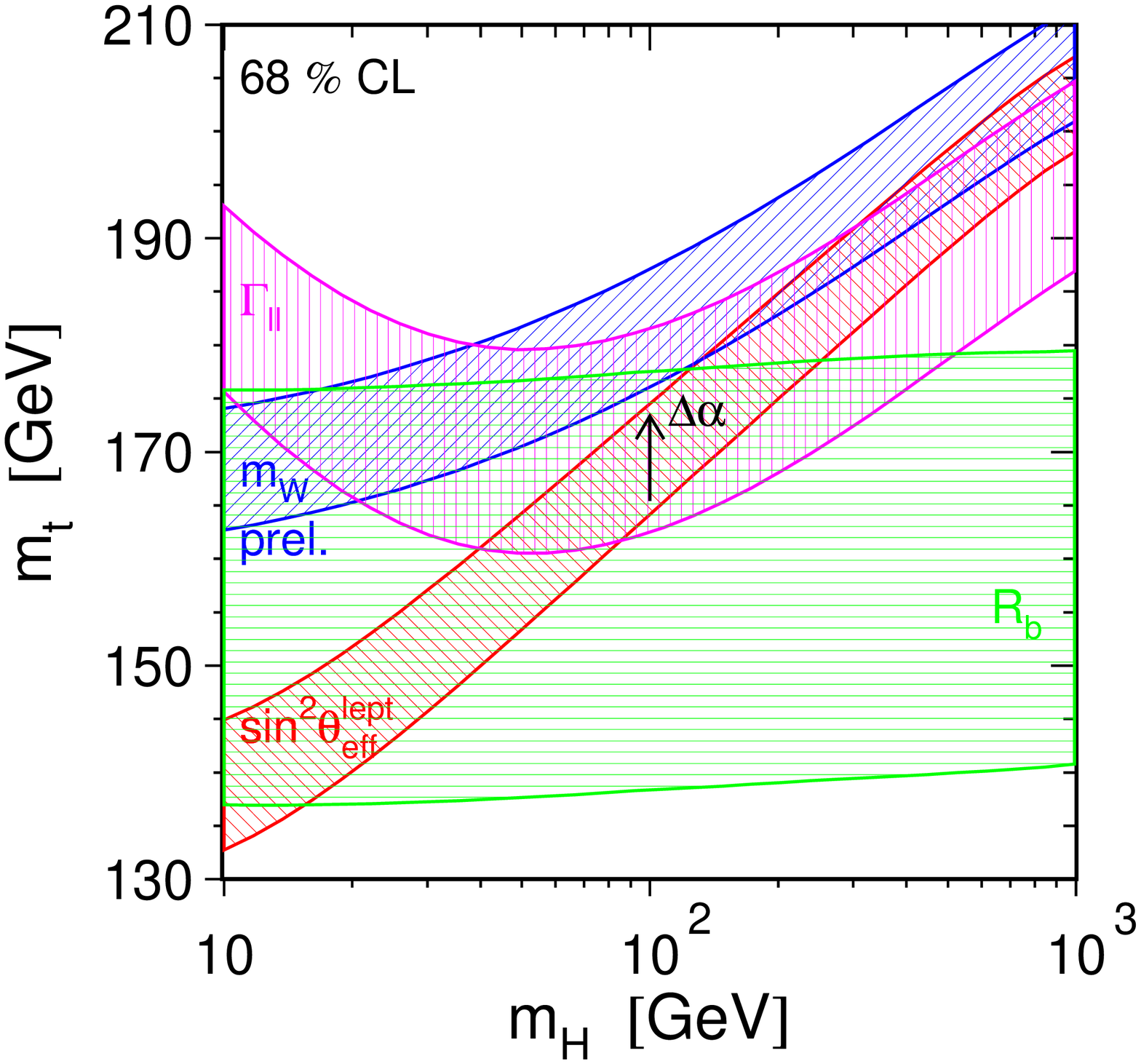,width=0.9\linewidth}} 
\caption[Constraints on $\Mt$ and $\MH$ from measurements of $\Rbz$,
$\Gll$, $\swsqeffl$ and $\MW$] {Constraints on $\Mt$ and $\MH$ from
measurements of $\Rbz$, $\Gll$, $\swsqeffl$ and $\MW$.  Each band
gives the $\pm1\sigma$ constraint from the indicated measurement.  The
parametric uncertainty due to the uncertainty in the hadronic vacuum
polarisation, $\dalhad=0.02758\pm0.00035$, is not included in the
width of these bands as it is small except for the $\swsqeffl$ band,
where the $\pm1\sigma$ uncertainty is indicated by the arrow labeled
$\Delta\alpha$.  The direct measurement of $\MW$ used here is
preliminary.}
\label{fig:msm:mh-mt-bands} 
\end{center}
\end{figure}

Relative to their measurement accuracy, four pseudo-observables are
particularly sensitive to the masses of the interesting $\SM$
particles, the top quark or the Higgs boson, while at the same time
are largely independent of QCD effects. These pseudo-observables are
$\Rbz$, $\Gll$, $\swsqeffl$ and $\MW$.  Each of these measurements
imposes a constraint on the size of electroweak radiative corrections,
which is graphically shown in Figure~\ref{fig:msm:mh-mt-bands} as a
band in the $(\MH,\Mt)$ plane.  Significant non-linearities occur in
these constraints over the allowed $\MH$ range.

Owing to the top-dependent vertex corrections as shown in
Figure~\ref{fig:b_vertex}, the quantity $\Rb$ is sensitive to $\Mt$,
while as a ratio of hadronic decay widths it is largely insensitive to
the other four $\SM$ input parameters, including the mass of the Higgs
boson, as shown in Figure~\ref{fig:coup:b-vertex}.  Within the $\SM$
framework, the measurement of $\Rbz$ therefore provides particularly
unambiguous information on $\Mt$.  If $\Rbz$ had been measured smaller
(i.e., its band shifted upwards in Figure~\ref{fig:msm:mh-mt-bands})
by a standard deviation, the indirect constraints on $\Mt$ and $\MH$
would both move toward higher values, along the almost parallel and
overlapping bands of the $\Gll$, $\swsqeffl$ and $\MW$ constraints.

The $\Gll$ band shown in Figure~\ref{fig:msm:mh-mt-bands} implies that
the preferred $\Mt$ exhibits a broad minimum around
$\MH\approx50~\GeV$.  In combination with the $\Rbz$ band preferring
an even lower value of $\Mt$, this results in an indirect
determination of $\Mt$ which is remarkably stable against variations
in $\swsqeffl$.  In contrast with the enhanced stability of the $\Mt$
determination, the favoured value of $\MH$ is very sensitive to
$\swsqeffl$.  It should also be noted that, of all the bands, only
$\swsqeffl$ is sensitive to the value of $\dalhad$, as indicated by
the arrow in Figure~\ref{fig:msm:mh-mt-bands}.

The effects of ZH production, or real Higgsstrahlung, are ignored
here, as well as in all results quoted in this paper. They are
negligible for $\MH>50~\GeV$. For $\MH<50~\GeV$, the rise of $\Mt$
with decreasing $\MH$ predicted by the $\Gll$ constraint band would be
somewhat suppressed, due to the fact that most, but not all, ZH events
where the $\Zzero$ decays to leptons would have been classed as
contributing to $\Ghad$ rather than $\Gll$.  Based on a detailed
analysis~\cite{Kawamoto:2004pi} it is concluded that, apart from the
determination of $\alfas$, Higgsstrahlung would not appreciably shift
the results of the $\SM$ analyses presented in
Section~\ref{sec:msm:msm}.

\begin{figure}[p]
\begin{center}
\mbox{\epsfig{file=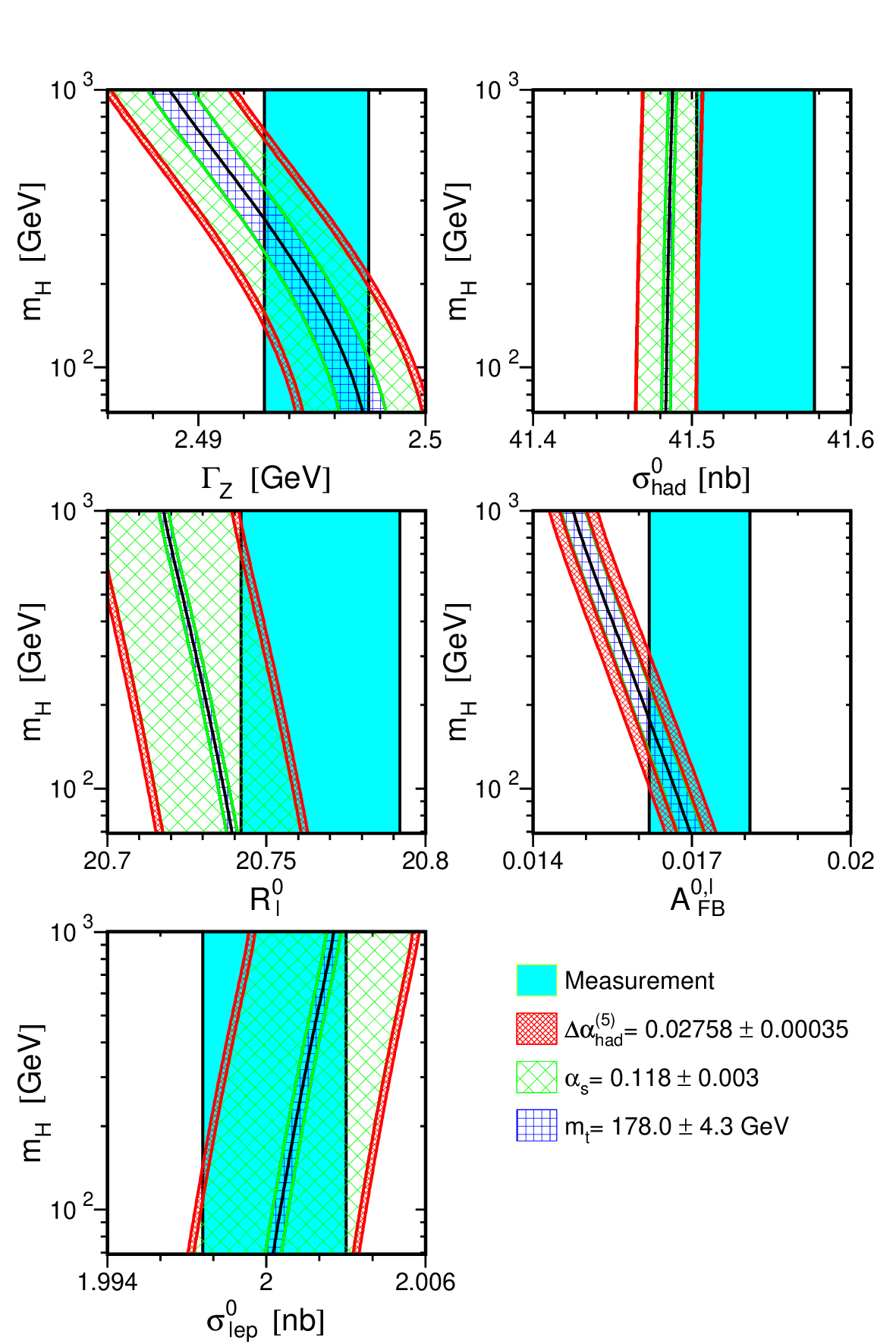,width=0.8\linewidth}} 
\caption[Higgs sensitivity of $\GZ$, $\shad$, $\Rl$, $\Afbzl$ and
$\swsqeffl(\Qfbhad)$] {Comparison of the LEP combined measurements of
  $\GZ$, $\shad$, $\Rl$, $\Afbzl$ and $\slept$ with the $\SM$
  prediction as a function of the mass of the Higgs boson. The
  measurement with its uncertainty is shown as the vertical band. The
  width of the $\SM$ band arises due to the uncertainties in
  $\dalhad$, $\alfmz$ and $\Mt$ in the ranges indicated.  The total
  width of the band is the linear sum of these uncertainties. }
\label{fig:msm:higgs-1} 
\end{center}
\end{figure}

\begin{figure}[p]
\begin{center}
\mbox{\epsfig{file=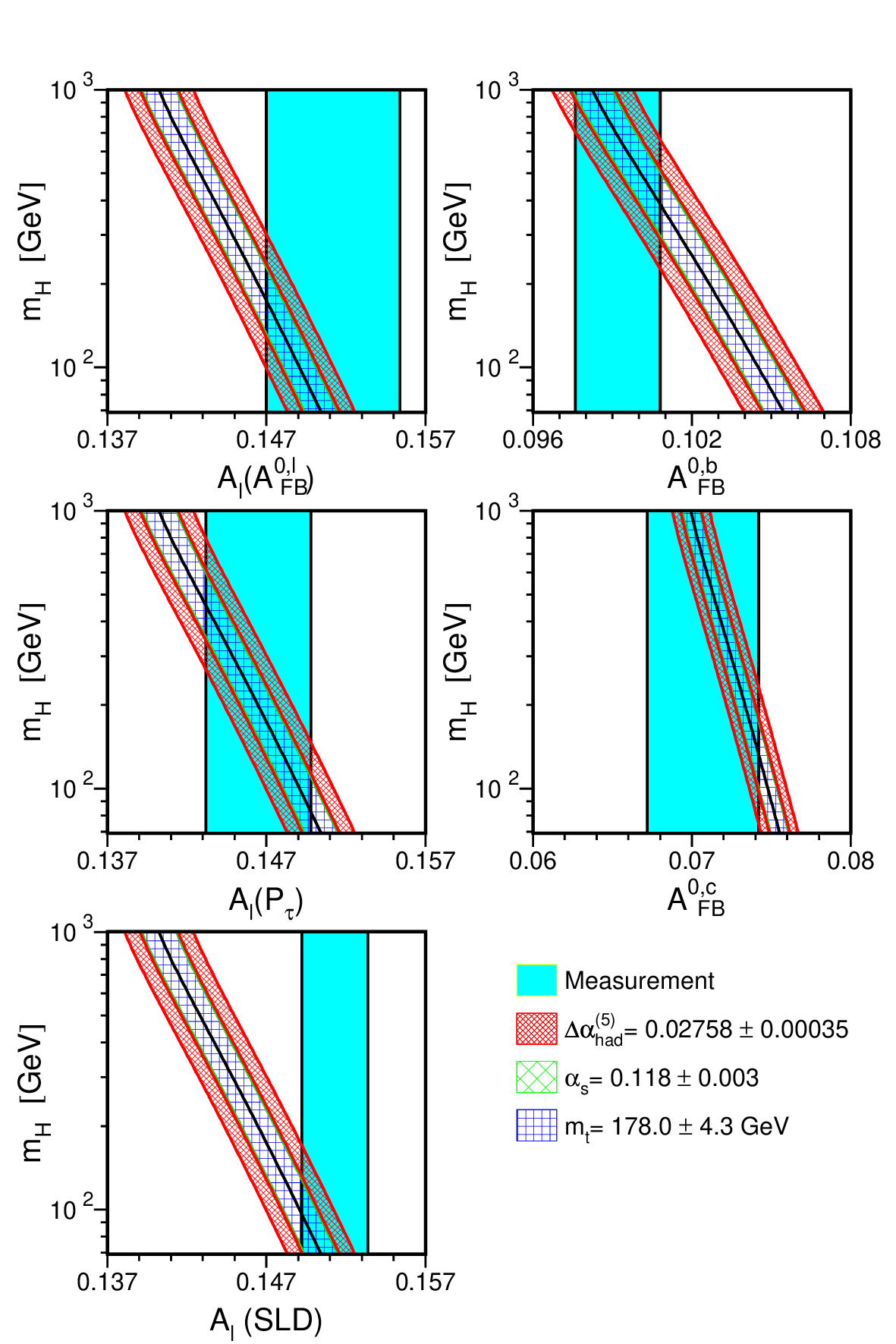,width=0.8\linewidth}} 
\caption[Higgs sensitivity of $\cAl(\Afbzl)$, $\cAl(P_\tau)$,
$\cAl$(SLD), $\Afbzb$ and $\Afbzc$] {Comparison of the LEP/SLD
  combined measurements of $\cAl(\Afbzl)$, $\cAl(P_\tau)$,
  $\cAl$(SLD), $\Afbzb$ and $\Afbzc$ with the $\SM$ prediction as a
  function of the mass of the Higgs boson.  The measurement with its
  uncertainty is shown as the vertical band. The width of the $\SM$
  band arises due to the uncertainties in $\dalhad$, $\alfmz$ and
  $\Mt$ in the ranges indicated.  The total width of the band is the
  linear sum of these uncertainties. }
\label{fig:msm:higgs-2} 
\end{center}
\end{figure}

\begin{figure}[p]
\begin{center}
\mbox{\epsfig{file=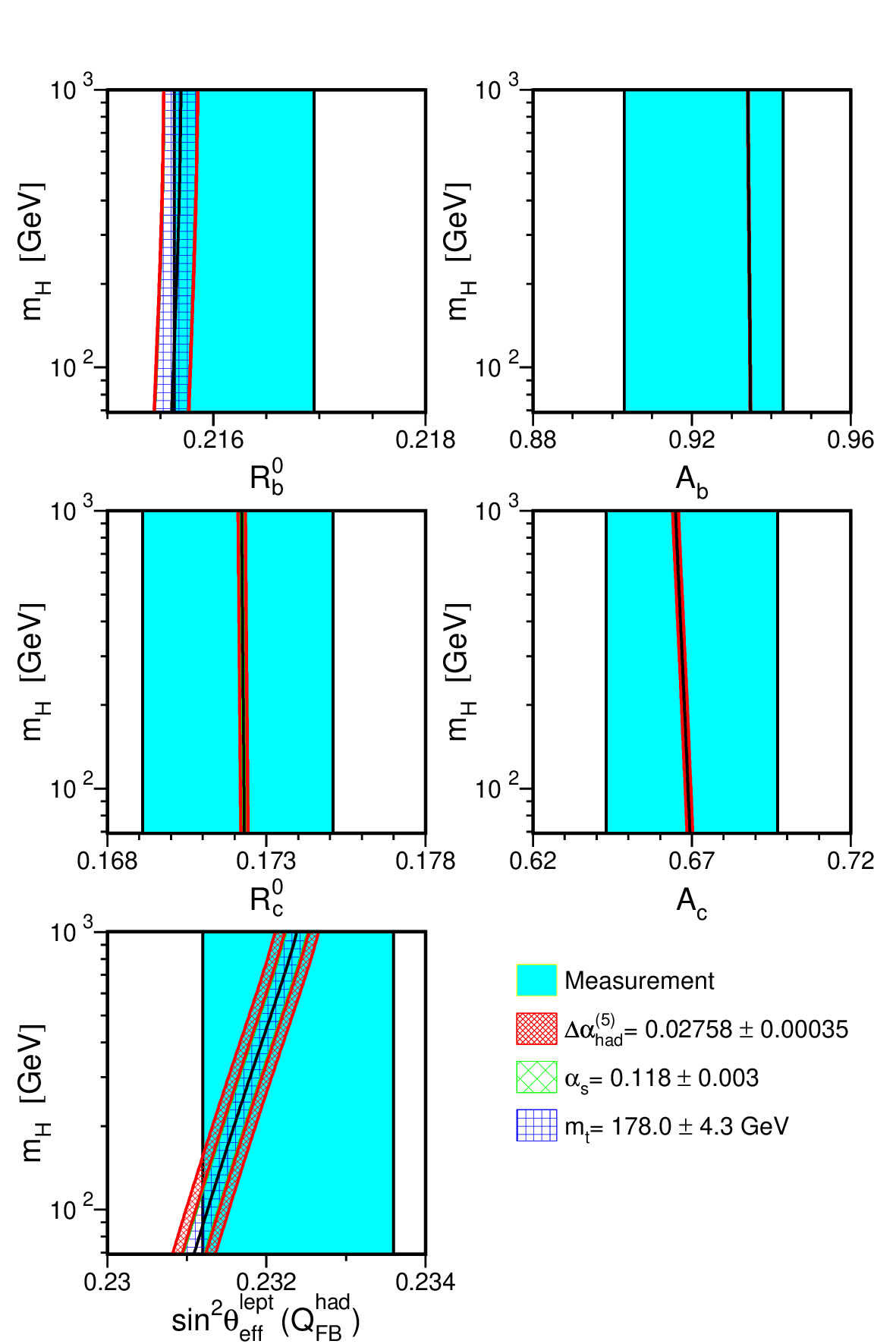,width=0.8\linewidth}} 
\caption[Higgs sensitivity of $\Rbz$, $\Rcz$, $\cAb$, $\cAc$ and $\swsqeffl(\Qfbhad)$] 
{Comparison of the LEP/SLD combined measurements of $\Rbz$, $\Rcz$,
  $\cAb$, $\cAc$ and $\swsqeffl(\Qfbhad)$ with the $\SM$ prediction as
  a function of the mass of the Higgs boson.  The measurement with its
  uncertainty is shown as the vertical band.  The width of the $\SM$
  band arises due to the uncertainties in $\dalhad$, $\alfmz$ and
  $\Mt$ in the ranges indicated.  The total width of the band is the
  linear sum of these uncertainties.  }
\label{fig:msm:higgs-3} 
\end{center}
\end{figure}

\begin{figure}[p]
\begin{center}
\mbox{\epsfig{file=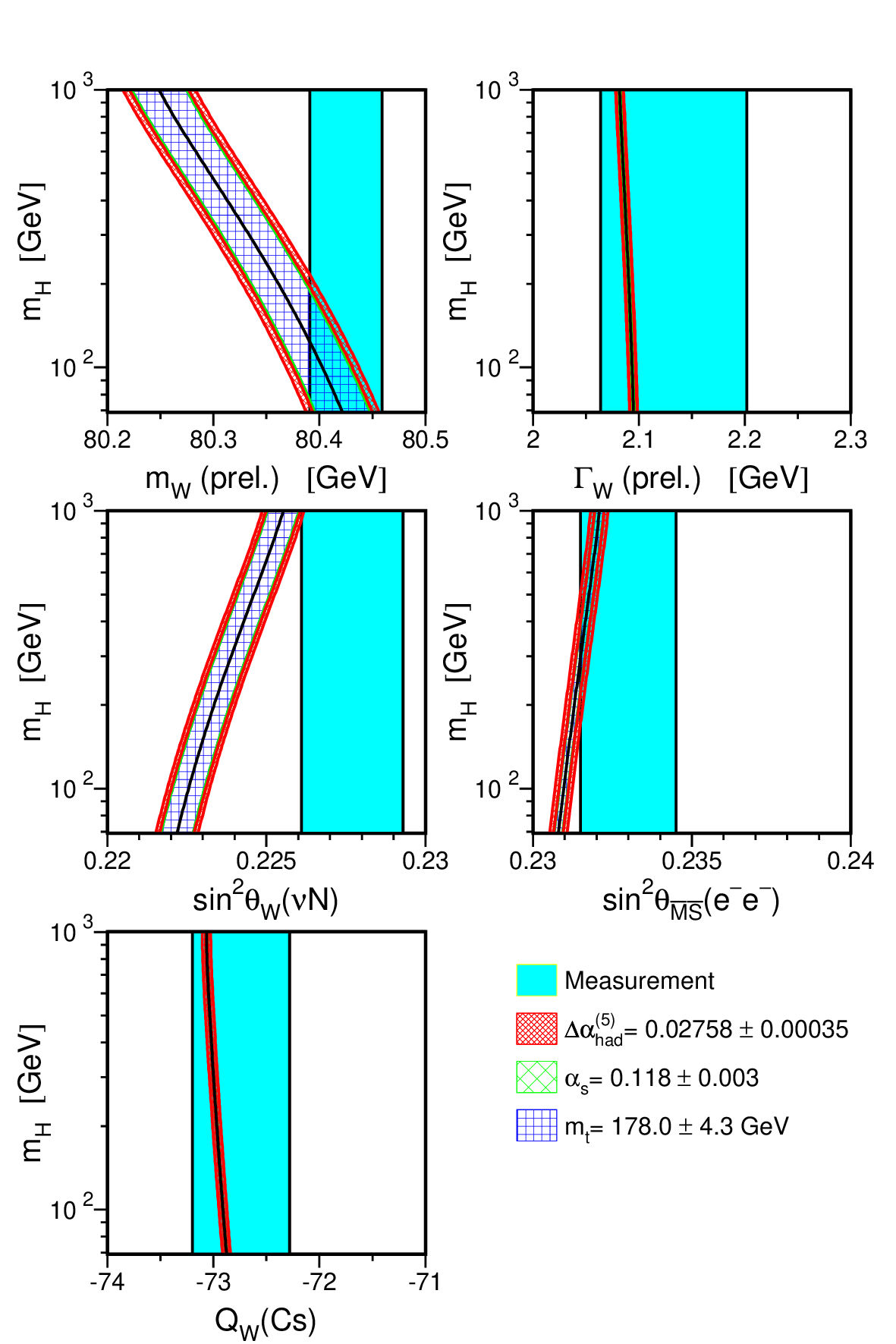,width=0.8\linewidth}} 
\caption[Higgs sensitivity of $\MW$, $\GW$, $\swsq$, $\swsqMSb$ and
$\QW$] {Comparison of the combined measurements of $\MW$ and $\GW$,
and the results from low $Q^2$ processes $\swsq~(\nu \mathrm{N})$,
$\swsqMSb~(\emem)$ and $\QW$ (APV) with the $\SM$ prediction as a
function of the mass of the Higgs boson.  The measurement with its
uncertainty is shown as the vertical band.  The width of the $\SM$
band arises due to the uncertainties in $\dalhad$, $\alfmz$ and $\Mt$
in the ranges indicated.  The total width of the band is the linear
sum of these uncertainties.  The direct measurements of $\MW$ and
$\GW$ used here are preliminary. }
\label{fig:msm:higgs-4} 
\end{center}
\end{figure}

The dependence of all pseudo-observables on the mass of the Higgs
boson within the framework of the $\SM$ is visualised in
Figures~\ref{fig:msm:higgs-1} to~\ref{fig:msm:higgs-4}, comparing the
experimental result with the value of the observable calculated within
the framework of the minimal $\SM$ as a function of the Higgs-boson
mass.  Non-linear effects, as already observed in
Figure~\ref{fig:msm:mh-mt-bands}, are clearly visible.

For the quantity $\swsqeffl$ determined in various asymmetry
measurements, it has already been shown in Figure~\ref{fig:coup:sef2}
that the parametric uncertainty on the $\SM$ prediction arising from
$\dalhad$ is non-negligible compared to the experimental uncertainty
of the average.  As a consequence, the uncertainty on the hadronic
vacuum polarisation is one of the limiting factors in the extraction
of the mass of the Higgs boson.  This situation underlines the
importance of further improved determinations of the hadronic vacuum
polarisation through measurements of the hadronic cross-section in
electron-positron annihilations at low centre-of-mass energies.
Compared to $\swsqeffl$, the W-boson mass is relatively less sensitive
to $\dalhad$ than to $\Mt$ and $\MH$, making $\MW$, measured at the
Tevatron and at \LEPII, an ideal observable to further reduce the
error on the prediction of the Higgs-boson mass.

\subsection{Theoretical Uncertainties}
\label{sec:msm:TU1}

Theoretical uncertainties in radiative corrections and the calculation
of pseudo-observables arise due to the fact that the perturbative
expansion is known and calculated only up to a finite order.  Many
theorists perform the various complicated calculations of radiative
corrections.  In order to make this work accessible to
experimentalists in a consistent way, the relevant calculations are
incorporated in computer programs such as TOPAZ0~\cite{\TOPAZref},
using the general minimal subtraction renormalisation scheme, and
ZFITTER~\cite{\ZFITTERref}, using the on-mass-shell renormalisation
scheme.  For the realistic observables, the measured cross-sections
and asymmetries, the following corrections are included in TOPAZ0 and
ZFITTER: up to $\calO(\alpha^2)$ and leading $\calO(\alpha^3)$
for initial-state QED radiation including pairs, $\calO(\alpha)$
for final-state QED radiation and QED initial-final state
interference, $\calO(\alfas^3)$ for final-state QCD radiation and
$\calO(\alpha\alfas)$ for mixed QED/QCD final-state radiation.
These corrections are needed to extract the pseudo-observables
discussed in this report from the realistic observables.  For the
calculation of the expectation for the extracted pseudo-observables
discussed in this report, the final-state corrections listed above are
also available for the Z decay widths.  Furthermore, complete one-loop
electroweak radiative corrections, re-summed leading one-loop
corrections and two-loop corrections up to $\calO(\alpha\alfas,
\alpha\alfas^2, \GF^2\Mt^4, \GF^2\Mt^2\MZ^2, \GF\Mt^2\alfas,
\GF\Mt^2\alfas^2)$ are included.  Overviews and summaries of radiative
corrections in Z-pole physics are given in
References~\citen{LEP1YR89VOL1, bib-PCLI, BardinPassarinoBook}, which
should be consulted for references to the original calculations.

Missing higher-order electroweak, strong and mixed corrections cause
the calculation of any observable to be incomplete and thus
approximate.  Ambiguities also arise due to the choice of
renormalisation schemes, re-summation schemes, momentum-transfer
scales in loop corrections, and schemes to implement the factorisation
of various corrections.  These ambiguities reflect and are of the same
order as the missing higher-order corrections.  The uncertainty on the
predicted observables due to these effects is thus estimated by
comparing results obtained using different calculations performed to
equivalent order~\cite{bib-PCLI,BP:98,PCP99,BCEHWWW01,Snowmass:2001}.
Recent developments in the calculation of electroweak radiative
corrections include the complete two-loop corrections for the mass of
the W boson~\cite{Twoloop-MW}, leading three-loop top-quark
contributions to the $\rho$ parameter~\cite{Threeloop-rho}, and
fermionic two-loop corrections for the effective electroweak mixing
angle~\cite{Twoloop-sin2teff}.  The remaining theoretical
uncertainties are estimated to be $\pm4~\MeV$ in
$\MW$~\cite{Twoloop-MW} and $\pm4.9\cdot10^{-5}$ in
$\swsqeffl$~\cite{Twoloop-sin2teff}, respectively.

The recent calculations and their associated theoretical uncertainties
are implemented in ZFITTER~6.42~\cite{\ZFITTERref} and used
here.\footnote{ The default flags of ZFITTER~6.42 are used, except for
setting {\tt AMT4=6} to access these latest electroweak radiative
corrections and setting {\tt ALEM=2} to take into account the
externally supplied value of $\dalhad$.  The effects of the
theoretical uncertainties in the calculations of $\MW$ and $\swsqeffl$
are simulated by changing the ZFITTER flags {\tt DMWW} and {\tt DSWW}
from their default value of {\tt 0} to {\tt$\tt\pm$1}. } Numerical
results for theoretical uncertainties calculated with ZFITTER are
reported in Table~\ref{tab:TU} for several pseudo-observables.  The
uncertainties due to missing higher-orders are in general small
compared to the leading parametric uncertainties, with the exception
of the effective electroweak mixing angle.  The latter uncertainty
dominates all other theoretical uncertainties in global $\SM$
analyses.

\subsubsection{QCD Uncertainties}
\label{sec:msm:QCD-TU}

The largest QCD correction in the calculation of Z-pole observables
arises through the final-state QCD radiation factor in quark-pair
production (Equation~\ref{eq:Gff}), modifying the decay width of the Z
into hadrons, $\Ghad$, and thus also the Z-pole observables $\Gtot$,
$\Rl$, $\shad$ and $\slept$, which depend trivially on $\Ghad$.  The
quark asymmetries also require significant QCD corrections.  The
theoretical uncertainty in the calculation of the observables related
to $\Ghad$ due to unknown higher-order QCD effects, and conversely in
the $\alfmz$ values extracted from measurements of these quantities,
is a subject of current discussion.  Estimates of the corresponding
theoretical uncertainty on $\alfmz$ extracted from these observables
vary from 0.0005 to 0.003~\cite{QCD:Kuhn:1996, QCD:Soper:1996,
QCD:Bethke:2000, Bethke:2004uy, Stenzel:2005sg}, a range which spans
the uncertainty on $\alfmz$ caused by the experimental errors on the
measured hadronic Z-pole observables.

Virtual quark loops with additional gluon exchange induce QCD
corrections to the propagators, which introduce an additional, but
much smaller $\alfas$ dependence in the calculation of each Z-pole
observable, mainly through $\rhof$ and $\kappaf$.  These two-loop
$\calO(\alpha\alfas)$ corrections are small but known to leading
order only, hence smaller but non-negligible additional theoretical
uncertainties on the prediction of any Z-pole observable arise.

This has several consequences: The extracted value of $\alfmz$ is
mainly given by the dependence of the hadronic Z-pole observables on
the final-state QCD radiation factor.  As it is the very same
final-state QCD correction factor entering the calculation of all
hadronic Z-pole observables, most theoretical QCD uncertainties are
fully correlated and affect the extracted $\alfmz$ value independent
of which observable is used.  Because of this strong correlation, the
extracted values of the other $\SM$ input parameters are largely
insensitive to the theoretical uncertainty due to unknown higher-order
QCD effects, as in the fit any such bias is effectively absorbed in
the fit value of $\alfmz$.\footnote{ Note that this theoretical
uncertainty would have to be known quantitatively and included
explicitly if external measurements of $\alfmz$ were included in the
analyses.  For the $\SM$ analyses presented here, however, this is not
necessary as external constraints on $\alfmz$, even without any
uncertainty, would not lead to reduced uncertainties on the other
$\SM$ input parameters. }

\section{Analysis Procedure}
\label{sec:msm:proc}

In order to determine the five relevant $\SM$ input parameters a
$\chi^2$ minimisation is performed using the program
MINUIT~\cite{MINUIT}.  The $\chi^2$ is calculated as usual by
comparing the measurements of Z-pole and other observables, their
errors and correlations including those discussed in
Chapter~\ref{chap:corr}, with the predictions calculated in the
framework of the $\SM$. The results combined in the previous chapters
under the hypothesis of lepton universality, which is inherent to the
$\SM$, are used for measurements of leptonic Z-pole observables.  All
are reported in Table~\ref{tab:msm:input}.  The predictions are
calculated as a function of the five $\SM$ input parameters by the
program ZFITTER, while TOPAZ0 and ZFITTER are used to calculate
theoretical uncertainties. Both programs include all relevant
electroweak radiative corrections.  All five $\SM$ input parameters
are allowed to vary in the fit, so that parametric uncertainties are
correctly treated and propagated.

This analysis procedure tests quantitatively how well the $\SM$ is
able to describe the complete set of all measurements with just one
value for each of the five $\SM$ input parameters.  For interpreting
the adequacy of this description, however, the large contribution to
the $\chi^2$ arising from the asymmetry measurements as discussed in
the previous chapter has to be taken into account.

In addition, the mass of the only particle of the $\SM$ which remains
without significant direct experimental evidence, the mass of the
Higgs boson, will be constrained.  For this determination, the
additional measurements presented in Section~\ref{sec:msm:add}, such
as the direct measurements of $\MW$ and $\Mt$ at \LEPII\ and the
Tevatron, are also included, in order to obtain the best precision.

In the case of those observables which are $\SM$ input parameters and
thus fit parameters, such as $\MZ$, $\dalhad$ and $\Mt$, special care
is needed when evaluating the performance of various measurements in
constraining the fitted mass of the Higgs boson.  As in general all
measurements carry information about all $\SM$ input parameters, a
shift of such a measurement by one standard deviation does not lead to
a shift of the fitted Higgs-boson mass given by the corresponding
fitted correlation coefficient. As for all other measurements, a fit
to the new set of measurements has to be performed.

\subsection{Treatment of Systematic Uncertainties}

As discussed in detail in the previous chapters, the experimental
measurements have associated uncertainties which are of both
statistical and systematic nature. Both sources are assumed to be and
are treated as Gaussian errors corresponding to a symmetric interval
around the central value with 68\% probability content. While this is
a valid model for statistical and many systematic errors, some
systematic uncertainties are derived from discrete tests, e.g.,
performing a Monte Carlo test with and without a certain option
affecting the event generation and detector simulation. For errors of
this type, a flat, box-like probability distribution, or any other,
could also be applicable. For the analyses presented in the following,
studies show that the central values of the fitted parameters are
affected only slightly by the particular choice of the probability
density function for such uncertainties.  A somewhat larger effect is
seen for the fitted uncertainty of the fitted parameters.  Since a box
of size $\pm\sigma$ has a spread of $\pm\sigma/\sqrt{3}$, the
uncertainties of the fitted parameters would decrease if such a model
were to be applied to these less tractable errors.  Thus the results
presented below are considered conservative.

Theoretical uncertainties due to missing higher order corrections as
discussed above are typically implemented by offering various choices
or options in the programs TOPAZ0 and ZFITTER when calculating
radiative corrections.  As these choices correspond to discrete
options (flags), they cannot be varied during a fit.  Rather, the
analysis is repeated with different flag settings.  The change in the
five extracted $\SM$ input parameters is taken as an estimate of the
theoretical uncertainty for the option studied.  The flags are varied
one by one and the fits are repeated. The maximum deviation of any
given flag change is taken as the theoretical uncertainty, thus
avoiding double counting due to correlated variations governed by
different flags.  Since this uncertainty is usually much smaller than
the uncertainty arising from the experimental uncertainties in the
measured Z-pole observables (Table~\ref{tab:TU}), it is not included
in the results presented in the following.  By far the largest
electroweak theoretical uncertainty affecting the determination of the
five $\SM$ input parameters, mainly the mass of the Higgs boson, is
that of the effective electroweak mixing angle.

\section{Standard Model Analyses}
\label{sec:msm:msm}

\subsection{Z-Pole Results}

Based on the electroweak observables measured at \LEPI\ and by SLD,
and presented before, a fit is performed to the hadronic vacuum
polarisation and the 14 Z-pole observables derived under the
assumption of lepton universality, in order to determine the five
input parameters of the $\SM$.  The result is reported in
Table~\ref{tab:msmfit-lep1sld}. A $\chidf$ of 16.0/10 is obtained,
corresponding to a probability of 9.9\%.  The largest contribution to
the $\chi^2$ arises from the asymmetry measurements as discussed in
Section~\ref{sec:coup:disc}.  The $\SM$ describes the complete set of
measurements with a unique set of values for the five $\SM$ input
parameters.

Tests show that the inclusion of a direct measurement of $\alfmz$, or
even fixing $\alfmz$, results in negligible improvements in the
determination of the other $\SM$ input parameters, since correlation
coefficients between $\alfmz$ and all other parameters are small.
Similarly, the cross-section scale, which depends directly on the
normalization of the luminosity measurement, decouples from other
$\SM$ input parameters.  The fit results are rather stable except for
a small shift in $\alfmz$ when the measurement containing the
cross-section normalisation, $\shad$, is dropped from the input
measurements.

\begin{table}[t]
\begin{center}
\renewcommand{\arraystretch}{1.25}
\begin{tabular}{|c||r@{$\pm$}l||rrrrr|}
\hline
Parameter & \multicolumn{2}{|c||}{Value} 
          & \multicolumn{5}{|c| }{Correlations} \\
          & \multicolumn{2}{|c||}{ }
          & {$\dalhad$} & {$\alfmz$} 
          & {$\MZ$} & {$\Mt$} & {$\LOGMH$}      \\
\hline
\hline
$\dalhad$   &$0.02759$&$0.00035$&$ 1.00$&$     $&$     $&$     $&$     $\\
$\alfmz$    &$0.1190 $&$0.0027 $&$-0.04$&$ 1.00$&$     $&$     $&$     $\\
$\MZ~[\GeV]$&$91.1874$&$0.0021 $&$-0.01$&$-0.03$&$ 1.00$&$     $&$     $\\
$\Mt~[\GeV]$&$173    $&$^{13}_{10}$
                                &$-0.03$&$ 0.19$&$-0.07$&$ 1.00$&$     $\\
$\LOGMH$    &$2.05   $&$^{0.43}_{0.34}$
                                &$-0.29$&$ 0.25$&$-0.02$&$ 0.89$&$ 1.00$\\
\hline
$\MH~[\GeV]$&$111    $&$^{190}_{60}$
                                &$-0.29$&$ 0.25$&$-0.02$&$ 0.89$&$ 1.00$\\
\hline
\end{tabular}
\caption[Results for $\SM$ input parameters from Z-pole measurements]
{Results for the five $\SM$ input parameters derived from a fit to the
Z-pole results and $\dalhad$.  The fit has a $\chidf$ of 16.0/10,
corresponding to a probability of 9.9\%.  See Section~\ref{sec:msm:TU}
for a discussion of the theoretical uncertainties not included
here. The results on $\MH$, obtained by exponentiating the fit results
on $\LOGMH$, are also shown.}
\label{tab:msmfit-lep1sld}
\end{center}
\end{table}

\subsubsection{Discussion}

The Z-pole data alone are not able to improve significantly on the
determination of $\dalhad$ compared to the direct determination
presented in Section~\ref{sec:msm:vacpol}.  The strong coupling
constant, $\alfmz$, mainly determined by the leptonic pole
cross-section $\slept=\shad/\Rl$ as discussed in
Sections~\ref{chap:partrafo} and~\ref{sec:msm:QCD-TU} and shown in
Figure~\ref{fig:msm:higgs-1}, is one of the most precise
determinations of this quantity and in good agreement with other
determinations~\cite{Bethke:2004uy} and the world
average~\cite{PDG2004}, but theoretical issues currently obscure the
appropriate theoretical uncertainty to assign in its interpretation,
as discussed in Section~\ref{sec:msm:TU1}.  A dedicated analysis
following the detailed prescription given in
Reference~\citen{Stenzel:2005sg} yields a theoretical uncertainty of
0.0010 on $\alfmz$ extracted from this set of Z-pole measurements.

The role of the mass of the Z boson is now changed from that of a
model-independent parameter, unrelated to the other pseudo-observables
except for defining the pole position in the extraction of the pole
observables, to that of a fundamental input parameter of the $\SM$
affecting the calculation of all pseudo-observables.  Because of its
high precision with respect to the other measurements, the uncertainty
on $\MZ$ remains unchanged.

The pole mass of the top quark is predicted with an accuracy of about
$12~\GeV$. This precise prediction for a fundamental particle of the
$\SM$ not directly accessible at the Z pole emphasises clearly the
predictive power of the $\SM$ as well as the precision of the
experimental results.

Despite the logarithmic dependence of the electroweak radiative
corrections on the mass of the Higgs boson, its value is nevertheless
predicted within a factor of about 2.  The value obtained shows the
self-consistency of the $\SM$ analysis presented here, as such an
analysis would be inconsistent and invalid for resulting Higgs-boson
masses too small, as discussed in Section~\ref{sec:intro_SM_remnants},
or close to or larger than $1~\TeV$.  The large correlation
coefficient of $\MH$ with $\Mt$ shows that the precision of the $\MH$
prediction will significantly improve when the direct measurement of
$\Mt$ is included, as will be shown in Section~\ref{sec:msm:higgs}.

Having determined the five $\SM$ input parameters as given in
Table~\ref{tab:msmfit-lep1sld}, the parameters discussed in
Section~\ref{sec:intro_ew} are then predicted to be:
\begin{equation}
\begin{array}{rclcrcl}
\swsq          & = &  0.22331 \pm 0.00062              &~~&
               &   &                                   \\[2mm]
\swsqeffl      & = &  0.23149 \pm 0.00016              &~~&
\kappa_\ell    & = &  1.0366  \pm 0.0025               \\[2mm]
\swsqeffb      & = &  0.23293 \pm^{0.00031}_{0.00025}  &~~&
\kappab        & = &  1.0431  \pm 0.0036               \\[2mm]
\rhol          & = &  1.00509 \pm^{0.00067}_{0.00081}  &~~&
-\Drw          & = &  0.0242  \pm 0.0021               \\[2mm]
\rhob          & = &  0.99426 \pm^{0.00079}_{0.00164}  &~~&
\Dr            & = &  0.0363  \pm 0.0019               \\[1mm]
\end{array}
\end{equation}
The quantities presented here are obtained from the same data set.
Hence they are correlated with the five $\SM$ input parameters and
cannot be used independently.  Predictions of many more observables
within the $\SM$ framework are reported in
Appendix~\ref{app:SM:preds}.

Besides the hadronic vacuum polarisation $\dalhad$, only results from
the Z-pole measurements, whose precision will not be improved in the
near future, are used up to this point.  The impact of the precision
measurements of $\Mt$, $\MW$ and $\GW$, as discussed in
Sections~\ref{sec:msm:add:top} and~\ref{sec:msm:add:W}, is considered
in the following.  Note that these results are expected to benefit
from new measurements in the near future.

\subsection{The Mass of the Top Quark and of the W Boson}
\label{sec:mtdirect}

The above indirect constraint on the pole mass of the top quark,
$\Mt=173^{+13}_{-10}~\GeV$ (Table~\ref{tab:msmfit-lep1sld}), can be
compared with the result of the direct measurement of $\Mt$ at Run-I
of the Tevatron, $\Mt=178.0\pm4.3~\GeV$~\cite{PP-MT:combination}.  The
indirect determination is in good agreement with the direct
measurement.  It is impressive to note that even before the discovery
of the top-quark in 1995, the then available set of electroweak
precision data allowed the mass of the top quark to be predicted
correctly as verified by its direct measurement obtained later, see
Section~\ref{sec:intro_impact}.

\begin{figure}[p]
\begin{center}
\mbox{\epsfig{file=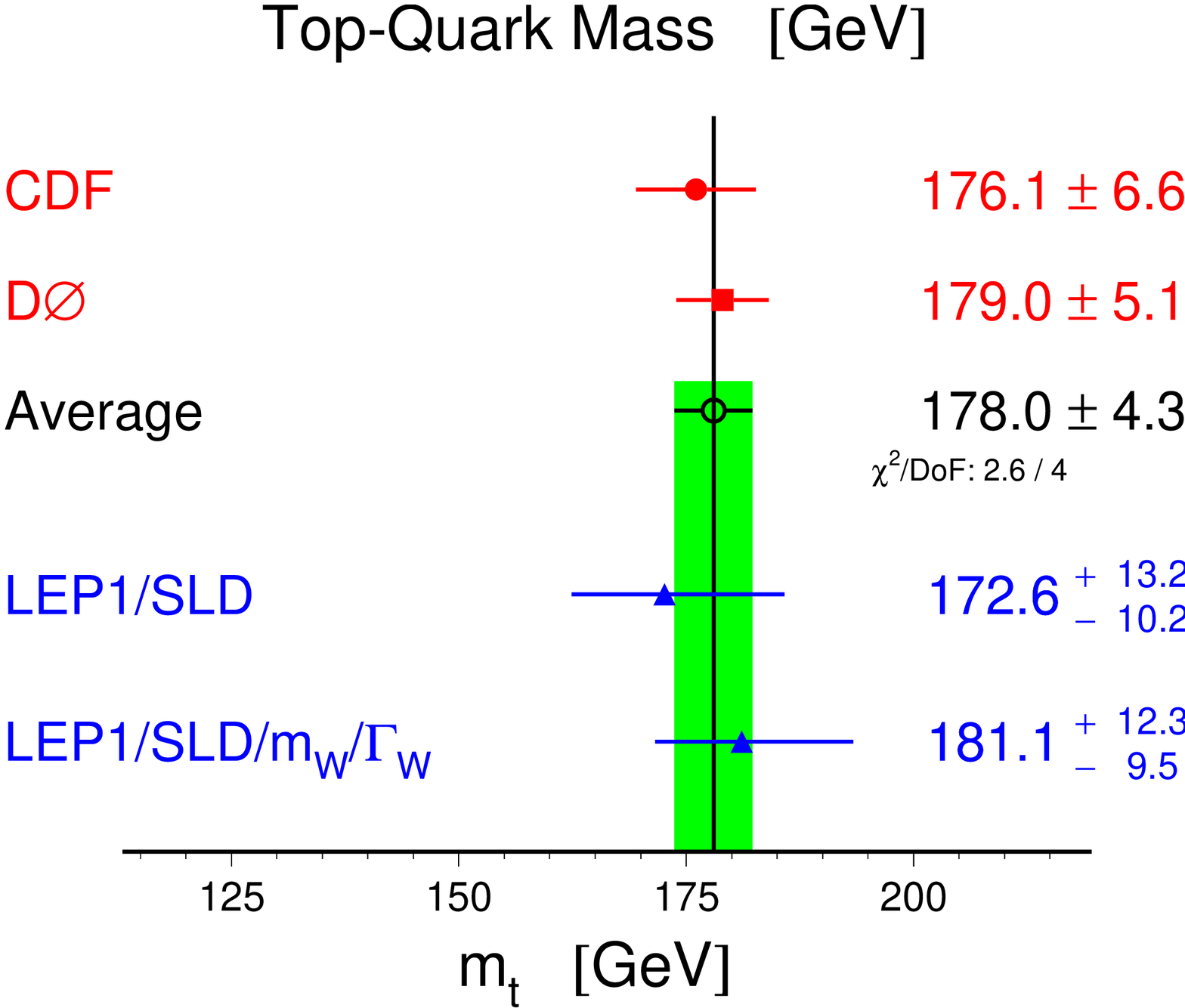,width=0.7\linewidth}} 
\caption[Comparison of results on the mass of the top quark] {Results
on the mass of the top quark. The direct measurements of $\Mt$ at
Run-I of the Tevatron (top) are compared with the indirect
determinations (bottom). }
\label{fig:msm:mt} 
\end{center}
\end{figure}

The accuracy of the indirect constraint on $\Mt$ is improved by
including the combined results on the W boson mass and width measured
at Run-I of the Tevatron and at \LEPII\ as presented in
Section~\ref{sec:msm:add:W}:
\begin{eqnarray}
\Mt & = & 181^{+12}_{-9}~\GeV \label{eq:sm:mt-1}\,.
\end{eqnarray}
It can therefore be seen that the direct measurement of the top-quark
mass is nearly three times as accurate as its indirect determination
within the framework of the $\SM$.  The different determinations of
$\Mt$ are compared in Figure~\ref{fig:msm:mt}.

Based on the results listed in Table~\ref{tab:msmfit-lep1sld}, the
prediction for the mass of the W boson is:
\begin{eqnarray}
\MW       & = & 80.363   \pm 0.032~\GeV \,,\label{eq:sm:mw-0}
\end{eqnarray}
which is in agreement at the level of 1.3 standard deviations with the
combined direct measurement of $\MW=80.425\pm0.034~\GeV$ as presented
in Section~\ref{sec:msm:add:W}.  

\begin{figure}[p]
\begin{center}
\mbox{\epsfig{file=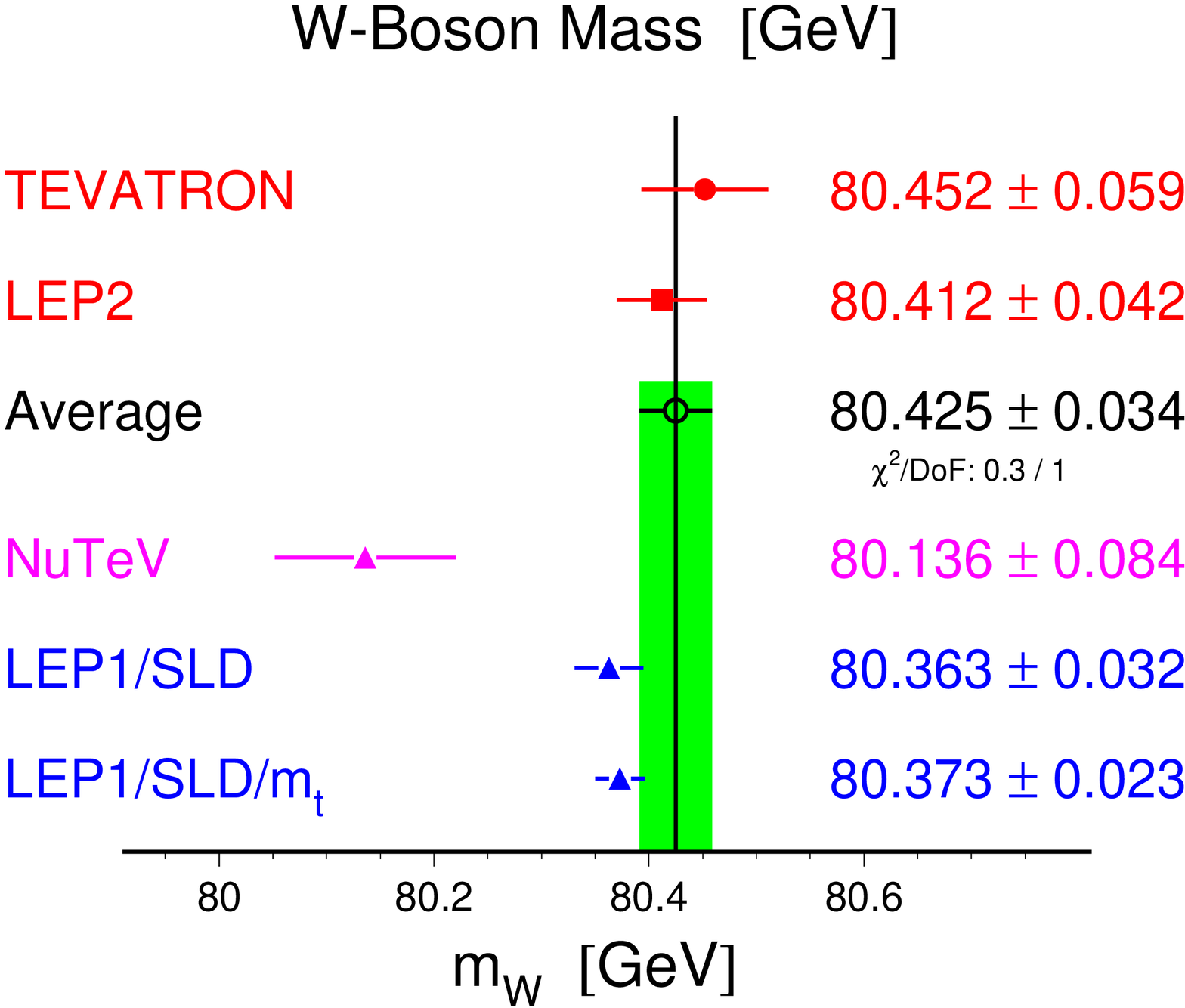,width=0.7\linewidth}} 
\caption[Comparison of results on the mass of the W boson] {Results on
the mass of the W boson, $\MW$. The direct measurements of $\MW$ at
\LEPII\ (preliminary) and at Run-I of the Tevatron (top) are compared
with the indirect determinations (bottom).  The NuTeV result
interpreted in terms of $\MW$ is shown separately. }
\label{fig:msm:mw} 
\end{center}
\end{figure}

The accuracy of the $\MW$ prediction is improved when the direct
measurement of the top-quark mass from Run-I of the Tevatron is
included:
\begin{eqnarray}
\MW       & = & 80.373   \pm 0.023~\GeV \,.\label{eq:sm:mw-1}
\end{eqnarray}
The indirect $\SM$ constraint on $\MW$ is therefore seen to be more
precise than the current direct measurements.  For a stringent test of
the $\SM$, the mass of the W boson should thus be measured directly to
an accuracy of $20~\MeV$ or better.  The different determinations of
$\MW$ are compared in Figure~\ref{fig:msm:mw}, also showing the NuTeV
result when interpreted as a measurement of $\MW$.

\subsection{The Mass of the Higgs Boson}
\label{sec:msm:higgs}

\begin{figure}[p]
\begin{center}
\mbox{\epsfig{file=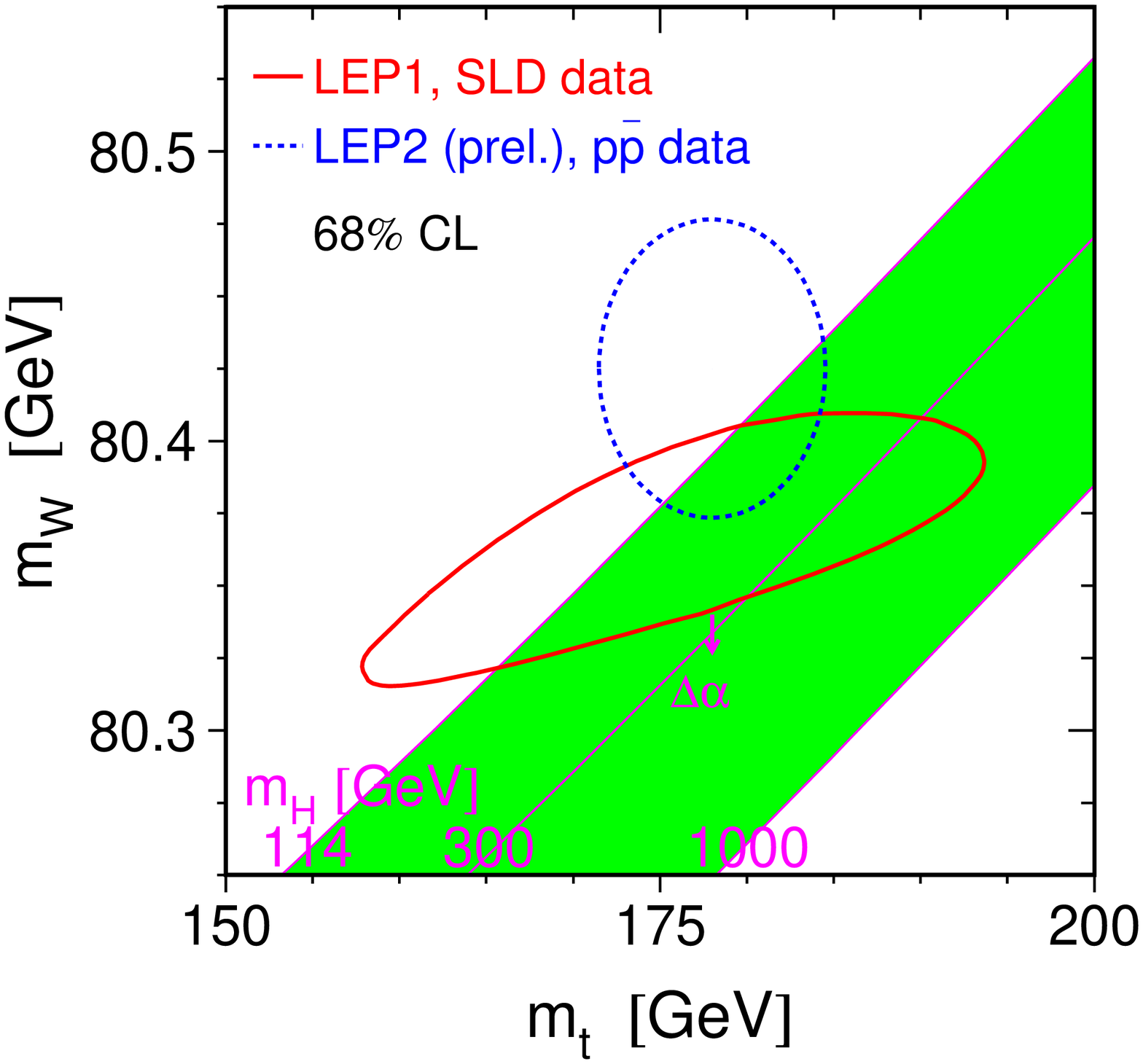,width=0.9\linewidth}} 
\caption[Comparison of top-quark and W-boson mass determinations]
{Contour curves of 68\% probability in the $(\Mt,\MW)$ plane.  The
shaded band shows the $\SM$ prediction based on the value for $\GF$
for various values of the Higgs-boson mass and fixed $\dalhad$;
varying the hadronic vacuum polarisation by
$\dalhad=0.02758\pm0.00035$ yields an additional uncertainty on the
$\SM$ prediction shown by the arrow labeled $\Delta\alpha$. The direct
measurement of $\MW$ used here is preliminary. }
\label{fig:msm:mt-mw} 
\end{center}
\end{figure}

The comparison between the indirect constraints and the direct
measurements of $\Mt$ and $\MW$ in the ($\Mt,\MW$) plane is shown in
Figure~\ref{fig:msm:mt-mw}.  The observed agreement is a crucial test
of the $\SM$.  Since the $\SM$ is so successful in predicting the
values of $\MW$ and $\Mt$, this type of analysis is now extended to
predict the mass of the Higgs boson.  As seen in the figure, both
contours prefer low values for the mass of the Higgs boson.

\begin{figure}[p]
\begin{center}
\vskip -1cm
\mbox{\epsfig{file=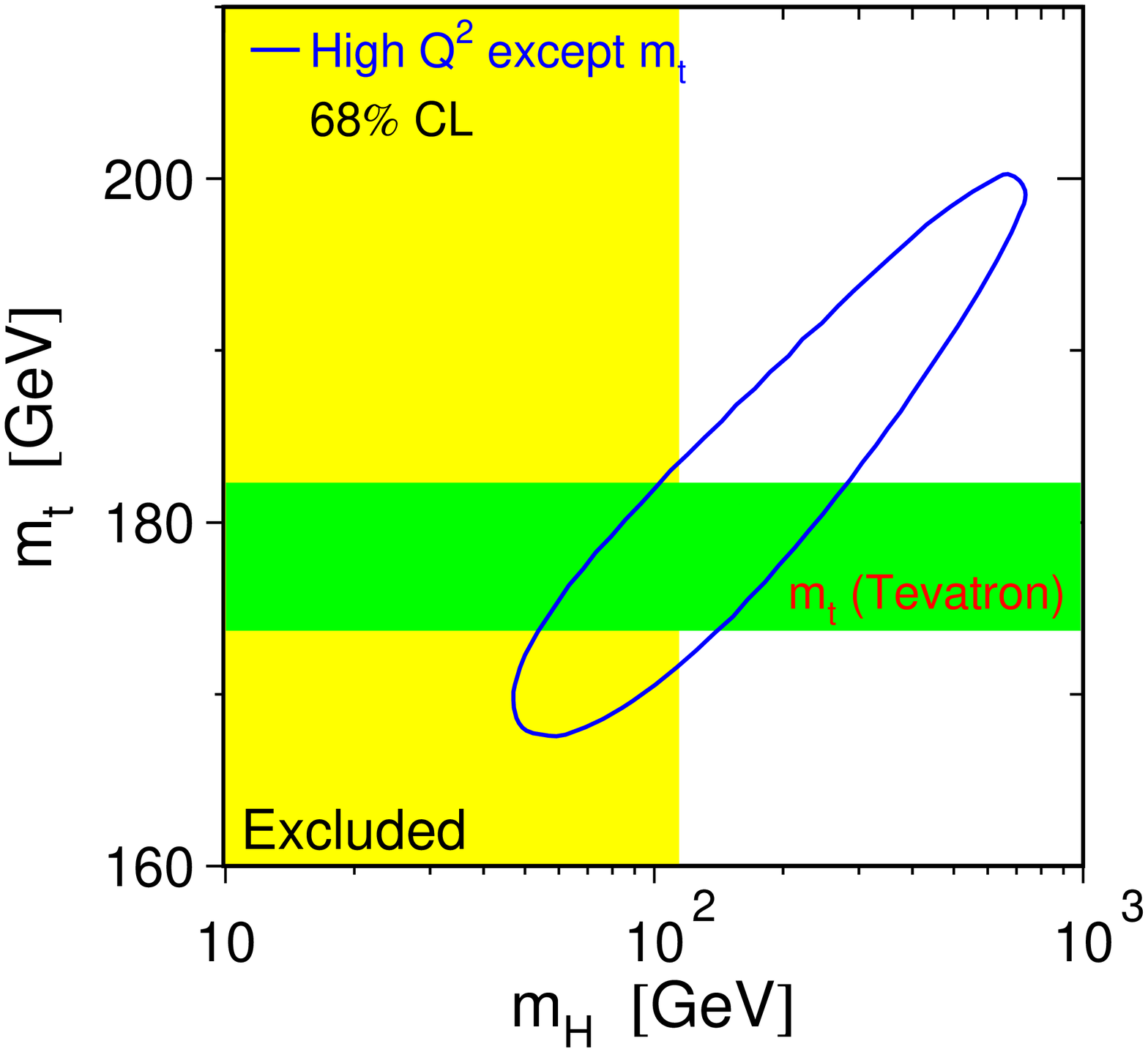,width=0.6\linewidth}} 
\vskip -1cm
\mbox{\epsfig{file=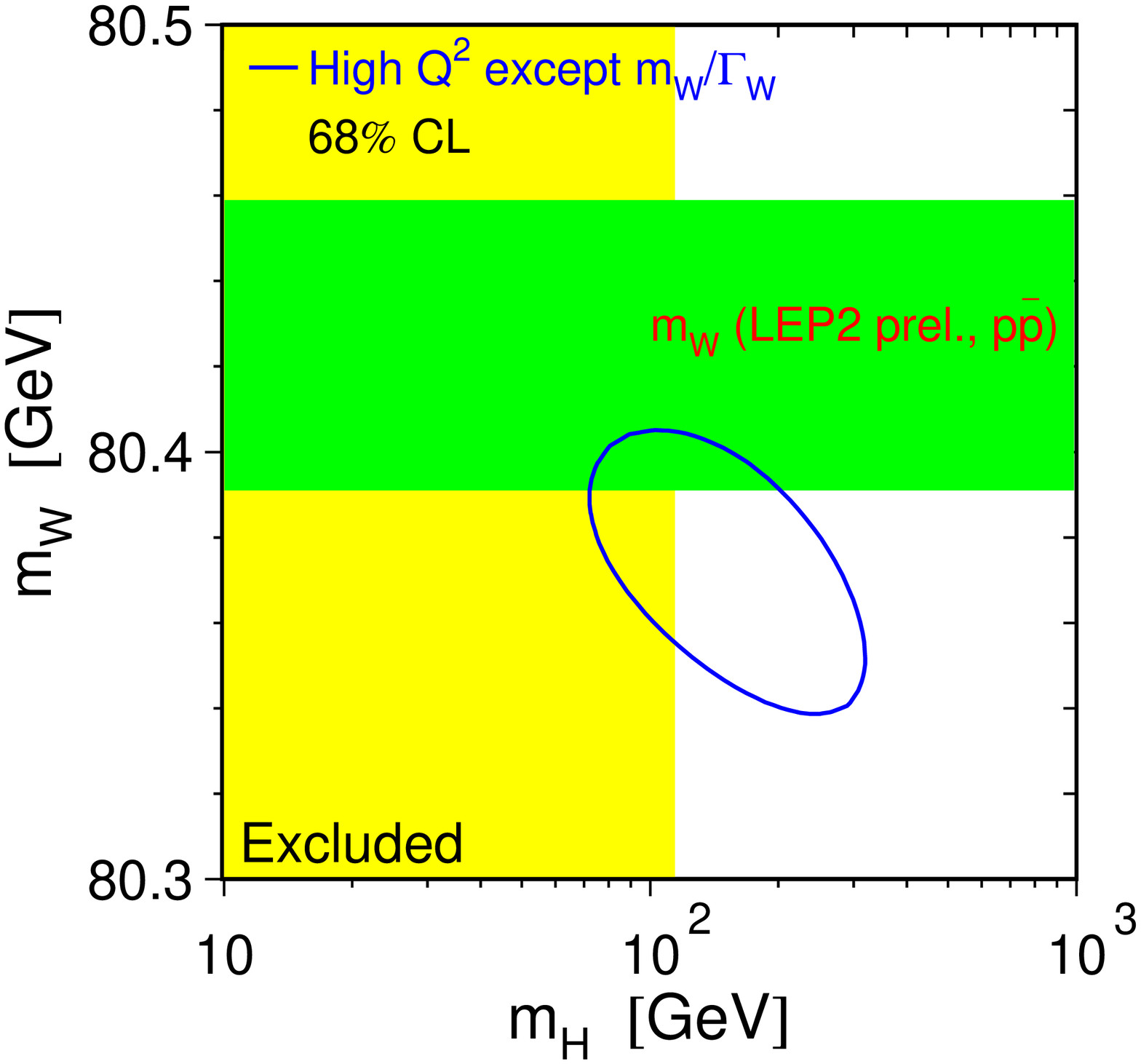,width=0.6\linewidth}}
\vskip -0.5cm
\caption[Constraints on the Higgs-boson mass from $\Mt$ and $\MW$]
{Contour curves of 68\% probability in (top) the $(\Mt,\MH)$ plane and
(bottom) the $(\MW,\MH)$ plane, based on all 18 measurements except
the direct measurement of $\Mt$ and the direct measurements of $\MW$
and $\GW$, respectively.  The direct measurements of these excluded
observables are shown as the horizontal bands of width $\pm1$ standard
deviation.  The vertical band shows the 95\% confidence level
exclusion limit on $\MH$ of $114.4~\GeV$ derived from the direct
search at \LEPII~\cite{LEPSMHIGGS}.  The direct measurements of $\MW$
and $\GW$ used here are preliminary.}
\label{fig:msm:mhmt-mhmw} 
\end{center}
\end{figure}

In order to obtain the most stringent constraint on the mass of the
$\SM$ Higgs boson, the analysis is performed using the hadronic vacuum
polarisation, the 14 Z-pole results, as well as the three additional
results measured in high-$Q^2$ interactions as discussed in
Section~\ref{sec:msm:add}, namely $\Mt$, $\MW$ and $\GW$, for a total
of 18 input measurements.  The relative importance of including the
direct measurements of $\Mt$ and $\MW$ in constraining $\MH$ is shown
in Figure~\ref{fig:msm:mhmt-mhmw}.  At the current level of
experimental precision, the direct measurement of $\Mt$ is more
important.  A measurement of $\MW$ with increased precision, however,
will become very valuable, especially in conjunction with an improved
$\Mt$ measurement.

The results are shown in Table~\ref{tab:msmfit-all}.  A $\chidf$ of
18.3/13 is obtained, corresponding to a probability of 15\%.  The
largest contribution to the $\chi^2$ is again caused by the asymmetry
measurements as discussed in Section~\ref{sec:coup:disc}.  Thus also
the complete set of measurements is accommodated by a single set of
values for the five $\SM$ input parameters.

\begin{figure}[p]
\begin{center}
\vskip -1cm
\mbox{\epsfig{file=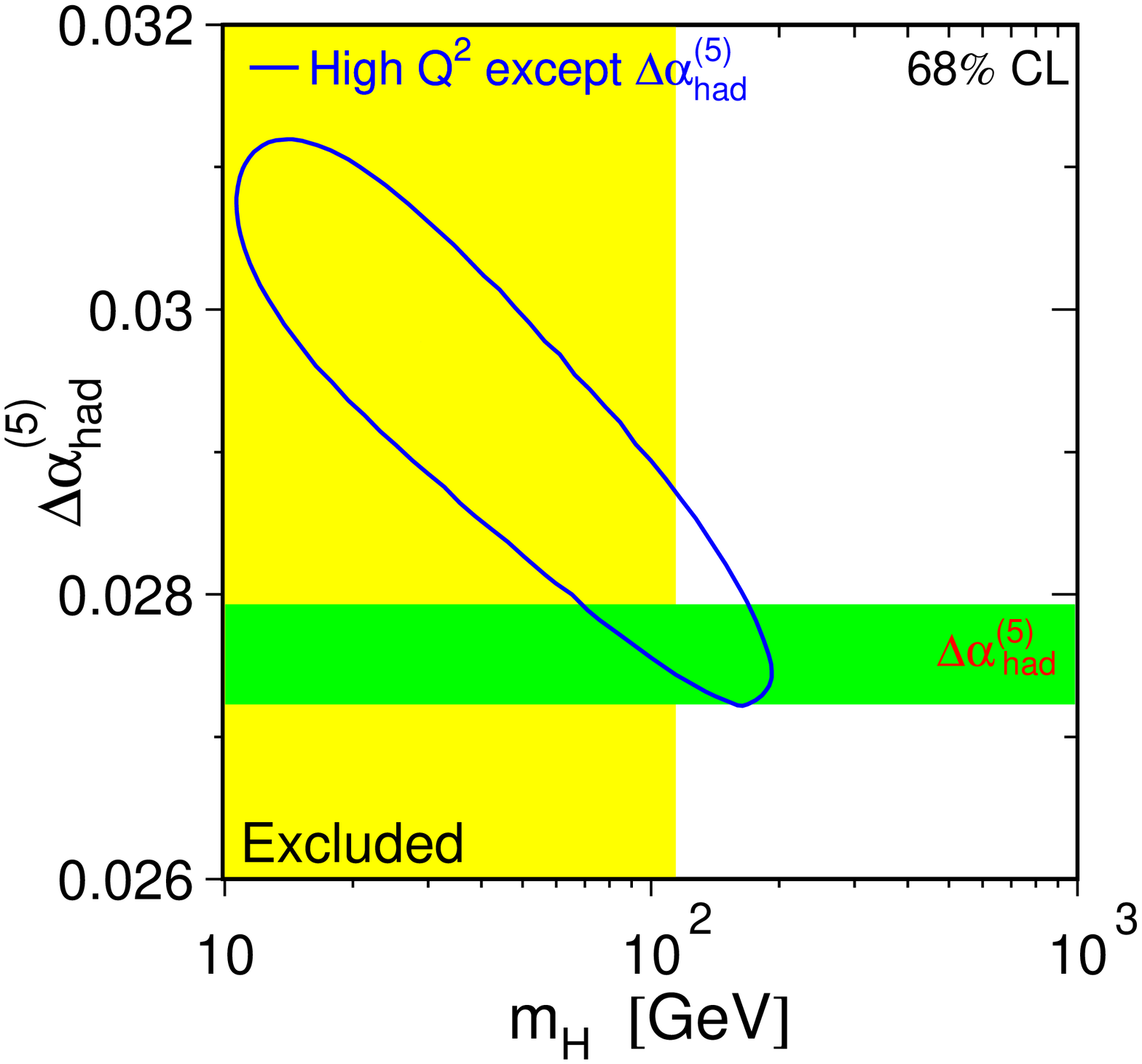,width=0.6\linewidth}}
\vskip -0.5cm
\caption[Constraints on the Higgs-boson mass from $\dalhad$] {Contour
curve of 68\% probability in the $(\dalhad,\MH)$ plane, based on all
18 measurements except the constraint on $\dalhad$.  The direct
measurements of the excluded observable is shown as the horizontal
bands of width $\pm1$ standard deviation.  The vertical band shows the
95\% confidence level exclusion limit on $\MH$ of $114.4~\GeV$ derived
from the direct search at \LEPII~\cite{LEPSMHIGGS}.  The direct
measurements of $\MW$ and $\GW$ used here are preliminary.}
\label{fig:msm:mhah} 
\end{center}
\end{figure}

Compared to the results shown in Table~\ref{tab:msmfit-lep1sld}, very
good agreement is observed.  The relative uncertainty on $\MH$
decreases by about a half, mainly due to the inclusion of the direct
measurements of $\Mt$ and $\MW$. A change of the measured top-quark
mass by one standard deviation, $4.3~\GeV$, changes the fitted
Higgs-boson mass by about 30\%, or 0.12 in $\LOGMH$.  The importance
of the external $\dalhad=0.02758\pm0.00035$ determination for the
constraint on $\MH$ is shown in Figure~\ref{fig:msm:mhah}.  Without
the external $\dalhad$ constraint, the fit results are
$\dalhad=0.0298^{+0.0010}_{-0.0017}$ and $\MH=29^{+77}_{-15}~\GeV$,
with a correlation of $-0.88$ between these two fit results.

\begin{figure}[p]
\begin{center}
\mbox{\epsfig{file=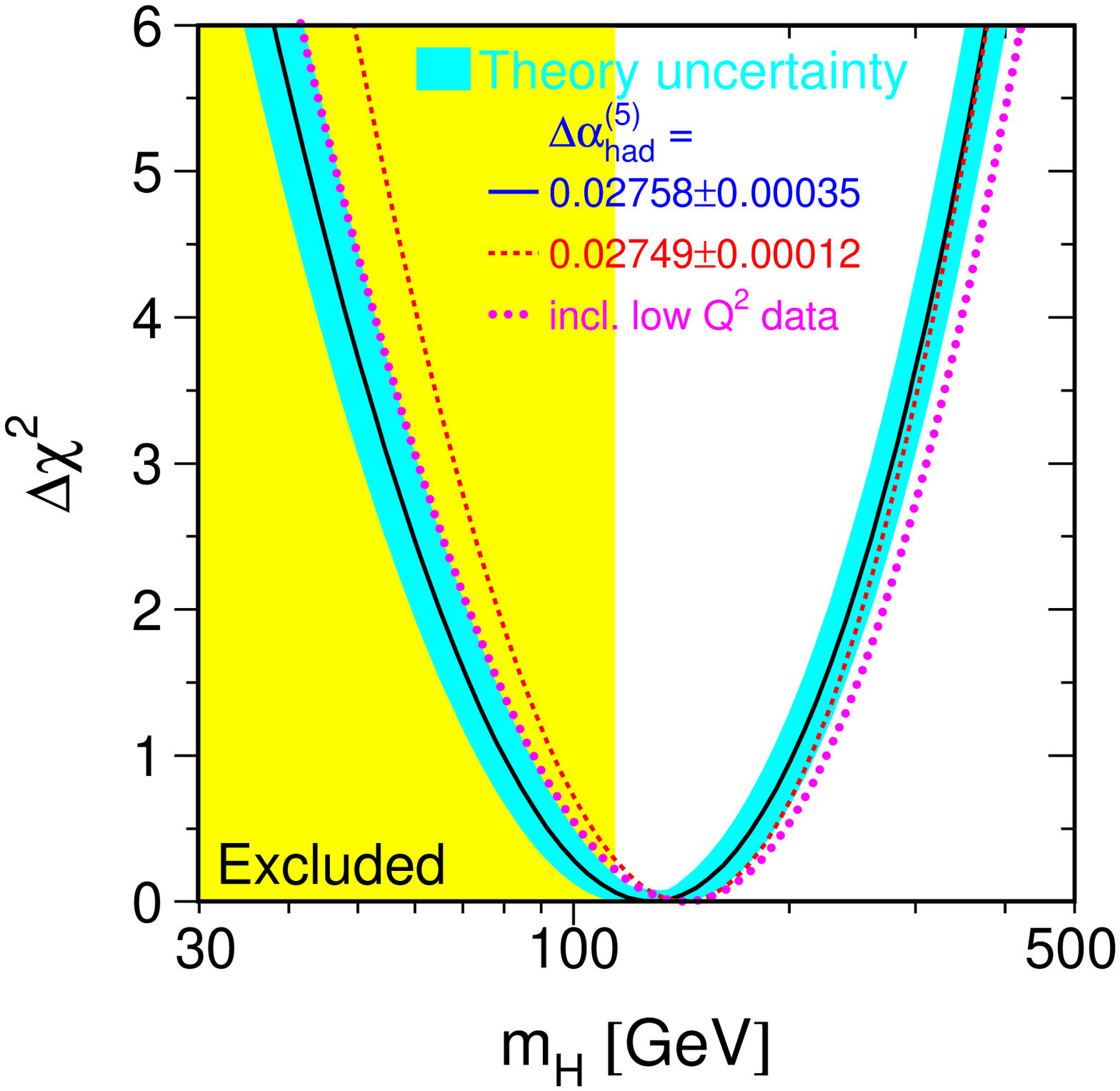,width=0.9\linewidth}} 
\caption[The ``blue-band'' plot $\Delta\chi^2(\MH)$]
{$\Delta\chi^2(\MH)=\chi^2_{min}(\MH)-\chi^2_{\min}$ as a function of
$\MH$.  The line is the result of the fit using all 18 results.  The
associated band represents the estimate of the theoretical uncertainty
due to missing higher-order corrections as discussed in
Section~\ref{sec:msm:TU}. The vertical band shows the 95\% confidence
level exclusion limit on $\MH$ of $114.4~\GeV$ derived from the direct
search at \LEPII~\cite{LEPSMHIGGS}. The dashed curve is the result
obtained using the theory-driven $\dalhad$ determination of
Equation~\ref{eq:dalhad:qcd}. The direct measurements of $\MW$ and
$\GW$ used here are preliminary.}
\label{fig:msmfit-chi2} 
\end{center}
\end{figure}

The $\Delta\chi^2(\MH)=\chi^2_{min}(\MH)-\chi^2_{\min}$ curve is shown
in Figure~\ref{fig:msmfit-chi2}.  The effect of the theoretical
uncertainties in the $\SM$ calculations due to missing higher-order
corrections as discussed in Section~\ref{sec:msm:TU} is shown by the
thickness of the shaded curve.  Including these errors, the one-sided
95\%~CL upper limit on $\LOGMH$, given at $\Delta\chi^2=2.7$, is:
\begin{eqnarray}
\LOGMH ~ < ~ 2.455     ~&~\hbox{or}~&~ 
\MH    ~ < ~ 285~\GeV \label{eq:sm:mh-exp}\,,
\end{eqnarray}
assuming a prior probability density flat in $\LOGMH$.\footnote{ 
Integrating the one-dimensional probability density function instead
of taking $\Delta\chi^2=2.7$, the upper limit at 95\% confidence level
is $280~\GeV$. In case a prior probability density flat in $\MH$ is
assumed, the upper limit at 95\% confidence level, calculated by
integration, increases to $337~\GeV$.}  In case the theory-driven
$\dalhad$ determination of Equation~\ref{eq:dalhad:qcd} is used, the
central value of $\MH$ increases while the uncertainty on $\MH$ is
reduced so that the upper limit changes only slightly.  These results
are clearly consistent with the 95\% confidence level lower limit on
$\MH$ of $114.4~\GeV$ based on the direct search performed at
\LEPII~\cite{LEPSMHIGGS}.\footnote{ The direct search limit can be
taken into account as follows: since the electroweak precision
observables are sensitive to $\LOGMH$, and the direct search exclusion
significance rises steeply with decreasing mass, the direct search
limit essentially constitutes a cut off in $\LOGMH$.  Renormalising
the probability content of the region $\MH>114~\GeV$ to 100\%, with
zero probability for $\MH<114~\GeV$, the 95\% confidence level upper
limit on the mass of the Higgs boson becomes: $\LOGMH < 2.485$
($\MH<306~\GeV$) for a prior probability density flat in $\LOGMH$, or
$\MH<353~\GeV$ for a prior probability density flat in $\MH$.  For the
calculation of both cases, the one-dimensional probability density is
integrated.}

\begin{table}[p]
\begin{center}
\renewcommand{\arraystretch}{1.25}
\begin{tabular}{|c||r@{$\pm$}l||rrrrr|}
\hline
Parameter & \multicolumn{2}{|c||}{Value} 
          & \multicolumn{5}{|c| }{Correlations} \\
          & \multicolumn{2}{|c||}{ }
          & {$\dalhad$} & {$\alfmz$} 
          & {$\MZ$} & {$\Mt$} & {$\LOGMH$}      \\
\hline
\hline
$\dalhad$   &$0.02767$&$0.00034$&$ 1.00$&$     $&$     $&$     $&$     $\\
$\alfmz$    &$0.1188 $&$0.0027 $&$-0.02$&$ 1.00$&$     $&$     $&$     $\\
$\MZ~[\GeV]$&$91.1874$&$0.0021 $&$-0.01$&$-0.02$&$ 1.00$&$     $&$     $\\
$\Mt~[\GeV]$&$178.5  $&$3.9    $&$-0.05$&$ 0.11$&$-0.03$&$ 1.00$&$     $\\
$\LOGMH$    &$2.11   $&$0.20   $&$-0.46$&$ 0.18$&$ 0.06$&$ 0.67$&$ 1.00$\\
\hline
$\MH~[\GeV]$&$129    $&$^{74}_{49}$
                                &$-0.46$&$ 0.18$&$ 0.06$&$ 0.67$&$ 1.00$\\
\hline
\end{tabular}
\caption[Results for $\SM$ input parameters] {Results for the five
$\SM$ input parameters derived from a fit to the Z-pole results and
$\dalhad$, plus $\Mt$, $\MW$, and $\GW$ from Tevatron Run-I and
\LEPII.  The fit has a $\chidf$ of 18.3/13, corresponding to a
probability of 15\%.  See Section~\ref{sec:msm:TU} for a discussion of
the theoretical uncertainties not included here. The results on $\MH$,
obtained by exponentiating the fit results on $\LOGMH$, are also
shown. The direct measurements of $\MW$ and $\GW$ used here are
preliminary. }
\label{tab:msmfit-all}
\end{center}
\end{table}

The lower limit on the mass of the $\SM$ Higgs boson from \LEPII\ is
derived from the non-observation of direct Higgs production.
Similarly, the upper mass limit given above arises from the
observation of radiative corrections that can be largely attributed to
the presence of a heavy top quark.  The inferred radiative corrections
are not significantly different from those expected when the Higgs
mass is at the electroweak scale, where the contributions to radiative
corrections from the Higgs boson are relatively small.

\section{Discussion}
\label{sec:msm:disc}

\begin{figure}[p]
\begin{center}
\mbox{\epsfig{file=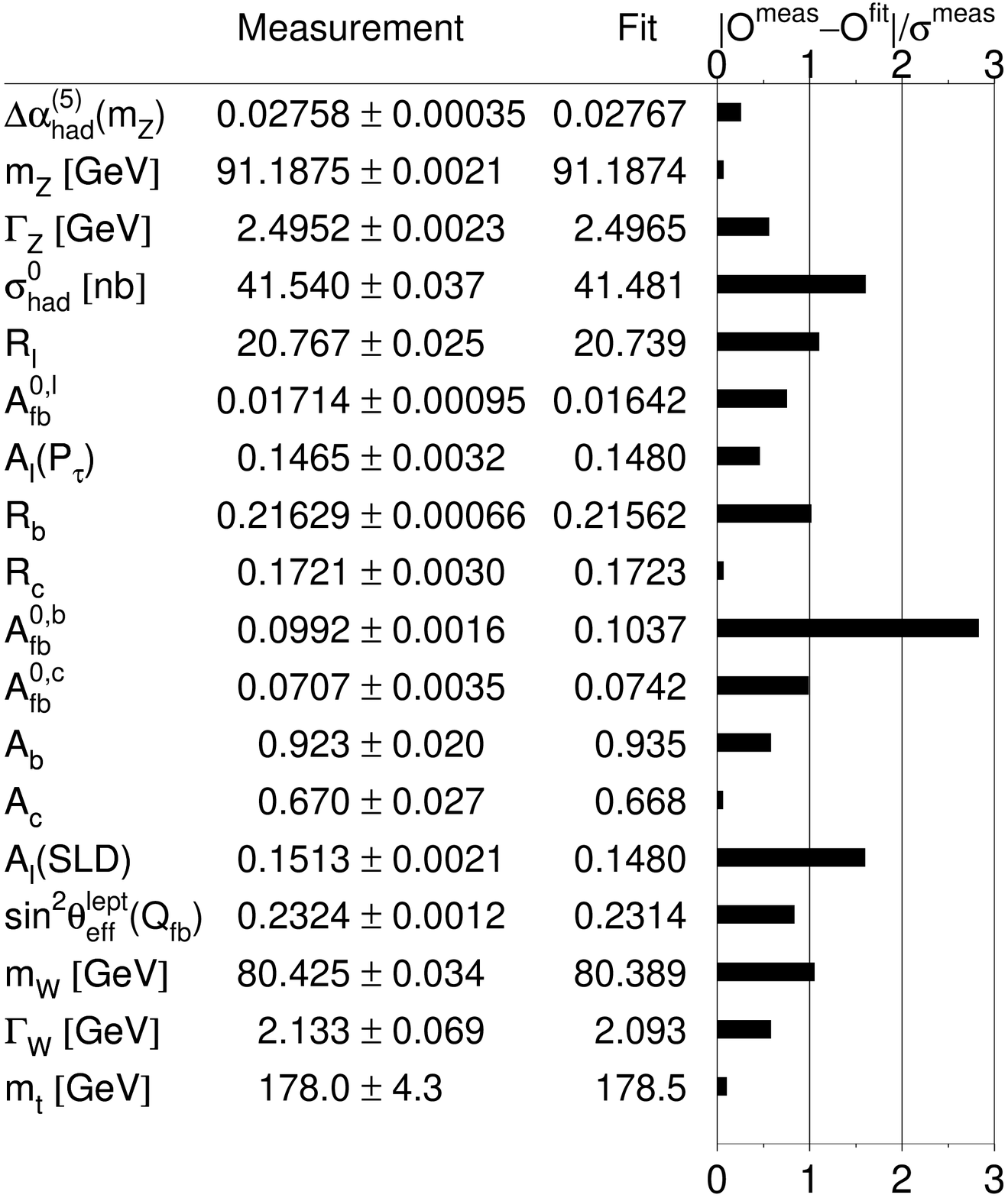,width=0.8\linewidth}} 
\caption[Measurements and pulls] {Comparison of the measurements with
the expectation of the $\SM$, calculated for the five $\SM$ input
parameter values in the minimum of the global $\chi^2$ of the fit.
Also shown is the pull of each measurement, where pull is defined as
the difference of measurement and expectation in units of the
measurement uncertainty. The direct measurements of $\MW$ and $\GW$
used here are preliminary.}
\label{fig:msm:pulls} 
\end{center}
\end{figure}

The global $\chidf$ of the $\SM$ fit is 18.3/13, corresponding to a
probability of 15\%.  Predictions for the individual measurements
entering this analysis and the resulting pulls contributing to the
global $\chi^2$ are reported in Table~\ref{tab:msm:input}.
Predictions of many other observables within the $\SM$ framework are
reported in Appendix~\ref{app:SM:preds}.  The pulls of the
measurements are also shown in Figure~\ref{fig:msm:pulls}.  Here, the
pull is defined as the difference between the measured and the
predicted value, in units of the measurement uncertainty, calculated
for the values of the five $\SM$ input parameters in the minimum of
the $\chi^2$.

\begin{table}[p]
\begin{center}
  \renewcommand{\arraystretch}{1.30}
\begin{tabular}{|ll||r|r||r|r|l|}
\hline
 && {Measurement with}  & {Systematic} & {Standard Model} & {Pull}  \\
 && {Total Error}       & {Error}      & {High-$Q^2$ Fit} & {    }  \\
\hline
\hline
& $\dalhad$~\cite{bib-BP05}
                & $0.02758\pm0.00035$   
                               & 0.00034              & $0.02767\pm0.00035$ & $0.3$ \\
\hline
\hline
&$\MZ$ [\GeV{}] & $91.1875\pm0.0021\pz$
                               &${}^{(a)}$0.0017$\pz$ & $91.1874\pm0.0021\pz$ & $0.1$ \\
&$\GZ$ [\GeV{}] & $2.4952 \pm0.0023\pz$
                               &${}^{(a)}$0.0012$\pz$ &  $2.4965\pm0.0015\pz$ & $0.6$ \\
&$\shad$ [nb]   & $41.540 \pm0.037\pzz$ 
                               &${}^{(a)}$0.028$\pzz$ &  $41.481\pm0.014\pzz$ & $1.6$ \\
&$\Rl$          & $20.767 \pm0.025\pzz$ 
                               &${}^{(a)}$0.007$\pzz$ &  $20.739\pm0.018\pzz$ & $1.1$ \\
&$\Afbzl$       & $0.0171 \pm0.0010\pz$ 
                               &${}^{(a)}$0.0003$\pz$ &  $0.01642\pm0.00024 $ & $0.8$ \\
+& correlation matrix &&&& \\[-2mm]
 & ~~Table~\ref{tab:lsafbresult} &&&& \\
\hline
&$\cAl~(P_\tau)$& $0.1465 \pm0.0033\pz$ & 0.0015$\pz$ & $0.1480\pm 0.0011\pz$ & $0.5$ \\
\hline
&$\cAl$~(SLD)   & $0.1513 \pm0.0021\pz$ & 0.0011$\pz$ & $0.1480\pm 0.0011\pz$ & $1.6$ \\
\hline
&$\Rbz{}$       & $0.21629\pm0.00066$   & 0.00050     & $0.21562\pm0.00013  $ & $1.0$ \\
&$\Rcz{}$       & $0.1721\pm0.0030\pz$  & 0.0019$\pz$ & $0.1723 \pm0.0001\pz$ & $0.1$ \\
&$\Afbzb{}$     & $0.0992\pm0.0016\pz$  & 0.0007$\pz$ & $0.1037\pm 0.0008\pz$ & $2.8$ \\
&$\Afbzc{}$     & $0.0707\pm0.0035\pz$  & 0.0017$\pz$ & $0.0742\pm 0.0006\pz$ & $1.0$ \\
&$\cAb$         & $0.923\pm 0.020\pzz$  & 0.013$\pzz$ & $0.9346\pm 0.0001\pz$ & $0.6$ \\
&$\cAc$         & $0.670\pm 0.027\pzz$  & 0.015$\pzz$ & $0.6683\pm 0.0005\pz$ & $0.1$ \\
+& correlation matrix &&&& \\[-2mm]
 & ~~Table~\ref{tab:14parcor} &&&& \\
\hline
&$\swsqeffl$
  ($\Qfbhad$)   & $0.2324\pm0.0012\pz$  & 0.0010$\pz$ & $0.23140\pm0.00014$   & $0.8$ \\
\hline
\hline
&$\Mt$ [\GeV{}] (Run-I~\cite{PP-MT:combination})
                & $178.0\pm4.3\pzz\pzz$ &3.3$\pzz\pzz$& $178.5\pm3.9\pzz\pzz$ & $0.1$ \\
\hline
&$\MW$ [\GeV{}]
                & $80.425\pm0.034\pzz$  &             & $80.389\pm0.019 \pzz$ & $1.1$ \\
&$\GW$ [\GeV{}]
                & $ 2.133\pm0.069\pzz$  &             & $2.093\pm0.002  \pzz$ & $0.6$ \\
+& correlation given in &&&& \\[-2mm]
 & ~~Section~\ref{sec:msm:add:W} &&&& \\
\hline
\end{tabular}\end{center}
\caption[Overview of results]{ Summary of measurements included in the
  analyses of the five $\SM$ input parameters.  The top 15 results are
  included in the Z-pole and the high-$Q^2$ fit, while the bottom
  three results are only used in the high-$Q^2$ fit.  The total errors
  in column 2 include the systematic errors listed in column 3.
  Although the systematic errors include both correlated and
  uncorrelated sources, the determination of the systematic part of
  each error is approximate.  The $\SM$ results in column 4 and the
  pulls (absolute value of the difference between measurement and fit
  in units of the total measurement error, see
  Figure~\ref{fig:msm:pulls}) in column 5 are derived from the $\SM$
  analysis of all 18 results, including also the correlations between
  results presented in Chapter~\ref{chap:corr}.  The direct
  measurements of $\MW$ and $\GW$ used here are preliminary. \\
  $^{(a)}$\small{Only common systematic errors are indicated.}\\ }
\label{tab:msm:input}
\end{table}

The largest contribution to the overall $\chi^2$, 2.8 standard
deviations, has already been discussed in Section~\ref{sec:coup:asym},
namely the b-quark forward-backward asymmetry measured at \LEPI.  Two
other measurements, the hadronic pole cross-section $\shad$ and the
left-right asymmetry measured by SLD, dominating $\cAl$(SLD), cause
pulls of 1.6 standard deviations. The pulls of all other measurements
are about one standard deviation or less.

Compared to the uncertainty of the measurements, $\shad$ exhibits only
a weak dependence on any of the five $\SM$ input parameters.  The
principal dependence of this quantity is on the number of light
neutrino generations, which is constant and equal to three in the
$\SM$.  The exclusion of $\shad$ affects the five fitted $\SM$ input
parameters only slightly.

\begin{figure}[p]
\begin{center}
\mbox{\epsfig{file=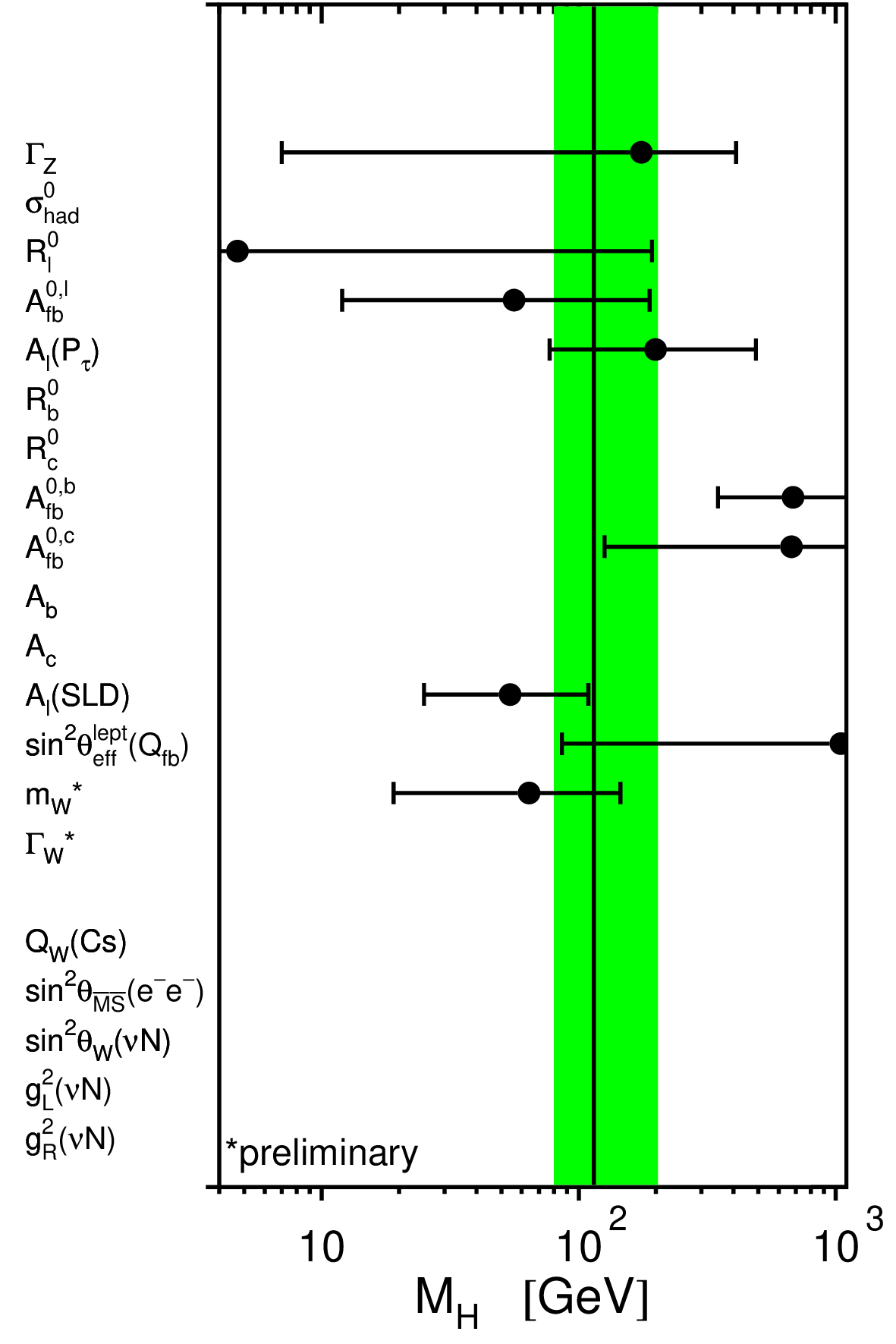,width=0.8\linewidth}} 
\caption[Higgs-boson mass constraints] {Constraints on the mass of the
Higgs boson from each pseudo-observable. The Higgs-boson mass and its
68\% CL uncertainty is obtained from a five-parameter $\SM$ fit to the
observable, constraining $\dalhad=0.02758\pm0.00035$,
$\alfmz=0.118\pm0.003$, $\MZ=91.1875\pm0.0021~\GeV$ and Tevatron Run-I
$\Mt=178.0\pm4.3~\GeV$. Only significant constraints are
shown. Because of these four common constraints the resulting
Higgs-boson mass values cannot be combined.  The shaded band denotes
the overall constraint on the mass of the Higgs boson derived from all
pseudo-observables reported in Table~\ref{tab:msmfit-all}.  The direct
measurements of $\MW$ and $\GW$ used in that analysis are
preliminary.}
\label{fig:msm:higgs} 
\end{center}
\end{figure}

The constraint on the Higgs-boson mass arising from each
pseudo-observable is shown in Figure~\ref{fig:msm:higgs}.  The
corresponding Higgs-boson mass is obtained from a five-parameter $\SM$
fit to the observable, constraining $\dalhad=0.02758\pm0.00035$,
$\alfmz=0.118\pm0.003$, $\MZ=91.1875\pm0.0021~\GeV$ and Tevatron
Run-I $\Mt=178.0\pm4.3~\GeV$.  The region of very low Higgs-boson
masses is approximate, since in that region the effect of the ZH
four-fermion process may become non-negligible, see
Section~\ref{sec:intro_SM_remnants}.

As far as the mass of the Higgs boson is concerned, $\Afbzb$ and
$\cAl$(SLD), both determining $\swsqeffl$, exhibit a high sensitivity,
but prefer a high and a low $\MH$ value, respectively, as shown in
Figure~\ref{fig:msm:higgs}.  Within the $\SM$ analysis, other
observables also prefer low values for $\MH$, such as the other
leptonic asymmetry measurements as well as the combined measurement of
the mass of the W boson.  Therefore the pull of the $\Afbzb$
measurement is enlarged compared to the $\swsqeffl$ combination
discussed in Section~\ref{sec:coup:sef2lept}, while the pull of
$\cAl$(SLD) is reduced.

Because of these considerations, it is interesting to repeat the $\SM$
analysis,
\begin{itemize}
\item  excluding $\cAl$(SLD):
\begin{eqnarray}
\MH=  175^{+99}_{-66}~\GeV & \mathrm{and} & \chidf=  14.6/12~(27\%)\,;
\end{eqnarray} 
\item  excluding $\Afbzb$(LEP):
\begin{eqnarray}
\MH=\pz76^{+54}_{-33}~\GeV & \mathrm{and} & \chidf=\pz9.7/12~(64\%)\,;
\end{eqnarray} 
\item  excluding both $\cAl$(SLD) and $\Afbzb$(LEP):
\begin{eqnarray}
\MH=  103^{+76}_{-48}~\GeV & \mathrm{and} & \chidf=\pz8.7/11~(65\%)\,.
\end{eqnarray} 
\end{itemize}
The largest changes in the results are observed in $\LOGMH$ for the
second case, removing $\Afbzb$(LEP), as expected when removing this
measurement with the largest pull and highly sensitive to $\MH$.
Given the size of the experimental errors on the b-quark couplings
compared to the size of their radiative corrections as expected in the
$\SM$, a potential explanation in terms of new physics phenomena would
require modifications of the right-handed b-quark coupling at the
level of the Born-term values, as already discussed in
Section~\ref{sec:coup:disc}.

The extracted constraints on the five input parameters of the $\SM$
can be used to calculate predictions for other measured electroweak
observables.  As an example, the atomic parity violation parameter of
cesium $\QWCs$, the electroweak mixing angle $\swsqMSb$ measured in
polarised M\o ller scattering, and the results from neutrino-nucleon
scattering on $\gnlq^2$ and $\gnrq^2$ by the NuTeV experiment, all
discussed in Section~\ref{sec:msm:add:lowQ2}, are considered.  The
measurements and predictions are compared in
Table~\ref{tab:msm:predict}.  While good agreement is found for
$\QWCs$, $\swsqMSb$ and $\gnrq^2$, the result on the left-handed
coupling $\gnlq^2$, measured nearly eight times more precisely than
the right-handed coupling $\gnrq^2$, and its prediction show a
discrepancy at the level of 3.0 standard deviations.  When the four
measurements are included in the fit input, the five fitted $\SM$
input parameters changed by less than 10\% of their error except for
the Higgs-boson mass: $\LOGMH$ increases by 0.04 corresponding to an
increase of 11\% in $\MH$, while the $\chidf$ increases to 28.9/17,
corresponding to a probability of only 3.5\%.

In the data collected at LEP in the year 2000, the last year of
\LEPII\ operation, tantalising hints for the production of the Higgs
boson were found~\cite{LEPSMHIGGS}.  However, combining all data of
all four LEP experiments, the observation, compared to the expected
background, corresponds to a significance of about 1.7 standard
deviations only, at a mass of about $115~\GeV$~\cite{LEPSMHIGGS}.  New
data is needed to confirm or exclude this indication and its
interpretation.  Such a value for $\MH$ is in good agreement with the
above values predicted within the $\SM$ based on the analysis of
precision electroweak observables.

\begin{table}[t]
\begin{center}
  \renewcommand{\arraystretch}{1.30}
\begin{tabular}{|ll||r||r|r|l|}
\hline
 && {Measurement with}  & {Standard Model} & {Pull}  \\
 && {Total Error}       & {High-$Q^2$ Fit} & {    }  \\
\hline
\hline
&APV~\cite{QWCs:theo:2003:new}
                &                        &                      &        \\
\hline
&$\QWCs$        & $-72.74\pm0.46\pzz\pz$ & $-72.942\pm0.037\pzz$& $ 0.4$ \\
\hline
\hline
&M\o ller~\cite{E158RunI+II+III}
                &                        &                      &        \\
\hline
&$\swsqMSb$     & $0.2330\pm0.0015\pz$   & $0.23111\pm0.00014$  & $1.3$  \\
\hline
\hline
&$\nu$N~\cite{bib-NuTeV-final}
                &                        &                      &        \\
\hline
&$\gnlq^2$      & $0.30005\pm0.00137$    & $0.30415\pm0.00023$  & $3.0$  \\
&$\gnrq^2$      & $0.03076\pm0.00110$    & $0.03014\pm0.00003$  & $0.6$  \\
\hline
\end{tabular}\end{center}
\caption[Overview of predictions]{ Summary of predictions for results
 obtained in low-$Q^2$ processes described in
 Section~\ref{sec:msm:add:lowQ2}, derived from the fit to all
 high-$Q^2$ data.  Good agreement is observed except for the $\nu$N
 result on $\gnlq^2$ measured by the NuTeV experiment.  The direct
 measurements of $\MW$ and $\GW$ used in the high-$Q^2$ fit are
 preliminary. }
\label{tab:msm:predict}
\end{table}

\chapter{Summary and Conclusions}
\label{chap:sum}

The four LEP experiments ALEPH, DELPHI, L3 and OPAL and the SLD
experiment at the SLC perform measurements in electron-positron
collisions at centre-of-mass energies corresponding to the mass of the
Z boson.  Accumulating about 18 million Z decays with excellent
detectors, the measurements are of unprecedented accuracy in
high-energy particle physics.  In particular, the pair-production of
charged leptons and heavy quarks as well as inclusive hadron
production are analysed by measuring production cross-sections and
various cross-section asymmetries. These measurements are presented in
Chapters~\ref{chap:lsafb} to~\ref{sec:lq} .

The measurements are used to determine the mass of the Z boson, its
decay widths and its couplings to the various fermion species, for
example:
\begin{eqnarray*}
\MZ       & = &           91.1875  \pm 0.0021~\GeV \\
\GZ       & = &            2.4952  \pm 0.0023~\GeV \\
\rho_\ell & = &            1.0050  \pm 0.0010      \\
\swsqeffl & = &            0.23153 \pm 0.00016     \,.
\end{eqnarray*}
The number of light neutrino species is determined to be
$2.9840\pm0.0082$, in good agreement with expectations based on the
three observed generations of fundamental fermions.  In general, the
uncertainties on the measured parameters have been reduced by two to
three orders of magnitude with respect to the experimental results
available before the startup of SLC and LEP.

In addition, the large and diverse set of precise measurements allows
many relations inspired by the Standard Model to be stringently tested
(Chapter~\ref{chap:Z+coup}) and the free parameters of the model to be
tightly constrained (Chapter~\ref{chap:MSM}).  Lepton universality of
the neutral weak current is established at the per-mille level.  The
masses of W boson and top quarks are predicted to be:
$\MW=80.363\pm0.032~\GeV$ and $\Mt=173^{+13}_{-10}~\GeV$, agreeing
well with the direct measurements of these quantities, successfully
testing the Standard Model at the level of its radiative corrections.

While most measurements are well accommodated, the various
measurements of forward-backward and polarised asymmetries, when
interpreted in terms of a single quantity, the leptonic effective
electroweak mixing angle, show a dispersion larger than expected with
a $\chidf$ of 11.8/5, corresponding to a probability of 3.7\%, for the
average value shown above.  Within the Standard-Model framework, the
large $\chidf$ is mainly caused by the measurement of the
forward-backward asymmetry in $\bb$ production, discussed in
Chapter~\ref{chap:MSM}.  This could be explained by new physics, for
example, modifying the right-handed b-quark coupling, or simply by a
fluctuation.  Further improvements on the precision of Z-pole
observables could come from a linear collider taking data at the
Z-pole.

Including the direct measurements of $\Mt$ and $\MW$, the mass of the
Standard-Model Higgs boson is predicted with a relative uncertainty of
about 50\% and found to be less than $285~\GeV$ at 95\% confidence
level.  These results are in good agreement with the lower limit of
$114~\GeV$ at 95\% confidence level obtained from direct searches at
\LEPII.  Overall, the $\SM$ is verified to be a good theory up to the
$100~\GeV$ scale.  The data impose very tight constraints on any new
physics beyond the $\SM$.  Any extended theory must be consistent with
the $\SM$ or one or more Higgs doublet models such as super-symmetry.

Analysing results measured in low-$Q^2$ interactions, the largest
deviation with respect to the expectation, at the level of three
standard deviations, is found for the left-handed neutrino-quark
effective coupling combination, as measured in neutrino-nucleon
scattering by the NuTeV collaboration.  Note that this measurement and
the forward-backward asymmetry in $\bb$ production measured at LEP are
completely independent experimentally, are of different precision, and
are extracted from different processes involving different fermion
species, at greatly differing scales of momentum transfer.  The
constraints on the Standard-Model parameters are rather stable when
the NuTeV measurement is included in the determination of
Standard-Model parameters, while this is not the case when the
measurement of the forward-backward asymmetry in $\bb$ production is
removed.

In the next years, improvements to three other important electroweak
observables will become available, namely the masses of the top quark
and of the W boson, and the hadronic vacuum polarisation.  Assuming
future world-average uncertainties of $2~\GeV$ on $\Mt$, $25~\MeV$ on
$\MW$, and $0.00010$ on $\dalhad$, the mass of the Higgs boson can be
predicted with a relative uncertainty of about 30\%.  The direct
observation of Higgs-boson production would eventually allow its mass
to be measured at the $\GeV$ level, allowing another crucial test of
the Standard Model through a comparison of the direct and indirect
results.

\appendix

\chapter{Author Lists}
\label{app:author-list}

The ALEPH, DELPHI, L3, OPAL and SLD Collaborations have provided the
inputs for the combined results presented in this Report.  The LEP
Electroweak Working Group and the SLD Heavy Flavour and Electroweak
Groups have performed the combinations.  The Working Groups consist of
members of the five Collaborations.  The lists of authors from the
Collaborations follow.

\clearpage

\section{The ALEPH Collaboration}

{

\newcommand{\aAachen}     { 1}
\newcommand{\aAnnecy}     { 2}
\newcommand{\aBarcelona}  { 3}
\newcommand{\aBari}       { 4}
\newcommand{\aBeijing}    { 5}
\newcommand{\aCERN}       { 6}
\newcommand{\aAubiere}    { 7}
\newcommand{\aCopenhagen} { 8}
\newcommand{\aAttiki}     { 9}
\newcommand{\aPalaiseau}  {10}
\newcommand{\aFirenze}    {11}
\newcommand{\aTallahasee} {12}
\newcommand{\aFrascati}   {13}
\newcommand{\aGlasgow}    {14}
\newcommand{\aOrem}       {15}
\newcommand{\aHeidelberg} {16}
\newcommand{\aLondon}     {17}
\newcommand{\aInnsbruck}  {18}
\newcommand{\aLancaster}  {19}
\newcommand{\aBelgium}    {20}
\newcommand{\aMainz}      {21}
\newcommand{\aMarseille}  {22}
\newcommand{\aMilano}     {23}
\newcommand{\aMunich}     {24}
\newcommand{\aParis}      {25}
\newcommand{\aPisa}       {26}
\newcommand{\aSurrey}     {27}
\newcommand{\aOxon}       {28}
\newcommand{\aYvette}     {29}
\newcommand{\aSantaCruz}  {30}
\newcommand{\aSheffield}  {31}
\newcommand{\aSiegen}     {32}
\newcommand{\aTrieste}    {33}
\newcommand{\aWashington} {34}
\newcommand{\aWisconsin}  {35}
\newcommand{\aZurich}     {36}

\tolerance=10000
\hbadness=5000
\raggedright
\raggedbottom
\sloppy

\noindent
S.\thinspace Schael,$\!^{\aAachen}$
\nopagebreak
R.\thinspace Barate,$\!^{\aAnnecy}$
R.\thinspace Bruneli\`ere,$\!^{\aAnnecy}$
D.\thinspace Buskulic,$\!^{\aAnnecy}$
I.\thinspace De\thinspace Bonis,$\!^{\aAnnecy}$
D.\thinspace Decamp,$\!^{\aAnnecy}$
P.\thinspace Ghez,$\!^{\aAnnecy}$
C.\thinspace Goy,$\!^{\aAnnecy}$
S.\thinspace J\'ez\'equel,$\!^{\aAnnecy}$
J.-P.\thinspace Lees,$\!^{\aAnnecy}$
% ...

\bigskip

\newcommand{\AlInst}[2]{\item[$^{#1}$] {#2}}

\begin{list}{A}{\itemsep=0pt plus 0pt minus 0pt\parsep=0pt plus 0pt minus 0pt
                \topsep=0pt plus 0pt minus 0pt}
\AlInst{\aAachen}{
Physikalisches Institut der RWTH-Aachen, D-52056 Aachen, Germany}
\AlInst{\aAnnecy}{
Laboratoire de Physique des Particules (LAPP), IN$^{2}$P$^{3}$-CNRS,
F-74019 Annecy-le-Vieux Cedex, France}
\AlInst{\aBarcelona}{
Institut de F\'{i}sica d'Altes Energies, Universitat Aut\`{o}noma
de Barcelona, E-08193 Bellaterra (Barcelona), Spain$^{a7}$}
\AlInst{\aBari}{
Dipartimento di Fisica, INFN Sezione di Bari, I-70126 Bari, Italy}
\AlInst{\aBeijing}{
Institute of High Energy Physics, Academia Sinica, Beijing, The People's
Republic of China$^{a8}$}
\AlInst{\aCERN}{
European Laboratory for Particle Physics (CERN), CH-1211 Geneva 23,
Switzerland}
\AlInst{\aAubiere}{
Laboratoire de Physique Corpusculaire, Universit\'e Blaise Pascal,
IN$^{2}$P$^{3}$-CNRS, Clermont-Ferrand, F-63177 Aubi\`{e}re, France}
\AlInst{\aCopenhagen}{
Niels Bohr Institute, 2100 Copenhagen, DK-Denmark$^{a9}$}
\AlInst{\aAttiki}{
Nuclear Research Center Demokritos (NRCD), GR-15310 Attiki, Greece}
\AlInst{\aPalaiseau}{
Laoratoire Leprince-Ringuet, Ecole Polytechnique, IN$^{2}$P$^{3}$-CNRS,
\mbox{F-91128} Palaiseau Cedex, France}
\AlInst{\aFirenze}{
Dipartimento di Fisica, Universit\`a di Firenze, INFN Sezione di Firenze,
I-50125 Firenze, Italy}
\AlInst{\aTallahasee}{
Supercomputer Computations Research Institute, Florida State University,
Tallahasee, FL-32306-4052, USA}
\AlInst{\aFrascati}{
Laboratori Nazionali dell'INFN (LNF-INFN), I-00044 Frascati, Italy}
\AlInst{\aGlasgow}{
Department of Physics and Astronomy, University of Glasgow,
Glasgow G12 8QQ,United Kingdom$^{a10}$}
\AlInst{\aOrem}{
Utah Valley State College, Orem, UT 84058, U.S.A.}
\AlInst{\aHeidelberg}{
Kirchhoff-Institut f\"ur Physik, Universit\"at Heidelberg, D-69120
Heidelberg, Germany$^{a16}$}
\AlInst{\aLondon}{
Department of Physics, Imperial College, London SW7 2BZ,
United Kingdom$^{a10}$}
\AlInst{\aInnsbruck}{
Institut f\"ur Experimentalphysik, Universit\"at Innsbruck, A-6020
Innsbruck, Austria$^{a18}$}
\AlInst{\aLancaster}{
Department of Physics, University of Lancaster, Lancaster LA1 4YB,
United Kingdom$^{a10}$}
\AlInst{\aBelgium}{
Institut de Physique Nucl\'eaire, D\'epartement de Physique,
Universit\'e Catholique de Louvain, 1348 Louvain-la-Neuve, Belgium}
\AlInst{\aMainz}{
Institut f\"ur Physik, Universit\"at Mainz, D-55099 Mainz, Germany$^{a16}$}
\AlInst{\aMarseille}{
Centre de Physique des Particules de Marseille, Univ M\'editerran\'ee,
IN$^{2}$P$^{3}$-CNRS, F-13288 Marseille, France}
\AlInst{\aMilano}{
Dipartimento di Fisica, Universit\`a di Milano e INFN Sezione di
Milano, I-20133 Milano, Italy.}
\AlInst{\aMunich}{
Max-Planck-Institut f\"ur Physik, Werner-Heisenberg-Institut,
D-80805 M\"unchen, Germany$^{a16}$}
\AlInst{\aParis}{
Laboratoire de l'Acc\'el\'erateur Lin\'eaire, Universit\'e de Paris-Sud,
IN$^{2}$P$^{3}$-CNRS, F-91898 Orsay Cedex, France}
\AlInst{\aPisa}{
Dipartimento di Fisica dell'Universit\`a, INFN Sezione di Pisa,
e Scuola Normale Superiore, I-56010 Pisa, Italy}
\AlInst{\aSurrey}{
Department of Physics, Royal Holloway \& Bedford New College,
University of London, Egham, Surrey TW20 OEX, United Kingdom$^{a10}$}
\AlInst{\aOxon}{
Particle Physics Dept., Rutherford Appleton Laboratory,
Chilton, Didcot, Oxon OX11 OQX, United Kingdom$^{a10}$}
\AlInst{\aYvette}{
CEA, DAPNIA/Service de Physique des Particules,
CE-Saclay, F-91191 Gif-sur-Yvette Cedex, France$^{a17}$}
\AlInst{\aSantaCruz}{
Institute for Particle Physics, University of California at
Santa Cruz, Santa Cruz, CA 95064, USA$^{a22}$}
\AlInst{\aSheffield}{
Department of Physics, University of Sheffield, Sheffield S3 7RH,
United Kingdom$^{a10}$}
\AlInst{\aSiegen}{
Fachbereich Physik, Universit\"at Siegen, D-57068 Siegen, Germany$^{a16}$}
\AlInst{\aTrieste}{
Dipartimento di Fisica, Universit\`a di Trieste e INFN Sezione di Trieste,
I-34127 Trieste, Italy}
\AlInst{\aWashington}{
Experimental Elementary Particle Physics, University of Washington, Seattle,
WA 98195 U.S.A.}
\AlInst{\aWisconsin}{
Department of Physics, University of Wisconsin, Madison, WI 53706,
USA$^{a11}$}
\AlInst{\aZurich}{
Institute for Particle Physics, ETH H\"onggerberg, 8093 Z\"urich,
Switzerland.}
\end{list}

\bigskip

\begin{list}{A}{\itemsep=0pt plus 0pt minus 0pt\parsep=0pt plus 0pt minus 0pt
                \topsep=0pt plus 0pt minus 0pt}
\AlInst{a1}{Also at CERN, 1211 Geneva 23, Switzerland.}
\AlInst{a2}{Now at Fermilab, PO Box 500, MS 352, Batavia, IL 60510, USA}
\AlInst{a3}{Also at Dipartimento di Fisica di Catania and INFN Sezione di
 Catania, 95129 Catania, Italy.}
\AlInst{a4}{Now at University of Florida, Department of Physics, Gainesville, Florida 32611-8440, USA}
\AlInst{a5}{Also IFSI sezione di Torino, INAF, Italy.}
\AlInst{a6}{Also at Groupe d'Astroparticules de Montpellier, Universit\'{e} de Montpellier II, 34095, Montpellier, France.}
\AlInst{a7}{Supported by CICYT, Spain.}
\AlInst{a8}{Supported by the National Science Foundation of China.}
\AlInst{a9}{Supported by the Danish Natural Science Research Council.}
\AlInst{a10}{Supported by the UK Particle Physics and Astronomy Research
Council.}
\AlInst{a11}{Supported by the US Department of Energy, grant
DE-FG0295-ER40896.}
\AlInst{a12}{Now at Departement de Physique Corpusculaire, Universit\'e de
Gen\`eve, 1211 Gen\`eve 4, Switzerland.}
\AlInst{a13}{Also at Department of Physics, Tsinghua University, Beijing, The People's Republic of China.}
\AlInst{a14}{Supported by the Leverhulme Trust.}
\AlInst{a15}{Permanent address: Universitat de Barcelona, 08208 Barcelona,
Spain.}
\AlInst{a16}{Supported by Bundesministerium f\"ur Bildung
und Forschung, Germany.}
\AlInst{a17}{Supported by the Direction des Sciences de la
Mati\`ere, C.E.A.}
\AlInst{a18}{Supported by the Austrian Ministry for Science and Transport.}
\AlInst{a19}{Now at SAP AG, 69185 Walldorf, Germany}
\AlInst{a20}{Now at Groupe d' Astroparticules de Montpellier, Universit\'e de Montpellier II, 34095 Montpellier, France.}
\AlInst{a21}{Now at BNP Paribas, 60325 Frankfurt am Mainz, Germany}
\AlInst{a22}{Supported by the US Department of Energy,
grant DE-FG03-92ER40689.}
\AlInst{a23}{Now at Institut Inter-universitaire des hautes Energies (IIHE), CP 230, Universit\'{e} Libre de Bruxelles, 1050 Bruxelles, Belgique}
\AlInst{a24}{Now at Dipartimento di Fisica e Tecnologie Relative, Universit\`a di Palermo, Palermo, Italy.}
\AlInst{a25}{Deceased.}
\AlInst{a26}{Now at SLAC, Stanford, CA 94309, U.S.A}
\AlInst{a27}{Now at CERN, 1211 Geneva 23, Switzerland}
\AlInst{a28}{Research Fellow of the Belgium FNRS}
\AlInst{a29}{Research Associate of the Belgium FNRS}
\AlInst{a30}{Now at Liverpool University, Liverpool L69 7ZE, United Kingdom}
\AlInst{a31}{Supported by the Federal Office for Scientific, Technical and Cultural Affairs through
the Interuniversity Attraction Pole P5/27}
\AlInst{a32}{Now at Henryk Niewodnicznski Institute of Nuclear Physics, Polish Academy of Sciences, Cracow, Poland}
\AlInst{a33}{Supported by the Commission of the European Communities, contract
ERBFMBICT982894}
\AlInst{a34}{Also Istituto di Fisica Generale, Universit\`a di Torino, 10125 Torino, Italy}

\end{list}

}

\clearpage

\section{The DELPHI Collaboration}

{

\newcommand{\DpName}[2]{\hbox{#1,$\!^{#2}$}}
\newcommand{\DpNameTwo}[3]{\hbox{#1,$\!^{{#2},{#3}}$}}
\newcommand{\DpNameThree}[4]{\hbox{#1,$\!^{{#2},{#3},{#4}}$}}
\newskip\Bigfill \Bigfill = 0pt plus 1000fill
\newcommand{\DpNameLast}[2]{\hbox{#1.$\!^{#2}$}\hspace{\Bigfill}}

\tolerance=10000
\hbadness=5000
\raggedright

\newcommand{\dAMES}{1}
\newcommand{\dAIM}{2}
\newcommand{\dATHENS}{3}
\newcommand{\dBERGEN}{4}
\newcommand{\dBOLOGNA}{5}
\newcommand{\dBRASIL}{6}
\newcommand{\dBRATISLAVA}{7}
\newcommand{\dCDF}{8}
\newcommand{\dCERN}{9}
\newcommand{\dCRN}{10}
\newcommand{\dDESY}{11}
\newcommand{\dDEMOKRITOS}{12}
\newcommand{\dFZU}{13}
\newcommand{\dGENOVA}{14}
\newcommand{\dGRENOBLE}{15}
\newcommand{\dHELSINKI}{16}
\newcommand{\dJINR}{17}
\newcommand{\dKARLSRUHE}{18}
\newcommand{\dKRAKOWone}{19}
\newcommand{\dKRAKOWtwo}{20}
\newcommand{\dLAL}{21}
\newcommand{\dLANCASTER}{22}
\newcommand{\dLIP}{23}
\newcommand{\dLIVERPOOL}{24}
\newcommand{\dGLASGOW}{25}
\newcommand{\dLPNHE}{26}
\newcommand{\dLUND}{27}
\newcommand{\dLYON}{28}
\newcommand{\dMARSEILLE}{29}
\newcommand{\dMILANO}{30}
\newcommand{\dMILANOtwo}{31}
\newcommand{\dNBI}{32}
\newcommand{\dNC}{33}
\newcommand{\dNIKHEF}{34}
\newcommand{\dNTUATHENS}{35}
\newcommand{\dOSLO}{36}
\newcommand{\dOVIEDO}{37}
\newcommand{\dOXFORD}{38}
\newcommand{\dPADOVA}{39}
\newcommand{\dRAL}{40}
\newcommand{\dROMAtwo}{41}
\newcommand{\dROMAtre}{42}
\newcommand{\dSACLAY}{43}
\newcommand{\dSANTANDER}{44}
\newcommand{\dSAPIENZA}{45}
\newcommand{\dSERPUKHOV}{46}
\newcommand{\dSLOVENIJA}{47}
\newcommand{\dSTOCKHOLM}{48}
\newcommand{\dTORINO}{49}
\newcommand{\dTORINOTH}{50}
\newcommand{\dTU}{51}
\newcommand{\dUFRJ}{52}
\newcommand{\dUPPSALA}{53}
\newcommand{\dVALENCIA}{54}
\newcommand{\dVIENNA}{55}
\newcommand{\dWARSZAWA}{56}
\newcommand{\dWUPPERTAL}{57}
\newcommand{\dAACHEN}{58}

\noindent
\DpName{J.\thinspace Abdallah}{\dLPNHE}
\DpName{P.\thinspace Abreu}{\dLIP}
\DpName{W.\thinspace Adam}{\dVIENNA}
\DpName{T.\thinspace Adye}{\dRAL}
\DpName{P.\thinspace Adzic}{\dDEMOKRITOS}
\DpName{I.\thinspace Ajinenko}{\dSERPUKHOV}
\DpName{T.\thinspace Albrecht}{\dKARLSRUHE}
\DpName{T.\thinspace Alderweireld}{\dAIM}
\DpName{G.D.\thinspace Alekseev}{\dJINR}
\DpName{R.\thinspace Alemany-Fernandez}{\dCERN}
% ...

\bigskip

\newcommand{\DpInst}[2]{\item[$^{#2}$] {#1}}

\begin{list}{A}{\itemsep=0pt plus 0pt minus 0pt\parsep=0pt plus 0pt minus 0pt
                \topsep=0pt plus 0pt minus 0pt}
\DpInst{Department of Physics and Astronomy, Iowa State
     University, Ames IA 50011-3160, USA
    }{\dAMES}
\DpInst{Physics Department, Universiteit Antwerpen,
     Universiteitsplein 1, B-2610 Antwerpen, Belgium \\
     \indent~~and IIHE, ULB-VUB,
     Pleinlaan 2, B-1050 Brussels, Belgium \\
     \indent~~and Facult\'e des Sciences,
     Univ. de l'Etat Mons, Av. Maistriau 19, B-7000 Mons, Belgium
    }{\dAIM}
\DpInst{Physics Laboratory, University of Athens, Solonos Str.
     104, GR-10680 Athens, Greece
    }{\dATHENS}
\DpInst{Department of Physics, University of Bergen,
     All\'egaten 55, NO-5007 Bergen, Norway
    }{\dBERGEN}
\DpInst{Dipartimento di Fisica, Universit\`a di Bologna and INFN,
     Via Irnerio 46, IT-40126 Bologna, Italy
    }{\dBOLOGNA}
\DpInst{Centro Brasileiro de Pesquisas F\'{\i}sicas, rua Xavier Sigaud 150,
     BR-22290 Rio de Janeiro, Brazil \\
     \indent~~and Depto. de F\'{\i}sica, Pont. Univ. Cat\'olica,
     C.P. 38071 BR-22453 Rio de Janeiro, Brazil \\
     \indent~~and Inst. de F\'{\i}sica, Univ. Estadual do Rio de Janeiro,
     rua S\~{a}o Francisco Xavier 524, Rio de Janeiro, Brazil
    }{\dBRASIL}
\DpInst{Comenius University, Faculty of Mathematics and Physics,
     Mlynska Dolina, SK-84215 Bratislava, Slovakia
    }{\dBRATISLAVA}
\DpInst{Coll\`ege de France, Lab. de Physique Corpusculaire, IN2P3-CNRS,
     FR-75231 Paris Cedex 05, France
    }{\dCDF}
\DpInst{CERN, CH-1211 Geneva 23, Switzerland
    }{\dCERN}
\DpInst{Institut de Recherches Subatomiques, IN2P3 - CNRS/ULP - BP20,
     FR-67037 Strasbourg Cedex, France
    }{\dCRN}
\DpInst{Now at DESY-Zeuthen, Platanenallee 6, D-15735 Zeuthen, Germany
    }{\dDESY}
\DpInst{Institute of Nuclear Physics, N.C.S.R. Demokritos,
     P.O. Box 60228, GR-15310 Athens, Greece
    }{\dDEMOKRITOS}
\DpInst{FZU, Inst. of Phys. of the C.A.S. High Energy Physics Division,
     Na Slovance 2, CZ-180 40, Praha 8, Czech Republic
    }{\dFZU}
\DpInst{Dipartimento di Fisica, Universit\`a di Genova and INFN,
     Via Dodecaneso 33, IT-16146 Genova, Italy
    }{\dGENOVA}
\DpInst{Institut des Sciences Nucl\'eaires, IN2P3-CNRS, Universit\'e
     de Grenoble 1, FR-38026 Grenoble Cedex, France
    }{\dGRENOBLE}
\DpInst{Helsinki Institute of Physics and Department of Physical Sciences,
     P.O. Box 64, FIN-00014 University of Helsinki,
     \indent~~Finland
    }{\dHELSINKI}
\DpInst{Joint Institute for Nuclear Research, Dubna, Head Post
     Office, P.O. Box 79, RU-101 000 Moscow, Russian Federation
    }{\dJINR}
\DpInst{Institut f\"ur Experimentelle Kernphysik,
     Universit\"at Karlsruhe, Postfach 6980, DE-76128 Karlsruhe,
     Germany
    }{\dKARLSRUHE}
\DpInst{Institute of Nuclear Physics PAN,Ul. Radzikowskiego 152,
     PL-31142 Krakow, Poland
    }{\dKRAKOWone}
\DpInst{Faculty of Physics and Nuclear Techniques, University of Mining
     and Metallurgy, PL-30055 Krakow, Poland
    }{\dKRAKOWtwo}
\DpInst{Universit\'e de Paris-Sud, Lab. de l'Acc\'el\'erateur
     Lin\'eaire, IN2P3-CNRS, B\^{a}t. 200, FR-91405 Orsay Cedex, France
    }{\dLAL}
\DpInst{School of Physics and Chemistry, University of Lancaster,
     Lancaster LA1 4YB, UK
    }{\dLANCASTER}
\DpInst{LIP, IST, FCUL - Av. Elias Garcia, 14-$1^{o}$,
     PT-1000 Lisboa Codex, Portugal
    }{\dLIP}
\DpInst{Department of Physics, University of Liverpool, P.O.
     Box 147, Liverpool L69 3BX, UK
    }{\dLIVERPOOL}
\DpInst{Dept. of Physics and Astronomy, Kelvin Building,
     University of Glasgow, Glasgow G12 8QQ
    }{\dGLASGOW}
\DpInst{LPNHE, IN2P3-CNRS, Univ.~Paris VI et VII, Tour 33 (RdC),
     4 place Jussieu, FR-75252 Paris Cedex 05, France
    }{\dLPNHE}
\DpInst{Department of Physics, University of Lund,
     S\"olvegatan 14, SE-223 63 Lund, Sweden
    }{\dLUND}
\DpInst{Universit\'e Claude Bernard de Lyon, IPNL, IN2P3-CNRS,
     FR-69622 Villeurbanne Cedex, France
    }{\dLYON}
\DpInst{Univ. d'Aix - Marseille II - CPP, IN2P3-CNRS,
     FR-13288 Marseille Cedex 09, France
    }{\dMARSEILLE}
\DpInst{Dipartimento di Fisica, Universit\`a di Milano and INFN-MILANO,
     Via Celoria 16, IT-20133 Milan, Italy
    }{\dMILANO}
\DpInst{Dipartimento di Fisica, Univ. di Milano-Bicocca and
     INFN-MILANO, Piazza della Scienza 2, IT-20126 Milan, Italy
    }{\dMILANOtwo}
\DpInst{Niels Bohr Institute, Blegdamsvej 17,
     DK-2100 Copenhagen {\O}, Denmark
    }{\dNBI}
\DpInst{IPNP of MFF, Charles Univ., Areal MFF,
     V Holesovickach 2, CZ-180 00, Praha 8, Czech Republic
    }{\dNC}
\DpInst{NIKHEF, Postbus 41882, NL-1009 DB
     Amsterdam, The Netherlands
    }{\dNIKHEF}
\DpInst{National Technical University, Physics Department,
     Zografou Campus, GR-15773 Athens, Greece
    }{\dNTUATHENS}
\DpInst{Physics Department, University of Oslo, Blindern,
     NO-0316 Oslo, Norway
    }{\dOSLO}
\DpInst{Dpto. Fisica, Univ. Oviedo, Avda. Calvo Sotelo
     s/n, ES-33007 Oviedo, Spain
    }{\dOVIEDO}
\DpInst{Department of Physics, University of Oxford,
     Keble Road, Oxford OX1 3RH, UK
    }{\dOXFORD}
\DpInst{Dipartimento di Fisica, Universit\`a di Padova and
     INFN, Via Marzolo 8, IT-35131 Padua, Italy
    }{\dPADOVA}
\DpInst{Rutherford Appleton Laboratory, Chilton, Didcot
     OX11 OQX, UK
    }{\dRAL}
\DpInst{Dipartimento di Fisica, Universit\`a di Roma II and
     INFN, Tor Vergata, IT-00173 Rome, Italy
    }{\dROMAtwo}
\DpInst{Dipartimento di Fisica, Universit\`a di Roma III and
     INFN, Via della Vasca Navale 84, IT-00146 Rome, Italy
    }{\dROMAtre}
\DpInst{DAPNIA/Service de Physique des Particules,
     CEA-Saclay, FR-91191 Gif-sur-Yvette Cedex, France
    }{\dSACLAY}
\DpInst{Instituto de Fisica de Cantabria (CSIC-UC), Avda.
     los Castros s/n, ES-39006 Santander, Spain
    }{\dSANTANDER}
\DpInst{Dipartimento di Fisica, Universit\`a degli Studi di Roma
     La Sapienza, Piazzale Aldo Moro 2, IT-00185 Rome, Italy
    }{\dSAPIENZA}
\DpInst{Inst. for High Energy Physics, Serpukov
     P.O. Box 35, Protvino, (Moscow Region), Russian Federation
    }{\dSERPUKHOV}
\DpInst{J. Stefan Institute, Jamova 39, SI-1000 Ljubljana, Slovenia
     and Laboratory for Astroparticle Physics,\\
     \indent~~and Nova Gorica Polytechnic, Kostanjeviska 16a, SI-5000 Nova Gorica, Slovenia, \\
     \indent~~and Department of Physics, University of Ljubljana,
     SI-1000 Ljubljana, Slovenia
    }{\dSLOVENIJA}
\DpInst{Fysikum, Stockholm University,
     Box 6730, SE-113 85 Stockholm, Sweden
    }{\dSTOCKHOLM}
\DpInst{Dipartimento di Fisica Sperimentale, Universit\`a di
     Torino and INFN, Via P. Giuria 1, IT-10125 Turin, Italy
    }{\dTORINO}
\DpInst{INFN,Sezione di Torino, and Dipartimento di Fisica Teorica,
     Universit\`a di Torino, Via P. Giuria 1, IT-10125 Turin, Italy
    }{\dTORINOTH}
\DpInst{Dipartimento di Fisica, Universit\`a di Trieste and
     INFN, Via A. Valerio 2, IT-34127 Trieste, Italy \\
     \indent~~and Istituto di Fisica, Universit\`a di Udine,
     IT-33100 Udine, Italy
    }{\dTU}
\DpInst{Univ. Federal do Rio de Janeiro, C.P. 68528
     Cidade Univ., Ilha do Fund\~ao
     BR-21945-970 Rio de Janeiro, Brazil
    }{\dUFRJ}
\DpInst{Department of Radiation Sciences, University of
     Uppsala, P.O. Box 535, SE-751 21 Uppsala, Sweden
    }{\dUPPSALA}
\DpInst{IFIC, Valencia-CSIC, and D.F.A.M.N., U. de Valencia,
     Avda. Dr. Moliner 50, ES-46100 Burjassot (Valencia), Spain
    }{\dVALENCIA}
\DpInst{Institut f\"ur Hochenergiephysik, \"Osterr. Akad.
     d. Wissensch., Nikolsdorfergasse 18, AT-1050 Vienna, Austria
    }{\dVIENNA}
\DpInst{Inst. Nuclear Studies and University of Warsaw, Ul.
     Hoza 69, PL-00681 Warsaw, Poland
    }{\dWARSZAWA}
\DpInst{Fachbereich Physik, University of Wuppertal, Postfach
     100 127, DE-42097 Wuppertal, Germany
    }{\dWUPPERTAL}
\DpInst{Now at I.Physikalisches Institut, RWTH Aachen,
          Sommerfeldstrasse 14, DE-52056 Aachen, Germany
    }{\dAACHEN}
\DpInst{Deceased}{\dagger}
\end{list}
}

\clearpage

\section{The L3 Collaboration}

{
\newcount\tutecount  \tutecount=0
\def\tutenum#1{\global\advance\tutecount by 1 \xdef#1{\the\tutecount}}
\def\tute#1{$^{#1}$}
\tutenum\aachen            %
\tutenum\alabama           %
\tutenum\basel             %
\tutenum\beijing           %
\tutenum\berlin            %
\tutenum\bologna           %
\tutenum\bucharest         %
\tutenum\budapest          %
\tutenum\caltech           %
\tutenum\cern              %
\tutenum\cmu               %
\tutenum\cyprus            %
\tutenum\debrecen          %
\tutenum\dublin            %
\tutenum\eth               %
\tutenum\florence          %
\tutenum\florida           %
\tutenum\geneva            %
\tutenum\hamburg           %
\tutenum\hefei             %
\tutenum\korea             %
\tutenum\lapp              %
\tutenum\lausanne          %
\tutenum\lecce             %
\tutenum\lsu               %
\tutenum\lyon              %
\tutenum\madrid            %
\tutenum\mich              %
\tutenum\milan             %
\tutenum\mit               %
\tutenum\moscow            %
\tutenum\naples            %
\tutenum\ne                %
\tutenum\nikhef            %
\tutenum\nymegen           %
\tutenum\panjab            %
\tutenum\perugia           %
\tutenum\peters            %
\tutenum\potenza           %
\tutenum\prince            %
\tutenum\psinst            %
\tutenum\purdue            %
\tutenum\riverside         %
\tutenum\rome              %
\tutenum\salerno           %
\tutenum\santiago          %
\tutenum\seft              %
\tutenum\sofia             %
\tutenum\taiwan            %
\tutenum\tata              %
\tutenum\tsinghua          %
\tutenum\ucsd              %
\tutenum\utrecht           %
\tutenum\wl                %
\tutenum\zeuthen           %

{
\tolerance=10000
\hbadness=5000
\raggedright
\def\r{\rlap,}
\noindent

M.\thinspace Acciarri\r\tute\milan\
P.\thinspace Achard\r\tute\geneva\
O.\thinspace Adriani\r\tute{\florence}\
M.\thinspace Aguilar-Benitez\r\tute\madrid\
J.\thinspace Alcaraz\r\tute{\madrid}\
G.\thinspace Alemanni\r\tute\lausanne\
J.\thinspace Allaby\r\tute\cern\
A.\thinspace Aloisio\r\tute\naples\
M.G.\thinspace Alviggi\r\tute\naples\
G.\thinspace Ambrosi\r\tute\geneva\
% ...

\bigskip

\begin{list}{A}{\itemsep=0pt plus 0pt minus 0pt\parsep=0pt plus 0pt minus 0pt
                \topsep=0pt plus 0pt minus 0pt}
\item[$^{\aachen}$]
 III. Physikalisches Institut, RWTH, D-52056 Aachen, Germany$^{\S}$
\item[$^{\alabama}$] University of Alabama, Tuscaloosa, AL 35486, USA
\item[$^{\basel}$] Institute of Physics, University of Basel, CH-4056 Basel,
     Switzerland
\item[$^{\beijing}$] Institute of High Energy Physics, IHEP,
  100039 Beijing, China$^{\triangle}$
\item[$^{\berlin}$] Humboldt University, D-10099 Berlin, FRG$^{\S}$
\item[$^{\bologna}$] University of Bologna and INFN-Sezione di Bologna,
     I-40126 Bologna, Italy
\item[$^{\bucharest}$] Institute of Atomic Physics and University of Bucharest,
     R-76900 Bucharest, Romania
\item[$^{\budapest}$] Central Research Institute for Physics of the
     Hungarian Academy of Sciences, H-1525 Budapest 114, Hungary$^{\ddag}$
\item[$^{\caltech}$] California Institute of Technology, Pasadena, CA 91125, USA
\item[$^{\cern}$] European Laboratory for Particle Physics, CERN,
     CH-1211 Geneva 23, Switzerland
\item[$^{\cmu}$] Carnegie Mellon University, Pittsburgh, PA 15213, USA
\item[$^{\cyprus}$] Department of Physics, University of Cyprus,
     Nicosia, Cyprus
\item[$^{\debrecen}$] KLTE-ATOMKI, H-4010 Debrecen, Hungary$^\P$
\item[$^{\dublin}$] UCD School of Physics, 
  University College Dublin, Belfield, Dublin 4, Ireland
\item[$^{\eth}$] Eidgen\"ossische Technische Hochschule, ETH Z\"urich,
     CH-8093 Z\"urich, Switzerland
\item[$^{\florence}$] INFN Sezione di Firenze and University of Florence,
     I-50125 Florence, Italy
\item[$^{\florida}$] Florida Institute of Technology, Melbourne, FL 32901, USA
\item[$^{\geneva}$] University of Geneva, CH-1211 Geneva 4, Switzerland
\item[$^{\hamburg}$] University of Hamburg, D-22761 Hamburg, Germany
\item[$^{\hefei}$] Chinese University of Science and Technology, USTC,
      Hefei, Anhui 230 029, China$^{\triangle}$
\item[$^{\korea}$]  The Center for High Energy Physics,
     Kyungpook National University, 702-701 Taegu, Republic of Korea
\item[$^{\lapp}$] Laboratoire d'Annecy-le-Vieux de Physique des Particules,
     LAPP,IN2P3-CNRS, BP 110, F-74941 Annecy-le-Vieux CEDEX, France
\item[$^{\lausanne}$] University of Lausanne, CH-1015 Lausanne, Switzerland
\item[$^{\lecce}$] INFN-Sezione di Lecce and Universit\'a Degli Studi di Lecce,
     I-73100 Lecce, Italy
\item[$^{\lsu}$] Louisiana State University, Baton Rouge, LA 70803, USA
\item[$^{\lyon}$] Institut de Physique Nucl\'eaire de Lyon,
     IN2P3-CNRS,Universit\'e Claude Bernard,
     F-69622 Villeurbanne, France
\item[$^{\madrid}$] Centro de Investigaciones Energ{\'e}ticas,
     Medioambientales y Tecnol\'ogicas, CIEMAT, E-28040 Madrid,
     Spain$^{\flat}$
\item[$^{\mich}$] University of Michigan, Ann Arbor, MI 48109, USA
\item[$^{\milan}$] INFN-Sezione di Milano, I-20133 Milan, Italy
\item[$^{\mit}$] Massachusetts Institute of Technology, Cambridge, MA 02139, USA
\item[$^{\moscow}$] Institute of Theoretical and Experimental Physics, ITEP,
     Moscow, Russia
\item[$^{\naples}$] INFN-Sezione di Napoli and University of Naples,
     I-80125 Naples, Italy
\item[$^{\ne}$] Northeastern University, Boston, MA 02115, USA
\item[$^{\nikhef}$] National Institute for High Energy Physics, NIKHEF,
     and University of Amsterdam, NL-1009 DB Amsterdam, The Netherlands
\item[$^{\nymegen}$] Radboud University and NIKHEF,
     NL-6525 ED Nijmegen, The Netherlands
\item[$^{\panjab}$] Panjab University, Chandigarh 160 014, India
\item[$^{\perugia}$] INFN-Sezione di Perugia and Universit\`a Degli
     Studi di Perugia, I-06100 Perugia, Italy
\item[$^{\peters}$] Nuclear Physics Institute, St. Petersburg, Russia
\item[$^{\potenza}$] INFN-Sezione di Napoli and University of Potenza,
     I-85100 Potenza, Italy
\item[$^{\prince}$] Princeton University, Princeton, NJ 08544, USA
\item[$^{\psinst}$] Paul Scherrer Institut, PSI, CH-5232 Villigen, Switzerland
\item[$^{\purdue}$] Purdue University, West Lafayette, IN 47907, USA
\item[$^{\riverside}$] University of Californa, Riverside, CA 92521, USA
\item[$^{\rome}$] INFN-Sezione di Roma and University of Rome, ``La Sapienza",
     I-00185 Rome, Italy
\item[$^{\salerno}$] University and INFN, Salerno, I-84100 Salerno, Italy
\item[$^{\santiago}$] Dept. de Fisica de Particulas Elementales, Univ. de Santiago,
     E-15706 Santiago de Compostela, Spain
\item[$^{\seft}$] SEFT, Research Institute for High Energy Physics, P.O. Box 9,
      SF-00014 Helsinki, Finland
\item[$^{\sofia}$] Bulgarian Academy of Sciences, Central Lab.~of
     Mechatronics and Instrumentation, BU-1113 Sofia, Bulgaria
\item[$^{\taiwan}$] National Central University, Chung-Li, Taiwan, China
\item[$^{\tata}$] Tata Institute of Fundamental Research, Mumbai (Bombay) 400 005, India
\item[$^{\tsinghua}$] Department of Physics, National Tsing Hua University,
      Taiwan, China
\item[$^{\ucsd}$] University of California, San Diego, CA 92093, USA
\item[$^{\utrecht}$] Utrecht University and NIKHEF, NL-3584 CB Utrecht,
     The Netherlands
\item[$^{\wl}$] World Laboratory, FBLJA  Project, CH-1211 Geneva 23, Switzerland
\item[$^{\zeuthen}$] DESY, D-15738 Zeuthen, Germany
\end{list}

\bigskip

\begin{list}{A}{\itemsep=0pt plus 0pt minus 0pt\parsep=0pt plus 0pt minus 0pt
                \topsep=0pt plus 0pt minus 0pt}
\item[$^{\S}$]  Supported by the German Bundesministerium
        f\"ur Bildung, Wissenschaft, Forschung und Technologie.
\item[$^{\ddag}$] Supported by the Hungarian OTKA fund under contract
numbers T019181, F023259 and T037350.
\item[$^{\P}$] Also supported by the Hungarian OTKA fund under contract
  number T026178.
\item[$^{\flat}$] Supported also by the Comisi\'on Interministerial de Ciencia y
        Tecnolog{\'\i}a.
\item[$^{\sharp}$] Also supported by CONICET and Universidad Nacional de La Plata,
        CC 67, 1900 La Plata, Argentina.
\item[$^{\triangle}$] Supported by the National Natural Science
  Foundation of China.
\end{list}
}

}

\clearpage

\section{The OPAL Collaboration}

{
\newcommand {\thachn}{1}
\newcommand {\ualbrta}{2}
\newcommand {\brkbck}{3}
\newcommand {\bham}{4}
\newcommand {\blgna}{5}
\newcommand {\bnn}{6}
\newcommand {\ubclmb}{7}
\newcommand {\brnlu}{8}
\newcommand {\bdapst}{9}
\newcommand {\cvdsh}{10}
\newcommand {\scly}{11}
\newcommand {\cernch}{12}
\newcommand {\fiuchi}{13}
\newcommand {\inrdbr}{14}
\newcommand {\dsyhbg}{15}
\newcommand {\duku}{16}
\newcommand {\ufrbg}{17}
\newcommand {\uhdlbg}{18}
\newcommand {\uinda}{19}
\newcommand {\umanchsr}{20}
\newcommand {\umlnd}{21}
\newcommand {\umntrl}{22}
\newcommand {\mpimnchn}{23}
\newcommand {\lmumnchn}{24}
\newcommand {\uorgn}{25}
\newcommand {\crltn}{26}
\newcommand {\crppo}{27}
\newcommand {\nrcotwa}{28}
\newcommand {\qmwcull}{29}
\newcommand {\rvsd}{30}
\newcommand {\clrcral}{31}
\newcommand {\tchnn}{32}
\newcommand {\tlavv}{33}
\newcommand {\tky}{34}
\newcommand {\ucll}{35}
\newcommand {\uvctr}{36}
\newcommand {\wzmn}{37}
\newcommand {\ylu}{38}
\newcommand {\hqasymt}{a}
\newcommand {\ncrns}{b}
\newcommand {\zbu}{c}
\newcommand {\ncrng}{d}
\newcommand {\zottaw}{e}
\newcommand {\zcern}{f}
\newcommand {\hqasymi}{g}
\newcommand {\zcolba}{h}
\newcommand {\ncrnn}{i}
\newcommand {\ncrnc}{j}
\newcommand {\charmd}{k}
\newcommand {\ncrne}{l}
\newcommand {\ncrnz}{m}
\newcommand {\ncrnx}{n}
\newcommand {\zerau}{o}
\newcommand {\zerlgn}{p}
\newcommand {\zfnal}{q}
\newcommand {\zfiu}{r}
\newcommand {\prxh}{s}
\newcommand {\ncrni}{t}
\newcommand {\ncrnb}{u}
\newcommand {\ncrnr}{v}
\newcommand {\ncrnu}{w}
\newcommand {\ncrnua}{x}
\newcommand {\ncrnq}{y}
\newcommand {\zklsrh}{z}
\newcommand {\ncrnh}{aa}
\newcommand {\zlyon}{ab}
\newcommand {\ncrnj}{ac}
\newcommand {\incchgb}{ad}
\newcommand {\zumass}{ae}
\newcommand {\ncrnp}{af}
\newcommand {\zumich}{ag}
\newcommand {\zedoh}{ah}
\newcommand {\ncrnf}{ai}
\newcommand {\znijm}{aj}
\newcommand {\znd}{ak}
\newcommand {\zprag}{al}
\newcommand {\zral}{am}
\newcommand {\ncrno}{an}
\newcommand {\incchgf}{ao}
\newcommand {\zrit}{ap}
\newcommand {\ncrnl}{aq}
\newcommand {\ncrnw}{ar}
\newcommand {\ncrna}{as}
\newcommand {\zutah}{at}
\newcommand {\zvict}{au}
\newcommand {\ncrnm}{av}
\newcommand {\zeth}{aw}
\newcommand {\hqasymk}{ax}
\newcommand {\puedu}{ay}
\newcommand {\lll}{az}
\newcommand {\laurnu}{ba}
\newcommand {\tkymu}{bb}
\newcommand {\harvardu}{bc}
\newcommand {\ncrnzz}{\dagger}
\tolerance=10000
\hbadness=5000
\raggedright
G.\thinspace Abbiendi,$\!^{\blgna}$
P.D.\thinspace Acton,$\!^{\brnlu}$
C.\thinspace Ainsley,$\!^{\cvdsh}$
P.F.\thinspace {\AA}kesson,$\!^{\bnn}$
G.\thinspace Alexander,$\!^{\tlavv}$
J.\thinspace Allison,$\!^{\umanchsr}$
P.P.\thinspace Allport,$\!^{\cvdsh}$
N.\thinspace Altekamp,$\!^{\cvdsh}$
P.\thinspace Amaral,$\!^{\fiuchi}$
K.\thinspace Ametewee,$\!^{\brnlu}$
% ...

\bigskip

\begin{list}{A}{\itemsep=0pt plus 0pt minus 0pt\parsep=0pt plus 0pt minus 0pt
                \topsep=0pt plus 0pt minus 0pt}

\item[$^{ \thachn}$]  Technische Hochschule Aachen, III Physikalisches Institut, Sommerfeldstrasse 26-28, D-52056 Aachen, Germany
\item[$^{\ualbrta}$]  University of Alberta,  Department of Physics, Edmonton AB T6G 2J1, Canada
\item[$^{\brkbck}$]  Birkbeck College, London, WC1E 7HV, UK
\item[$^{\bham}$]  School of Physics and Astronomy, University of Birmingham, Birmingham B15 2TT, UK
\item[$^{ \blgna}$]  Dipartimento di Fisica dell' Universit\`a di Bologna and INFN, I-40126 Bologna, Italy
\item[$^{ \bnn}$]  Physikalisches Institut, Universit\"at Bonn, D-53115 Bonn, Germany
\item[$^{\ubclmb}$]  University of British Columbia, Department of Physics, Vancouver BC V6T 1Z1, Canada
\item[$^{\brnlu}$]  Brunel University, Uxbridge, Middlesex UB8 3PH, UK
\item[$^{\bdapst}$]  Research Institute for Particle and Nuclear Physics, H-1525 Budapest, P O  Box 49, Hungary
\item[$^{ \cvdsh}$]  Cavendish Laboratory, Cambridge CB3 0HE, UK
\item[$^{\scly}$]  CEA, DAPNIA/SPP, CE-Saclay, F-91191 Gif-sur-Yvette, France
\item[$^{ \cernch}$]  CERN, European Organisation for Nuclear Research, CH-1211 Geneva 23, Switzerland
\item[$^{  \fiuchi}$]  Enrico Fermi Institute and Department of Physics, University of Chicago, Chicago IL 60637, USA
\item[$^{\inrdbr}$]  Institute of Nuclear Research, H-4001 Debrecen, P O  Box 51, Hungary
\item[$^{\dsyhbg}$]  Universit\"at Hamburg/DESY, II Institut f\"ur Experimental Physik, Notkestrasse 85, D-22607 Hamburg, Germany
\item[$^{\duku}$]  Duke University, Department of Physics, Durham, NC 27708-0305, USA
\item[$^{\ufrbg}$]  Fakult\"at f\"ur Physik, Albert Ludwigs Universit\"at, D-79104 Freiburg, Germany
\item[$^{\uhdlbg}$]  Physikalisches Institut, Universit\"at Heidelberg, D-69120 Heidelberg, Germany
\item[$^{\uinda}$]  Indiana University, Department of Physics, Swain Hall West 117, Bloomington IN 47405, USA
\item[$^{\umanchsr}$]  Department of Physics, Schuster Laboratory, The University, Manchester M13 9PL, UK
\item[$^{\umlnd}$]  Department of Physics, University of Maryland, College Park, MD 20742, USA
\item[$^{\umntrl}$]  Laboratoire de Physique Nucl\'eaire, Universit\'e de Montr\'eal, Montr\'eal, Quebec H3C 3J7, Canada
\item[$^{\mpimnchn}$]  Max-Planck-Institute f\"ur Physik, F\"ohring Ring 6, 80805 M\"unchen, Germany
\item[$^{\lmumnchn}$]  University Ludwigs-Maximilians-Universit\"at M\"unchen, Sektion Physik, Am Coulombwall 1, D-85748 Garching, Germany
\item[$^{\uorgn}$]  University of Oregon, Department of Physics, Eugene OR 97403, USA
\item[$^{ \crltn}$]  Ottawa-Carleton Institute for Physics, Department of Physics, Carleton University, Ottawa, Ontario K1S 5B6, Canada
\item[$^{ \crppo}$]  Centre for Research in Particle Physics, Carleton University, Ottawa, Ontario K1S 5B6, Canada
\item[$^{\nrcotwa}$]  National Research Council of Canada, Herzberg Institute of Astrophysics, Ottawa, Ontario K1A 0R6, Canada
\item[$^{\qmwcull}$]  Queen Mary and Westfield College, University of London, London E1 4NS, UK
\item[$^{  \rvsd}$]  Department of Physics, University of California, Riverside CA 92521, USA
\item[$^{\clrcral}$]  CLRC Rutherford Appleton Laboratory, Chilton, Didcot, Oxfordshire OX11 0QX, UK
\item[$^{ \tchnn}$]  Department of Physics, Technion-Israel Institute of Technology, Haifa 32000, Israel
\item[$^{ \tlavv}$]  Department of Physics and Astronomy, Tel Aviv University, Tel Aviv 69978, Israel
\item[$^{ \tky}$]  International Centre for Elementary Particle Physics and Department of Physics, University of Tokyo, Tokyo 113-0033, and Kobe University, Kobe 657-8501, Japan
\item[$^{\ucll}$]  University College London, London WC1E 6BT, UK
\item[$^{\uvctr}$]  University of Victoria, Department of Physics, P O Box 3055, Victoria BC V8W 3P6, Canada
\item[$^{\wzmn}$]  Particle Physics Department, Weizmann Institute of Science, Rehovot 76100, Israel
\item[$^{\ylu}$]  Yale University, Department of Physics, New Haven, CT 06520, USA

\end{list}

\bigskip

\begin{list}{A}{\itemsep=0pt plus 0pt minus 0pt\parsep=0pt plus 0pt minus 0pt
                \topsep=0pt plus 0pt minus 0pt}

\item[$^{\hqasymt}$]  now at RWTH Aachen, Germany %
\item[$^{\ncrns}$]  now at University of Antwerpen, Physics Department,B-2610 Antwerpen, Belgium %
\item[$^{\zbu}$]  now at Boston University, Boston, USA %
\item[$^{\ncrng}$]  and Research Institute for Particle and Nuclear Physics, Budapest, Hungary %
\item[$^{\zottaw}$]  now at Department of Physics, Carleton University, Ottawa, ON, K1S 5B6, Canada %
\item[$^{\zcern}$]  now at CERN, 1211 Geneva 23, Switzerland %
\item[$^{\hqasymi}$]  and CERN, PH Department, 1211 Geneva 23, Switzerland %
\item[$^{\zcolba}$]  now at Columbia University, New York, New York, USA %
\item[$^{\ncrnn}$]  now at University of Mining and Metallurgy, Cracow, Poland %
\item[$^{\ncrnc}$]  and Institute of Nuclear Research, Debrecen, Hungary %
\item[$^{\charmd}$]  and Department of Experimental Physics, Lajos Kossuth University, Debrecen, Hungary %
\item[$^{\ncrne}$]  and Department of Experimental Physics, University of Debrecen, Hungary %
\item[$^{\ncrnz}$]  now at DESY, Notkestrasse 85, D-22607 Hamburg, Germany %
\item[$^{\ncrnx}$]  now at DESY Zeuthen, Platanenallee 6, D-15738 Zeuthen, Germany %
\item[$^{\zerau}$]  now at Embry-Riddle Aeronautical University, Prescott, Arizona 86301, USA  %
\item[$^{\zerlgn}$]  now at Friedrich-Alexander-Universit\"at Erlangen-N\"urnberg, 91054 Erlangen, Germany  %
\item[$^{\zfnal}$]  now at Fermi National Accelerator Laboratory, Batavia, Illinois, USA  %
\item[$^{\zfiu}$]  now at Florida International University, Miami, Florida, USA  %
\item[$^{\prxh}$]  on leave from Research Institute for Computer Peripherals, Hangzhou, China %
\item[$^{\ncrni}$]  now at Dept. Physics, University of Illinois at Urbana-Champaign, USA %
\item[$^{\ncrnb}$]  now at University of Iowa, Dept of Physics and Astronomy, Iowa, USA %
\item[$^{\ncrnr}$]  now at IEKP Universit\"at Karlsruhe, Germany %
\item[$^{\ncrnu}$]  now at High Energy Accelerator Research Organisation (KEK), Tsukuba, Ibaraki, Japan %
\item[$^{\ncrnua}$]  and High Energy Accelerator Research Organisation (KEK), Tsukuba, Ibaraki, Japan %
\item[$^{\ncrnq}$]  now at IPHE Universit\'e de Lausanne, CH-1015 Lausanne, Switzerland %
\item[$^{\zklsrh}$]  now at Institut f\"ur Experimentelle Kernphysik, Universit\"at Karlsruhe, Karlsruhe, Germany %
\item[$^{\ncrnh}$]  now at University of Liverpool, Dept of Physics, Liverpool L69 3BX, UK %
\item[$^{\zlyon}$]  now at IPNL, Universit\'e Claude Bernard Lyon-1, Villeurbanne, France %
\item[$^{\ncrnj}$]  and Manchester University Manchester, M13 9PL, Manchester, UK %
\item[$^{\incchgb}$]  now at CPP Marseille, Facult\'e des Sciences de Luminy, Marseille, France %
\item[$^{\zumass}$]  now at University of Massachusetts, Amherst, Massachusetts 01003-4525 USA %
\item[$^{\ncrnp}$]  now at The University of Melbourne, Victoria, Australia %
\item[$^{\zumich}$]  now at The University of Michigan, Ann Arbor, Michigan, USA %
\item[$^{\zedoh}$]  now at MPI f\"ur Physik, 80805 M\"unchen, Germany %
\item[$^{\ncrnf}$]  and MPI f\"ur Physik, 80805 M\"unchen, Germany %
\item[$^{\znijm}$]  now at IMAPP, Radboud University Nijmegen, Toernooiveld 1, 6525 ED Nijmegen, The Netherlands %
\item[$^{\znd}$]  now at University of Notre Dame, Notre Dame, Indiana 46556-5670 USA %
\item[$^{\zprag}$]  now at Institute of Physics, Academy of Sciences of the Czech Republic, Prague, Czech Republic %
\item[$^{\zral}$]  now at CLRC Rutherford Appleton Laboratory, Chilton, Didcot, Oxfordshire OX11 0QX, UK %
\item[$^{\ncrno}$]  now at University of California, San Diego, USA %
\item[$^{\incchgf}$]  and Shinshu University, Matsumoto 390, Japan %
\item[$^{\zrit}$]  now at Royal Institute of Technology, Stockholm, Sweden %
\item[$^{\ncrnl}$]  now at University of Toronto, Dept of Physics, Toronto, Canada %
\item[$^{\ncrnw}$]  now at TRIUMF, Vancouver V6T 2A3, Canada %
\item[$^{\ncrna}$]  and at TRIUMF, Vancouver V6T 2A3, Canada %
\item[$^{\zutah}$]  now at The University of Utah, Salt Lake City, Utah, USA %
\item[$^{\zvict}$]  now at University of Victoria, Department of Physics, Victoria BC V8W 3P6, Canada %
\item[$^{\ncrnm}$]  now at Bergische Universit\"at, Wuppertal, Germany %
\item[$^{\zeth}$]  now at Eidgenossische Technische Hochschule Zurich (ETH), Zurich, Switzerland %
\item[$^{\hqasymk}$] now at University of Kansas, Dept of Physics and Astronomy, Lawrence, KS 66045, USA %
\item[$^{\puedu}$] now at Purdue University, West Lafayette, IN 47907, USA %
\item[$^{\lll}$] now at Lawrence Livermore National Lab, California, USA %
\item[$^{\laurnu}$] now at Laurentian University, Ontario, Canada %
\item[$^{\tkymu}$] now at Tokyo Metropolitan University, Tokyo, Japan %
\item[$^{\harvardu}$] now at Harvard university, Cambridge, MA, USA %
\item[$^{\ncrnzz}$]  deceased

\end{list}

\bigskip

In addition to the support staff at our own
institutions we are pleased to acknowledge the  \\
Department of Energy, USA, \\
National Science Foundation, USA, \\
Particle Physics and Astronomy Research Council, UK, \\
Natural Sciences and Engineering Research Council, Canada, \\
Israel Science Foundation, administered by the Israel
Academy of Science and Humanities, \\
Benoziyo Center for High Energy Physics,\\
Japanese Ministry of Education, Culture, Sports, Science and
Technology (MEXT) and a grant under the MEXT International
Science Research Program,\\
Japanese Society for the Promotion of Science (JSPS),\\
German Israeli Bi-national Science Foundation (GIF), \\
Bundesministerium f\"ur Bildung und Forschung, Germany, \\
National Research Council of Canada, \\
Hungarian Foundation for Scientific Research, OTKA T-038240,
and T-042864,\\
The NWO/NATO Fund for Scientific Research, the Netherlands.\\

}

\clearpage

\section{The SLD Collaboration}

{
\def\iADEL{$^{1}$}
\def\iAOMORI{$^{2}$}
\def\iBRUN{$^{3}$}
\def\iBOLO{$^{4}$}
\def\iBU{$^{5}$}
\def\iCALT{$^{6}$}
\def\iCIN{$^{7}$}
\def\iCOLO{$^{8}$}
\def\iCOLUM{$^{9}$}
\def\iCSU{$^{10}$}
\def\iFERR{$^{11}$}
\def\iFRAS{$^{12}$}
\def\iILL{$^{13}$}
\def\iJHU{$^{14}$}
\def\iLBL{$^{15}$}
\def\iMASS{$^{16}$}
\def\iMISSI{$^{17}$}
\def\iMIT{$^{18}$}
\def\iMOSCOW{$^{19}$}
\def\iNAGO{$^{20}$}
\def\iOREG{$^{21}$}
\def\iOXF{$^{22}$}
\def\iPAD{$^{23}$}
\def\iPERU{$^{24}$}
\def\iPISA{$^{25}$}
\def\iRAL{$^{26}$}
\def\iRUTG{$^{27}$}
\def\iSLAC{$^{28}$}
\def\iSOGA{$^{29}$}
\def\iSOONG{$^{30}$}
\def\iTENN{$^{31}$}
\def\iTOHO{$^{32}$}
\def\iUCSB{$^{33}$}
\def\iUCSC{$^{34}$}
\def\iVAND{$^{35}$}
\def\iWASH{$^{36}$}
\def\iWISC{$^{37}$}
\def\iYALE{$^{38}$}

\tolerance=10000
\hbadness=5000
\raggedright

\mbox{Kenji\thinspace Abe,$\!$\iNAGO}
\mbox{Koya\thinspace Abe,$\!$\iTOHO}
\mbox{T.\thinspace Abe,$\!$\iSLAC}
\mbox{I.\thinspace Abt,$\!$\iILL}
\mbox{P.D.\thinspace Acton,$\!$\iBRUN}
\mbox{I.\thinspace Adam,$\!$\iSLAC}
\mbox{G.\thinspace Agnew,$\!$\iBRUN}
\mbox{T.\thinspace Akagi,$\!$\iSLAC}
\mbox{H.\thinspace Akimoto,$\!$\iSLAC}
\mbox{N.J.\thinspace Allen,$\!$\iBRUN}
\mbox{W.W.\thinspace Ash,$\!$\iSLAC$^\dagger$}
\mbox{D.\thinspace Aston,$\!$\iSLAC}
\mbox{N.\thinspace Bacchetta,$\!$\iPAD}
\mbox{K.G.\thinspace Baird,$\!$\iMASS}
\mbox{C.\thinspace Baltay,$\!$\iYALE}
\mbox{H.R.\thinspace Band,$\!$\iWISC}
\mbox{M.B.\thinspace Barakat,$\!$\iYALE}
\mbox{G.J.\thinspace Baranko,$\!$\iCOLO}
\mbox{O.\thinspace Bardon,$\!$\iMIT}
\mbox{T.L.\thinspace Barklow,$\!$\iSLAC}
\mbox{G.L.\thinspace Bashindzhagian,$\!$\iMOSCOW}
\mbox{R.\thinspace Battiston,$\!$\iPERU}
\mbox{J.M.\thinspace Bauer,$\!$\iMISSI}
\mbox{A.O.\thinspace Bazarko,$\!$\iCOLUM}
\mbox{A.\thinspace Bean,$\!$\iSLAC}
\mbox{G.\thinspace Bellodi,$\!$\iOXF}
\mbox{R.\thinspace Ben-David,$\!$\iYALE}
\mbox{A.C.\thinspace Benvenuti,$\!$\iBOLO}
\mbox{R.\thinspace Berger,$\!$\iSLAC}                                           \mbox{M.\thinspace Biasini,$\!$\iPERU}
\mbox{T.\thinspace Bienz,$\!$\iSLAC}
\mbox{G.M.\thinspace Bilei,$\!$\iPERU}
\mbox{D.\thinspace Bisello,$\!$\iPAD}
\mbox{G.\thinspace Blaylock,$\!$\iMASS}
\mbox{J.R.\thinspace Bogart,$\!$\iSLAC}
\mbox{B.\thinspace Bolen,$\!$\iMISSI}
\mbox{T.\thinspace Bolton,$\!$\iCOLUM}
\mbox{G.R.\thinspace Bower,$\!$\iSLAC}
\mbox{J.E.\thinspace Brau,$\!$\iOREG}
\mbox{M.\thinspace Breidenbach,$\!$\iSLAC}
\mbox{W.M.\thinspace Bugg,$\!$\iTENN}
\mbox{D.\thinspace Burke,$\!$\iSLAC}
\mbox{T.H.\thinspace Burnett,$\!$\iWASH}
\mbox{P.N.\thinspace Burrows,$\!$\iOXF}
\mbox{W.\thinspace Busza,$\!$\iMIT}
\mbox{A.\thinspace Calcaterra,$\!$\iFRAS}
\mbox{D.O.\thinspace Caldwell,$\!$\iUCSB}
\mbox{B.\thinspace Camanzi,$\!$\iRAL}
\mbox{M.\thinspace Carpinelli,$\!$\iPISA}
\mbox{J.\thinspace Carr,$\!$\iCOLO}
\mbox{R.\thinspace Cassell,$\!$\iSLAC}
\mbox{R.\thinspace Castaldi,$\!$\iPISA}
\mbox{A.\thinspace Castro,$\!$\iPAD}
\mbox{M.\thinspace Cavalli-Sforza,$\!$\iUCSC}
\mbox{G.B.\thinspace Chadwick,$\!$\iSLAC}
\mbox{A.\thinspace Chou,$\!$\iSLAC}
\mbox{E.\thinspace Church,$\!$\iWASH}
\mbox{R.\thinspace Claus,$\!$\iSLAC}
\mbox{H.O.\thinspace Cohn,$\!$\iTENN}
\mbox{J.A.\thinspace Coller,$\!$\iBU}
\mbox{M.R.\thinspace Convery,$\!$\iSLAC}
\mbox{V.\thinspace Cook,$\!$\iWASH}
\mbox{R.\thinspace Cotton,$\!$\iBRUN}
\mbox{R.F.\thinspace Cowan,$\!$\iMIT}
\mbox{P.A.\thinspace Coyle,$\!$\iUCSC}
\mbox{D.G.\thinspace Coyne,$\!$\iUCSC}
\mbox{G.\thinspace Crawford,$\!$\iSLAC}
\mbox{A.\thinspace D'Oliveira,$\!$\iCIN}
\mbox{C.J.S.\thinspace Damerell,$\!$\iRAL}
\mbox{M.\thinspace Daoudi,$\!$\iSLAC}
\mbox{S.\thinspace Dasu,$\!$\iWISC}
\mbox{N.\thinspace de Groot,$\!$\iRAL}
\mbox{R.\thinspace de Sangro,$\!$\iFRAS}
\mbox{P.\thinspace De Simone,$\!$\iFRAS}
\mbox{S.\thinspace De Simone,$\!$\iFRAS}
\mbox{R.\thinspace Dell'Orso,$\!$\iPISA}
\mbox{P.J.\thinspace Dervan,$\!$\iBRUN}
\mbox{M.\thinspace Dima,$\!$\iCSU}
\mbox{D.N.\thinspace Dong,$\!$\iMIT}
\mbox{M.\thinspace Doser,$\!$\iSLAC}
\mbox{P.Y.C.\thinspace Du,$\!$\iTENN}
\mbox{R.\thinspace Dubois,$\!$\iSLAC}
\mbox{J.E.\thinspace Duboscq,$\!$\iUCSB}
\mbox{G.\thinspace Eigen,$\!$\iCALT}
\mbox{B.I.\thinspace Eisenstein,$\!$\iILL}
\mbox{R.\thinspace Elia,$\!$\iSLAC}
\mbox{E.\thinspace Erdos,$\!$\iCOLO}
\mbox{I.\thinspace Erofeeva,$\!$\iMOSCOW}
\mbox{V.\thinspace Eschenburg,$\!$\iMISSI}
\mbox{E.\thinspace Etzion,$\!$\iWISC}
\mbox{S.\thinspace Fahey,$\!$\iCOLO}
\mbox{D.\thinspace Falciai,$\!$\iFRAS}
\mbox{C.\thinspace Fan,$\!$\iCOLO}
\mbox{J.P.\thinspace Fernandez,$\!$\iUCSC}
\mbox{M.J.\thinspace Fero,$\!$\iMIT}
\mbox{K.\thinspace Flood,$\!$\iMASS}
\mbox{R.\thinspace Frey,$\!$\iOREG}
\mbox{J.I.\thinspace Friedman,$\!$\iMIT}
\mbox{K.\thinspace Furuno,$\!$\iOREG}
\mbox{E.L.\thinspace Garwin,$\!$\iSLAC}
\mbox{T.\thinspace Gillman,$\!$\iRAL}
\mbox{G.\thinspace Gladding,$\!$\iILL}
\mbox{S.\thinspace Gonzalez,$\!$\iMIT}
\mbox{G.D.\thinspace Hallewell,$\!$\iSLAC}
\mbox{E.L.\thinspace Hart,$\!$\iTENN}
\mbox{J.L.\thinspace Harton,$\!$\iCSU}
\mbox{A.\thinspace Hasan,$\!$\iBRUN}
\mbox{Y.\thinspace Hasegawa,$\!$\iTOHO}
\mbox{K.\thinspace Hasuko,$\!$\iTOHO}
\mbox{S.\thinspace Hedges,$\!$\iBU}
\mbox{S.S.\thinspace Hertzbach,$\!$\iMASS}
\mbox{M.D.\thinspace Hildreth,$\!$\iSLAC}
\mbox{D.G.\thinspace Hitlin,$\!$\iCALT}
\mbox{A.\thinspace Honma,$\!$\iSLAC}
\mbox{J.\thinspace S.\thinspace Huber,$\!$\iOREG}
\mbox{M.E.\thinspace Huffer,$\!$\iSLAC}
\mbox{E.W.\thinspace Hughes,$\!$\iSLAC}
\mbox{X.\thinspace Huynh,$\!$\iSLAC}
\mbox{H.\thinspace Hwang,$\!$\iOREG}
\mbox{M.\thinspace Iwasaki,$\!$\iOREG}
\mbox{Y.\thinspace Iwasaki,$\!$\iTOHO}
\mbox{J.M.\thinspace Izen,$\!$\iILL}
\mbox{D.J.\thinspace Jackson,$\!$\iRAL}
\mbox{P.\thinspace Jacques,$\!$\iRUTG}
\mbox{J.A.\thinspace Jaros,$\!$\iSLAC}
\mbox{Z.Y.\thinspace Jiang,$\!$\iSLAC}
\mbox{A.S.\thinspace Johnson,$\!$\iSLAC}
\mbox{J.R.\thinspace Johnson,$\!$\iWISC}
\mbox{R.A.\thinspace Johnson,$\!$\iCIN}
\mbox{T.\thinspace Junk,$\!$\iSLAC}
\mbox{R.\thinspace Kajikawa,$\!$\iNAGO}
\mbox{M.\thinspace Kalelkar,$\!$\iRUTG}
\mbox{Y.A.\thinspace Kamyshkov,$\!$\iTENN}
\mbox{H.J.\thinspace Kang,$\!$\iRUTG}
\mbox{I.\thinspace Karliner,$\!$\iILL}
\mbox{H.\thinspace Kawahara,$\!$\iSLAC}
\mbox{M.H.\thinspace Kelsey,$\!$\iSLAC}
\mbox{H.W.\thinspace Kendall,$\!$\iMIT$^\dagger$}
\mbox{Y.D.\thinspace Kim,$\!$\iSOGA}
\mbox{M.\thinspace King,$\!$\iSLAC}
\mbox{R.\thinspace King,$\!$\iSLAC}
\mbox{R.R.\thinspace Kofler,$\!$\iMASS}
\mbox{N.M.\thinspace Krishna,$\!$\iSLAC}
\mbox{Y.\thinspace Kwon,$\!$\iSLAC}
\mbox{J.F.\thinspace Labs,$\!$\iSLAC}
\mbox{R.S.\thinspace Kroeger,$\!$\iMISSI}
\mbox{M.\thinspace Langston,$\!$\iOREG}
\mbox{A.\thinspace Lath,$\!$\iMIT}
\mbox{J.A.\thinspace Lauber,$\!$\iCOLO}
\mbox{D.W.G.\thinspace Leith,$\!$\iSLAC}
\mbox{V.\thinspace Lia,$\!$\iMIT}
\mbox{C.\thinspace Lin,$\!$\iMASS}
\mbox{M.X.\thinspace Liu,$\!$\iUCSC}
\mbox{M.\thinspace Loreti,$\!$\iPAD}
\mbox{A.\thinspace Lu,$\!$\iUCSB}
\mbox{H.L.\thinspace Lynch,$\!$\iSLAC}
\mbox{J.\thinspace Ma,$\!$\iWASH}
\mbox{G.\thinspace Mancinelli,$\!$\iRUTG}
\mbox{S.\thinspace Manly,$\!$\iYALE}
\mbox{G.\thinspace Mantovani,$\!$\iPERU}
\mbox{T.W.\thinspace Markiewicz,$\!$\iSLAC}
\mbox{T.\thinspace Maruyama,$\!$\iSLAC}
\mbox{H.\thinspace Masuda,$\!$\iSLAC}
\mbox{E.\thinspace Mazzucato,$\!$\iFERR}
\mbox{J.F.\thinspace McGowan,$\!$\iILL}
\mbox{A.K.\thinspace McKemey,$\!$\iBRUN}
\mbox{B.T.\thinspace Meadows,$\!$\iCIN}
\mbox{R.\thinspace Messner,$\!$\iSLAC}
\mbox{P.M.\thinspace Mockett,$\!$\iWASH}
\mbox{K.C.\thinspace Moffeit,$\!$\iSLAC}
\mbox{T.B.\thinspace Moore,$\!$\iYALE}
\mbox{M.\thinspace Morii,$\!$\iSLAC}
\mbox{B.\thinspace Mours,$\!$\iSLAC}
\mbox{D.\thinspace Muller,$\!$\iSLAC}
\mbox{G.\thinspace Mueller,$\!$\iSLAC}
\mbox{V.\thinspace Murzin,$\!$\iMOSCOW}
\mbox{T.\thinspace Nagamine,$\!$\iTOHO}
\mbox{S.\thinspace Narita,$\!$\iTOHO}
\mbox{U.\thinspace Nauenberg,$\!$\iCOLO}
\mbox{H.\thinspace Neal,$\!$\iYALE}
\mbox{G.\thinspace Nesom,$\!$\iOXF}
\mbox{M.\thinspace Nussbaum,$\!$\iCIN$^\dagger$}
\mbox{Y.\thinspace Ohnishi,$\!$\iNAGO}
\mbox{N.\thinspace Oishi,$\!$\iNAGO}
\mbox{D.\thinspace Onoprienko,$\!$\iTENN}
\mbox{L.S.\thinspace Osborne,$\!$\iMIT}
\mbox{R.S.\thinspace Panvini,$\!$\iVAND$^\dagger$}
\mbox{C.H.\thinspace Park,$\!$\iSOONG}
\mbox{H.\thinspace Park,$\!$\iOREG}
\mbox{T.J.\thinspace Pavel,$\!$\iSLAC}
\mbox{I.\thinspace Peruzzi,$\!$\iFRAS}
\mbox{L.\thinspace Pescara,$\!$\iPAD}
\mbox{M.\thinspace Piccolo,$\!$\iFRAS}
\mbox{L.\thinspace Piemontese,$\!$\iFERR}
\mbox{E.\thinspace Pieroni,$\!$\iPISA}
\mbox{K.T.\thinspace Pitts,$\!$\iOREG}
\mbox{R.J.\thinspace Plano,$\!$\iRUTG}
\mbox{R.\thinspace Prepost,$\!$\iWISC}
\mbox{C.Y.\thinspace Prescott,$\!$\iSLAC}
\mbox{G.\thinspace Punkar,$\!$\iUCSB}
\mbox{J.\thinspace Quigley,$\!$\iMIT}
\mbox{B.N.\thinspace Ratcliff,$\!$\iSLAC}
\mbox{K.\thinspace Reeves,$\!$\iSLAC}
\mbox{T.W.\thinspace Reeves,$\!$\iVAND}
\mbox{J.\thinspace Reidy,$\!$\iMISSI}
\mbox{P.L.\thinspace Reinertsen,$\!$\iUCSC}
\mbox{P.E.\thinspace Rensing,$\!$\iSLAC}
\mbox{L.S.\thinspace Rochester,$\!$\iSLAC}
\mbox{J.E.\thinspace Rothberg,$\!$\iWASH}
\mbox{P.C.\thinspace Rowson,$\!$\iSLAC}
\mbox{J.J.\thinspace Russell,$\!$\iSLAC}
\mbox{O.H.\thinspace Saxton,$\!$\iSLAC}
\mbox{T.\thinspace Schalk,$\!$\iUCSC}
\mbox{R.H.\thinspace Schindler,$\!$\iSLAC}
\mbox{U.\thinspace Schneekloth,$\!$\iMIT}
\mbox{B.A.\thinspace Schumm,$\!$\iUCSC}
\mbox{J.\thinspace Schwiening,$\!$\iSLAC}
\mbox{A.\thinspace Seiden,$\!$\iUCSC}
\mbox{S.\thinspace Sen,$\!$\iYALE}
\mbox{V.V.\thinspace Serbo,$\!$\iSLAC}
\mbox{L.\thinspace Servoli,$\!$\iPERU}
\mbox{M.H.\thinspace Shaevitz,$\!$\iCOLUM}
\mbox{J.T.\thinspace Shank,$\!$\iBU}
\mbox{G.\thinspace Shapiro,$\!$\iLBL$^\dagger$}
\mbox{D.J.\thinspace Sherden,$\!$\iSLAC}
\mbox{K.D.\thinspace Shmakov,$\!$\iTENN}
\mbox{C.\thinspace Simopoulos,$\!$\iSLAC}
\mbox{N.B.\thinspace Sinev,$\!$\iOREG}
\mbox{S.R.\thinspace Smith,$\!$\iSLAC}
\mbox{M.B.\thinspace Smy,$\!$\iCSU}
\mbox{J.A.\thinspace Snyder,$\!$\iYALE}
\mbox{M.D.\thinspace Sokoloff,$\!$\iCIN}
\mbox{H.\thinspace Staengle,$\!$\iCSU}
\mbox{A.\thinspace Stahl,$\!$\iSLAC}
\mbox{P.\thinspace Stamer,$\!$\iRUTG}
\mbox{H.\thinspace Steiner,$\!$\iLBL}
\mbox{R.\thinspace Steiner,$\!$\iADEL}
\mbox{M.G.\thinspace Strauss,$\!$\iMASS}
\mbox{D.\thinspace Su,$\!$\iSLAC}
\mbox{F.\thinspace Suekane,$\!$\iTOHO}
\mbox{A.\thinspace Sugiyama,$\!$\iNAGO}
\mbox{A.\thinspace Suzuki,$\!$\iNAGO}
\mbox{S.\thinspace Suzuki,$\!$\iNAGO}
\mbox{M.\thinspace Swartz,$\!$\iJHU}
\mbox{A.\thinspace Szumilo,$\!$\iWASH}
\mbox{T.\thinspace Takahashi,$\!$\iSLAC}
\mbox{F.E.\thinspace Taylor,$\!$\iMIT}
\mbox{J.J.\thinspace Thaler,$\!$\iILL}
\mbox{J.\thinspace Thom,$\!$\iSLAC}
\mbox{E.\thinspace Torrence,$\!$\iMIT}
\mbox{A.I.\thinspace Trandafir,$\!$\iMASS}
\mbox{J.D.\thinspace Turk,$\!$\iYALE}
\mbox{T.\thinspace Usher,$\!$\iSLAC}
\mbox{J.\thinspace Va'vra,$\!$\iSLAC}
\mbox{C.\thinspace Vannini,$\!$\iPISA}
\mbox{E.\thinspace Vella,$\!$\iWASH}
\mbox{J.P.\thinspace Venuti,$\!$\iVAND}
\mbox{R.\thinspace Verdier,$\!$\iMIT}
\mbox{P.G.\thinspace Verdini,$\!$\iPISA}
\mbox{D.L.\thinspace Wagner,$\!$\iCOLO}
\mbox{S.R.\thinspace Wagner,$\!$\iCOLO}
\mbox{A.P.\thinspace Waite,$\!$\iSLAC}
\mbox{S.\thinspace Walston,$\!$\iOREG}
\mbox{J.\thinspace Wang,$\!$\iSLAC}
\mbox{S.J.\thinspace Watts,$\!$\iBRUN}
\mbox{A.W.\thinspace Weidemann,$\!$\iTENN}
\mbox{E.R.\thinspace Weiss,$\!$\iWASH}
\mbox{J.S.\thinspace Whitaker,$\!$\iBU}
\mbox{S.L.\thinspace White,$\!$\iTENN}
\mbox{F.J.\thinspace Wickens,$\!$\iRAL}
\mbox{D.A.\thinspace Williams,$\!$\iUCSC}
\mbox{D.C.\thinspace Williams,$\!$\iUCSC}
\mbox{S.H.\thinspace Williams,$\!$\iSLAC}
\mbox{S.\thinspace Willocq,$\!$\iMASS}
\mbox{R.J.\thinspace Wilson,$\!$\iCSU}
\mbox{W.J.\thinspace Wisniewski,$\!$\iSLAC}
\mbox{J.L.\thinspace Wittlin,$\!$\iMASS}
\mbox{M.\thinspace Woods,$\!$\iSLAC}
\mbox{G.B.\thinspace Word,$\!$\iRUTG}
\mbox{T.R.\thinspace Wright,$\!$\iWISC}
\mbox{J.\thinspace Wyss,$\!$\iPAD}
\mbox{R.K.\thinspace Yamamoto,$\!$\iMIT}
\mbox{J.M.\thinspace Yamartino,$\!$\iMIT}
\mbox{X.Q.\thinspace Yang,$\!$\iOREG}
\mbox{J.\thinspace Yashima,$\!$\iTOHO}
\mbox{S.J.\thinspace Yellin,$\!$\iUCSB}
\mbox{C.C.\thinspace Young,$\!$\iSLAC}
\mbox{H.\thinspace Yuta\unskip.\iAOMORI}
\mbox{G.\thinspace Zapalac,$\!$\iWISC}
\mbox{R.W.\thinspace Zdarko,$\!$\iSLAC}
\mbox{C.\thinspace Zeitlin,$\!$\iOREG}
\mbox{J.\thinspace Zhou.$\!$\iOREG}

\bigskip

\begin{list}{A}{\itemsep=0pt plus 0pt minus 0pt\parsep=0pt plus 0pt minus 0pt
                \topsep=0pt plus 0pt minus 0pt}
\item[\iADEL]
  Adelphi University, Garden City, New York, 11530,
\item[\iAOMORI]
  Aomori University, Aomori, 030 Japan,
\item[\iBRUN]
  Brunel University, Uxbridge, Middlesex, UB8 3PH United Kingdom,
\item[\iBOLO]
  INFN Sezione di Bologna, I-40126, Bologna, Italy,
\item[\iBU]
  Boston University, Boston, Massachusetts 02215,
\item[\iCALT]
  California Institute of Technology, Pasadena, California, 91125,
\item[\iCIN]
  University of Cincinnati, Cincinnati, Ohio, 45221,
\item[\iCOLO]
  University of Colorado, Boulder, Colorado 80309,
\item[\iCSU]
  Colorado State University, Ft. Collins, Colorado 80523,
\item[\iFERR]
  INFN Sezione di Ferrara and Universita di Ferrara, I-44100 Ferrara, Italy,
\item[\iFRAS]
  INFN Laboratori Nazionali di Frascati, I-00044 Frascati, Italy,
\item[\iILL]
  University of Illinois, Urbana, Illinois, 61801,
\item[\iJHU]
  Johns Hopkins University,  Baltimore, Maryland 21218-2686,
\item[\iLBL]
  Lawrence Berkeley National Laboratory, University of California, Berkeley, California 94720,
\item[\iMASS]
  University of Massachusetts, Amherst, Massachusetts 01003,
\item[\iMISSI]
  University of Mississippi, University, Mississippi 38677,
\item[\iMIT]
  Massachusetts Institute of Technology, Cambridge, Massachusetts 02139,
\item[\iMOSCOW]
  Institute of Nuclear Physics, Moscow State University, 119899 Moscow, Russia,
\item[\iNAGO]
  Nagoya University, Chikusa-ku, Nagoya, 464 Japan,
\item[\iOREG]
  University of Oregon, Eugene, Oregon 97403,
\item[\iOXF]
  Oxford University, Oxford, OX1 3RH, United Kingdom,
\item[\iPAD]
  INFN Sezione di Padova and Universita di Padova, I-35100 Padova, Italy,
\item[\iPERU]
  INFN Sezione di Perugia and Universita di Perugia, I-06100 Perugia, Italy,
\item[\iPISA]
  INFN Sezione di Pisa and Universita di Pisa, I-56100 Pisa, Italy,
\item[\iRAL]
  Rutherford Appleton Laboratory, Chilton, Didcot, Oxon OX11 0QX United Kingdom,
\item[\iRUTG]
  Rutgers University, Piscataway, New Jersey 08855,
\item[\iSLAC]
  Stanford Linear Accelerator Center, Stanford University, Stanford, California
94309,
\item[\iSOGA]
  Sogang University, Seoul, Korea,
\item[\iSOONG]
  Soongsil University, Seoul, Korea 156-743,
\item[\iTENN]
  University of Tennessee, Knoxville, Tennessee 37996,
\item[\iTOHO]
  Tohoku University, Sendai, 980 Japan,
\item[\iUCSB]
  University of California at Santa Barbara, Santa Barbara, California 93106,
\item[\iUCSC]
  University of California at Santa Cruz, Santa Cruz, California 95064,
\item[\iVAND]
  Vanderbilt University, Nashville,Tennessee 37235,
\item[\iWASH]
  University of Washington, Seattle, Washington 98105,
\item[\iWISC]
  University of Wisconsin, Madison,Wisconsin 53706,
\item[\iYALE]
  Yale University, New Haven, Connecticut 06511.
\item[$^\dagger$] Deceased.

\end{list}

\bigskip

This work was supported by the U.S. Department of Energy and National
Science Foundation, the UK Particle Physics and Astronomy Research
Council, the Istituto Nazionale di Fisica Nucleare of Italy and the
Japan-US Cooperative Research Project on High Energy Physics.
}

\chapter{Heavy-Flavour Fit including Off-Peak Asymmetries}
\label{sec:hqappfit}
The full 18 parameter fit to the LEP and SLD data including the
off-peak asymmetries gave the following
results:

\begin{eqnarray*}
  \Rbz    &=& 0.21628   \pm  0.00066\\
  \Rcz    &=& 0.1722    \pm  0.0031 \\
  \Abl    &=& 0.0560    \pm  0.0066 \\
  \Acl    &=&-0.018     \pm  0.013  \\
  \Abp    &=& 0.0982    \pm  0.0017 \\
  \Acp    &=& 0.0635    \pm  0.0036 \\
  \Abh    &=& 0.1125    \pm  0.0055 \\
  \Ach    &=& 0.125     \pm  0.011  \\
  \cAb    &=& 0.924     \pm  0.020  \\
  \cAc    &=& 0.669     \pm  0.027  \\
  \Brbl   &=& 0.1070    \pm  0.0022 \\
  \Brbclp &=& 0.0802    \pm  0.0018 \\
  \Brcl   &=& 0.0971    \pm  0.0032 \\
  \chiM   &=& 0.1250    \pm  0.0039 \\
  \fDp    &=& 0.235     \pm  0.016  \\
  \fDs    &=& 0.126     \pm  0.026  \\
  \fcb    &=& 0.092     \pm  0.022  \\
  \PcDst  &=& 0.1622    \pm  0.0048 \,
\end{eqnarray*}
with a $\chi^2/$d.o.f.{} of  $48/(105-18)$. The corresponding correlation
matrix is given in Table~\ref{tab:18parcor}.
The energies for the  peak$-$2, peak and peak+2 results are respectively
89.55 \GeV{}, 91.26 \GeV{} and 92.94 \GeV.
Note that the asymmetry results shown here are not the pole
asymmetries.%
\begin{table}[p]
\begin{center}
\begin{sideways}
\begin{minipage}[b]{\textheight}
\begin{center}
\footnotesize
\begin{tabular}{|l||rrrrrrrrrrrrrrrrrr|}
\hline
&\makebox[0.45cm]{$1)$}
&\makebox[0.45cm]{$2)$}
&\makebox[0.45cm]{$3)$}
&\makebox[0.45cm]{$4)$}
&\makebox[0.45cm]{$5)$}
&\makebox[0.45cm]{$6)$}
&\makebox[0.45cm]{$7)$}
&\makebox[0.45cm]{$8)$}
&\makebox[0.45cm]{$9)$}
&\makebox[0.45cm]{$10)$}
&\makebox[0.45cm]{$11)$}
&\makebox[0.45cm]{$12)$}
&\makebox[0.45cm]{$13)$}
&\makebox[0.45cm]{$14)$}
&\makebox[0.45cm]{$15)$}
&\makebox[0.45cm]{$16)$}
&\makebox[0.45cm]{$17)$}
&\makebox[0.45cm]{$18)$}\\
&\makebox[0.45cm]{\Rb}
&\makebox[0.45cm]{\Rc}
&\makebox[0.45cm]{$\Abb$}
&\makebox[0.45cm]{$\Acc$}
&\makebox[0.45cm]{$\Abb$}
&\makebox[0.45cm]{$\Acc$}
&\makebox[0.45cm]{$\Abb$}
&\makebox[0.45cm]{$\Acc$}
&\makebox[0.45cm]{\cAb}
&\makebox[0.45cm]{\cAc}
&\makebox[0.45cm]{$\BR$}
&\makebox[0.45cm]{$\BR$}
&\makebox[0.45cm]{$\BR$}
&\makebox[0.45cm]{\chiM}
&\makebox[0.45cm]{$\fDp$}
&\makebox[0.45cm]{$\fDs$}
&\makebox[0.45cm]{$f(c_{bar.})$}
&\makebox[0.45cm]{$P$}\\
&
&
&\makebox[0.45cm]{$(-2)$}
&\makebox[0.45cm]{$(-2)$}
&\makebox[0.45cm]{(pk)}
&\makebox[0.45cm]{(pk)}
&\makebox[0.45cm]{$(+2)$}
&\makebox[0.45cm]{$(+2)$}
&
&
&\makebox[0.45cm]{$(1)$}
&\makebox[0.45cm]{$(2)$}
&\makebox[0.45cm]{$(3)$}
&
&
&
&
&\\
\hline\hline
 1) & $ 1.00$ &         &         &         &         &         &         
&         &         &         &         &         &         &         
&         &         &         &         \\ 
 2) & $-0.18$ & $ 1.00$ &         &         &         &         &         
&         &         &         &         &         &         &         
&         &         &         &         \\ 
 3) & $-0.02$ & $ 0.01$ & $ 1.00$ &         &         &         &         
&         &         &         &         &         &         &         
&         &         &         &         \\ 
 4) & $ 0.00$ & $ 0.01$ & $ 0.13$ & $ 1.00$ &         &         &         
&         &         &         &         &         &         &         
&         &         &         &         \\ 
 5) & $-0.10$ & $ 0.03$ & $ 0.03$ & $ 0.01$ & $ 1.00$ &         &         
&         &         &         &         &         &         &         
&         &         &         &         \\ 
 6) & $ 0.07$ & $-0.06$ & $ 0.00$ & $ 0.02$ & $ 0.15$ & $ 1.00$ &         
&         &         &         &         &         &         &         
&         &         &         &         \\ 
 7) & $-0.04$ & $ 0.01$ & $ 0.01$ & $ 0.01$ & $ 0.08$ & $ 0.02$ & $ 1.00$ 
&         &         &         &         &         &         &         
&         &         &         &         \\ 
 8) & $ 0.03$ & $-0.04$ & $ 0.00$ & $ 0.01$ & $ 0.02$ & $ 0.15$ & $ 0.13$ 
& $ 1.00$ &         &         &         &         &         &         
&         &         &         &         \\ 
 9) & $-0.08$ & $ 0.04$ & $ 0.01$ & $ 0.00$ & $ 0.06$ & $-0.02$ & $ 0.02$ 
& $-0.01$ & $ 1.00$ &         &         &         &         &         
&         &         &         &         \\ 
10) & $ 0.04$ & $-0.06$ & $ 0.00$ & $ 0.00$ & $ 0.01$ & $ 0.04$ & $ 0.00$ 
& $ 0.02$ & $ 0.11$ & $ 1.00$ &         &         &         &         
&         &         &         &         \\ 
11) & $-0.08$ & $ 0.05$ & $ 0.00$ & $ 0.01$ & $ 0.00$ & $ 0.18$ & $ 0.00$ 
& $ 0.07$ & $-0.02$ & $ 0.02$ & $ 1.00$ &         &         &         
&         &         &         &         \\ 
12) & $-0.03$ & $-0.01$ & $ 0.00$ & $-0.02$ & $-0.05$ & $-0.23$ & $-0.03$ 
& $-0.08$ & $ 0.02$ & $-0.04$ & $-0.24$ & $ 1.00$ &         &         
&         &         &         &         \\ 
13) & $-0.01$ & $-0.29$ & $ 0.00$ & $ 0.02$ & $ 0.00$ & $-0.21$ & $ 0.00$ 
& $-0.14$ & $ 0.03$ & $-0.02$ & $ 0.00$ & $ 0.10$ & $ 1.00$ &         
&         &         &         &         \\ 
14) & $ 0.00$ & $ 0.02$ & $ 0.01$ & $ 0.02$ & $ 0.11$ & $ 0.08$ & $ 0.03$ 
& $ 0.02$ & $ 0.06$ & $ 0.00$ & $ 0.29$ & $-0.23$ & $ 0.16$ & $ 1.00$ 
&         &         &         &         \\ 
15) & $-0.15$ & $-0.10$ & $ 0.00$ & $ 0.00$ & $ 0.01$ & $-0.03$ & $ 0.01$ 
& $-0.02$ & $ 0.00$ & $ 0.00$ & $ 0.04$ & $ 0.02$ & $ 0.00$ & $ 0.02$ 
& $ 1.00$ &         &         &         \\ 
16) & $-0.03$ & $ 0.13$ & $ 0.00$ & $ 0.00$ & $ 0.00$ & $-0.02$ & $ 0.00$ 
& $-0.01$ & $ 0.00$ & $ 0.00$ & $ 0.01$ & $ 0.00$ & $-0.01$ & $-0.01$ 
& $-0.40$ & $ 1.00$ &         &         \\ 
17) & $ 0.11$ & $ 0.17$ & $ 0.00$ & $ 0.00$ & $-0.01$ & $ 0.04$ & $ 0.00$ 
& $ 0.02$ & $ 0.00$ & $ 0.00$ & $-0.02$ & $-0.01$ & $-0.02$ & $ 0.00$ 
& $-0.24$ & $-0.49$ & $ 1.00$ &         \\ 
18) & $ 0.13$ & $-0.43$ & $ 0.00$ & $ 0.00$ & $-0.02$ & $ 0.04$ & $-0.01$ 
& $ 0.02$ & $-0.02$ & $ 0.02$ & $-0.01$ & $ 0.01$ & $ 0.13$ & $ 0.00$ 
& $ 0.08$ & $-0.06$ & $-0.14$ & $ 1.00$ \\ 
\hline
\end{tabular}
\normalsize
\end{center}
\caption[The correlation matrix for the set of the 18 heavy flavour
  parameters.]{
  The correlation matrix for the set of the 18 heavy flavour
  parameters. $\BR(1)$, $\BR(2)$ and $\BR(3)$ 
  denote $\Brbl$, $\Brbclp$ and $\Brcl$
  respectively, $P$ denotes $\PcDst$.  }
\label{tab:18parcor}
\end{minipage}
\end{sideways}
\end{center}
\end{table}

\clearpage

\chapter{The Measurements used in the Heavy Flavour Averages}

\label{sec:hqappendix}

In Tables~\ref{tab:Rbinp} to~\ref{tab:RcPcDstinp} the results used in
the combination are listed.  In each case an indication of the dataset
used and the type of analysis is given.  The values of centre-of-mass
energy are given where relevant.  In each table, following the number
quoted in the referenced publication, the corrected value of each
measurement is given. For these values all external input parameters
as detailed in Section~\ref{sec:hqinputs} are used. In addition all
other fit parameters that affect the measurement in question via
explicit dependencies or correlations with simultaneous measurements
are fixed to the results of the LEP/SLD combination.  The corrected
value is followed by the statistical error, the internal systematic,
the systematic error common to more than one measurement, the effect
of a $\pm 1 \sigma$ change in all the other averaged parameters on the
value used in the average for this measurement, the total systematic
error, and the total error.

Contributions to the common systematic error quoted here are from
any physics source that is potentially common between the different
experiments. Detector systematics that are common between different analyses
of the same experiment are considered internal.
\begin{table}[p]
\begin{center}
\begin{tabular}{|l||c|c|c|c|c|}
 \hline
 & \mca{1}{ALEPH} & \mca{1}{DELPHI} & \mca{1}{L3} & \mca{1}{OPAL} & \mca{1}{SLD} \\ \hline 
 & 92-95&92-95&94-95&92-95&93-98 \\ 
 & \tmcite{ref:alife}&\tmcite{ref:drb}&\tmcite{ref:lrbmixed}&\tmcite{ref:omixed}&\tmcite{Abe:2005nq} \\ \hline \hline 
Published \Rbz{} & 0.2159&0.21634&0.2174&0.2178&0.2159 \\ \hline \hline
Corrected \Rbz{} & 0.2158&0.21643&0.2166&0.2176&0.2158 \\ \hline \hline
Statistical & 0.0009&0.00067&0.0013&0.0011&0.0009 \\ \hline 
Internal Systematic & 0.0007&0.00038&0.0014&0.0009&0.0005 \\ 
Common Systematic & 0.0006&0.00039&0.0018&0.0008&0.0005 \\  
Other Param. Sys. & 0.0001&0.00014&0.0010&0.0004&0.0002 \\ \hline 
Total Systematic & 0.0009&0.00056&0.0025&0.0012&0.0008 \\ \hline \hline
Total Error & 0.0013&0.00087&0.0028&0.0017&0.0012 \\ \hline 
\end{tabular}
\end{center}
\caption[The measurements of \Rbz{}.] {The measurements of \Rbz{}. 
  All measurements use a lifetime tag enhanced by other features like 
  invariant mass cuts or high $p_T$ leptons. 
 }
\label{tab:Rbinp}
\end{table}
\begin{table}[p]
\begin{center}
\begin{sideways}
\begin{minipage}[b]{\textheight}
\begin{center}
\begin{tabular}{|l||c|c|c|c|c|c|c|c|}
 \hline
 & \mca{3}{ALEPH} & \mca{2}{DELPHI} & \mca{2}{OPAL} & \mca{1}{SLD} \\ \hline 
 & 91-95&91-95&92-95&92-95&92-95&91-94&90-95&93-97\\ 
 & D-meson&c-count&lepton&c-count&D-meson&c-count&D-meson&D-meson \\  
 &   & (result) &  & (result) & (result) & (result) & (result) &   \\  
 & \tmcite{ref:arcd}&\tmcite{ref:arcc}&\tmcite{ref:arcd}&\tmcite{ref:drcc}&\tmcite{ref:drcd,ref:drcc}&\tmcite{ref:orcc}&\tmcite{ref:orcd}&\tmcite{Abe:2005nq} \\ \hline \hline 
Published \Rcz{} & 0.169&0.174&0.168&0.169&0.161&0.167&0.180&0.1744 \\ \hline \hline
Corrected \Rcz{} & 0.168&0.174&0.169&0.169&0.161&0.164&0.177&0.1741 \\ \hline \hline
Statistical & 0.008&0.005&0.006&0.005&0.010&0.012&0.010&0.0031 \\ \hline 
Internal Systematic & 0.008&0.006&0.004&0.005&0.006&0.013&0.010&0.0010 \\ 
Common Systematic & 0.003&0.009&0.004&0.008&0.006&0.010&0.006&0.0016 \\  
Other Param. Sys. & 0.000&0.000&0.005&0.000&0.001&0.000&0.000&0.0004 \\ \hline 
Total Systematic & 0.008&0.011&0.008&0.009&0.009&0.016&0.012&0.0020 \\ \hline \hline
Total Error & 0.012&0.012&0.010&0.010&0.014&0.020&0.015&0.0037 \\ \hline 
\end{tabular}
\end{center}
\caption[The measurements of \Rcz{}.] {The measurements of \Rcz{}. 
 ``c-count'' denotes the determination of \Rcz{} from the sum of production rates of weakly decaying charmed hadrons. 
 ``D-meson'' denotes any single/double tag 
 analysis using exclusive and/or inclusive  $D$ meson reconstruction. 
 The columns with the mention ``(result)'' are not directly used in the  
 global average, only the corresponding measurements 
 (\PcDst, \RcfDp, \RcfDs, \RcfLc, \RcfDz{} and \RcPcDst{} see tables
 \ref{tab:PcDstinp}-\ref{tab:RcPcDstinp}) are included.  }
\label{tab:Rcinp}
\end{minipage}
\end{sideways}
\end{center}
\end{table}
\begin{table}[p]
\begin{center}
\begin{sideways}
\begin{minipage}[b]{\textheight}
\begin{center}
\begin{tabular}{|l||c|c|c|c|c|c|c|c|c|c|c|}
 \hline
 & \mca{4}{ALEPH} & \mca{3}{DELPHI} & \mca{1}{L3} & \mca{3}{OPAL} \\ \hline 
 & 91-95 & 91-95 & 91-95&91-95&91-95&92-95&92-00&90-95&91-00&90-00&90-95 \\ 
 & lepton & lepton & lepton&jet&lepton&D-meson&multi&lepton&jet&lepton&D-meson \\  
 & \tmcite{\alasy} & \tmcite{\alasy} & \tmcite{\alasy}&\tmcite{ref:ajet}&\tmcite{\dlasy}&\tmcite{ref:ddasy}&\tmcite{ref:dnnasy}&\tmcite{\llasy}&\tmcite{ref:ojet}&\tmcite{ref:olasy}&\tmcite{ref:odsac} \\ \hline \hline 
\roots\ (\GeV) & 88.38 & 89.38 & 90.21&89.47&89.434&89.434&89.449&89.50&89.50&89.51&89.49 \\ \hline 
Published \Abl & -13.1 & 5.5 & -0.4&4.4&6.7&5.7&6.3&6.1&5.8&4.7&-9. \\ \hline \hline
Corrected \Abl & \mca{3}{5.2} &4.6&6.4&4.8&6.6&6.3&6.0&5.2&-5. \\ \hline \hline
Statistical & \mca{3}{1.8} &1.2&2.2&7.3&1.4&2.9&1.5&1.8&10. \\ \hline 
Internal Systematic & \mca{3}{0.1} &0.0&0.2&0.7&0.2&0.3&0.1&0.1&2. \\ 
Common Systematic & \mca{3}{0.1} &0.0&0.1&0.2&0.0&0.2&0.0&0.1&1. \\  
Other Param. Sys. & \mca{3}{0.1} &0.1&0.2&0.8&0.0&0.1&0.1&0.2&1. \\ \hline 
Total Systematic & \mca{3}{0.2} &0.1&0.3&1.0&0.2&0.4&0.1&0.2&2. \\ \hline \hline
Total Error & \mca{3}{1.8} &1.2&2.2&7.4&1.4&3.0&1.5&1.8&11. \\ \hline 
\end{tabular}
\end{center}
\caption[The measurements of \Abl.] {The measurements of \Abl. 
The ``Corrected'' values are quoted at $\sqrt{s} = 89.55 \, \GeV$. 
All numbers are given in \%. 
 }
\label{tab:Ablinp}
\end{minipage}
\end{sideways}
\end{center}
\end{table}
\begin{table}[p]
\begin{center}
\begin{sideways}
\begin{minipage}[b]{\textheight}
\begin{center}
\begin{tabular}{|l||c|c|c|c|c|c|c|c|}
 \hline
 & \mca{4}{ALEPH} & \mca{2}{DELPHI} & \mca{2}{OPAL} \\ \hline 
 & 91-95 & 91-95 & 91-95&91-95&91-95&92-95&90-00&90-95 \\ 
 & lepton & lepton & lepton&D-meson&lepton&D-meson&lepton&D-meson \\  
 & \tmcite{\alasy} & \tmcite{\alasy} & \tmcite{\alasy}&\tmcite{ref:adsac}&\tmcite{\dlasy}&\tmcite{ref:ddasy}&\tmcite{ref:olasy}&\tmcite{ref:odsac} \\ \hline \hline 
\roots\ (\GeV) & 88.38 & 89.38   & 90.21&89.37&89.434&89.434&89.51&89.49 \\ \hline 
Published \Acl & -12.4 & -2.3 & -0.3&-1.0&3.1&-5.0&-6.8&3.9 \\ \hline \hline
Corrected \Acl & \mca{3}{-1.5} &0.2&3.5&-4.4&-6.2&2.5 \\ \hline \hline
Statistical & \mca{3}{2.4} &4.3&3.4&3.6&2.5&4.9 \\ \hline 
Internal Systematic & \mca{3}{0.2} &0.9&0.4&0.3&0.9&0.8 \\ 
Common Systematic & \mca{3}{0.1} &0.1&0.1&0.1&0.1&0.3 \\  
Other Param. Sys. & \mca{3}{0.1} &0.2&0.2&0.1&0.3&0.1 \\ \hline 
Total Systematic & \mca{3}{0.2} &0.9&0.4&0.4&0.9&0.8 \\ \hline \hline
Total Error & \mca{3}{2.4} &4.4&3.5&3.6&2.7&5.0 \\ \hline 
\end{tabular}
\end{center}
\caption[The measurements of \Acl.] {The measurements of \Acl. 
The ``Corrected'' values are quoted at $\sqrt{s} = 89.55 \, \GeV$. 
All numbers are given in \%. 
 }
\label{tab:Aclinp}
\end{minipage}
\end{sideways}
\end{center}
\end{table}
\begin{table}[p]
\begin{center}
\begin{sideways}
\begin{minipage}[b]{\textheight}
\begin{center}
\begin{tabular}{|l||c|c|c|c|c|c|c|c|c|c|}
 \hline
 & \mca{2}{ALEPH} & \mca{3}{DELPHI} & \mca{2}{L3} & \mca{3}{OPAL} \\ \hline 
 & 91-95&91-95&91-95&92-95&92-00&91-95&90-95&91-00&90-00&90-95 \\ 
 & lepton&jet&lepton&D-meson&multi&jet&lepton&jet&lepton&D-meson \\  
 & \tmcite{\alasy}&\tmcite{ref:ajet}&\tmcite{\dlasy}&\tmcite{ref:ddasy}&\tmcite{ref:dnnasy}&\tmcite{ref:ljet}&\tmcite{\llasy}&\tmcite{ref:ojet}&\tmcite{ref:olasy}&\tmcite{ref:odsac} \\ \hline \hline 
\roots\ (\GeV) & 91.21&91.23&91.26&91.235&91.231&91.24&91.26&91.26&91.25&91.24 \\ \hline 
Published \Abp & 9.52&10.00&10.04&7.6&9.58&9.3&9.80&9.77&9.72&9.4 \\ \hline \hline
Corrected \Abp & 9.98&10.03&10.15&7.9&9.67&9.3&9.66&9.71&9.77&9.7 \\ \hline \hline
Statistical & 0.40&0.27&0.55&1.9&0.32&1.0&0.65&0.36&0.40&2.6 \\ \hline 
Internal Systematic & 0.07&0.10&0.17&0.5&0.15&0.5&0.27&0.15&0.07&2.1 \\ 
Common Systematic & 0.10&0.02&0.16&0.6&0.04&0.2&0.16&0.08&0.13&0.3 \\  
Other Param. Sys. & 0.12&0.05&0.10&0.2&0.03&0.1&0.12&0.05&0.10&0.2 \\ \hline 
Total Systematic & 0.17&0.12&0.25&0.8&0.15&0.6&0.33&0.18&0.18&2.1 \\ \hline \hline
Total Error & 0.44&0.29&0.60&2.1&0.35&1.2&0.73&0.40&0.44&3.4 \\ \hline 
\end{tabular}
\end{center}
\caption[The measurements of \Abp.] {The measurements of \Abp. 
The ``Corrected'' values are quoted at $\sqrt{s} = 91.26 \, \GeV$. 
All numbers are given in \%. 
 }
\label{tab:Abpinp}
\end{minipage}
\end{sideways}
\end{center}
\end{table}
\begin{table}[p]
\begin{center}
\begin{tabular}{|l||c|c|c|c|c|c|c|}
 \hline
 & \mca{2}{ALEPH} & \mca{2}{DELPHI} & \mca{1}{L3} & \mca{2}{OPAL} \\ \hline 
 & 91-95&91-95&91-95&92-95&90-95&90-00&90-95 \\ 
 & lepton&D-meson&lepton&D-meson&lepton&lepton&D-meson \\  
 & \tmcite{\alasy}&\tmcite{ref:adsac}&\tmcite{\dlasy}&\tmcite{ref:ddasy}&\tmcite{\llasy}&\tmcite{ref:olasy}&\tmcite{ref:odsac} \\ \hline \hline 
\roots\ (\GeV) & 91.21&91.22&91.26&91.235&91.24&91.25&91.24 \\ \hline 
Published \Acp & 6.45&6.3&6.3&6.59&7.8&5.68&6.3 \\ \hline \hline
Corrected \Acp & 6.62&6.3&6.2&6.49&8.2&5.70&6.5 \\ \hline \hline
Statistical & 0.56&0.9&0.9&0.93&3.0&0.54&1.2 \\ \hline 
Internal Systematic & 0.24&0.2&0.5&0.26&1.7&0.19&0.5 \\ 
Common Systematic & 0.22&0.2&0.2&0.07&0.6&0.22&0.3 \\  
Other Param. Sys. & 0.20&0.0&0.2&0.03&0.7&0.20&0.0 \\ \hline 
Total Systematic & 0.38&0.3&0.6&0.27&1.9&0.36&0.6 \\ \hline \hline
Total Error & 0.68&0.9&1.1&0.97&3.6&0.65&1.3 \\ \hline 
\end{tabular}
\end{center}
\caption[The measurements of \Acp.] {The measurements of \Acp. 
The ``Corrected'' values are quoted at $\sqrt{s} = 91.26 \, \GeV$. 
All numbers are given in \%. 
 }
\label{tab:Acpinp}
\end{table}
\begin{table}[p]
\begin{center}
\begin{sideways}
\begin{minipage}[b]{\textheight}
\begin{center}
\begin{tabular}{|l||c|c|c|c|c|c|c|c|c|c|c|}
 \hline
 & \mca{4}{ALEPH} & \mca{3}{DELPHI} & \mca{1}{L3} & \mca{3}{OPAL} \\ \hline 
 & 91-95 & 91-95 & 91-95&91-95&91-95&92-95&92-00&90-95&91-00&90-00&90-95 \\ 
 & lepton & lepton & lepton&jet&lepton&D-meson&multi&lepton&jet&lepton&D-meson \\  
 & \tmcite{\alasy} & \tmcite{\alasy} & \tmcite{\alasy}&\tmcite{ref:ajet}&\tmcite{\dlasy}&\tmcite{ref:ddasy}&\tmcite{ref:dnnasy}&\tmcite{\llasy}&\tmcite{ref:ojet}&\tmcite{ref:olasy}&\tmcite{ref:odsac} \\ \hline \hline 
\roots\ (\GeV) & 92.05 & 92.94 & 93.90&92.95&92.990&92.990&92.990&93.10&92.91&92.95&92.95 \\ \hline 
Published \Abh & 11.1 & 10.4 & 13.8&11.72&11.2&8.8&10.4&13.7&12.2&10.3&-2.1 \\ \hline \hline
Corrected \Abh & \mca{3}{11.1} &11.69&11.4&8.6&10.4&13.7&12.2&10.1&-0.2 \\ \hline \hline
Statistical & \mca{3}{1.4} &0.98&1.8&6.2&1.2&2.4&1.2&1.5&8.7 \\ \hline 
Internal Systematic & \mca{3}{0.2} &0.11&0.1&0.9&0.3&0.3&0.2&0.1&2.0 \\ 
Common Systematic & \mca{3}{0.2} &0.02&0.1&0.5&0.0&0.2&0.1&0.2&1.2 \\  
Other Param. Sys. & \mca{3}{0.2} &0.12&0.2&0.5&0.0&0.1&0.1&0.2&0.7 \\ \hline 
Total Systematic & \mca{3}{0.3} &0.16&0.3&1.1&0.3&0.4&0.2&0.3&2.4 \\ \hline \hline
Total Error & \mca{3}{1.5} &0.99&1.8&6.3&1.2&2.4&1.3&1.5&9.0 \\ \hline 
\end{tabular}
\end{center}
\caption[The measurements of \Abh.] {The measurements of \Abh. 
The ``Corrected'' values are quoted at $\sqrt{s} = 92.94 \, \GeV$. 
All numbers are given in \%. 
 }
\label{tab:Abhinp}
\end{minipage}
\end{sideways}
\end{center}
\end{table}
\begin{table}[p]
\begin{center}
\begin{sideways}
\begin{minipage}[b]{\textheight}
\begin{center}
\begin{tabular}{|l||c|c|c|c|c|c|c|c|}
 \hline
 & \mca{4}{ALEPH} & \mca{2}{DELPHI} & \mca{2}{OPAL} \\ \hline 
 & 91-95 & 91-95 & 91-95&91-95&91-95&92-95&90-00&90-95 \\ 
 & lepton & lepton & lepton&D-meson&lepton&D-meson&lepton&D-meson \\  
 & \tmcite{\alasy} & \tmcite{\alasy} & \tmcite{\alasy}&\tmcite{ref:adsac}&\tmcite{\dlasy}&\tmcite{ref:ddasy}&\tmcite{ref:olasy}&\tmcite{ref:odsac} \\ \hline \hline 
\roots\ (\GeV) & 92.05 & 92.94 & 93.90  &92.96&92.990&92.990&92.95&92.95 \\ \hline 
Published \Ach & 10.6 & 11.9 & 12.1&11.0&11.0&11.8&14.6&15.8 \\ \hline \hline
Corrected \Ach & \mca{3}{11.9} &10.9&10.9&11.4&14.9&14.6 \\ \hline \hline
Statistical & \mca{3}{2.0} &3.3&2.8&3.1&2.0&4.0 \\ \hline 
Internal Systematic & \mca{3}{0.3} &0.7&0.4&0.5&0.5&0.7 \\ 
Common Systematic & \mca{3}{0.3} &0.1&0.2&0.1&0.2&0.5 \\  
Other Param. Sys. & \mca{3}{0.3} &0.2&0.3&0.1&0.4&0.1 \\ \hline 
Total Systematic & \mca{3}{0.6} &0.7&0.6&0.6&0.7&0.9 \\ \hline \hline
Total Error & \mca{3}{2.1} &3.4&2.8&3.1&2.1&4.1 \\ \hline 
\end{tabular}
\end{center}
\caption[The measurements of \Ach.] {The measurements of \Ach. 
The ``Corrected'' values are quoted at $\sqrt{s} = 92.94 \, \GeV$. 
All numbers are given in \%. 
 }
\label{tab:Achinp}
\end{minipage}
\end{sideways}
\end{center}
\end{table}
\begin{table}[p]
\begin{center}
\begin{tabular}{|l||c|c|c|c|}
 \hline
 & \mca{4}{SLD} \\ \hline 
 & 93-98&93-98&94-95&96-98 \\ 
 & lepton&jet&$K^{\pm}$&$K$+vertex \\  
 & \tmcite{\SLDacbl}&\tmcite{\SLDabj}&\tmcite{\SLDabk}&\tmcite{ref:SLD_vtxasy} \\ \hline \hline 
\roots\ (\GeV) & 91.28&91.28&91.28&91.28 \\ \hline 
Published \cAb & 0.919&0.907&0.86&0.919 \\ \hline \hline
Corrected \cAb & 0.939&0.907&0.86&0.917 \\ \hline \hline
Statistical & 0.030&0.020&0.09&0.018 \\ \hline 
Internal Systematic & 0.018&0.023&0.10&0.017 \\ 
Common Systematic & 0.009&0.003&0.01&0.003 \\  
Other Param. Sys. & 0.011&0.001&0.00&0.002 \\ \hline 
Total Systematic & 0.023&0.024&0.10&0.017 \\ \hline \hline
Total Error & 0.037&0.031&0.13&0.025 \\ \hline 
\end{tabular}
\end{center}
\caption{The measurements of \cAb. 
 }
\label{tab:cAbinp}
\end{table}
\begin{table}[p]
\begin{center}
\begin{tabular}{|l||c|c|c|}
 \hline
 & \mca{3}{SLD} \\ \hline 
 & 93-98&93-98&96-98 \\ 
 & lepton&D-meson&$K$+vertex \\  
 & \tmcite{\SLDacbl}&\tmcite{ref:SLD_ACD}&\tmcite{ref:SLD_vtxasy} \\ \hline \hline 
\roots\ (\GeV) & 91.28&91.28&91.28 \\ \hline 
Published \cAc & 0.583&0.688&0.673 \\ \hline \hline
Corrected \cAc & 0.587&0.689&0.674 \\ \hline \hline
Statistical & 0.055&0.035&0.029 \\ \hline 
Internal Systematic & 0.045&0.020&0.023 \\ 
Common Systematic & 0.022&0.004&0.002 \\  
Other Param. Sys. & 0.017&0.001&0.002 \\ \hline 
Total Systematic & 0.053&0.021&0.023 \\ \hline \hline
Total Error & 0.076&0.041&0.037 \\ \hline 
\end{tabular}
\end{center}
\caption{The measurements of \cAc. 
 }
\label{tab:cAcinp}
\end{table}
\clearpage
\begin{table}[p]
\begin{center}
\begin{tabular}{|l||c|c|c|c|c|c|}
 \hline
 & \mca{1}{ALEPH} & \mca{1}{DELPHI} & \mca{2}{L3} & \mca{2}{OPAL} \\ \hline 
 & 91-95&94-95&94-95&92&92-95&92-95 \\ 
 & multi&multi&multi&multi&multi&multi \\  
 & \tmcite{ref:abl}&\tmcite{ref:dbl}&\tmcite{ref:lrbmixed}&\tmcite{ref:lbl}&\tmcite{ref:obl}&\tmcite{ref:obl} \\ \hline \hline 
Published \Brbl & 10.70&10.70&10.16&10.68&10.78&10.96 \\ \hline \hline
Corrected \Brbl & 10.74&10.70&10.26&10.82&\mca{2}{10.86}  \\ \hline \hline
Statistical & 0.10&0.14&0.09&0.11&\mca{2}{0.09}  \\ \hline 
Internal Systematic & 0.15&0.14&0.16&0.36&\mca{2}{0.21}  \\ 
Common Systematic & 0.23&0.43&0.31&0.22&\mca{2}{0.19}  \\  
Other Param. Sys. & 0.03&0.07&0.03&0.09&\mca{2}{0.02}  \\ \hline 
Total Systematic & 0.28&0.45&0.35&0.43&\mca{2}{0.29}  \\ \hline \hline
Total Error & 0.29&0.48&0.36&0.45&\mca{2}{0.30}  \\ \hline 
\end{tabular}
\end{center}
\caption[The measurements of \Brbl.] {The measurements of \Brbl. 
 All numbers are given in \%. 
 }
\label{tab:Brblinp}
\end{table}
\begin{table}[p]
\begin{center}
\begin{tabular}{|l||c|c|c|c|}
 \hline
 & \mca{1}{ALEPH} & \mca{1}{DELPHI} & \mca{2}{OPAL} \\ \hline 
 & 91-95&94-95&92-95&92-95 \\ 
 & multi&multi&multi&multi \\  
 & \tmcite{ref:abl}&\tmcite{ref:dbl}&\tmcite{ref:obl}&\tmcite{ref:obl} \\ \hline \hline 
Published \Brbclp & 8.18&7.98&8.37&8.17 \\ \hline \hline
Corrected \Brbclp & 8.11&7.98&\mca{2}{8.42}  \\ \hline \hline
Statistical & 0.15&0.22&\mca{2}{0.15}  \\ \hline 
Internal Systematic & 0.18&0.16&\mca{2}{0.22}  \\ 
Common Systematic & 0.15&0.22&\mca{2}{0.32}  \\  
Other Param. Sys. & 0.05&0.04&\mca{2}{0.04}  \\ \hline 
Total Systematic & 0.24&0.27&\mca{2}{0.39}  \\ \hline \hline
Total Error & 0.29&0.35&\mca{2}{0.42}  \\ \hline 
\end{tabular}
\end{center}
\caption[The measurements of \Brbclp.] {The measurements of \Brbclp. 
 All numbers are given in \%. 
 }
\label{tab:Brbclpinp}
\end{table}
\begin{table}[p]
\begin{center}
\begin{tabular}{|l||c|c|}
 \hline
 & \mca{1}{DELPHI} & \mca{1}{OPAL} \\ \hline 
 & 92-95&90-95 \\ 
 & $D$+lepton&$D$+lepton \\  
 & \tmcite{ref:drcd}&\tmcite{ref:ocl} \\ \hline \hline 
Published \Brcl & 9.58&9.5 \\ \hline \hline
Corrected \Brcl & 9.67&9.6 \\ \hline \hline
Statistical & 0.42&0.6 \\ \hline 
Internal Systematic & 0.24&0.5 \\ 
Common Systematic & 0.13&0.4 \\  
Other Param. Sys. & 0.01&0.0 \\ \hline 
Total Systematic & 0.27&0.7 \\ \hline \hline
Total Error & 0.50&0.9 \\ \hline 
\end{tabular}
\end{center}
\caption[The measurements of \Brcl.] {The measurements of \Brcl. 
 All numbers are given in \%. 
 }
\label{tab:Brclinp}
\end{table}
\begin{table}[p]
\begin{center}
\begin{tabular}{|l||c|c|c|c|}
 \hline
 & \mca{1}{ALEPH} & \mca{1}{DELPHI} & \mca{1}{L3} & \mca{1}{OPAL} \\ \hline 
 & 91-95&94-95&90-95&90-00 \\ 
 & multi&multi&lepton&lepton \\  
 & \tmcite{\alasy}&\tmcite{ref:dbl}&\tmcite{\llasy}&\tmcite{ref:olasy} \\ \hline \hline 
Published \chiM & 0.1196&0.127&0.1192&0.1312 \\ \hline \hline
Corrected \chiM & 0.1199&0.127&0.1199&0.1318 \\ \hline \hline
Statistical & 0.0049&0.013&0.0066&0.0046 \\ \hline 
Internal Systematic & 0.0021&0.005&0.0023&0.0015 \\ 
Common Systematic & 0.0040&0.003&0.0026&0.0037 \\  
Other Param. Sys. & 0.0012&0.001&0.0016&0.0016 \\ \hline 
Total Systematic & 0.0047&0.006&0.0038&0.0043 \\ \hline \hline
Total Error & 0.0068&0.014&0.0076&0.0063 \\ \hline 
\end{tabular}
\end{center}
\caption{The measurements of \chiM. 
 }
\label{tab:chiMinp}
\end{table}
\begin{table}[p]
\begin{center}
\begin{tabular}{|l||c|c|}
 \hline
 & \mca{1}{DELPHI} & \mca{1}{OPAL} \\ \hline 
 & 92-95&90-95 \\ 
 & D-meson&D-meson \\  
 & \tmcite{ref:drcd}&\tmcite{ref:orcd} \\ \hline \hline 
Published \PcDst & 0.174&0.1516 \\ \hline \hline
Corrected \PcDst & 0.174&0.1546 \\ \hline \hline
Statistical & 0.010&0.0038 \\ \hline 
Internal Systematic & 0.004&0.0045 \\ 
Common Systematic & 0.001&0.0050 \\  
Other Param. Sys. & 0.000&0.0021 \\ \hline 
Total Systematic & 0.004&0.0070 \\ \hline \hline
Total Error & 0.011&0.0080 \\ \hline 
\end{tabular}
\end{center}
\caption{The measurements of \PcDst. 
 }
\label{tab:PcDstinp}
\end{table}
\begin{table}[p]
\begin{center}
\begin{tabular}{|l||c|c|c|}
 \hline
 & \mca{1}{ALEPH} & \mca{1}{DELPHI} & \mca{1}{OPAL} \\ \hline 
 & 91-95&92-95&91-94 \\ 
 & c-count&c-count&c-count \\  
 & \tmcite{ref:arcc}&\tmcite{ref:drcc}&\tmcite{ref:orcc} \\ \hline \hline 
Published \RcfDp & 0.0409&0.0384&0.0393 \\ \hline \hline
Corrected \RcfDp & 0.0402&0.0386&0.0386 \\ \hline \hline
Statistical & 0.0014&0.0014&0.0056 \\ \hline 
Internal Systematic & 0.0012&0.0012&0.0026 \\ 
Common Systematic & 0.0029&0.0025&0.0028 \\  
Other Param. Sys. & 0.0012&0.0008&0.0015 \\ \hline 
Total Systematic & 0.0033&0.0029&0.0041 \\ \hline \hline
Total Error & 0.0036&0.0032&0.0069 \\ \hline 
\end{tabular}
\end{center}
\caption{The measurements of \RcfDp. 
 }
\label{tab:RcfDpinp}
\end{table}
\begin{table}[p]
\begin{center}
\begin{tabular}{|l||c|c|c|}
 \hline
 & \mca{1}{ALEPH} & \mca{1}{DELPHI} & \mca{1}{OPAL} \\ \hline 
 & 91-95&92-95&91-94 \\ 
 & c-count&c-count&c-count \\  
 & \tmcite{ref:arcc}&\tmcite{ref:drcc}&\tmcite{ref:orcc} \\ \hline \hline 
Published \RcfDs & 0.0199&0.0213&0.0161 \\ \hline \hline
Corrected \RcfDs & 0.0206&0.0213&0.0158 \\ \hline \hline
Statistical & 0.0036&0.0018&0.0048 \\ \hline 
Internal Systematic & 0.0011&0.0009&0.0007 \\ 
Common Systematic & 0.0047&0.0048&0.0037 \\  
Other Param. Sys. & 0.0003&0.0004&0.0006 \\ \hline 
Total Systematic & 0.0048&0.0049&0.0038 \\ \hline \hline
Total Error & 0.0060&0.0052&0.0061 \\ \hline 
\end{tabular}
\end{center}
\caption{The measurements of \RcfDs. 
 }
\label{tab:RcfDsinp}
\end{table}
\begin{table}[p]
\begin{center}
\begin{tabular}{|l||c|c|c|}
 \hline
 & \mca{1}{ALEPH} & \mca{1}{DELPHI} & \mca{1}{OPAL} \\ \hline 
 & 91-95&92-95&91-94 \\ 
 & c-count&c-count&c-count \\  
 & \tmcite{ref:arcc}&\tmcite{ref:drcc}&\tmcite{ref:orcc} \\ \hline \hline 
Published \RcfBar & 0.0169&0.0170&0.0107 \\ \hline \hline
Corrected \RcfBar & 0.0155&0.0170&0.0089 \\ \hline \hline
Statistical & 0.0017&0.0040&0.0065 \\ \hline 
Internal Systematic & 0.0005&0.0014&0.0008 \\ 
Common Systematic & 0.0038&0.0040&0.0028 \\  
Other Param. Sys. & 0.0004&0.0004&0.0005 \\ \hline 
Total Systematic & 0.0039&0.0043&0.0030 \\ \hline \hline
Total Error & 0.0042&0.0058&0.0072 \\ \hline 
\end{tabular}
\end{center}
\caption{The measurements of \RcfBar. 
 }
\label{tab:RcfLcinp}
\end{table}
\begin{table}[p]
\begin{center}
\begin{tabular}{|l||c|c|c|}
 \hline
 & \mca{1}{ALEPH} & \mca{1}{DELPHI} & \mca{1}{OPAL} \\ \hline 
 & 91-95&92-95&91-94 \\ 
 & c-count&c-count&c-count \\  
 & \tmcite{ref:arcc}&\tmcite{ref:drcc}&\tmcite{ref:orcc} \\ \hline \hline 
Published \RcfDz & 0.0961&0.0927&0.1013 \\ \hline \hline
Corrected \RcfDz & 0.0966&0.0929&0.1027 \\ \hline \hline
Statistical & 0.0031&0.0027&0.0080 \\ \hline 
Internal Systematic & 0.0036&0.0026&0.0033 \\ 
Common Systematic & 0.0042&0.0024&0.0038 \\  
Other Param. Sys. & 0.0018&0.0019&0.0016 \\ \hline 
Total Systematic & 0.0058&0.0040&0.0053 \\ \hline \hline
Total Error & 0.0066&0.0048&0.0095 \\ \hline 
\end{tabular}
\end{center}
\caption{The measurements of \RcfDz. 
 }
\label{tab:RcfDzinp}
\end{table}
\begin{table}[p]
\begin{center}
\begin{tabular}{|l||c|c|}
 \hline
 & \mca{1}{DELPHI} & \mca{1}{OPAL} \\ \hline 
 & 92-95&90-95 \\ 
 &  D-meson &D-meson \\  
 & \tmcite{ref:drcc}&\tmcite{ref:orcd} \\ \hline \hline 
Published \RcPcDst & 0.0283&0.0272 \\ \hline \hline
Corrected \RcPcDst & 0.0284&0.0271 \\ \hline \hline
Statistical & 0.0007&0.0005 \\ \hline 
Internal Systematic & 0.0008&0.0008 \\ 
Common Systematic & 0.0006&0.0010 \\  
Other Param. Sys. & 0.0009&0.0001 \\ \hline 
Total Systematic & 0.0013&0.0013 \\ \hline \hline
Total Error & 0.0015&0.0014 \\ \hline 
\end{tabular}
\end{center}
\caption{The measurements of \RcPcDst. 
 }
\label{tab:RcPcDstinp}
\end{table}

\clearpage

\chapter{Limits on Non-Standard Z Decays}
\label{sec:NP-Z-decays}

Numerical limits on possible contributions to Z final states from
sources beyond the Standard Model ($\SM$) are obtained by taking the
difference between the widths in Table~\ref{tab:width} or the
branching fractions of Table~\ref{tab:brfrac}, and the corresponding
$\SM$ predictions, as is shown in Table~\ref{tab:xsm}.

Decays of Z-Bosons into non-$\SM$ particles with observable final
states identical to the $\SM$ ones would result in different selection
efficiencies, and therefore these limits must be treated with
care. Extra contributions to the total width or to the invisible
width, however, are safe in this respect.

In order to calculate the upper limit for such contributions,
parametric errors on the $\SM$ prediction are added in quadrature to
the experimental errors. The unknown value of the Higgs boson mass is
taken into account by choosing its value within the range of
114~\GeV~\cite{LEPSMHIGGS} to 1000~\GeV\ such that the $\SM$
prediction is minimal, \ie, either $\MH=114~\GeV$ for the leptonic
branching fractions or $\MH=1000~\GeV$ for all other quantities listed
in the first column of Table~\ref{tab:xsm}.  The values assumed for
$\Mt$ and $\alqed$ are those of Table~\ref{tab:msm:input}, while for
$\alfmz$ a value with an enlarged error of $0.118\pm0.003$ is
chosen. The best description of the Z-pole results is obtained by
using the value derived from the full $\SM$ fit of
Table~\ref{tab:msmfit-all}, but this value of $\alfmz$ would be
affected by contributions from physics beyond the $\SM$ to hadronic Z
decays and can therefore not be used here.  The enlarged error takes
into account the uncertainties involved when applying an external
value of $\alfmz$ to hadronic Z decays, as is discussed in
Section~\ref{sec:msm:QCD-TU}.  Clearly the derived limits on Z decays
involving hadrons depend on the choice of $\alfmz$, while limits on
extra contributions to the leptonic widths are almost insensitive to
it.  All branching fractions, however, depend on the choice of
$\alfmz$, due to the their strong correlations arising from
constraining their sum to be equal to unity.

The 95\%~CL upper limits on extra, non-$\SM$ contributions to the Z
widths and branching fractions derived in the way described above are
summarised in Table~\ref{tab:xsm}; these limits are of Bayesian type
assuming zero probability below the minimal $\SM$ prediction and a
uniform prior probability above.

\begin{table}[htb] \begin{center}
\renewcommand{\arraystretch}{1.15}
\begin{tabular} {|l||r|r||r|r|}
\hline %
Z-decay to:
  &  ~~$\Delta\Gamma_x$~[\MeV] & min. $\Gamma_{\SM}$~[\MeV] &
     ~~$\Delta$B$_x~[\%]$~~         & min. B$_{\SM}~[\%]$ \\
\hline %
\hline %
 \multicolumn{5}{|c|}{Without Lepton Universality} \\
\hline %
$         \ff$     &  11.4    &  2488.7$\pz\pm$1.9$\pz$    &
                       --- $\pzz$    &     --- $\pzz$               \\
$         \qq$   &    14.6    & 1736.6$\pz\pm$1.8$\pz$ &
                    0.35$\pz$ & 69.777$\pz\pm$0.021$\pz$ \\
$         \ee$      & 0.32    & 83.82$\pm$0.04&
                      0.0075 & 3.3664$\pm$0.0023 \\
$         \mumu$    & 0.49    & 83.82$\pm$0.04 &
                      0.014 & 3.3664$\pm$0.0023 \\ 
$         \tautau$  & 0.82    & 83.63$\pm$0.04 &
                      0.025 & 3.3588$\pm$0.0023 \\
$         \bb$      & 5.3     & 374.6$\pz\pm$0.4$\pz$ &
                      0.17$\pz$ &   15.051$\pz\pm$0.012$\pz$ \\
$         \cc$      & 11.4    &  299.1$\pz\pm$0.4$\pz$ &
                       0.43$\pz$ &  12.017$\pz\pm$0.008$\pz$ \\
$         \inv$     &  3.1    &  500.9$\pz\pm$0.2$\pz$   &
                      0.11$\pz$  & 20.104$\pz\pm$0.014$\pz$\\
\hline %
\hline %
    \multicolumn{5}{|c|}{With Lepton Universality} \\
\hline %
$         \ff$                 & 11.4  &  2488.7$\pz\pm$1.9$\pz$  &    
                           --- $\pzz$  &    --- $\pzz$        \\
$         \qq$                &  12.2   & 1736.6$\pz\pm$1.8$\pz$ &
                                0.23$\pz$ & 69.777$\pz\pm$0.021$\pz$ \\
$         \ee+\mumu+\tautau$  & 0.97      & 251.27$\pm$0.12&
                                0.018     & 10.0916$\pm$0.0069 \\
$         \bb$                &  4.8        & 374.6$\pz\pm$0.4$\pz$ &
                                 0.15$\pz$ & 15.051$\pz\pm$0.012$\pz$  \\
$         \cc$                &  11.0      & 299.1$\pz\pm$0.4$\pz$  &
                                 0.42$\pz$ & 12.017$\pz\pm$0.008$\pz$  \\
$         \inv$               &  2.0 & 500.9$\pz\pm$0.2$\pz$ &
                                0.062 & 20.104$\pz\pm$0.014$\pz$ \\
\hline %
\end{tabular} 
\caption[Limits on non-$\SM$ widths and branching fractions]
{\label{tab:xsm} 95\%~CL limits on non-$\SM$ contributions to the $\Zzero$
  widths ($\Delta\Gamma_x$, second column) and branching fractions 
  ($\Delta$B$_x$, fourth column) derived from the results of 
  Tables~\ref{tab:width} and~\ref{tab:brfrac}. 
  The minimal $\SM$ predictions for the widths and
  branching fractions with their parametric uncertainties 
  arising from the errors in $\Mt$, $\alfmz$ and $\alqed$
  are shown in the third and fifth columns, respectively.
  Note that there are correlations among the experimental and theoretical
  errors, and therefore the limits must not be used simultaneously.}
\end{center} 
\end{table}

\clearpage

\chapter{Tests of Electroweak Radiative Corrections}
\label{sec:msm:eps-stu}

\section{Parametrisations}

As discussed in Section~\ref{sec:intro_ew}, the expected structure of
electroweak radiative corrections in the Standard Model ($\SM$) shows
contributions quadratic in the fermion masses and only logarithmic in
the Higgs-boson mass.  It has been studied how the small Higgs-mass
dependence can be disentangled from the large top-quark mass
dependence. For this purpose, four new effective parameters,
$\epsilon_1$, $\epsilon_2$, $\epsilon_3$ and $\epsilon_b$ are
introduced~\cite{\epsilonpar}.  They are defined such that they vanish
in the approximation when only effects due to pure QED and QCD are
taken into account. In terms of the auxiliary quantities $\stzsq$,
defined in Equation~\ref{eq:sm:s20}, and $\Delta\kappa'$, relating
$\swsqeffl$ to $\stzsq$ analogously to Equation~\ref{eq:sin}:
\begin{eqnarray}
\swsqeffl & = & (1+\Delta\kappa')\stzsq \label{eq:sm:eps-0}\,,
\end{eqnarray}
the $\epsilon$ parameters are given by:
\begin{eqnarray}
\epsilon_1 & = & \Delta\rho 
\label{eq:mi:eps-1}\\
\epsilon_2 & = & 
\ctzsq\Delta\rho + \frac{\stzsq}{\ctzsq-\stzsq}\Drw-2\stzsq\Delta\kappa' 
\label{eq:mi:eps-2}\\
\epsilon_3 & = & 
\ctzsq\Delta\rho + (\ctzsq-\stzsq)\Delta\kappa'  
\label{eq:mi:eps-3}\\
\epsilon_b & = & \frac{1}{2}\Delta\rho_b 
\label{eq:mi:eps-b}\,.
\end{eqnarray}
Within the $\SM$ the leading contributions in terms of $\Mt$ and $\MH$
are:
\begin{eqnarray}
\epsilon_1 & = & 
\frac{3\GF\Mt^2}{8\sqrt{2}\pi^2}-
\frac{3\GF\MW^2}{4\sqrt{2}\pi^2}\twsq\ln\frac{\MH}{\MZ} 
+\ldots \label{eq:sm:eps-1}\\
\epsilon_2 & = & -\frac{\GF\MW^2}{2\sqrt{2}\pi^2}\ln\frac{\Mt}{\MZ} 
+\ldots \label{eq:sm:eps-2}\\
\epsilon_3 & = & 
\frac{\GF\MW^2}{12\sqrt{2}\pi^2}\ln\frac{\MH}{\MZ}-
\frac{\GF\MW^2}{ 6\sqrt{2}\pi^2}\ln\frac{\Mt}{\MZ} 
+\ldots \label{eq:sm:eps-3}\\
\epsilon_b & = & 
-\frac{\GF\Mt^2}{4\sqrt{2}\pi^2} 
+\ldots \label{eq:sm:eps-b}\,.
\end{eqnarray}
Note that comparing to the equations given in
Section~\ref{sec:intro_ew}, the argument of the natural logarithm is
$\MH/\MZ$ rather than $\MH/\MW$.  The difference is of subleading
order.

The $\epsilon$ parameters separate electroweak radiative corrections
in quadratic $\Mt$ effects and logarithmic $\MH$ effects.  Such a
rearrangement is also useful in the search for new physics effects in
precision measurements.  Another commonly used description is based on
the so-called $STU$ parameters~\cite{\STUpar}, extended by an
additional parameter, $\gamma_b$, for the b-quark sector~\cite{PDG98}.
Approximate linear relations between these two sets of parameters
exist:
\begin{eqnarray}
S        & \simeq &           +  \epsilon_3\frac{4\stzsq}{\alqed} - c_S 
\label{eq:mi:stu-s} \\
T        & \simeq & \phantom{\pm}\epsilon_1\frac{1      }{\alqed} - c_T 
\label{eq:mi:stu-t} \\
U        & \simeq &           -  \epsilon_2\frac{4\stzsq}{\alqed} - c_U
\label{eq:mi:stu-u} \\
\gamma_b & \simeq &        2.29  \epsilon_b                  - c_\gamma 
\label{eq:mi:stu-b} \,.
\end{eqnarray}
In the literature, these parameters are in fact defined as shifts
relative to a fixed set of $\SM$ values $c_i$, $i=S,T,U,\gamma$, so
that $S=T=U=\gamma_b=0$ at that point.  Thus these parameters measure
deviations from the electroweak radiative corrections expected in the
$\SM$, in particular new physics effects in oblique electroweak
corrections, \ie, those entering through vacuum polarisation diagrams.
For numerical results presented in the following, we use as the fixed
subtraction point the values corresponding to: $\dalhad=0.02758$,
$\alfmz=0.118$, $\MZ=91.1875~\GeV$, $\Mt=175~\GeV$, $\MH=150~\GeV$.
Predictions of these parameters within the $\SM$ framework are
reported in Appendix~\ref{app:SM:preds}.

\section{Results}

The formulae listed above and in Chapter~\ref{sec:intro} are combined
to express the measured quantities in terms of the $\epsilon$ or $STU$
parameters, and the latter are then determined as usual in a
$\chi^2$-fit to the measurements.  In both analyses, the largest
contribution to the $\chi^2$ arises from the asymmetry measurements as
discussed in Section~\ref{sec:coup:rho-sef2}.  Note that the
experimental results on light quark flavours presented in
Appendix~\ref{sec:lqappendix} are not used.

The Z-pole measurements performed by SLD and at \LEPI\ constrain the
parameters $\epsilon_1~(T)$, $\epsilon_3~(S)$ and
$\epsilon_b~(\gamma_b)$.  Given these, the measurements of the W-boson
mass or of the on-shell electroweak mixing angle are solely
determining $\epsilon_2~(U)$.  The other additional measurements
discussed in Section~\ref{sec:msm:add} are not included here as they
can be expressed in terms of neither the $\epsilon$ nor the $STU$
parameters without additional assumptions.  Because of its explicit
$\Mt$ and $\MH$ dependence, the measurement of $\swsq$ by NuTeV cannot
be included.

\begin{figure}[p]
\begin{center}
\mbox{\epsfig{file=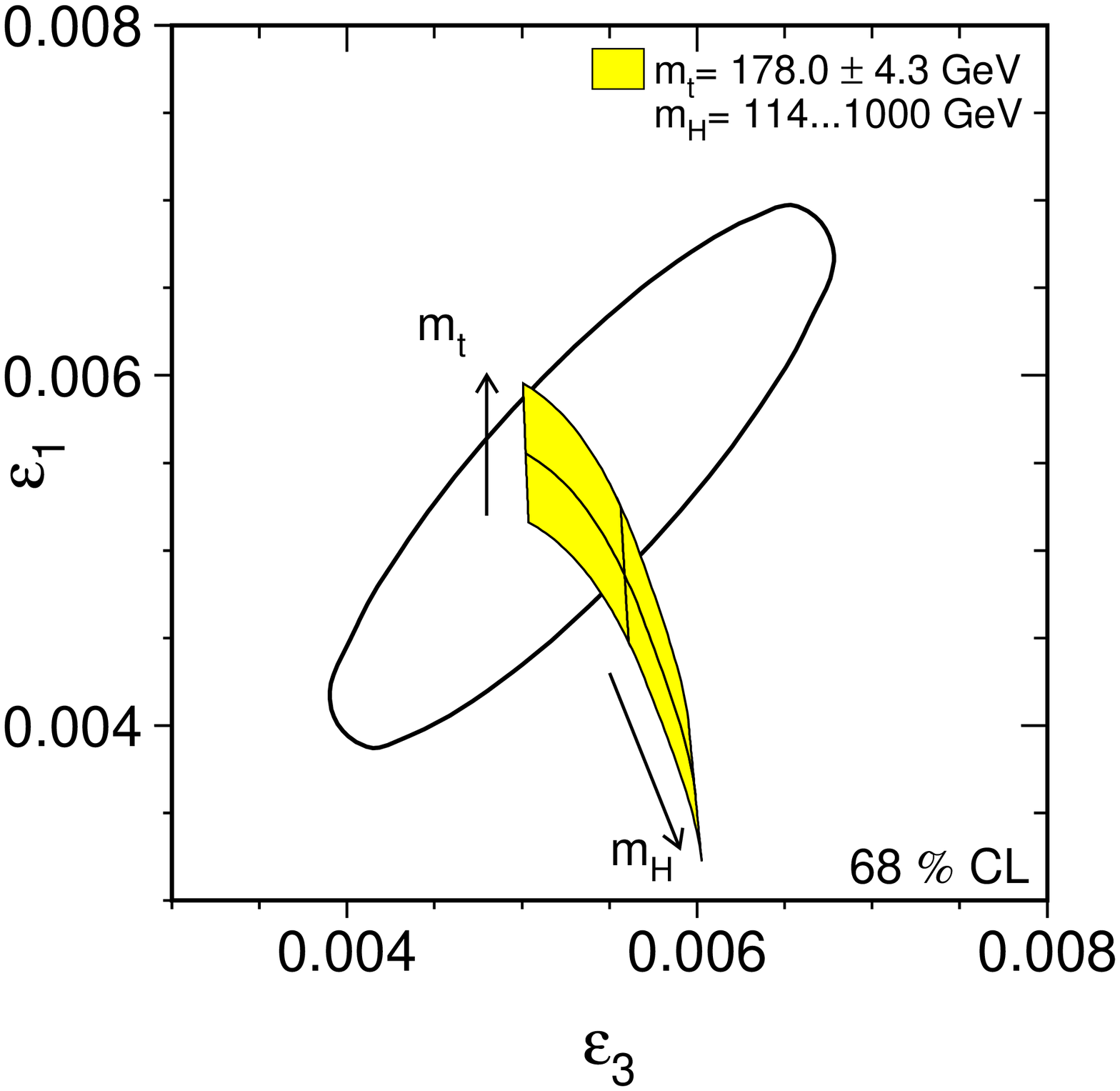,width=0.9\linewidth}} 
\caption[Results on $\epsilon$ parameters] { Contour curve of 68\%
probability in the ($\epsilon_3,\epsilon_1$) plane. The shaded region
shows the predictions within the $\SM$ for $\Mt=178.0\pm4.3~\GeV$
(Tevatron Run-I) and $\MH=300^{+700}_{-186}~\GeV$, for a fixed
hadronic vacuum polarisation of $\dalhad=0.02758$.  The direct
measurement of $\MW$ used here is preliminary. }
\label{fig:epsfit} 
\end{center}
\end{figure}

The results of the fit of all $\epsilon$ parameters to all LEP and SLD
results including the measurements of the W-boson mass are reported in
Table~\ref{tab:epsfit-all}, and are shown as a contour curve in the
($\epsilon_3,\epsilon_1$) plane in Figure~\ref{fig:epsfit}.  All
$\epsilon$ parameters are significantly different from zero, in
particular the case for $\epsilon_2$ determined by the W-boson mass,
showing again that genuine electroweak radiative corrections beyond
the running of $\alpha$ and $\alfas$ are observed. The allowed region
in $\epsilon$-parameter space overlaps with the region expected in the
$\SM$ for a light Higgs boson.  Despite the $\cAb$ result discussed
before, the extracted value for $\epsilon_b$ agrees with the $\SM$
expectation because of the strong constraint given by the $\Rbz$
result.

\begin{figure}[p]
\begin{center}
\mbox{\epsfig{file=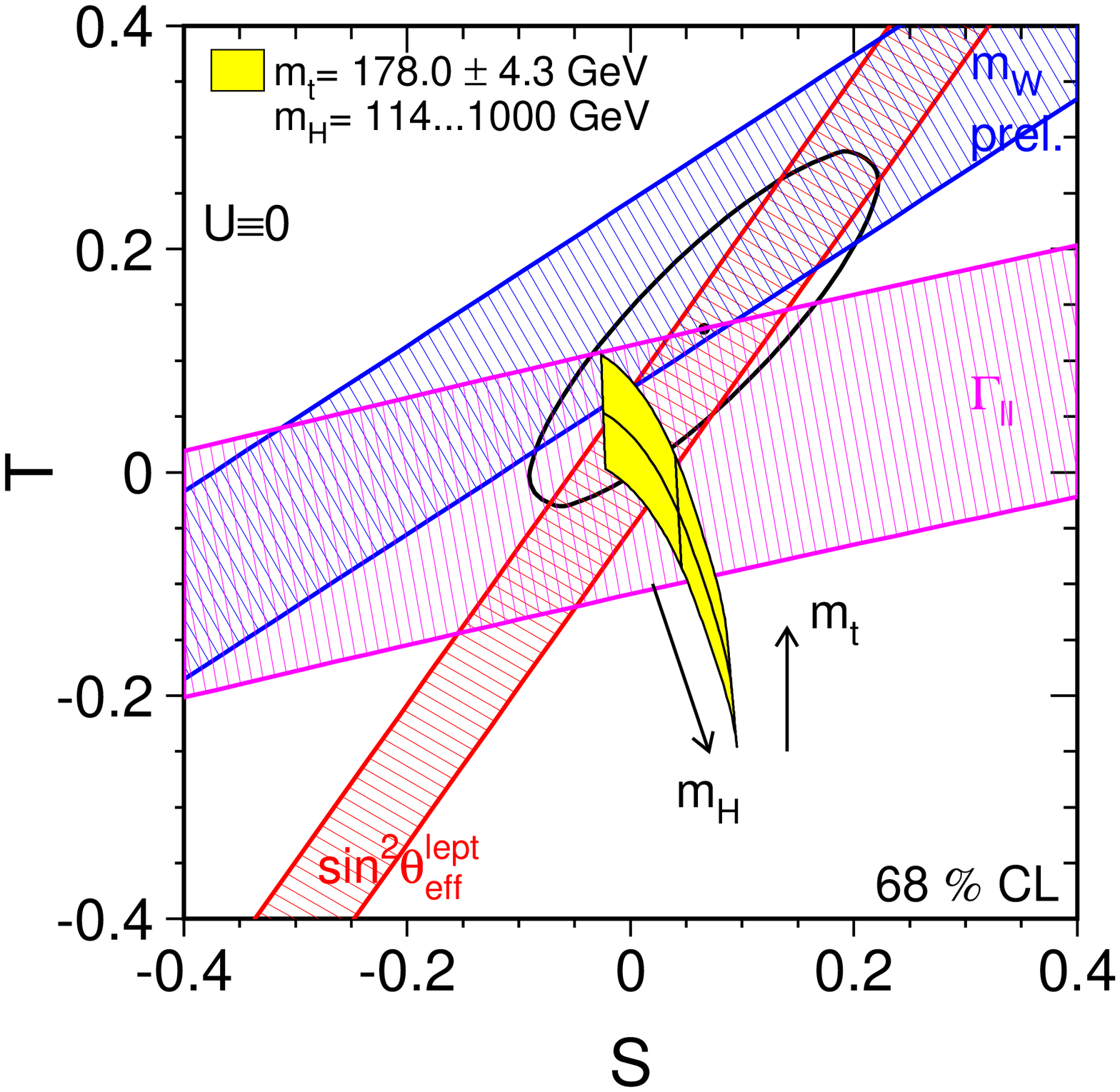,width=0.9\linewidth}} 
\caption[Results on $STU$ parameters] { Contour curve of 68\%
probability in the ($T,S$) plane.  Also shown are $\pm1$ standard
deviation bands corresponding to the measurements of $\Gll$, $\MW$ and
$\swsqeffl$.  The shaded region shows the predictions within the $\SM$
for $\Mt=178.0\pm4.3~\GeV$ (Tevatron Run-I) and
$\MH=300^{+700}_{-186}~\GeV$, for a fixed hadronic vacuum polarisation
of $\dalhad=0.02758$.  The $\SM$ reference point at which all $STU$
parameters vanish is chosen to be: $\dalhad=0.02758$, $\alfmz=0.118$,
$\MZ=91.1875~\GeV$, $\Mt=175~\GeV$, $\MH=150~\GeV$.  The constraint
$U=0$ is always applied. The direct measurement of $\MW$ used here is
preliminary.}
\label{fig:stufit} 
\end{center}
\end{figure}

The results of the fit of the $STU$ parameters to the same data set
are shown in Table~\ref{tab:stufit-all-u=0}.  The constraint $U=0$ is
imposed, as the mass and width of the W boson is the only measurement
sensitive to $U$ and models with deviations in $U$ constitute a more
severe deviation from the $\SM$ symmetry framework than implied by $S$
and $T$.  In the ($T,S$) plane, the overall result as well as bands
corresponding to the most precise measurements are shown in
Figure~\ref{fig:stufit}.  The $STU$ analyses show that there are no
large unexpected electroweak radiative corrections, as the values of
the $STU$ parameters are in agreement with zero.

\begin{table}[t]
\begin{center}
\renewcommand{\arraystretch}{1.25}
\begin{tabular}{|l||r@{$\pm$}l||rrrrrrr|}
\hline
Parameter & \multicolumn{2}{|c||}{Value} 
          & \multicolumn{7}{|c| }{Correlations} \\
          & \multicolumn{2}{|c||}{ }
          & $\dalhad$ & $\alfmz$ & $\MZ$ 
          & $\epsilon_1$ & $\epsilon_2$ & $\epsilon_3$ & $\epsilon_{b}$ \\
\hline
\hline
$\dalhad$     &$0.02758$&$0.00035$& $ 1.00$&$     $&$     $&$     $&$     $&$     $&$     $\\
$\alfmz $     &$ 0.1185$&$0.0039 $& $ 0.00$&$ 1.00$&$     $&$     $&$     $&$     $&$     $\\
$\MZ~[\GeV]$  &$91.1873$&$0.0021 $& $ 0.00$&$ 0.02$&$ 1.00$&$     $&$     $&$     $&$     $\\
$\epsilon_1$  &$+0.0054$&$0.0010 $& $ 0.00$&$-0.37$&$-0.11$&$ 1.00$&$     $&$     $&$     $\\
$\epsilon_2$  &$-0.0089$&$0.0012 $& $ 0.06$&$-0.25$&$-0.03$&$ 0.60$&$ 1.00$&$     $&$     $\\
$\epsilon_3$ &$+0.00534$&$0.00094$& $-0.31$&$-0.28$&$-0.06$&$ 0.86$&$ 0.40$&$ 1.00$&$     $\\
$\epsilon_{b}$&$-0.0050$&$0.0016 $& $ 0.00$&$-0.63$&$ 0.00$&$ 0.00$&$-0.01$&$ 0.02$&$ 1.00$\\
\hline
\end{tabular}
\caption[Results on all $\epsilon$ parameters] {Results on the
$\epsilon$ parameters including their correlations derived from a fit
to all $\LEPI$ and SLD measurements and including the combined
preliminary measurement of the W-boson mass. The $\chidf$ has a value
of 15.7/9, corresponding to a probability of 7.2\%.  }
\label{tab:epsfit-all}
\end{center}
\end{table}

\begin{table}[t]
\begin{center}
\renewcommand{\arraystretch}{1.25}
\begin{tabular}{|l||r@{$\pm$}l||rrrrrr|}
\hline
Parameter & \multicolumn{2}{|c||}{Value} 
          & \multicolumn{6}{|c| }{Correlations} \\
          & \multicolumn{2}{|c||}{ }
          & $\dalhad$ & $\alfmz$ & $\MZ$ 
          & $S$       & $T$      & $\gamma_{b}$ \\
\hline
\hline
$\dalhad$     &$0.02760$&$0.00035$& $ 1.00$&$     $&$     $&$     $&$     $&$     $\\
$\alfmz $     &$ 0.1174$&$0.0038 $& $ 0.02$&$ 1.00$&$     $&$     $&$     $&$     $\\
$\MZ~[\GeV]$  &$91.1872$&$0.0021 $& $ 0.00$&$ 0.01$&$ 1.00$&$     $&$     $&$     $\\
$S$           &$+0.07  $&$0.10   $& $-0.36$&$-0.20$&$-0.05$&$ 1.00$&$     $&$     $\\
$T$           &$+0.13  $&$0.10   $& $-0.05$&$-0.28$&$-0.11$&$ 0.85$&$ 1.00$&$     $\\
$\gamma_{b}$  &$+0.0014$&$0.0038 $& $ 0.00$&$-0.66$&$ 0.00$&$ 0.02$&$ 0.01$&$ 1.00$\\
\hline
\end{tabular}
\caption[Results on the three Z-pole $STU\gamma_b$ parameters]
{Results on the $STU\gamma_b$ parameters including their correlations
derived from a fit to all $\LEPI$ and SLD measurements and including
the combined preliminary measurement of the W-boson mass.  The
parameter $U$ is fixed to 0.  The $\chidf$ has a value of 17.1/10,
corresponding to a probability of 7.2\%.  }
\label{tab:stufit-all-u=0}
\end{center}
\end{table}

\clearpage

\chapter{Results using Light Flavour Hadronic Events}
\label{sec:lqappendix}

Measurements using tagged samples of specific light flavour quarks
(up, down or strange) are summarised here, together with information
on the partial widths of the $\Zzero$ to up-type and down-type quarks
in hadronic $\Zzero$ decays inferred from the observed rate of direct
photon production.  With some extra assumptions, these results are
then used to make checks of light flavour couplings.

\section{Asymmetry Measurements}

\begin{figure}[htb]\begin{center}
\mbox{\epsfig{file=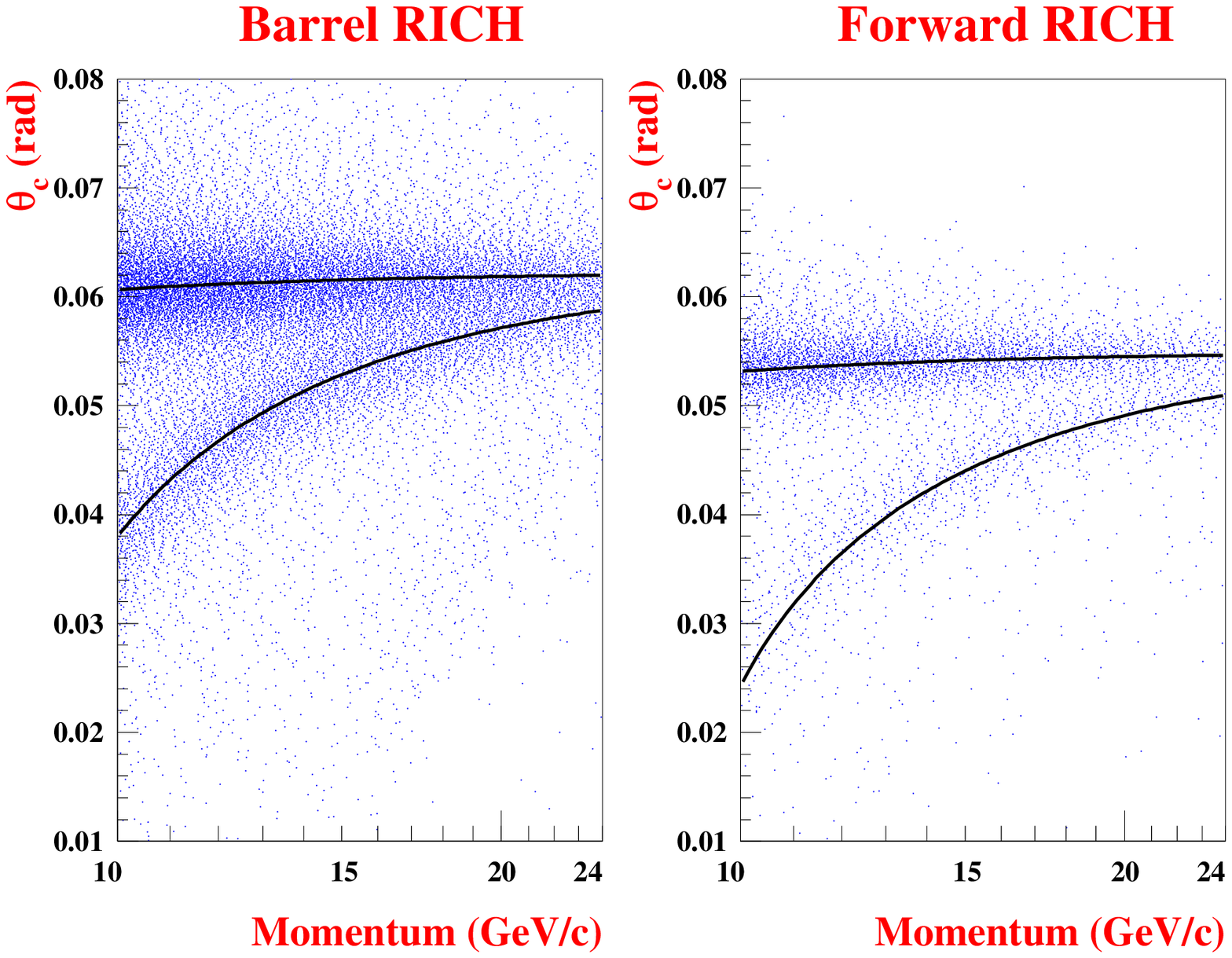,width=\textwidth}}
\end{center}
\caption[Performance of the DELPHI RICH]
{\label{fig:lqdrich} For a sample of tracks in DELPHI 1994 data, the
  reconstructed average Cherenkov angle in the gaseous radiator of
  (left) the barrel RICH and (right) the forward RICH as a function of
  the particle momentum.  The two solid lines show the Cherenkov
  angle for the pion (upper) and kaon (lower) hypotheses.  }
\end{figure}

The first measurement of the strange quark forward-backward asymmetry
was made by DELPHI~\cite{ref:dsfirst}, using 1992 data, and
identifying strange quark events from kaons in the Ring Imaging
Cherenkov detectors (RICH).  The measurement was then updated with the
full 1992-1995 data set~\cite{ref:dstrange}.  The Barrel RICH covers
$40^{\circ}<\theta<140^{\circ}$, and was used for the full
dataset. The Forward RICH covers $15^{\circ}<\theta<35^{\circ}$ plus
$145^{\circ}<\theta<165^{\circ}$, and was used for the 1994--1995
data.  Kaons with momenta between 10 and $24~\GeV$ are selected in the
RICH detectors, with an average identification efficiency of 53\%
(42\%) in the Barrel (Forward) region. At least two photoelectrons had
to be identified in the ring, and the angle of the ring with respect
to the track direction had to be consistent with the theoretical
expectation for kaons within 2.5 standard deviations, and at least 2
standard deviations away from the pion expectation.  The distribution
of Cherenkov angle as a function of momentum is shown in
Figure~\ref{fig:lqdrich}.  The quark direction is taken to be the
event thrust axis, signed according to the charge of the identified
kaon. The strange fraction of the sample selected by the kaon tag is
43\%.  For events in the barrel region, which overlap with the
micro-vertex acceptance, bottom and charm quark events are removed by
a requirement on the event b-tagging probability, which increases the
strange fraction to 55\% and reduces the dependence of the result on
modelling kaon production in heavy quark decays.

The asymmetry of the selected event sample is a linear combination of
five quark forward-backward asymmetries, weighted by the fraction of
that flavour and a flavour dependent charge dilution factor, as in
Equation~\ref{eq:afbsum}.  The asymmetry of the selected sample is
estimated by a $\chi^2$ fit to the asymmetry in bins of $\cos \theta$
of the event thrust axis, signed by the charge of the kaon. The sample
asymmetry is corrected for background, dominated by misidentified
pions.  This correction depends on the polar angle of the kaon
candidate.  The s-quark asymmetry is then evaluated, taking into
account the fraction of each quark flavour in the kaon-tagged sample,
and the probability that the charge is correctly tagged for each
flavour.  Corrections for QED radiation and QCD effects are also made.
The analysis is somewhat model dependent, in that it assumes the
Standard Model ($\SM$) prediction for production fractions for each
flavour, and for non-strange asymmetries, taken either from
ZFITTER~\cite{\ZFITTERref} or from LEP combined measurements. The
analysis also relies on the Monte Carlo simulation to compute the
efficiencies and dilutions for each flavour. However, the explicit
dependence on the other flavour asymmetries can be included in the
result, which is:
\begin{eqnarray}
 \Afbzs      &=&   0.1008 \pm 0.0113 \pm 0.0036
                 - 0.0210  ( \Afbzc - 0.0709)/ 0.0709 \nonumber\\
             & & + 0.0121  ( \Afbzd - 0.1031)/ 0.1031             
                 + 0.0115  ( \Afbzu - 0.0736)/ 0.0736 \,,
\label{app:afb0s}
\end{eqnarray}
where the first error is statistical and the second systematic.  The
dependence on the b-quark forward-backward asymmetry is a factor 10
smaller and has been neglected. The quoted systematic error in the 
original publication of 0.0040
included an uncertainty for the measured c-quark
asymmetry, which is replaced by the explicit dependence here.

OPAL~\cite{ref:olight} has also measured light quark asymmetries,
using the full 1990-1995 data-set, and high-momentum stable particles
as a tag for light flavour events.  Their approach is quite different
from that of DELPHI, aiming for the minimum model dependence. The tag
method uses the fact that the leading particle in a jet tends to carry
the quantum numbers of the primary quark, and that the decay of c- and
b-hadrons does not usually yield very high momentum stable particles.
Identified $\pi^{\pm}$, $\mathrm{K}^{\pm}$,
$\mathrm{p(\overline{p})}$, $\Kzeros$ or $\Lambda
(\overline{\Lambda})$ hadrons with momentum, $p_h$, satisfying $2 p_h
/\sqrt{s} > 0.5$ are selected.  Charged protons, pions and kaons are
identified from the $\mathrm{d}E/\mathrm{d} x$ measured in the OPAL
jet chamber, while $\Kzeros$ and $\Lambda (\overline{\Lambda})$
are selected by reconstructing their decay vertex and mass cuts.  Only
events where the polar angle of the thrust axis satisfies $|\cos
\theta |<0.8$ are considered, and after all selection cuts about 110
thousand tagged hemispheres are retained out of 4.3 million events.
The purities range from 89.5\% for pions to 59\% for protons.

With the 5 different tags, the analysis uses a system of 5 single and
15 double tag equations to derive the light flavour composition of the
tagged hemispheres directly from data (see Section~\ref{sec:hq} for a
description of the double tag method).  The unknowns are the 15
$\eta^h_{\mathrm{q}}$, the fractions of hemispheres of flavour q
tagged by hadron $h$, and the three light flavour partial widths
$R_{\mathrm{q}}$, plus one hemisphere correlation coefficient 
which is assumed to
be the same for all tagging hadrons and flavours.  The small heavy
quark fractions are measured from data from a b-tagged sample for
b-quarks, and from Monte Carlo simulation using measured uncertainties
on their properties for c-quark events.  To solve the system of
equations, it is then also assumed that $\Rd = \Rs \equiv \Rds$, and
that a few hadronisation symmetries are valid, for example
$\eta_{\mathrm{d}}^{\pi^{\pm}} = \eta_{\mathrm{u}}^{\pi^{\pm}} $.  In
order to measure the forward-backward asymmetry, the charge tagging
probabilities are also measured from the double tagged events, and it
is assumed that $\Afbzd = \Afbzs \equiv \Afbzds$.

The OPAL results are
\begin{eqnarray}
\Afbzds &=&   0.072 \pm 0.035 \pm 0.011              
            - 0.0119 ( \Afbzc - 0.0722)/ 0.0722 \label{app:afb0d}\,, \\
\Afbzu  &=&   0.044 \pm 0.067 \pm 0.018                
            - 0.0334 ( \Afbzc - 0.0722) /0.0722 \label{app:afb0u}\,.
\end{eqnarray}
The correlation between the two results is +91\%.  The correlation is
positive because although the quark asymmetries have the same sign,
the up and down-type quarks have opposite charge. The asymmetry for a
given tag particle is therefore of opposite sign if the leading
particle includes an up-type quark compared to a down-type quark.
These pole asymmetries include corrections of $+0.004$ which have been
applied to the measured \Ads\ and \Auu\ to account for QCD and ISR
effects.  The dependence on the c-quark forward-backward asymmetry has
been quoted explicitly, and the results have negligible dependence on
other $\SM$ parameters.  Correlated systematic uncertainties with
other measurements are also very small.

\begin{figure}[t]\begin{center}
\mbox{\epsfig{file=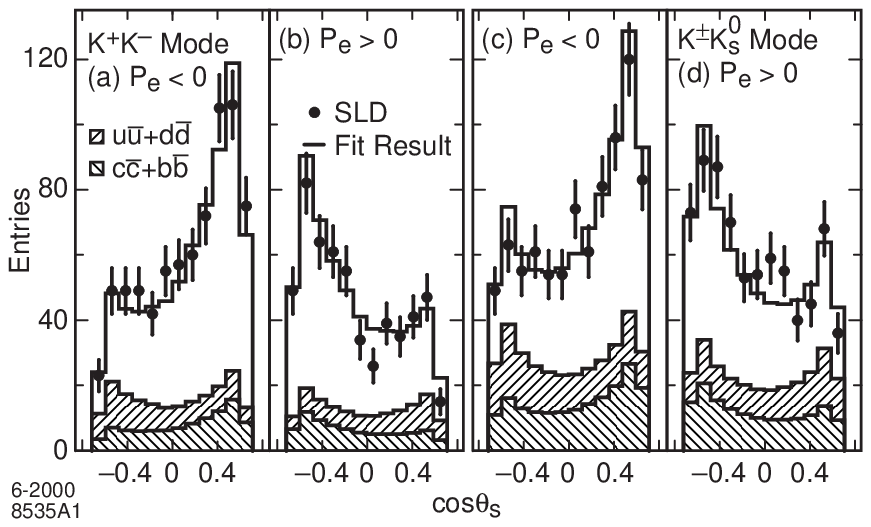,width=0.9\textwidth}}
\end{center}
\caption[SLD measurement of $\cAs$]
{\label{fig:lqsldas} Measured s-quark polar angle distributions (dots)
  for selected SLD events in the (a),(b) $\mathrm{K^+K^-}$ and (c),(d)
  $\mathrm{K^{\pm}\Kzeros}$ modes, produced with (a),(c) left- and
  (b),(d) right-polarised electron beams. The histograms represent the
  result of a simultaneous fit to the four distributions, and the
  upper (lower) hatched areas indicate the estimated $\uu + \dd~(\cc +
  \bb)$ backgrounds.}
\end{figure}

SLD have published a measurement of the strange quark coupling
parameter, \cAs, from the left-right forward-backward asymmetry of
events tagged by the presence in each hemisphere of a high momentum
$\mathrm{K}^{\pm}$ or $\Kzeros$~\cite{ref:sldas}.  The
measurement uses the full sample of 550,000 $\Zzero$ decays recorded
in 1993--1998.  Charged kaons with momentum above $9~\GeV$ are
identified by the Cherenkov Ring Imaging Detector (CRID), with
efficiency (purity) of 48\% (91.5\%).  Neutral kaons with momentum
above $5~\GeV$ are reconstructed from the decay $\Kzeros
\rightarrow \pi^+ \pi^-$ with an efficiency (purity) of 24\% (90.7\%).
Background from kaons from heavy flavour events is suppressed by
identifying B and D decay vertices.  Requiring a strange tag in both
hemispheres further suppresses the $\uu+\dd$ events. The thrust axis
is used to estimate the s-quark production angle, with the charge
identified from a $\mathrm{K}^{\pm}$ in one hemisphere, which must be
opposite to either a $\mathrm{K}^{\mp}$ or a $\Kzeros$. For the
two tagging cases, 1290 and 1580 events are selected, with $\ssb$
purities of 73\% and 60\% respectively.  The corresponding analysing
powers are 0.95 and 0.70, where analysing power is defined as
$(N_r-N_w)/(N_r+N_w)$, and $N_r$ $(N_w)$ is the number of events where
the thrust axis was signed correctly (incorrectly).  The asymmetry is
derived from a simultaneous maximum likelihood fit to the
distributions shown in Figure~\ref{fig:lqsldas}, taking into account
contributions from each flavour.  As in the case of the OPAL
measurement, this analysis is designed to be self-calibrating as much
as possible.  The analysing powers and the $\uu + \dd$ backgrounds are
constrained from the data, by examining the relative rates of single
and multi-tagged hemispheres.

The result of the fit is
\begin{eqnarray}
\cAs & = & 0.895 \pm 0.066 \pm 0.062       
           - 0.1258  (\cAc - 0.641) / 0.641 \nonumber \\
     &   & + 0.0275 (\cAd - 0.935) / 0.935              
           + 0.0558 (\cAu - 0.666)/ 0.666  \nonumber \\
     &   & + 0.0150 (\Rc - 0.1735) /0.1735                 
           + 0.2291 ((\Ru+\Rd)/\Rs - 1.780)/ 1.780 \label{eq:SLD:As}\, , 
\end{eqnarray} 
where the dependence on other electroweak parameters is given
explicitly. The dependences on $\cAb$ and $\Rb$ are negligible. Due to
strong correlations, the dependence on the light quark partial widths
is given in terms of $(\Ru+\Rd)/\Rs$ as a convenient approximation.

Common systematic uncertainties between any of these light quark
results and the measurements in the heavy flavour sector, for example
from QCD corrections, or the SLC electron beam polarisation, can
safely be neglected for all these results, since the total statistical
and systematic errors are relatively much larger. Correlations between
the light-quark results are also small, in particular because the OPAL
and SLD results rely on data to constrain systematic uncertainties.

\section{Partial Width Measurements}

The OPAL analysis described above~\cite{ref:olight}, using 1990-1995
data, and high-momentum stable particles as a light-flavour tag, also
gives measurements of the ratios:
\begin{eqnarray} 
        \frac{ \Ru }{ \Rd + \Ru + \Rs } 
& = & 1-\frac{2\Rds}{ \Rd + \Ru + \Rs} 
~ = ~ 0.258 \pm 0.031 \pm 0.032 \label{app:Ru}\,,
\end{eqnarray}
where $\Rds = \Rd = \Rs$.

In addition, DELPHI~\cite{ref:dphoton}, L3~\cite{ref:lphoton} and
OPAL~\cite{Abbiendi:2003ke} have used the rate of hadronic events with
identified direct photons, interpreted as final-state radiation from
quarks, to access effective couplings defined as:
\begin{eqnarray} 
\cu &=& 4\, (\gvu^2 + \gau^2 + \gvc^2 + \gac^2 ) /2 \\
\cd &=& 4\, (\gvd^2 + \gad^2 + \gvs^2 + \gas^2  + \gvb^2 + \gab^2 ) /3 \,,
\end{eqnarray}
which are proportional to the up-type or down-type partial widths.
The measured quantity is the partial width of hadronic events with an
isolated photon, which is given by
\begin{equation}
\Gamma(\Zzero \ra \gamma + {\mathrm{jets}}) ( \ycut )
 ~=~ \frac {h}{9} \frac {\alpha} {2 \pi} F(\ycut) \Sqqg \,,
\end{equation}
where $ h = 3 \GF \MZ^3 / 24 \pi \sqrt{2} $, $F(\ycut)$ expresses the
theoretical matrix element calculation for the rate of
$\mathrm{qq\gamma}$ events as a function of the jet resolution
parameter $\ycut$, and $\Sqqg$ is a function of the effective
couplings. The matrix elements are known to $\calO(\alpha \alfas)$.
The values of $\alfas$ used to evaluate the matrix elements
and their uncertainty
partly reflect the lack of a higher order calculation.
(see for example~\cite{ref:ophoton}).
The couplings contribute as:
\begin{equation}
\label{eq:sqqg}
S_{\mathrm{qq\gamma}} ~=~ 8 \cu + (3-\epsilon) \cd \,.
\end{equation}
This reflects the relative strengths of the up and down-type quark
couplings to the photon.  The quantity $\epsilon$ takes into account
the b-quark mass, and is also expected to depend on the jet resolution
as discussed below. The analyses combine this with the total hadronic
width of the $\Zzero$, which can be expressed as
\begin{equation}
\label{eq:ghadsqq}
\Ghad ~=~ h \left[ 1 + \frac {\alfas} {\pi} + 
1.41 \left( \frac  {\alfas} {\pi} \right)^2 -
12.8 \left( \frac  {\alfas} {\pi} \right)^3
\right]\Sqq \,,
\end{equation}
and
\begin{equation}
\label{eq:sqq}
\Sqq ~=~ 2 \cu + 3 \cd \,.
\end{equation}
In this case the QCD correction is known to third order in $\alfas$.
The Equations~\ref{eq:sqqg} and~\ref{eq:sqq} can be solved to give the
effective couplings, $\cu$ and $\cd$.  In this paper we find a LEP
combined value for $\Sqqg$ and use this in the following section with
all the other Z lineshape information to investigate quark couplings.
Although ALEPH have also investigated prompt photon
production~\cite{ref:aphoton}, the collaboration chose not to
interpret these QCD studies in terms of electroweak couplings.

Experimentally, the photon is identified in hadronic events as an
isolated calorimeter cluster, with no associated track. DELPHI and
OPAL use shower shape variables to reduce the background from light
neutral meson decays such as $\pi^0 \rightarrow \gamma \gamma$.  The
other dominant background is from initial-state radiation. This is
reduced by restricting the analysis to the central region of the
detector. The event samples and the photon selection criteria are
outlined in Table~\ref{tab:lqphoton}.

\begin{table}[htb]
\begin{center}
\renewcommand{\arraystretch}{1.1}
\begin{tabular}{|l||c|c|c|}
\hline
& DELPHI & L3 & OPAL \\
\hline\hline
Data set: & & & \\
Years      & 1991--93 & 1990--91 & 1990--95 \\
Multihadron events & 1.5 M & 320 k & 3.0 M \\
\hline
Photon selection: & & & \\
$\theta_\gamma$ in range & $25-155^\circ$ & $45-135^{\circ}$ & 
                $ | \cos \theta_{\gamma} | < 0.72 $\\
 $E_{\gamma}$ satisfies & $>5.5$~\GeV & $>5.0$~\GeV & $>7.0$~\GeV  \\
Isolation half angle & $20^{\circ}, E>500$~\MeV 
& $15^{\circ}, E>500$~\MeV 
& $0.235{\mathrm{rad}}, E>500$~\MeV \\
\hline
Jet scheme:& Durham, $\ycut=0.02$ & JADE, $\ycut=0.05$ & JADE, $\ycut=0.08$ \\
Photon--jet  &   same           & $\gamma$ $20^{\circ}$ from jet & same \\
\hline
\end{tabular}
\caption[Comparison of direct photon analyses]
{Comparison of direct photon analyses. The jet finding schemes and
  resolution parameters are those chosen for the central value of the
  electroweak couplings by each experiment.}
\label{tab:lqphoton}
\end{center}
\end{table}

The particles in the event excluding the photon are grouped into jets
using some jet resolution parameter $\ycut$.
The jet finding is then extended to include the photon 
using the same jet resolution
parameter (DELPHI, OPAL), and the event is only retained if the
photon is not merged with a jet. In the case of L3, an 
angular separation between the photon and the jets is required.
The rate of isolated photons therefore depends on the jet
resolution parameter that has been chosen. The rate as a function of
$\ycut$ is used in various QCD studies, but one working point is
chosen for the calculation of electroweak parameters of relevance
here.  The rates are corrected for detector and fragmentation effects,
and for the geometric acceptance. When compared with the predictions
of matrix element calculations they yield a measurement of $\Sqqg$.

The correction to account for the b-quark mass was estimated by L3 to
be $\epsilon = 0.2 \pm 0.1$.  However this correction should depend on
the effective mass of the photon-jet system. No correction was used by
OPAL, while DELPHI adopted the same correction as L3. However,
in the OPAL and DELPHI analyses,
the effective mass of the photon-jet system is constrained to be
about an order of magnitude larger than for L3, and the relative
impact of the b-quark mass should be much smaller.  For this reason,
the correction has been used here for the L3 result only.

The published values for $\Sqqg$ with the error categories chosen by
the three experiments are as follows: for DELPHI,
\begin{equation}
 \Sqqg ~=~ 11.71 \pm 0.43 \pm 0.78 \pm 0.50 \pm 0.25 
^{+1.07}_{-1.78} \,,
\end{equation}
where the errors account for statistics, experimental effects, theory,
$\alfas$ and the $y_{cut}$ range respectively; for L3
\begin{equation}
 \Sqqg ~=~ 11.88 \pm 1.17 \pm 0.09 \pm 0.63  \,,
\end{equation}
where the errors represent statistical and experimental effects,
hadronisation and variations of the photon-jet collinearity cut; and
for OPAL,
\begin{equation}
\Sqqg ~=~ 13.74 \pm 0.30 \pm 0.27 ^{+0.12}_{-0.04}  \,,
\end{equation}
where the the first error is statistical,
the second is systematic and the third comes from
the uncertainty in evaluating
the matrix element $F(\ycut)$.

The OPAL result uses more data, and controls the experimental 
uncertainties by fitting the distribution of the shower
shape variable for the rate of neutral hadrons
misidentified as photons. This result dominates the average.
The OPAL uncertainties 
due to hadronisation or fragmentation, $\alfas^{(1)}$
and theory amount to 0.25. The uncertainty in the
DELPHI and L3 measurements due to these effects
are estimated to be 0.60. These uncertainties are
treated as fully correlated.
An additional common uncertainty of 0.36 due to 
possible common experimental effects is estimated for 
DELPHI and L3. 
These common uncertainties are used to calculate the
off-diagonal terms in the covariance matrix relating the three
measurements of $\Sqqg$, which are combined using a $\chi^2$ fit based
on the heavy-flavour averaging procedure. The average is
very insensitive to variations in the assumptions about 
correlated uncertainties.
The value of $\epsilon$ was
set equal to zero for the DELPHI and OPAL results, and constrained to
$\epsilon=0.2\pm0.1$ for the L3 result. The average value of
$\Sqqg^0\equiv 8\cu + 3\cd$ was then found to be:
\begin{equation}
\Sqqg^0~ \equiv~ 8\cu + 3\cd ~=~ 13.67 \pm 0.42 \,.
\label{eq:sqqgresult}
\end{equation}
This result is uncorrelated with the Z width, and is used in the
following section to infer information on quark couplings.

\section{Comparison with Standard Model Expectations}

The $\SM$ analysis presented in Table~\ref{tab:msm:input} predicts the
following values for the observables discussed above:
\begin{eqnarray}
\Afbzs ~ = ~ \Afbzd             & = &  0.1039  \pm 0.0008   \\[1mm]
\Afbzu                          & = &  0.0742  \pm 0.0006   \\[1mm]
\cAs                            & = &  0.9357  \pm 0.0001   \\[1mm]
\frac{ \Ru }{ \Rd + \Ru + \Rs } & = &  0.2816  \pm 0.0001   \\[1mm]
\Sqqg^0                         & = &  13.677  \pm 0.005\pz \,.
\end{eqnarray}
The agreement is good.

\section{Z Boson Properties and Effective Couplings}

The properties of the Z boson and effective couplings of the neutral
weak current are now determined for all five quark flavours.  In
contrast to the procedure adopted in Chapter~\ref{chap:Z+coup}, the
hadronic partial width is no longer an independent parameter but is
calculated as:
\begin{eqnarray}
\Ghad & = & \Gdd+\Guu+\Gss+\Gcc+\Gbb\,.
\label{app:gz_hadr}
\end{eqnarray}

Since there are not sufficiently many different light-quark flavour
pseudo-observables measured to disentangle u, d and s quarks
completely, an assumption is made in the extraction of
pseudo-observables such as partial widths or effective coupling
constants for light quarks: quark universality is imposed for the two
down-type light-quark flavours, so that $\mathrm{s}\equiv\mathrm{d}$
for all pseudo-observables relating to s and d quarks.

As reported in the previous sections, the experimental results on
pseudo-observables for light quarks depend explicitly on the values of
pseudo-observables for other quark flavours.  In order to treat these
dependencies correctly, the global analyses presented in
Sections~\ref{chap:partrafo} and~\ref{chap:coupling} are extended to
include and treat light quark flavours.  For simplicity, the
cross-section ratios $R_{\mathrm{q}}$ for light quarks u, d and s are
interpreted directly as ratios of partial widths, $\Rqz$; this is
justified as the difference $\Rq-\Rqz$ is negligible relative to the
experimental uncertainties of the light quark measurements.

\subsection{Z-Boson Decay Widths and Branching Fractions}

Following the analysis in Section~\ref{chap:partrafo} and including
the ratio of partial widths
$\Ruz/(\Rdz+\Ruz+\Rsz)=\Guu/(\Gdd+\Guu+\Gss)$, Equation~\ref{app:Ru},
the partial Z decay widths and branching fractions for all five quark
flavours are determined.  The results for the heavy quark flavours b
and c are nearly unchanged from those shown in Tables~\ref{tab:width}
and~\ref{tab:brfrac}, and are not repeated here.  The results for the
light quark flavours are reported in Tables~\ref{app:lq:gz}
and~\ref{app:lq:bz} for partial Z decay widths and Z branching
fractions, respectively.  The strong anti-correlation between the
light-quark partial widths and between their branching fractions
arises through Equation~\ref{app:gz_hadr} from the precisely measured
b, c and inclusive hadronic partial Z decay widths.

\begin{table}[htbp]
\begin{center}
\renewcommand{\arraystretch}{1.25}
\begin{tabular}{|c||r@{$\pm$}l||rr|}
\hline
Parameter & \multicolumn{2}{|c||}{Value} 
          & \multicolumn{2}{|c| }{Correlations} \\
          & \multicolumn{2}{|c||}{$[\MeV]$} & {$\Gss=\Gdd$} & {$\Guu$} \\
\hline
\hline
$\Gss=\Gdd$ & $396$ & $ 24$ & $ 1.00$ & $     $ \\
$\Guu$      & $275$ & $ 48$ & $-0.99$ & $ 1.00$ \\
\hline
\end{tabular}
\caption[Partial $\Zzero$ widths]
{ Partial $\Zzero$ decay widths and error correlation coefficients for
  light quarks. }
\label{app:lq:gz}
\end{center}
\end{table}

\begin{table}[htbp]
\begin{center}
\renewcommand{\arraystretch}{1.25}
\begin{tabular}{|c||r@{$\pm$}l||rr|}
\hline
Parameter & \multicolumn{2}{|c||}{Value} 
          & \multicolumn{2}{|c| }{Correlations} \\
          & \multicolumn{2}{|c||}{[\%]} & 
{$B(\Zzero\to\ssb)=B(\Zzero\to\dd)$} & 
{$B(\Zzero\to\uu)$} \\
\hline
\hline
$B(\Zzero\to\ssb)=B(\Zzero\to\dd)$ & $15.9$  & $1.0$  & $ 1.00$ & $     $ \\
$B(\Zzero\to\uu)$                 & $11.0$  & $1.9$  & $-0.99$ & $ 1.00$ \\
\hline
\end{tabular}
\caption[$\Zzero$ branching fractions]
{$\Zzero$ branching fractions and error correlation coefficients for
  light quarks. }
\label{app:lq:bz}
\end{center}
\end{table}

In order to test quark universality in Z decays quantitatively, the
ratios of the quark partial widths or equivalently the ratio of the
quark branching fractions are calculated for up-type quarks and for
down-type quarks. The results are:
\begin{eqnarray}
\frac{\Gdd}{\Gbb} & = & 
\frac{B(\Zzero\to\dd)}{B(\Zzero\to\bb)}
                  ~ = ~ 1.049\pm0.064    \\
\frac{\Guu}{\Gcc} & = & 
\frac{B(\Zzero\to\uu)}{B(\Zzero\to\cc)}
                  ~ = ~ 0.92\pz\pm0.16\pz \,,
\end{eqnarray}
with a correlation of $-0.98$.  In both cases, good agreement with
unity is observed.  Assuming quark universality, quark mass effects
and $\SM$ b-specific vertex corrections are expected to decrease
$\Gbb$ and $B(\Zzero\to\bb)$ by about 1.9\% relative to the light
down-type quark flavour d; this is also shown in
Figure~\ref{fig:coup:b-vertex}.

\subsection{Effective Couplings of the Neutral Weak Current}

In addition to the measurement of the partial width ratio,
Equation~\ref{app:Ru}, by including the four measurements of the
various light quark asymmetries, Equations~\ref{app:afb0s}
to~\ref{eq:SLD:As}, and the measurement of $\Sqqg^0$,
Equation~\ref{eq:sqqgresult}, the analyses presented in
Section~\ref{chap:coupling} are extended to determine effective
couplings of all five quark flavours.  The results for the heavy quark
flavours b and c are nearly unchanged from those shown in
Tables~\ref{tab:coup:aq}, \ref{tab:coup:gq} and~\ref{tab:coup:rsq},
and are not repeated here.  The asymmetry parameters $\cAs=\cAd$ and
$\cAu$, derived from the various asymmetry measurements, are listed in
Table~\ref{app:coup:aq}.  The flavour-dependent effective coupling
constants as well as the $\rhoq$ parameters and the effective
electroweak mixing angles $\swsqeffq$ for light quarks are reported in
Tables~\ref{app:coup:gq} and~\ref{app:coup:rsq}, respectively.  Good
agreement with the $\SM$ expectation is observed in all cases.

\begin{table}[p]
\begin{center}
\renewcommand{\arraystretch}{1.25}
\begin{tabular}{|c||r@{$\pm$}l||rr|}
\hline
Parameter & \multicolumn{2}{|c||}{Value} 
          & \multicolumn{2}{|c|}{Correlations} \\
          & \multicolumn{2}{|c||}{ }
          & {$\cAs=\cAd$} & {$\cAu$} \\
\hline
\hline
$\cAs=\cAd$ & $0.902$ & $0.087$ & $ 1.00$ & $     $ \\
$\cAu$      & $0.82 $ & $0.32 $ & $ 0.61$ & $ 1.00$ \\
\hline
\end{tabular}
\caption[Results on the asymmetry parameter $\cAq$ for light quarks]
{Results on the asymmetry parameters $\cAq$ for light quarks.  Note
that since $\cAf=2r/(1+r^2)$ with $r=\gvf/\gaf$, values $|\cAf|>1$ are
unphysical. The combination has a $\chidf$ of 4.9/5, corresponding to
a probability of 43\%.}
\label{app:coup:aq}
\end{center}
\end{table}

\begin{table}[p]
\begin{center}
\renewcommand{\arraystretch}{1.25}
\begin{tabular}{|c||r@{$\pm$}l||rrrr|}
\hline
Parameter & \multicolumn{2}{|c||}{Value} 
          & \multicolumn{4}{|c| }{Correlations} \\
          & \multicolumn{2}{|c||}{ }
          & {$\gas=\gad$} & {$\gau$} & {$\gvs=\gvd$} & {$\gvu$} \\
\hline
\hline
$\gas=\gad$& $-0.52$ & $^{0.05}_{0.03}$ &$ 1.00$&$     $&$     $&$     $\\
$\gau$     & $+0.47$ & $^{0.05}_{0.33}$ &$-0.43$&$ 1.00$&$     $&$     $\\
$\gvs=\gvd$& $-0.33$ & $^{0.05}_{0.07}$ &$-0.92$&$ 0.59$&$ 1.00$&$     $\\
$\gvu$     & $+0.24$ & $^{0.28}_{0.11}$ &$ 0.61$&$-0.91$&$-0.61$&$ 1.00$\\
\hline
\hline
Parameter & \multicolumn{2}{|c||}{Value} 
          & \multicolumn{4}{|c| }{Correlations} \\
          & \multicolumn{2}{|c||}{ }
          & {$\gls=\gld$} & {$\glu$} & {$\grs=\grd$} & {$\gru$} \\
\hline
\hline
$\gls=\gld$& $-0.423$   & $0.012$          &$ 1.00$&$     $&$     $&$     $\\
$\glu$     & $+0.356$   & $0.035$          &$-0.13$&$ 1.00$&$     $&$     $\\
$\grs=\grd$& $+0.10\pz$ & $^{0.04}_{0.06}$ &$ 0.72$&$-0.59$&$ 1.00$&$     $\\
$\gru$     & $-0.11\pz$ & $^{0.30}_{0.07}$ &$-0.51$&$+0.84$&$-0.60$&$ 1.00$\\
\hline
\end{tabular}
\caption[Results on the effective coupling constants for light quarks]
{Results on the effective coupling constants $\gaq$ and $\gvq$ as well
as $\glq$ and $\grq$ for light quarks. Because of the non-parabolic
nature of the $\chi^2$ being minimised, the above error matrices are
approximate.  The combination has a $\chidf$ of 5.4/7, corresponding
to a probability of 62\%.}
\label{app:coup:gq}
\end{center}
\end{table}

\begin{table}[p]
\begin{center}
\renewcommand{\arraystretch}{1.25}
\begin{tabular}{|c||r@{$\pm$}l||rrrr|}
\hline
Parameter & \multicolumn{2}{|c||}{Value} 
          & \multicolumn{4}{|c| }{Correlations} \\
          & \multicolumn{2}{|c||}{ }
          & {$\rhos    =\rhod$}     & {$\rhou$} 
          & {$\swsqeffs=\swsqeffd$} & {$\swsqeffu$} \\
\hline
\hline
$\rhos=\rhod$        &$1.09$ &$^{0.12}_{0.21}$&$ 1.00$&$     $&$     $&$     $\\
$\rhou$              &$0.88$ &$^{0.20}_{0.37}$&$ 0.42$&$ 1.00$&$     $&$     $\\
$\swsqeffs=\swsqeffd$&$0.28$ &$^{0.08}_{0.16}$&$ 0.96$&$ 0.55$&$ 1.00$&$     $\\
$\swsqeffu$          &$0.18$ &$^{0.09}_{0.18}$&$ 0.58$&$ 0.94$&$ 0.61$&$ 1.00$\\
\hline
\end{tabular}
\caption[Results on $\rhoq$ and $\swsqeffq$ for light quarks] {Results
on the $\rhoq$ parameter and the effective electroweak mixing angle
$\swsqeffq$ for light quarks.  Because of the non-parabolic nature of
the $\chi^2$ being minimised as a function of $\rhoq$ and $\swsqeffq$,
the above error matrix is approximate. Note that $\swsqeffu\ge0$ has
to be enforced as a boundary condition in the calculation of the
errors. The combination has a $\chidf$ of 5.4/7, corresponding to a
probability of 62\%.}
\label{app:coup:rsq}
\end{center}
\end{table}

\begin{figure}[htbp]
\begin{center}
\mbox{\epsfig{file=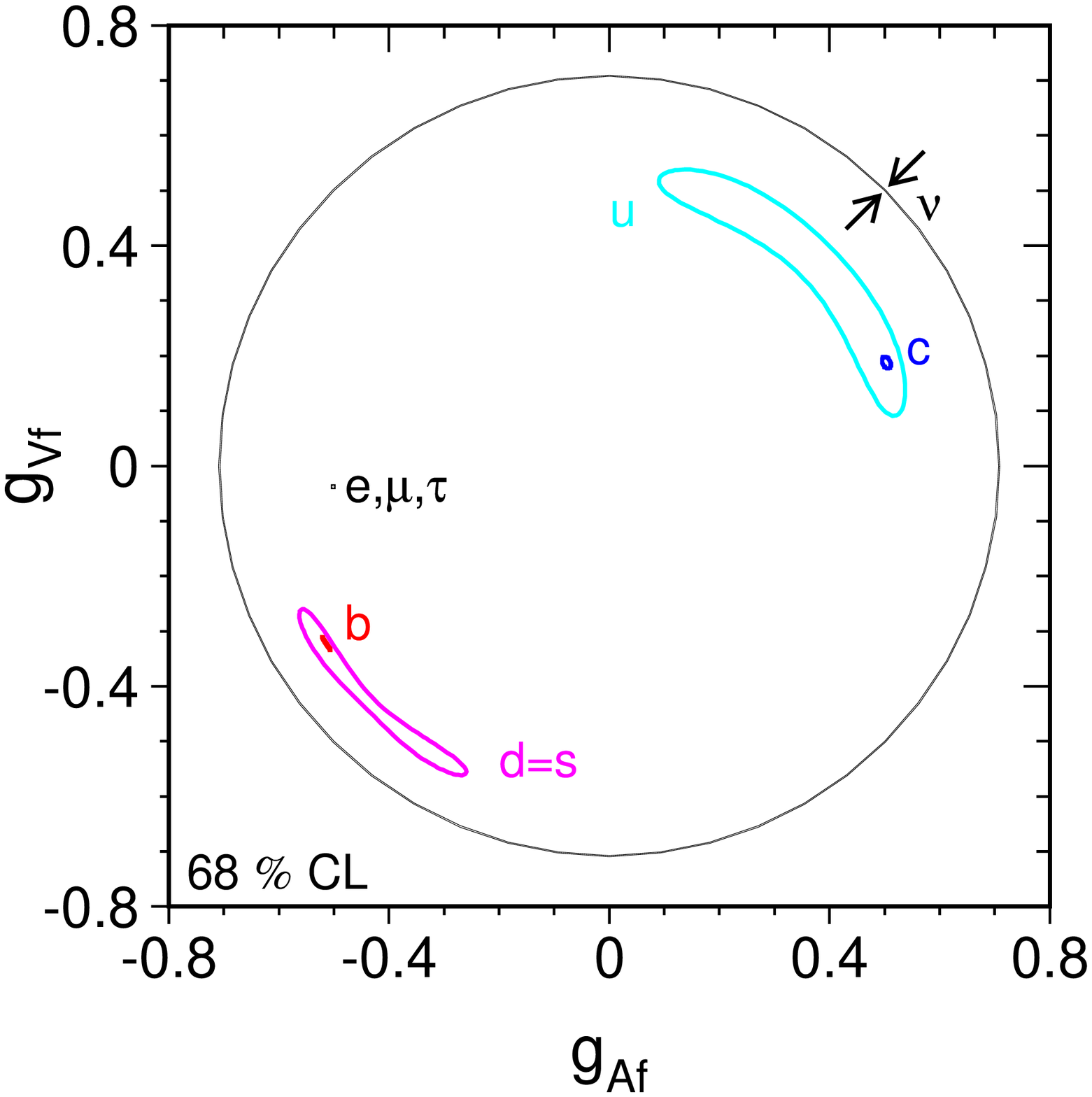,width=0.9\linewidth}} 
\caption[Comparison of the effective coupling constants] { Comparison
of the effective vector and axial-vector coupling constants for
fermions. For the light-quark contours (u and d=s), a second solution
exists, mirroring the contour curves at the origin.  The allowed area
for neutrinos, assuming three generations of neutrinos with identical
vector couplings and identical axial-vector couplings, is bounded by
circles centred at the origin since the invisible partial width
constrains the sum of the squares of the effective couplings only. }
\label{app:coup:gf} 
\end{center}
\end{figure}

The results on the effective coupling constants are also shown
graphically in Figure~\ref{app:coup:gf}.  Note that the lepton and
heavy-quark regions are expanded in Figures~\ref{fig:coup:gl}
and~\ref{fig:coup:gq}.  Since the energy dependence of the
forward-backward asymmetry of light quarks is not measured, the
ambiguity $\gvq\leftrightarrow\gaq$ arises for the light quarks,
corresponding to a mirror symmetry of the contour curves in the
$(\gvq,\gaq)$ plane along the diagonal $\gvq=\gaq$.  Each light quark
contour includes and connects both regions since the light-quark
measurements are not precise enough to exclude $\gvq=\gaq$ with
sufficient significance.  In addition, since $\cAq$ determines only
the relative sign between $\gvq$ and $\gaq$, there is also an
inversion symmetry about the origin,
$(\gvq,\gaq)\leftrightarrow(-\gvq,-\gaq)$, of the contour curves for
light quarks.  The corresponding mirror solutions are not shown in
Figure~\ref{app:coup:gf}.  Effective couplings for u and d quarks are
also measured in electron-proton collisions at HERA~\cite{HERAud} and
in proton-antiproton collisions at the Tevatron~\cite{CDFII-Afb},
albeit less precisely.

The leptonic effective electroweak mixing angle determined from the
four light-quark asymmetry measurements is:
\begin{eqnarray}
\swsqeffl & = & 0.2320\pm0.0021\,,
\end{eqnarray}
dominated by DELPHI measurement of $\Afbzs$, Equation~\ref{app:afb0s},
where the combination has a $\chidf$ of 1.5/3, corresponding to a
probability of 68\%.  For the determination of $\swsqeffl$, the
parametric dependence of the SLD $\cAs$ result,
Equation~\ref{eq:SLD:As}, on the value of the light-quark partial
width ratio $(\Ru+\Rd)/\Rs$ is neglected.  This result is in good
agreement with all the determinations of this quantity presented in
Section~\ref{sec:coup:sef2lept}.

\clearpage

\chapter{Standard Model Predictions}
\label{app:SM:preds}

Constraints on the free parameters of the Standard Model ($\SM$)
derived from the precision electroweak measurements are discussed in
Chapter~\ref{chap:MSM}. Based on these analyses, $\SM$ predictions for
all pseudo-observables are calculated and reported in
Tables~\ref{app:SM:pred:1} to~\ref{app:SM:pred:7}. Two sets are
calculated, for the Z-pole $\SM$ analysis shown in
Table~\ref{tab:msmfit-lep1sld} as well as for the high-$Q^2$ $\SM$
analysis shown in Table~\ref{tab:msmfit-all}. The improvements in the
accuracy of many predictions is clearly visible. Note that all
predictions are derived from the same sets of observables, hence they
are correlated with the five $\SM$ input parameters and cannot be used
independently.

The uncertainties quoted on the $\SM$ predictions include only the
parametric uncertainties arising from the fit to the set of
measurements, hence do not take additional theoretical uncertainties
(Section~\ref{sec:msm:TU1}) or phase-space uncertainties arising from
uncertainties in the masses of fermions other than the top quark into
account.

\renewcommand{\nb}{\mathrm{nb}}
\newcommand{\MSbar}{\mathrm{\overline{MS}}}

\newcommand{\Dalhad}[1]{\Delta         \alpha^{(5)}_{\mathrm{had}}(#1)}
\newcommand{\DALhad}[1]{\Delta\overline\alpha^{(5)}_{\mathrm{had}}(#1)}
\newcommand{\alphas}[1]{       \alpha_{\mathrm{S}}(#1)}
\newcommand{\ALphas}[1]{\overline\alpha_{\mathrm{S}}(#1)}

\newcommand{\Mv}[1]{m_{\mathrm{#1}}}
\newcommand{\MV}[1]{\overline{m}_{\mathrm{#1}}}
\newcommand{\Mf}[1]{m_{\mathrm{#1}}}
\newcommand{\MF}[1]{\overline{m}_{\mathrm{#1}}}
\newcommand{\Ms}[1]{m_{\mathrm{#1}}}
\newcommand{\MS}[1]{\overline{m}_{\mathrm{#1}}}

\newcommand{\ro}[1]{\rho_{\mathrm{#1}}}
\newcommand{\ka}[1]{\kappa_{\mathrm{#1}}}
\newcommand{\eps}[1]{\epsilon_{\mathrm{#1}}}
\newcommand{\stu}[1]{#1+c_{#1}}
\newcommand{\sts}[1]{#1}
\newcommand{\stw}[2]{\sin^2\theta^{\mathrm{#1}}_{\mathrm{#2}}}
\newcommand{\sTw}[2]{\sin^2\overline\theta^{\mathrm{#1}}_{\mathrm{#2}}}
\newcommand{\Del}[1]{\Delta\mathrm{#1}}
\newcommand{\Gam}[2]{\Gamma^{\mathrm{#1}}_{\mathrm{#2}}}
\newcommand{\GAM}[2]{\overline{\Gamma}^{\mathrm{#1}}_{\mathrm{#2}}}
\newcommand{\RGam}[3]{\Gam{#1}{#2}/\Gam{#1}{#3}}
\newcommand{\Bf}[2]{B(\mathrm{#1}\to\mathrm{#2})}
\newcommand{\g}[1]{g_{\mathrm{#1}}}
\newcommand{\cA}[2]{\mathcal{A}^{\mathrm{#1}}_{\mathrm{#2}}}
\newcommand{\cP}[2]{\mathcal{P}^{\mathrm{#1}}_{\mathrm{#2}}}
\newcommand{\sig}[2]{\sigma^{\mathrm{#1}}_{\mathrm{#2}}}
\newcommand{\R}[2]{R^{\mathrm{#1}}_{\mathrm{#2}}}
\newcommand{\A}[2]{A^{\mathrm{#1}}_{\mathrm{#2}}}
\newcommand{\smat}[3]{#1^{\mathrm{#2}}_{\mathrm{#3}}}
\newcommand{\VCKM}[1]{V_{\mathrm{#1}}}
\newcommand{\Ciq}[2]{C_{\mathrm{#1}\mathrm{#2}}}
\renewcommand{\QW}[1]{Q_{\mathrm{W}}(\mathrm{#1})}

\newcommand{\calsym}[3]{$#1$ & $#2$ & $\,\pm\, #3$}
\newcommand{\calasy}[4]{$#1$ & $#2$ & $\,\pm\,^{#3}_{#4}$}
\newcommand{\colsym}[3]{$#2$ & $\,\pm\, #3$}
\newcommand{\colasy}[4]{$#2$ & $\,\pm\,^{#3}_{#4}$}

\begin{table}[ht]
\begin{center}
\renewcommand{\arraystretch}{1.2}
\begin{tabular}{|l||r@{}l||r@{}l|}
\hline
 Observable & \multicolumn{2}{|c||}{Standard Model}
            & \multicolumn{2}{|c|}{Standard Model} \\
            & \multicolumn{2}{|c||}{Z-pole Fit    } 
            & \multicolumn{2}{|c|}{High-$Q^2$ Fit} \\
\hline
\hline
 \calsym{\Dalhad{\Mv{Z}}}{0.02759}{0.00035} & \colsym{\Dalhad{\Mv{Z}}}{0.02767}{0.00034} \\
 \calsym{\alphas{\Mv{Z}}}{0.1190}{0.0028} & \colsym{\alphas{\Mv{Z}}}{0.1188}{0.0027} \\
 \calsym{\Mv{Z}~[\GeV]}{91.1874}{0.0021} & \colsym{\Mv{Z}~[\GeV]}{91.1874}{0.0021} \\
 \calasy{\Mf{t}~[\GeV]}{173}{13}{10} & \colsym{\Mf{t}~[\GeV]}{178.5}{3.9} \\
 \calasy{\Ms{H}~[\GeV]}{111}{190}{60} & \colasy{\Ms{H}~[\GeV]}{129}{74}{49} \\
\hline
 \calasy{\LOGMH}{2.05}{0.43}{0.34} & \colsym{\LOGMH}{2.11}{0.20} \\
\hline
\hline
 \calsym{\Mv{W}~[\GeV]}{80.363}{0.032} & \colsym{\Mv{W}~[\GeV]}{80.389}{0.019} \\
 \calsym{\sTw{\MSbar}{W}}{0.23120}{0.00016} & \colsym{\sTw{\MSbar}{W}}{0.23111}{0.00014} \\
 \calsym{\stw{}{W}}{0.22331}{0.00062} & \colsym{\stw{}{W}}{0.22281}{0.00036} \\
 \calsym{\Del{r}}{0.0363}{0.0019} & \colsym{\Del{r}}{0.0348}{0.0011} \\
 \calsym{\Del{r_w}}{-0.0242}{0.0021} & \colsym{\Del{r_w}}{-0.0259}{0.0013} \\
\hline
\end{tabular}
\caption[$\SM$ predictions for pseudo-observables]{$\SM$ predictions
for pseudo-observables derived from the Z-pole and the high-$Q^2$
$\SM$ fits. The matrices of correlation coefficients for the five
$\SM$ input parameters are shown in Tables~\ref{tab:msmfit-lep1sld}
and~\ref{tab:msmfit-all}, respectively.}
\label{app:SM:pred:1}
\end{center}
\end{table}

\begin{table}[ht]
\begin{center}
\renewcommand{\arraystretch}{1.2}
\begin{tabular}{|l||r@{}l||r@{}l|}
\hline
 Observable & \multicolumn{2}{|c||}{Standard Model}
            & \multicolumn{2}{|c|}{Standard Model} \\
            & \multicolumn{2}{|c||}{Z-pole Fit    } 
            & \multicolumn{2}{|c|}{High-$Q^2$ Fit} \\
\hline
\hline           
 \calsym{\Gam{W}{tot}~[\GeV]}{2.0910}{0.0029} & \colsym{\Gam{W}{tot}~[\GeV]}{2.0930}{0.0017} \\
 \calsym{\Gam{W}{had}~[\GeV]}{1.4119}{0.0022} & \colsym{\Gam{W}{had}~[\GeV]}{1.4132}{0.0014} \\
 \calsym{\Gam{W}{e\nu}~[\GeV]}{0.22636}{0.00028} & \colsym{\Gam{W}{e\nu}~[\GeV]}{0.22659}{0.00015} \\
 \calsym{\Gam{W}{\mu\nu}~[\GeV]}{0.22636}{0.00028} & \colsym{\Gam{W}{\mu\nu}~[\GeV]}{0.22659}{0.00015} \\
 \calsym{\Gam{W}{\tau\nu}~[\GeV]}{0.22636}{0.00028} & \colsym{\Gam{W}{\tau\nu}~[\GeV]}{0.22659}{0.00015} \\
 \calsym{\Gam{Z}{tot}~[\GeV]}{2.4956}{0.0019} & \colsym{\Gam{Z}{tot}~[\GeV]}{2.4965}{0.0015} \\
 \calsym{\Gam{Z}{had}~[\GeV]}{1.7423}{0.0016} & \colsym{\Gam{Z}{had}~[\GeV]}{1.7427}{0.0015} \\
 \calasy{\Gam{Z}{ee}~[\GeV]}{0.083987}{0.000058}{0.000068} & \colsym{\Gam{Z}{ee}~[\GeV]}{0.084031}{0.000027} \\
 \calasy{\Gam{Z}{\mu\mu}~[\GeV]}{0.083986}{0.000058}{0.000068} & \colsym{\Gam{Z}{\mu\mu}~[\GeV]}{0.084030}{0.000027} \\
 \calasy{\Gam{Z}{\tau\tau}~[\GeV]}{0.083796}{0.000058}{0.000067} & \colsym{\Gam{Z}{\tau\tau}~[\GeV]}{0.083840}{0.000026} \\
 \calsym{\Gam{Z}{dd}~[\GeV]}{0.38299}{0.00036} & \colsym{\Gam{Z}{dd}~[\GeV]}{0.38317}{0.00026} \\
 \calsym{\Gam{Z}{ss}~[\GeV]}{0.38299}{0.00036} & \colsym{\Gam{Z}{ss}~[\GeV]}{0.38317}{0.00026} \\
 \calasy{\Gam{Z}{bb}~[\GeV]}{0.37602}{0.00054}{0.00074} & \colsym{\Gam{Z}{bb}~[\GeV]}{0.37577}{0.00037} \\
 \calsym{\Gam{Z}{uu}~[\GeV]}{0.30016}{0.00046} & \colsym{\Gam{Z}{uu}~[\GeV]}{0.30035}{0.00035} \\
 \calsym{\Gam{Z}{cc}~[\GeV]}{0.30010}{0.00047} & \colsym{\Gam{Z}{cc}~[\GeV]}{0.30028}{0.00036} \\
 \calasy{\Gam{Z}{inv}~[\GeV]}{0.50162}{0.00034}{0.00040} & \colsym{\Gam{Z}{inv}~[\GeV]}{0.50185}{0.00014} \\
 \calsym{\RGam{Z}{inv}{\ell}}{5.97258}{0.00057} & \colsym{\RGam{Z}{inv}{\ell}}{5.97227}{0.00050} \\
 \calsym{\RGam{Z}{\ell}{\nu}}{0.502296}{0.000048} & \colsym{\RGam{Z}{\ell}{\nu}}{0.502322}{0.000042} \\
 \calsym{\RGam{Z}{\nu}{\ell}}{1.99086}{0.00019} & \colsym{\RGam{Z}{\nu}{\ell}}{1.99076}{0.00017} \\
 \calsym{\Bf{W}{had}}{0.67523}{0.00019} & \colsym{\Bf{W}{had}}{0.67522}{0.00019} \\
 \calsym{\Bf{W}{e\nu}}{0.108255}{0.000065} & \colsym{\Bf{W}{e\nu}}{0.108260}{0.000063} \\
 \calsym{\Bf{W}{\mu\nu}}{0.108255}{0.000065} & \colsym{\Bf{W}{\mu\nu}}{0.108260}{0.000063} \\
 \calsym{\Bf{W}{\tau\nu}}{0.108255}{0.000065} & \colsym{\Bf{W}{\tau\nu}}{0.108260}{0.000063} \\
 \calsym{\Bf{Z}{had}}{0.69812}{0.00019} & \colsym{\Bf{Z}{had}}{0.69807}{0.00018} \\
 \calsym{\Bf{Z}{ee}}{0.033653}{0.000022} & \colsym{\Bf{Z}{ee}}{0.033659}{0.000020} \\
 \calsym{\Bf{Z}{\mu\mu}}{0.033653}{0.000022} & \colsym{\Bf{Z}{\mu\mu}}{0.033659}{0.000020} \\
 \calsym{\Bf{Z}{\tau\tau}}{0.033577}{0.000022} & \colsym{\Bf{Z}{\tau\tau}}{0.033583}{0.000020} \\
 \calasy{\Bf{Z}{dd}}{0.153464}{0.000042}{0.000030} & \colsym{\Bf{Z}{dd}}{0.153484}{0.000014} \\
 \calasy{\Bf{Z}{ss}}{0.153464}{0.000042}{0.000030} & \colsym{\Bf{Z}{ss}}{0.153484}{0.000014} \\
 \calasy{\Bf{Z}{bb}}{0.15067}{0.00025}{0.00033} & \colsym{\Bf{Z}{bb}}{0.15052}{0.00010} \\
 \calasy{\Bf{Z}{uu}}{0.120273}{0.000110}{0.000087} & \colsym{\Bf{Z}{uu}}{0.120307}{0.000069} \\
 \calasy{\Bf{Z}{cc}}{0.120248}{0.000113}{0.000091} & \colsym{\Bf{Z}{cc}}{0.120282}{0.000073} \\
 \calsym{\Bf{Z}{inv}}{0.20100}{0.00013} & \colsym{\Bf{Z}{inv}}{0.20102}{0.00012} \\
\hline
\end{tabular}
\caption[$\SM$ predictions for pseudo-observables]{$\SM$ predictions
for pseudo-observables derived from the Z-pole and the high-$Q^2$
$\SM$ fits.}
\label{app:SM:pred:2}
\end{center}
\end{table}

\begin{table}[ht]
\begin{center}
\renewcommand{\arraystretch}{1.2}
\begin{tabular}{|l||r@{}l||r@{}l|}
\hline
 Observable & \multicolumn{2}{|c||}{Standard Model}
            & \multicolumn{2}{|c|}{Standard Model} \\
            & \multicolumn{2}{|c||}{Z-pole Fit    } 
            & \multicolumn{2}{|c|}{High-$Q^2$ Fit} \\
\hline
\hline           
 \calasy{\g{A\nu}}{0.50199}{0.00017}{0.00020} & \colsym{\g{A\nu}}{0.502112}{0.000069} \\
 \calasy{\g{Ae}}{-0.50127}{0.00020}{0.00017} & \colsym{\g{Ae}}{-0.501389}{0.000068} \\
 \calasy{\g{A\mu}}{-0.50127}{0.00020}{0.00017} & \colsym{\g{A\mu}}{-0.501389}{0.000068} \\
 \calasy{\g{A\tau}}{-0.50127}{0.00020}{0.00017} & \colsym{\g{A\tau}}{-0.501389}{0.000068} \\
 \calasy{\g{Ad}}{-0.50168}{0.00020}{0.00017} & \colsym{\g{Ad}}{-0.501803}{0.000068} \\
 \calasy{\g{As}}{-0.50168}{0.00020}{0.00017} & \colsym{\g{As}}{-0.501803}{0.000068} \\
 \calasy{\g{Ab}}{-0.49856}{0.00041}{0.00020} & \colasy{\g{Ab}}{-0.49844}{0.00013}{0.00010} \\
 \calasy{\g{Au}}{0.50144}{0.00017}{0.00020} & \colsym{\g{Au}}{0.501562}{0.000068} \\
 \calasy{\g{Ac}}{0.50144}{0.00017}{0.00020} & \colsym{\g{Ac}}{0.501562}{0.000068} \\
 \calasy{\g{V\nu}}{0.50199}{0.00017}{0.00020} & \colsym{\g{V\nu}}{0.502112}{0.000069} \\
 \calsym{\g{Ve}}{-0.03712}{0.00032} & \colsym{\g{Ve}}{-0.03730}{0.00028} \\
 \calsym{\g{V\mu}}{-0.03712}{0.00032} & \colsym{\g{V\mu}}{-0.03730}{0.00028} \\
 \calsym{\g{V\tau}}{-0.03712}{0.00032} & \colsym{\g{V\tau}}{-0.03730}{0.00028} \\
 \calsym{\g{Vd}}{-0.34699}{0.00017} & \colsym{\g{Vd}}{-0.34714}{0.00012} \\
 \calsym{\g{Vs}}{-0.34699}{0.00017} & \colsym{\g{Vs}}{-0.34714}{0.00012} \\
 \calasy{\g{Vb}}{-0.34372}{0.00049}{0.00028} & \colasy{\g{Vb}}{-0.34360}{0.00018}{0.00016} \\
 \calsym{\g{Vu}}{0.19204}{0.00023} & \colsym{\g{Vu}}{0.19221}{0.00020} \\
 \calsym{\g{Vc}}{0.19204}{0.00023} & \colsym{\g{Vc}}{0.19221}{0.00020} \\
 \calasy{\g{L\nu}}{0.50199}{0.00017}{0.00020} & \colsym{\g{L\nu}}{0.502112}{0.000069} \\
 \calsym{\g{Le}}{-0.26919}{0.00020} & \colsym{\g{Le}}{-0.26935}{0.00016} \\
 \calsym{\g{L\mu}}{-0.26919}{0.00020} & \colsym{\g{L\mu}}{-0.26935}{0.00016} \\
 \calsym{\g{L\tau}}{-0.26919}{0.00020} & \colsym{\g{L\tau}}{-0.26935}{0.00016} \\
 \calasy{\g{Ld}}{-0.42434}{0.00018}{0.00016} & \colsym{\g{Ld}}{-0.424470}{0.000087} \\
 \calasy{\g{Ls}}{-0.42434}{0.00018}{0.00016} & \colsym{\g{Ls}}{-0.424470}{0.000087} \\
 \calasy{\g{Lb}}{-0.42114}{0.00045}{0.00024} & \colasy{\g{Lb}}{-0.42102}{0.00015}{0.00013} \\
 \calsym{\g{Lu}}{0.34674}{0.00017} & \colsym{\g{Lu}}{0.34688}{0.00012} \\
 \calsym{\g{Lc}}{0.34674}{0.00017} & \colsym{\g{Lc}}{0.34688}{0.00012} \\
 \calasy{\g{Re}}{0.23208}{0.00016}{0.00018} & \colsym{\g{Re}}{0.23204}{0.00013} \\
 \calasy{\g{R\mu}}{0.23208}{0.00016}{0.00018} & \colsym{\g{R\mu}}{0.23204}{0.00013} \\
 \calasy{\g{R\tau}}{0.23208}{0.00016}{0.00018} & \colsym{\g{R\tau}}{0.23204}{0.00013} \\
 \calasy{\g{Rd}}{0.077345}{0.000053}{0.000061} & \colsym{\g{Rd}}{0.077333}{0.000044} \\
 \calasy{\g{Rs}}{0.077345}{0.000053}{0.000061} & \colsym{\g{Rs}}{0.077333}{0.000044} \\
 \calasy{\g{Rb}}{0.077420}{0.000052}{0.000061} & \colsym{\g{Rb}}{0.077417}{0.000040} \\
 \calsym{\g{Ru}}{-0.15470}{0.00011} & \colsym{\g{Ru}}{-0.154677}{0.000087} \\
 \calsym{\g{Rc}}{-0.15470}{0.00011} & \colsym{\g{Rc}}{-0.154677}{0.000087} \\
\hline
\end{tabular}
\caption[$\SM$ predictions for pseudo-observables]{$\SM$ predictions
for pseudo-observables derived from the Z-pole and the high-$Q^2$
$\SM$ fits.}
\label{app:SM:pred:3}
\end{center}
\end{table}

\begin{table}[ht]
\begin{center}
\renewcommand{\arraystretch}{1.2}
\begin{tabular}{|l||r@{}l||r@{}l|}
\hline
 Observable & \multicolumn{2}{|c||}{Standard Model}
            & \multicolumn{2}{|c|}{Standard Model} \\
            & \multicolumn{2}{|c||}{Z-pole Fit    } 
            & \multicolumn{2}{|c|}{High-$Q^2$ Fit} \\
\hline
\hline           
 \calsym{\cA{}{e}}{0.1473}{0.0012} & \colsym{\cA{}{e}}{0.1480}{0.0011} \\
 \calsym{\cA{}{\mu}}{0.1473}{0.0012} & \colsym{\cA{}{\mu}}{0.1480}{0.0011} \\
 \calsym{\cA{}{\tau}}{0.1473}{0.0012} & \colsym{\cA{}{\tau}}{0.1480}{0.0011} \\
 \calsym{\cA{}{d}}{0.93569}{0.00010} & \colsym{\cA{}{d}}{0.935748}{0.000088} \\
 \calsym{\cA{}{s}}{0.93569}{0.00010} & \colsym{\cA{}{s}}{0.935748}{0.000088} \\
 \calasy{\cA{}{b}}{0.93462}{0.00016}{0.00020} & \colsym{\cA{}{b}}{0.934588}{0.000100} \\
 \calsym{\cA{}{u}}{0.66798}{0.00055} & \colsym{\cA{}{u}}{0.66829}{0.00048} \\
 \calsym{\cA{}{c}}{0.66798}{0.00055} & \colsym{\cA{}{c}}{0.66829}{0.00048} \\
 \calsym{\cP{\tau}{e}}{0.1473}{0.0012} & \colsym{\cP{\tau}{e}}{0.1480}{0.0011} \\
 \calsym{\cP{\tau}{\tau}}{0.1473}{0.0012} & \colsym{\cP{\tau}{\tau}}{0.1480}{0.0011} \\
 \calsym{\sig{0}{had}~[\nb]}{41.476}{0.015} & \colsym{\sig{0}{had}~[\nb]}{41.481}{0.014} \\
 \calsym{\sig{0}{e}~[\nb]}{1.9994}{0.0026} & \colsym{\sig{0}{e}~[\nb]}{2.0001}{0.0024} \\
 \calsym{\sig{0}{\mu}~[\nb]}{1.9993}{0.0026} & \colsym{\sig{0}{\mu}~[\nb]}{2.0001}{0.0024} \\
 \calsym{\sig{0}{\tau}~[\nb]}{1.9948}{0.0026} & \colsym{\sig{0}{\tau}~[\nb]}{1.9956}{0.0024} \\
 \calsym{\R{0}{e}}{20.744}{0.019} & \colsym{\R{0}{e}}{20.739}{0.018} \\
 \calsym{\R{0}{\mu}}{20.745}{0.019} & \colsym{\R{0}{\mu}}{20.740}{0.018} \\
 \calsym{\R{0}{\tau}}{20.792}{0.019} & \colsym{\R{0}{\tau}}{20.786}{0.018} \\
 \calasy{\R{0}{d}}{0.219824}{0.000093}{0.000082} & \colsym{\R{0}{d}}{0.219868}{0.000055} \\
 \calasy{\R{0}{s}}{0.219824}{0.000093}{0.000082} & \colsym{\R{0}{s}}{0.219868}{0.000055} \\
 \calasy{\R{0}{b}}{0.21583}{0.00033}{0.00045} & \colsym{\R{0}{b}}{0.21562}{0.00013} \\
 \calasy{\R{0}{u}}{0.17228}{0.00015}{0.00011} & \colsym{\R{0}{u}}{0.172341}{0.000063} \\
 \calasy{\R{0}{c}}{0.17225}{0.00016}{0.00012} & \colsym{\R{0}{c}}{0.172305}{0.000068} \\
 \calsym{\A{0,e}{FB}}{0.01627}{0.00027} & \colsym{\A{0,e}{FB}}{0.01642}{0.00024} \\
 \calsym{\A{0,\mu}{FB}}{0.01627}{0.00027} & \colsym{\A{0,\mu}{FB}}{0.01642}{0.00024} \\
 \calsym{\A{0,\tau}{FB}}{0.01627}{0.00027} & \colsym{\A{0,\tau}{FB}}{0.01642}{0.00024} \\
 \calsym{\A{0,d}{FB}}{0.10335}{0.00088} & \colsym{\A{0,d}{FB}}{0.10385}{0.00078} \\
 \calsym{\A{0,s}{FB}}{0.10335}{0.00088} & \colsym{\A{0,s}{FB}}{0.10385}{0.00078} \\
 \calsym{\A{0,b}{FB}}{0.10324}{0.00088} & \colsym{\A{0,b}{FB}}{0.10373}{0.00077} \\
 \calsym{\A{0,u}{FB}}{0.07378}{0.00068} & \colsym{\A{0,u}{FB}}{0.07417}{0.00060} \\
 \calsym{\A{0,c}{FB}}{0.07378}{0.00068} & \colsym{\A{0,c}{FB}}{0.07417}{0.00060} \\
\hline
\end{tabular}
\caption[$\SM$ predictions for pseudo-observables]{$\SM$ predictions
for pseudo-observables derived from the Z-pole and the high-$Q^2$
$\SM$ fits.}
\label{app:SM:pred:4}
\end{center}
\end{table}

\begin{table}[ht]
\begin{center}
\renewcommand{\arraystretch}{1.2}
\begin{tabular}{|l||r@{}l||r@{}l|}
\hline
 Observable & \multicolumn{2}{|c||}{Standard Model}
            & \multicolumn{2}{|c|}{Standard Model} \\
            & \multicolumn{2}{|c||}{Z-pole Fit    } 
            & \multicolumn{2}{|c|}{High-$Q^2$ Fit} \\
\hline
\hline
 \calasy{\ro{\nu}}{1.00799}{0.00068}{0.00081} & \colsym{\ro{\nu}}{1.00847}{0.00028} \\
 \calasy{\ro{e}}{1.00509}{0.00067}{0.00081} & \colsym{\ro{e}}{1.00556}{0.00027} \\
 \calasy{\ro{\mu}}{1.00509}{0.00067}{0.00081} & \colsym{\ro{\mu}}{1.00556}{0.00027} \\
 \calasy{\ro{\tau}}{1.00509}{0.00067}{0.00081} & \colsym{\ro{\tau}}{1.00556}{0.00027} \\
 \calasy{\ro{d}}{1.00675}{0.00067}{0.00081} & \colsym{\ro{d}}{1.00723}{0.00027} \\
 \calasy{\ro{s}}{1.00675}{0.00067}{0.00081} & \colsym{\ro{s}}{1.00723}{0.00027} \\
 \calasy{\ro{b}}{0.99426}{0.00079}{0.00164} & \colasy{\ro{b}}{0.99376}{0.00040}{0.00052} \\
 \calasy{\ro{u}}{1.00578}{0.00067}{0.00081} & \colsym{\ro{u}}{1.00626}{0.00027} \\
 \calasy{\ro{c}}{1.00578}{0.00067}{0.00081} & \colsym{\ro{c}}{1.00626}{0.00027} \\
 \calsym{\stw{\nu}{eff}}{0.23111}{0.00016} & \colsym{\stw{\nu}{eff}}{0.23102}{0.00014} \\
 \calsym{\stw{e}{eff}}{0.23149}{0.00016} & \colsym{\stw{e}{eff}}{0.23140}{0.00014} \\
 \calsym{\stw{\mu}{eff}}{0.23149}{0.00016} & \colsym{\stw{\mu}{eff}}{0.23140}{0.00014} \\
 \calsym{\stw{\tau}{eff}}{0.23149}{0.00016} & \colsym{\stw{\tau}{eff}}{0.23140}{0.00014} \\
 \calsym{\stw{d}{eff}}{0.23126}{0.00016} & \colsym{\stw{d}{eff}}{0.23117}{0.00014} \\
 \calsym{\stw{s}{eff}}{0.23126}{0.00016} & \colsym{\stw{s}{eff}}{0.23117}{0.00014} \\
 \calasy{\stw{b}{eff}}{0.23293}{0.00031}{0.00025} & \colsym{\stw{b}{eff}}{0.23298}{0.00016} \\
 \calsym{\stw{u}{eff}}{0.23138}{0.00016} & \colsym{\stw{u}{eff}}{0.23129}{0.00014} \\
 \calsym{\stw{c}{eff}}{0.23138}{0.00016} & \colsym{\stw{c}{eff}}{0.23129}{0.00014} \\
 \calsym{\ka{\nu}}{1.0349}{0.0026} & \colsym{\ka{\nu}}{1.0368}{0.0012} \\
 \calsym{\ka{e}}{1.0366}{0.0026} & \colsym{\ka{e}}{1.0385}{0.0012} \\
 \calsym{\ka{\mu}}{1.0366}{0.0026} & \colsym{\ka{\mu}}{1.0385}{0.0012} \\
 \calsym{\ka{\tau}}{1.0366}{0.0026} & \colsym{\ka{\tau}}{1.0385}{0.0012} \\
 \calsym{\ka{d}}{1.0356}{0.0026} & \colsym{\ka{d}}{1.0375}{0.0012} \\
 \calsym{\ka{s}}{1.0356}{0.0026} & \colsym{\ka{s}}{1.0375}{0.0012} \\
 \calsym{\ka{b}}{1.0431}{0.0036} & \colsym{\ka{b}}{1.0456}{0.0015} \\
 \calsym{\ka{u}}{1.0361}{0.0026} & \colsym{\ka{u}}{1.0381}{0.0012} \\
 \calsym{\ka{c}}{1.0361}{0.0026} & \colsym{\ka{c}}{1.0381}{0.0012} \\
\hline
\end{tabular}
\caption[$\SM$ predictions for pseudo-observables]{$\SM$ predictions
for pseudo-observables derived from the Z-pole and the high-$Q^2$
$\SM$ fits.}
\label{app:SM:pred:5}
\end{center}
\end{table}

\begin{table}[ht]
\begin{center}
\renewcommand{\arraystretch}{1.2}
\begin{tabular}{|l||r@{}l||r@{}l|}
\hline
 Observable & \multicolumn{2}{|c||}{Standard Model}
            & \multicolumn{2}{|c|}{Standard Model} \\
            & \multicolumn{2}{|c||}{Z-pole Fit    } 
            & \multicolumn{2}{|c|}{High-$Q^2$ Fit} \\
\hline
\hline           
 \calasy{\eps{1}}{0.00506}{0.00066}{0.00080} & \colsym{\eps{1}}{0.00553}{0.00027} \\
 \calasy{\eps{2}}{-0.00746}{0.00018}{0.00021} & \colsym{\eps{2}}{-0.007612}{0.000100} \\
 \calasy{\eps{3}}{0.00502}{0.00053}{0.00076} & \colasy{\eps{3}}{0.00511}{0.00027}{0.00039} \\
 \calasy{\eps{b}}{-0.00540}{0.00079}{0.00108} & \colsym{\eps{b}}{-0.00588}{0.00032} \\
 \calasy{\stu{S}}{0.598}{0.064}{0.090} & \colasy{\stu{S}}{0.608}{0.032}{0.046} \\
 \calasy{\stu{T}}{0.653}{0.085}{0.103} & \colsym{\stu{T}}{0.713}{0.035} \\
 \calasy{\stu{U}}{0.889}{0.026}{0.022} & \colasy{\stu{U}}{0.907}{0.013}{0.011} \\
 \calasy{\stu{\gamma_b}}{-0.0124}{0.0018}{0.0025} & \colsym{\stu{\gamma_b}}{-0.01347}{0.00073} \\
 \calasy{\sts{S}}{-0.025}{0.064}{0.090} & \colasy{\sts{S}}{-0.014}{0.032}{0.046} \\
 \calasy{\sts{T}}{-0.010}{0.085}{0.103} & \colsym{\sts{T}}{0.051}{0.035} \\
 \calasy{\sts{U}}{-0.001}{0.026}{0.022} & \colasy{\sts{U}}{0.017}{0.013}{0.011} \\
 \calasy{\sts{\gamma_b}}{0.0005}{0.0018}{0.0025} & \colsym{\sts{\gamma_b}}{-0.00066}{0.00073} \\
 \calsym{\RGam{Z}{\mu}{e}}{0.99999198984}{0.00000000084} & \colsym{\RGam{Z}{\mu}{e}}{0.99999199026}{0.00000000075} \\
 \calsym{\RGam{Z}{\tau}{e}}{0.99773494}{0.00000024} & \colsym{\RGam{Z}{\tau}{e}}{0.99773506}{0.00000021} \\
 \calasy{\RGam{Z}{d}{b}}{1.0185}{0.0025}{0.0019} & \colsym{\RGam{Z}{d}{b}}{1.01971}{0.00077} \\
 \calasy{\RGam{Z}{s}{b}}{1.0185}{0.0025}{0.0019} & \colsym{\RGam{Z}{s}{b}}{1.01971}{0.00077} \\
 \calsym{\RGam{Z}{u}{c}}{1.000205}{0.000036} & \colsym{\RGam{Z}{u}{c}}{1.000207}{0.000035} \\
 \calasy{\g{Ad}/\g{Ab}}{1.00626}{0.00110}{0.00080} & \colsym{\g{Ad}/\g{Ab}}{1.00675}{0.00032} \\
 \calasy{\g{As}/\g{Ab}}{1.00626}{0.00110}{0.00080} & \colsym{\g{As}/\g{Ab}}{1.00675}{0.00032} \\
 \calasy{\g{Vd}/\g{Vb}}{1.0095}{0.0016}{0.0012} & \colsym{\g{Vd}/\g{Vb}}{1.01028}{0.00048} \\
 \calasy{\g{Vs}/\g{Vb}}{1.0095}{0.0016}{0.0012} & \colsym{\g{Vs}/\g{Vb}}{1.01028}{0.00048} \\
 \calsym{\MV{Z}~[\GeV]}{91.1532}{0.0021} & \colsym{\MV{Z}~[\GeV]}{91.1532}{0.0021} \\
 \calsym{\GAM{Z}{tot}~[\GeV]}{2.4947}{0.0019} & \colsym{\GAM{Z}{tot}~[\GeV]}{2.4956}{0.0015} \\
\hline
\end{tabular}
\caption[$\SM$ predictions for pseudo-observables]{$\SM$ predictions
for pseudo-observables derived from the Z-pole and the high-$Q^2$
$\SM$ fits.}
\label{app:SM:pred:6}
\end{center}
\end{table}

\begin{table}[ht]
\begin{center}
\renewcommand{\arraystretch}{1.2}
\begin{tabular}{|l||r@{}l||r@{}l|}
\hline
 Observable & \multicolumn{2}{|c||}{Standard Model}
            & \multicolumn{2}{|c|}{Standard Model} \\
            & \multicolumn{2}{|c||}{Z-pole Fit    } 
            & \multicolumn{2}{|c|}{High-$Q^2$ Fit} \\
\hline
\hline           
\calsym{\Ciq{1}{u}}{-0.18883}{0.00024} & \colsym{\Ciq{1}{u}}{-0.18902}{0.00018} \\
 \calasy{\Ciq{1}{d}}{0.34105}{0.00030}{0.00027} & \colsym{\Ciq{1}{d}}{0.34126}{0.00013} \\
 \calsym{\Ciq{2}{u}}{-0.03751}{0.00027} & \colsym{\Ciq{2}{u}}{-0.03762}{0.00023} \\
 \calsym{\Ciq{2}{d}}{0.02351}{0.00025} & \colsym{\Ciq{2}{d}}{0.02358}{0.00021} \\
 \calasy{\QW{Cs}}{-72.926}{0.065}{0.072} & \colsym{\QW{Cs}}{-72.942}{0.037} \\
 \calsym{\QW{Tl}}{-116.40}{0.11} & \colsym{\QW{Tl}}{-116.432}{0.056} \\
 \calasy{\QW{Pb}}{-118.30}{0.10}{0.12} & \colsym{\QW{Pb}}{-118.332}{0.057} \\
 \calasy{\QW{Bi}}{-119.22}{0.10}{0.12} & \colsym{\QW{Bi}}{-119.246}{0.057} \\
 \calsym{\g{{\nu}Lud}^2}{0.30381}{0.00043} & \colsym{\g{{\nu}Lud}^2}{0.30415}{0.00022} \\
 \calsym{\g{{\nu}Rud}^2}{0.030126}{0.000057} & \colsym{\g{{\nu}Rud}^2}{0.030142}{0.000031} \\
 \calsym{S_{qq}}{6.7502}{0.0030} & \colasy{S_{qq}}{6.7526}{0.0021}{0.0024} \\
 \calsym{S_{qq\gamma}}{13.6700}{0.0085} & \colsym{S_{qq\gamma}}{13.6768}{0.0052} \\
 \calasy{\Gam{Z}{u}/\Gam{Z}{u+d+s}}{0.28154}{0.00014}{0.00012} & \colsym{\Gam{Z}{u}/\Gam{Z}{u+d+s}}{0.28157}{0.00010} \\
 \calasy{\Gam{Z}{d}/\Gam{Z}{u+d+s}}{0.359231}{0.000058}{0.000069} & \colsym{\Gam{Z}{d}/\Gam{Z}{u+d+s}}{0.359216}{0.000052} \\
\hline
\end{tabular}
\caption[$\SM$ predictions for pseudo-observables]{$\SM$ predictions
for pseudo-observables derived from the Z-pole and the high-$Q^2$
$\SM$ fits.}
\label{app:SM:pred:7}
\end{center}
\end{table}

\bibliographystyle{PhysRep}

\bibliography{physrep}

\begin{mcbibliography}{100}

\bibitem{Hasert:1973ff}
Gargamelle Neutrino Collaboration, F.~J. Hasert {\it et~al.},
\newblock  Phys. Lett. {\bf B46}  (1973) 138\relax
\relax
\bibitem{Arnison:1983rp}
UA1 Collaboration, G. Arnison {\it et~al.},
\newblock  Phys. Lett. {\bf B122}  (1983) 103\relax
\relax
\bibitem{Banner:1983jy}
UA2 Collaboration, M. Banner {\it et~al.},
\newblock  Phys. Lett. {\bf B122}  (1983) 476\relax
\relax
\bibitem{Arnison:1983mk}
UA1 Collaboration, G. Arnison {\it et~al.},
\newblock  Phys. Lett. {\bf B126}  (1983) 398\relax
\relax
\bibitem{Bagnaia:1983zx}
UA2 Collaboration, P. Bagnaia {\it et~al.},
\newblock  Phys. Lett. {\bf B129}  (1983) 130\relax
\relax
\bibitem{Glashow:1961tr}
S.~L. Glashow,
\newblock  Nucl. Phys. {\bf 22}  (1961) 579\relax
\relax
\bibitem{Weinberg:1967tq}
S. Weinberg,
\newblock  Phys. Rev. Lett. {\bf 19}  (1967) 1264\relax
\relax
\bibitem{Salam:1968rm}
A. Salam,
\newblock  Weak and Electromagnetic Interactions, p. 367, in Elementary
  Particle Theory, Proceedings of the 1968 Nobel Symposium, ed. N. Svartholm,
\newblock  (Almquist and Wiksells, Stockholm, 1968)\relax
\relax
\bibitem{Veltman:1968ki}
M. Veltman,
\newblock  Nucl. Phys. {\bf B7}  (1968) 637\relax
\relax
\bibitem{'tHooft:1971rn}
G. 't~Hooft,
\newblock  Nucl. Phys. {\bf B35}  (1971) 167\relax
\relax
\bibitem{'tHooft:1972fi}
G. 't~Hooft and M. Veltman,
\newblock  Nucl. Phys. {\bf B44}  (1972) 189\relax
\relax
\bibitem{'tHooft:1972ue}
G. 't~Hooft and M. Veltman,
\newblock  Nucl. Phys. {\bf B50}  (1972) 318\relax
\relax
\bibitem{LEP}
{\em {\it LEP Design Report}, CERN-LEP/84-01},
\newblock  Internal report, CERN, 1984,
\newblock  \\The main features of LEP have been reviewed by:\relax
\relax
\bibitem{Myers:1990sk}
S. Myers and E. Picasso,
\newblock  Contemp. Phys. {\bf 31}  (1990) 387--403\relax
\relax
\bibitem{Brandt:2000xk}
D. Brandt {\it et~al.},
\newblock  Rept. Prog. Phys. {\bf 63}  (2000) 939--1000,
\newblock  \\A useful retrospective view of the accelerator is presented
  in:\relax
\relax
\bibitem{Assmann:2002th}
R. Assmann, M. Lamont, and S. Myers,
\newblock  Nucl. Phys. Proc. Suppl. {\bf 109B}  (2002) 17--31\relax
\relax
\bibitem{SLC}
{\em {\it SLAC Linear Collider Conceptual Design Report}, SLAC-R-229},
\newblock  Internal report, SLAC, 1980\relax
\relax
\bibitem{Decamp:1990jr}
ALEPH Collaboration, D. Decamp {\it et~al.},
\newblock  Nucl. Instrum. Meth. {\bf A294}  (1990) 121--178\relax
\relax
\bibitem{Buskulic:1995wz}
ALEPH Collaboration, D. Buskulic {\it et~al.},
\newblock  Nucl. Instrum. Meth. {\bf A360}  (1995) 481--506\relax
\relax
\bibitem{Aarnio:1991vx}
DELPHI Collaboration, P. Aarnio {\it et~al.},
\newblock  Nucl. Instrum. Meth. {\bf A303}  (1991) 233--276\relax
\relax
\bibitem{Abreu:1996uz}
DELPHI Collaboration, P. Abreu {\it et~al.},
\newblock  Nucl. Instrum. Meth. {\bf A378}  (1996) 57--100\relax
\relax
\bibitem{L3:1990kx}
L3 Collaboration, B. Adeva {\it et~al.},
\newblock  Nucl. Instrum. Meth. {\bf A289}  (1990) 35--102\relax
\relax
\bibitem{Acciarri:1994yk}
M. Acciarri {\it et~al.},
\newblock  Nucl. Instrum. Meth. {\bf A351}  (1994) 300--312\relax
\relax
\bibitem{Chemarin:1994bp}
M. Chemarin {\it et~al.},
\newblock  Nucl. Instrum. Meth. {\bf A349}  (1994) 345--355\relax
\relax
\bibitem{Adam:1996fj}
A. Adam {\it et~al.},
\newblock  Nucl. Instrum. Meth. {\bf A383}  (1996) 342--366\relax
\relax
\bibitem{Ahmet:1991eg}
OPAL Collaboration, K. Ahmet {\it et~al.},
\newblock  Nucl. Instrum. Meth. {\bf A305}  (1991) 275--319\relax
\relax
\bibitem{Allport:1993kp}
OPAL Collaboration, P.~P. Allport {\it et~al.},
\newblock  Nucl. Instrum. Meth. {\bf A324}  (1993) 34--52\relax
\relax
\bibitem{Allport:1994ec}
OPAL Collaboration, P.~P. Allport {\it et~al.},
\newblock  Nucl. Instrum. Meth. {\bf A346}  (1994) 476--495\relax
\relax
\bibitem{Anderson:1994ve}
OPAL Collaboration, B.~E. Anderson {\it et~al.},
\newblock  IEEE Trans. Nucl. Sci. {\bf 41}  (1994) 845--852\relax
\relax
\bibitem{Abrams:1989aw}
Mark-II Collaboration, G.~S. Abrams {\it et~al.},
\newblock  Phys. Rev. Lett. {\bf 63}  (1989) 724\relax
\relax
\bibitem{SLD-CDC}
SLD Collaboration, M.~J. Fero {\it et~al.},
\newblock  Nucl. Instrum. Meth. {\bf A367}  (1995) 111\relax
\relax
\bibitem{SLD-LAC}
SLD Collaboration, D. Axen {\it et~al.},
\newblock  Nucl. Instrum. Meth. {\bf A328}  (1993) 472\relax
\relax
\bibitem{SLD-CRID}
SLD Collaboration, K. Abe {\it et~al.},
\newblock  Nucl. Instrum. Meth. {\bf A343}  (1994) 74\relax
\relax
\bibitem{SLD-LUM}
SLD Collaboration, S.~C. Berridge {\it et~al.},
\newblock  IEEE Trans. Nucl. Sci. {\bf 39}  (1992) 1242\relax
\relax
\bibitem{SLD-VXD}
SLD Collaboration, K. Abe {\it et~al.},
\newblock  Nucl. Instrum. Meth. {\bf A400}  (1997) 287\relax
\relax
\bibitem{SLD-WIC}
SLD Collaboration, A.~C. Benvenuti {\it et~al.},
\newblock  Nucl. Instrum. Meth. {\bf A276}  (1989) 94\relax
\relax
\bibitem{bib-MZpaper}
{Working Group on LEP Energy (L.{\,}Arnaudon {\etal})},
\newblock  Phys. Lett. {\bf B307}  (1993) 187--193\relax
\relax
\bibitem{bib-ECAL92}
{Working Group on LEP Energy, L.{\,}Arnaudon {\etal}},
\newblock  {\em The Energy Calibration of LEP in 1992},
\newblock  Preprint CERN SL/93-21 (DI), CERN, 1993\relax
\relax
\bibitem{bib-ECAL93}
{Working Group on LEP Energy (R.{\,}Assmann {\etal})},
\newblock  Z. Phys. {\bf C66}  (1995) 567--582\relax
\relax
\bibitem{bib-ECAL95}
R. Assmann {\it et~al.},
\newblock  Eur. Phys. J. {\bf C6}  (1999) 187--223\relax
\relax
\bibitem{bib-transpol}
A.~A. Sokolov and I.~M. Ternov,
\newblock  Phys. Dokl. {\bf 8}  (1964) 1203--1205\relax
\relax
\bibitem{JETSET}
T. Sjostrand,
\newblock  Comput. Phys. Commun. {\bf 82}  (1994) 74--90,
\newblock  (JETSET)\relax
\relax
\bibitem{HERWIG}
G. Marchesini {\it et~al.},
\newblock  Comput. Phys. Commun. {\bf 67}  (1992) 465--508,
\newblock  (HERWIG)\relax
\relax
\bibitem{ARIADNE}
L. Lonnblad,
\newblock  Comput. Phys. Commun. {\bf 71}  (1992) 15--31,
\newblock  (ARIADNE)\relax
\relax
\bibitem{KORALZ}
{S. Jadach, B.F.L. Ward and Z. W{\c{a}s}},
\newblock  Comput. Phys. Commun. {\bf 79}  (1994) 503,
\newblock  (KORALZ 4.0)\relax
\relax
\bibitem{KK}
{S. Jadach, B.F.L. Ward and Z. W{\c{a}s}},
\newblock  Comput. Phys. Commun. {\bf 130}  (2000) 260,
\newblock  (KK Monte Carlo)\relax
\relax
\bibitem{BABAMC}
{F.A. Berends, R. Kleiss and W. Hollik},
\newblock  Nucl. Phys. {\bf B304}  (1988) 712,
\newblock  (BABAMC)\relax
\relax
\bibitem{BHLUMI4}
{S. Jadach, W. Placzek, E. Richter-W{\c{a}s}, B.F.L. Ward and Z. W{\c{a}s}},
\newblock  Comput. Phys. Commun. {\bf 102}  (1997) 229,
\newblock  (BHLUMI 4.04)\relax
\relax
\bibitem{GEANT}
R. Brun {\it et~al.},
\newblock  {\em GEANT3},
\newblock  Preprint CERN DD/EE/84-1, CERN, 1987,
\newblock  details of its implementation may be found in the individual
  detector references,~\cite{\ALEPHdet,\DELPHIdet,\Ldet,\OPALdet}\relax
\relax
\bibitem{Ross:1975fq}
D.~A. Ross and M.~J.~G. Veltman,
\newblock  Nucl. Phys. {\bf B95}  (1975) 135\relax
\relax
\bibitem{Veltman:1977kh}
M.~J.~G. Veltman,
\newblock  Nucl. Phys. {\bf B123}  (1977) 89\relax
\relax
\bibitem{Ross:1973fp}
D.~A. Ross and J.~C. Taylor,
\newblock  Nucl. Phys. {\bf B51}  (1973) 125--144\relax
\relax
\bibitem{Sirlin:1980nh}
A. Sirlin,
\newblock  Phys. Rev. {\bf D22}  (1980) 971--981\relax
\relax
\bibitem{Burgers:LEP1YR89VOL1}
G. Burgers and F. Jegerlehner, in {\em Z Physics At Lep 1. Proceedings,
  Workshop, Geneva, Switzerland, September 4-5, 1989. Vol. 1: Standard
  Physics}, CERN 89-08, ed. G. Altarelli, R. Kleiss, and C. Verzegnassi,
  (CERN, Geneva, Switzerland, 1989), p.~55\relax
\relax
\bibitem{Jegerlehner:1991dq}
F. Jegerlehner, in {\em Testing the Standard Model - TASI-90}, proceedings:
  Theoretical Advanced Study Institute in Elementary Particle Physics, Boulder,
  Colo., Jun 3-27, 1990, ed. M. Cvetic and P. Langacker,  (World Scientific,
  Singapore, 1991), p. 916\relax
\relax
\bibitem{Kawamoto:2004pi}
T. Kawamoto and R.~G. Kellogg,
\newblock  Phys. Rev. {\bf D69}  (2004) 113008\relax
\relax
\bibitem{Montagna:1993py}
G. Montagna {\it et~al.},
\newblock  Nucl. Phys. {\bf B401}  (1993) 3--66\relax
\relax
\bibitem{Montagna:1993ai}
G. Montagna {\it et~al.},
\newblock  Comput. Phys. Commun. {\bf 76}  (1993) 328--360\relax
\relax
\bibitem{Montagna:1996ja}
G. Montagna {\it et~al.},
\newblock  Comput. Phys. Commun. {\bf 93}  (1996) 120--126\relax
\relax
\bibitem{Montagna:1998kp}
G. Montagna {\it et~al.},
\newblock  Comput. Phys. Commun. {\bf 117}  (1999) 278--289,
\newblock  updated to include initial state pair radiation (G. Passarino, priv.
  comm.)\relax
\relax
\bibitem{Bardin:1989di}
D.~Y. Bardin {\it et~al.},
\newblock  Z. Phys. {\bf C44}  (1989) 493\relax
\relax
\bibitem{Bardin:1990tq}
D.~Y. Bardin {\it et~al.},
\newblock  Comput. Phys. Commun. {\bf 59}  (1990) 303--312\relax
\relax
\bibitem{Bardin:1991fu}
D.~Y. Bardin {\it et~al.},
\newblock  Nucl. Phys. {\bf B351}  (1991) 1--48\relax
\relax
\bibitem{Bardin:1991de}
D.~Y. Bardin {\it et~al.},
\newblock  Phys. Lett. {\bf B255}  (1991) 290--296\relax
\relax
\bibitem{Bardin:1992jc}
D.~Y. Bardin {\it et~al.},
\newblock  {\em ZFITTER: An Analytical program for fermion pair production in
  ${\ee}$ annihilation},
\newblock  Eprint arXiv:hep-ph/9412201, 1992\relax
\relax
\bibitem{Bardin:1999yd}
D.~Y. Bardin {\it et~al.},
\newblock  Comput. Phys. Commun. {\bf 133}  (2001) 229--395,
\newblock  updated with results from~\cite{Arbuzov}\relax
\relax
\bibitem{Kobel:2000aw}
{Two Fermion Working Group, M. Kobel, {\it et al.}},
\newblock  {\em Two-fermion production in electron positron collisions},
\newblock  Eprint hep-ph/0007180, 2000\relax
\relax
\bibitem{Arbuzov:2005ma}
A.~B. Arbuzov {\it et~al.},
\newblock  {\em ZFITTER: a semi-analytical program for fermion pair production
  in e+e- annihilation, from version 6.21 to version 6.42},
\newblock  Eprint hep-ph/0507146, 2005\relax
\relax
\bibitem{BardinPassarinoBook}
D. Bardin and G. Passarino,
\newblock  The standard model in the making: Precision study of the electroweak
  interactions,
\newblock  (Clarendon, Oxford, UK, 1999)\relax
\relax
\bibitem{Berends:LEP1YR89VOL1}
F. Berends {\it et~al.}, in {\em Z Physics At Lep 1. Proceedings, Workshop,
  Geneva, Switzerland, September 4-5, 1989. Vol. 1: Standard Physics}, CERN
  89-08, ed. G. Altarelli, R. Kleiss, and C. Verzegnassi,  (CERN, Geneva,
  Switzerland, 1989), p.~89\relax
\relax
\bibitem{Grassi:2000dz}
P.~A. Grassi, B.~A. Kniehl, and A. Sirlin,
\newblock  Phys. Rev. Lett. {\bf 86}  (2001) 389--392\relax
\relax
\bibitem{Sirlin:1991rt}
A. Sirlin,
\newblock  Phys. Lett. {\bf B267}  (1991) 240--242\relax
\relax
\bibitem{Sirlin:1991fd}
A. Sirlin,
\newblock  Phys. Rev. Lett. {\bf 67}  (1991) 2127--2130\relax
\relax
\bibitem{bib-PCLI-QCD}
K. Chetyrkin {\it et~al.}, in {\em Reports of the working group on precision
  calculations for the Z resonance}, CERN 95-03, ed. D. Bardin, W. Hollik, and
  G. Passarino,  (CERN, Geneva, Switzerland, 1995), p. 175\relax
\relax
\bibitem{Czarnecki:1996ei}
A. Czarnecki and J.~H. Kuhn,
\newblock  Phys. Rev. Lett. {\bf 77}  (1996) 3955--3958\relax
\relax
\bibitem{Harlander:1997zb}
R. Harlander, T. Seidensticker, and M. Steinhauser,
\newblock  Phys. Lett. {\bf B426}  (1998) 125--132\relax
\relax
\bibitem{Blondel:1988gp}
A. Blondel {\it et~al.},
\newblock  Nucl. Phys. {\bf B304}  (1988) 438\relax
\relax
\bibitem{Boehm:LEP1YR89VOL1}
M. B{\"o}hm and W. Hollik, in {\em Z Physics At Lep 1. Proceedings, Workshop,
  Geneva, Switzerland, September 4-5, 1989. Vol. 1: Standard Physics}, CERN
  89-08, ed. G. Altarelli, R. Kleiss, and C. Verzegnassi,  (CERN, Geneva,
  Switzerland, 1989), pp. 203--234\relax
\relax
\bibitem{LEPSMHIGGS}
{ALEPH, DELPHI, L3, and OPAL Collaborations},
\newblock  Phys. Lett. {\bf B565}  (2003) 61--75\relax
\relax
\bibitem{BLUE:1988}
L. Lyons, D. Gibaut, and P. Clifford,
\newblock  Nucl. Instrum. Meth. {\bf A270}  (1988) 110\relax
\relax
\bibitem{BLUE:2003}
A. Valassi,
\newblock  Nucl. Instrum. Meth. {\bf A500}  (2003) 391--405\relax
\relax
\bibitem{Decamp:1990ky}
ALEPH Collaboration, D. Decamp {\it et~al.},
\newblock  Z. Phys. {\bf C48}  (1990) 365--392\relax
\relax
\bibitem{Decamp:1992aj}
ALEPH Collaboration, D. Decamp {\it et~al.},
\newblock  Z. Phys. {\bf C53}  (1992) 1--20\relax
\relax
\bibitem{Buskulic:1993gu}
ALEPH Collaboration, D. Buskulic {\it et~al.},
\newblock  Z. Phys. {\bf C60}  (1993) 71--82\relax
\relax
\bibitem{Buskulic:1994ea}
ALEPH Collaboration, D. Buskulic {\it et~al.},
\newblock  Z. Phys. {\bf C62}  (1994) 539--550\relax
\relax
\bibitem{Barate:1999ce}
ALEPH Collaboration, R. Barate {\it et~al.},
\newblock  Eur. Phys. J. {\bf C14}  (2000) 1--50\relax
\relax
\bibitem{Abreu:1991wj}
DELPHI Collaboration, P. Abreu {\it et~al.},
\newblock  Nucl. Phys. {\bf B367}  (1991) 511--574\relax
\relax
\bibitem{Abreu:1994wg}
DELPHI Collaboration, P. Abreu {\it et~al.},
\newblock  Nucl. Phys. {\bf B417}  (1994) 3--57\relax
\relax
\bibitem{Abreu:1994ds}
DELPHI Collaboration, P. Abreu {\it et~al.},
\newblock  Nucl. Phys. {\bf B418}  (1994) 403--427\relax
\relax
\bibitem{Abreu:2000mh}
DELPHI Collaboration, P. Abreu {\it et~al.},
\newblock  Eur. Phys. J. {\bf C16}  (2000) 371--405\relax
\relax
\bibitem{Adeva:1991jh}
L3 Collaboration, B. Adeva {\it et~al.},
\newblock  Z. Phys. {\bf C51}  (1991) 179--204\relax
\relax
\bibitem{Adriani:1993gk}
L3 Collaboration, O. Adriani {\it et~al.},
\newblock  Phys. Rept. {\bf 236}  (1993) 1--146\relax
\relax
\bibitem{Acciarri:1994gx}
L3 Collaboration, M. Acciarri {\it et~al.},
\newblock  Z. Phys. {\bf C62}  (1994) 551--576\relax
\relax
\bibitem{Acciarri:2000ai}
L3 Collaboration, M. Acciarri {\it et~al.},
\newblock  Eur. Phys. J. {\bf C16}  (2000) 1--40\relax
\relax
\bibitem{Alexander:1991qw}
OPAL Collaboration, G. Alexander {\it et~al.},
\newblock  Z. Phys. {\bf C52}  (1991) 175--208\relax
\relax
\bibitem{Acton:1993yc}
OPAL Collaboration, P.~D. Acton {\it et~al.},
\newblock  Z. Phys. {\bf C58}  (1993) 219--238\relax
\relax
\bibitem{Akers:1994is}
OPAL Collaboration, R. Akers {\it et~al.},
\newblock  Z. Phys. {\bf C61}  (1994) 19--34\relax
\relax
\bibitem{Abbiendi:2000hu}
OPAL Collaboration, G. Abbiendi {\it et~al.},
\newblock  Eur. Phys. J. {\bf C19}  (2001) 587--651\relax
\relax
\bibitem{bib-detsimo}
OPAL Collaboration, J. Allison {\it et~al.},
\newblock  Nucl. Instrum. Meth. {\bf A317}  (1992) 47--74\relax
\relax
\bibitem{Field:1996dk}
J.~H. Field and T. Riemann,
\newblock  Comput. Phys. Commun. {\bf 94}  (1996) 53--87,
\newblock  (BHAGENE3)\relax
\relax
\bibitem{Jadach:1997nk}
S. Jadach, W. Placzek, and B.~F.~L. Ward,
\newblock  Phys. Lett. {\bf B390}  (1997) 298--308\relax
\relax
\bibitem{UNIBAB}
H. Anlauf {\it et~al.},
\newblock  Comput. Phys. Commun. {\bf 79}  (1994) 466--486,
\newblock  (UNIBAB)\relax
\relax
\bibitem{FERMISV}
{J. Hilgart, R. Kleiss and F. Le Diberder},
\newblock  Comput. Phys. Commun. {\bf 75}  (1993) 191,
\newblock  (FERMISV)\relax
\relax
\bibitem{grc4f}
J. Fujimoto {\it et~al.},
\newblock  Comput. Phys. Commun. {\bf 100}  (1997) 128--156,
\newblock  (GRC4f)\relax
\relax
\bibitem{Bederede:1995pc}
For the ALEPH Collaboration, D. Bederede {\it et~al.},
\newblock  Nucl. Instrum. Meth. {\bf A365}  (1995) 117--134\relax
\relax
\bibitem{Brock:1996ty}
For the L3 Collaboration, I.~C. Brock {\it et~al.},
\newblock  Nucl. Instrum. Meth. {\bf A381}  (1996) 236--266\relax
\relax
\bibitem{Abbiendi:1999zx}
OPAL Collaboration, G. Abbiendi {\it et~al.},
\newblock  Eur. Phys. J. {\bf C14}  (2000) 373--425\relax
\relax
\bibitem{ALIBABA}
{W. Beenakker, F.A. Berends and S.C. van der Marck},
\newblock  Nucl. Phys. {\bf B349}  (1991) 323--368,
\newblock  (ALIBABA)\relax
\relax
\bibitem{bib-EPOL}
LEP Polarization Collaboration, L. Arnaudon {\it et~al.},
\newblock  Phys. Lett. {\bf B284}  (1992) 431--439\relax
\relax
\bibitem{bib-EPOL2}
L. Arnaudon {\it et~al.},
\newblock  Z. Phys. {\bf C66}  (1995) 45--62\relax
\relax
\bibitem{bib-firstPOL}
L. Knudsen {\it et~al.},
\newblock  Phys. Lett. {\bf B270}  (1991) 97--104\relax
\relax
\bibitem{bib-polarimeter}
M. Placidi and R. Rossmanith,
\newblock  Nucl. Instr. Meth. {\bf A274}  (1989) 79\relax
\relax
\bibitem{Arbuzov}
A.~B. Arbuzov,
\newblock  JHEP {\bf 07}  (2001) 043\relax
\relax
\bibitem{bib-JEG2}
S. Eidelman and F. Jegerlehner,
\newblock  Z. Phys. {\bf C67}  (1995) 585--602\relax
\relax
\bibitem{bib-BP05}
H. Burkhardt and B. Pietrzyk,
\newblock  Phys. Rev. {\bf D72}  (2005) 057501\relax
\relax
\bibitem{bib-tierror1}
W. Beenakker and G. Passarino,
\newblock  Phys. Lett. {\bf B425}  (1998) 199\relax
\relax
\bibitem{bib-tierror2}
W. Beenakker and G. Passarino, private communication\relax
\relax
\bibitem{BHLUMI061}
B.~F.~L. Ward {\it et~al.},
\newblock  Phys. Lett. {\bf B450}  (1999) 262--266\relax
\relax
\bibitem{Montagna:1998vb}
G. Montagna {\it et~al.},
\newblock  Nucl. Phys. {\bf B547}  (1999) 39--59\relax
\relax
\bibitem{Montagna:1999eu}
G. Montagna {\it et~al.},
\newblock  Phys. Lett. {\bf B459}  (1999) 649--652\relax
\relax
\bibitem{Berends88}
{F.A. Berends, W.L. van Neerven and G.J.H. Burgers},
\newblock  Nucl. Phys. {\bf B297}  (1988) 429,
\newblock  erratum: B304 (1988) 921\relax
\relax
\bibitem{KF}
E.~A. Kuraev and V.~S. Fadin,
\newblock  Sov. J. Nucl. Phys. {\bf 41}  (1985) 466--472\relax
\relax
\bibitem{Montagna}
{G. Montagna, O. Nicrosini and F. Piccinini},
\newblock  Phys. Lett. {\bf B406}  (1997) 243--248\relax
\relax
\bibitem{Skrzypek92}
M. Skrzypek,
\newblock  Acta Phys. Polon. {\bf B23}  (1992) 135\relax
\relax
\bibitem{JSW}
{S. Jadach, M. Skrzypek and B.F.L. Ward},
\newblock  Phys. Lett. {\bf B257}  (1991) 173--178\relax
\relax
\bibitem{YFS}
{D.R. Yennie, S.C. Frautschi and H. Suura},
\newblock  Ann. Phys. {\bf 13}  (1961) 379--452\relax
\relax
\bibitem{JPS}
{S. Jadach, M. Skrzypek and B. Pietrzyk},
\newblock  Phys. Lett. {\bf B456}  (1999) 77\relax
\relax
\bibitem{Martinez:1991ta}
M. Martinez {\it et~al.},
\newblock  Z. Phys. {\bf C49}  (1991) 645--656\relax
\relax
\bibitem{Martinez:1995kp}
M. Martinez and F. Teubert,
\newblock  Z. Phys. {\bf C65}  (1995) 267--276,
\newblock  updated with results summarized in~\cite{JPS}
  and~\cite{bib-PCLI}\relax
\relax
\bibitem{bib-ifi}
{S. Jadach, B. Pietrzyk, E. Tournefier, B.F.L. Ward and Z. W{\c{a}s}},
\newblock  Phys. Lett. {\bf B465}  (1999) 254\relax
\relax
\bibitem{ISPP88}
{B.A. Kniehl, M. Krawczyk, J.H. K{\"u}hn and R.G. Stuart},
\newblock  Phys. Lett. {\bf B209}  (1988) 337\relax
\relax
\bibitem{JSMpairs}
{S. Jadach, M. Skrzypek and M. Martinez},
\newblock  Phys. Lett. {\bf B280}  (1992) 129--136\relax
\relax
\bibitem{PCP99}
D.~Y. Bardin, M. Gr{\"u}newald, and G. Passarino,
\newblock  {\em Precision calculation project report},
\newblock  Eprint arXiv:hep-ph/9902452, 1999\relax
\relax
\bibitem{bib-smat1}
A. Leike, T. Riemann, and J. Rose,
\newblock  Phys. Lett. {\bf B273}  (1991) 513--518\relax
\relax
\bibitem{bib-smat2}
T. Riemann,
\newblock  Phys. Lett. {\bf B293}  (1992) 451--456\relax
\relax
\bibitem{bib-smat3}
S. Kirsch and T. Riemann,
\newblock  Comp. Phys. Commun. {\bf 88}  (1995) 89--108\relax
\relax
\bibitem{Adriani:1993gkS}
L3 Collaboration, O. Adriani {\it et~al.},
\newblock  Phys. Rept. {\bf 236}  (1993) 1--146\relax
\relax
\bibitem{Acciarri:2000aiS}
L3 Collaboration, M. Acciarri {\it et~al.},
\newblock  Eur. Phys. J. {\bf C16}  (2000) 1--40\relax
\relax
\bibitem{Abbiendi:2000huS}
OPAL Collaboration, G. Abbiendi {\it et~al.},
\newblock  Eur. Phys. J. {\bf C19}  (2001) 587--651\relax
\relax
\bibitem{TOPAZ}
TOPAZ Collaboration, K. Miyabayashi {\it et~al.},
\newblock  Phys. Lett. {\bf B347}  (1995) 171\relax
\relax
\bibitem{VENUS}
VENUS Collaboration, K. Yusa {\it et~al.},
\newblock  Phys. Lett. {\bf B447}  (1999) 167\relax
\relax
\bibitem{bib-LEP2smata}
ALEPH Collaboration, D. Busculic {\it et~al.},
\newblock  Z.Phys. {\bf C 71}  (1996) 179\relax
\relax
\bibitem{bib-LEP2smatd}
DELPHI Collaboration, P. Abreu {\it et~al.},
\newblock  Eur. Phys. J. {\bf C11}  (1999) 383--407\relax
\relax
\bibitem{bib-LEP2smatl1}
L3 Collaboration, O. Adriani {\it et~al.},
\newblock  Phys. Lett. {\bf B315}  (1993) 494--502\relax
\relax
\bibitem{bib-LEP2smatl2}
L3 Collaboration, M. Acciarri {\it et~al.},
\newblock  Phys. Lett. {\bf B479}  (2000) 101--117\relax
\relax
\bibitem{bib-LEP2smatl3}
L3 Collaboration, M. Acciarri {\it et~al.},
\newblock  Phys. Lett. {\bf B489}  (2000) 93--101\relax
\relax
\bibitem{bib-LEP2smato}
OPAL Collaboration, K. Ackerstaff {\it et~al.},
\newblock  Eur. Phys. J. {\bf C2}  (1998) 441--472\relax
\relax
\bibitem{PDG2004}
{Particle Data Group, S. Eidelman, {\it et al.}},
\newblock  Phys. Lett. {\bf B592}  (2004) 1\relax
\relax
\bibitem{Abe:1993sh}
SLD Collaboration, K. Abe {\it et~al.},
\newblock  Phys. Rev. Lett. {\bf 70}  (1993) 2515--2520\relax
\relax
\bibitem{Abe:1994wx}
SLD Collaboration, K. Abe {\it et~al.},
\newblock  Phys. Rev. Lett. {\bf 73}  (1994) 25--29\relax
\relax
\bibitem{Abe:1997nj}
SLD Collaboration, K. Abe {\it et~al.},
\newblock  Phys. Rev. Lett. {\bf 78}  (1997) 2075--2079\relax
\relax
\bibitem{Abe:2000dq}
SLD Collaboration, K. Abe {\it et~al.},
\newblock  Phys. Rev. Lett. {\bf 84}  (2000) 5945--5949\relax
\relax
\bibitem{Maruyama:1992dv}
T. Maruyama {\it et~al.},
\newblock  Phys. Rev. {\bf B46}  (1992) 4261--4264\relax
\relax
\bibitem{Maruyama:1991mk}
T. Maruyama {\it et~al.},
\newblock  Phys. Rev. Lett. {\bf 66}  (1991) 2376--2379\relax
\relax
\bibitem{Nakanishi:1991}
T. Nakanishi {\it et~al.},
\newblock  Phys. Lett. {\bf A158}  (1991) 345--349\relax
\relax
\bibitem{ref:sld-arctests}
T. Limberg, P. Emma, and R. Rossmanith,
\newblock  {\em The North Arc of the SLC as a spin rotator},
\newblock  Internal Report SLAC-PUB-6210, SLAC, 1993,
\newblock  Presented at 1993 Particle Accelerator Conference (PAC 93),
  Washington, DC, 17-20 May 1993\relax
\relax
\bibitem{King:1994spin}
R.~C. King,
\newblock  A precise measurement of the left-right asymmetry of Z boson
  production at the SLAC linear collider,
\newblock  Ph.D. thesis, Stanford Univ., Sep 1994,
\newblock  SLAC-0452, pp.40-46\relax
\relax
\bibitem{Lipps:1954}
F.~W. Lipps and H.~A. Tolhoek,
\newblock  Physica {\bf 20}  (1954) 85--98\relax
\relax
\bibitem{ref:sld-pockel}
A. Lath,
\newblock  A Precise measurement of the left-right cross-section asymmetry in Z
  boson production,
\newblock  Ph.D. thesis, Mass. Inst. of Tech., Sep 1994,
\newblock  SLAC-0454, pp.93-96\relax
\relax
\bibitem{ref:sld-EGS4}
W.~R. Nelson, H. Hirayama, and D.~W.~O. Rogers,
\newblock  {\em THE EGS4 CODE SYSTEM},
\newblock  Internal Report SLAC-0265, SLAC, Dec 1985\relax
\relax
\bibitem{Torrence:1997bdMC}
E.~C. Torrence,
\newblock  Search for anomalous couplings in the decay of polarized Z bosons to
  tau lepton pairs,
\newblock  Ph.D. thesis, Mass. Inst. of Tech., Jun 1997,
\newblock  SLAC-R-0509, pp.191-214\relax
\relax
\bibitem{ref:sld-pgc}
R.~C. Field {\it et~al.},
\newblock  IEEE Trans. Nucl. Sci. {\bf 45}  (1998) 670--675\relax
\relax
\bibitem{Berridge:1998wy}
S.~C. Berridge {\it et~al.},
\newblock  {\em A quartz fiber / tungsten calorimeter for the Compton
  polarimeter at SLAC},
\newblock  Internal Report SLAC-REPRINT-1998-024, SLAC, 1998,
\newblock  Prepared for 13th International Symposium on High-Energy Spin
  Physics (SPIN 98), Protvino, Russia, 8-12 Sep 1998\relax
\relax
\bibitem{Onoprienko:2000qfc}
D.~V. Onoprienko,
\newblock  Precise measurement of the left-right asymmetry in ${\Zzero}$ boson
  production by ${\ee}$ collisions: Electron beam polarization measurement with
  the quartz fiber calorimeter,
\newblock  Ph.D. thesis, SLAC and Tennessee U., Aug 2000,
\newblock  SLAC-R-556, pp.53-100\relax
\relax
\bibitem{ref:sld-morrisQED}
M.~L. Swartz,
\newblock  Phys. Rev. {\bf D58}  (1998) 014010\relax
\relax
\bibitem{Levi:1989tn}
M.~E. Levi {\it et~al.},
\newblock  {\em Precision Synchrotron Radiation Detectors},
\newblock  Internal Report SLAC-PUB-4921, SLAC, 1989,
\newblock  Presented at IEEE Particle Accelerator Conf., Chicago, Ill., Mar
  20-23, 1989\relax
\relax
\bibitem{Kent:1989zc}
J. Kent {\it et~al.},
\newblock  {\em Precision Measurements of the SLC Beam Energy},
\newblock  Internal Report SLAC-PUB-4922, SLAC, 1989,
\newblock  Presented at IEEE Particle Accelerator Conf., Chicago, Ill., Mar
  20-23, 1989\relax
\relax
\bibitem{ref:sld-zpeak}
D.~V. Onoprienko,
\newblock  Precise measurement of the left-right asymmetry in ${\Zzero}$ boson
  production by ${\ee}$ collisions: Electron beam polarization measurement with
  the quartz fiber calorimeter,
\newblock  Ph.D. thesis, SLAC and Tennessee U., Aug 2000,
\newblock  SLAC-R-556, pp.25-31\relax
\relax
\bibitem{Ben-David:1994es}
R.~J. Ben-David,
\newblock  The First measurement of the left-right cross-section asymmetry in Z
  boson production,
\newblock  Ph.D. thesis, Yale Univ., May 1994,
\newblock  UMI-94-33702, pp.69-92\relax
\relax
\bibitem{Park:1993es}
H. Park,
\newblock  A Measurement of the left-right cross-section asymmetry in
  ${\Zzero}$ production with polarized ${\ee}$ collisions,
\newblock  Ph.D. thesis, Oregon Univ., Dec 1993,
\newblock  SLAC-435, pp.61-95, 130-150\relax
\relax
\bibitem{ref:sld-extinct}
R.~D. Elia,
\newblock  Measurement of the left-right asymmetry in Z boson production by
  electron - positron collisions,
\newblock  Ph.D. thesis, Stanford Univ., Apr 1993,
\newblock  SLAC-0429, pp.123-126\relax
\relax
\bibitem{ref:sld-levchuk}
M. Swartz {\it et~al.},
\newblock  Nucl. Instrum. Meth. {\bf A363}  (1995) 526--537\relax
\relax
\bibitem{ref:sld-Mott}
G. Mulhollan {\it et~al.},
\newblock  {\em A derivative standard for polarimeter calibration},
\newblock  Internal Report SLAC-PUB-7325, SLAC, 1995,
\newblock  Talk given at 16th IEEE Particle Accelerator Conference (PAC 95) and
  International Conference on High-energy Accelerators (IUPAP), Dallas, Texas,
  1-5 May 1995\relax
\relax
\bibitem{ref:sld-posipol}
H.~R. Band, P.~C. Rowson, and T.~R. Wright,
\newblock  {\em The Positron Polarization (POSPOL) Experiments: T-419},
\newblock  Internal Report SLD-note 268, SLD, 2000,
\newblock
  http://www-sldnt.slac.stanford.edu/sldbb/SLDNotes/sld-note{\%}20268.pdf\relax
\relax
\bibitem{ref:sld-al2000}
SLD Collaboration, K. Abe {\it et~al.},
\newblock  Phys. Rev. Lett. {\bf 86}  (2001) 1162--1166\relax
\relax
\bibitem{ref:sld-al1997}
SLD Collaboration, K. Abe {\it et~al.},
\newblock  Phys. Rev. Lett. {\bf 79}  (1997) 804--808\relax
\relax
\bibitem{ref:sld-dmiba}
M. Martinez and R. Miquel,
\newblock  Z. Phys. {\bf C53}  (1992) 115--126\relax
\relax
\bibitem{Eberhard:1989ve}
P.~H. Eberhard {\it et~al.}, in {\em Z Physics At Lep 1. Proceedings, Workshop,
  Geneva, Switzerland, September 4-5, 1989. Vol. 1: Standard Physics}, CERN
  89-08, ed. G. Altarelli, R. Kleiss, and C. Verzegnassi,  (CERN, Geneva,
  Switzerland, 1989), pp. 235--265\relax
\relax
\bibitem{Hagiwara:1990fn}
K. Hagiwara, A.~D. Martin, and D. Zeppenfeld,
\newblock  Phys. Lett. {\bf B235}  (1990) 198--202\relax
\relax
\bibitem{Davier:1993nw}
M. Davier {\it et~al.},
\newblock  Phys. Lett. {\bf B306}  (1993) 411--417\relax
\relax
\bibitem{Tsai:1971vv}
Y.-S. Tsai,
\newblock  Phys. Rev. {\bf D4}  (1971) 2821,
\newblock  [Erratum-ibid.\ D {\bf 13} (1976) 771]\relax
\relax
\bibitem{ALEPHTAU}
ALEPH Collaboration, A. Heister {\it et~al.},
\newblock  Eur. Phys. J. {\bf C20}  (2001) 401--430\relax
\relax
\bibitem{Buskulic:1996vx}
ALEPH Collaboration, D. Buskulic {\it et~al.},
\newblock  Z. Phys. {\bf C69}  (1996) 183--194\relax
\relax
\bibitem{Buskulic:1993vk}
ALEPH Collaboration, D. Buskulic {\it et~al.},
\newblock  Z. Phys. {\bf C59}  (1993) 369--386\relax
\relax
\bibitem{Decamp:1991vz}
ALEPH Collaboration, D. Decamp {\it et~al.},
\newblock  Phys. Lett. {\bf B265}  (1991) 430--444\relax
\relax
\bibitem{DELPHITAU}
DELPHI Collaboration, P. Abreu {\it et~al.},
\newblock  Eur. Phys. J. {\bf C14}  (2000) 585--611\relax
\relax
\bibitem{Abreu:1995ku}
DELPHI Collaboration, P. Abreu {\it et~al.},
\newblock  Z. Phys. {\bf C67}  (1995) 183--202\relax
\relax
\bibitem{L3TAU}
L3 Collaboration, M. Acciarri {\it et~al.},
\newblock  Phys. Lett. {\bf B429}  (1998) 387--398\relax
\relax
\bibitem{Acciarri:1994qt}
L3 Collaboration, M. Acciarri {\it et~al.},
\newblock  Phys. Lett. {\bf B341}  (1994) 245--256\relax
\relax
\bibitem{Adriani:1992zn}
L3 Collaboration, O. Adriani {\it et~al.},
\newblock  Phys. Lett. {\bf B294}  (1992) 466--478\relax
\relax
\bibitem{OPALTAU}
OPAL Collaboration, G. Abbiendi {\it et~al.},
\newblock  Eur. Phys. J. {\bf C21}  (2001) 1--21\relax
\relax
\bibitem{Alexander:1996ha}
OPAL Collaboration, G. Alexander {\it et~al.},
\newblock  Z. Phys. {\bf C72}  (1996) 365--375\relax
\relax
\bibitem{Akers:1995db}
OPAL Collaboration, R. Akers {\it et~al.},
\newblock  Z. Phys. {\bf C65}  (1995) 1--16\relax
\relax
\bibitem{Alexander:1991am}
OPAL Collaboration, G. Alexander {\it et~al.},
\newblock  Phys. Lett. {\bf B266}  (1991) 201--217\relax
\relax
\bibitem{Smith:1977rr}
J. Smith, J.~A.~M. Vermaseren, and J. G.~Grammer,
\newblock  Phys. Rev. {\bf D15}  (1977) 3280\relax
\relax
\bibitem{Berends:1984sd}
F.~A. Berends, P.~H. Daverveldt, and R. Kleiss,
\newblock  Phys. Lett. {\bf B148}  (1984) 489\relax
\relax
\bibitem{Berends:1985gf}
F.~A. Berends, P.~H. Daverveldt, and R. Kleiss,
\newblock  Nucl. Phys. {\bf B253}  (1985) 441\relax
\relax
\bibitem{PDG98}
{Particle Data Group, C. Caso {\it et al.}},
\newblock  Eur. Phys. J. {\bf C3}  (1998) 1--794\relax
\relax
\bibitem{PDG2000}
{Particle Data Group, D. E. Groom {\it et al.}},
\newblock  Eur. Phys. J. {\bf C15}  (2000) 1--878\relax
\relax
\bibitem{Jadach:1990mz}
S. Jadach, J.~H. Kuhn, and Z. W{\c{a}s},
\newblock  Comput. Phys. Commun. {\bf 64}  (1990) 275--299,
\newblock  (TAUOLA)\relax
\relax
\bibitem{Jadach:1993hs}
S. Jadach {\it et~al.},
\newblock  Comput. Phys. Commun. {\bf 76}  (1993) 361--380,
\newblock  (TAUOLA: Version 2.4)\relax
\relax
\bibitem{PHOTOS}
E. Barbiero, B. van Eijk, and Z. W{\c{a}}s,
\newblock  Comput. Phys. Commun. {\bf 66}  (1991) 115,
\newblock  CERN-TH 7033/93, (PHOTOS)\relax
\relax
\bibitem{Decker}
R. Decker and M. Finkemeier,
\newblock  Phys. Rev. {\bf D48}  (1993) 4203\relax
\relax
\bibitem{Finkemeier}
M. Finkemeier,
\newblock  Radiative corrections to the decay ${\tpinu}$,
\newblock  Ph.D. thesis, University of Karlsruhe, Feb 1994\relax
\relax
\bibitem{ref:atag}
ALEPH Collaboration, D. Buskulic {\it et~al.},
\newblock  Phys. Lett. {\bf B313}  (1993) 535--548\relax
\relax
\bibitem{ref:lrbmixed}
L3 Collaboration, M. Acciarri {\it et~al.},
\newblock  Eur. Phys. J. {\bf C13}  (2000) 47--61\relax
\relax
\bibitem{ref:omixed}
OPAL Collaboration, G. Abbiendi {\it et~al.},
\newblock  Eur. Phys. J. {\bf C8}  (1999) 217--239\relax
\relax
\bibitem{SLD_ZVTOP}
D.~J. Jackson,
\newblock  Nucl. Instrum. Meth. {\bf A388}  (1997) 247--253\relax
\relax
\bibitem{ref:drb}
DELPHI Collaboration, P. Abreu {\it et~al.},
\newblock  Eur. Phys. J. {\bf C10}  (1999) 415--442\relax
\relax
\bibitem{ref:dbtag}
DELPHI Collaboration, J. Abdallah {\it et~al.},
\newblock  Eur. Phys. J. {\bf C32}  (2004) 185--208\relax
\relax
\bibitem{ref:aimp}
ALEPH Collaboration, R. Barate {\it et~al.},
\newblock  Phys. Lett. {\bf B401}  (1997) 150--162\relax
\relax
\bibitem{ref:orcd}
OPAL Collaboration, K. Ackerstaff {\it et~al.},
\newblock  Eur. Phys. J. {\bf C1}  (1998) 439--459\relax
\relax
\bibitem{ref:arcc}
ALEPH Collaboration, R. Barate {\it et~al.},
\newblock  Eur. Phys. J. {\bf C16}  (2000) 597--611\relax
\relax
\bibitem{ref:drcd}
DELPHI Collaboration, P. Abreu {\it et~al.},
\newblock  Eur. Phys. J. {\bf C12}  (2000) 209--224\relax
\relax
\bibitem{Abe:2005nq}
SLD Collaboration, K. Abe {\it et~al.},
\newblock  Phys. Rev. {\bf D71}  (2005) 112004\relax
\relax
\bibitem{fernando}
P. Billoir {\it et~al.},
\newblock  Nucl. Instrum. Meth. {\bf A360}  (1995) 532--558\relax
\relax
\bibitem{ref:alife}
ALEPH Collaboration, R. Barate {\it et~al.},
\newblock  Phys. Lett. {\bf B401}  (1997) 163--175\relax
\relax
\bibitem{ref:arcd}
ALEPH Collaboration, R. Barate {\it et~al.},
\newblock  Eur. Phys. J. {\bf C4}  (1998) 557--570\relax
\relax
\bibitem{ref:drcc}
DELPHI Collaboration, P. Abreu {\it et~al.},
\newblock  Eur. Phys. J. {\bf C12}  (2000) 225--241\relax
\relax
\bibitem{ref:orcc}
OPAL Collaboration, G. Alexander {\it et~al.},
\newblock  Z. Phys. {\bf C72}  (1996) 1--16\relax
\relax
\bibitem{ref:ocl}
OPAL Collaboration, G. Abbiendi {\it et~al.},
\newblock  Eur. Phys. J. {\bf C8}  (1999) 573--584\relax
\relax
\bibitem{ourpap}
{LEP Heavy Flavor Working Group, D. Abbaneo {\it et al.}},
\newblock  Eur. Phys. J. {\bf C4}  (1998) 185--191\relax
\relax
\bibitem{ref:alasy}
ALEPH Collaboration, A. Heister {\it et~al.},
\newblock  Eur. Phys. J. {\bf C24}  (2002) 177--191\relax
\relax
\bibitem{ref:ajet}
ALEPH Collaboration, A. Heister {\it et~al.},
\newblock  Eur. Phys. J. {\bf C22}  (2001) 201--215\relax
\relax
\bibitem{Abreu:1995vn}
DELPHI Collaboration, P. Abreu {\it et~al.},
\newblock  Z. Phys. {\bf C65}  (1995) 569--586\relax
\relax
\bibitem{Abdallah:2003gp}
DELPHI Collaboration, J. Abdallah {\it et~al.},
\newblock  Eur. Phys. J. {\bf C34}  (2004) 109--125\relax
\relax
\bibitem{ref:llasy1}
L3 Collaboration, O. Adriani {\it et~al.},
\newblock  Phys. Lett. {\bf B292}  (1992) 454--462\relax
\relax
\bibitem{ref:llasy2}
L3 Collaboration, M. Acciarri {\it et~al.},
\newblock  Phys. Lett. {\bf B448}  (1999) 152--162\relax
\relax
\bibitem{ref:olasy}
OPAL Collaboration, G. Abbiendi {\it et~al.},
\newblock  Phys. Lett. {\bf B577}  (2003) 18--36\relax
\relax
\bibitem{ref:ljet}
L3 Collaboration, M. Acciarri {\it et~al.},
\newblock  Phys. Lett. {\bf B439}  (1998) 225--236\relax
\relax
\bibitem{ref:ojet}
OPAL Collaboration, G. Abbiendi {\it et~al.},
\newblock  Phys. Lett. {\bf B546}  (2002) 29--47\relax
\relax
\bibitem{ref:dnnasy}
DELPHI Collaboration, J. Abdallah {\it et~al.},
\newblock  Eur. Phys. J. {\bf C40}  (2005) 1--25\relax
\relax
\bibitem{ref:adsac}
ALEPH Collaboration, R. Barate {\it et~al.},
\newblock  Phys. Lett. {\bf B434}  (1998) 415--425\relax
\relax
\bibitem{ref:ddasy}
DELPHI Collaboration, P. Abreu {\it et~al.},
\newblock  Eur. Phys. J. {\bf C10}  (1999) 219--237\relax
\relax
\bibitem{ref:odsac}
OPAL Collaboration, G. Alexander {\it et~al.},
\newblock  Z. Phys. {\bf C73}  (1997) 379--395\relax
\relax
\bibitem{ref:SLD_AQL}
SLD Collaboration, K. Abe {\it et~al.},
\newblock  Phys. Rev. Lett. {\bf 88}  (2002) 151801\relax
\relax
\bibitem{Abe:1999hb}
SLD Collaboration, K. Abe {\it et~al.},
\newblock  Phys. Rev. Lett. {\bf 83}  (1999) 3384--3389\relax
\relax
\bibitem{ref:SLD_ACD}
SLD Collaboration, K. Abe {\it et~al.},
\newblock  Phys. Rev. {\bf D63}  (2001) 032005\relax
\relax
\bibitem{Abe:1998wk}
SLD Collaboration, K. Abe {\it et~al.},
\newblock  Phys. Rev. Lett. {\bf 81}  (1998) 942--946\relax
\relax
\bibitem{Abe:2002fs}
SLD Collaboration, K. Abe {\it et~al.},
\newblock  Phys. Rev. Lett. {\bf 90}  (2003) 141804\relax
\relax
\bibitem{ref:SLD_vtxasy}
SLD Collaboration, K. Abe {\it et~al.},
\newblock  Phys. Rev. Lett. {\bf 94}  (2005) 091801\relax
\relax
\bibitem{ref:SLD_ABK1}
SLD Collaboration, K. Abe {\it et~al.},
\newblock  Phys. Rev. Lett. {\bf 83}  (1999) 1902--1907\relax
\relax
\bibitem{ref:abl}
ALEPH Collaboration, A. Heister {\it et~al.},
\newblock  Eur. Phys. J. {\bf C22}  (2002) 613--626\relax
\relax
\bibitem{ref:dbl}
DELPHI Collaboration, P. Abreu {\it et~al.},
\newblock  Eur. Phys. J. {\bf C20}  (2001) 455--478\relax
\relax
\bibitem{ref:lbl}
L3 Collaboration, M. Acciarri {\it et~al.},
\newblock  Z. Phys. {\bf C71}  (1996) 379--390\relax
\relax
\bibitem{ref:obl}
OPAL Collaboration, G. Abbiendi {\it et~al.},
\newblock  Eur. Phys. J. {\bf C13}  (2000) 225--240\relax
\relax
\bibitem{pet}
C. Peterson {\it et~al.},
\newblock  Phys. Rev. {\bf D27}  (1983) 105\relax
\relax
\bibitem{epsb_results_1}
ALEPH Collaboration, D. Buskulic {\it et~al.},
\newblock  Phys. Lett. {\bf B357}  (1995) 699--714\relax
\relax
\bibitem{epsb_results_2}
ALEPH Collaboration, D. Buskulic {\it et~al.},
\newblock  Z. Phys. {\bf C62}  (1994) 179--198\relax
\relax
\bibitem{epsb_results_3}
DELPHI Collaboration, P. Abreu {\it et~al.},
\newblock  Z. Phys. {\bf C66}  (1995) 323--340\relax
\relax
\bibitem{epsb_results_4}
OPAL Collaboration, G. Alexander {\it et~al.},
\newblock  Phys. Lett. {\bf B364}  (1995) 93--106\relax
\relax
\bibitem{epsb_results_5}
OPAL Collaboration, R. Akers {\it et~al.},
\newblock  Z. Phys. {\bf C60}  (1993) 199--216\relax
\relax
\bibitem{collins}
P.~D.~B. Collins and T.~P. Spiller,
\newblock  J. Phys. {\bf G11}  (1985) 1289\relax
\relax
\bibitem{kart}
V.~G. Kartvelishvili, A.~K. Likhoded, and V.~A. Petrov,
\newblock  Phys. Lett. {\bf B78}  (1978) 615\relax
\relax
\bibitem{epsc_dstar_1}
ALEPH Collaboration, D. Buskulic {\it et~al.},
\newblock  Z. Phys. {\bf C62}  (1994) 1--14\relax
\relax
\bibitem{epsc_dstar_2}
DELPHI Collaboration, P. Abreu {\it et~al.},
\newblock  Z. Phys. {\bf C59}  (1993) 533,
\newblock  erratum: Z. Phys. C65 (1995) 709\relax
\relax
\bibitem{epsc_dstar_3}
OPAL Collaboration, R. Akers {\it et~al.},
\newblock  Z. Phys. {\bf C67}  (1995) 27--44\relax
\relax
\bibitem{gbb_meas_a}
ALEPH Collaboration, R. Barate {\it et~al.},
\newblock  Phys. Lett. {\bf B434}  (1998) 437--450\relax
\relax
\bibitem{gbb_meas_di}
DELPHI Collaboration, P. Abreu {\it et~al.},
\newblock  Phys. Lett. {\bf B405}  (1997) 202--214\relax
\relax
\bibitem{gbb_meas_dii}
DELPHI Collaboration, P. Abreu {\it et~al.},
\newblock  Phys. Lett. {\bf B462}  (1999) 425--439\relax
\relax
\bibitem{gbb_meas_o}
OPAL Collaboration, G. Abbiendi {\it et~al.},
\newblock  Eur. Phys. J. {\bf C18}  (2001) 447--460\relax
\relax
\bibitem{gbb_meas_s}
SLD Collaboration, K. Abe {\it et~al.},
\newblock  Phys. Lett. {\bf B507}  (2001) 61--69\relax
\relax
\bibitem{gcc_l}
L3 Collaboration, M. Acciarri {\it et~al.},
\newblock  Phys. Lett. {\bf B476}  (2000) 243--255\relax
\relax
\bibitem{gcc_o}
OPAL Collaboration, G. Abbiendi {\it et~al.},
\newblock  Eur. Phys. J. {\bf C13}  (2000) 1--13\relax
\relax
\bibitem{gsplit}
A. Giammanco, in Gluon splitting into heavy flavours at the Z peak.
  Proceedings, 12th International Workshop, DIS 2004, Strbske Pleso, Slovakia,
  April 14-18, 2004. Vol. 1 + 2, ed. D. Bruncko, J. Ferencei, and P. Strizenec,
   (Inst. Exp. Phys. SAS, Kosice, Slovak Republic, 2004), pp. 783--786\relax
\relax
\bibitem{bmultid}
DELPHI Collaboration, P. Abreu {\it et~al.},
\newblock  Phys. Lett. {\bf B425}  (1998) 399--412\relax
\relax
\bibitem{bmultil}
L3 Collaboration, M. Acciarri {\it et~al.},
\newblock  Phys. Lett. {\bf B416}  (1998) 220--232\relax
\relax
\bibitem{bmultio}
OPAL Collaboration, R. Akers {\it et~al.},
\newblock  Z. Phys. {\bf C61}  (1994) 209--222\relax
\relax
\bibitem{ctopo}
MARK-III Collaboration, D. Coffman {\it et~al.},
\newblock  Phys. Lett. {\bf B263}  (1991) 135--140\relax
\relax
\bibitem{cleo_lc}
CLEO Collaboration, D.~E. Jaffe {\it et~al.},
\newblock  Phys. Rev. {\bf D62}  (2000) 072005\relax
\relax
\bibitem{cleospe}
CLEO Collaboration, S. Henderson {\it et~al.},
\newblock  Phys. Rev. {\bf D45}  (1992) 2212--2231\relax
\relax
\bibitem{accmm}
G. Altarelli {\it et~al.},
\newblock  Nucl. Phys. {\bf B208}  (1982) 365\relax
\relax
\bibitem{isgw}
N. Isgur {\it et~al.},
\newblock  Phys. Rev. {\bf D39}  (1989) 799\relax
\relax
\bibitem{delcocl}
DELCO Collaboration, W. Bacino {\it et~al.},
\newblock  Phys. Rev. Lett. {\bf 43}  (1979) 1073\relax
\relax
\bibitem{mark3cl}
MARK-III Collaboration, R.~M. Baltrusaitis {\it et~al.},
\newblock  Phys. Rev. Lett. {\bf 54}  (1985) 1976\relax
\relax
\bibitem{ref:lephf}
{ALEPH, DELPHI, L3, OPAL Collaborations},
\newblock  Nucl. Instrum. Meth. {\bf A378}  (1996) 101--115\relax
\relax
\bibitem{cleobd}
CLEO Collaboration, D. Bortoletto {\it et~al.},
\newblock  Phys. Rev. {\bf D45}  (1992) 21--35\relax
\relax
\bibitem{cleoupver1}
CLEO Collaboration, T.~E. Coan {\it et~al.},
\newblock  Phys. Rev. Lett. {\bf 80}  (1998) 1150--1155\relax
\relax
\bibitem{cleoupver2}
CLEO Collaboration, D. Gibaut {\it et~al.},
\newblock  Phys. Rev. {\bf D53}  (1996) 4734--4746\relax
\relax
\bibitem{cleoupver3}
CLEO Collaboration, R. Ammar {\it et~al.},
\newblock  Phys. Rev. {\bf D55}  (1997) 13--18\relax
\relax
\bibitem{aleph2d}
ALEPH Collaboration, R. Barate {\it et~al.},
\newblock  Eur. Phys. J. {\bf C4}  (1998) 387--407\relax
\relax
\bibitem{btotaua}
ALEPH Collaboration, R. Barate {\it et~al.},
\newblock  Eur. Phys. J. {\bf C19}  (2001) 213--227\relax
\relax
\bibitem{btotaud}
DELPHI Collaboration, P. Abreu {\it et~al.},
\newblock  Phys. Lett. {\bf B496}  (2000) 43--58\relax
\relax
\bibitem{btotaula}
L3 Collaboration, M. Acciarri {\it et~al.},
\newblock  Phys. Lett. {\bf B332}  (1994) 201--208\relax
\relax
\bibitem{btotaulb}
L3 Collaboration, M. Acciarri {\it et~al.},
\newblock  Z. Phys. {\bf C71}  (1996) 379--390\relax
\relax
\bibitem{paolo}
P. Nason and C. Oleari,
\newblock  Phys. Lett. {\bf B407}  (1997) 57--60\relax
\relax
\bibitem{lampe}
G. Altarelli and B. Lampe,
\newblock  Nucl. Phys. {\bf B391}  (1993) 3--22\relax
\relax
\bibitem{neerv}
V. Ravindran and W.~L. van Neerven,
\newblock  Phys. Lett. {\bf B445}  (1998) 214--222\relax
\relax
\bibitem{sey}
S. Catani and M.~H. Seymour,
\newblock  JHEP {\bf 9907}  (1999) 023\relax
\relax
\bibitem{hfasycor}
A. Freitas and K. M{\"o}nig,
\newblock  Eur. Phys. J. {\bf C40}  (2005) 493--496\relax
\relax
\bibitem{ALEPHcharge1996}
ALEPH Collaboration, D. Buskulic {\it et~al.},
\newblock  Z. Phys. {\bf C71}  (1996) 357--378\relax
\relax
\bibitem{DELPHIcharge}
DELPHI Collaboration, P. Abreu {\it et~al.},
\newblock  Phys. Lett. {\bf B277}  (1992) 371--382\relax
\relax
\bibitem{OPALcharge}
OPAL Collaboration, P.~D. Acton {\it et~al.},
\newblock  Phys. Lett. {\bf B294}  (1992) 436--450\relax
\relax
\bibitem{Buskulic:1996tx}
ALEPH Collaboration, D. Buskulic {\it et~al.},
\newblock  Phys. Lett. {\bf B388}  (1996) 419--430\relax
\relax
\bibitem{Acciarri:1998iw}
L3 Collaboration, M. Acciarri {\it et~al.},
\newblock  Phys. Lett. {\bf B450}  (1999) 281--293\relax
\relax
\bibitem{Abbiendi:2002vu}
OPAL Collaboration, G. Abbiendi {\it et~al.},
\newblock  Phys. Lett. {\bf B544}  (2002) 259--273\relax
\relax
\bibitem{Barate:1999qx}
ALEPH Collaboration, R. Barate {\it et~al.},
\newblock  Eur. Phys. J. {\bf C12}  (2000) 183--207\relax
\relax
\bibitem{Abreu:2000ap}
DELPHI Collaboration, P. Abreu {\it et~al.},
\newblock  Phys. Lett. {\bf B485}  (2000) 45--61\relax
\relax
\bibitem{Adriani:1993ca}
L3 Collaboration, O. Adriani {\it et~al.},
\newblock  Phys. Lett. {\bf B306}  (1993) 187--196\relax
\relax
\bibitem{Abbiendi:2003dh}
OPAL Collaboration, G. Abbiendi {\it et~al.},
\newblock  Eur. Phys. J. {\bf C33}  (2004) 173--212\relax
\relax
\bibitem{Amaldi:1987fu}
U. Amaldi {\it et~al.},
\newblock  Phys. Rev. {\bf D36}  (1987) 1385\relax
\relax
\bibitem{bib-Gmu-1}
T. van Ritbergen and R.~G. Stuart,
\newblock  Phys. Rev. Lett. {\bf 82}  (1999) 488--491\relax
\relax
\bibitem{bib-Gmu-2}
T. van Ritbergen and R.~G. Stuart,
\newblock  Nucl. Phys. {\bf B564}  (2000) 343--390\relax
\relax
\bibitem{bib-Gmu-3}
M. Steinhauser and T. Seidensticker,
\newblock  Phys. Lett. {\bf B467}  (1999) 271--278\relax
\relax
\bibitem{bib-alphalept}
M. Steinhauser,
\newblock  Phys. Lett. {\bf B429}  (1998) 158--161\relax
\relax
\bibitem{bib-Burk}
H. Burkhardt and B. Pietrzyk,
\newblock  Phys. Lett. {\bf B356}  (1995) 398--403\relax
\relax
\bibitem{BES_01}
BES Collaboration, J.~Z. Bai {\it et~al.},
\newblock  Phys. Rev. Lett. {\bf 88}  (2002) 101802\relax
\relax
\bibitem{CMD_03}
CMD-2 Collaboration, R.~R. Akhmetshin {\it et~al.},
\newblock  Phys. Lett. {\bf B578}  (2004) 285--289\relax
\relax
\bibitem{KLOE_04}
KLOE Collaboration, A. Aloisio {\it et~al.},
\newblock  Phys. Lett. {\bf B606}  (2005) 12--24\relax
\relax
\bibitem{bib-Swartz}
M.~L. Swartz,
\newblock  Phys. Rev. {\bf D53}  (1996) 5268--5282\relax
\relax
\bibitem{bib-Zeppe}
A.~D. Martin and D. Zeppenfeld,
\newblock  Phys. Lett. {\bf B345}  (1995) 558--563\relax
\relax
\bibitem{bib-Alemany}
R. Alemany, M. Davier, and A. Hocker,
\newblock  Eur. Phys. J. {\bf C2}  (1998) 123--135\relax
\relax
\bibitem{bib-Davier}
M. Davier and A. Hocker,
\newblock  Phys. Lett. {\bf B419}  (1998) 419--431\relax
\relax
\bibitem{bib-alphaKuhn}
J.~H. Kuhn and M. Steinhauser,
\newblock  Phys. Lett. {\bf B437}  (1998) 425--431\relax
\relax
\bibitem{bib-jeger99}
F. Jegerlehner, in Proceedings, 4th International Symposium, RADCOR'98, ed. J.
  Sola,  (World Scientific, Singapore, Sep 1999), p.~75\relax
\relax
\bibitem{bib-Erler}
J. Erler,
\newblock  Phys. Rev. {\bf D59}  (1999) 054008\relax
\relax
\bibitem{bib-ADMartin}
A.~D. Martin, J. Outhwaite, and M.~G. Ryskin,
\newblock  Phys. Lett. {\bf B492}  (2000) 69--73\relax
\relax
\bibitem{bib-Troconiz-Yndurain}
J.~F. de~Troconiz and F.~J. Yndurain,
\newblock  Phys. Rev. {\bf D65}  (2002) 093002\relax
\relax
\bibitem{bib-Hagiwara:2003}
K. Hagiwara {\it et~al.},
\newblock  Phys. Rev. {\bf D69}  (2004) 093003\relax
\relax
\bibitem{bib-Troconiz-Yndurain-2004}
J.~F. de~Troconiz and F.~J. Yndurain,
\newblock  Phys. Rev. {\bf D71}  (2005) 073008\relax
\relax
\bibitem{CDF-top:1994}
CDF Collaboration, F. Abe {\it et~al.},
\newblock  Phys. Rev. Lett. {\bf 73}  (1994) 225--231\relax
\relax
\bibitem{CDF-top:1995}
CDF Collaboration, F. Abe {\it et~al.},
\newblock  Phys. Rev. Lett. {\bf 74}  (1995) 2626--2631\relax
\relax
\bibitem{D0-top:1995}
{D\O} Collaboration, S. Abachi {\it et~al.},
\newblock  Phys. Rev. Lett. {\bf 74}  (1995) 2632--2637\relax
\relax
\bibitem{Mtop1-CDF-di-l-PRLa}
CDF Collaboration, F. Abe {\it et~al.},
\newblock  Phys. Rev. Lett. {\bf 80}  (1998) 2779--2784\relax
\relax
\bibitem{Mtop1-CDF-di-l-PRLb}
CDF Collaboration, F. Abe {\it et~al.},
\newblock  Phys. Rev. Lett. {\bf 82}  (1999) 271--276\relax
\relax
\bibitem{Mtop1-CDF-di-l-PRLb-E}
CDF Collaboration, F. Abe {\it et~al.},
\newblock  Erratum: Phys. Rev. Lett. {\bf 82}  (1999) 2808--2809\relax
\relax
\bibitem{Mtop1-CDF-l+j-PRL}
CDF Collaboration, F. Abe {\it et~al.},
\newblock  Phys. Rev. Lett. {\bf 80}  (1998) 2767--2772\relax
\relax
\bibitem{Mtop1-CDF-l+j-PRD}
CDF Collaboration, T. Affolder {\it et~al.},
\newblock  Phys. Rev. {\bf D63}  (2001) 032003\relax
\relax
\bibitem{Mtop1-CDF-all-j-PRL}
CDF Collaboration, F. Abe {\it et~al.},
\newblock  Phys. Rev. Lett. {\bf 79}  (1997) 1992--1997\relax
\relax
\bibitem{D0-top:prl-ll}
{D\O} Collaboration, B. Abbott {\it et~al.},
\newblock  Phys. Rev. Lett. {\bf 80}  (1998) 2063--2068\relax
\relax
\bibitem{D0-top:prd-ll}
{D\O} Collaboration, B. Abbott {\it et~al.},
\newblock  Phys. Rev. {\bf D60}  (1999) 052001\relax
\relax
\bibitem{D0-top:prl-lj}
{D\O} Collaboration, S. Abachi {\it et~al.},
\newblock  Phys. Rev. Lett. {\bf 79}  (1997) 1197--1202\relax
\relax
\bibitem{D0-top:prd-lj}
{D\O} Collaboration, B. Abbott {\it et~al.},
\newblock  Phys. Rev. {\bf D58}  (1998) 052001\relax
\relax
\bibitem{Mtop1-D0-l+j-new1}
{D\O} Collaboration, V.~M. Abazov {\it et~al.},
\newblock  Nature {\bf 429}  (2004) 638--642\relax
\relax
\bibitem{Mtop1-D0-l+j-new2}
{D\O} Collaboration, V.~M. Abazov {\it et~al.},
\newblock  {\em New measurement of the top quark mass in lepton + jets t anti-t
  events at {D\O}},
\newblock  Eprint hep-ex/0407005, 2004\relax
\relax
\bibitem{Mtop1-D0-all-j-PRL}
{D\O} Collaboration, V.~M. Abazov {\it et~al.},
\newblock  Phys. Lett. {\bf B606}  (2005) 25--33\relax
\relax
\bibitem{PP-MT:combination}
{The CDF and {D\O} Collaborations, and the Tevatron Electroweak Working Group},
\newblock  {\em Combination of CDF and {D\O} results on the top-quark mass},
\newblock  Eprint hep-ex/0404010, 2004\relax
\relax
\bibitem{UA1-MW}
UA1 Collaboration, C. Albajar {\it et~al.},
\newblock  Z. Phys. {\bf C44}  (1989) 15\relax
\relax
\bibitem{UA2-MW}
UA2 Collaboration, J. Alitti {\it et~al.},
\newblock  Phys. Lett. {\bf B276}  (1992) 354--364\relax
\relax
\bibitem{CDF-MW}
CDF Collaboration, T. Affolder {\it et~al.},
\newblock  Phys. Rev. {\bf D64}  (2001) 52001--39\relax
\relax
\bibitem{CDF-MW-PRL95}
CDF Collaboration, F. Abe {\it et~al.},
\newblock  Phys. Rev. Lett. {\bf 75}  (1995) 11--16\relax
\relax
\bibitem{CDF-MW-PRD95}
CDF Collaboration, F. Abe {\it et~al.},
\newblock  Phys. Rev. {\bf D52}  (1995) 4784--4827\relax
\relax
\bibitem{CDF-MW-PRL90}
CDF Collaboration, F. Abe {\it et~al.},
\newblock  Phys. Rev. Lett. {\bf 65}  (1990) 2243--2246\relax
\relax
\bibitem{CDF-MW-PRD90}
CDF Collaboration, F. Abe {\it et~al.},
\newblock  Phys. Rev. {\bf D43}  (1991) 2070--2093\relax
\relax
\bibitem{D0-MW:central}
{D\O} Collaboration, B. Abbott {\it et~al.},
\newblock  Phys. Rev. Lett. {\bf 80}  (1998) 3008\relax
\relax
\bibitem{D0-MW:endcap}
{D\O} Collaboration, B. Abbott {\it et~al.},
\newblock  Phys. Rev. Lett. {\bf 84}  (2000) 222--227\relax
\relax
\bibitem{D0-MW:edge}
{D\O} Collaboration, V.~M. Abazov {\it et~al.},
\newblock  Phys. Rev. {\bf D66}  (2002) 012001\relax
\relax
\bibitem{D0-MW:large}
{D\O} Collaboration, B. Abbott {\it et~al.},
\newblock  Phys. Rev. {\bf D62}  (2000) 092006\relax
\relax
\bibitem{CDF-GW}
CDF Collaboration, T. Affolder {\it et~al.},
\newblock  Phys. Rev. Lett. {\bf 85}  (2000) 3347--3352\relax
\relax
\bibitem{D0-GW}
{D\O} Collaboration, V.~M. Abazov {\it et~al.},
\newblock  Phys. Rev. {\bf D66}  (2002) 032008\relax
\relax
\bibitem{PP-MW-GW:combination}
{The CDF Collaboration, the D\O\ Collaboration, and the Tevatron Electroweak
  Working Group},
\newblock  Phys. Rev. {\bf D70}  (2004) 092008\relax
\relax
\bibitem{ALEPH-MW}
ALEPH Collaboration, R. Barate {\it et~al.},
\newblock  Eur. Phys. J. {\bf C17}  (2000) 241--261\relax
\relax
\bibitem{DELPHI-MW}
DELPHI Collaboration, P. Abreu {\it et~al.},
\newblock  Phys. Lett. {\bf B511}  (2001) 159--177\relax
\relax
\bibitem{L3-MW}
L3 Collaboration, M. Acciarri {\it et~al.},
\newblock  Phys. Lett. {\bf B454}  (1999) 386--398\relax
\relax
\bibitem{OPAL-MW}
OPAL Collaboration, G. Abbiendi {\it et~al.},
\newblock  Phys. Lett. {\bf B507}  (2001) 29--46\relax
\relax
\bibitem{LEP-MW:combination}
{The LEP Collaborations ALEPH, DELPHI, L3, OPAL, the LEP Electroweak Working
  Group, and the SLD Electroweak and Heavy Flavour Groups},
\newblock  {\em A combination of preliminary electroweak measurements and
  constraints on the standard model},
\newblock  Eprint hep-ex/0412015, CERN, 2004\relax
\relax
\bibitem{QWCs:exp:1}
C.~S. Wood {\it et~al.},
\newblock  Science {\bf 275}  (1997) 1759\relax
\relax
\bibitem{QWCs:exp:2}
S.~C. Bennett and C.~E. Wieman,
\newblock  Phys. Rev. Lett. {\bf 82}  (1999) 2484--2487\relax
\relax
\bibitem{QWTl:exp:1}
N.~H. Edwards {\it et~al.},
\newblock  Phys. Rev. Lett. {\bf 74}  (1995) 2654--2657\relax
\relax
\bibitem{QWTl:exp:2}
P.~A. Vetter {\it et~al.},
\newblock  Phys. Rev. Lett. {\bf 74}  (1995) 2658--2661\relax
\relax
\bibitem{QWCs:theo:2003:new}
J.~S.~M. Ginges and V.~V. Flambaum,
\newblock  Phys. Rept. {\bf 397}  (2004) 63--154\relax
\relax
\bibitem{QW:MSM}
D.~Y. Bardin {\it et~al.},
\newblock  Eur. Phys. J. {\bf C22}  (2001) 99--104\relax
\relax
\bibitem{E158RunI}
SLAC E158 Collaboration, P. Anthony {\it et~al.},
\newblock  Phys. Rev. Lett. {\bf 92}  (2004) 181602\relax
\relax
\bibitem{E158RunI+II+III}
SLAC E158 Collaboration, P.~L. Anthony {\it et~al.},
\newblock  Phys. Rev. Lett. {\bf 95}  (2005) 081601\relax
\relax
\bibitem{PaschosWolfenstein}
E.~A. Paschos and L. Wolfenstein,
\newblock  Phys. Rev. {\bf D7}  (1973) 91--95\relax
\relax
\bibitem{bib-NuTeV-final}
NuTeV Collaboration, G.~P. Zeller {\it et~al.},
\newblock  Phys. Rev. Lett. {\bf 88}  (2002) 091802,
\newblock  erratum: 90 (2003) 239902\relax
\relax
\bibitem{Zeller:2002du}
NuTeV Collaboration, G.~P. Zeller {\it et~al.},
\newblock  Phys. Rev. {\bf D65}  (2002) 111103\relax
\relax
\bibitem{McFarland:2003jw}
K.~S. McFarland and S.-O. Moch,
\newblock  {\em Conventional physics explanations for the NuTeV ${\swsq}$},
\newblock  Eprint hep-ph/0306052, 2003\relax
\relax
\bibitem{Olness:2003wz}
F. Olness {\it et~al.},
\newblock  Eur. Phys. J. {\bf C40}  (2005) 145--156\relax
\relax
\bibitem{Kretzer:2003wy}
S. Kretzer {\it et~al.},
\newblock  Phys. Rev. Lett. {\bf 93}  (2004) 041802\relax
\relax
\bibitem{Martin:2004dh}
A.~D. Martin {\it et~al.},
\newblock  Eur. Phys. J. {\bf C39}  (2005) 155--161\relax
\relax
\bibitem{KMcF:NuInt01}
K.~S. McFarland {\it et~al.},
\newblock  Nucl. Phys. Proc. Suppl. {\bf 112}  (2002) 226--233\relax
\relax
\bibitem{LEP1YR89VOL1}
Ed. G. Altarelli, R. Kleiss, and C. Verzegnassi,
\newblock  {\em Z Physics At Lep 1. Proceedings, Workshop, Geneva, Switzerland,
  September 4-5, 1989. Vol. 1: Standard Physics},
\newblock  (CERN, Geneva, Switzerland, 1989),
\newblock  Yellow Report CERN 89-08\relax
\relax
\bibitem{bib-PCLI}
D. Bardin {\it et~al.}, in {\em Reports of the working group on precision
  calculations for the Z resonance}, CERN 95-03, ed. D. Bardin, W. Hollik, and
  G. Passarino,  (CERN, Geneva, Switzerland, 1995), pp. 7--162\relax
\relax
\bibitem{BP:98}
D.~Y. Bardin and G. Passarino,
\newblock  {\em Upgrading of precision calculations for electroweak
  observables},
\newblock  Eprint hep-ph/9803425, 1998\relax
\relax
\bibitem{BCEHWWW01}
U. Baur {\it et~al.},
\newblock  eConf {\bf C010630}  (2001) P122,
\newblock  hep-ph/0111314\relax
\relax
\bibitem{Snowmass:2001}
{The Snowmass Working Group on Precision Electroweak Measurements},
\newblock  eConf {\bf C010630}  (2001) P1WG1,
\newblock  hep-ph/0202001\relax
\relax
\bibitem{Twoloop-MW}
M. Awramik {\it et~al.},
\newblock  Phys. Rev. {\bf D69}  (2004) 053006\relax
\relax
\bibitem{Threeloop-rho}
M. Faisst {\it et~al.},
\newblock  Nucl. Phys. {\bf B665}  (2003) 649--662\relax
\relax
\bibitem{Twoloop-sin2teff}
M. Awramik {\it et~al.},
\newblock  Phys. Rev. Lett. {\bf 93}  (2004) 201805\relax
\relax
\bibitem{QCD:Kuhn:1996}
K.~G. Chetyrkin, J.~H. Kuhn, and A. Kwiatkowski,
\newblock  Phys. Rept. {\bf 277}  (1996) 189--281\relax
\relax
\bibitem{QCD:Soper:1996}
D.~E. Soper and L.~R. Surguladze,
\newblock  Phys. Rev. {\bf D54}  (1996) 4566--4577\relax
\relax
\bibitem{QCD:Bethke:2000}
S. Bethke,
\newblock  J. Phys. {\bf G26}  (2000) R27\relax
\relax
\bibitem{Bethke:2004uy}
S. Bethke,
\newblock  Nucl. Phys. Proc. Suppl. {\bf 135}  (2004) 345--352\relax
\relax
\bibitem{Stenzel:2005sg}
H. Stenzel,
\newblock  JHEP {\bf 07}  (2005) 0132\relax
\relax
\bibitem{MINUIT}
F. James and M. Roos,
\newblock  Comp. Phys. Commun. {\bf 10}  (1975) 343\relax
\relax
\bibitem{Altarelli:1991}
G. Altarelli and R. Barbieri,
\newblock  Phys. Lett. {\bf B253}  (1991) 161--167\relax
\relax
\bibitem{Altarelli:1992a}
G. Altarelli, R. Barbieri, and S. Jadach,
\newblock  Nucl. Phys. {\bf B369}  (1992) 3--32\relax
\relax
\bibitem{Altarelli:1992b}
G. Altarelli, R. Barbieri, and S. Jadach,
\newblock  Nucl. Phys. {\bf B376}  (1992) 444\relax
\relax
\bibitem{Altarelli:1993a}
G. Altarelli, R. Barbieri, and F. Caravaglios,
\newblock  Nucl. Phys. {\bf B405}  (1993) 3--23\relax
\relax
\bibitem{Altarelli:1993b}
G. Altarelli, R. Barbieri, and F. Caravaglios,
\newblock  Phys. Lett. {\bf B314}  (1993) 357--363\relax
\relax
\bibitem{Altarelli:1995}
G. Altarelli, R. Barbieri, and F. Caravaglios,
\newblock  Phys. Lett. {\bf B349}  (1995) 145--154\relax
\relax
\bibitem{Altarelli:1997}
G. Altarelli, R. Barbieri, and F. Caravaglios,
\newblock  Int. J. Mod. Phys. {\bf A13}  (1998) 1031\relax
\relax
\bibitem{Peskin-Takeuchi:1990}
M.~E. Peskin and T. Takeuchi,
\newblock  Phys. Rev. Lett. {\bf 65}  (1990) 964--967\relax
\relax
\bibitem{Peskin-Takeuchi:1992}
M.~E. Peskin and T. Takeuchi,
\newblock  Phys. Rev. {\bf D46}  (1992) 381--409\relax
\relax
\bibitem{ref:dsfirst}
DELPHI Collaboration, P. Abreu {\it et~al.},
\newblock  Z. Phys. {\bf C67}  (1995) 1--14\relax
\relax
\bibitem{ref:dstrange}
DELPHI Collaboration, P. Abreu {\it et~al.},
\newblock  Eur. Phys. J. {\bf C14}  (2000) 613--631\relax
\relax
\bibitem{ref:olight}
OPAL Collaboration, K. Ackerstaff {\it et~al.},
\newblock  Z. Phys. {\bf C76}  (1997) 387--400\relax
\relax
\bibitem{ref:sldas}
SLD Collaboration, K. Abe {\it et~al.},
\newblock  Phys. Rev. Lett. {\bf 85}  (2000) 5059--5063\relax
\relax
\bibitem{ref:dphoton}
DELPHI Collaboration, P. Abreu {\it et~al.},
\newblock  Z. Phys. {\bf C69}  (1995) 1--14\relax
\relax
\bibitem{ref:lphoton}
L3 Collaboration, O. Adriani {\it et~al.},
\newblock  Phys. Lett. {\bf B301}  (1993) 136--144\relax
\relax
\bibitem{Abbiendi:2003ke}
OPAL Collaboration, G. Abbiendi {\it et~al.},
\newblock  Phys. Lett. {\bf B586}  (2004) 167--182\relax
\relax
\bibitem{ref:ophoton}
OPAL Collaboration, P.~D. Acton {\it et~al.},
\newblock  Z. Phys. {\bf C58}  (1993) 405--418\relax
\relax
\bibitem{ref:aphoton}
ALEPH Collaboration, D. Buskulic {\it et~al.},
\newblock  Z. Phys. {\bf C57}  (1993) 17--36\relax
\relax
\bibitem{HERAud}
H1 Collaboration, A. Aktas {\it et~al.},
\newblock  Phys. Lett. {\bf B632}  (2006) 35--42\relax
\relax
\bibitem{CDFII-Afb}
CDF Collaboration, D. Acosta {\it et~al.},
\newblock  Phys. Rev. {\bf D71}  (2005) 052002\relax
\relax
\end{mcbibliography}

\end{document}